\documentclass[twocolumn,aps,prd,superscriptaddress,preprintnumbers,nofootinbib,10pt]{revtex4-2}

\usepackage[a4paper,margin=1.5cm]{geometry}

\usepackage{amsfonts}
\usepackage[tbtags]{amsmath}
\usepackage{amssymb}
\usepackage{color}
\usepackage{courier}
\usepackage{dsfont}
\usepackage{euscript}
\usepackage{epsfig}
\usepackage{epstopdf}
\usepackage{float}
\usepackage{graphics}
\usepackage{graphicx}
\usepackage{hyperref}
\usepackage[utf8]{inputenc}
\usepackage{latexsym}
\usepackage{MnSymbol}
\usepackage{pifont}
\usepackage{physics}
\usepackage{slashed}
\usepackage[normalem]{ulem}
\usepackage{url}
\usepackage{verbatim}
\usepackage{widetable}
\usepackage{dcolumn}
\usepackage{bm}
\usepackage{siunitx}
\usepackage{xspace}
\usepackage{dsfont}
\usepackage{newtxtext}
\usepackage{adjustbox}
\usepackage{caption}
\usepackage{subcaption}
\usepackage{tikz}
\usepackage{tikz-feynman}
\tikzfeynmanset{compat=1.1.0}

\usetikzlibrary{arrows.meta,positioning}

\usepackage{ragged2e}
\usepackage{enumitem}

\usetikzlibrary{positioning, shapes.geometric, arrows.meta, shadows, fit, backgrounds}

\definecolor{darkblue}{RGB}{31,78,121}
\definecolor{lightblue}{RGB}{174,198,207}
\definecolor{darkorange}{RGB}{214,96,77}
\definecolor{lightorange}{RGB}{253,187,132}
\definecolor{darkgreen}{RGB}{44,127,84}
\definecolor{lightgreen}{RGB}{161,215,106}

\makeatletter
\long\def\@makecaption#1#2{%
  \vskip\abovecaptionskip
  \parbox{\columnwidth}{%
    \small
    \setlength{\parindent}{0pt}%
    \setlength{\parskip}{0pt}%
    \noindent\justifying
    #1.\quad #2\par
  }%
  \vskip\belowcaptionskip}
\makeatother

\usepackage[noabbrev]{cleveref}
\hypersetup{
    colorlinks=true,
    linkcolor=blue,
    citecolor=blue,
    urlcolor=blue
}

\makeatletter
\long\def\@makecaption#1#2{%
  \vskip\abovecaptionskip
  \parbox{0.95\linewidth}{
    \small
    \setlength{\parindent}{0pt}
    \setlength{\parskip}{0pt}
    \noindent\justifying
    #1.\quad #2\par
  }
  \vskip\belowcaptionskip}
\makeatother

\crefname{equation}{Eq.}{Eqs.}
\crefname{figure}{Fig.}{Figs.}
\crefname{section}{Sec.}{Sections}
\crefname{table}{Table}{Tables}

\newcommand{\arepoTwo}{\textsc{arepo-2}\xspace}
\newcommand{\arepoOne}{\textsc{arepo-1}\xspace}
\newcommand{\arepo}{\textsc{arepo}\xspace}
\newcommand{\gizmo}{\textsc{gizmo}\xspace}

\newcommand{\Msun}{{\rm M_\odot}}

\newcommand{\kpc}{{\rm kpc}}

\newcommand{\Gyr}{{\rm Gyr}}
\newcommand{\kms}{{\rm km\, s^{-1}}}
\newcommand{\cpm}{{\rm cm^{2}\, g^{-1}}}

\DeclareSIUnit\erg{erg}
\newcommand{\mnras}{MNRAS}
\newcommand{\aap}{A\&A}
\newcommand{\apjl}{ApJL}
\newcommand{\apjs}{ApJS}

\newcommand{\jcap}{JCAP}
\newcommand{\physrep}{Phys.\ Rep.}

\graphicspath{{./Figures/}}
\newbox{\ORCIDicon}
\sbox{\ORCIDicon}{\large\includegraphics[width=0.8em]{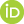}}

\begin{document}

\preprint{Physical Review D (PRD)}

\title{Self-Interacting Dark Matter: From Particle Physics Theories to Structure Formation Simulations}

\author{Martin Rosenlyst\,\href{https://orcid.org/0000-0003-1245-7748}{\usebox{\ORCIDicon}}}
\email{mrosen12@mit.edu}
\affiliation{Physics Department and Kavli Institute for Astrophysics and Space Research, Massachusetts Institute of Technology, Cambridge, MA 02139, USA}

\author{Vinh Tran\,\href{https://orcid.org/0009-0003-6068-6921}{\usebox{\ORCIDicon}}}
\affiliation{Department of Physics, California Institute of Technology, Pasadena, CA 91125, USA}
\affiliation{Physics Department and Kavli Institute for Astrophysics and Space Research, Massachusetts Institute of Technology, Cambridge, MA 02139, USA}

\author{Oliver Zier\,\href{https://orcid.org/0000-0003-1811-8915}{\usebox{\ORCIDicon}}}
\affiliation{Center for Astrophysics | Harvard \& Smithsonian, 60 Garden St, Cambridge, MA 02138, USA}

\author{Xuejian Shen\,\href{https://orcid.org/0000-0002-6196-823X}{\usebox{\ORCIDicon}}}
\affiliation{Center for Astrophysics | Harvard \& Smithsonian, 60 Garden St, Cambridge, MA 02138, USA}
\affiliation{Physics Department and Kavli Institute for Astrophysics and Space Research, Massachusetts Institute of Technology, Cambridge, MA 02139, USA}

\author{Mark Vogelsberger\,\href{https://orcid.org/0000-0001-8593-7692}{\usebox{\ORCIDicon}}}
\affiliation{Physics Department and Kavli Institute for Astrophysics and Space Research, Massachusetts Institute of Technology, Cambridge, MA 02139, USA}

\author{Thejs Brinckmann\,\href{https://orcid.org/0000-0002-1492-5181}{\usebox{\ORCIDicon}}}
\affiliation{CP3-Origins, University of Southern Denmark, Campusvej 55, 5230 Odense M, Denmark}

\author{Mads T. Frandsen\,\href{https://orcid.org/0000-0003-2061-562X}{\usebox{\ORCIDicon}}}
\affiliation{CP3-Origins, University of Southern Denmark, Campusvej 55, 5230 Odense M, Denmark}

\author{Jes\'us Zavala\,\href{https://orcid.org/0000-0003-4442-908X}{\usebox{\ORCIDicon}}}
\affiliation{Center for Astrophysics and Cosmology, Science Institute, University of Iceland, Dunhagi 5, 107 Reykjavik, Iceland}

\author{Frederik Holt Dihver Christensen\,\href{https://orcid.org/0009-0000-8260-5828}{\usebox{\ORCIDicon}}}
\affiliation{CP3-Origins, University of Southern Denmark, Campusvej 55, 5230 Odense M, Denmark}

\author{Mikkel Theiss Kristensen\,\href{https://orcid.org/0000-0002-4802-0452}{\usebox{\ORCIDicon}}}
\affiliation{CP3-Origins, University of Southern Denmark, Campusvej 55, 5230 Odense M, Denmark}

\date{\today}

\begin{abstract} 
We establish a unified theory-to-simulation framework for self-interacting dark matter (SIDM) that integrates three components: (i) a comprehensive, self-contained treatment of its particle-physics foundations and a broad range of model classes; (ii) a systematic, quantitative mapping from microscopic dark-sector physics to structure-formation simulation inputs through explicit expressions and numerical prescriptions; and (iii) the development, implementation, and validation of a broad suite of SIDM algorithms and a dedicated DM decay module in the moving-mesh code \arepoTwo. Under (i), we detail the particle content, symmetries, interactions, production mechanisms, and constraints of a broad landscape of elementary and composite SIDM models, and present representative viable benchmark models. Under (ii), the mapping specifies cosmological initial conditions, internal-state masses and populations, reaction networks, velocity- and angle-dependent cross sections, dissipative energy loss, and decay rates and recoil kinematics. We then apply this mapping to a broad range of SIDM model classes. Under (iii), we implement and validate SIDM algorithms with constant and velocity-dependent cross sections, covering isotropic and anisotropic scattering, multi-state elastic and inelastic reactions, and dissipative processes, alongside a dedicated DM decay module in \arepoTwo. This part builds on the novel SIDM infrastructure developed in our previous companion paper. For angle-dependent scattering, we implement anisotropic Monte Carlo (aMC), frequent small-angle drag--diffusion (fSIDM), and a hybrid algorithm that combines fSIDM with rare large-angle aMC events and achieves speed-ups of up to $\sim100\times$ over aMC. We validate the implementations through beam-lattice deflection, thermalization and decay tests, isolated-halo gravothermal core-collapse simulations, and cosmological box and zoom-in simulations. The results recover analytical predictions and published benchmarks and demonstrate agreement across numerical schemes. Our theory-to-simulation framework serves as a cross-disciplinary reference and a practical bridge between SIDM particle physics and simulations of nonlinear structure formation.
\end{abstract}

\maketitle

\makeatletter
\def\l@subsubsection#1#2{}
\makeatother

\tableofcontents
\thispagestyle{empty}

\section{Introduction}
\label{sec:introduction}

Evidence for dark matter (DM) spans a remarkable range of scales, from the dynamics of galaxy clusters~\citep{Zwicky:1933ren} and galaxy rotation curves~\citep{Rubin:1983tro} to X-ray observations of the intracluster gas~\citep{Briel1992Coma}
and gravitational lensing in merging galaxy clusters, such as the Bullet Cluster~\citep{Clowe2006Bullet}.
Complementary constraints from the cosmic microwave
background~\citep{Smoot1992COBE,Planck2018Parameters}
and large-scale structure~\citep{Eisenstein2005BAO}
establish that DM constitutes most of the matter density in the Universe. In the concordance $\Lambda$CDM cosmology, a cold and effectively collisionless DM component (CDM) drives hierarchical structure formation~\citep{White1978,Blumenthal1984,Davis1985} and provides the gravitational backbone within which galaxies assemble. Despite this empirical success, the microscopic identity
of DM remains unknown
~\citep[e.g.,][]{BertoneHooper2018HistoryDM,Cirelli:2024ssz}. Gravity-only CDM simulations predict a rich hierarchy of gravitationally bound DM haloes and subhaloes spanning many orders of magnitude in mass~\citep[e.g.,][]{Springel2008Aquarius,Diemand2008ViaLacteaII,Boylan-Kolchin:2009alo}. Virialized haloes typically exhibit approximately universal, centrally cuspy density profiles~\citep{Navarro_1996a,navarro1997universal}. For conventional CDM candidates, such as weakly interacting massive particles (WIMPs), the primordial free-streaming cut-off can occur at extremely small scales, allowing the hierarchy in principle to continue down to approximately Earth-mass microhaloes~\citep[e.g.,][]{Diemand2005EarthMassMicrohalos,Bringmann:2009vf,wang2020universal,zheng2024abundance}.

Although the CDM model has been widely successful on large scales~\citep{Springel_2005n}, comparisons between predictions and nearby galaxies have historically revealed a number of apparent small-scale discrepancies. Early examples include the `cusp--core''~\citep[e.g.,][]{Flores_1994,Moore_1994} and `missing satellites'' problems~\citep[e.g.,][]{Klypin1999MissingSatellites,Moore_1999a}, followed by the ``too big to fail'' problem~\citep[e.g.,][]{Boylan-Kolchin_2011,Boylan-Kolchin_2012,Tollerud_2014}, and the observed diversity of dwarf-galaxy rotation curves~\citep[e.g.,][]{Oman_2015,Kamada:2016euw}, as well as the range of inner DM densities inferred for dwarf satellites~\citep{Fattahi:2018ioj,10.1093/mnras/sty2505,Zavala:2019sjk,Kaplinghat2019} Neither the observational status nor the interpretation of these tensions is straightforward~\citep[see, e.g.,][]{Bullock_2017}. The apparent deficit of satellites, for example, has been reduced substantially by the discovery of increasingly faint Milky Way satellites~\citep{Bechtol2015,Drlica2015} together with corrections for survey incompleteness and the destruction of subhaloes by the Galactic baryonic component~\citep{Kim2018}. More generally, feedback, star formation and the gravitational influence of baryons can reshape both galaxies and their DM haloes, potentially alleviating several of the classical small-scale tensions~\citep[e.g.,][]{Governato_2010,PontzenGovernato2012CuspCore,Sawala2016APOSTLE,Wetzel2016,GarrisonKimmel2019,Sales:2022ich}. Phenomenological models have also explored galactic rotation curves and Tully--Fisher relations using effective DM profiles and equations of state~\citep{Petersen:2019mfm,Oikonomou:2025bsi,Oikonomou:2026zhu}.

These questions nevertheless stimulated extensive exploration of alternatives to standard CDM~\citep[e.g.,][]{Hogan2000,Spergel_1999,Bode:2000gq,Dalcanton2001,Buckley2018}. A complementary motivation comes from particle physics. No particle in the Standard Model of particle physics
(SM) can account for the observed DM abundance, and
extensive direct and indirect detection and
accelerator-based searches have so far yielded no
definitive evidence for the particle nature of DM
~\citep[e.g.,][]{Bertone:2004pz,Bertone2010,Cirelli:2024ssz,Kahlhoefer:2017dnp,Aprile2018,Schumann:2019eaa,Meng2021,Aalbers2023}. These null results leave open the possibility that DM participates in non-gravitational interactions within a hidden dark sector, with potential signatures in cosmic structure.

\setcounter{page}{1}

Self-interacting dark matter (SIDM) provides a particularly well-motivated realization of this possibility~\citep[e.g.,][]{Carlson:1992fn,deLaix1995,Firmani2000,Spergel_1999}. In the simplest scenario, DM particles undergo elastic self-scattering characterized by a velocity-independent cross section per unit mass. Collisional heat transport then drives the inner halo toward an approximately isothermal configuration and can transform a central cusp into a lower-density core~\citep[e.g.,][]{Yoshida2000,Dave_2000,Colin2002,Vogelsberger_2012,rocha2013cosmological,Elbert2015SIDMCores}. Cross sections around $\sigma/m_{\rm DM}\sim0.1$--$10\,\cpm$ can produce substantial effects at the characteristic velocities of dwarf galaxies~\citep[e.g.,][]{Zavala:2012us,Kaplinghat:2015aga,Tulin:2017ara}. SIDM has consequently been studied extensively as a possible explanation for the internal structure and diversity of dwarf galaxies, satellite systems and DM haloes more generally~\citep{Bullock_2017,Tulin_2018,Adhikari_2022}.

A constant self-interaction cross section that produces appreciable effects on dwarf- and galaxy-scale haloes is not necessarily compatible with constraints at the much larger characteristic velocities of galaxy clusters. Cluster density profiles and shapes, as well as the dynamics of merging systems, place constraints on elastic SIDM at substantially larger characteristic velocities~\citep[e.g.,][]{Miralda-Escude_2002,Randall:2008ppe,Peter2013SIDMHaloShapes,Harvey_2015,Shen2022esidm}. Recent analyses using strong lensing together with stellar kinematics and the dynamics of merging double-radio-relic clusters further probe this regime~\citep{ODonnell:2025pkw,Jee2026}, while other cluster studies have derived upper limits reaching the $\sim0.1\,\cpm$ level~\citep{Elbert_2016,Sagunski_2020,Robertson:2018anx,Andrade_2021,Eckert_2022}. The large separation between dwarf-, galaxy- and cluster-scale velocities therefore provides strong motivation for velocity-dependent cross sections~\citep{Kaplinghat_2015}. Such behaviour is not merely phenomenological: it arises naturally when DM interacts through a light force carrier, for which a Yukawa-like potential can enhance scattering at low velocities while suppressing it at cluster scales~\citep[e.g.,][]{Feng:2009hw,Buckley_2010,Loeb_2011,Tulin_2013}.

This connection to particle physics considerably broadens the SIDM paradigm. Hidden-sector realizations may contain new Abelian or non-Abelian forces, elementary or composite DM, several internal states, particle--antiparticle asymmetries and additional light degrees of freedom~\citep[e.g.,][]{Ackerman2009,Arkani-Hamed:2008hhe,Feng:2009mn,Loeb_2011,Cyr-Racine_2012,Tulin_2013,Cline:2013pca,Boddy:2014yra}. Their self-interactions can be short- or long-range, elastic or inelastic, conservative or dissipative, approximately isotropic or strongly anisotropic, and weakly or strongly dependent on relative velocity. Composite sectors can additionally contain a spectrum of dark hadrons, atoms or molecules connected through several reaction and decay channels. SIDM therefore extends beyond a simple phenomenological modification of collisionless DM characterized by a single constant cross section and can arise from a broad range of microscopic dark-sector realizations whose particle-physics properties may influence nonlinear cosmic structure.

The central difficulty is that particle theories and structure-formation simulations are formulated in different variables. A microscopic model is specified by its fields, symmetries, particle masses, couplings and interaction terms. An $N$-body simulation instead requires the non-relativistic quantities that determine the motion and interactions of numerical particles. In the simplest elastic limit, this reduction may yield only a constant $\sigma/m_{\rm DM}$. More general models require velocity-dependent cross sections, angular scattering distributions, multiple internal-state masses and abundances, allowed reaction channels and their thresholds, energy exchange in endothermic or exothermic reactions, dissipative energy losses, decay rates and recoil kinematics. Dark-sector processes operating before nonlinear structure formation can also modify the matter transfer function and therefore the cosmological initial conditions. Translating the microscopic parameter space into this set of simulation inputs is thus an essential step between particle-physics model building and astrophysical predictions.

The need to preserve the relevant microscopic information is particularly clear for light-mediated SIDM. When the mediator mass is smaller than the DM mass, scattering can become not only velocity dependent but also strongly anisotropic, approaching a forward-peaked Rutherford-like form in the long-range limit~\citep{Feng:2009hw,Buckley_2010,Loeb_2011,Tulin_2013}. Halo evolution, however, is determined by the transport of momentum and energy rather than simply by the total number of collisions. Anisotropic interactions are therefore often approximated through isotropic scattering matched to an angle-weighted transfer or viscosity cross section~\citep{Colquhoun_2020,Yang_2022D,Sabarish_2024}. Such mappings can reproduce important aspects of isolated and approximately equilibrated haloes, but they necessarily compress the full differential cross section, ${\rm d}\sigma/{\rm d}\Omega(v_{\rm rel},\theta)$, into a lower-dimensional description. The angular information can matter in strongly anisotropic or far-from-equilibrium systems, including satellite disruption, gravothermal evolution and cluster mergers~\citep{Kahlhoefer_2015,Kahlhoefer_2014,Robertson_2017b,Fischer_2021a,Fischer_2021b,Fischer_2022,Fischer_2023,Fischer_2024a,Ragagnin_2024,Sabarish_2024}. Connecting microscopic models to these systems therefore motivates simulations that can retain the full velocity and angular dependence when required.

Recent observational findings have renewed interest in SIDM, but with the focus shifting towards a regime opposite to that explored in much of the early work discussed above. Strong-lensing analyses of galaxy clusters have reported substructures that appear unusually concentrated relative to standard CDM expectations~\citep{Meneghetti2020,Ragagnin2022,Meneghetti2023,Natarajan2026}. On galaxy scales, gravitational imaging has detected a small number of individual low-mass perturbers, including a recently reported object with a mass of order $10^{6}\,\Msun$~\citep{Vegetti2010,Vegetti2012,Powell2025}. Analyses of the earlier detections inferred concentrations above typical CDM expectations~\citep{Minor2021,Ballard2024,Despali2025,Enzi2025}, although translating lensing measurements into virial masses and concentrations requires extrapolation beyond the region directly constrained by the data. When expressed in quantities more directly probed by lensing, the earlier systems are unusual but are not necessarily individually incompatible with CDM; their interpretation can additionally depend on luminosity constraints~\citep{Tajalli2025}. The smallest and most recently reported perturber, by contrast, has been argued to be difficult to reconcile with both cold and warm DM while being compatible with a gravothermally collapsed SIDM halo~\citep{Vegetti2026}. Cold stellar streams provide a complementary route to the same mass regime. Perturbations in thin Milky Way streams can encode the density structure of otherwise invisible subhaloes. The GD-1 spur and gaps were reported in~\citep{PriceWhelanBonaca2018GD1}, while subsequent dynamical analyses inferred a compact perturber and examined its possible connection to core-collapsed SIDM substructure~\citep{Bonaca2019,Zhang2024GD1SIDM}. These observations have therefore generated a new wave of interest in SIDM models in which self-interactions do not merely lower central densities but instead drive a subset of haloes toward much higher central densities.

The relevant mechanism is gravothermal core collapse. Because self-gravitating systems have negative heat capacity, the transport of heat from the central region to the outer halo through DM self-interactions causes the core to lose energy, contract and become progressively hotter and denser. This runaway evolution is closely related to the instability known as gravothermal catastrophe from globular-cluster dynamics~\citep{Lynden1968}. At sufficiently large cross sections ($\sigma/m_{\rm DM}\gtrsim 10\,\cpm$), dwarf-scale haloes can enter this regime within a Hubble time and reach central densities orders of magnitude above their collisionless counterparts~\citep[e.g.,][]{Balberg2002,Koda_2011,Pollack:2014rja,Turner2021}. This possibility has implications for dense substructure, the diversity of dwarf-galaxy rotation curves and the population of dense Milky Way satellites~\citep{Zavala:2019sjk,2021MNRAS503920C,Meshveliani:2022rih,Yu2025,Kong2025}. Core collapse in sufficiently massive and concentrated high-redshift haloes has also been explored as a route to DM-seeded massive black holes, with possible connections to the early formation of massive quasars~\citep{Pollack:2014rja,Xiao2021,Feng2021,Shen2025,Sarkar:2026xtb} and to the faint active galactic nuclei observed by JWST as ``Little Red Dots''~\citep{Shen2025,Shen2026,Jiang2026,Roberts2026}.

The diversity of astrophysical systems accessible to SIDM can potentially probe different aspects of the underlying dark-sector microphysics. Dwarf galaxies, stellar streams, satellites, massive galaxies and clusters span widely different characteristic velocities, densities and dynamical environments, making them sensitive not only to the strength and velocity dependence of self-interactions, but potentially also to their angular structure, inelastic thresholds, dissipative energy loss and the presence of multiple or metastable DM states. Connecting such observables to microscopic particle physics, however, ultimately requires nonlinear simulations. Formally, collisional DM evolution is described by the Boltzmann equation with a collision operator determined by the underlying interaction, but solving this equation directly across the dynamic range of cosmological structure formation is prohibitively expensive. The most widely used numerical approach instead embeds stochastic Monte-Carlo scattering in $N$-body codes, with pairwise interaction probabilities chosen to reproduce the appropriate collision rate in the continuum limit~\citep[e.g.,][]{Yoshida2000,Burkert_2000,Dave_2000,Vogelsberger_2012,rocha2013cosmological,Robertson2017SIDM,Koda_2011,Fry_2015,Robertson_2017b,Banerjee_2020}. These methods have enabled cosmological SIDM calculations over a wide range of scales, but the numerical demands become particularly severe during gravothermal collapse, where rapidly increasing central densities and interaction rates require accurate time integration, sufficient spatial and mass resolution, and careful control of numerical parameters~\citep[e.g.,][]{palubski2024numerical,Mace2026}.

Strongly forward-peaked scattering introduces a different numerical difficulty. The total collision rate can be extremely high even when individual encounters produce negligible deflections, making an event-by-event Monte-Carlo treatment inefficient. In the frequent small-angle limit, the collective effect of these encounters can instead be described through an effective drag force and transverse momentum diffusion~\citep{Kahlhoefer_2014,Fischer_2021a}. Interactions containing both a large population of small deflections and a non-negligible tail of large-angle events motivate hybrid approaches that combine the frequent- and rare-scattering limits~\citep{Arido:2024hfk}. High-density regions introduce another complication because particles may interact multiple times within a single timestep. In massively parallel calculations, these events require specialized communication strategies to avoid violations of conservation laws~\citep{Robertson2017SIDM,Fischer_2021a,Fischer_2024a,Robertson_2017b}. Multi-state, inelastic and dissipative reactions add further requirements because both the available reaction channels and the post-interaction kinematics depend on state identity, relative velocity and energy exchange.

Baryonic physics provides an additional layer between microscopic DM interactions and observations. In massive galaxies, baryons can substantially alter the gravitational environment and therefore the evolution and observable signatures of SIDM~\citep[e.g.,][]{Despali2019InterplaySIDMBaryons}. This motivates hydrodynamical calculations in which galaxy formation and DM self-interactions are evolved self-consistently. SIDM has consequently been incorporated into cosmological galaxy-formation programmes including EAGLE-SIDM~\citep{Robertson2018,Robertson2021,Forouhar2022}, BAHAMAS-SIDM~\citep{Robertson_2019}, TANGO-SIDM~\citep{Correa2022,Correa2025} and the \textsc{AIDA}-TNG suite~\citep{Despali2025AIDATNG}, as well as high-resolution simulations of individual dwarf- and Milky-Way-mass systems using FIRE~\citep{robles2017sidm,Sameie2021,Vargya2022} and Lyra~\citep{Gutcke2025}. Together, these efforts demonstrate that interpreting SIDM phenomenology increasingly requires algorithms that are not only accurate, but sufficiently efficient and modular to operate within modern galaxy-formation simulations.

Several of these calculations, including \textsc{AIDA}-TNG and Lyra, employ the moving-mesh code \arepo and build on its original Monte-Carlo SIDM implementation~\citep{Vogelsberger_2012,vogelsberger2019evaporating}. In \arepoOne, the SIDM neighbour search is coupled to the gravitational tree. This architecture complicates the use of subsequent gravity-solver optimizations and makes extensions to more general dark-sector interactions increasingly cumbersome. Our companion work~\cite{Zier:2026} introduces a new SIDM infrastructure in \arepoTwo that separates the interaction solver from the gravity tree, employs a dedicated DM neighbour-search tree and a hypercube-based communication scheme that natively supports multiple interactions per timestep, and exposes a modular interface for generic two-body dark-sector processes.

The purpose of the present work is to build the particle-physics and numerical framework on top of this infrastructure. We develop a unified theory-to-simulation description of SIDM in which representative contact-like, light-mediator, resonant, vector, asymmetric, multi-state inelastic and composite dark-sector models are first organized in terms of their microscopic ingredients. We discuss their particle content, symmetries, masses and couplings, the associated scattering, annihilation and decay processes, mechanisms for producing the DM abundance, and the relevant cosmological, astrophysical and experimental constraints. In the spirit of the Effective Theory of Structure Formation (ETHOS) framework~\citep{Cyr-Racine:2015ihg,Vogelsberger:2015gpr}, we establish an explicit mapping from the microscopic particle-physics parameter space, $\mathcal{P}_{\rm PP}$, to the simulation inputs, $\mathcal{P}_{\rm SIM}$, extending the description of nonlinear dark-sector dynamics to a broader range of interactions. Depending on the model, these inputs include modifications to the initial conditions, internal-state populations and reaction networks, velocity- and angle-dependent cross sections, inelastic and dissipative energy exchange, and decay rates and kinematics.

Building on the new \arepoTwo infrastructure~\citep{Zier:2026}, we implement methods for isotropic and anisotropic self-scattering with constant or velocity-dependent cross sections, including strongly forward-peaked interactions, as well as multi-state and inelastic reactions, dissipative scattering and DM decays. The angular framework includes explicit anisotropic Monte-Carlo scattering, a frequent small-angle description and a hybrid treatment that evolves the cumulative effect of small deflections through drag--diffusion while retaining explicit rare large-angle scattering. We validate the resulting implementations with controlled deflection, decay and thermalization tests, isolated gravothermal-collapse tests, and full cosmological box and zoom-in simulations. The resulting framework is intended to provide a common interface between microscopic dark-sector theories, the simulation inputs derived from them and their nonlinear astrophysical signatures. The following section gives a guide to the particle-physics, mapping and numerical components of the paper.

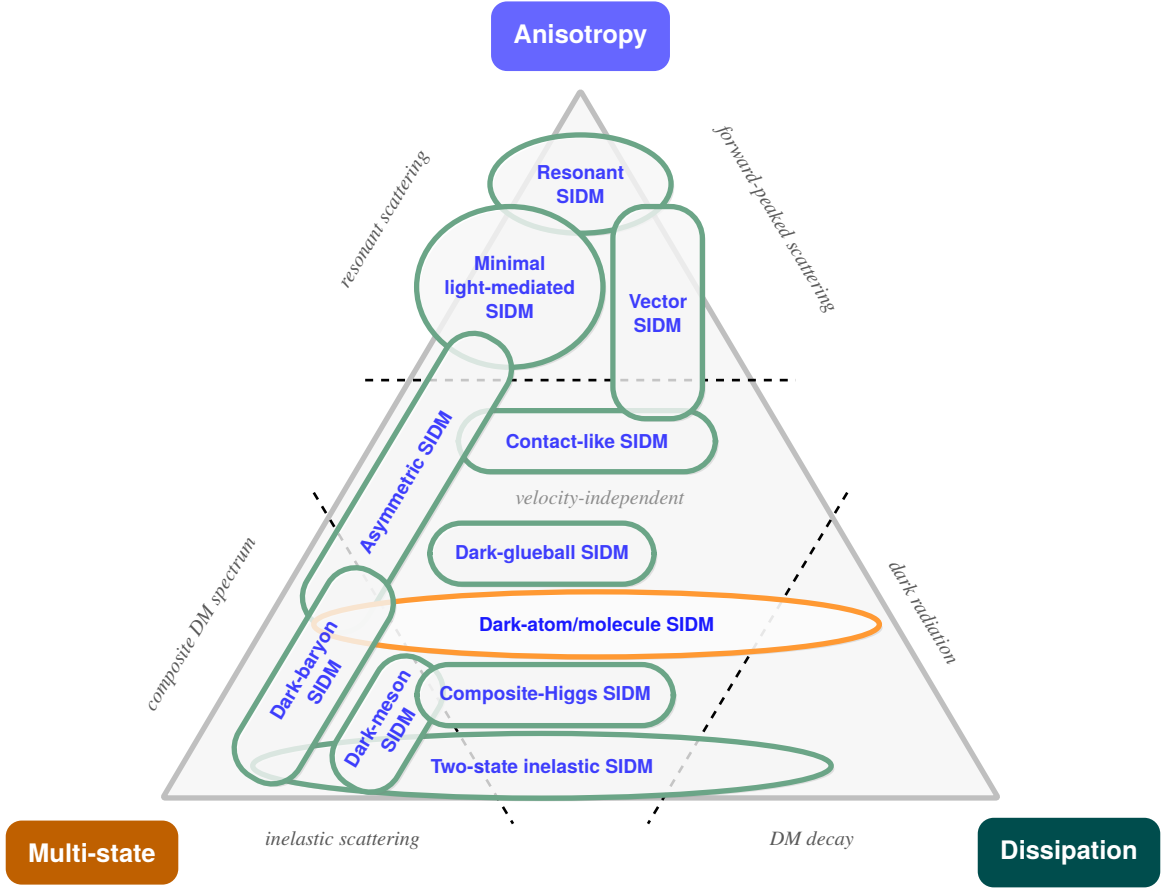
\begin{figure*}[t]
\centering
\begin{tikzpicture}[scale=0.85, transform shape,
  font=\sffamily,
  corner/.style={draw=none, rounded corners=6pt, inner sep=10pt, text=white, font=\bfseries\sffamily\large, align=center},
  bubble/.style={draw=darkgreen!70, line width=2pt, fill=white, fill opacity=0.75, rounded corners=12pt, inner sep=7pt, font=\small\sffamily, align=center, drop shadow={shadow xshift=0.3ex, shadow yshift=-0.3ex, opacity=0.2}},
  bubbleyellow/.style={draw=orange!80, line width=2pt, fill=white, fill opacity=0.9, rounded corners=12pt, inner sep=7pt, font=\small\sffamily, align=center, drop shadow={shadow xshift=0.3ex, shadow yshift=-0.3ex, opacity=0.2}},
  bubblen/.style={draw=darkred!70, line width=2pt, fill=white, fill opacity=0.9, rounded corners=12pt, inner sep=7pt, font=\small\sffamily, align=center, drop shadow={shadow xshift=0.3ex, shadow yshift=-0.3ex, opacity=0.2}},
  manif/.style={font=\itshape\small, text=black!65},
  manifc/.style={font=\itshape\small, text=black!45}
]

\coordinate (A) at (0,6.5);
\coordinate (B) at (-6.5,-4.5);
\coordinate (C) at (6.5,-4.5);

\fill[black!3] (A) -- (B) -- (C) -- cycle;
\draw[line width=2pt, draw=black!25, rounded corners=1pt] (A) -- (B) -- (C) -- cycle;

\draw[dashed, line width=1pt, black] (-3.35,2.0) -- (3.35,2.0);   
\draw[dashed, line width=1pt, black] (-4.15,0.25) -- (-1.05,-4.9);  
\draw[dashed, line width=1pt, black] (4.15,0.25) -- (1.05,-4.9);    

\node[corner, fill=blue!60] at ($(A)+(0,0.85)$) {Anisotropy};
\node[corner, fill=orange!75!black] at ($(B)+(-1.1,-0.85)$) {Multi-state};
\node[corner, fill=teal!60!black] at ($(C)+(1.1,-0.85)$) {Dissipation};

\node[manifc] at (0.3,0.15) {velocity-independent};

\node[manif, rotate=59]  at (-3.05,4.5)  {resonant scattering};
\node[manif, rotate=-59] at (3.05,4.5)   {forward-peaked scattering};
\node[manif, rotate=59]  at (-5.9,-1.8) {composite DM spectrum};
\node[manif]             at (-3.7,-5.15) {inelastic scattering};
\node[manif, rotate=-59] at (5.35,-1.6)  {dark radiation};
\node[manif]             at (3.6,-5.15)  {DM decay};

\node[bubble, minimum width=4cm, minimum height=0.95cm] at (0.1,1.05) {\textbf{\hyperref[sec:contact-like]{\parbox{3cm}{Contact-like SIDM}}}};

\node[bubble, ellipse, minimum width=2.1cm, minimum height=1.4cm] at (0.0,5.05) {\textbf{\hyperref[sec:Resonant_SIDM]{\parbox{1.5cm}{Resonant SIDM}}}};

\node[bubble, ellipse, inner sep=0pt, minimum width=0.5cm, minimum height=2.5cm, align=center] at (-1.1,3.45) {\textbf{\hyperref[sec:light-mediator]{\parbox{2.05cm}{\centering Minimal light-mediated SIDM}}}};

\node[bubble, minimum width=0.8cm, minimum height=3.3cm, align=center, text width=0.9cm] at (1.2,3.05) {\textbf{\hyperref[sec:Vector_SIDM]{\parbox{0.9cm}{\centering Vector SIDM}}}};

\node[
  bubble,
  minimum width=5.2cm,
  minimum height=1.1cm,
  rotate=59,
  transform shape
] (inelasticbubble) at (-2.7,0.4)
{\textbf{\textcolor{blue}{\parbox[c]{3cm}{\centering Asymmetric SIDM}}}};

\node[
  rotate=59,
  transform shape,
  inner sep=0pt,
  outer sep=0pt,
  text opacity=0
] (inelasticbubble) at (-3.5,0.5)
{\hyperref[sec:Asymmetric_SIDM]{\rule{1.0cm}{1.5cm}}};

\node[bubble, minimum width=3cm, minimum height=0.95cm] at (-0.6,-0.7) {\textbf{\hyperref[sec:dark_glueballs]{\parbox{3cm}{Dark-glueball SIDM}}}};

\node[bubble, ellipse, minimum width=9cm, minimum height=0.8cm] at (-0.6,-4) {\textbf{\hyperref[sec: inelastic_sidm_models]{\parbox{4.5cm}{\centering Two-state inelastic SIDM}}}};

\node[bubbleyellow, ellipse, minimum width=8.8cm, minimum height=0.1cm] at (0.25,-1.8) {\textbf{\hyperref[sec:dark_atoms]{\parbox{4.5cm}{Dark-atom/molecule SIDM}}}};

\node[
  bubble,
  minimum width=3.2cm,
  minimum height=1.1cm,
  rotate=59,
  transform shape,
  align=center
] at (-4.15,-2.6)
{\textbf{\textcolor{blue}{\parbox[c]{3.2cm}{\centering Dark-baryon\\ SIDM}}}};

\node[
  rotate=59,
  transform shape,
  inner sep=0pt,
  outer sep=0pt,
  text opacity=0
] at (-5.15,-2.6)
{\hyperref[sec:dark_baryons]{\rule{1.1cm}{1.1cm}}};

\node[
  bubble,
  minimum width=1.75cm,
  minimum height=1.1cm,
  rotate=59,
  transform shape,
  align=center
] at (-3,-3.35)
{\textbf{\textcolor{blue}{\parbox[c]{1.75cm}{\centering Dark-meson\\ SIDM}}}};

\node[
  rotate=59,
  transform shape,
  inner sep=0pt,
  outer sep=0pt,
  text opacity=0
] at (-4,-3.35)
{\hyperref[sec:dark_mesons]{\rule{1.1cm}{1.1cm}}};

\node[bubble, minimum width=3.5cm, minimum height=0.95cm] at (-0.55,-2.9) {\textbf{\hyperref[sec:dark_mesons]{\parbox{3.5cm}{Composite-Higgs SIDM}}}};

\end{tikzpicture}
\caption{
Schematic three-axis map of the particle-physics SIDM model classes discussed in Chapter~\ref{sec:pp_to_simulation_inputs}. The axes represent
anisotropy of the scattering, internal multi-state structure and dissipative processes, and the dashed lines mark qualitative thresholds along them. The central region within all three thresholds represents approximately isotropic, elastic, single-state SIDM with a velocity-independent cross section. Crossing any threshold indicates the onset of the corresponding additional feature, together with velocity-dependent scattering where applicable. Each bubble represents a model class and spans the qualitatively different realizations that may occur within it. A bubble crossing one or more thresholds therefore indicates that only a subset of its realizations exhibits the corresponding characteristics, and an individual realization may exhibit one or more of them. Overlapping bubbles indicate that the model classes are not mutually exclusive and can share realizations. The positions, sizes, threshold crossings and overlaps are illustrative only. Italic labels indicate representative microscopic mechanisms or structures. Green bubbles denote model classes that can, in principle, be simulated with our present \arepoTwo implementations, whereas yellow denotes classes that are only partly captured.
}
\label{fig:map_pp_to_effective}
\end{figure*}

\section{Guide for the busy reader}
\label{sec:guide_busy_reader}

This paper develops a unified theory-to-simulation framework for SIDM, connecting the particle-physics description of dark-sector models to the simulation inputs required for structure-formation simulations and to their numerical implementation in \arepoTwo, as summarized in \cref{fig:pp_to_simulation_inputs}. Chapters~\ref{sec: sidm_pp_models} and~\ref{sec:pp_to_simulation_inputs} develop the particle-physics toolbox, establish the theory-to-simulation mapping and apply it to representative dark-sector model classes. Building on this mapping, Chapters~\ref{sec: implementation} and~\ref{sec: verification tests} present and validate the corresponding numerical algorithms, summarized in \cref{fig:sidm_flowchart}.

\medskip
\noindent\textbf{From particle physics to simulation inputs.}
In Chapter~\ref{sec: sidm_pp_models}, we introduce the particle-physics toolbox from which SIDM models can be constructed. We first define the scope and taxonomy of the SIDM model realizations considered in this work (\cref{sec:scope_and_taxonomy}) and summarize the cross-section quantities most relevant for structure formation (\cref{sec:relevant_cross_section}). We then introduce the dark-sector particle content, symmetries, masses and couplings at Lagrangian level, together with the resulting scattering, annihilation and decay processes, DM-production mechanisms and associated cosmological, astrophysical and experimental constraints (\cref{sec: pp_terminology}). We then use minimal light-mediated SIDM model class as a benchmark to demonstrate how complementary probes constrain the microscopic parameter space (\cref{sec:interplay_constraints}), and show how broader dark-sector realizations span a wide range of viable SIDM phenomenology (\cref{sec:reopen_parameter_space}). As microscopic baselines, we consider elementary SIDM consisting of a single DM species with either contact-like self-interactions or self-interactions mediated by a single dark-sector mediator, as well as composite DM whose low-energy self-interactions arise from underlying strong dynamics. These constructions can be extended by additional dark-sector degrees of freedom or further ingredients such as a DM asymmetry.

In Chapter~\ref{sec:pp_to_simulation_inputs}, we establish the mapping from this microscopic particle physics, $\mathcal{P}_{\rm PP}$, to the corresponding simulation inputs, $\mathcal{P}_{\rm SIM}$. The simulation inputs comprise the initial conditions (\cref{sec:initial_conditions}), internal states and their initial populations (\cref{sec:internal_state_structure}), reaction networks (\cref{sec:reaction_network}), dissipative energy loss (\cref{sec:dissipative_energy_loss}), differential and angle-weighted cross sections (\cref{sec:cross_sections}), and decay rates and kinematics (\cref{sec:decay_inputs}). This framework provides the interface between microscopic dark-sector theories and structure-formation simulations and, while SIDM is our primary focus, also applies to dark-sector scenarios beyond the SIDM paradigm. Later in Chapter~\ref{sec:pp_to_simulation_inputs}, we apply this mapping to representative dark-sector model classes constructed from the particle-physics toolbox of Chapter~\ref{sec: sidm_pp_models}. We consider contact-like and minimal light-mediator SIDM (Secs.~\ref{sec:contact-like} and~\ref{sec:light-mediator}), resonant SIDM (\cref{sec:Resonant_SIDM}), vector SIDM (\cref{sec:Vector_SIDM}), asymmetric SIDM (\cref{sec:Asymmetric_SIDM}), multi-state inelastic SIDM (\cref{sec: inelastic_sidm_models}) and composite dark sectors (\cref{sec:composite_SIDM}). Related non-SIDM scenarios, including decaying DM, warm DM and ultra-light bosonic DM, are briefly discussed in the context of modified initial conditions (\cref{sec:initial_conditions}). Each class name identifies the particle-physics feature used to organize the discussion rather than a property that uniquely characterizes the theory; models may therefore belong to several classes at once. For example, a composite dark-baryon model may also be asymmetric and resonant. For each class, we determine the corresponding simulation inputs, assess which aspects can be simulated with the new \arepoTwo implementations summarized in \cref{sec:implemented_in_arepo}, and discuss the astrophysical systems in which the relevant DM properties can be probed most effectively (\cref{sec:simulation_diagnostics_DM}).

\medskip
\noindent\textbf{Numerical implementations and validation.}
In Chapter~\ref{sec: implementation}, we build on the new Monte Carlo-based SIDM module for the moving-mesh code \arepoTwo introduced in our companion paper \cite{Zier:2026}, which we briefly summarize in \cref{sec:SIDM_module}. This module combines (i) a dedicated neighbour-search tree for DM particles with (ii) a new hypercube-based communication scheme that naturally handles multiple scattering events per particle within a single timestep. Together, these improvements yield a substantial speed-up over legacy SIDM implementations, bringing the runtime close to that of CDM simulations, particularly during the late stages of gravothermal collapse. The module is designed to be readily extended to arbitrary two-body DM interactions through a simplified programming interface, without requiring detailed knowledge of the underlying parallelization scheme. Building on this framework, we implement a range of SIDM algorithms, including velocity-dependent scattering (\cref{sec:imp_velocity_dep_SIDM}), anisotropic Monte Carlo scattering (\cref{sec:anisotropic_MC}), frequent small-angle scattering (\cref{sec: fsidm}), the angular-hybrid SIDM scheme (\cref{sec:asidm}), multi-state and inelastic reactions (\cref{sec:imp_inelastic_SIDM}) and dissipative scattering (\cref{sec:imp_dissipative_SIDM}), as well as our new decaying-DM module (\cref{sec:imp_decaying_DM}). The corresponding timestep criteria are discussed in 
\cref{sec: time-stepping criterion}.

In Chapter~\ref{sec: verification tests}, we validate these implementations with deflection tests (\cref{sec: deflection tests}), decaying-DM tests (\cref{sec:dDM_box_test}), thermalization tests (\cref{sec: thermalization tests}), isolated halo evolution (\cref{sec: core collapse tests}) and cosmological and zoom-in simulations (Secs.~\ref{sec:cosmological_DMO_simulations} and \ref{sec: cosmological zoom-in simulations}) spanning a range of systems and dynamical regimes. Together, these implementations provide the numerical framework for evolving DM structure using the simulation inputs and for studying the corresponding SIDM model classes.

\bigskip
Finally, in Chapter~\ref{sec:discussion}, we discuss the main advances presented in this paper, including the new algorithms introduced relative to \arepoOne, the numerical properties and challenges of the implementations, and further directions for SIDM simulations. We conclude in Chapter~\ref{sec: conclusions}, while the appendices collect technical details of the angular velocity updates and the angular hybrid SIDM scattering scheme.

\section{Particle-physics ingredients of self-interacting dark matter}
\label{sec: sidm_pp_models}

Throughout this work, we use self-interacting DM (SIDM) as an umbrella term for DM scenarios in which non-gravitational interactions among DM particles or internal DM states can modify their late-time dynamics and the formation of cosmic structure. SIDM therefore does not correspond to a unique microscopic theory, and a large number of particle-physics realizations have been proposed in the literature~\citep[e.g.,][]{Spergel_1999,Tulin_2018,Adhikari_2022}. In this chapter, we characterize representative elementary and composite SIDM models in terms of the common particle-physics ingredients from which they are constructed: their particle content, symmetries, mass spectrum and interaction structure. Elementary dark sectors contain DM and mediator degrees of freedom that are fundamental fields, whereas in composite dark sectors the relevant low-energy degrees of freedom emerge from an underlying confining theory. In either case, these microscopic ingredients determine the allowed scattering, annihilation and decay processes and thereby the resulting dark-sector phenomenology.

A viable SIDM model must also admit a consistent cosmological history. We therefore discuss the origin of the DM and mediator masses, the stability of the DM candidate and mechanisms for generating its present-day abundance. We also consider the cosmological, astrophysical and laboratory constraints that determine which regions of the resulting microscopic parameter space remain viable. Together, these elements provide the particle-physics foundation for the representative SIDM model classes considered in Chapter~\ref{sec:pp_to_simulation_inputs}.

\medskip
\noindent
\textbf{Chapter overview.}
This chapter develops the particle-physics foundation for the theory-to-simulation framework introduced in the next chapter. We first define the scope and taxonomy of the SIDM models considered here (\cref{sec:scope_and_taxonomy}) and clarify the differential and angle-weighted cross sections relevant for structure formation (\cref{sec:relevant_cross_section}). We then provide a particle-physics primer (\cref{sec: pp_terminology}), progressing from fields and gauge interactions (\cref{sec:fields_and_particles,sec:gauge_interactions}) through DM mass generation, stability and relic abundance (\cref{sec:DM_mass_stability_abundance}), the Lagrangian and Feynman-diagram description (\cref{sec:lagrangian,sec:feynman_diagrams}), and both renormalizable and non-perturbative composite interactions (\cref{sec:renormalizable_interactions,sec:composite_dark_sectors_non_pert_ints}). We then connect these microscopic interactions to self-scattering and cross sections (\cref{sec: pp_ints_to_SM_self_interactions,sec:interactions_to_cross_sections}), before discussing DM production and experimental searches (\cref{sec:DM_production_mechanisms,sec:DM_search_experiments}). Finally, we synthesize the cosmological, astrophysical and particle-physics constraints on minimal light-mediated SIDM and show how broader, well-motivated dark-sector constructions open a diverse range of viable SIDM phenomenology (\cref{sec:interplay_constraints,sec:reopen_parameter_space}).

\subsection{Scope and taxonomy of SIDM models}
\label{sec:scope_and_taxonomy}

Throughout this chapter, we use representative particle-physics realizations of SIDM as microscopic baselines for organizing the discussion. For elementary DM, we consider a single DM species whose self-interactions arise from contact interactions or, when kinematically relevant, through the exchange of a dark-sector mediator particle. As an alternative microscopic baseline, we consider composite DM, where the low-energy DM states are bound states of elementary constituents and their self-interactions originate from the confining dynamics. These constructions can be supplemented by additional DM states or other dark-sector degrees of freedom, as well as by further ingredients such as a DM asymmetry. Minimal light-mediated SIDM models, consisting of a single DM species interacting through a single dark-sector mediator, provide one particularly useful benchmark. While some of these simplest realizations are subject to strong combined constraints, they occupy only a small subset of the much broader SIDM theory space, which includes a wide range of viable and well-motivated elementary and composite dark-sector constructions.

At the level relevant for structure formation, the resulting SIDM realizations can be organized along three physically distinct but complementary axes, illustrated schematically in \cref{fig:map_pp_to_effective}. Throughout this discussion, we use \emph{DM states} to denote massive dark-sector states that constitute the DM population during structure formation, including sufficiently long-lived excited states, while \textit{dark-sector} \textit{degrees} \textit{of} \textit{freedom} refers more broadly to additional states such as mediators or dark radiation that need not themselves constitute DM:

\begin{enumerate}[leftmargin=*, labelsep=0.5em]
\item \textbf{Angular structure (anisotropy):} from approximately isotropic scattering to strongly angle-dependent scattering, including forward-peaked or resonant scattering. Strongly anisotropic regimes typically also exhibit a pronounced velocity dependence of the cross section.

\item \textbf{Internal-state structure (multi-state):} from a single DM state to scenarios in which multiple DM states are populated and connected by elastic or inelastic reactions, including endothermic and exothermic transitions and possible decays of excited states. Such state structure can arise naturally, for example, from a rich composite-DM spectrum of bound states.

\item \textbf{Dissipation:} from conservative effective DM interactions, which preserve the energy carried by the interacting DM degrees of freedom, to dissipative effective interactions in which energy is transferred out of the DM subsystem to other dark-sector degrees of freedom, for example through the emission of dark radiation.
\end{enumerate}

In \cref{fig:map_pp_to_effective}, the SIDM model classes considered later in Chapter~\ref{sec:pp_to_simulation_inputs} are placed schematically according to these three characteristics. The classes are not mutually exclusive: a given microscopic realization may occupy several regions of this classification and exhibit more than one of the corresponding features. The connection between these characteristics and the quantities required for structure-formation simulations is developed systematically in Chapter~\ref{sec:pp_to_simulation_inputs}.

\subsection{Cross sections relevant for structure formation}
\label{sec:relevant_cross_section}

We first discuss the quantities derived from the differential scattering cross section that enter structure-formation simulations. At the microscopic level, the fundamental quantity is the differential cross section, whose dynamical impact is commonly summarized by angle-integrated cross sections.

The differential cross section,
\begin{equation}
\frac{\mathrm{d}\sigma}{\mathrm{d}\Omega}(\theta,v_{\rm rel}),
\end{equation}
encodes the full angular and velocity dependence of the two-body interaction. This is the quantity that should ideally be implemented in simulations whenever the goal is to capture the complete scattering physics~\cite{Robertson_2017b,Banerjee_2020}. In practice, however, cosmological $N$-body simulations often adopt an effective isotropic description that approximates how self-interactions redistribute momentum and energy within haloes~\cite{Vogelsberger_2012,rocha2013cosmological,Robertson_2017b}.

For isotropic scattering, the total cross section adequately characterizes both the interaction rate and its dynamical impact. For anisotropic interactions, particularly strongly forward-peaked scattering generated by long-range or light-mediator forces, the total cross section can instead substantially overestimate the dynamical importance of self-interactions because small-angle deflections transfer little momentum and energy~\cite{Buckley_2010,Tulin_2013,Kahlhoefer_2014}.

Since halo evolution is governed by the redistribution of momentum and energy rather than by the raw number of scatterings, it is useful to introduce angle-weighted cross sections,
\begin{equation}
\sigma_X
=
\int \mathrm{d}\Omega\,
w_X(\theta)\,
\frac{\mathrm{d}\sigma}{\mathrm{d}\Omega},
\label{eq:weighted_cross_sections}
\end{equation}
where the weight $w_X(\theta)$ is chosen to emphasize the aspect of the scattering most relevant for the macroscopic problem. The most common choices are~\cite{Tulin:2013teo,Boddy:2016bbu,Tulin_2018,Kahlhoefer:2017umn,Robertson:2016qef,Yang:2022hkm,Fischer:2025rky}:

\begin{enumerate}[leftmargin=*, labelsep=0.5em]

\item \textbf{Total cross section}, $\sigma_{\rm tot}$ (overall scattering rate):
\[
w_{\rm tot}(\theta) = 1 .
\]
This counts all events equally, independent of deflection angle, and thus measures only the total interaction rate. It is adequate for isotropic interactions, but in forward-dominated models it strongly overestimates the dynamical impact, since most scatterings barely alter particle trajectories.

\item \textbf{Transfer cross section}, $\sigma_{\rm T}$ (momentum transfer for distinguishable particles)~\cite{Feng:2009hw,Tulin_2013}:
\[
w_{\rm T}(\theta) = 1 - \cos\theta .
\]
In the center-of-mass frame the longitudinal momentum change of one particle is $\Delta p = m_{\rm DM} v_{\rm rel}(1-\cos\theta)/2$, so the factor $(1-\cos\theta)$ directly measures the fractional forward momentum transfer. Small-angle deflections, for which $(1-\cos\theta)\simeq \theta^2/2$, are strongly suppressed. The transfer cross section therefore quantifies the efficiency of momentum redistribution (effective drag) inside haloes.

\item \textbf{Modified transfer cross section}, $\sigma_{\widetilde{\rm T}}$ (momentum transfer for identical particles)~\cite{Kahlhoefer_2014,Kahlhoefer:2017umn,Robertson_2017b}:
\[
w_{\widetilde{\rm T}}(\theta) = 2\,\left(1 - |\cos\theta|\right) .
\]
For identical particles, the $t$- and $u$-channel amplitudes interfere and backward scattering ($\theta \to \pi$) is physically equivalent to forward scattering. The absolute value treats forward and backward deflections symmetrically, preventing double counting of equivalent momentum exchange. This is the appropriate generalization of $\sigma_{\rm T}$ for identical particles.

\item \textbf{Viscosity cross section}, $\sigma_{\rm V}$ (energy and transverse momentum transfer, for both distinguishable and identical particles)~\cite{Tulin_2013,Cline_2014a,Boddy:2016bbu}:
\[
w_{\rm V}(\theta) = \frac{3}{2}\,\left(1 - \cos^{2}\theta\right) .
\]
Here $(1-\cos^2\theta)=\sin^2\theta$ is proportional to the squared transverse momentum transfer, suppressing both forward and backward deflections, which exchange little transverse momentum. The viscosity cross section thus measures the rate of velocity isotropization and thermal conduction -- the efficiency of energy redistribution rather than mere momentum loss.
\end{enumerate}

\noindent
In the isotropic limit, these definitions all coincide,
\begin{equation}
\sigma_{\rm tot}
=
\sigma_{\rm T}
=
\sigma_{\widetilde{\rm T}}
=
\sigma_{\rm V},
\end{equation}
so the precise choice is immaterial. This equivalence does not hold for anisotropic scattering.

Among these angle-weighted cross sections, the viscosity cross section is often a useful single-number proxy for elastic SIDM structure formation~\cite{Yang:2022hkm}. The macroscopic evolution of an SIDM halo -- core formation, gravothermal evolution and eventual core collapse -- is controlled primarily by heat transport, i.e.\ the redistribution of random kinetic energy by DM self-scattering between halo regions with different velocity dispersions, rather than by the total scattering rate~\cite{Balberg2002,Koda_2011}. Because both forward ($\theta\to0$) and backward ($\theta\to\pi$) scattering contribute little to this heat transport, the viscosity cross section $\sigma_{\rm V}$, with its $\sin^2\theta$ weighting, captures the relevant physics better than the transfer cross section $\sigma_{\rm T}$, weighted by $1-\cos\theta$. Comparisons with simulations that sample the full angular dependence of $\mathrm{d}\sigma/\mathrm{d}\Omega$ show that viscosity-based matching can reproduce halo evolution more accurately than transfer-cross-section matching for strongly anisotropic elastic scattering~\cite{Yang:2022hkm}. Nevertheless, the accuracy of any single-number proxy depends on the interaction model and the observable being considered. For \emph{inelastic} and \emph{dissipative} scatterings, the situation is more involved because changes in the internal state and the associated energy loss or gain can significantly affect halo evolution even for forward scattering~\cite{vogelsberger2019evaporating,Shen:2021frv}. A single elastic transport proxy is therefore generally insufficient, and these processes are more faithfully described by the full channel-dependent differential and total interaction rates. More generally, the most complete approach, adopted in this work, is to implement the full differential cross section directly~\cite{Robertson_2017b,Banerjee_2020}, since any angle-integrated scalar proxy necessarily discards part of the angular information.

For \emph{inelastic} and \emph{dissipative} scatterings, the situation is more involved because changes in the internal state and the associated energy loss or gain can significantly affect halo evolution even for small angular deflections~\cite{vogelsberger2019evaporating,Shen:2021frv}. A single elastic transport proxy is therefore generally insufficient. Instead, each inelastic or dissipative reaction channel must be characterized by its channel-dependent differential cross section, $\mathrm{d}\sigma_{\mathcal R}/\mathrm{d}\Omega(v_{\rm rel},\theta)$, whose angular integral determines the corresponding interaction cross section $\sigma_{\mathcal R}(v_{\rm rel})$ and hence the reaction rate, while its angular dependence determines the distribution of scattering outcomes.

\subsection{Particle-physics terminology and microscopic ingredients}
\label{sec: pp_terminology}

In this section, we provide a concise primer on the particle-physics terminology and microscopic ingredients used throughout this work. Rather than providing a comprehensive review of DM model building, we introduce the particle-physics concepts needed to understand how SIDM models are constructed and how their microscopic properties are translated into structure-formation simulation inputs. Readers already familiar with particle physics and SIDM phenomenology may skip this section.

\subsubsection{Fields and particles}
\label{sec:fields_and_particles}

In quantum field theory, each particle species is associated with an underlying quantum field, and particles correspond to excitations of that field. For our purposes, a field can simply be regarded as the mathematical object used to describe a particle species, while the Lagrangian introduced below specifies its dynamics, mass and interactions.

\begin{table*}[t]
\centering
\begin{tabular}{lcccc}
\hline
\textbf{Field} & \textbf{Symbol} & \textbf{Spin} & \textbf{Elementary SM analogue} & \textbf{Composite SM analogue} \\
\hline
Scalar DM & $s$ & $0$ & Higgs boson & Scalar mesons (e.g.\ $f_0$ states) \\
Fermionic DM & $\chi$ & $1/2$ & Quarks, leptons (e.g.\ electron) & Baryons (e.g.\ proton, neutron) \\
Vector DM & $\chi_\mu$ & $1$ & Gauge bosons (e.g.\ $W^\pm$, $Z^0$) & Vector mesons (e.g.\ $\rho$, $\omega$) \\
\hline
Scalar mediator & $\phi$ & $0$ & Higgs boson & Scalar mesons (e.g.\ $f_0$ states) \\
Vector mediator & $\phi_\mu$ & $1$ & Photon, $W^\pm$, $Z^0$, gluons & Vector mesons (e.g.\ $\rho$, $\omega$) \\
\hline
\end{tabular}
\caption{Summary of the representative DM and dark-mediator fields used in the explicit microscopic realizations considered in this work, including their notation and spin, together with elementary and composite analogues in the Standard Model of particle physics (SM) and its low-energy bound-state spectrum. The spin classification is more general than the specific parity assignments adopted here; in particular, spin-$0$ and spin-$1$ states may also occur as pseudoscalar and axial-vector states, respectively.}
\label{tab:dm_fields}
\end{table*}

In this work, we consider DM candidates with spin $0$, $1/2$ and $1$. Their microscopic properties are further determined by their gauge and discrete quantum numbers and by the symmetries of the underlying dark sector. Spin alone does not fully specify the nature of a particle. In particular, spin-$0$ states can have scalar or pseudoscalar character, while spin-$1$ states can have vector or axial-vector character. These distinctions describe how the corresponding fields transform under parity, i.e.\ under the spatial inversion $\vec{x}\rightarrow-\vec{x}$: a scalar is unchanged whereas a pseudoscalar changes sign, with analogous parity distinctions between vector and axial-vector fields. Familiar spin-$0$ examples are the scalar Higgs boson and the pseudoscalar pions in the Standard Model of particle physics (SM)\footnote{The Standard Model of particle physics is the well-tested theory describing the known elementary particles and their electromagnetic, weak and strong interactions; see the foundational works of~\cite{Glashow:1961tr,Weinberg:1967tq,Salam:1968rm} for the electroweak theory, \cite{Englert:1964et,Higgs:1964pj,Guralnik:1964eu} for the Higgs mechanism, and~\cite{GrossWilczek:1973,Politzer:1973} for the quantum chromodynamics of the strong nuclear force.}.

In the explicit microscopic realizations considered below, we focus on scalar spin-$0$ DM, denoted by $s$, fermionic spin-$1/2$ DM, denoted by $\chi$, and vector spin-$1$ DM, denoted by $\chi_\mu$. A spin-$0$ particle carries no intrinsic angular momentum; a fermion carries half-integer spin, like the electron, quarks or composite baryons such as the proton and neutron; and a spin-$1$ particle carries one unit of intrinsic angular momentum.

Self-interactions between DM particles can arise in several ways: from contact-like interactions, which appear effectively point-like; from the exchange of a mediator particle that acts as a dark-sector force carrier, analogous to how the photon mediates the electromagnetic force; or from the internal dynamics of composite DM states built from elementary constituents. Representative contact and mediator-exchange interactions are illustrated in \cref{fig:selfint_realizations_all}, while examples of composite self-interactions are shown in \cref{fig:selfint_composite_realizations}. Mediators may themselves be scalar particles (denoted $\phi$) or vector particles (denoted $\phi_\mu$). A scalar mediator carries spin $0$, whereas a vector mediator carries spin $1$. In this sense, DM self-interactions can be viewed as arising from a new ``dark force,'' in close analogy with familiar interactions in the SM. The particle content and notation used throughout this work are summarized in Table~\ref{tab:dm_fields}.

The spin of the DM particle and the type of mediator constrain the possible interactions and therefore shape the velocity and angular dependence of the differential cross section, $\mathrm{d}\sigma/\mathrm{d}\Omega(\theta,v_{\rm rel})$, which in turn governs the impact of self-interactions on cosmic structure formation.

\subsubsection{Gauge interactions (forces)}
\label{sec:gauge_interactions}

Having introduced the relevant fields and mediators, we now turn to gauge interactions, one of the most common microscopic origins of DM self-interactions. A natural possibility is that DM self-interactions arise from a \emph{gauge theory} (for a detailed introduction, see~\cite{Bohm:2001yx}), in which the interactions follow from an underlying symmetry principle. Electromagnetism provides the simplest SM example: it is associated with a $U(1)$ gauge symmetry and is mediated by the photon. If DM carries a conserved \emph{dark charge} under a new Abelian $U(1)_D$ gauge symmetry, it interacts through the corresponding spin-$1$ mediator $\phi_\mu$ introduced above, often called a \emph{dark photon}~\cite{Okun:1982xi,Holdom:1985ag}. More generally, the gauge group may be non-Abelian, such as $SU(2)_D$ or $SU(3)_D$, in which case DM self-interactions can be mediated by self-interacting non-Abelian gauge bosons~\cite{Gross:2015cwa,Ko:2016fcd,Tulin:2017ara}. A non-Abelian gauge theory is specified not only by its gauge group but also by its matter content and representations, which together determine the renormalization-group evolution of the gauge coupling. The balance between gauge and matter degrees of freedom can determine whether the theory remains perturbative at low energies or becomes strongly coupled and confines, in which case the relevant low-energy degrees of freedom are composite states rather than the underlying elementary fields. Fermionic dark sectors may additionally contain Yukawa interactions with scalar fields, so many microscopic realizations are more generally gauge--Yukawa theories; here, however, we organize the discussion according to the gauge dynamics most directly responsible for the self-interactions.

At an intuitive level, a gauge interaction arises when a symmetry of the fields is required to hold locally, i.e.\ independently at each point in spacetime. Maintaining this local symmetry requires the introduction of a gauge field, whose associated gauge boson mediates the interaction. For an unbroken $U(1)_D$ symmetry, the corresponding dark photon is massless. If the symmetry is spontaneously broken, for example by a dark Higgs field acquiring a non-zero vacuum expectation value, the gauge boson acquires a mass while remaining the mediator of the dark interaction. The mediator can therefore be light without being massless, with its mass setting the interaction range and playing a central role in the velocity and angular dependence of DM self-scattering.

The simplest case is an Abelian gauge symmetry $U(1)$, corresponding to local 
phase rotations of a field, $\chi \to e^{i\alpha(x)}\chi$, where $\alpha$ is 
allowed to vary in space and time. Consistency then requires a new gauge field; in electromagnetism this is the photon, while in the dark sector DM charged under $U(1)_D$ interacts via exchange of a dark photon $\phi_\mu$. The key physical point is that the exchange of a \emph{light} vector mediator can render DM--DM scattering long-range and strongly dependent on velocity and scattering angle, in close analogy to Rutherford scattering in  electromagnetism~\cite{Feng:2009hw,Buckley_2010,Loeb_2011,Tulin_2013}. This is 
one of the standard microscopic routes to velocity- and angle-dependent SIDM.

If DM carries a conserved \emph{dark charge} under a new Abelian $U(1)_D$ gauge symmetry, it interacts through the corresponding spin-$1$ mediator $\phi_\mu$ introduced above, often called a \emph{dark photon}~\cite{Okun:1982xi,Holdom:1985ag}. More generally, the gauge group may be non-Abelian, such as $SU(2)_D$ or $SU(3)_D$~\cite{YangMills:1954}. Unlike an Abelian $U(1)$ symmetry, non-Abelian gauge transformations do not generally commute, and the corresponding gauge bosons themselves carry the gauge charge and can therefore interact with one another. In the SM, this structure underlies the weak and strong interactions. In a dark sector, such non-Abelian gauge bosons can mediate DM self-interactions~\cite{Buckley:2009in,Ko:2014nha,Boddy:2014yra,Tulin:2017ara}. It is important, however, to distinguish two qualitatively different regimes of such a non-Abelian dark sector, since the \emph{same} gauge group $SU(N)_D$ can realize either picture depending on its coupling strength. If the gauge coupling remains \emph{weak} (perturbative, possibly with Higgs-like, massive mediators), the non-Abelian gauge bosons remain perturbative force carriers and can mediate self-interactions between DM states carrying the corresponding dark charge. These DM states may themselves be elementary or composite, with the latter possibility arising, for example, from additional confining dynamics. The mediators are analogous to the $W^\pm$ and $Z^0$ bosons of the weak interaction in the SM. For simplicity, we denote all such vector mediators collectively by $\phi_\mu$, irrespective of the specific gauge group. If, on the other hand, the coupling becomes \emph{strong} at a confinement scale $\Lambda_D$, the fundamental constituents are confined and the low energy DM states are instead \emph{composite} objects, as we discuss next.

A particularly rich possibility arises when the non-Abelian dynamics confines~\cite{Kribs:2016cew,Cline:2013zca,Boddy:2014yra,Hochberg:2014kqa}. In the SM, the non-Abelian $SU(3)$ gauge interaction of quantum chromodynamics (QCD), mediated by gluons, confines at low energies and binds elementary quarks and gluons into composite hadrons, including mesons, baryons and glueballs. The same confinement mechanism can operate in a dark sector with a non-Abelian gauge group $G_D$, for example $SU(N)_D$. If the gauge-field contribution dominates sufficiently over the matter-field contribution to the running of the coupling, the $G_D$ interaction becomes strongly coupled and confines at a scale $\Lambda_D$. The elementary dark constituents then bind into composite \emph{dark hadrons}, in analogy with QCD. Such composite states can naturally possess sizeable self-interactions, a rich spectrum of bound states and multiple interaction channels, as discussed in detail in \cref{sec:composite_SIDM}.

If the dark sector contains attractive but non-confining interactions, its constituents may form weakly bound states in close analogy to atomic physics in the SM. In particular, an attractive dark $U(1)_D$ interaction mediated by a light dark photon can give rise to \emph{dark atoms}~\cite{Kaplan:2009de,CyrRacine:2012fz,Cline:2013pca} and, in more complex scenarios, \emph{dark molecules}~\cite{Cline:2013pca,Gurian:2021qhk}. Such bound states can themselves possess sizeable self-interactions, much as neutral hydrogen atoms scatter efficiently at low energies through their underlying electromagnetic constituents~\cite{Krstic:1999}. These possibilities further broaden the range of effective interactions relevant for structure formation.

A related realization of dark $U(1)_D$ gauge interactions is found in \emph{partially interacting DM} (PIDM), where only a subdominant fraction of DM has significant self-interactions~\cite{Fan:2013yva}. In the dissipative \emph{double-disk DM} scenario, a massless dark photon enables radiative cooling of this component, potentially leading to the formation of a dark disk~\cite{Fan:2013tia}.

In summary, gauge theories provide a broad microscopic framework for SIDM. The simplest realization is an Abelian $U(1)_D$ interaction mediated by a dark photon. More general perturbative dark sectors may involve non-Abelian gauge interactions and, in the presence of scalar fields and fermionic matter, Yukawa interactions, giving rise to gauge--Yukawa theories. If the non-Abelian gauge dynamics instead becomes strongly coupled at low energies, confinement produces composite dark states with potentially sizeable self-interactions. Attractive but non-confining gauge interactions can additionally form weakly bound states such as dark atoms and molecules, while combinations of these structures are also possible. These possibilities closely parallel familiar structures in the SM, including electromagnetic, weak, strong and Yukawa interactions, but realized within the dark sector.

\subsubsection{DM mass generation, stability and relic abundance}
\label{sec:DM_mass_stability_abundance}

A microscopic SIDM model must explain the origin
of the DM mass, its stability on cosmological
timescales, and how the observed relic
abundance~\cite{Planck2018Parameters} is produced
in the early Universe
~\cite{Bertone:2004pz,Feng:2010gw,Cirelli:2024ssz}.

\medskip
\noindent
\textbf{Origin of the DM mass.} In the SM, most elementary particles acquire their masses through the \emph{Higgs mechanism}~\cite{Englert:1964et,Higgs:1964pj,Guralnik:1964eu}: electroweak symmetry breaking gives the Higgs field a non-zero vacuum expectation value, and particles coupled to this field thereby acquire mass. A dark sector can mirror this structure. A dark sector can acquire mass through Higgs-sector interactions in several ways. For scalar DM, a renormalizable Higgs-portal coupling can directly connect the DM field to the SM Higgs sector~\cite{Silveira:1985rk,Patt:2006fw}. For gauge-singlet fermionic DM, no analogous renormalizable direct Higgs portal exists; Higgs interactions typically proceed through an additional scalar mediator, such as a dark Higgs, that mixes with the SM Higgs, or through higher-dimensional operators. Independently, the dark sector may contain its own Higgs field whose non-zero vacuum expectation value generates masses for dark-sector particles, including DM states and dark gauge bosons, for example giving a $U(1)_D$ gauge mediator a finite mass~\cite{Holdom:1985ag}.

Second, mass can arise predominantly from \emph{strong dynamics}, as in QCD, where most of the nucleon mass is generated by the strong interaction rather than by the Higgs-generated quark masses, as confirmed by lattice-QCD calculations~\cite{BMW:2008jgk}. An analogous mechanism can operate in a confining dark gauge theory: for constituent masses not much larger than the confinement scale, the masses of the resulting composite DM states are set primarily by the dynamical dark confinement scale $\Lambda_D$, rather than by the constituent masses themselves.

Third, the hierarchy of particle masses can be controlled by exact or approximate \emph{symmetries}. For fermions, \emph{chiral symmetries} can protect small masses against large quantum corrections~\cite{tHooft:1979rat}; the electron in the SM provides a familiar example. Scalar particles are not protected by chiral symmetries, but their masses may instead be stabilized by other symmetries or dynamical structures. Examples include \emph{supersymmetry}~\cite{Martin:1997ns}, \emph{shift symmetries} as for pseudo-Nambu--Goldstone bosons~\cite{Weinberg:1978kz,Wilczek:1977pj}, and \emph{compositeness}, where the scalar is a bound state of a new strong interaction~\cite{Kaplan:1983fs,Kaplan:1983sm}. Keeping scalar masses light and theoretically stable usually requires additional structure. Examples include \emph{supersymmetry}~\cite{Martin:1997ns}, which relates fermions and bosons such that their quantum contributions to scalar masses cancel; \emph{shift symmetries}, under which the scalar can be shifted by a constant without changing the physics, forbidding a large mass term, as for pseudo-Nambu--Goldstone bosons~\cite{Weinberg:1978kz,Wilczek:1977pj}; and \emph{compositeness}, in which the scalar is not elementary but a bound state of a new strong interaction, so that its mass is generated dynamically~\cite{Kaplan:1983fs,Kaplan:1983sm}.
Finally, these mechanisms can also be combined. In \emph{composite Higgs} models~\cite{Kaplan:1983fs,Agashe:2004rs}, the Higgs itself is not elementary but a bound state of a new strong interaction, so that mass generation emerges from strong dynamics rather than from a fundamental scalar. 

SIDM models can therefore realize DM mass generation through Higgs-like mechanisms, strong dynamics or combinations thereof, with the resulting mass scales constrained by the symmetries and dynamics of the dark sector. The mass-generation mechanism and symmetry structure establish the characteristic DM and mediator mass scales (e.g.\ $m_s$, $m_\chi$, $m_{\chi_\mu}$, $m_\phi$ and $m_{\phi_\mu}$). These scales, together with the interaction strengths, determine the self-scattering regime relevant for structure formation. In particular, the mediator mass relative to the characteristic momentum transfer sets the interaction range: heavy mediators lead to effectively short-range, approximately isotropic scattering, whereas sufficiently light mediators can produce long-range, velocity- and angle-dependent interactions.

\medskip
\noindent
\textbf{Origin of DM stability.} 
DM must be sufficiently stable on cosmological timescales~\cite{Audren:2014bca,Poulin:2016nat}, 
its lifetime greatly exceeding the age of the Universe. Several mechanisms can ensure such longevity; for a comprehensive discussion 
we refer the reader to~\cite{Hambye:2010zb}.

\emph{Absolute stability through symmetries.} A common mechanism for ensuring DM stability is an exact symmetry under which the DM candidate carries a non-trivial charge. The simplest example is a discrete $Z_2$ symmetry, under which the DM particle is ``odd'' while SM particles are ``even''~\cite{Hambye:2010zb}.\footnote{Under a $Z_2$ symmetry, fields transform as either $s\rightarrow+s$ (even) or $s\rightarrow-s$ (odd). Conservation of the $Z_2$ charge implies that the lightest odd particle cannot decay solely into even states and is therefore stable. More generally, stability may be protected by $Z_3$ or higher $Z_N$ symmetries~\cite{Ma:2007gq,Belanger:2012vp}, under which a field with discrete charge $q$ transforms as $s\rightarrow e^{2\pi i q/N}s$. The lightest state carrying a non-trivial conserved $Z_N$ charge is then protected against decay into neutral states.} A well-known realization is $R$-parity in supersymmetric theories, for which the lightest supersymmetric particle can be stable~\cite{Goldberg:1983nd,Ellis:1983ew,Jungman:1995df}. DM stability may also arise if DM is the lightest charged state under a conserved \emph{continuous global symmetry}, for example a $U(1)$ symmetry analogous to electric-charge conservation.\footnote{A $U(1)$ symmetry allows the field to be multiplied by a common complex phase, $s \rightarrow e^{i\alpha}s$, without changing the physics, where $\alpha$ is constant for a global $U(1)$ and spacetime-dependent for a local (gauged) $U(1)$. The symmetry implies conservation of an associated charge, much like electric charge.}

\emph{Accidental and emergent stability.} The symmetries protecting DM need not be imposed specifically for this purpose, but may emerge from the field content and interaction structure of the theory. A familiar SM example is baryon number, which arises as an accidental global symmetry of the renormalizable SM Lagrangian and renders the proton extremely long-lived. Analogously, a confining dark gauge theory may possess an accidental \emph{dark baryon number} that stabilizes the lightest dark baryon~\cite{Antipin:2015xia,Mitridate:2017oky}. Stability may also follow from a residual symmetry that remains unbroken after spontaneous symmetry breaking, thereby protecting the lightest state carrying the corresponding charge~\cite{Krauss:1988zc}.

\emph{Approximate symmetries and small violations.} The symmetries ensuring DM stability need not be exact: small violations are acceptable provided they do not reduce the DM lifetime below the age of the Universe. For instance, DM might carry a tiny electric charge (so-called millicharged DM~\cite{Holdom:1985ag,Davidson:2000hf}), permitting extremely weak interactions with ordinary photons. If this charge is small enough, such interactions neither induce rapid decay nor cause excessive energy loss, and remain consistent with observations.

\emph{Long-lived but unstable dark matter.} Even without an exact protecting symmetry, DM can be cosmologically long-lived if its decay rate is sufficiently suppressed. A well-known example is the keV-scale sterile neutrino, whose decay into SM particles is suppressed by both its small mass and its tiny mixing with active neutrinos~\cite{Dodelson:1993je,Boyarsky:2018tvu}.

In summary, DM stability can arise from symmetries directly tied to the interaction structure of the model, including exact, residual or accidental symmetries, but may also be imposed explicitly. Alternatively, sufficiently long lifetimes can result from kinematic or dynamical suppression of decay rates. The essential requirement is that the DM candidate remains sufficiently long-lived on cosmological timescales.

\medskip
\noindent
\textbf{DM relic abundance.} Beyond stability, a viable DM candidate must also reproduce the relic abundance observed today, $\Omega_{\mathrm{DM}}h^2 \approx 0.12$~\cite{Planck2018Parameters}. Several production mechanisms can account for how the correct cosmic DM density arose in the early Universe; in \cref{sec:DM_production_mechanisms} we discuss four representative cases in more detail.

In the simplest and most widely studied scenario -- \emph{thermal freeze-out}~\cite{Lee:1977ua,Scherrer:1985zt,Griest:1990kh,Kolb:1990vq} -- DM was initially in thermal equilibrium with the hot plasma shortly after the Big Bang. As the Universe expanded and cooled, interactions became less frequent, and once the annihilation rate dropped below the expansion rate the DM number density effectively ``froze out.'' The residual abundance is then set primarily by the annihilation cross section: stronger annihilation leaves fewer particles, weaker annihilation a larger relic density. In the alternative \emph{freeze-in} scenario~\cite{Hall:2009bx,Elahi:2014fsa,Bernal:2017kxu}, DM never reaches thermal equilibrium; coupling only feebly to the SM, it is slowly produced from the hot plasma, with its abundance gradually accumulating until production ceases as the Universe cools. A further possibility is \emph{reannihilation}~\cite{Chu:2011be,Hambye:2019dwd}, in which DM is first produced through a feeble portal interaction but then annihilates efficiently within the dark sector, further depleting its abundance before freeze-out completes. Another well-motivated case is \emph{asymmetric} DM~\cite{Nussinov:1985xr,Kaplan:2009ag,Davoudiasl:2012uw,DuttaBanik:2020vfr,Barman:2021ost,Cui:2020dly,Falkowski:2011xh,Patel:2022xyv,Biswas:2018sib,Narendra:2018vfw,Nagata:2016knk,Arina:2011cu,Arina:2012fb,Narendra:2017uxl,Narendra:2019cyt,Mahapatra:2023dbr,Borah:2022ask}; see also the reviews~\cite{Petraki:2013wwa,Zurek:2013wia}. Here the present-day density arises from an initial particle--antiparticle asymmetry, analogous to the baryon asymmetry of ordinary matter: once a dark-sector asymmetry is generated, the symmetric component annihilates away, leaving a net excess of DM particles. This scenario is especially compelling because the observed density ($\rho_{\mathrm{DM}} \approx 5\,\rho_{\mathrm{B}}$) is of the same order as the visible matter density, hinting that the two abundances may share a common origin~\cite{Kaplan:2009ag,Zurek:2013wia}.

\medskip
More generally, the same symmetries that protect DM from rapid decay also constrain its allowed interactions, shape how its mass is generated and determine which production mechanisms are viable in the early Universe. In this sense, self-interactions, mass generation, stability and relic production are not independent theoretical choices, but different consequences of the same microscopic framework. The symmetry structure and particle spectrum of the dark sector therefore play a central organizing role: they dictate which interaction terms are permitted, which are forbidden and how the theory behaves at high energies. To make these connections explicit, we now turn to the Lagrangian description, in which these principles are implemented in a precise and systematic way.

\subsubsection{The Lagrangian}
\label{sec:lagrangian}

The microscopic particle properties are encoded in the Lagrangian density, $\mathcal{L}$. Observable quantities such as differential cross sections, decay rates, lifetimes, annihilation rates, branching ratios and production rates all follow from $\mathcal{L}$~\cite{Peskin:1995ev,Weinberg:1995mt}. The quantum dynamics is described by the path integral~\cite{Feynman:1948ur},
\begin{equation}
\label{eq:path_integral}
\mathcal{Z} = \int \mathcal{D}\Phi \;
e^{iS[\Phi]},
\end{equation}
defined in terms of the action
\begin{equation}
\label{eq:action}
S = \int \mathrm{d}^4 x \, \mathcal{L},
\end{equation}
where $\Phi$ denotes all fields in the theory. The path integral sums over field configurations weighted by the phase $e^{iS[\Phi]}$, from which one derives correlation functions, scattering amplitudes and, ultimately, cross sections and decay rates.

A generic Lagrangian contains
\begin{enumerate}
\item \emph{Kinetic terms}, which describe propagation;
\item \emph{Mass terms}, which set the rest masses;
\item \emph{Interaction terms}, which encode scattering, annihilation and decay.
\end{enumerate}
Below we construct these ingredients for scalar ($s$) and fermionic ($\chi$) DM, focusing on the interaction structures relevant for SIDM, and we extend this to vector DM ($\chi_\mu$) in \cref{sec:Vector_SIDM}.

A central organizing principle is \emph{renormalizability}~\cite{Dyson:1949bp,tHooft:1972tcz,Wilson:1973jj,Peskin:1995ev}. In natural units $\hbar=c=1$, all quantities are measured in powers of energy. Since the action in \cref{eq:action} is dimensionless and $[\mathrm{d}^4x]=-4$, the Lagrangian density has mass dimension $[\mathcal{L}]=4$ in four spacetime dimensions. The basic field dimensions used here are
\begin{equation}
\begin{aligned}
[s] = 1,  \qquad  [\phi] = 1, & \qquad  [\chi_\mu] = 1,  \qquad [\phi_\mu]  = 1, \\ [\chi] = \tfrac{3}{2}, 
 \qquad 
&[m] = 1,  \qquad
[\partial_\mu] = 1.
\end{aligned}
\end{equation} Renormalizable Lagrangian terms are those with total mass dimension less than or equal to four. Loosely speaking, these are the interactions whose strength does not grow with energy, so that the theory remains predictive up to arbitrarily high energies: a finite number of measured parameters (masses and couplings) suffices to compute any process. Before writing down examples, we note two further requirements. First, the Lagrangian must respect the symmetries that stabilize DM. For example, a complex scalar usually appears as $s^\dagger s$, while a fermion appears as $\bar{\chi}\chi$, where $\bar{\chi}\equiv \chi^\dagger\gamma^0$ is the Dirac adjoint.\footnote{The Dirac adjoint $\bar{\chi}=\chi^\dagger\gamma^0$ is needed to build Lorentz-invariant fermion bilinears such as $\bar{\chi}\chi$. Here $\gamma^\mu$ ($\mu=0,1,2,3$) are the constant $4\times4$ Dirac (gamma) matrices, which act on the four-component spinor $\chi$. For a complex scalar, $s$ and $s^\dagger$ correspond to the particle and antiparticle degrees of freedom.} These combinations preserve the assumed dark charge or symmetry. Second, Lorentz invariance requires that all spacetime indices be contracted. In $\partial_\mu s^\dagger \partial^\mu s$ and $\bar{\chi}\gamma^\mu\chi\,\phi_\mu$, for instance, the repeated indices $\mu=0,1,2,3$ are summed, making each term a Lorentz scalar.

Examples of renormalizable terms built from $s$, $\chi$, $\phi$ and $\phi_\mu$ that satisfy these requirements include
\begin{align}
\begin{aligned} \label{eq:renomalizable_terms}
&\partial_\mu s^\dagger \partial^\mu s, 
\qquad
s^\dagger s, 
\qquad
(s^\dagger s)(s^\dagger s),
\qquad 
\\[6pt] &(s^\dagger s)\phi,
\qquad
(s^\dagger s)\phi^2, \qquad
i\left(s^\dagger\partial^\mu s
-(\partial^\mu s^\dagger)s\right)\phi_\mu,
\qquad \\[6pt]
& \bar{\chi} i \gamma^\mu \partial_\mu \chi,
\qquad
\bar{\chi}\chi,
\qquad
\bar{\chi}\chi \phi,
\qquad
\bar{\chi}\gamma^\mu\chi \phi_\mu .
\end{aligned}
\end{align}
Each has total mass dimension four or smaller and may appear in a fundamental four-dimensional Lagrangian if permitted by the symmetries.

Lagrangian terms with mass dimension greater than four are called \emph{higher-dimensional operators}. To keep the action dimensionless, their coefficients must carry inverse powers of some energy scale, $1/\Lambda^{n}$, and they therefore signal an \emph{effective field theory} -- a description valid only below the cutoff scale $\Lambda$, above which new particles or interactions must appear to complete the theory~\cite{Weinberg:1978kz,Georgi:1993mps}. Such non-renormalizable terms typically cause scattering amplitudes to grow with energy; at sufficiently high energies these amplitudes become too large, indicating that the low-energy description is incomplete. Such operators can therefore be viewed as remnants of a more complete high-energy theory.

For example,
\[
(\partial_\mu s^\dagger \partial^\mu s)(s^\dagger s)
\quad \text{and} \quad
(\bar{\chi}\chi)(\bar{\chi}\chi)
\]
are dimension-six operators and must be suppressed by a heavy scale $\Lambda$, e.g.
\[
\frac{1}{\Lambda^2}(\partial_\mu s^\dagger \partial^\mu s)(s^\dagger s)
\quad \text{or} \quad
\frac{1}{\Lambda^2}(\bar{\chi}\chi)(\bar{\chi}\chi).
\]
Such terms are useful as effective interactions, but indicate that the theory is valid only below a characteristic energy scale $\Lambda$.

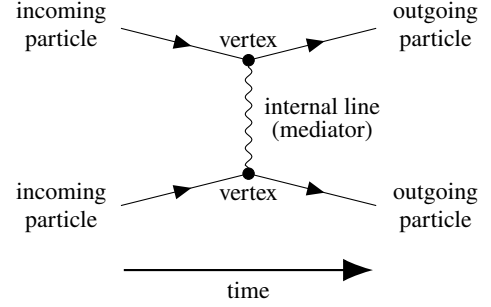
\begin{figure}[t]
\centering
\begin{tikzpicture}[scale=0.75]

\begin{feynman}
  \vertex[dot] (a) at (0,1.0) {};
  \vertex[dot] (b) at (0,-1.0) {};

  \vertex (i1) at (-2.4, 1.6) {};
  \vertex (i2) at (-2.4,-1.6) {};
  \vertex (f1) at ( 2.4, 1.6) {};
  \vertex (f2) at ( 2.4,-1.6) {};

  \diagram*{
    (i1) -- [fermion] (a) -- [fermion] (f1),
    (i2) -- [fermion] (b) -- [fermion] (f2),
    (a)  -- [photon] (b),
  };
\end{feynman}

\node[font=\small, align=center] at (-3.3, 1.6) {incoming\\particle};
\node[font=\small, align=center] at (-3.3,-1.6) {incoming\\particle};

\node[font=\small, align=center] at ( 3.3, 1.6) {outgoing\\particle};
\node[font=\small, align=center] at ( 3.3,-1.6) {outgoing\\particle};

\node[font=\small, align=center] at (1.3,0.2) {internal line};
\node[font=\small, align=center] at (1.3,-0.25) {(mediator)};

\node[font=\small] at (0,1.35) {vertex};
\node[font=\small] at (0,-1.35) {vertex};

\draw[-{Latex[length=4mm]}, thick] (-2.2,-2.7) -- (2.2,-2.7);
\node[font=\small] at (0,-3.05) {time};

\end{tikzpicture}
\caption{Illustrative Feynman diagram showing a self-scattering process mediated by vector exchange. External lines represent incoming and outgoing particles, the dots indicate interaction vertices and the wavy internal line denotes the exchanged mediator. By convention, time flows from left to right.}
\label{fig:feynman_legend_example}
\end{figure}

\subsubsection{Feynman diagrams}
\label{sec:feynman_diagrams}

The interaction structures introduced below can be represented graphically by \emph{Feynman diagrams}~\cite{Feynman:1949zx}, which provide an intuitive bookkeeping of how particles propagate and interact. The elementary vertices corresponding to each interaction term are shown in \cref{fig:sidm_vertices}, and combine into complete processes such as the vector-mediated self-scattering example in \cref{fig:feynman_legend_example}. Further realizations of DM self-interaction processes are collected in \cref{fig:selfint_realizations_all}.

The diagrammatic conventions used throughout this work are illustrated in \cref{fig:feynman_legend_example}, where time flows from left to right. External lines denote incoming and outgoing particles, vertices denote interaction terms in the Lagrangian, and internal lines denote mediators exchanged between vertices. Scalar particles (such as $s$ or $\phi$) are drawn as dashed lines, fermions (such as $\chi$) as solid lines with arrows, and vector particles (such as $\chi_\mu$ or $\phi_\mu$) as wavy lines. The arrow on a fermion line distinguishes particles from antiparticles: $\chi$ carries an arrow pointing forward in time, whereas $\bar{\chi}$ carries the opposite arrow.\footnote{An antiparticle can be represented mathematically as a particle propagating backward in time, but this is a formal device rather than literal backward time evolution.} In this way, Feynman diagrams compactly encode the interaction terms and the physical processes generated by the Lagrangian.

\subsubsection{Renormalizable interactions and fundamental mediator models}
\label{sec:renormalizable_interactions}

\begin{figure}[t]
\centering
\begin{subfigure}[t]{0.62\columnwidth}
\centering
\begin{tikzpicture}
\begin{feynman}
  \vertex (a) at (0,0);

  \vertex (s1) at (-1.2, 0.9) {$s$};
  \vertex (s2) at (-1.2,-0.9) {$s$};
  \vertex (s3) at ( 1.2, 0.9) {$s$};
  \vertex (s4) at ( 1.2,-0.9) {$s$};

  \diagram*{
    (s1) -- [scalar] (a),
    (s2) -- [scalar] (a),
    (s3) -- [scalar] (a),
    (s4) -- [scalar] (a),
  };
\end{feynman}

\node[font=\scriptsize, above=3pt] at (a) {$\lambda$};

\end{tikzpicture}
\caption{Scalar self-interaction ($\lambda \, s^4$).}
\label{fig:vertex_s4}
\end{subfigure}

\begin{subfigure}[t]{0.99\columnwidth}
\centering
\begin{tikzpicture}

\begin{scope}[shift={(0,0)}]
\begin{feynman}
  \vertex (b) at (0,0);

  \vertex (s1) at (-1.2, 0.9) {$s$};
  \vertex (s2) at (-1.2,-0.9) {$s$};
  \vertex (phi) at ( 1.2, 0.0) {$\phi$};

  \diagram*{
    (s1) -- [scalar] (b),
    (s2) -- [scalar] (b),
    (phi) -- [scalar] (b),
  };
\end{feynman}
\node[font=\scriptsize, above=3pt] at (b) {$\mu$};
\end{scope}

\begin{scope}[shift={(4.5,0)}]
\begin{feynman}
  \vertex (c) at (0,0);

  \vertex (s1) at (-1.2, 0.9) {$s$};
  \vertex (s2) at (-1.2,-0.9) {$s$};
  \vertex (phi1) at ( 1.2, 0.9) {$\phi$};
  \vertex (phi2) at ( 1.2,-0.9) {$\phi$};

  \diagram*{
    (s1) -- [scalar] (c),
    (s2) -- [scalar] (c),
    (phi1) -- [scalar] (c),
    (phi2) -- [scalar] (c),
  };
\end{feynman}
\node[font=\scriptsize, above=3pt] at (c) {$\kappa $};
\end{scope}

\end{tikzpicture}
\caption{Scalar–mediator interactions ($\mu \, s^2\phi$ and $\kappa\, s^2\phi^2$).}
\label{fig:scalar_vertices}
\end{subfigure}

\vspace{2mm}

\begin{subfigure}[t]{0.49\columnwidth}
\centering
\begin{tikzpicture}
\begin{feynman}
  \vertex (v) at (0,0);

  \vertex (s1) at (-1.2, 0.9) {$s$};
  \vertex (s2) at (-1.2,-0.9) {$s^\dagger$};
  \vertex (phi)  at ( 1.2, 0.0) {$\phi_\mu$};

  \diagram*{
    (s1) -- [scalar] (v),
    (s2) -- [scalar] (v),
    (phi)  -- [photon]  (v),
  };
\end{feynman}

\node[font=\scriptsize, above=3pt] at (v) {$g$};

\end{tikzpicture}
\caption{Scalar--vector interaction
$ig\left(s^\dagger\partial^\mu s
-(\partial^\mu s^\dagger)s\right)\phi_\mu$.}
\label{fig:vertex_scalar_vector}
\end{subfigure}

\vspace{2mm}

\begin{subfigure}[t]{0.49\columnwidth}
\centering
\begin{tikzpicture}
\begin{feynman}
  \vertex (v) at (0,0);

  \vertex (chi1) at (-1.2, 0.9) {$\chi$};
  \vertex (chi2) at (-1.2,-0.9) {$\bar{\chi}$};
  \vertex (phi)  at ( 1.2, 0.0) {$\phi$};

  \diagram*{
    (chi1) -- [fermion] (v),
    (chi2) -- [anti fermion] (v),
    (phi)  -- [scalar]  (v),
  };
\end{feynman}

\node[font=\scriptsize, above=3pt] at (v) {$y$};

\end{tikzpicture}
\caption{Fermion–scalar (Yukawa) interaction ($y\,\bar{\chi}\chi\,\phi$).}
\label{fig:vertex_yukawa}
\end{subfigure}
\hfill
\begin{subfigure}[t]{0.49\columnwidth}
\centering
\begin{tikzpicture}
\begin{feynman}
  \vertex (v) at (0,0);

  \vertex (chi1) at (-1.2, 0.9) {$\chi$};
  \vertex (chi2) at (-1.2,-0.9) {$\bar{\chi}$};
  \vertex (phi)  at ( 1.2, 0.0) {$\phi_\mu$};

  \diagram*{
    (chi1) -- [fermion] (v),
    (chi2) -- [anti fermion] (v),
    (phi)  -- [photon]  (v),
  };
\end{feynman}

\node[font=\scriptsize, above=3pt] at (v) {$g$};

\end{tikzpicture}
\caption{Fermion–vector interaction ($g\,\bar{\chi}\gamma^\mu\chi\,\phi_\mu$).}
\label{fig:vertex_vector}
\end{subfigure}

\caption{Basic renormalizable interactions used to build microscopic SIDM models for both scalar and fermionic DM. }
\label{fig:sidm_vertices}
\end{figure}
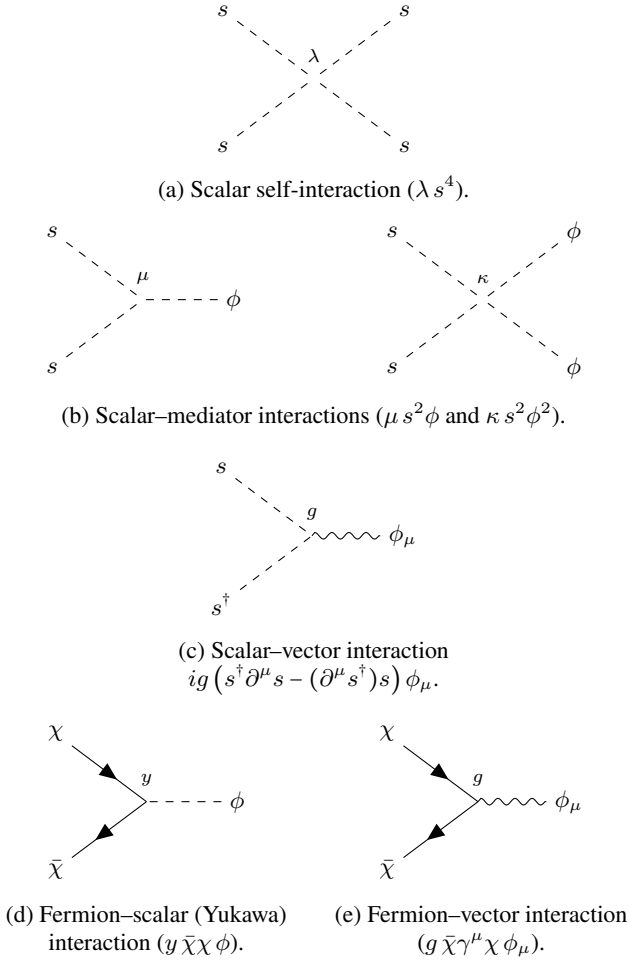

Having fixed the rules for constructing $\mathcal{L}$, we now collect the renormalizable interaction terms available for scalar and fermionic DM, together with the corresponding Feynman vertices shown in \cref{fig:sidm_vertices}.

\medskip
\noindent
\textbf{Scalar dark matter.}
A free scalar field $s$ (spin $0$), i.e. one without interactions, is described by
\begin{equation}
\mathcal{L}_{\rm free} =
\begin{cases}
\frac{1}{2}\,\partial_\mu s\,\partial^\mu s
- \frac{1}{2}\, m_s^2\, s^2 
& \text{(real scalar)} \\[6pt]
\partial_\mu s^\dagger \partial^\mu s
- m_s^2\, s^\dagger s 
& \text{(complex scalar)}
\end{cases}
\end{equation}
in which the first term is the \emph{kinetic term}, built from the field derivatives $\partial_\mu s$ and governing how the field propagates through spacetime, while the second is the \emph{mass term}, quadratic in $s$ and fixing the rest mass $m_s$ of the corresponding particle. A real scalar satisfies $s^\dagger=s$ and carries a single real degree of freedom, hence the normalization factor $1/2$; a complex scalar instead has two real degrees of freedom and is treated together with its conjugate, without this factor. Terms quadratic in the fields determine the propagators (the lines of a Feynman diagram), while terms with three or more fields give the interaction vertices. The relevant self-interactions and mediator couplings are:

\begin{itemize}[leftmargin=*, labelsep=0.5em]

\item \emph{Quartic self-interaction.}
The simplest self-coupling is
\begin{equation}
\label{eq:quartic_coupling}
\mathcal{L}_{\rm int} \supset
\begin{cases}
-\dfrac{\lambda}{4!}\, s^4, 
& \text{(real scalar)} \\[6pt]
-\lambda\, (s^\dagger s)^2, 
& \text{(complex scalar)}
\end{cases}
\end{equation}
where the real-scalar case carries the symmetry factor $1/4!$ and $(s^\dagger s)^2$ is the simplest neutral complex-scalar combination. With dimensionless coupling $\lambda$, these terms drive contact-like elastic scattering, $ss\rightarrow ss$ for real scalar DM (\cref{fig:vertex_s4}) and $s s^\dagger \rightarrow s s^\dagger$ for complex scalar DM.

\item \emph{Scalar mediator couplings.}
Scalar DM may instead interact through the exchange of a scalar mediator $\phi$, via
\begin{equation}
\mathcal{L}_{\rm int} \supset
\begin{cases}
\dfrac{\mu}{2}\, s^2 \phi & \text{(real scalar DM)} \\[6pt]
\mu\, (s^\dagger s)\phi & \text{(complex scalar DM)}
\end{cases}
\label{eq:scalar_mediator_couplings_mu}
\end{equation}
and
\begin{equation}
\mathcal{L}_{\rm int} \supset
\begin{cases}
\dfrac{\kappa}{4}\, s^2 \phi^2 & \text{(real scalar DM)} \\[6pt]
\kappa\, (s^\dagger s)\phi^2 & \text{(complex scalar DM)} 
\end{cases}
\label{eq:scalar_mediator_couplings_kappa}
\end{equation}
shown in \cref{fig:scalar_vertices}. Individually, each such vertex only couples DM to the mediator and does not by itself constitute a self-interaction. Rather, as we will see below, two such vertices can combine, for instance through the exchange of a $\phi$ mediator between two DM particles, to generate DM self-scattering.

\item \emph{Vector mediator couplings.}
If scalar DM carries a dark charge, the covariant derivative generates a coupling to a vector mediator $\phi_\mu$,
\begin{equation}
\label{eq:vector_int}
\mathcal{L}_{\rm int} \supset
ig\,\left(
s^\dagger \partial^\mu s
- (\partial^\mu s^\dagger)s
\right)\phi_\mu .
\end{equation}
Since a real scalar cannot carry a $U(1)$ charge, this minimal coupling exists only for complex scalar DM. This term is illustrated in \cref{fig:vertex_scalar_vector}, where it corresponds to the exchange of a light $\phi_\mu$ mediator.

\end{itemize}

\medskip
\noindent
\textbf{Fermionic dark matter.}
A free fermionic field $\chi$ (spin $1/2$) is governed by
\begin{equation} \label{eq: fermionic free Lagrangian}
\mathcal{L}_{\rm free}
= \bar{\chi} i \gamma^\mu \partial_\mu \chi - m_\chi \bar{\chi} \chi ,
\end{equation}
where the first term is kinetic and the second fixes the rest mass $m_\chi$. Its interactions parallel those of the scalar case:

\begin{itemize}[leftmargin=*, labelsep=0.5em]\item \emph{Scalar (Yukawa) coupling.}
Coupling to a scalar mediator $\phi$ takes the Yukawa form
\begin{equation}
\mathcal{L}_{\rm int} \supset y \, \bar{\chi}\chi \, \phi , \label{eq:yukawa_coupling}
\end{equation}
illustrated in \cref{fig:vertex_yukawa}. This vertex alone does not constitute a self-interaction; rather, two such vertices must combine through the exchange of a $\phi$ between two $\chi$ particles to generate self-scattering.\item \emph{Vector coupling.}
Coupling to a vector mediator $\phi_\mu$ takes the form
\begin{equation} \label{eq:vector_exchange}
\mathcal{L}_{\rm int} \supset g \, \bar{\chi}\gamma^\mu \chi \, \phi_\mu ,
\end{equation}
illustrated in \cref{fig:vertex_vector}. 
\end{itemize} 

As we will discuss in \cref{sec: pp_ints_to_SM_self_interactions}, for light mediators $\phi$ or $\phi_\mu$, the exchange process yields self-interactions characterized by velocity and angle dependence.

\begin{figure}[t]
\centering

\begin{subfigure}[t]{0.48\columnwidth}
\centering
\begin{tikzpicture}
\begin{feynman}
  \vertex (m) at (0,0);
  
  \vertex (phi) at (-1.0, 0.0) {$\phi_\mu$};
  \vertex (gam) at ( 1.0, 0.0) {$\gamma_\mu$};

  \diagram*{
    (phi) -- [photon] (m),
    (m)   -- [photon] (gam),
  };
\end{feynman}

\draw[thick] ($(m)+(-0.10,0.10)$) -- ($(m)+(0.10,-0.10)$);
\draw[thick] ($(m)+(-0.10,-0.10)$) -- ($(m)+(0.10,0.10)$);
\node[below=3pt of m] {\small $\epsilon_\gamma$};
\end{tikzpicture}
\caption{Photon portal}
\label{fig:portal_photon}
\end{subfigure}
\hfill
\begin{subfigure}[t]{0.48\columnwidth}
\centering
\begin{tikzpicture}
\begin{feynman}
  \vertex (m) at (0,0);
  
  \vertex (phi) at (-1.0, 0.0) {$\phi_\mu$};
  \vertex (Z)   at ( 1.0, 0.0) {$Z^\mu$};

  \diagram*{
    (phi) -- [photon] (m),
    (m)   -- [photon] (Z),
  };
\end{feynman}

\draw[thick] ($(m)+(-0.10,0.10)$) -- ($(m)+(0.10,-0.10)$);
\draw[thick] ($(m)+(-0.10,-0.10)$) -- ($(m)+(0.10,0.10)$);
\node[below=3pt of m] {\small $\epsilon_Z$};
\end{tikzpicture}
\caption{$Z$-boson portal}
\label{fig:portal_Z}
\end{subfigure}

\vspace{3mm}

\begin{subfigure}[t]{0.48\columnwidth}
\centering
\begin{tikzpicture}
\begin{feynman}
  \vertex (m) at (0,0);
  
  \vertex (phi) at (-1.0, 0.0) {$\phi$};
  \vertex (h)   at ( 1.0, 0.0) {$h$};

  \diagram*{
    (phi) -- [scalar] (m),
    (m)   -- [scalar] (h),
  };
\end{feynman}

\draw[thick] ($(m)+(-0.10,0.10)$) -- ($(m)+(0.10,-0.10)$);
\draw[thick] ($(m)+(-0.10,-0.10)$) -- ($(m)+(0.10,0.10)$);
\node[below=3pt of m] {\small $\epsilon_{h}$};
\end{tikzpicture}
\caption{Higgs portal (mediator)}
\label{fig:portal_Higgs_mediator}
\end{subfigure}

\vspace{3mm}

\begin{subfigure}[t]{0.48\columnwidth}
\centering
\begin{tikzpicture}
\begin{scope}[shift={(4.5,0)}]
\begin{feynman}
  \vertex (c) at (0,0);

  \vertex (s1) at (-1.2, 0.9) {$s$};
  \vertex (s2) at (-1.2,-0.9) {$s$};
  \vertex (phi1) at ( 1.2, 0.9) {$h$};
  \vertex (phi2) at ( 1.2,-0.9) {$h$};

  \diagram*{
    (s1) -- [scalar] (c),
    (s2) -- [scalar] (c),
    (phi1) -- [scalar] (c),
    (phi2) -- [scalar] (c),
  };
\end{feynman}
\node[font=\scriptsize, above=3pt] at (c) {$\lambda_{sh} $};
\end{scope}

\end{tikzpicture}
\caption{Higgs portal (DM)}
\label{fig:portal_Higgs_DM}
\end{subfigure}

\caption{
Four renormalizable portal interactions connecting the dark sector to the SM: (\emph{a}) kinetic mixing between dark photon $\phi_\mu$ and photon $\gamma_\mu$, (\emph{b}) mass mixing between dark vector $\phi_\mu$ and $Z$ boson, (\emph{c}) scalar mixing between dark scalar mediator $\phi$ and Higgs boson $h$, and (\emph{d}) quartic coupling between scalar DM $s$ and Higgs boson $h$ with coupling constant $\lambda_{sh}$. Portal mixings in (\emph{a})--(\emph{c}) are indicated by crosses. These are allowed renormalizable portals given the symmetries of both sectors.
}
\label{fig:portal_couplings}
\end{figure}
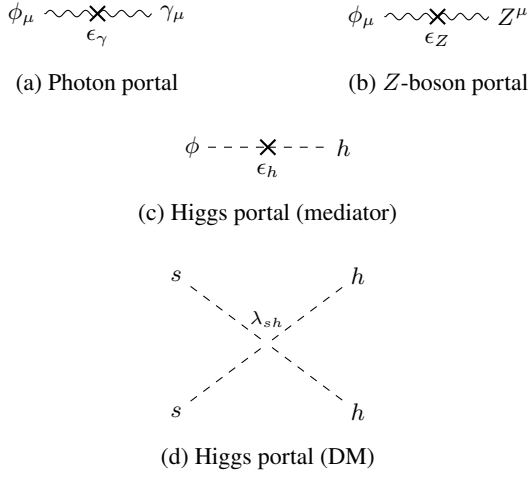

\medskip
\noindent
\textbf{Portal interactions connecting the dark sector to the SM.}
The interactions discussed above act entirely within the dark sector. A connection to the SM can arise through additional \emph{portal interactions}, allowing the dark-sector particles to be produced or detected through SM processes. At the renormalizable level, the possible interactions are strongly constrained by Lorentz and SM gauge symmetries. For the mediator and DM fields considered here, the most relevant portals are:

\begin{itemize}[leftmargin=*, labelsep=0.5em]
\item \emph{Higgs portal.}
A scalar mediator can couple to the SM Higgs sector and thereby mix with the physical Higgs boson after electroweak symmetry breaking. For a real scalar mediator $\phi$, the resulting Higgs--mediator mixing can be parametrized as
\begin{equation}
\mathcal{L}_{\rm portal} \supset \epsilon_h\, m_h^2\, \phi \, h ,
\label{eq:higgs_portal_mediator}
\end{equation}
where $\epsilon_h$ denotes the Higgs--mediator mixing parameter. Scalar DM can similarly couple directly to the Higgs sector. In terms of the physical Higgs field, we write
\begin{equation}
\mathcal{L}_{\rm portal} \supset
\begin{cases}
\dfrac{1}{4}\,\lambda_{sh}\, s^2\, h^2 & \text{(real scalar DM)} \\[6pt]
\lambda_{sh}\, (s^\dagger s)\, h^2 & \text{(complex scalar DM)}
\end{cases}
\label{eq:higgs_portal_dm}
\end{equation}
where $\lambda_{sh}$ denotes the corresponding scalar-DM Higgs-portal coupling~\cite{Patt:2006fw,March-Russell:2008lng,Ahlers:2008qc,Andreas:2008xy,Arina:2010wv,Kahlhoefer:2017umn}.

\item \emph{Vector (photon) portal.}
For a vector mediator $\phi_\mu$, the dark and visible sectors can be connected through kinetic mixing with the SM electromagnetic field,
\begin{equation}
\label{eq:photon_portal}
\mathcal{L}_{\rm portal}
\supset
\frac{\epsilon_\gamma}{2}\,(\partial_\mu\phi_\nu-\partial_\nu\phi_\mu)F^{\mu\nu},
\end{equation}
where $F_{\mu\nu}$ is the electromagnetic field-strength tensor and $\epsilon_\gamma$ parametrizes the kinetic mixing~\cite{Holdom:1985ag,Foot:2004pa,Feldman:2006wd,Pospelov:2008jd,Arkani-Hamed:2008hhe}.

\item \emph{$Z$-boson portal.}
A vector mediator may also couple to the neutral weak
sector through mass mixing with the $Z$ boson,
\begin{equation}
\label{eq:Z_portal}
\mathcal{L}_{\rm portal}
\supset
\epsilon_Z\,m_Z^2\,\phi_\mu Z^\mu,
\end{equation}
where $m_Z$ is the $Z$-boson mass and
$\epsilon_Z$ is a dimensionless parameter
characterizing the mass mixing
~\cite{Babu:1997st,Davoudiasl:2012ag,Davoudiasl:2013aya}.
\end{itemize}

Together, these constitute the renormalizable interactions allowed by the assumed symmetries that connect the dark and visible sectors. In the corresponding Feynman diagrams (\cref{fig:portal_couplings}), the mass and kinetic mixings are indicated by crosses. Through these portals, dark-sector particles acquire small couplings to visible matter: Higgs-portal mixing connects dark scalars to SM particles that obtain their masses from the Higgs field, while kinetic mixing allows a dark photon to interact with electrically charged SM particles. The same couplings can also allow dark-sector particles to decay into visible states, for example $\phi_\mu \to e^+e^-$ or $\phi \to \bar{f}f$, thereby providing direct experimental signatures. Consequently, the portal strengths $\epsilon_\gamma$, $\epsilon_Z$, $\epsilon_h$ and $\lambda_{sh}$ are constrained both by requirements associated with reproducing the observed DM relic abundance (\cref{sec:DM_production_mechanisms}) and by a broad range of experimental, cosmological and astrophysical probes discussed in \cref{sec:DM_search_experiments}.

\begin{figure*}
    \centering    \includegraphics[width=1.5\columnwidth]{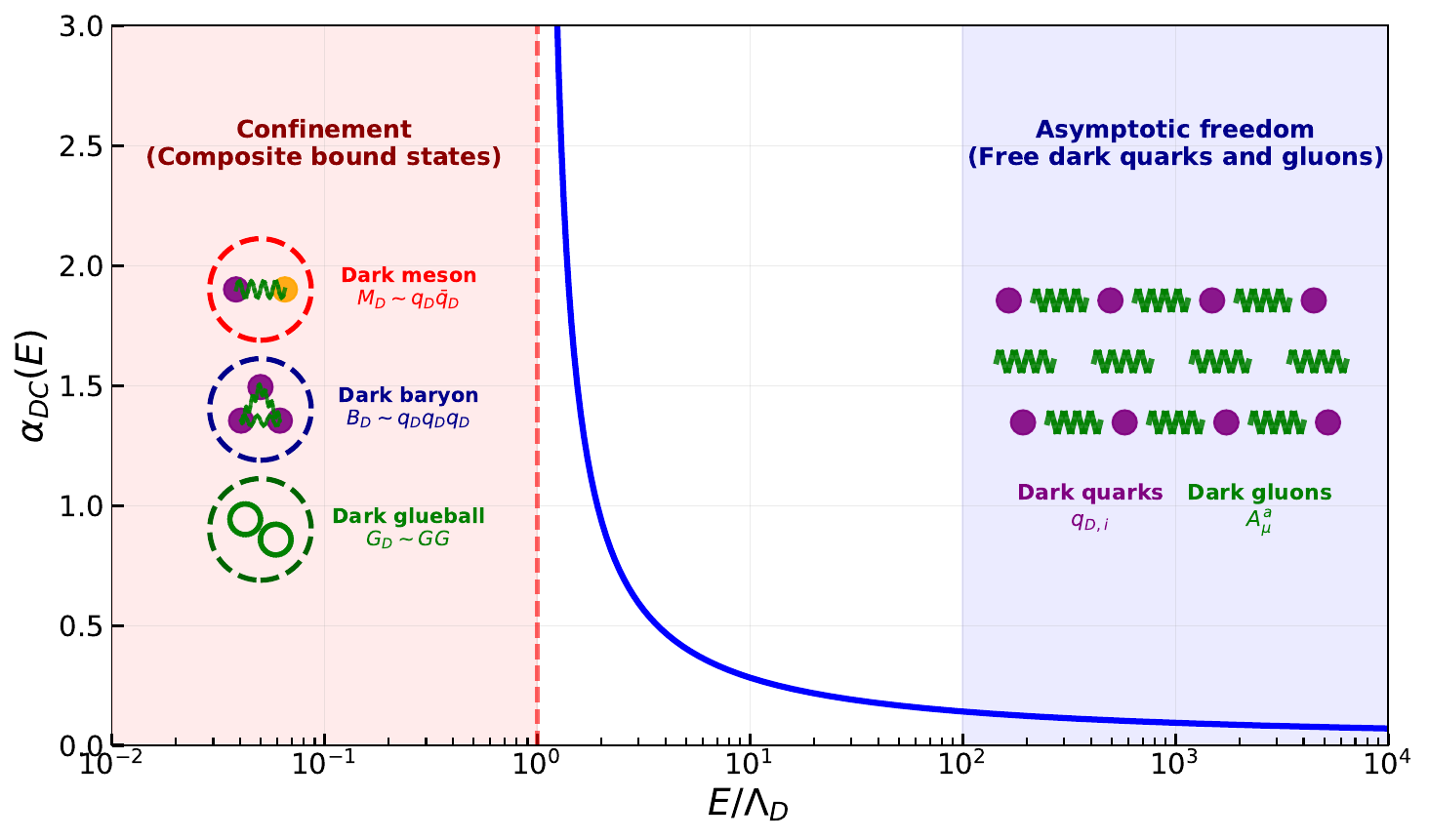}
\caption{
Running of the dark gauge coupling $\alpha_{\rm DC}(E)$ as a function of energy scale $E$ relative to the confinement scale $\Lambda_D$.
For $E \gg \Lambda_D$, dark quarks $q_{D,i}$ and dark gluons $A^a_\mu$ behave as weakly interacting elementary particles.
Near $\Lambda_D$ the coupling becomes strong and the theory confines.
Below this scale the relevant degrees of freedom are composite bound states such as dark mesons ($M_D \sim q_D\bar{q}_D$), dark baryons ($B_D \sim q_Dq_Dq_D$) and dark glueballs ($G_D \sim GG$), in direct analogy with ordinary QCD below $\Lambda_{\rm QCD} \approx 200$ MeV.
}
\label{fig:dark_sector_running_coupling}
\end{figure*}

\subsubsection{Non-perturbative interactions and composite dark sectors}
\label{sec:composite_dark_sectors_non_pert_ints}

The renormalizable interactions above treat DM and its mediators as \emph{elementary} fields -- point-like particles appearing directly in the microscopic Lagrangian. This is a natural starting point, but not the only one. Many familiar particles are instead \emph{composite}, i.e.\ bound states of more fundamental constituents: in the SM, protons and neutrons are baryons made of quarks, and pions and other mesons are quark--antiquark states, all held together by the strong interaction. An analogous possibility arises in the dark sector, where the particle constituting DM may itself be a composite bound state of more fundamental dark-sector constituents, rather than an elementary field~\cite{Kribs:2016cew,Cline:2013zca,Boddy:2014yra,Hochberg:2014dra}.

This possibility is especially natural in theories with a new dark gauge interaction. Just as ordinary quarks interact through the non-Abelian gauge theory of quantum chromodynamics (QCD), dark-sector fermions may interact through their own gauge force, often called \emph{dark QCD}~\cite{Bai:2010qg,Hochberg:2014dra,Kribs:2016cew}. When this force grows strong at low energies, the elementary fields are no longer the natural way to describe the physics. Instead, they bind together into composite, gauge-invariant states. A composite dark sector is therefore not an ad hoc complication, but the expected low-energy behavior of a confining gauge theory, in direct analogy with ordinary hadron physics.

\medskip
\noindent
\textbf{From fundamental constituents to bound states.}
A minimal microscopic realization begins with a set of dark fermions, denoted by $q_{D,i}$, which are often referred to as \emph{dark quarks}. These fields carry a dark gauge charge and interact through a dark gauge force associated with a symmetry group such as\footnote{Here $G_{\rm DC}$ denotes the dark-sector gauge symmetry. The groups $SU(N)$, $SO(N)$ and $Sp(N)$ are continuous Lie groups that determine the allowed interactions among the dark fields and the structure of the resulting bound states. In particular, $SU(N)$ is the closest analogue of the QCD gauge group $SU(3)$, leading to confinement and composite states such as dark mesons and dark baryons, while $SO(N)$ and $Sp(N)$ have different group-theoretic properties and can give rise to qualitatively different spectra.}
\begin{equation}\label{eq:strong_dark_gauge_group}
G_{\rm DC} = SU(N)_{\rm DC}, \qquad SO(N)_{\rm DC}, \qquad Sp(N)_{\rm DC}.
\end{equation}
This construction is directly analogous to QCD, where quarks carry color charge and interact via the strong force. The subscript ``DC'' denotes ``dark color.'' The choice of gauge group determines how the dark quarks interact, how they combine into gauge-invariant composite states, and how the corresponding force carriers -- the dark gauge bosons -- self-interact~\cite{Kribs:2016cew,Hochberg:2014kqa}.

Schematically, the renormalizable dark-sector Lagrangian reads
\begin{align} 
\begin{aligned}
\mathcal{L}_{\rm DC}
=&
-\frac{1}{4} G^a_{\mu\nu} G^{a\,\mu\nu}
+ \sum_i \bar{q}_{D,i}(i\gamma^\mu D_\mu-m_i)q_{D,i}
\\ &+ \mathcal{L}_{\rm dark\text{-}int},
\label{eq:dark_composite_lagrangian}
\end{aligned} 
\end{align}
where $G^a_{\mu\nu}$ is the dark gauge field-strength tensor, $m_i$ the dark-quark masses and $D_\mu=\partial_\mu+i g_{\rm DC} T^a A^a_\mu$ the covariant derivative. The first term describes the propagation and self-interactions of the dark gauge bosons, the second the dark quarks and their coupling to the gauge field, while $\mathcal{L}_{\rm dark\text{-}int}$ may contain portal, Yukawa or scalar couplings, depending on the model. This structure closely mirrors that of QCD.

The crucial feature is that the gauge coupling $g_{\rm DC}$ need not stay small at all energies. In asymptotically free non-Abelian theories it grows towards low energies~\cite{GrossWilczek:1973,Politzer:1973}, and below the confinement scale $\Lambda_D$ the interaction becomes non-perturbative: the elementary fields $q_D$ and $A^a_\mu$ are no longer the correct low-energy degrees of freedom and instead bind into composite states. This is exactly what occurs in QCD, whose low-energy spectrum consists of hadrons rather than free quarks and gluons. A confining dark sector therefore generically predicts a spectrum of composite dark hadrons~\cite{Kribs:2016cew}, as illustrated schematically in \cref{fig:dark_sector_running_coupling}.

\begin{table*}[t]
\centering
\small
\setlength{\tabcolsep}{8pt}
\renewcommand{\arraystretch}{1.25}
\begin{tabular}{llll}
\hline
\textbf{Microscopic ingredient} & \textbf{Symbol / structure} & \textbf{Dark-sector interpretation} & \textbf{Visible-sector analogue} \\
\hline
Dark constituent fermion & $q_{D,i}$ & Dark quark & Quark \\
Dark gauge boson & $A_\mu^a$ & Dark gluon & Gluon \\
Dark gauge group & $SU(N)_{\rm DC},\, SO(N)_{\rm DC},\, Sp(N)_{\rm DC}$ & Dark gauge symmetry & QCD color group \\
\hline
Dark meson & $M_D \sim q_D \bar{q}_D$ & Composite boson & Pion, kaon, $\rho$ meson \\
Dark baryon & $B_D \sim q_D q_D \cdots q_D$ & Composite fermion or boson & Proton, neutron \\
Dark glueball & $\mathcal{O}_{\rm GB} \sim \mathrm{Tr}(G_{\mu\nu}G^{\mu\nu})$ & Pure gauge bound boson & Glueball \\
Dark pion (pNGB) & $\pi_D \sim q_D \bar{q}_D \ \ (\text{pNGB}) $ & Light composite boson & QCD pion \\
Weakly bound state & $(p_D e_D)_{\rm bound}$ & Dark atom / molecule & Hydrogen atom / molecule \\
\hline
\end{tabular}
\caption{Microscopic ingredients and bound states in composite dark-sector models, together with their visible-sector analogues.}
\label{tab:composite_dark_sector_overview}
\end{table*}

\medskip
\noindent
\textbf{Composite spectrum.}
The most common composite states in QCD-like dark sectors are direct analogues of ordinary hadrons:
\begin{itemize}[leftmargin=*, labelsep=0.5em]
\item \emph{Dark mesons}~\cite{Hochberg:2014kqa,Kribs:2016cew},
\begin{equation}
M_D \sim q_D \bar{q}_D ,
\end{equation}
are bosonic bound states of a dark quark and antiquark, analogous to pions, kaons or vector mesons.

\item \emph{Dark baryons}~\cite{Antipin:2015xia,Mitridate:2017oky},
\begin{equation}
B_D \sim q_D q_D \cdots q_D ,
\end{equation}
contain a number of constituents fixed by the gauge group: in an $SU(N)_{\rm DC}$ theory a baryon is built from $N$ dark quarks contracted into a gauge singlet, just as an ordinary $SU(3)$ baryon consists of three quarks. Dark baryons are the analogues of protons and neutrons.

\item \emph{Dark glueballs}~\cite{Boddy:2014yra,Soni:2016gzf} are bound states of the dark force carriers alone, rather than of matter particles. Associated with operators such as
\begin{equation}
\mathcal{O}_{\rm GB} \sim \mathrm{Tr}(G_{\mu\nu}G^{\mu\nu}),
\end{equation}
they are ``self-bound'' states of the dark force, analogous to the glueballs predicted in ordinary QCD.
\end{itemize}

Even a microscopically simple theory can thus generate a rich low-energy spectrum -- a general feature of strong dynamics, whereby a few fundamental fields are reorganized into many composite states.

\medskip
\noindent
\textbf{Symmetries, mass scales and stability.}
The spectrum of a confining dark sector is determined not only by the gauge group in \cref{eq:strong_dark_gauge_group}, but also by the fermion representations, the flavor content---that is, the number and species of dark fermions with otherwise identical gauge quantum numbers---and the hierarchy between the constituent masses and the confinement scale. Of particular interest is the chiral regime, $m_i\ll\Lambda_D$, where the dark-quark masses are small compared with the scale of the strong dynamics. The theory then possesses an approximate global flavor symmetry that is spontaneously broken by the formation of a condensate,
\begin{equation}
\langle \bar{q}_D q_D \rangle \neq 0.
\end{equation}
For $N_F$ Dirac flavors, the symmetry-breaking pattern depends on whether the fermion representation under the confining gauge group is complex, real or pseudoreal:
\begin{equation}
\label{eq:global_symmetry_breaking}
\begin{aligned}
\text{complex:}\quad & SU(N_F)_L\times SU(N_F)_R \rightarrow SU(N_F)_V,\\ \
\text{real:}\quad & SU(2N_F)\rightarrow SO(2N_F),\ \\
\text{pseudoreal:}\quad & SU(2N_F)\rightarrow Sp(2N_F).
\end{aligned}
\end{equation}
The broken generators give rise to Nambu--Goldstone modes, which become light pseudo-Nambu--Goldstone bosons once the dark-quark masses and other explicit symmetry-breaking interactions are included. These states are the dark-sector analogues of the QCD pions and are commonly referred to as \emph{dark pions}~\cite{Hochberg:2014dra,Hochberg:2014kqa}.

The hierarchy between $m_i$ and $\Lambda_D$ also determines the composite mass spectrum. In the chiral regime, generic non-Goldstone hadrons have masses set by the confinement scale, $m_{\rm comp}\sim\Lambda_D$, whereas the pseudo-Nambu--Goldstone states can be parametrically lighter because their masses are controlled by explicit chiral-symmetry breaking. In the opposite regime, $m_i\gg\Lambda_D$, the masses of constituent bound states are instead dominated by the constituent masses, $m_{\rm comp}\simeq\sum_i m_i+E_{\rm bind}$. Composite dark sectors can therefore span qualitatively different spectra, from light dark pions accompanied by confinement-scale resonances to heavy constituent-dominated bound states.

The same symmetries can also determine which composite states are stable. A conserved dark baryon number, for example, renders the lightest state carrying this charge stable, providing a natural DM candidate~\cite{Antipin:2015xia,Mitridate:2017oky}. Dark mesons or pions may likewise be stable or long-lived when protected by exact or approximate flavor symmetries, by discrete symmetries that survive the symmetry breaking in \cref{eq:global_symmetry_breaking}, or when their decays are kinematically forbidden. The symmetry structure of the confining theory therefore determines both the low-energy spectrum and which of its composite states can constitute DM.

\medskip
\noindent
\textbf{Resolving compositeness and effective interactions.}
Composite particles do not always reveal their internal structure: whether they behave as extended or point-like objects depends on the momentum transfer of the interaction. If the process probes length scales much larger than the bound-state size -- equivalently, momentum transfers much smaller than its inverse, $q \ll \Lambda_{\rm comp} \sim R^{-1}$ -- the composite state behaves as point-like.\footnote{In QCD-like dark sectors the compositeness scale is set by confinement, $\Lambda_{\rm comp} \sim \Lambda_D$. For weakly bound states such as dark atoms or molecules these scales can differ.} Its interactions are then captured by the same renormalizable operators introduced earlier (cf.\ \cref{eq:renomalizable_terms}), now with fields that may represent either elementary or composite degrees of freedom. The renormalizable Lagrangian thus provides a universal low-energy description, independent of the microscopic origin of the particles.

At higher momentum transfer, $q \sim \Lambda_{\rm comp}$, the internal structure begins to be resolved, and the point-particle description must be supplemented by finite-size effects such as form factors or higher-derivative interactions. These are captured systematically by an effective-field-theory expansion~\cite{Georgi:1993mps},
\begin{equation} \label{eq:effective_Lagrangian_expansion}
\mathcal{L}_{\rm eff}
=
\mathcal{L}_{\rm ren}
+
\sum_{n>4}
\frac{C_n}{\Lambda_{\rm comp}^{\,n-4}}\,\mathcal{O}_n ,
\end{equation}
where $\mathcal{L}_{\rm ren}$ collects the renormalizable terms of mass dimension $\leq 4$ (cf.\ \cref{eq:renomalizable_terms}). The operators $\mathcal{O}_n$ involve extra fields and/or derivatives -- probing shorter distances inside the composite object -- while the coefficients $C_n$ encode their strength and depend on the underlying theory. Their suppression by powers of $\Lambda_{\rm comp}$ keeps these corrections small for $q \ll \Lambda_{\rm comp}$; as $q$ approaches $\Lambda_{\rm comp}$ they grow in importance, signalling the onset of compositeness, and for $q \gg \Lambda_{\rm comp}$ the effective description breaks down and must be replaced by the microscopic theory.

The appearance of operators with $[\mathcal{L}] > 4$ here does \emph{not} signal a breakdown of the fundamental theory; it merely reflects that the low-energy description is written in terms of composite degrees of freedom. Even a renormalizable microscopic theory naturally generates such higher-dimensional operators in its low-energy effective description, encoding the internal structure of its bound states. Composite dark sectors thus concretely illustrate how a simple renormalizable theory can produce a hierarchy of effective interactions, with point-like behavior emerging as an approximation and deviations controlled by $\Lambda_{\rm comp}$.

\medskip
\noindent
\textbf{Weakly bound composite states.}
Composite dark sectors need not be strongly coupled in the QCD-like sense; even perturbative interactions can form bound states when attractive forces are present. The simplest example is a \emph{dark atom}~\cite{Kaplan:2009de,CyrRacine:2012fz,Cline:2013pca}, formed from a dark proton $p_D$ and dark electron $e_D$ of opposite charge under a dark $U(1)_D$ force mediated by a dark vector $\phi_\mu$. A schematic Lagrangian is
\begin{equation}
\mathcal{L}_{D}
=
\bar{p}_D(i\gamma^\mu D_\mu+m_{p_D})p_D
+
\bar{e}_D(i\gamma^\mu D_\mu+m_{e_D})e_D,
\end{equation}
with covariant derivative
\begin{equation}
D_\mu = \partial_\mu + i g Q_D \phi_\mu ,
\end{equation}
where the constituents carry opposite dark charges $Q_D=\pm 1$. This is the dark analogue of the proton--electron system in atomic physics: the composite object is weakly bound rather than hadronic, but the underlying idea is the same -- the observed DM may be a bound state rather than an elementary field. More elaborate possibilities include dark molecules and dark chemistry~\cite{Cline:2013pca,Gurian:2021qhk}.

\medskip
In summary, composite dark sectors provide a broad class of well-motivated extensions of elementary SIDM. Renormalizable gauge theories with fundamental dark constituents can generate a rich low-energy spectrum of dark mesons, baryons, glueballs and pions, as well as more weakly bound states such as dark atoms and molecules (see \cref{tab:composite_dark_sector_overview}). The resulting spectrum and interaction structure are controlled by the gauge symmetry, matter content, constituent masses and confinement or binding scale. At momentum transfers well below the compositeness scale, these states can often be treated as effective point-like degrees of freedom, whereas at higher momentum transfer their internal structure becomes relevant through form factors and higher-dimensional interactions. Composite dark sectors therefore connect simple microscopic gauge theories to a much richer range of effective DM self-interactions relevant for structure formation.

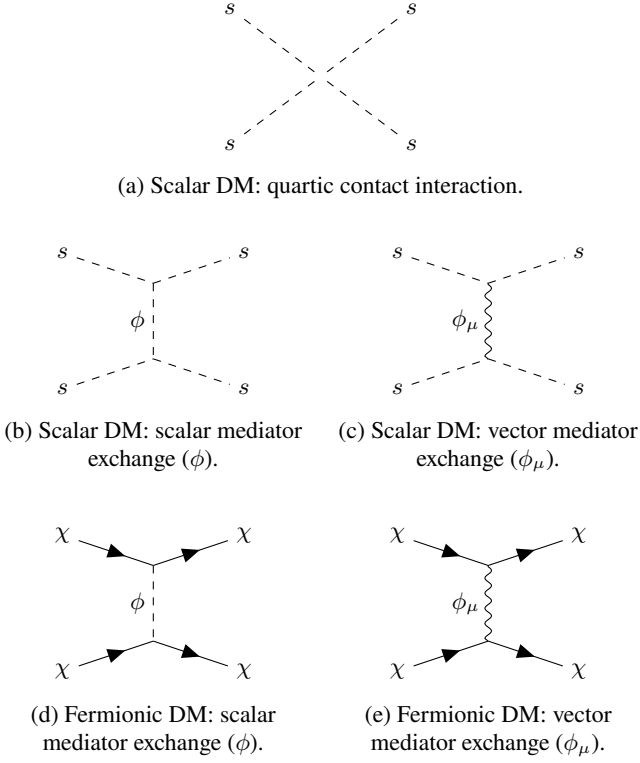
\begin{figure}[t]
\centering

\begin{subfigure}[t]{0.62\columnwidth}
\centering
\begin{tikzpicture}
\begin{feynman}
  \vertex (a) at (0,0);

  \vertex (s1) at (-1.2, 0.9) {$s$};
  \vertex (s2) at (-1.2,-0.9) {$s$};
  \vertex (s3) at ( 1.2, 0.9) {$s$};
  \vertex (s4) at ( 1.2,-0.9) {$s$};

  \diagram*{
    (s1) -- [scalar] (a),
    (s2) -- [scalar] (a),
    (s3) -- [scalar] (a),
    (s4) -- [scalar] (a),
  };
\end{feynman}
\end{tikzpicture}
\caption{Scalar DM: quartic contact interaction.}
\label{fig:selfint_scalar_contact}
\end{subfigure}

\vspace{4mm}

\begin{subfigure}[t]{0.49\columnwidth}
\centering
\begin{tikzpicture}
\begin{feynman}
  \vertex (a) at (0,0.5);
  \vertex (b) at (0,-0.5);

  \vertex (i1) at (-1.2, 0.9) {$s$};
  \vertex (i2) at (-1.2,-0.9) {$s$};
  \vertex (f1) at ( 1.2, 0.9) {$s$};
  \vertex (f2) at ( 1.2,-0.9) {$s$};

  \diagram*{
    (i1) -- [scalar] (a) -- [scalar] (f1),
    (i2) -- [scalar] (b) -- [scalar] (f2),
    (a)  -- [scalar, edge label'=$\phi$] (b),
  };
\end{feynman}
\end{tikzpicture}
\caption{Scalar DM: scalar mediator exchange ($\phi$).}
\label{fig:selfint_scalar_phi}
\end{subfigure}
\hfill
\begin{subfigure}[t]{0.49\columnwidth}
\centering
\begin{tikzpicture}
\begin{feynman}
  \vertex (a) at (0,0.5);
  \vertex (b) at (0,-0.5);

  \vertex (i1) at (-1.2, 0.9) {$s$};
  \vertex (i2) at (-1.2,-0.9) {$s$};
  \vertex (f1) at ( 1.2, 0.9) {$s$};
  \vertex (f2) at ( 1.2,-0.9) {$s$};

  \diagram*{
    (i1) -- [scalar] (a) -- [scalar] (f1),
    (i2) -- [scalar] (b) -- [scalar] (f2),
    (a)  -- [photon, edge label'=$\phi_\mu$] (b),
  };
\end{feynman}
\end{tikzpicture}
\caption{Scalar DM: vector mediator exchange ($\phi_\mu$).}
\label{fig:selfint_scalar_vec}
\end{subfigure}

\vspace{5mm}

\begin{subfigure}[t]{0.49\columnwidth}
\centering
\begin{tikzpicture}
\begin{feynman}
  \vertex (a) at (0,0.5);
  \vertex (b) at (0,-0.5);

  \vertex (i1) at (-1.2, 0.9) {$\chi$};
  \vertex (i2) at (-1.2,-0.9) {$\chi$};
  \vertex (f1) at ( 1.2, 0.9) {$\chi$};
  \vertex (f2) at ( 1.2,-0.9) {$\chi$};

  \diagram*{
    (i1) -- [fermion] (a) -- [fermion] (f1),
    (i2) -- [fermion] (b) -- [fermion] (f2),
    (a)  -- [scalar, edge label'=$\phi$] (b),
  };
\end{feynman}
\end{tikzpicture}
\caption{Fermionic DM: scalar mediator exchange ($\phi$).}
\label{fig:selfint_fermion_phi}
\end{subfigure}
\hfill
\begin{subfigure}[t]{0.49\columnwidth}
\centering
\begin{tikzpicture}
\begin{feynman}
  \vertex (a) at (0,0.5);
  \vertex (b) at (0,-0.5);

  \vertex (i1) at (-1.2, 0.9) {$\chi$};
  \vertex (i2) at (-1.2,-0.9) {$\chi$};
  \vertex (f1) at ( 1.2, 0.9) {$\chi$};
  \vertex (f2) at ( 1.2,-0.9) {$\chi$};

  \diagram*{
    (i1) -- [fermion] (a) -- [fermion] (f1),
    (i2) -- [fermion] (b) -- [fermion] (f2),
    (a)  -- [photon, edge label'=$\phi_\mu$] (b),
  };
\end{feynman}
\end{tikzpicture}
\caption{Fermionic DM: vector mediator exchange ($\phi_\mu$).}
\label{fig:selfint_fermion_vec}
\end{subfigure}

\caption{Representative microscopic realizations generating DM self-interactions for scalar DM ($s$) and fermionic DM ($\chi$): direct contact interactions and self-scattering induced by scalar or vector mediator exchange.}
\label{fig:selfint_realizations_all}
\end{figure}

\subsubsection{From microscopic interactions to DM self-interactions}
\label{sec: pp_ints_to_SM_self_interactions}

The renormalizable interaction terms introduced in \cref{fig:sidm_vertices}
determine how DM particles interact at the microscopic level.
From these couplings, one can construct the self-scattering processes relevant for astrophysical systems.
Representative examples are shown in \cref{fig:selfint_realizations_all}.

\medskip
\noindent
\textbf{Elementary contact interactions and mediator exchange.}
For scalar DM $s$, self-interactions may arise in three distinct ways.
First, a direct quartic term $s^4$ generates a contact interaction in which four scalar fields meet at a single point,
as shown in \cref{fig:selfint_scalar_contact}.
This is the simplest realization of scalar self-scattering, corresponding to a local process $ss \rightarrow ss$ without mediator exchange.
Second, if the scalar field couples to another scalar mediator $\phi$ through a term such as $s^2\phi$
(cf.~\cref{fig:scalar_vertices}), self-scattering can occur through $\phi$ exchange, as illustrated in \cref{fig:selfint_scalar_phi}.
The interaction is then mediated by the propagation of $\phi$ rather than being point-like.
If the mediator is light, the interaction can become long-range and velocity-dependent.
Third, if the scalar couples to a vector mediator $\phi_\mu$
(cf.~\cref{fig:vertex_scalar_vector}), exchange of this vector field produces the process shown in \cref{fig:selfint_scalar_vec}.
As in the scalar-mediator case, the scattering properties depend on the mediator mass and coupling structure.

For fermionic DM $\chi$, there is no renormalizable four-fermion contact interaction at dimension four, i.e.\ no fundamental $\chi^4$-type term in the Lagrangian.
Self-scattering therefore arises through mediator exchange.
A scalar mediator coupled via $\bar{\chi}\chi\,\phi$
(cf.~\cref{fig:vertex_yukawa}) generates the diagram shown in \cref{fig:selfint_fermion_phi},
while a vector mediator coupled through $\bar{\chi}\gamma^\mu\chi\,\phi_\mu$
(cf.~\cref{fig:vertex_vector}) produces the process in \cref{fig:selfint_fermion_vec}.
As in the scalar case, the angular and velocity dependence of the cross section is controlled by the mediator mass.

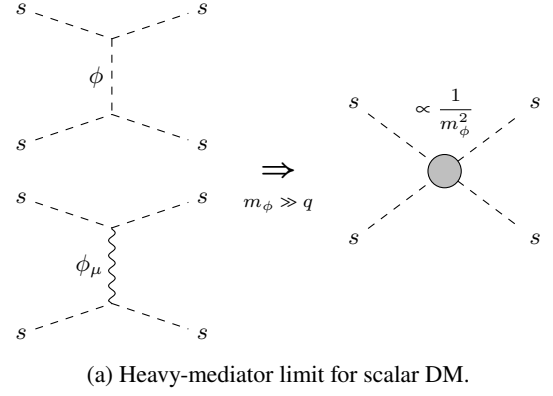
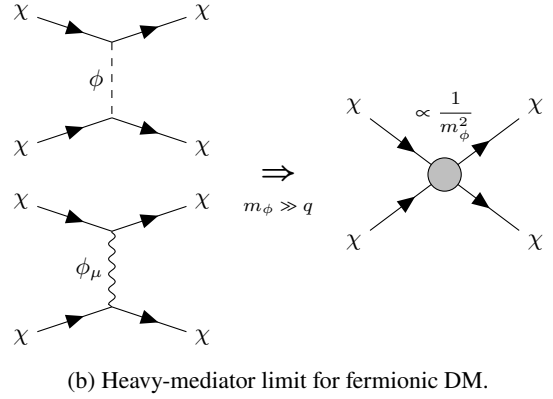
\begin{figure}[t]
\centering

\begin{subfigure}[t]{0.99\columnwidth}
\centering
\begin{tikzpicture}

\begin{scope}[shift={(0,1.25)}]
\begin{feynman}
  \vertex (a) at (0,0.5);
  \vertex (b) at (0,-0.5);

  \vertex (i1) at (-1.2, 0.9) {$s$};
  \vertex (i2) at (-1.2,-0.9) {$s$};
  \vertex (f1) at ( 1.2, 0.9) {$s$};
  \vertex (f2) at ( 1.2,-0.9) {$s$};

  \diagram*{
    (i1) -- [scalar] (a) -- [scalar] (f1),
    (i2) -- [scalar] (b) -- [scalar] (f2),
    (a)  -- [scalar, edge label'=$\phi$] (b),
  };
\end{feynman}
\end{scope}

\begin{scope}[shift={(0,-1.25)}]
\begin{feynman}
  \vertex (a) at (0,0.5);
  \vertex (b) at (0,-0.5);

  \vertex (i1) at (-1.2, 0.9) {$s$};
  \vertex (i2) at (-1.2,-0.9) {$s$};
  \vertex (f1) at ( 1.2, 0.9) {$s$};
  \vertex (f2) at ( 1.2,-0.9) {$s$};

  \diagram*{
    (i1) -- [scalar] (a) -- [scalar] (f1),
    (i2) -- [scalar] (b) -- [scalar] (f2),
    (a)  -- [photon, edge label'=$\phi_\mu$] (b),
  };
\end{feynman}
\end{scope}

\node at (2.2,0) {\Large $\Rightarrow$};
\node[font=\scriptsize] at (2.2,-0.45) {$m_\phi \gg q$};

\begin{scope}[shift={(4.4,0)}]
\begin{feynman}
  \vertex (c) at (0,0);

  \vertex (i1) at (-1.2, 0.9) {$s$};
  \vertex (i2) at (-1.2,-0.9) {$s$};
  \vertex (f1) at ( 1.2, 0.9) {$s$};
  \vertex (f2) at ( 1.2,-0.9) {$s$};

  \diagram*{
    (i1) -- [scalar] (c) -- [scalar] (f1),
    (i2) -- [scalar] (c) -- [scalar] (f2),
  };
\end{feynman}

\fill[black!25] (0,0) circle (0.22);
\draw (0,0) circle (0.22);
\node[font=\scriptsize, above=3pt] at (0,0.22) {$\displaystyle \propto \frac{1}{m_\phi^{2}}$};
\end{scope}

\end{tikzpicture}
\caption{Heavy-mediator limit for scalar DM.}
\label{fig:heavy_limit_scalar}
\end{subfigure}

\vspace{5mm}

\begin{subfigure}[t]{0.99\columnwidth}
\centering
\begin{tikzpicture}

\begin{scope}[shift={(0,1.25)}]
\begin{feynman}
  \vertex (a) at (0,0.5);
  \vertex (b) at (0,-0.5);

  \vertex (i1) at (-1.2, 0.9) {$\chi$};
  \vertex (i2) at (-1.2,-0.9) {$\chi$};
  \vertex (f1) at ( 1.2, 0.9) {$\chi$};
  \vertex (f2) at ( 1.2,-0.9) {$\chi$};

  \diagram*{
    (i1) -- [fermion] (a) -- [fermion] (f1),
    (i2) -- [fermion] (b) -- [fermion] (f2),
    (a)  -- [scalar, edge label'=$\phi$] (b),
  };
\end{feynman}
\end{scope}

\begin{scope}[shift={(0,-1.25)}]
\begin{feynman}
  \vertex (a) at (0,0.5);
  \vertex (b) at (0,-0.5);

  \vertex (i1) at (-1.2, 0.9) {$\chi$};
  \vertex (i2) at (-1.2,-0.9) {$\chi$};
  \vertex (f1) at ( 1.2, 0.9) {$\chi$};
  \vertex (f2) at ( 1.2,-0.9) {$\chi$};

  \diagram*{
    (i1) -- [fermion] (a) -- [fermion] (f1),
    (i2) -- [fermion] (b) -- [fermion] (f2),
    (a)  -- [photon, edge label'=$\phi_\mu$] (b),
  };
\end{feynman}
\end{scope}

\node at (2.2,0) {\Large $\Rightarrow$};
\node[font=\scriptsize] at (2.2,-0.45) {$m_\phi \gg q$};

\begin{scope}[shift={(4.4,0)}]
\begin{feynman}
  \vertex (c) at (0,0);

  \vertex (i1) at (-1.2, 0.9) {$\chi$};
  \vertex (i2) at (-1.2,-0.9) {$\chi$};
  \vertex (f1) at ( 1.2, 0.9) {$\chi$};
  \vertex (f2) at ( 1.2,-0.9) {$\chi$};

  \diagram*{
    (i1) -- [fermion] (c) -- [fermion] (f1),
    (i2) -- [fermion] (c) -- [fermion] (f2),
  };
\end{feynman}

\fill[black!25] (0,0) circle (0.22);
\draw (0,0) circle (0.22);
\node[font=\scriptsize, above=3pt] at (0,0.22) {$\displaystyle \propto \frac{1}{m_\phi^{2}}$};
\end{scope}

\end{tikzpicture}
\caption{Heavy-mediator limit for fermionic DM.}
\label{fig:heavy_limit_fermion}
\end{subfigure}

\caption{Heavy-mediator limit. When the mediator mass is much larger than the typical momentum transfer, $m_\phi \gg q$, mediator exchange cannot be resolved dynamically and reduces to an effective contact interaction. The same logic applies to both scalar and fermionic DM.}
\label{fig:heavy_mediator_limit_both}
\end{figure}

\medskip
\noindent
\textbf{Heavy-mediator limit and effective contact interactions.}
Renormalizability restricts the fundamental UV interaction terms to operators of mass dimension four or less, but the physics relevant for DM haloes usually occurs at energies far below the mass scale of possible mediators.
In this regime, an effective-field-theory description is often more appropriate.
As illustrated in \cref{fig:heavy_limit_scalar,fig:heavy_limit_fermion}, if a mediator with mass $m_\phi$ is much heavier than the typical momentum transfer in haloes, $m_\phi^2 \gg q^2$, it cannot be resolved dynamically and may be integrated out.
The exchange diagrams on the left-hand side of the two subfigures then reduce to local contact interactions, represented by the four-point interactions on the right-hand side.
The resulting low-energy theory contains higher-dimensional operators encoding the heavy mediator in a point-like manner.

For scalar DM (cf.~\cref{fig:heavy_limit_scalar}), if the scalar couples to a heavy mediator through
$\frac{\mu}{2} (s^\dagger s)\phi$, integrating out $\phi$ generates an effective quartic interaction
\begin{equation} \label{eq:four_scalar_int}
\frac{\mu^2}{m_\phi^2}\, (s^\dagger s)^2 ,
\end{equation}
which induces local scattering $ss^\dagger \rightarrow ss^\dagger$.
Similarly, a heavy vector mediator produces an effective short-range interaction suppressed by $1/m_\phi^2$.
In all cases, the heavy mediator appears as a contact interaction at low energies.

For fermionic DM (cf.~\cref{fig:heavy_limit_fermion}), integrating out a heavy mediator for $m_\phi^2 \gg q^2$ yields local four-fermion interactions of the schematic form
\begin{equation} \label{eq:four_fermion_int}
\frac{g^2}{m_\phi^2}
(\bar{\chi}\gamma^\mu \chi)(\bar{\chi}\gamma_\mu \chi),
\end{equation}
or, for a scalar mediator,
\begin{equation} \label{eq:four_fermion_int2}
\frac{y^2}{m_\phi^2}
(\bar{\chi}\chi)(\bar{\chi}\chi).
\end{equation}
Both operators describe point-like self-scattering, $\chi\chi \rightarrow \chi\chi$, valid at momentum transfers much smaller than the mediator mass.

Such effective operators provide a controlled parametrization of short-range SIDM.
At halo velocities, they typically generate approximately isotropic and only weakly velocity-dependent cross sections.
The effective field theory (EFT) remains valid as long as $q^2 \ll m_\phi^2$.
Deviations at higher energies signal the breakdown of the EFT and the onset of the full mediator dynamics shown on the left-hand side of \cref{fig:heavy_limit_scalar,fig:heavy_limit_fermion}.

Together, \cref{fig:selfint_realizations_all} and \cref{fig:heavy_mediator_limit_both}
show how a small set of renormalizable interaction terms gives rise to different microscopic realizations of self-interacting DM.
Whether the resulting interaction is short-range and approximately isotropic, or long-range and forward-peaked, is primarily determined by the mediator mass and the underlying particle-physics structure.

\medskip
\noindent
\textbf{Non-perturbative and resonant self-interactions.}
The examples above describe SIDM through contact interactions or exchange of elementary mediators.
Strong velocity-dependent self-interactions can also arise without a light mediator, for example from non-perturbative dynamics or near-threshold resonances in the two-body scattering amplitude.

Resonant enhancement is a generic feature of two-body dynamics and can occur in theories with elementary DM as well as in composite dark sectors.
If the two-body system admits a quasi-bound state or resonance close to threshold\footnote{A quasi-bound state is a temporary state in which two particles stay close together for a short time during a collision, before separating again. It behaves like a short-lived bound state and appears as a resonance in the scattering process.}, the scattering amplitude is enhanced at low relative velocity, producing a pronounced velocity dependence.
This mechanism is familiar from hadron, nuclear and atomic physics, where it is described by the Breit--Wigner formalism \cite{Breit:1936zzb}, and has been emphasized in SIDM as a model-independent origin of velocity-dependent scattering \cite{Tulin:2013teo,Braaten:2013tza,Chu:2018fzy,Chu:2019awd}.
In particular, resonant SIDM (RSIDM) provides a framework in which velocity-dependent cross sections arise without requiring a light mediator \cite{Chu:2018fzy}.
Strongly interacting composite dark matter offers an independent realization, where the relevant dynamics is governed by bound-state structure, residual forces and near-threshold states rather than elementary mediator exchange \cite{Cline:2013zca}.

At low energies, such systems are naturally described by an effective theory of two-body scattering.
Independently of the microscopic origin, the leading interactions can be expressed through local operators such as
\begin{equation}
\mathcal{L}_{\rm eff} \supset 
C_4 (s^\dagger s)^2
+ \frac{C_6}{\Lambda^2} 
(\partial_\mu s^\dagger \partial^\mu s)(s^\dagger s),
\end{equation}
for scalar DM, and
\begin{equation}
\mathcal{L}_{\rm eff} \supset 
\frac{C_6}{\Lambda^2} (\bar{\chi}\chi)(\bar{\chi}\chi)
+ \frac{C_6'}{\Lambda^2} (\bar{\chi}\gamma^\mu\chi)(\bar{\chi}\gamma_\mu\chi),
\end{equation}
for fermionic DM.
These are the leading terms in the expansion of \cref{eq:effective_Lagrangian_expansion}.
Here $\Lambda$ denotes the high-energy scale at which the microscopic physics is resolved.
Depending on the scenario, it may correspond to the mass of a heavy mediator, $\Lambda \sim m_{\phi},m_{\phi_\mu}$, as in \cref{eq:four_scalar_int,eq:four_fermion_int,eq:four_fermion_int2}; the compositeness scale of a bound state, $\Lambda \sim \Lambda_D$; or the cutoff scale of the effective theory.

While these operators describe point-like scattering at low energies, $q \ll \Lambda$, their coefficients $C_n$ encode non-trivial physics such as bound states, resonances and strong dynamics.
As $q$ approaches $\Lambda$, higher-dimensional operators become increasingly important, signalling the breakdown of the point-particle approximation and the onset of the microscopic description.
Near a resonance, the effective description can be reorganized in terms of a small number of scattering parameters, such as the scattering length and effective range.
This effective-range expansion provides a universal parametrization of low-energy scattering largely independent of the underlying particle-physics model.

\begin{figure}[t]
\centering

\begin{subfigure}[t]{0.48\columnwidth}
\centering
\begin{tikzpicture}
\begin{feynman}
  \vertex (c) at (0,0);

  \vertex (i1) at (-1.5, 0.9) {$M_D$};
  \vertex (i2) at (-1.5,-0.9) {$M_D$};
  \vertex (f1) at ( 1.5, 0.9) {$M_D$};
  \vertex (f2) at ( 1.5,-0.9) {$M_D$};

  \diagram*{
    (i1) -- [scalar] (c) -- [scalar] (f1),
    (i2) -- [scalar] (c) -- [scalar] (f2),
  };

  \draw[fill=white] (c) circle (0.30);
  \node at (c) {\small $\Lambda_D$};
\end{feynman}
\end{tikzpicture}
\caption{Dark meson scattering via short-range composite dynamics.}
\label{fig:selfint_composite_mesons}
\end{subfigure}
\hfill
\begin{subfigure}[t]{0.48\columnwidth}
\centering
\begin{tikzpicture}
\begin{feynman}
  \vertex (a) at (0,0.6);
  \vertex (b) at (0,-0.6);

  \vertex (i1) at (-1.5, 0.9) {$B_D$};
  \vertex (i2) at (-1.5,-0.9) {$B_D$};
  \vertex (f1) at ( 1.5, 0.9) {$B_D$};
  \vertex (f2) at ( 1.5,-0.9) {$B_D$};

  \diagram*{
    (i1) -- [fermion] (a) -- [fermion] (f1),
    (i2) -- [fermion] (b) -- [fermion] (f2),
    (a) -- [scalar, edge label'=\small $\pi_D$] (b),
  };
\end{feynman}
\end{tikzpicture}
\caption{Dark baryon scattering via residual forces, e.g.\ dark-pion exchange.}
\label{fig:selfint_composite_baryons}
\end{subfigure}

\vspace{4mm}

\begin{subfigure}[t]{0.65\columnwidth}
\centering
\begin{tikzpicture}
\begin{feynman}
  \vertex (c) at (0,0);

  \vertex (i1) at (-1.6, 0.9) {$X_D$};
  \vertex (i2) at (-1.6,-0.9) {$X_D$};
  \vertex (f1) at ( 1.6, 0.9) {$X_D$};
  \vertex (f2) at ( 1.6,-0.9) {$X_D$};

  \diagram*{
    (i1) -- [scalar] (c) -- [scalar] (f1),
    (i2) -- [scalar] (c) -- [scalar] (f2),
  };

  \draw[fill=white] (c) circle (0.35);
  \node at (c) {$R_D$};
\end{feynman}
\end{tikzpicture}
\caption{Resonantly enhanced scattering via a near-threshold composite state $R_D$.}
\label{fig:selfint_composite_resonance}
\end{subfigure}

\caption{
Representative self-interaction channels in composite dark sectors.
Here $M_D$ denotes a generic dark meson, $B_D$ a dark baryon and $X_D$ a generic composite DM state.
Unlike the elementary-mediator realizations in \cref{fig:selfint_realizations_all}, these interactions arise from non-perturbative composite dynamics.
Panel (a) shows short-range dark-meson scattering generated by strong dynamics at the confinement/compositeness scale.
Panel (b) shows dark-baryon scattering induced by residual forces between composite states, analogous to pion-mediated nuclear interactions.
Panel (c) shows resonantly enhanced scattering through a near-threshold composite state $R_D$, which can generate strong velocity dependence even without a light elementary mediator.
}
\label{fig:selfint_composite_realizations}
\end{figure}
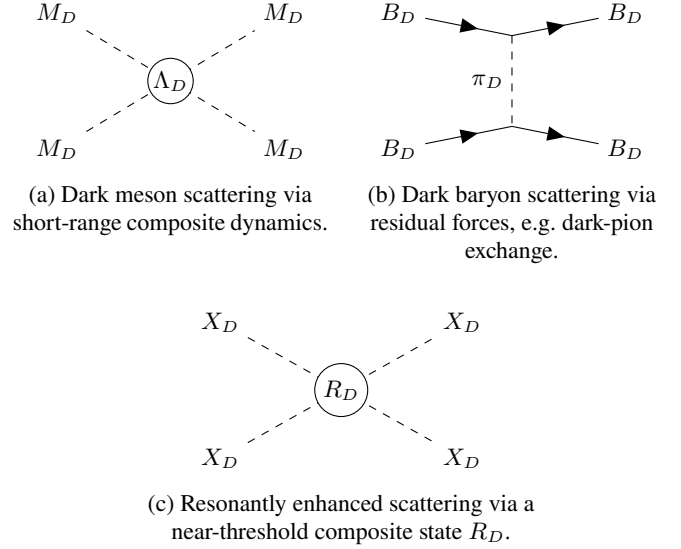

\medskip
\noindent
\textbf{Composite dynamics.}
A well-motivated realization of non-perturbative SIDM arises in composite dark sectors, where the low-energy degrees of freedom are bound states rather than elementary particles.
The microscopic origin of the self-interaction is then not the exchange of a new elementary force carrier, but the strong dynamics that bind the constituents.
The corresponding processes are illustrated in \cref{fig:selfint_composite_realizations}.

In \cref{fig:selfint_composite_mesons}, two dark mesons $M_D$ scatter through short-range interactions generated by confining dynamics at the scale $\Lambda_D$.At sufficiently low energies, this interaction
can be described by a local quartic operator,
\begin{equation}
\mathcal{L}_{\rm eff}
\supset \lambda_{M_D}(M_D^\dagger M_D)^2,
\end{equation}
where $C_{M_D}$ is a dimensionless effective
coupling determined by the underlying strong dynamics.
This is closely analogous to hadron--hadron scattering in QCD and arises naturally in composite DM scenarios \cite{Cline:2013zca}.

In \cref{fig:selfint_composite_baryons}, dark baryons $B_D$ interact through residual forces between composite states, in analogy with nucleon--nucleon interactions in QCD.
These forces are typically mediated by lighter composite states such as dark pions $\pi_D$, producing a finite-range interaction between heavy particles.
As in nuclear physics, the scattering amplitude can be enhanced by large scattering lengths or near-threshold bound states, providing a natural origin of large self-interactions~\cite{Cline:2022leq}.

Finally, \cref{fig:selfint_composite_resonance} illustrates resonantly enhanced scattering of DM particles $X_D$ through a near-threshold intermediate state $R_D$.
Here $X_D$ denotes the DM particle, which may be elementary or composite depending on the model.
The scattering amplitude is dominated by the intermediate state, producing a Breit--Wigner--type enhancement of the cross section at low velocities \cite{Breit:1936zzb}.
Such resonant behavior can arise both in composite systems and in models with elementary degrees of freedom, highlighting the generality of the mechanism \cite{Tulin:2013teo,Braaten:2013tza,Chu:2018qrm,Chu:2018fzy}.

Composite dark sectors therefore provide a concrete framework in which large and velocity-dependent self-interactions arise naturally.
At low energies, these interactions can still be described within the effective expansion of \cref{eq:effective_Lagrangian_expansion}, but the coefficients $C_n$ encode confinement, residual forces and bound-state structure rather than perturbative mediator exchange.
This shows how a common EFT framework can encompass many microscopic SIDM realizations, while the underlying physics encoded in the coefficients can differ substantially.

\subsubsection{From interactions to cross sections}
\label{sec:interactions_to_cross_sections}

The microscopic interaction terms introduced above dictate how DM particles scatter at the quantum level. For phenomenology and structure formation, however, the relevant quantities are not the interactions themselves but the
\emph{scattering amplitudes} and \emph{cross sections} they generate. The
connection proceeds through the chain
\begin{equation}
\mathcal{L}
\;\rightarrow\;
\mathcal{M}
\;\rightarrow\;
\frac{\mathrm{d}\sigma}{\mathrm{d}\Omega}(\theta, v_\mathrm{rel})
\;\rightarrow\;
\sigma_{X}(v_\mathrm{rel}),
\label{eq:L_to_sigma_chain}
\end{equation}
which links a given Lagrangian $\mathcal{L}$ to the angle-weighted cross sections
$\sigma_{X}(v_\mathrm{rel})$ (defined in \cref{eq:weighted_cross_sections}) that serve as simulation inputs. Each arrow represents a well-defined step: the Lagrangian fixes the invariant amplitude $\mathcal{M}$, the amplitude determines the differential cross section $\mathrm{d}\sigma/\mathrm{d}\Omega$, and angular integration with appropriate
weighting yields the velocity-dependent cross sections $\sigma_{X}$. Explicit cross-section expressions for the various self-interactions
mentioned above are collected in \cref{sec:cross_sections}.

\medskip
\noindent
\textbf{From the Lagrangian to the scattering amplitude.}
In quantum field theory, scattering processes are encoded in the path integral of \cref{eq:path_integral}, whose perturbative expansion is organized in terms of the Feynman diagrams introduced in \cref{sec:feynman_diagrams}. These diagrams translate the interaction terms of $\mathcal{L}$ into contributions to the scattering amplitude $\mathcal{M}$. Each diagram is assembled from vertex factors, propagators, and the spinor or polarization structures of the external states, with overall momentum conservation enforced by a delta function. The full amplitude is obtained by summing all diagrams consistent with the given
external states. Feynman diagrams thus provide a direct route from $\mathcal{L}$ to $\mathcal{M}$, from which all observables follow through $|\mathcal{M}|^2$.

\medskip
\noindent
\textbf{Differential cross section and kinematics.}
For elastic $2\to2$ scattering in the center-of-mass frame, the differential
cross section is
\begin{equation}
\frac{\mathrm{d}\sigma}{\mathrm{d}\Omega}
=
\frac{1}{64\pi^2 s}
\frac{|\vec{p}_f|}{|\vec{p}_i|}\,
|\mathcal{M}|^2 ,
\end{equation}
where $s=(p_1+p_2)^2$ is the squared center-of-mass energy and $\vec{p}_{i,f}$
the initial and final three-momenta. For elastic scattering of equal masses
$|\vec{p}_f|=|\vec{p}_i|$, and in the non-relativistic limit
$s\simeq 4m_{\rm DM}^2$, so that
\begin{equation}
\frac{\mathrm{d}\sigma}{\mathrm{d}\Omega}
\simeq
\frac{|\mathcal{M}|^2}{256\pi^2 m_{\rm DM}^2}.
\end{equation}
The angular and velocity dependence of
$\mathrm{d}\sigma/\mathrm{d}\Omega(\theta,v_{\rm rel})$ therefore follows
entirely from that of $\mathcal{M}$, which in turn depends on the momentum
transfer
\begin{equation}
q^2 = 2\mu_{\mathrm{DM}}^2 v_{\rm rel}^2 (1-\cos\theta),
\end{equation}
with $\mu_{\mathrm{DM}}$ the reduced mass. Large deflection angles correspond to
large momentum transfer, and forward scattering to small $q$.

\medskip
\noindent
\textbf{Mediator exchange and momentum dependence.}
For tree-level mediator exchange, the amplitude contains a propagator
\begin{equation}
\mathcal{M} \sim \frac{1}{q^2 + m_{\rm med}^2}.
\end{equation}
When the mediator is much lighter than the typical momentum transfer,
$m_{\rm med}\ll \mu_{\mathrm{DM}} v_{\rm rel}$, the propagator grows at small
$q$, strongly enhancing forward scattering; in position space this corresponds
to a long-range Yukawa (Coulomb-like) potential. In the opposite,
heavy-mediator limit $m_{\rm med}\gg \mu_{\mathrm{DM}} v_{\rm rel}$, the
propagator reduces to a constant $\sim 1/m_{\rm med}^2$, giving the
approximately isotropic, velocity-independent contact interaction discussed in
\cref{fig:heavy_mediator_limit_both}. Derivative couplings introduce additional
powers of momentum and, since the characteristic momentum scale is
$\mu_{\mathrm{DM}} v_{\rm rel}$, translate into explicit velocity dependence of
$|\mathcal{M}|^2$. The mapping from the Lagrangian structure to the momentum
dependence of $\mathcal{M}$ thus determines whether self-scattering is isotropic
or forward-peaked, and weakly or strongly velocity dependent.

\medskip
\noindent
\textbf{Low-energy universality and effective-range description.}
In most SIDM applications the scattering occurs at low velocities,
$v_{\rm rel}\sim 10^{-4}$--$10^{-2}\,c$, i.e.\ at momentum transfers well below
the underlying particle-physics scales. The process is then insensitive to
microscopic details and can be captured by a partial-wave expansion in which the
$s$-wave dominates, giving an isotropic amplitude~\cite{Chu:2019awd}
\begin{equation}
f(k) = \frac{1}{k\cot\delta_0 - i k},
\qquad
\frac{\mathrm{d}\sigma}{\mathrm{d}\Omega} = |f(k)|^2 ,
\end{equation}
with $k=\mu_{\mathrm{DM}} v_{\rm rel}$ and $\delta_0$ the $s$-wave phase shift.
The velocity dependence is encoded in the effective-range expansion
$k\cot\delta_0 = -1/a + \tfrac12 r_e k^2 + \cdots$, giving
\begin{equation} \label{eq:effective_range_cross}
\sigma(v) \approx
\frac{4\pi}{\left(-\tfrac{1}{a} + \tfrac{1}{2} r_e k^2 \right)^2 + k^2}.
\end{equation}
Here the scattering length $a$ sets the zero-velocity cross section,
$\sigma\to 4\pi a^2$, while the effective range $r_e$ captures the finite
spatial extent of the interaction. A large $|a|\gg r_e$ signals a shallow bound
state or near-threshold resonance and strongly enhances the low-velocity cross
section -- the model-independent origin of the resonant velocity dependence
introduced above (cf.\ \cref{fig:selfint_composite_realizations})%
~\cite{Tulin:2013teo,Braaten:2013tza,Chu:2018qrm,Chu:2018fzy}. Since very
different microscopic models can share the same $(a,r_e)$, this parametrization
makes explicit why SIDM halo phenomenology is often degenerate with respect to
the underlying particle-physics realization. Angular dependence re-appears only
once higher partial waves or long-range forces contribute.

\medskip
\noindent
\textbf{Effective cross sections for simulations.}
For structure formation the full differential cross section is rarely used
directly. Instead, one integrates it with the angular weights of
\cref{eq:weighted_cross_sections} to obtain the effective cross sections
$\sigma_X(v_{\rm rel})$ -- most importantly the transfer cross section
$\sigma_{\rm T}$ and viscosity cross section $\sigma_{\rm V}$ -- which quantify
the efficiency of momentum and energy transfer in a halo. It is these
velocity-dependent weighted cross sections, obtained through the chain of
\cref{eq:L_to_sigma_chain}, that constitute the actual simulation inputs and
determine the macroscopic impact of SIDM on halo structure and cosmic
evolution.

\medskip

Having established the particle-physics building blocks of a dark sector, we 
now turn to how these same ingredients determine the cosmological abundance of 
DM. The DM-production mechanism is not an independent addition to the theory, 
but is dictated by, for example, the same couplings and masses that set the self-interaction cross section. It therefore links the microphysics directly to the relic density.

\subsubsection{DM production mechanisms}
\label{sec:DM_production_mechanisms}

A viable SIDM model must explain not only the present-day self-interaction phenomenology, but also how the observed relic abundance, $\Omega_{\rm DM}h^2=0.1196\pm0.0031$, was generated in the early Universe~\cite{Planck_2020b}. The production mechanism connects the microscopic parameter space---including masses ${m_i}$, couplings ${g_i}$ and portal strengths ${\epsilon_\gamma,\epsilon_Z,\epsilon_h,\lambda_{sh}}$---to the present-day DM density and the thermal history of the dark sector. In particular, the portal interactions introduced in \cref{eq:photon_portal,eq:Z_portal,eq:higgs_portal_mediator,eq:higgs_portal_dm} and illustrated in \cref{fig:portal_couplings} determine the degree of thermal contact between the dark and visible sectors, while simultaneously providing targets for the experimental searches discussed in \cref{sec:DM_search_experiments}.

We consider four representative production mechanisms relevant for SIDM: thermal freeze-out, freeze-in, reannihilation and asymmetric production. These mechanisms correspond to qualitatively different thermal histories and select different regions of the underlying particle-physics parameter space.

\medskip
\noindent
\textbf{Thermal freeze-out.}
Thermal freeze-out is the most widely studied DM production mechanism~\cite{Lee:1977ua,Scherrer:1985zt,Griest:1990kh,Kolb:1990vq}. In this scenario, DM particles are initially in thermal equilibrium with the SM plasma. As the Universe expands and cools, the interaction rate $\Gamma_{\rm int}$ eventually falls below the Hubble rate $H(T)$, causing DM to decouple. After decoupling, the DM number density is effectively frozen, and subsequent cosmic expansion sets the present-day relic abundance.

The freeze-out process is governed by the competition between annihilation and expansion. The Boltzmann equation for the DM number density $n_{\rm DM}$ is
\begin{equation}
\frac{dn_{\rm DM}}{dt} + 3H n_{\rm DM} = -\langle \sigma_{\rm ann} v \rangle \left(n_{\rm DM}^2 - n_{\rm DM,eq}^2\right),
\label{eq:boltzmann_freezeout}
\end{equation}
where $H$ is the Hubble parameter, $\langle \sigma_{\rm ann} v \rangle$ is the thermally averaged annihilation cross section and $n_{\rm DM,eq}$ is the equilibrium number density. The right-hand side describes the net difference between annihilation and pair production, while $3H n_{\rm DM}$ accounts for dilution by cosmic expansion.

\begin{figure*}[t]
\centering
\includegraphics[width=1.30\columnwidth]{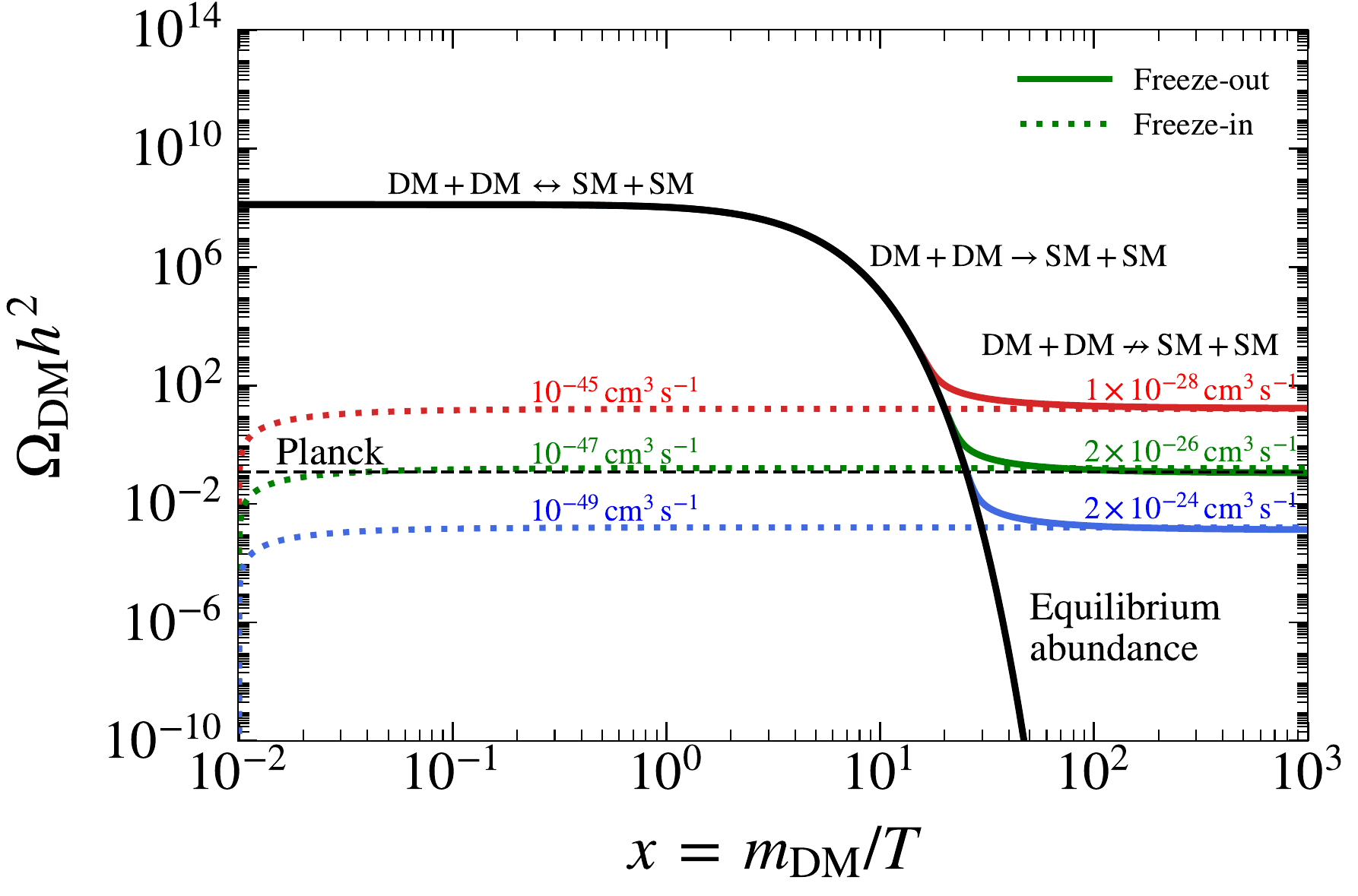}
\caption{
Evolution of the DM relic density $\Omega_{\rm DM}h^2$ as a function of $x \equiv m_{\rm DM}/T$ for freeze-out (colored solid curves) and freeze-in (colored dotted curves). The black curve shows the equilibrium abundance, and the black dashed line indicates the observed Planck value $\Omega_{\rm DM}h^2 = 0.1196\pm 0.0031$~\cite{Planck_2020b}. \emph{Freeze-out}: DM begins in thermal equilibrium (${\rm DM} + {\rm DM} \leftrightarrow {\rm SM} + {\rm SM}$, $x \ll 1$), tracks the equilibrium curve, and freezes out once the annihilation rate drops below the Hubble rate at $x_{\rm fo} \sim 20$--$30$. The abundance then approaches a constant plateau (${\rm DM} + {\rm DM} \nrightarrow {\rm SM} + {\rm SM}$, $x \gtrsim 30$). Larger $\langle\sigma_{\rm ann} v\rangle$ gives smaller final densities, $\Omega_{\rm DM} h^2 \propto 1/\langle\sigma_{\rm ann} v\rangle$, while the WIMP value $\langle\sigma_{\rm ann} v\rangle \approx 2 \times 10^{-26}~{\rm cm^3\,s^{-1}}$ (green curve) reproduces the observed abundance. \emph{Freeze-in}: DM is never in equilibrium and is produced gradually from the SM bath (${\rm SM} + {\rm SM} \to {\rm DM} + {\rm DM}$) through feeble portal interactions. 
The final abundance scales as $\Omega_{\rm DM} h^2 \propto \epsilon^2$, where $\epsilon$ denotes the portal coupling, in contrast to the inverse dependence on the annihilation cross section in freeze-out.
}
\label{fig:freezeout_freezein}
\end{figure*}

At high temperatures ($T \gg m_{\rm DM}$), DM is relativistic and remains in equilibrium through rapid annihilation and pair production, ${\rm DM} + {\rm DM} \leftrightarrow {\rm SM} + {\rm SM}$ (see \cref{fig:freezeout_freezein}, left region). Once $T \lesssim m_{\rm DM}$, the equilibrium density becomes exponentially suppressed, $n_{\rm DM,eq} \propto e^{-m_{\rm DM}/T}$, corresponding to $x \equiv m_{\rm DM}/T \gtrsim 1$ in \cref{fig:freezeout_freezein}. The black curve shows this equilibrium abundance, which the solid freeze-out curves follow at early times. When $\Gamma_{\rm ann} = n_{\rm DM} \langle \sigma_{\rm ann} v \rangle$ drops below $H$, pair production from the SM bath becomes inefficient and the process becomes effectively irreversible: ${\rm DM} + {\rm DM} \to {\rm SM} + {\rm SM}$ (middle region around $x \sim 10$--$30$). The DM density then departs from equilibrium and begins to freeze out. The freeze-out temperature $T_{\rm fo}$ is determined by
\begin{equation}
\Gamma_{\rm ann}(T_{\rm fo}) \sim H(T_{\rm fo}),
\end{equation}
which typically occurs at $x_{\rm fo} \equiv m_{\rm DM}/T_{\rm fo} \sim 20$--$30$. At later times, annihilation becomes negligible because the DM density is too diluted for particles to find collision partners, ${\rm DM} + {\rm DM} \nrightarrow {\rm SM} + {\rm SM}$ (right region for $x \gtrsim 30$), and the comoving number density approaches the constant plateau shown by the solid colored curves.

Solving \cref{eq:boltzmann_freezeout} in the freeze-out regime, with $Y \equiv n_{\rm DM}/s$ and $x = m_{\rm DM}/T$, gives~\cite{Kolb:1990vq,Gondolo:1990dk}
\begin{equation}
\Omega_{\rm DM} h^2 \simeq \frac{1.04 \times 10^9~{\rm GeV^{-1}}}{M_{\rm Pl} \sqrt{g_\ast(T_{\rm fo})}} \frac{x_{\rm fo}}{\langle \sigma_{\rm ann} v \rangle},
\label{eq:relic_abundance_freezeout_explicit}
\end{equation}
where $M_{\rm Pl} = 1.22 \times 10^{19}~{\rm GeV}$, $g_\ast(T_{\rm fo})$ is the effective number of relativistic degrees of freedom at freeze-out, typically $g_\ast \sim 100$, and $x_{\rm fo} \sim 20$--$30$. Hence,
$\Omega_{\rm DM} h^2 \propto 1/\langle \sigma_{\rm ann} v \rangle$:
larger annihilation rates deplete DM more efficiently before freeze-out and produce smaller present-day densities. This behavior is visible in \cref{fig:freezeout_freezein}: the solid curves with larger $\langle\sigma_{\rm ann} v\rangle$ (blue, $2 \times 10^{-24}~{\rm cm^3\,s^{-1}}$) freeze out to lower final abundances, while weaker cross sections (red, $10^{-28}~{\rm cm^3\,s^{-1}}$) yield higher relic densities.

For the canonical thermal relic cross section $\langle \sigma_{\rm ann} v \rangle \sim 2 \times 10^{-26}~{\rm cm^3\,s^{-1}}$ (green solid curve in \cref{fig:freezeout_freezein}), \cref{eq:relic_abundance_freezeout_explicit} gives $\Omega_{\rm DM} h^2 \sim 0.12$, in agreement with the Planck value indicated by the horizontal black dashed line~\cite{Planck_2020b}. This coincidence motivates the \emph{WIMP paradigm}~\cite{Jungman:1995df,Bertone:2004pz}: particles with weak-scale masses ($m_{\rm DM} \sim 100~{\rm GeV}$) and electroweak-strength interactions ($\langle \sigma_{\rm ann} v \rangle \sim \alpha_W^2/m_{\rm DM}^2 \sim 10^{-26}~{\rm cm^3\,s^{-1}}$) naturally reproduce the observed relic abundance. Examples include neutralinos in supersymmetric theories~\cite{Goldberg:1983nd,Ellis:1983ew} and Kaluza-Klein states in extra-dimensional models~\cite{Servant:2002aq,Cheng:2002ej}. However, null results from direct detection experiments and collider searches increasingly challenge the simplest WIMP scenarios, motivating alternative production mechanisms and interaction strengths~\cite{Arcadi:2017kky}.

\begin{figure}[t]
\centering

\begin{subfigure}[t]{0.49\columnwidth}
\centering
\begin{tikzpicture}
\begin{feynman}
  \vertex (a) at (-0.35,0);
  \vertex (m) at ( 0.35,0);
  \vertex (b) at ( 1.05,0);

  \vertex (i1) at (-1.4, 0.8) {$\chi$};
  \vertex (i2) at (-1.4,-0.8) {$\bar{\chi}$};

  \vertex (f1) at ( 2.0, 0.8) {$f$};
  \vertex (f2) at ( 2.0,-0.8) {$\bar{f}$};

  \diagram*{
    (i1) -- [fermion] (a),
    (i2) -- [anti fermion] (a),
    (a)  -- [photon, edge label=$\phi_\mu$] (m),
    (m)  -- [photon, edge label=$\gamma_\mu$] (b),
    (b)  -- [fermion] (f1),
    (f2) -- [fermion] (b),
  };
\end{feynman}

\draw[thick] ($(m)+(-0.10,0.10)$) -- ($(m)+(0.10,-0.10)$);
\draw[thick] ($(m)+(-0.10,-0.10)$) -- ($(m)+(0.10,0.10)$);
\node[below=2pt of m] {\small $\epsilon_\gamma$};
\end{tikzpicture}
\caption{Freeze-out via photon portal: $\chi\bar{\chi} \to \phi_\mu^\ast/\gamma^\ast \to f\bar{f}$.}
\label{fig:freezeout_photon}
\end{subfigure}
\hfill
\begin{subfigure}[t]{0.49\columnwidth}
\centering
\begin{tikzpicture}
\begin{feynman}
  \vertex (a) at (-0.35,0);
  \vertex (m) at ( 0.35,0);
  \vertex (b) at ( 1.05,0);

  \vertex (i1) at (-1.4, 0.8) {$s$};
  \vertex (i2) at (-1.4,-0.8) {$s $};

  \vertex (f1) at ( 2.0, 0.8) {$f$};
  \vertex (f2) at ( 2.0,-0.8) {$\bar{f}$};

  \diagram*{
    (i1) -- [scalar] (a),
    (i2) -- [scalar] (a),
    (a)  -- [scalar, edge label=$\phi$] (m),
    (m)  -- [scalar, edge label=$h$] (b),
    (b)  -- [fermion] (f1),
    (f2) -- [fermion] (b),
  };
\end{feynman}

\draw[thick] ($(m)+(-0.10,0.10)$) -- ($(m)+(0.10,-0.10)$);
\draw[thick] ($(m)+(-0.10,-0.10)$) -- ($(m)+(0.10,0.10)$);
\node[below=2pt of m] {\small $\epsilon_h$};
\end{tikzpicture}
\caption{Freeze-out via Higgs portal: $s s \to \phi^\ast/h^\ast \to f\bar{f}$.}
\label{fig:freezeout_higgs}
\end{subfigure}

\vspace{4mm}

\begin{subfigure}[t]{0.49\columnwidth}
\centering
\begin{tikzpicture}
\begin{feynman}
  \vertex (a) at (-0.35,0);
  \vertex (m) at ( 0.35,0);
  \vertex (b) at ( 1.05,0);

  \vertex (i1) at (-1.4, 0.8) {$f$};
  \vertex (i2) at (-1.4,-0.8) {$\bar{f}$};

  \vertex (f1) at ( 2.0, 0.8) {$\chi$};
  \vertex (f2) at ( 2.0,-0.8) {$\bar{\chi}$};

  \diagram*{
    (i1) -- [fermion] (a),
    (i2) -- [anti fermion] (a),
    (a)  -- [photon, edge label=$\gamma_\mu$] (m),
    (m)  -- [photon, edge label=$\phi_\mu$] (b),
    (b)  -- [fermion] (f1),
    (f2) -- [fermion] (b),
  };
\end{feynman}

\draw[thick] ($(m)+(-0.10,0.10)$) -- ($(m)+(0.10,-0.10)$);
\draw[thick] ($(m)+(-0.10,-0.10)$) -- ($(m)+(0.10,0.10)$);
\node[below=2pt of m] {\small $\epsilon_\gamma$};
\end{tikzpicture}
\caption{Freeze-in via photon portal:
$f\bar{f} \to \gamma^\ast/\phi_\mu^\ast \to \chi\bar{\chi}$. }
\label{fig:freezein_photon}
\end{subfigure}
\hfill
\begin{subfigure}[t]{0.49\columnwidth}
\centering
\begin{tikzpicture}
\begin{feynman}
  \vertex (a) at (-0.35,0);
  \vertex (m) at ( 0.35,0);
  \vertex (b) at ( 1.05,0);

  \vertex (i1) at (-1.4, 0.8) {$f$};
  \vertex (i2) at (-1.4,-0.8) {$\bar{f}$};

  \vertex (f1) at ( 2.0, 0.8) {$s$};
  \vertex (f2) at ( 2.0,-0.8) {$s$};

  \diagram*{
    (i1) -- [fermion] (a),
    (i2) -- [anti fermion] (a),
    (a)  -- [scalar, edge label=$h$] (m),
    (m)  -- [scalar, edge label=$\phi$] (b),
    (b)  -- [scalar] (f1),
    (f2) -- [scalar] (b),
  };
\end{feynman}

\draw[thick] ($(m)+(-0.10,0.10)$) -- ($(m)+(0.10,-0.10)$);
\draw[thick] ($(m)+(-0.10,-0.10)$) -- ($(m)+(0.10,0.10)$);
\node[below=2pt of m] {\small $\epsilon_h$};
\end{tikzpicture}
\caption{Freeze-in via Higgs portal: $f\bar{f} \to h^\ast/\phi^\ast \to s s$.}
\label{fig:freezein_higgs}
\end{subfigure}

\vspace{4mm}

\begin{subfigure}[t]{0.49\columnwidth}
\centering
\begin{tikzpicture}
\begin{feynman}
  \vertex (a) at (0, 0.5);
  \vertex (b) at (0,-0.5);

  \vertex (i1) at (-1.2, 1.0) {$\chi$};
  \vertex (i2) at (-1.2,-1.0) {$\bar{\chi}$};
  \vertex (f1) at ( 1.2, 1.0) {$\phi_\mu$};
  \vertex (f2) at ( 1.2,-1.0) {$\phi_\mu$};

  \diagram*{
    (i1) -- [fermion] (a) -- [photon] (f1),
    (i2) -- [anti fermion] (b) -- [photon] (f2),
    (a)  -- [fermion, edge label'=$\chi$] (b),
  };
\end{feynman}
\end{tikzpicture}
\caption{Reannihilation (dark photon): $\chi\bar{\chi} \to \phi_\mu\phi_\mu$.}
\label{fig:reannihilation_photon}
\end{subfigure}
\hfill
\begin{subfigure}[t]{0.49\columnwidth}
\centering
\begin{tikzpicture}
\begin{feynman}
  \vertex (a) at (0, 0.5);
  \vertex (b) at (0,-0.5);

  \vertex (i1) at (-1.2, 1.0) {$s$};
  \vertex (i2) at (-1.2,-1.0) {$s$};
  \vertex (f1) at ( 1.2, 1.0) {$\phi$};
  \vertex (f2) at ( 1.2,-1.0) {$\phi$};

  \diagram*{
    (i1) -- [scalar] (a) -- [scalar] (f1),
    (i2) -- [scalar] (b) -- [scalar] (f2),
    (a)  -- [scalar, edge label'=$s$] (b),
  };
\end{feynman}
\end{tikzpicture}
\caption{Reannihilation (scalar): $s s \to \phi\phi$.}
\label{fig:reannihilation_scalar}
\end{subfigure}

\caption{
Representative Feynman diagrams for DM production in light-mediator models. \emph{First and second rows}: freeze-out and freeze-in through photon portal interactions for fermionic DM and Higgs portal interactions for scalar DM. \emph{Third row}: reannihilation into dark sector mediators, either dark photons $\phi_\mu$ (left) or dark scalars $\phi$ (right). The portal couplings $\epsilon_\gamma$ and $\epsilon_h$ that enable experimental searches (\cref{sec:DM_search_experiments}) also govern production and annihilation in the early Universe. Portal mixings are indicated by crosses.
}
\label{fig:production_mechanisms}
\end{figure}
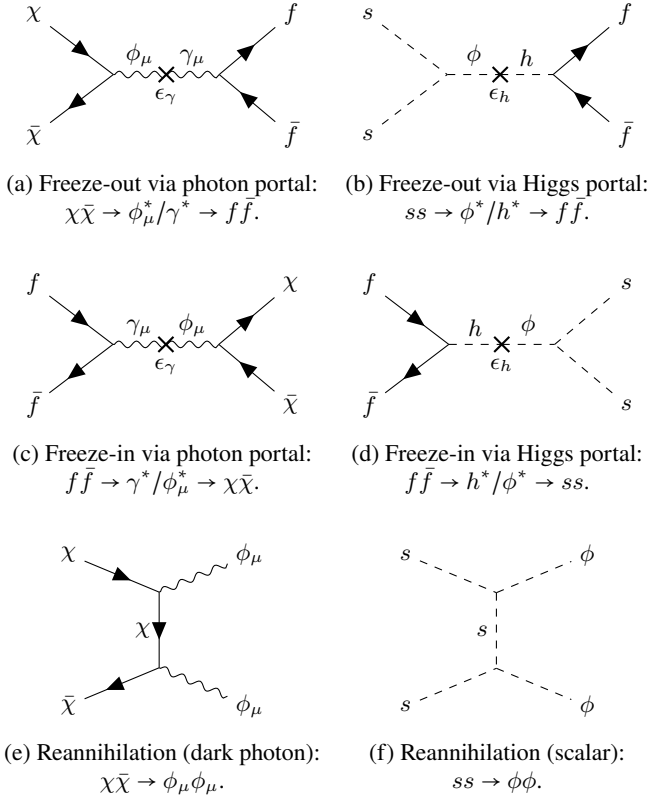

In light-mediator SIDM models, annihilation proceeds through the same mediators that generate self-interactions. For fermionic DM $\chi$ coupled to a vector mediator $\phi_\mu$ through $g\,\bar{\chi}\gamma^\mu\chi\,\phi_\mu$ (\cref{eq:vector_exchange}), the process $\chi\bar{\chi} \to \phi_\mu^\ast /\gamma_\mu^\ast \to f\bar{f}$ is shown in \cref{fig:freezeout_photon}. The dark photon mixes with the SM photon through kinetic mixing $\epsilon_\gamma$ (\cref{eq:photon_portal}), enabling annihilation into SM fermions $f$. For scalar DM $s$ coupled to a scalar mediator $\phi$ through $\mu\, s^2\phi$ (\cref{eq:scalar_mediator_couplings_mu}), the process $s s \to \phi^\ast /h^\ast  \to f\bar{f}$ proceeds through Higgs portal mixing $\epsilon_h$ (\cref{eq:higgs_portal_mediator}), as illustrated in \cref{fig:freezeout_higgs}. Analogous annihilation channels exist for the other SIDM model combinations discussed in this chapter, including fermionic DM with scalar mediators and scalar DM with vector mediators.

The interplay between self-interactions and annihilation is central to light-mediator SIDM phenomenology. The same DM coupling $\alpha_{\mathrm{DM}}$, mediator mass $m_{\rm med}$ and DM mass $m_{\rm DM}$ that determine $\sigma/m$ also control $\langle \sigma_{\rm ann} v \rangle$. For portal-mediated annihilation into SM final states in the limit $m_{\rm med}, m_f \ll m_{\rm DM}$, the thermally averaged cross section can be written schematically as
\begin{equation}
\langle \sigma_{\rm ann} v \rangle \sim \frac{\alpha_{\rm DM} \, \epsilon^2 \, \alpha_{\rm SM}}{m_{\rm DM}^2}  ,
\end{equation} 
where $\epsilon$ is the portal coupling ($\epsilon_\gamma$ for the photon portal or $\epsilon_h$ for the Higgs portal) and $\alpha_{\rm SM}$ is the relevant SM coupling. This correlation constrains the viable parameter space: large self-interactions, driven by large $\alpha_{\rm DM}$, tend to enhance annihilation and can overdeplete DM during freeze-out unless the DM mass or portal coupling is adjusted~\cite{Tulin:2013teo,Kaplinghat:2015aga}. Conversely, models that reproduce the relic abundance through freeze-out make specific predictions for self-scattering, linking early-Universe cosmology to late-time structure formation.

\medskip
\noindent
\textbf{Freeze-in production.}
Freeze-in occurs when the portal couplings are so weak that DM never reaches thermal equilibrium with the SM plasma~\cite{Hall:2009bx,Elahi:2014fsa,Bernal:2017kxu}. DM is instead gradually produced from the hot SM bath through rare scattering or decay processes involving the portal interactions shown in \cref{fig:portal_couplings}. The relevant channels depend on the DM and mediator types introduced in \cref{sec: pp_terminology}. For fermionic DM $\chi$ coupled through the photon portal (\cref{eq:photon_portal}), SM fermions produce DM pairs through $f\bar{f} \to \gamma^\ast/\phi_\mu^\ast \to \chi\bar{\chi}$ (\cref{fig:freezein_photon}). For scalar DM $s$ coupled through the Higgs portal (\cref{eq:higgs_portal_mediator}), $f\bar{f} \to h^\ast/\phi^\ast \to ss^\dagger$ produces DM through Higgs or scalar mediator exchange (\cref{fig:freezein_higgs}). Other SIDM model combinations have analogous portal-driven production channels.

The evolution of the freeze-in abundance is shown by the dotted curves in \cref{fig:freezeout_freezein}. Production is most efficient when the SM bath is hot and dense. As the Universe cools, the number density of energetic SM particles capable of producing DM becomes exponentially suppressed, and production ceases. The abundance therefore saturates at late times. This contrasts with freeze-out, where the DM abundance initially tracks its decreasing equilibrium value before interactions become inefficient and the comoving abundance freezes to an approximately constant value.

Because $\epsilon \ll 1$, the production rate remains far below the Hubble expansion rate and DM never equilibrates with the SM. The Boltzmann equation takes the form
\begin{equation}
\frac{dn_{\rm DM}}{dt} + 3H n_{\rm DM} = \langle \sigma_{\rm prod} v \rangle \, n_{\rm SM}^2,
\label{eq:boltzmann_freezein}
\end{equation}
where $\langle \sigma_{\rm prod} v \rangle$ is the thermally averaged production cross section and $n_{\rm SM}$ is the number density of the relevant SM particles. Unlike \cref{eq:boltzmann_freezeout}, this equation has no depletion term because DM never approaches equilibrium and annihilation back to the SM is negligible.

Freeze-in has the opposite coupling dependence to freeze-out. Whereas larger couplings in freeze-out increase annihilation and reduce the final abundance, freeze-in gives a relic density proportional to the interaction strength. Integrating \cref{eq:boltzmann_freezein} gives the schematic estimate
\begin{equation}
\Omega_{\rm DM} h^2 \sim
\frac{m_{\rm DM}s_0}{\rho_c/h^2}
\,\epsilon^2 \,\alpha_{\rm DM} \,\alpha_{\rm SM}
\,\frac{M_{\rm Pl}}{m_{\rm DM}},
\label{eq:relic_abundance_freezein}
\end{equation}
where $s_0 \simeq 2.89\times10^3\,\mathrm{cm}^{-3}$
is the present-day entropy density of the Universe,
and $\rho_c=3H_0^2/(8\pi G)$ is the present-day
critical energy density, with $G$ denoting Newton's
gravitational constant and
$H_0=100\,h\,\mathrm{km\,s^{-1}\,Mpc^{-1}}$
the present-day Hubble expansion rate.
Here $h$ is the dimensionless Hubble parameter,
such that $\rho_c/h^2 \simeq
1.05\times10^{-5}\,\mathrm{GeV\,cm^{-3}}$. Here $\epsilon$ denotes the relevant portal coupling ($\epsilon_\gamma$, $\epsilon_Z$ or $\epsilon_h$), while $\alpha_{\rm DM}$ and $\alpha_{\rm SM}$ are the dark-sector and SM fine structure constants. Stronger portal interactions produce more DM and hence a larger present-day abundance. Reproducing $\Omega_{\rm DM}h^2 = 0.1196\pm 0.0031$ typically requires $\epsilon \sim 10^{-12}$--$10^{-9}$~\cite{Chu:2011be}, depending on the masses and couplings.

Freeze-out and freeze-in also differ in when the abundance is set. In freeze-out, DM decouples at $x_{\rm fo} \sim 20$--$30$, whereas freeze-in production is typically most efficient at $T \sim m_{\rm DM}$ or above, depending on the DM--mediator hierarchy. This can lead to different thermal histories for the dark sector and different modifications to the linear matter power spectrum, as discussed in \cref{sec:initial_conditions}.

Freeze-in is especially motivated when DM stability follows from an approximate or accidental symmetry rather than an exact conservation law~\cite{Hambye:2010zb}. Sufficiently suppressed portal couplings can produce the correct abundance while leaving DM effectively invisible to direct and indirect detection experiments. Such models remain viable SIDM candidates if self-interactions within the dark sector are strong~\cite{Chu:2011be}. The portal coupling therefore provides a complementary constraint on the representative SIDM model classes considered later in Chapter~\ref{sec:pp_to_simulation_inputs}.

\medskip
\noindent
\textbf{Reannihilation.}
Reannihilation is a hybrid production scenario in which DM is first produced through portal interactions, but later undergoes a second phase of efficient annihilation within the dark sector~\cite{Carlson:1992fn,Drees:1992am,Drees:1993bu,Chu:2011be,Hambye:2019dwd}. This can occur when DM annihilation into dark sector states remains faster than the Hubble rate after chemical decoupling from the SM plasma, for example due to Sommerfeld enhancement or continued thermal equilibrium within the dark sector.

If the dark sector has a temperature $T_{\rm DM}$ different from the SM temperature $T_{\rm SM}$, DM can continue to annihilate efficiently into dark sector mediators. Portal production from the SM bath acts as a source term, while dark sector annihilation provides a depletion term, leading to a quasi-static equilibrium (QSE) abundance~\cite{Chu:2011be}. The Boltzmann equation is
\begin{equation}
\frac{dn_{\rm DM}}{dt} + 3H n_{\rm DM} = \langle \sigma_{\rm portal} v \rangle \, n_{\rm SM}^2 - \langle \sigma_{\rm DS} v \rangle \, n_{\rm DM}^2,
\label{eq:boltzmann_reannihilation}
\end{equation}
where $\langle \sigma_{\rm portal} v \rangle$ is the portal-mediated production cross section and $\langle \sigma_{\rm DS} v \rangle$ is the annihilation cross section into dark sector states. Balancing the two terms gives $n_{\rm DM}^{\rm QSE} \sim \sqrt{\langle \sigma_{\rm portal} v \rangle / \langle \sigma_{\rm DS} v \rangle} \, n_{\rm SM}^{\rm eq}$. The final relic abundance is approximately~\cite{Chu:2011be}
\begin{equation}
\Omega_{\rm DM} h^2 \sim
\frac{m_{\rm DM}s_0}{\rho_c/h^2}
\,\frac{x_f}{\langle \sigma_{\rm DS} v \rangle
\,m_{\rm DM}\,M_{\rm Pl}},
\label{eq:relic_abundance_reannihilation}
\end{equation}
where $x_f \equiv m_{\rm DM}/T_f$ is determined by
\begin{equation}
x_f \sim \log[\langle \sigma_{\rm eff} v \rangle \, m_{\rm DM} \, M_{\rm Pl}]
\label{eq:xf_reannihilation}
\end{equation}
with $\langle \sigma_{\rm eff} v \rangle = \sqrt{\langle \sigma_{\rm portal} v \rangle \langle \sigma_{\rm DS} v \rangle}$. Thus, the relic abundance depends inversely on the annihilation strength into dark sector states, but only logarithmically on the portal coupling through $x_f$.

In the reannihilation regime, the effective freeze-out
parameter $x_f$ can be considerably smaller than
in standard freeze-out. Consequently, reproducing
the observed relic abundance requires a smaller
dark-sector annihilation cross section, corresponding
to a weaker dark-sector coupling $\alpha_{\rm DM}$
for otherwise fixed model parameters
~\cite{Chu:2011be}. At fixed portal coupling, increasing the dark-sector annihilation rate reduces the final DM abundance.

Although less commonly invoked than freeze-out or freeze-in, reannihilation can be important in light-mediator SIDM models, especially when Sommerfeld enhancement increases annihilation rates at low velocities and the dark sector maintains $T_{\rm DM} < T_{\rm SM}$ but remains sufficiently warm. Dominant channels include $\chi\bar{\chi} \to \phi\phi$ for scalar mediators, $\chi\bar{\chi} \to \phi_\mu\phi_\mu$ for dark photon mediators (see \cref{fig:reannihilation_photon,fig:reannihilation_scalar}), and bound-state formation $\chi\bar{\chi} \to (\chi\bar{\chi})_{\rm bound} + \phi/\phi_\mu$ when the mediator is light~\cite{Petraki:2015hla,Cirelli:2015bda}. 

\medskip
\noindent
\textbf{Asymmetric dark matter.}
Asymmetric DM scenarios provide a qualitatively different production mechanism~\cite{Nussinov:1985xr,Kaplan:2009ag,Petraki:2013wwa,Zurek:2013wia}. Here, the present-day DM abundance originates from a primordial particle--antiparticle asymmetry in the dark sector, analogous to the baryon asymmetry in the SM. This framework is motivated by the observed relation $\Omega_{\rm DM} \approx 5\,\Omega_{\rm B}$~\cite{Planck_2020b}, which suggests that the two asymmetries may share a common origin.

The initial asymmetry between DM particles $\chi$ and antiparticles $\bar{\chi}$ can arise from dark-sector interactions satisfying the Sakharov conditions (baryon-number violation, C and CP violation, and departure from thermal equilibrium)~\cite{Sakharov:1967dj}, or it can be transferred from the SM baryon asymmetry through higher-dimensional operators or portal interactions~\cite{Kaplan:2009ag,Shelton:2010ta,Davoudiasl:2012uw}. Once established, the symmetric component annihilates away through processes such as $\chi\bar{\chi} \to f\bar{f}$ via portal interactions (\cref{fig:freezeout_photon,fig:freezeout_higgs}), leaving an excess of $\chi$ over $\bar{\chi}$. 
Neglecting inverse production, the Boltzmann equation
for the residual symmetric component is
\begin{equation}
\frac{dn_{\rm sym}}{dt}+3Hn_{\rm sym}
=
-\langle\sigma_{\rm portal}v\rangle
n_{\rm sym}(n_{\rm sym}+\Delta n),
\end{equation}
where $n_{\rm sym}=\min(n_\chi,n_{\bar{\chi}})$
is the number density of the less abundant species,
and $\Delta n=|n_\chi-n_{\bar{\chi}}|$ denotes the
particle--antiparticle number-density asymmetry.
In the absence of asymmetry-violating processes,
$\dot{\Delta n}+3H\Delta n=0$, so the comoving
asymmetry is conserved.
The final relic abundance is set by the initial asymmetry,
\begin{equation}
\Omega_{\rm DM} h^2 \sim \frac{m_{\rm DM}}{m_p} \frac{n_{\rm DM} - n_{\bar{\chi}}}{n_p - n_{\bar{p}}} \, \Omega_{\rm B} h^2,
\end{equation}
where $m_p$ is the proton mass and $n_p - n_{\bar{p}}$ is the baryon asymmetry. If the DM and baryon asymmetries are comparable, $n_{\rm DM} - n_{\bar{\chi}} \sim n_p - n_{\bar{p}}$, then $m_{\rm DM} \sim 5\,{\rm GeV}$ reproduces the observed density ratio~\cite{Kaplan:2009ag}.

Asymmetric SIDM has important phenomenological consequences. Since late-time annihilation is suppressed by the absence of antiparticles, asymmetric SIDM generically evades the stringent indirect detection constraints that apply to symmetric freeze-out scenarios~\cite{Cui:2020dly,Barman:2021ost}. However, if ADM particles can scatter with baryons and become gravitationally captured by stars, their suppressed annihilation allows them to accumulate and transport energy within stellar interiors, potentially altering stellar structure and evolution, particularly in DM-rich environments~\cite{Iocco:2012wk}. The relic abundance is also decoupled from the annihilation rate, allowing large self-interactions without necessarily underproducing DM. This differs from symmetric freeze-out scenarios, where large self-interactions are often correlated with large annihilation rates and therefore require mass or coupling adjustments~\cite{Buckley:2009in,Loeb:2010gj}.

Asymmetric DM can also allow efficient energy loss through inelastic processes. If the dark sector contains excited states or bound states, as in the dark atom models discussed in \cref{sec:dark_atoms}, energy released in inelastic collisions or radiative transitions is not recovered through annihilation. This can enhance dissipation and lead to more efficient core formation in haloes compared to symmetric SIDM~\cite{Cline:2013zca,Cline_2014a,Cyr-Racine:2013fsa}. The asymmetry-generation mechanism therefore provides complementary constraints on the representative SIDM model classes considered later in Chapter~\ref{sec:pp_to_simulation_inputs}.

\begin{figure*}[t]
\centering
\includegraphics[width=0.42\textwidth]{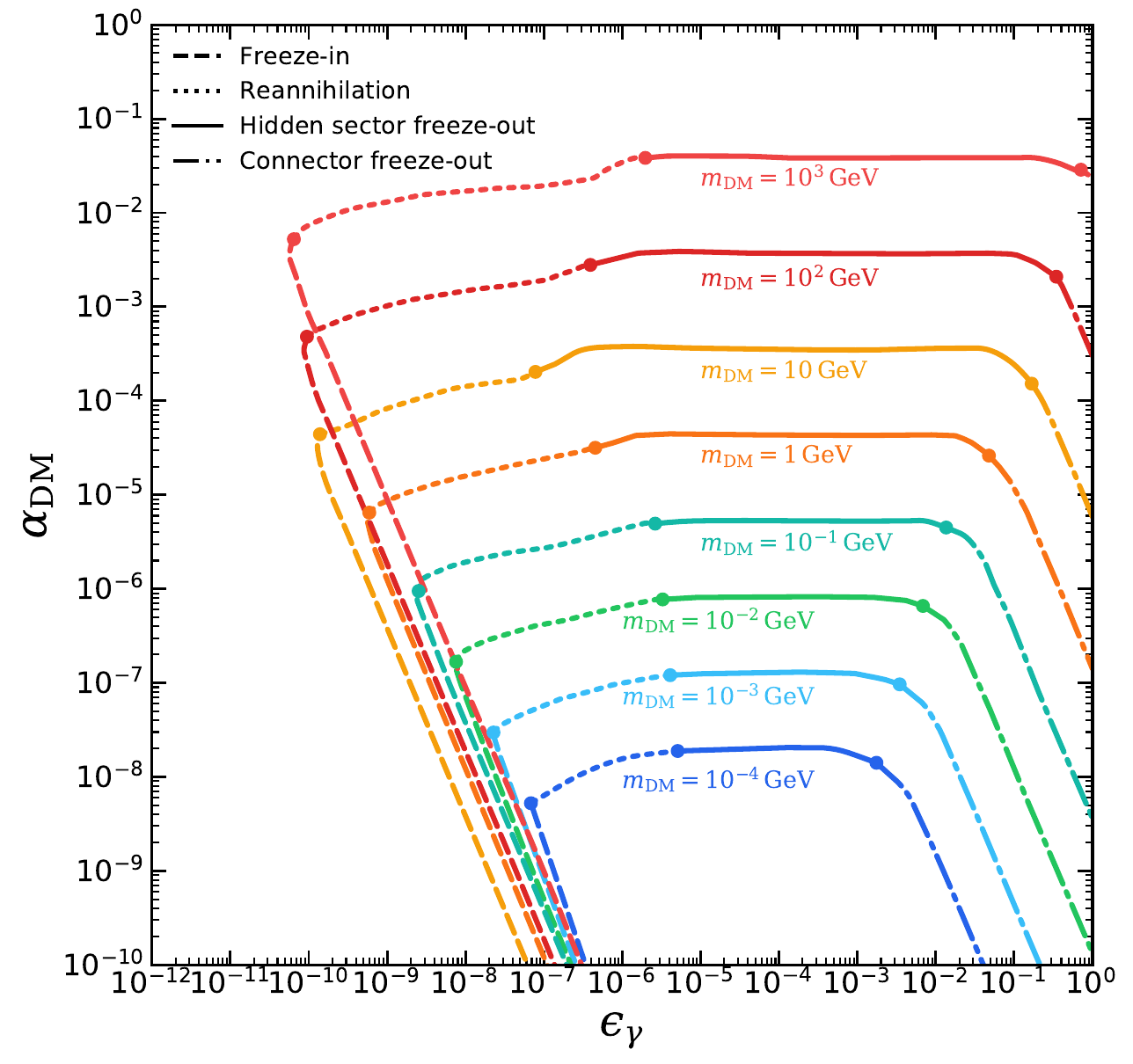}
\quad\quad
\includegraphics[width=0.47\textwidth]{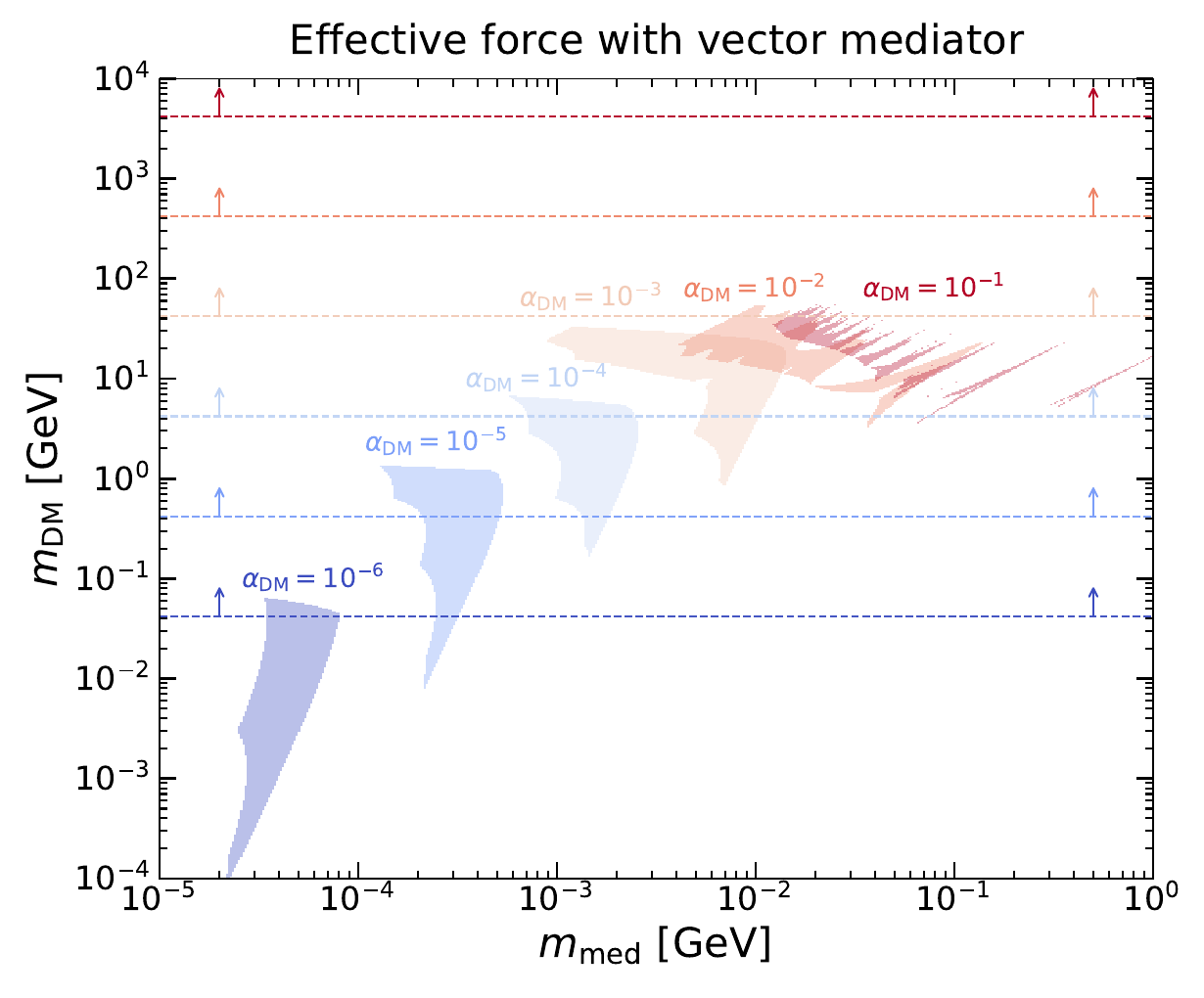}

\vspace{0.5em}

\includegraphics[width=0.47\textwidth]{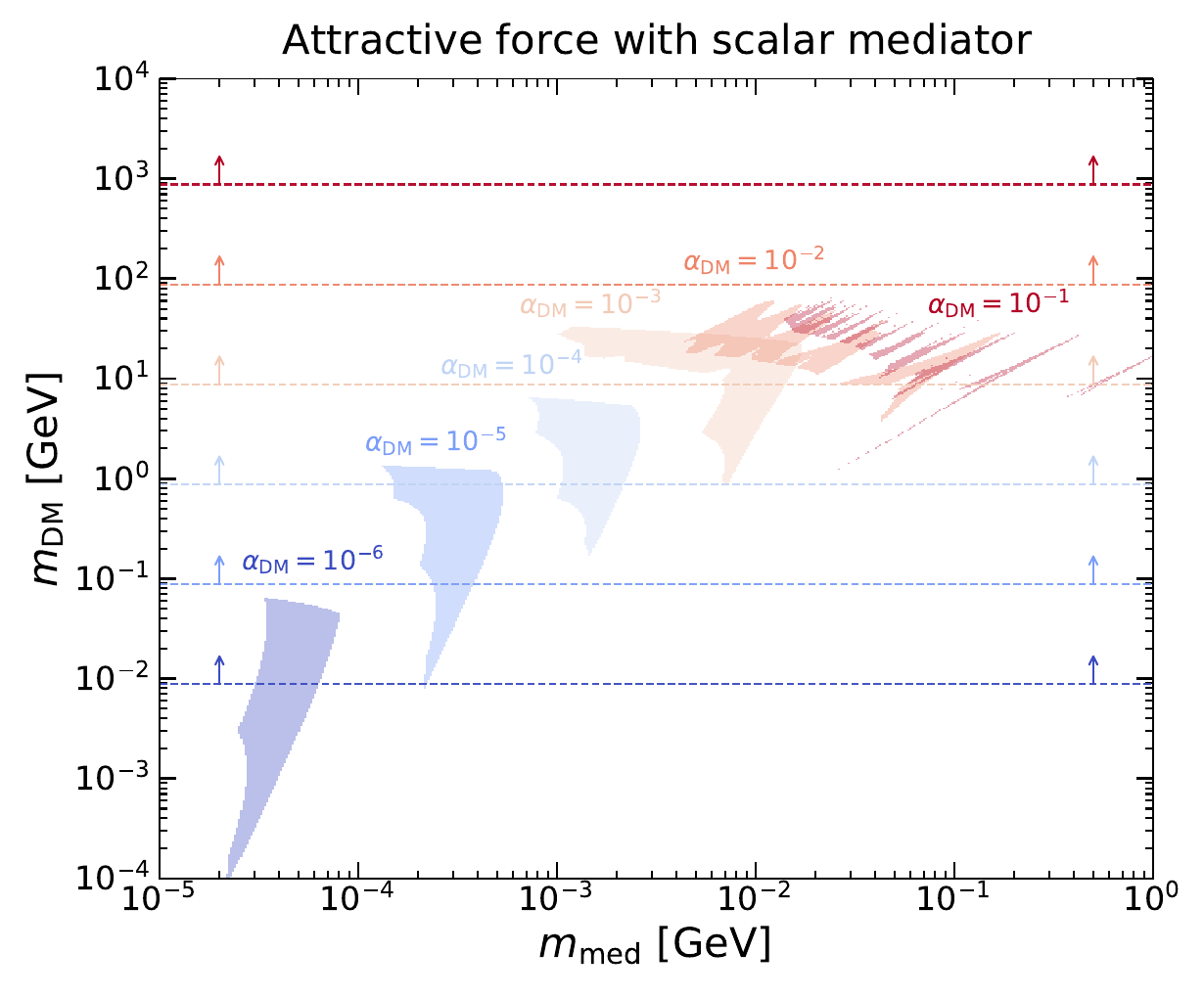}
\includegraphics[width=0.47\textwidth]{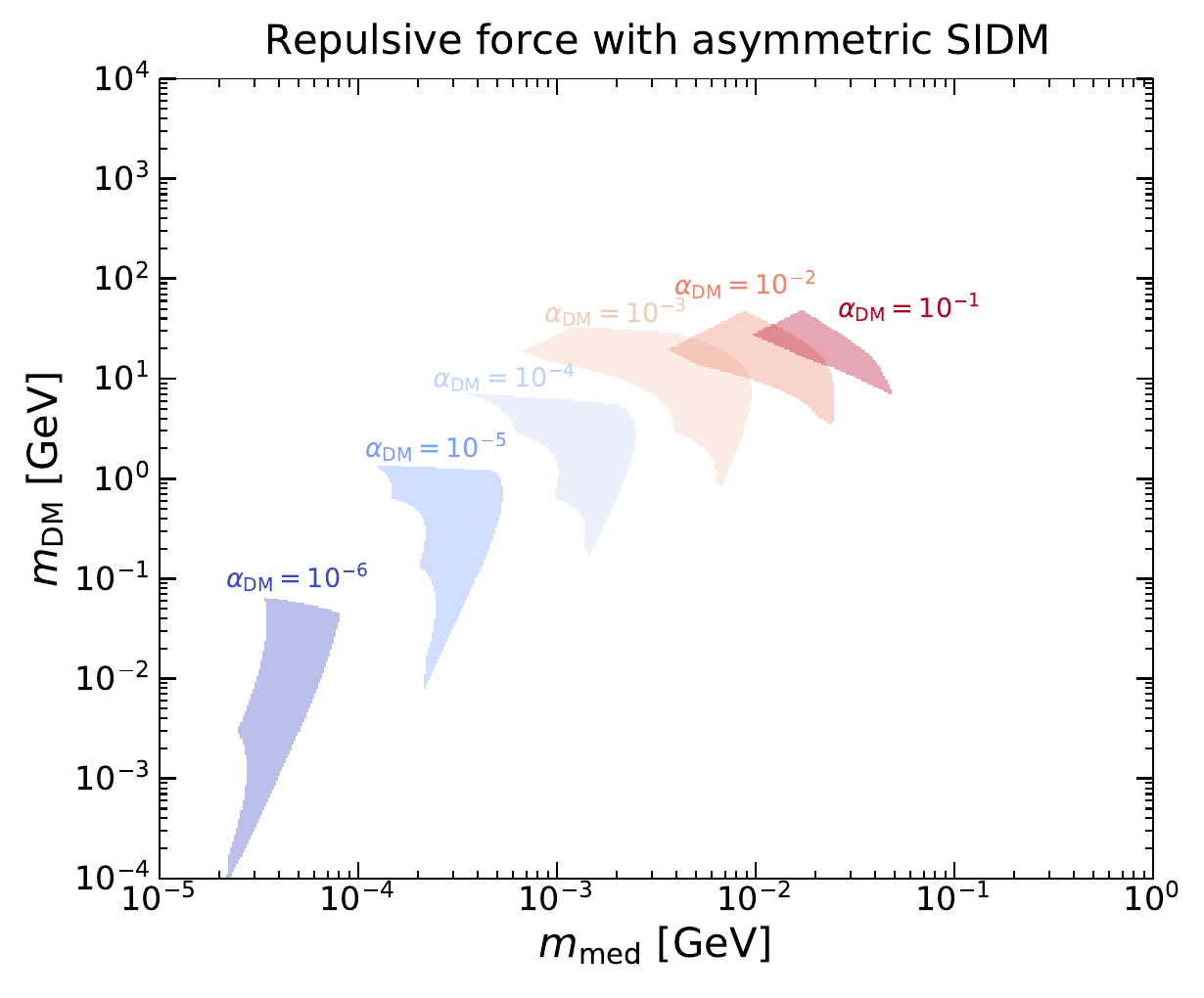}
\caption{
Relic-abundance and self-interaction constraints for representative light-mediator and asymmetric SIDM scenarios. \emph{Upper left:} Relic-density contours for the kinetic mixing portal in \cref{eq:photon_portal}, showing $\Omega_{\mathrm{DM}}h^2\approx0.12$ in the $(\alpha_{\mathrm{DM}},\epsilon_\gamma)$ plane for different DM masses. Inside the clock-shaped contours, DM is overproduced; outside, it is underproduced. The boundaries distinguish freeze-in (dashed), reannihilation (dotted), hidden-sector freeze-out (solid) and connector freeze-out (dot-dashed). \emph{Upper right:} Regions of the $(m_{\mathrm{DM}},m_{\mathrm{med}})$ plane compatible with the inferred astrophysical self-interaction constraints summarized in \cref{fig:cross_sections}, assuming an effective vector-mediated interaction with attractive and repulsive channels. \emph{Lower left:} Corresponding regions for an attractive scalar-mediated interaction through the Higgs portal in \cref{eq:higgs_portal_mediator}. \emph{Lower right:} Regions for asymmetric SIDM with a repulsive vector-mediated interaction, where the relic abundance is set by a primordial DM asymmetry. In the upper-right and lower-left panels, the colored dashed lines indicate the minimum DM mass required to obtain the observed relic abundance through the considered production mechanisms. The fragmented allowed regions at larger couplings in the attractive Yukawa channels arise from resonances and antiresonances in the self-scattering cross section~\cite{Tulin:2013teo}.
}
\label{fig:effective_force_vector}
\end{figure*}

\medskip
\noindent
\textbf{Self-interaction and relic-abundance constraints.} We now combine relic-density and self-interaction constraints in \cref{fig:effective_force_vector}. The upper-left panel presents the kinetic mixing portal, where the clock-shaped contours reproduce $\Omega_{\mathrm{DM}}h^2\approx0.12$ for different DM masses. The four production regimes differ in their degree of thermal contact between the dark and visible sectors: freeze-in occurs without equilibration, hidden-sector freeze-out proceeds mainly within the dark sector, connector freeze-out maintains thermal contact with the SM, and reannihilation involves dark-sector depletion following feeble portal production. The remaining panels show which microscopic parameters can also yield the required late-time self-interactions.

The upper-right panel of \cref{fig:effective_force_vector} shows the parameter regions satisfying the inferred astrophysical self-interaction constraints in \cref{fig:cross_sections} for an effective vector-mediated interaction. The cross section includes both attractive ($\chi\bar{\chi}$) and repulsive ($\chi\chi$ and $\bar{\chi}\bar{\chi}$) channels, averaged as $(\sigma_T^{+}+\sigma_T^{-})/2$, corresponding to the two signs of the Yukawa potential in \cref{eq:yukawa_potential_generic}. At larger couplings, resonances and antiresonances in the attractive channel produce narrow, fragmented allowed regions~\cite{Tulin:2013teo}. The colored dashed lines impose the additional relic-abundance requirements for the production mechanisms considered. These constraints allow simultaneous agreement with the observed DM abundance and astrophysically relevant self-interactions for $\alpha_{\mathrm{DM}}<10^{-3}$ in the vector-mediator scenario shown.

The lower-left panel shows the corresponding constraints for an attractive scalar-mediated interaction through the Higgs portal in \cref{eq:higgs_portal_mediator}, corresponding to the negative sign in the Yukawa potential. The attractive interaction can exhibit non-perturbative resonances and antiresonances, producing the fragmented allowed regions at larger couplings. Compared with the vector-mediated scenario, the scalar-mediated model permits simultaneous agreement with the considered relic-abundance and self-interaction constraints for couplings approaching $\alpha_{\mathrm{DM}}\sim10^{-2}$.

The lower-right panel instead illustrates asymmetric SIDM with a repulsive vector-mediated force. Because the surviving DM population consists predominantly of one species, self-interactions are dominated by $\chi\chi$ scattering, corresponding to the positive sign of the Yukawa potential. The relic abundance is set by a primordial particle--antiparticle asymmetry rather than symmetric freeze-out, so the relic-density boundaries shown for the other scenarios do not apply. The symmetric component must nevertheless be depleted efficiently, for example through portal-mediated annihilation, as constrained by the production regimes in the upper-left panel. A scalar-mediated asymmetric realization would instead retain attractive self-interactions, as in the lower-left panel, without the same symmetric-production bounds. This separation between relic abundance and self-interaction strength can in principle accommodate couplings approaching $\alpha_{\mathrm{DM}}\sim1$, although the perturbative description is then no longer reliable and additional cosmological and experimental constraints remain relevant.

\medskip
The relic abundance and late-time self-interaction phenomenology are therefore controlled by the same underlying particle-physics parameters, including the DM and mediator masses, dark-sector couplings and portal interactions with the SM. The portal couplings are particularly important because they govern the thermal connection between the dark and visible sectors while also determining the strength of many experimental signatures. We therefore turn next to the experimental and observational constraints that further restrict the viable SIDM parameter space.

\subsubsection{DM search experiments}
\label{sec:DM_search_experiments}

\begin{figure*}[t]
\centering

\begin{subfigure}[t]{0.32\textwidth}
\centering
\begin{tikzpicture}
\begin{feynman}
  \vertex (a) at (0, 0.7);
  \vertex (m) at (0, 0.0);
  \vertex (b) at (0,-0.7);

  \vertex (i1) at (-1.2, 1.0) {$\chi$};
  \vertex (f1) at ( 1.2, 1.0) {$\chi$};

  \vertex (i2) at (-1.2,-1.0) {$N$};
  \vertex (f2) at ( 1.2,-1.0) {$N$};

  \diagram*{
    (i1) -- [fermion] (a) -- [fermion] (f1),
    (i2) -- [fermion] (b) -- [fermion] (f2),
    (a) -- [photon, edge label'=$\phi_\mu$] (m),
    (m) -- [photon, edge label'=$\gamma_\mu$] (b),
  };
\end{feynman}

\draw[thick] ($(m)+(-0.10,0.10)$) -- ($(m)+(0.10,-0.10)$);
\draw[thick] ($(m)+(-0.10,-0.10)$) -- ($(m)+(0.10,0.10)$);
\node[right=2pt of m] {\small $\epsilon_\gamma$};
\end{tikzpicture}
\caption{Direct detection: $\chi N \rightarrow \chi N$.}
\label{fig:darkphoton_direct}
\end{subfigure}
\hfill
\begin{subfigure}[t]{0.32\textwidth}
\centering
\begin{tikzpicture}
\begin{feynman}
  \vertex (a) at (-0.35,0);
  \vertex (m) at ( 0.35,0);
  \vertex (b) at ( 1.05,0);

  \vertex (i1) at (-1.4, 0.8) {$\chi$};
  \vertex (i2) at (-1.4,-0.8) {$\bar\chi$};

  \vertex (f1) at ( 2.0, 0.8) {$e^-$};
  \vertex (f2) at ( 2.0,-0.8) {$e^+$};

  \diagram*{
    (i1) -- [fermion] (a),
    (i2) -- [anti fermion] (a),
    (a)  -- [photon, edge label=$\phi_\mu$] (m),
    (m)  -- [photon, edge label=$\gamma_\mu$] (b),
    (b)  -- [fermion] (f1),
    (f2) -- [fermion] (b),
  };
\end{feynman}

\draw[thick] ($(m)+(-0.10,0.10)$) -- ($(m)+(0.10,-0.10)$);
\draw[thick] ($(m)+(-0.10,-0.10)$) -- ($(m)+(0.10,0.10)$);
\node[below=2pt of m] {\small $\epsilon_\gamma$};
\end{tikzpicture}
\caption{Indirect detection: $\chi\bar\chi \rightarrow \phi_\mu^\ast/\gamma^\ast \rightarrow e^+e^-$.}
\label{fig:darkphoton_indirect}
\end{subfigure}
\hfill
\begin{subfigure}[t]{0.32\textwidth}
\centering
\begin{tikzpicture}
\begin{feynman}
  \vertex (a) at (-0.35,0);
  \vertex (m) at ( 0.35,0);
  \vertex (b) at ( 1.05,0);  \vertex (i1) at (-1.4, 0.8) {$q$};
  \vertex (i2) at (-1.4,-0.8) {$\bar q$};  \vertex (f1) at ( 2.0, 0.8) {$\chi$};
  \vertex (f2) at ( 2.0,-0.8) {$\bar\chi$};  \diagram*{
    (i1) -- [fermion] (a),
    (i2) -- [anti fermion] (a),
    (a)  -- [photon, edge label=$\gamma_\mu$] (m),
    (m)  -- [photon, edge label=$\phi_\mu$] (b),
    (b)  -- [fermion] (f1),
    (f2) -- [fermion] (b),
  };
\end{feynman}\draw[thick] ($(m)+(-0.10,0.10)$) -- ($(m)+(0.10,-0.10)$);
\draw[thick] ($(m)+(-0.10,-0.10)$) -- ($(m)+(0.10,0.10)$);
\node[below=2pt of m] {\small $\epsilon_\gamma$};
\end{tikzpicture}
\caption{Collider production: $q\bar q \rightarrow \gamma^\ast/\phi_\mu^\ast \rightarrow \chi\bar\chi$.}
\label{fig:darkphoton_collider}
\end{subfigure}
\caption{
\textbf{DM search experiments}. Representative search channels for fermionic DM $\chi$ interacting through a dark vector $\phi_\mu$ that couples to the SM via kinetic mixing with the photon, parametrized by $\epsilon_\gamma$. \emph{Left}: direct detection through DM--nucleus scattering. \emph{Middle}: indirect detection through DM annihilation into SM charged particles. \emph{Right}: collider production of DM pairs in proton--proton collisions via quark constituents. Analogous channels arise for other portal interactions, such as the Higgs and $Z$-boson portals. Portal mixings are indicated by crosses.
}
\label{fig:DM_search_experiments}
\end{figure*}
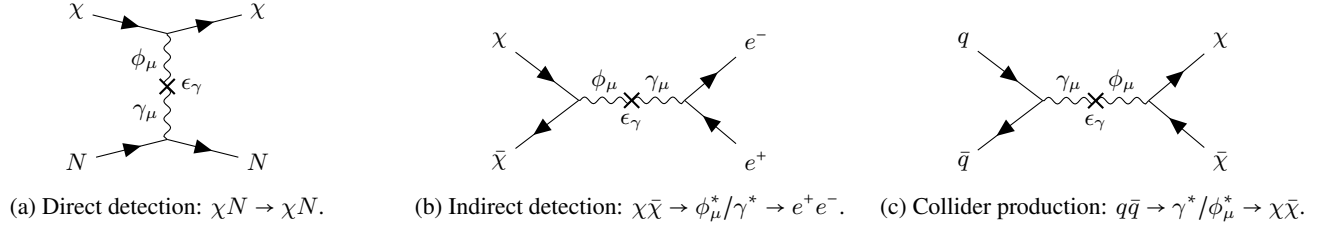

Many SIDM models generate sizeable and velocity-dependent self-interactions through a mediator that is much lighter than the DM, often at the MeV scale~\cite{Buckley:2009in,Feng:2009hw,Feng:2009mn,Loeb:2010gj,Kaplinghat:2015aga,Tulin:2013teo}. The mediator mass and dark-sector coupling determine the strength and velocity dependence of the DM self-scattering, while any coupling of the mediator to the SM particles opens complementary experimental and observational probes. Even a feeble portal interaction can therefore connect the self-interaction phenomenology relevant for structure formation to direct and indirect detection, collider and beam-dump searches, and cosmological or astrophysical constraints~\cite{Kaplinghat:2013yxa,Boddy:2014yra,Alexander:2016aln}. This complementarity is illustrated schematically in \cref{fig:DM_search_experiments}: the dark-sector interaction controls DM self-scattering, while the portal determines how the mediator communicates with the visible sector.

The portal interactions introduced in \cref{eq:photon_portal,eq:Z_portal,eq:higgs_portal_mediator,eq:higgs_portal_dm} specify this connection to the SM. Vector mediators can couple through photon kinetic mixing, parametrized by $\epsilon_\gamma$, or through mixing with the $Z$ boson, parametrized by $\epsilon_Z$, whereas scalar mediators can couple through the Higgs portal, with strength $\epsilon_h$; scalar DM may additionally interact with the Higgs through contact terms such as $\lambda_{sh}$. These couplings, indicated by crosses in \cref{fig:portal_couplings,fig:DM_search_experiments}, control the experimental visibility of the dark sector. Depending on the model, they determine DM--SM scattering rates, mediator production and decay rates, and the visible products of dark-sector annihilation or decay. They can also influence the thermal history and production mechanism of the dark sector discussed in \cref{sec:DM_production_mechanisms}.

Importantly, the interaction strength required for SIDM phenomenology and the portal coupling probed by experiments play distinct roles. The self-interaction regions shown in \cref{fig:effective_force_vector} are determined by the dark-sector masses and couplings under the assumption that the portal interaction does not significantly modify DM self-scattering. Self-interaction phenomenology alone therefore does not require a sizeable coupling to the SM. Cosmology and laboratory observations, however, can impose complementary constraints on this otherwise independent portal. If the coupling is too weak, a long-lived mediator can survive into cosmologically sensitive epochs and be constrained by Big Bang Nucleosynthesis (BBN), the Cosmic Microwave Background (CMB), or other limits on late energy injection. Increasing the portal coupling can shorten the mediator lifetime, but simultaneously enhances the signals targeted by direct-detection, indirect-detection, collider and fixed-target experiments. The viable parameter space is therefore determined by the simultaneous requirements of the desired DM self-interactions, an acceptable cosmological history and consistency with laboratory searches~\cite{Kaplinghat:2013yxa,Boddy:2014yra,Hambye:2019tjt}. This interplay is particularly restrictive in minimal light-mediator SIDM models, where the same small set of portal parameters controls both mediator phenomenology and the connection to the visible sector.

\medskip
\noindent
\textbf{Cosmological and astrophysical searches.}
Light mediators in SIDM models are strongly constrained by cosmology and astrophysics~\cite{Kaplinghat:2013yxa,Boddy:2014yra,Fradette:2014sza,Adams:1998nr,Chen:2003gz,Zhang:2007zzh,Galli:2009zc,Slatyer:2009yq}. A first class of constraints comes from the CMB. If DM annihilates during recombination at $z \simeq 1100$ into electromagnetically interacting particles, or into light mediators that later decay into such states, the injected energy can ionize neutral hydrogen, broaden the last-scattering surface and suppress the observed CMB temperature-anisotropy correlations~\cite{Slatyer:2015jla,Bringmann:2016din,Cirelli:2016rnw}. Energy injection at earlier epochs ($5\times10^4\lesssim z\lesssim2\times10^6$) can also produce so-called $\mu$-type spectral distortions of the nearly perfect blackbody spectrum of the CMB, providing complementary constraints on Sommerfeld-enhanced DM annihilation, particularly near resonances~\cite{Zavala:2009mi}. This gives an upper bound on the annihilation rate at recombination. The constraint is especially severe for vector-mediated models, where annihilation is typically s-wave and can be strongly Sommerfeld enhanced at the small velocities relevant at recombination. For scalar-mediated models, annihilation is typically p-wave suppressed, making the CMB bound much weaker.

A second class of CMB constraints applies to the mediator itself. If the mediator is still abundant at recombination and decays visibly, its decay products can inject electromagnetic energy and distort the CMB anisotropy spectrum~\cite{Poulin:2016anj,Slatyer:2012yq}. If it survives until after neutrino decoupling, it can reheat the photon bath but not the neutrinos, thereby modifying the effective number of relativistic degrees of freedom, $N_{\rm eff}$. Since \emph{Planck} constrains this quantity to $2.66 < N_{\rm eff} < 3.33$ at $2\sigma$~\cite{Planck:2018vyg}, this bounds the mediator lifetime and abundance~\cite{Hufnagel:2018bjp}. For even longer lifetimes, mediator decays are also constrained by X-ray observations~\cite{Essig:2013goa,Boddy:2015efa,Riemer-Sorensen:2015kqa}.

Comparable constraints arise from BBN. If the mediator decays during or after nucleosynthesis, it can photodissociate light nuclei, inject entropy or modify the Hubble expansion rate while the primordial light-element abundances are being established~\cite{Hufnagel:2018bjp,Sarkar:1998gx,Kirkman:2003uv}. These bounds depend on the mediator abundance and lifetime, on whether the mediator decouples while relativistic, and on whether its decay products are electromagnetic or invisible.

Astrophysical observations provide additional constraints. In particular, SN1987A can probe MeV-scale mediators because such particles may be produced efficiently in the hot and dense supernova core and contribute to its cooling~\cite{Kazanas:2014mca,Chang:2016ntp,Mahoney:2017jqk,Chang:2018rso}. In the vector portal case, dark photons can be produced through their mixing with ordinary photons; in the scalar portal case, light scalars can be emitted from the core and shorten the observed neutrino burst. Although supernova bounds have larger systematic uncertainties than cosmological constraints, they can still remove additional regions of light-mediator parameter space~\cite{Kazanas:2014mca,Chang:2016ntp,Mahoney:2017jqk,Chang:2018rso}.

\medskip
\noindent
\textbf{Direct detection.}
DM models can also be constrained by particle-physics searches. Direct detection experiments search for DM scattering off nuclei in terrestrial detectors. An incoming DM particle can transfer momentum to a nucleus, producing a recoil signal. Comparing the observed recoil spectrum to expected backgrounds then yields upper bounds on the DM--nucleus scattering cross section.

In SIDM models with portal interactions, the same mediator that generates DM self-interactions can mediate elastic scattering between DM and ordinary matter, $\chi N \rightarrow \chi N$, as illustrated in \cref{fig:darkphoton_direct}. This links dark-sector microphysics directly to laboratory observables. In light-mediator models, the interaction is not contact-like, and the recoil spectrum is typically more strongly peaked at low recoil energies than in heavy-mediator WIMP scenarios~\cite{Fornengo:2011sz,Li:2014vza,DelNobile:2015uua}. Existing limits from LUX, SuperCDMS, PandaX, XENON1T and LZ already exclude significant parts of minimal portal SIDM parameter space~\cite{LUX:2016ggv,PandaX-II:2017hlx,XENON:2018voc,LZ:2024zvo}.

For Higgs-portal scalar mediators, direct-detection constraints are especially severe because the mediator decay width to electrons is suppressed by the small electron Yukawa coupling, whereas the DM--nucleon interaction proceeds through the Higgs coupling to nucleons and is not suppressed to the same degree. The region where the scalar mediator decays early enough to satisfy BBN while evading direct detection is therefore extremely small. In the minimal scalar scenario, only a narrow region around $m_{\rm med}\sim 1~{\rm MeV}$ and $m_{\rm DM}\sim 0.5~{\rm GeV}$ remains viable, and this region is further reduced by supernova constraints~\cite{Kaplinghat:2013yxa,Hufnagel:2018bjp,Hambye:2019tjt}. According to the lower-left panel of
\cref{fig:effective_force_vector},
this remaining region can still yield
astrophysically relevant velocity-dependent
self-interactions for
$\alpha_\phi\sim5\times10^{-5}$.

The corresponding tension is milder in the vector case because both the dark-photon decay width and the direct-detection rate scale with the same kinetic-mixing parameter. However, the minimal vector-mediator scenario faces a different problem: for fermionic DM, annihilation into light vector mediators is typically s-wave, and the recombination-era annihilation rate is strongly enhanced by the Sommerfeld effect. This leads to powerful CMB constraints, which exclude the simplest thermal vector model across essentially all phenomenologically relevant parameter space when the mediator decays visibly~\cite{Kaplinghat:2013yxa,Boddy:2014yra,Hambye:2019tjt}. By contrast, scalar-mediated annihilation into light scalars is typically p-wave, which weakens late-time annihilation bounds and helps reopen part of the scalar parameter space despite stronger direct-detection tension.

\begin{figure*}
\centering
\resizebox{0.7\textwidth}{!}{%
\begin{tikzpicture}[
    x=1cm,
    y=1cm,
    >={Stealth[length=3.0mm,width=2.3mm]},
    box/.style={
        rounded corners=3mm,
        line width=1.5pt,
        fill=white,
        inner sep=0pt
    }
]

\definecolor{sidmblue}{RGB}{18,48,235}
\definecolor{prodgreen}{RGB}{0,74,13}
\definecolor{cosmopurple}{RGB}{76,13,157}
\definecolor{exptred}{RGB}{246,20,17}
\definecolor{exptdarkred}{RGB}{180,0,0}

\node[
    box,
    draw=sidmblue,
    minimum width=6.55cm,
    minimum height=3.30cm,
    anchor=north west
] (selfbox) at (0.20,10.00) {};

\node[
    text=sidmblue,
    font=\bfseries\fontsize{14.5}{16}\selectfont,
    anchor=north
] at (3.475,9.82)
{Self-interactions};

\node[
    anchor=north west,
    text width=5.90cm,
    align=left,
    font=\fontsize{7.75}{12}\selectfont
] at (0.45,9.12) {%
$\bullet$~~Small-scale structure may require\\[-0.02cm]
\hspace*{0.41cm}sizeable SIDM cross sections\\[0.19cm]
$\bullet$~~This favors sufficiently large $\alpha_{\rm DM}$\\[0.19cm]
$\bullet$~~Constrains $m_{\rm DM}$, $m_{\rm med}$ and $\alpha_{\rm DM}$
};

\node[
    box,
    draw=prodgreen,
    minimum width=6.75cm,
    minimum height=3.30cm,
    anchor=north west
] (prodbox) at (8.05,10.00) {};

\node[
    text=prodgreen,
    font=\bfseries\fontsize{14.5}{16}\selectfont,
    anchor=north
] at (11.425,9.82)
{DM production};

\node[
    anchor=north west,
    text width=6.15cm,
    align=left,
    font=\fontsize{7.75}{12}\selectfont
] at (8.32,9.12) {%
$\bullet$~~Must reproduce the observed relic abundance\\[0.19cm]
$\bullet$~~Freeze-out / freeze-in / reannihilation /\\[-0.02cm]
\hspace*{0.41cm}asymmetric DM / \ldots\\[0.19cm]
$\bullet$~~Further restricts $m_{\rm DM}$, $m_{\rm med}$,
$\alpha_{\rm DM}$ and $\epsilon_{\rm DM}$
};

\node[
    box,
    draw=black,
    minimum width=6.82cm,
    minimum height=2.18cm,
    anchor=north west
] (centralbox) at (4.18,5.90) {};

\node[
    anchor=north,
    align=center,
    font=\bfseries\fontsize{14.5}{15.5}\selectfont
] at (7.59,5.67)
{Minimal light-mediated\\[-0.03cm]SIDM};

\node[
    anchor=north,
    align=center,
    font=\fontsize{8.5}{11}\selectfont
] at (7.59,4.55)
{Key parameters: $m_{\rm DM}$, $m_{\rm med}$,
$\alpha_{\rm DM}$ and $\epsilon_{\rm DM}$};

\node[
    box,
    draw=cosmopurple,
    minimum width=5.65cm,
    minimum height=2.68cm,
    anchor=south west
] (cosmobox) at (0.20,0.15) {};

\node[
    text=cosmopurple,
    font=\bfseries\fontsize{13.5}{15}\selectfont,
    anchor=north
] at (3.025,2.62)
{Cosmology \& astrophysics};

\node[
    anchor=north west,
    text width=5.10cm,
    align=left,
    font=\fontsize{7.75}{11.5}\selectfont
] at (0.43,1.94) {%
$\bullet$~~BBN, CMB, and supernova bounds\\[0.17cm]
$\bullet$~~Mediator should decay sufficiently early\\[0.17cm]
$\bullet$~~This requires $\epsilon_{\rm DM}$ not too small
};

\node[
    box,
    draw=exptred,
    minimum width=5.80cm,
    minimum height=2.68cm,
    anchor=south west
] (expbox) at (9.02,0.15) {};

\node[
    text=exptred,
    font=\bfseries\fontsize{13}{14.5}\selectfont,
    anchor=north
] at (11.92,2.62)
{Particle-physics experiments};

\node[
    anchor=north west,
    text width=5.30cm,
    align=left,
    font=\fontsize{7.75}{11.5}\selectfont
] at (9.23,1.94) {%
$\bullet$~~Direct detection, indirect detection, colliders\\[0.17cm]
$\bullet$~~Typically require $\epsilon_{\rm DM}$ to be small\\[0.17cm]
$\bullet$~~Creates tension with cosmology
};

\draw[
    ->,
    sidmblue,
    line width=1.6pt
] (4.82,6.70) -- (5.27,5.98);

\node[
    text=sidmblue,
    font=\itshape\fontsize{9}{11}\selectfont,
    anchor=south
] at (3.28,6.13)
{favours larger $\alpha_{\rm DM}$};

\draw[
    ->,
    prodgreen,
    line width=1.6pt
] (10.08,6.70) -- (9.66,5.98);

\node[
    text=prodgreen,
    font=\itshape\fontsize{9}{11}\selectfont,
    anchor=south
] at (11.52,6.13)
{fixes relic density};

\draw[
    ->,
    cosmopurple,
    line width=1.6pt
] (4.96,2.83) -- (5.34,3.66);

\node[
    text=cosmopurple,
    font=\itshape\fontsize{9}{11}\selectfont,
    anchor=south
] at (2.83,3.02)
{needs mediator decay};

\draw[
    ->,
    exptred,
    line width=1.6pt
] (9.89,2.83) -- (9.48,3.66);

\node[
    text=exptred,
    font=\itshape\fontsize{9}{11}\selectfont,
    anchor=south
] at (12.18,3.02)
{limits portal coupling};

\node[
    text=exptdarkred,
    align=center,
    font=\bfseries\fontsize{8.5}{11.5}\selectfont
] at (7.44,1.55)
{Competing\\requirements\\on $\epsilon_{\rm DM}$};

\draw[
    ->,
    exptdarkred,
    line width=1.6pt
] (6.5,1.55) -- (5.92,1.55);

\draw[
    ->,
    exptdarkred,
    line width=1.6pt
] (8.37,1.55) -- (8.94,1.55);

\end{tikzpicture}
}
\caption{\centering
Schematic overview of combined SIDM constraints, illustrated with minimal light-mediated SIDM.
}
\label{fig:light_mediator_SIDM_constraints}
\end{figure*}
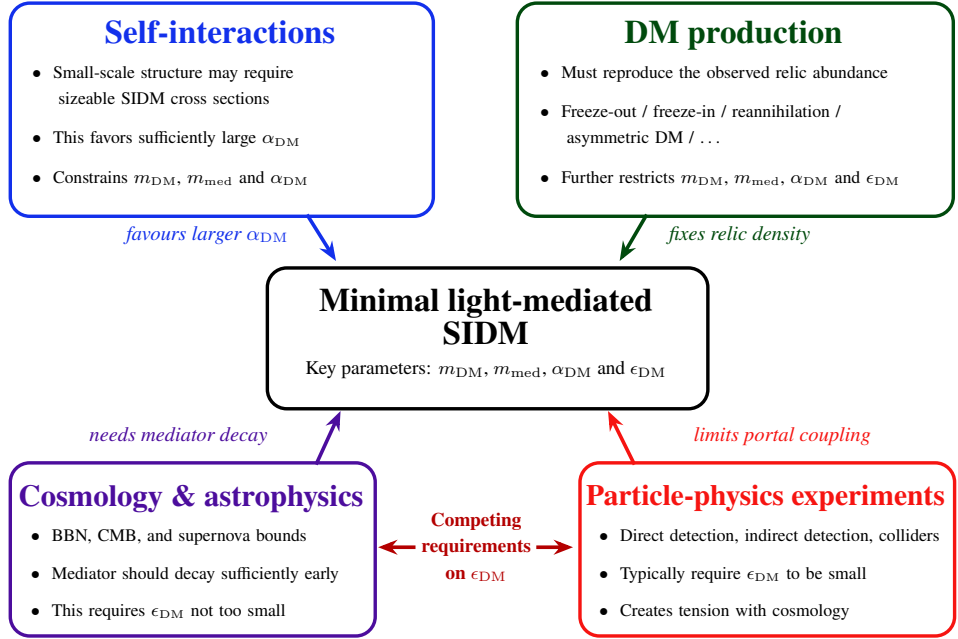

\medskip
\noindent
\textbf{Indirect detection.} 
Indirect detection searches for SM particles produced when DM annihilates or decays in astrophysical environments such as the Milky Way or dwarf galaxies. These experiments measure fluxes of photons, electrons, positrons or other particles and compare them to astrophysical backgrounds to constrain the DM annihilation rate.

In light-mediator SIDM models, DM often annihilates into mediator states that subsequently decay into SM particles~\cite{Kaplinghat:2013yxa,Kaplinghat:2015gha}, e.g.\ electron--positron pairs from $\chi\bar\chi \rightarrow \phi_\mu^\ast/\gamma^\ast \rightarrow e^+e^-$ as shown in \cref{fig:darkphoton_indirect}. A mediator much lighter than the DM particle can significantly enhance the late-time annihilation rate through the Sommerfeld effect, especially at the small DM velocities relevant today~\cite{Sommerfeld:1931qaf,Hisano:2004ds,Arkani-Hamed:2008hhe,Cassel:2009wt,Slatyer:2009vg,Feng:2010zp,Blum:2016nrz,Tulin:2013teo}.

This enhancement is particularly important for s-wave annihilation, where the tree-level cross section is velocity-independent and can generate large indirect-detection signals. For p-wave annihilation, the rate is naively suppressed by $v^2$ and would usually be negligible today. However, with a large DM--mediator mass hierarchy, Sommerfeld enhancement can partially compensate for this suppression, leading to an effective scaling $\propto 1/v$. The signal strength also depends on the astrophysical environment: typical DM velocities are $v_{\rm MW}\sim 10^{-3}$ in the Milky Way and $v_{\rm DG}\sim 10^{-5}$ in dwarf galaxies, so Sommerfeld effects can be much larger in dwarfs.

Current constraints from Fermi-LAT, AMS-02, Planck and related probes therefore strongly limit the annihilation cross section in SIDM models~\cite{Chen:2003gz,Galli:2009zc,Slatyer:2009yq,
Madhavacheril:2013cna,Lopez-Honorez:2013cua,Bringmann:2016din,Cirelli:2016rnw}. The bounds depend on mediator spin, portal structure, dark-sector thermal history and whether the DM is symmetric or asymmetric. In particular, asymmetric SIDM, discussed in \cref{sec:Asymmetric_SIDM}, suppresses late-time annihilation signals and can weaken indirect-detection constraints.

\medskip
\noindent
\textbf{Collider constraints.} 
Collider experiments probe DM by attempting to produce it directly in high-energy collisions. Since DM interacts only weakly with the SM, it typically escapes the detector and produces missing transverse momentum. Representative processes include $e^- e^+\to \gamma^\ast/\phi_\mu^\ast\to\chi\bar\chi$ in electron--positron colliders and $q\bar q\to \gamma^\ast/\phi_\mu^\ast\to\chi\bar\chi$ in proton--proton colliders, as shown in \cref{fig:darkphoton_collider}. If the light mediator is radiated and decays into SM particles, additional visible signatures such as lepton jets or displaced vertices can arise~\cite{Birkedal:2004xn,Feng:2005gj,Goodman:2010yf,Bai:2010hh,Buschmann:2015awa,ATLAS:2018txj,CMS:2019ctu}. For very weak portal couplings, fixed-target, beam-dump and low-energy $e^+e^-$ experiments provide complementary searches for mediator decays or missing-energy signatures~\cite{Batell:2009yf,Essig:2009nc,Alexander:2016aln}.

In scalar portal models, collider constraints are often driven by invisible Higgs decays, $h \to \phi\phi$. LHC measurements constrain the invisible branching ratio to $\mathrm{BR}(h\to \mathrm{inv.}) \lesssim 0.19$ at 95\% C.L., corresponding to $\Gamma_{\rm inv} \lesssim 0.5$--$1~\mathrm{MeV}$~\cite{CMS:2018yfx,ATLAS:2016neq}. This typically implies $\lambda_{\phi h} \lesssim 0.01$. For vector mediators, collider and fixed-target experiments constrain $\epsilon_\gamma$. For example, LHCb searches for dark photon decays such as $\gamma'\to \mu^+\mu^-$ exclude $\epsilon_\gamma \sim 10^{-5}$--$10^{-4}$ for $m_{\gamma'} \sim 200$--$300~\mathrm{MeV}$~\cite{LHCb:2019vmc}, with additional constraints from many other experiments~\cite{Jaeckel:2010ni,Berezhiani:2000gw,McDermott:2010pa}.

\subsection{Interplay of constraints on SIDM}
\label{sec:interplay_constraints}

Minimal light-mediated SIDM provides a particularly transparent benchmark for illustrating how astrophysical, cosmological and particle-physics information jointly constrains the underlying dark-sector parameters. The different probes are highly complementary: self-interactions, relic production, cosmological evolution and laboratory searches constrain different combinations of $m_{\rm DM}$, $m_{\rm med}$, $\alpha_{\rm DM}$ and the portal coupling $\epsilon_{\rm DM}$. Their interplay therefore provides a powerful way to identify which microscopic realizations are viable. The main requirements are:

\begin{itemize}[leftmargin=*, labelsep=0.5em]
\item \emph{Self-interactions:} astrophysical observations of small-scale structure can require sizable DM self-interactions and therefore a sufficiently large DM--mediator coupling $\alpha_{\mathrm{DM}}$. The corresponding constraints on the mediator and DM masses and couplings are shown in \cref{fig:effective_force_vector}, and are set in part by constraints on the self-interaction cross section inferred from a range of astrophysical systems, summarized in \cref{fig:cross_sections} and taken from~\cite{Kaplinghat:2015aga,Sagunski:2020spe,2021MNRAS503920C};
\item \emph{DM production:} the observed relic abundance must be reproduced through a viable production mechanism -- freeze-out, freeze-in, reannihilation or asymmetric production (see \cref{sec:DM_production_mechanisms}). Together with the self-interaction requirement above, this further constrains the allowed DM and mediator masses and couplings, as shown in \cref{fig:effective_force_vector};
\item \emph{Cosmology and astrophysics:} if light mediators are appreciably populated in the early Universe, BBN, CMB and supernova bounds can require them to decay sufficiently early and to avoid excessive late-time energy injection. In scenarios where this depletion proceeds through decays into SM particles, this typically imposes a lower bound on the portal coupling $\epsilon_{\mathrm{DM}}$ (see \cref{sec:DM_production_mechanisms});
\item \emph{Particle-physics experiments:} direct detection, indirect detection and collider searches constrain the same portal couplings $\epsilon_{\mathrm{DM}}$ to be small (see \cref{sec:DM_search_experiments}).
\end{itemize}

\noindent
This interplay is summarized schematically in \cref{fig:light_mediator_SIDM_constraints}. In minimal light-mediated SIDM, the self-interaction phenomenology primarily constrains $m_{\rm DM}$, $m_{\rm med}$ and $\alpha_{\rm DM}$, while the DM production history can additionally depend on the portal coupling $\epsilon_{\rm DM}$. Cosmological and particle-physics constraints further restrict these parameters, particularly $\epsilon_{\rm DM}$. Together, these complementary requirements sharply delineate the viable parameter space and reveal the correlations among the underlying particle-physics parameters. Similar interplay arises in many other SIDM model classes.

In the simplest Higgs-portal SIDM model with a
scalar mediator, for example, combining all
constraints leaves at most a narrow viable
window around $m_{\rm med}\sim1~{\rm MeV}$
and $m_{\rm DM}\sim0.5~{\rm GeV}$, where
$\alpha_\phi\sim5\times10^{-5}$ yields
astrophysically relevant velocity-dependent self-interactions (lower-left panel of
\cref{fig:effective_force_vector}),
while the relic density depends on the
production mechanism discussed in
\cref{sec:DM_production_mechanisms}
~\cite{Hufnagel:2018bjp,Hambye:2019tjt}.
 For vector mediators with kinetic mixing, the direct-detection tension is milder, but such scenarios are typically excluded by CMB constraints because annihilation into light mediators proceeds through an unsuppressed $s$-wave process~\cite{Kaplinghat:2013yxa,Boddy:2014yra,Bringmann:2016din,Cirelli:2016rnw,Hambye:2019tjt}. More generally, minimal thermal light-mediated SIDM models are either excluded or only marginally viable once all constraints are imposed together.

\subsection{A broad landscape of viable SIDM models}
\label{sec:reopen_parameter_space}

The strong constraints discussed above apply primarily to the most minimal thermal light-mediator realizations, rather than to SIDM as a broader particle-physics framework. A wide range of theoretically well-motivated dark-sector constructions naturally evade these restrictions and open substantial regions of viable parameter space, including modified thermal histories, asymmetric and inelastic sectors, additional dark states and composite dynamics. The most important possibilities are:

\begin{itemize}[leftmargin=*, labelsep=0.5em]
\item \emph{Modified thermal history:} 
A colder dark sector ($T_{\mathrm{DM}} < T_{\mathrm{SM}}$), or a subdominant annihilation into light mediators, reduces the required DM--mediator coupling and thereby weakens the cosmological, direct- and indirect-detection constraints~\cite{Chu:2016pew,Hambye:2019tjt}.

\item \emph{$p$-wave annihilation:} 
Velocity-suppressed annihilation relaxes the CMB and late-time indirect-detection bounds. This is one reason why scalar-mediated models can remain viable where the corresponding vector-mediated models are excluded~\cite{Bringmann:2016din,Cirelli:2016rnw,Hambye:2019tjt}.

\item \emph{Mediator depletion or hidden-sector dynamics:} The mediator abundance can be reduced through decays into additional dark-sector states, or through alternative annihilation channels, weakening the BBN and CMB constraints~\cite{Duerr:2018mbd,Ma:2017ucp,Hambye:2019tjt}.

\item \emph{Neutrino-dominated decays:} 
If the mediator decays predominantly into neutrinos, the bounds from electromagnetic energy injection at the CMB epoch are relaxed. This can be realized, e.g., through mass mixing with the $Z$ boson or in a gauged $L_\mu - L_\tau$ scenario~\cite{Bringmann:2016din,Kamada:2018zxi,Kamada:2018kmi,Garani:2019fpa}.

\item \emph{Suppressed direct detection:} 
Spin-dependent, velocity-suppressed, pseudoscalar or axial interactions can reduce the direct-detection signal, although often at the cost of modifying the 
self-interaction phenomenology~\cite{Kahlhoefer:2017umn,Agrawal:2020lea}.

\item \emph{Inelastic DM:}
Off-diagonal couplings between nearly degenerate states can kinematically suppress direct-detection signals while maintaining sizable self-interactions~\cite{Schutz:2014nka,Zhang:2016dck,Blennow:2016gde}. We discuss this model class in \cref{sec: inelastic_sidm_models}.

\item \emph{Asymmetric DM:} 
If the relic abundance is set by a primordial asymmetry rather than by thermal freeze-out, the link between annihilation and abundance is broken. This weakens the CMB and indirect-detection constraints and can reopen large regions of parameter space, 
including purely repulsive interactions (lower-right panel of \cref{fig:effective_force_vector})~\cite{Baldes:2017gzu,Hambye:2019tjt}. We discuss this model class in \cref{sec:Asymmetric_SIDM}.

\item \emph{Gravitational DM production:}
In contrast to the production mechanisms discussed in \cref{sec:DM_production_mechanisms}, DM may be produced gravitationally during or after inflation, including through graviton-mediated reheating processes, allowing negligibly small portal couplings to the SM and thereby decoupling the relic abundance from constraints on $\epsilon_{\rm DM}$. The dark-sector coupling $\alpha_{\rm DM}$ can still remain large enough to generate sizable self-interactions, although the viability depends on the inflationary/reheating history and on any accompanying light mediator~\cite{Ema:2018ucl,Ema:2019yrd,Mambrini:2021zpp}.
\end{itemize}

\noindent
Further possibilities include small DM-to-mediator mass ratios~\cite{Chu:2016pew}, stable or long-lived mediators depleted by hidden-sector annihilations~\cite{Duerr:2018mbd,Hambye:2019tjt}, and extended dark sectors with multiple states and interactions, see e.g.~\cite{Hambye:2019tjt}. Composite DM provides an alternative microscopic route to viable SIDM, in which sizeable self-interactions arise from the dynamics of bound states rather than from the exchange of an elementary light mediator. We discuss various composite SIDM model classes in \cref{sec:composite_SIDM}. Thus, as discussed further later in Chapter~\ref{sec:pp_to_simulation_inputs}, where we consider a range of representative dark-sector model classes, minimal light-mediated SIDM models are highly constrained, whereas relatively simple extensions and qualitatively distinct realizations such as composite SIDM can reopen substantial regions of viable parameter space and give rise to additional phenomenology that independently motivates these scenarios.

\medskip
The probes discussed above constrain complementary parts of the same underlying particle-physics framework. Laboratory searches primarily test the portal interactions connecting the dark and visible sectors, while astrophysical observations probe the strength and structure of DM self-interactions. Connecting these microscopic properties to structure formation, however, requires translating the particle-physics model into the quantities that govern the simulated DM dynamics. In the next chapter, we formulate this theory-to-simulation mapping and apply it to representative SIDM model classes.

\section{Mapping particle physics to simulations: framework and application}
\label{sec:pp_to_simulation_inputs}

Connecting a microscopic dark-sector theory to a structure-formation simulation requires more than specifying a cross section. The particle-physics model determines a coupled set of scattering rates, angular distributions, internal-state transitions, energy exchange, dissipation and decay processes, each of which must be translated into quantities that can be evaluated consistently during the dynamical evolution of DM. We therefore construct a unified theory-to-simulation framework that identifies the simulation-level quantities required to represent this microphysics and provides a systematic procedure for deriving them from the underlying particle theory. We denote this mapping schematically as
\begin{equation}
\mathcal{P}_{\rm PP}\rightarrow\mathcal{P}_{\rm SIM},
\end{equation}
where $\mathcal{P}_{\rm PP}$ contains the microscopic particle content, masses, couplings and interaction structure, while $\mathcal{P}_{\rm SIM}$ contains the quantities required for the numerical evolution. Importantly, the mapping can be degenerate: microscopically distinct theories may yield nearly identical simulation inputs over the velocities and physical conditions probed by a given astrophysical system.

In this chapter, we first establish the general structure of $\mathcal{P}_{\rm SIM}$ and derive how its components follow from $\mathcal{P}_{\rm PP}$. We then apply this framework systematically to representative SIDM model classes constructed from the particle-physics ingredients developed in Chapter~\ref{sec: sidm_pp_models}, explicitly translating their microscopic parameter spaces into simulation inputs and the corresponding \arepoTwo\ implementations.

\begin{figure*}[t]
    \centering
    \begin{tikzpicture}[
        node distance = 0.6cm and 2cm,
        mainbox/.style = {
            draw=darkblue, line width=1.2pt, rounded corners=4pt, 
            fill=blue!8, 
            drop shadow={shadow xshift=1.5pt, shadow yshift=-1.5pt, opacity=0.3},
            minimum width=3.5cm, minimum height=0.9cm,
            font=\normalsize\bfseries, align=center
        },
        subbox/.style = {
            draw=darkgreen!80!black, line width=0.8pt, rounded corners=3pt, 
            fill=lightgreen!30,
            minimum width=3.3cm, minimum height=0.65cm,
            font=\footnotesize, align=center
        },
        parambox/.style = {
            draw=darkorange, line width=0.8pt, rounded corners=3pt, 
            fill=lightorange!20,
            minimum width=3.5cm, minimum height=0.7cm,
            font=\footnotesize, align=center
        },
        algobox/.style = {
            draw=darkblue, line width=0.8pt, rounded corners=3pt, 
            fill=blue!8,
            minimum width=2.8cm, minimum height=0.65cm,
            font=\footnotesize, align=center
        },
        arrow/.style = {-Stealth, line width=1.2pt, darkblue},
        smallarrow/.style = {-Stealth, line width=0.6pt, darkblue},
        greenarrow/.style = {-Stealth, line width=1.2pt, darkgreen!80!black},
        greenline/.style = {line width=1.2pt, darkgreen!80!black},
        smallgreenarrow/.style = {-Stealth, line width=0.6pt, darkgreen!80!black},
        tinygreenarrow/.style = {-Stealth, line width=0.5pt, darkgreen!80!black},
        orangearrow/.style = {-Stealth, line width=0.8pt, darkorange},
        label/.style = {font=\scriptsize\itshape, text=darkblue},
        greenlabel/.style = {font=\scriptsize\itshape, text=darkgreen!80!black},
        bluelabel/.style = {font=\scriptsize\itshape, text=darkblue},
        orangelabel/.style = {font=\scriptsize\itshape, text=darkorange}
    ]
    
    \node[mainbox, fill=lightorange!30, draw=darkorange] (pp)
    {Particle Physics Model\\[1pt]\scriptsize (Lagrangian: Fields, symmetries \& interactions)};

    \node[parambox, below=0.4cm of pp] (params)
    {\textbf{Model Parameters}\\[1pt] Masses, couplings and \\ characteristic scales};

    \node[parambox, below=0.3cm of params] (feynman)
    {\textbf{Microscopic Processes}\\[1pt] Scattering, annihilation and decay};
    
    \draw[arrow, darkorange] (pp) -- (params);
    \draw[smallarrow, darkorange] (params) -- (feynman);
    
    \node[parambox, below=0.6cm of feynman] (constraints)
    {\textbf{External Constraints}\\[1pt]  Relic abundance, cosmology,\\astrophysics and laboratory searches};
    
    \draw[orangearrow] (feynman) -- (constraints);
    
    \node[mainbox, fill=lightgreen!40, draw=darkgreen!80!black, below left=1.3cm and 1.5cm of constraints] (ic)
    {Initial Conditions\\[1pt]\scriptsize (Linear evolution)};
    
    \node[mainbox, fill=lightgreen!40, draw=darkgreen!80!black, below right=1.3cm and 1.5cm of constraints] (nle)
    {Nonlinear Evolution\\[1pt]\scriptsize (Halo dynamics)};
    
    \draw[arrow, darkorange] (constraints.south) -- ++(0,-0.5) -| (ic.north);
    \draw[arrow, darkorange] (constraints.south) -- ++(0,-0.5) -| (nle.north);
    
    \coordinate (orange_split) at ($(constraints.south) + (0,-0.5)$);
    
    \node[orangelabel, below=0.05cm of orange_split]
    {\normalsize From theory to simulation inputs};
    
    \node[subbox, below=0.6cm of ic] (boltzmann)
    {\textbf{Boltzmann Solver}\\[1pt] \textsc{CAMB} / \textsc{CLASS}};
    
    \node[subbox, below=0.25cm of boltzmann] (transfer)
    {\textbf{Transfer function}\\[1pt] $T(k)$};
    
    \node[subbox, below=0.25cm of transfer] (pk)
    {\textbf{Power spectrum}\\[1pt] $P(k)$};
    
    \node[subbox, below=0.25cm of pk] (icgen)
    {\textbf{IC Generator}\\[1pt] \textsc{MUSIC} / \textsc{N-GenIC}};
    
    \node[subbox, below=0.25cm of icgen] (icfiles)
    {\textbf{IC file} \\[1pt] name.hdf5 / name.ics};
    
    \draw[greenarrow] (ic) -- (boltzmann);
    \draw[smallgreenarrow] (boltzmann) -- (transfer);
    \draw[smallgreenarrow] (transfer) -- (pk);
    \draw[smallgreenarrow] (pk) -- (icgen);
    \draw[smallgreenarrow] (icgen) -- (icfiles);
    
\node[subbox, below=0.6cm of nle] (states)
{\textbf{Mass splittings \& state fractions}\\[1pt] 
$\Delta_{i}$ and $f_{i}^{\rm init}$};

\node[subbox, below=0.25cm of states] (network)
{\textbf{Reaction network}\\[1pt] 
$\mathcal{R}=\{ij \to kl\}$};

\node[subbox, below=0.25cm of network] (dissipation)
{\textbf{Dissipative energy-loss fractions}\\[1pt]
$f_{\rm loss}^{\mathcal{R}}(v_{ij})$};

\node[subbox, below=0.25cm of dissipation] (cross)
{\textbf{Cross sections}\\[1pt] 
$\displaystyle\frac{d\sigma_\mathcal{R}}{d\Omega}(v_{\rm rel},\theta)$, 
$\sigma_X^\mathcal{R}(v_{\rm rel})$};

\node[subbox, below=0.25cm of cross] (decay)
{\textbf{Decay rates \& velocity kicks}\\[1pt] 
$\Gamma $ and $v_{\rm kick}^{\rm decay}$};

\draw[greenarrow] (nle) -- (states);

\begin{scope}[on background layer]
    \node[draw=darkorange!70, dashed, line width=1.0pt, rounded corners=6pt, fill=lightorange!10, 
          fit=(params)(feynman), inner sep=6pt] {};
    \node[draw=darkorange!70, dashed, line width=1.0pt, rounded corners=6pt, fill=lightorange!10, 
          fit=(constraints), inner sep=8pt] {};
    \node[draw=darkgreen!70!black, dashed, line width=1.0pt, rounded corners=6pt, fill=lightgreen!12, 
          fit=(states)(network)(dissipation)(cross)(decay), inner sep=6pt] {};
    \node[draw=darkgreen!70!black, dashed, line width=1.0pt, rounded corners=6pt, fill=lightgreen!12, 
          fit=(boltzmann)(transfer)(pk)(icgen)(icfiles), inner sep=6pt] {};
\end{scope}
    
    \coordinate (midpoint) at ($(icfiles.south)!0.5!(decay.south)$);
    
    \coordinate (pipeline_center) at ($(ic.south)!0.5!(nle.south) + (0,-2.5)$);
    
    \node[greenlabel] at (pipeline_center) {\normalsize \parbox{4cm}{\centering Theory-to-simulation pipeline}};
    
    \draw[tinygreenarrow] (pipeline_center) ++(-1.7,0.25) -- ++(-0.7,0);
    \draw[tinygreenarrow] (pipeline_center) ++(1.7,0.25) -- ++(0.6,0);
    
    \coordinate (horizontal_level) at ($(midpoint) + (0,-0.85)$);
    
    \coordinate (green_split) at (horizontal_level);
    
    \node[bluelabel, above=0.05cm of green_split]
    {\normalsize Simulation inputs into \arepoTwo};
    
    \node[mainbox, below=0.5cm of horizontal_level, xshift=-4cm, minimum width=7cm, minimum height=0.7cm, fill=blue!8] (newsidm)
    {\parbox{7.5cm}{\centering\normalsize\bfseries Our SIDM Module\\[1pt]\scriptsize (Dedicated DM tree, hypercube communication \& modular interface)}};
    
    \node[mainbox, below=0.5cm of horizontal_level, xshift=4cm, minimum width=7cm, minimum height=0.7cm, fill=blue!8] (newddm)
    {\parbox{7.5cm}{\centering\normalsize\bfseries Our Decaying DM Module\\[1pt]\scriptsize (State-based decaying DM)}};
    
    \draw[greenline] (icfiles.south) -- (icfiles.south |- horizontal_level);
    \draw[greenline] (decay.south) -- (decay.south |- horizontal_level);
    \draw[greenline] (icfiles.south |- horizontal_level) -- (decay.south |- horizontal_level);
    \draw[greenarrow] (newsidm.north |- horizontal_level) -- (newsidm.north);
    \draw[greenarrow] (newddm.north |- horizontal_level) -- (newddm.north);
    
    \coordinate (modules_midpoint) at ($(newsidm.south)!0.5!(newddm.south)$);
    
    \node[mainbox, below=0.75cm of modules_midpoint, minimum width=7cm, fill=blue!8] (algos)
    {Dark Matter Algorithms\\[1pt]\scriptsize (New \arepoTwo implementations)};
    
    \draw[arrow] (newsidm.south) -- ++(0,-0.3) -| (algos.north);
    \draw[arrow] (newddm.south) -- ++(0,-0.3) -| (algos.north);
    
    \node[algobox, below=0.45cm of algos, xshift=-3.75cm] (algo1)
    {\textbf{Angular scattering}\\[1pt]  
    \textbf{\hyperref[sec:anisotropic_MC]{aMC}} / \textbf{\hyperref[sec: fsidm]{fSIDM}} / \textbf{\hyperref[sec:asidm]{aSIDM}}};
    
    \node[algobox, right=0.3cm of algo1] (algo2)
    {\textbf{Velocity-independent scattering} \\[1pt] \textbf{\hyperref[sec:mc_velocity_dep]{MC}}};
    
    \node[algobox, right=0.3cm of algo2] (algo3)
    {\textbf{Velocity-dependent scattering} \\[1pt] \textbf{\hyperref[sec:mc_velocity_dep]{vd-MC}}};
    
    \node[algobox, below=0.3cm of algo1] (algo4)
    {\textbf{Dissipative scattering} \\[1pt] \textbf{\textbf{\hyperref[sec:imp_dissipative_SIDM]{dSIDM}}} / \textbf{\hyperref[sec:imp_dissipative_SIDM]{vd-dSIDM}}};
    
    \node[algobox, below=0.3cm of algo2] (algo5)
    {\textbf{Multi-state / inelastic scattering} \\[1pt] \textbf{\hyperref[sec:imp_inelastic_SIDM]{iSIDM}}};
    
    \node[algobox, below=0.3cm of algo3] (algo6)
    {\textbf{Decaying dark matter} \\[1pt] \textbf{\hyperref[sec:imp_decaying_DM]{dDM}}};
    
    \begin{scope}[on background layer]
        \node[draw=darkblue!70, dashed, line width=1.0pt, rounded corners=6pt, fill=lightblue!15, 
              fit=(algo1)(algo2)(algo3)(algo4)(algo5)(algo6), inner sep=8pt] {};
    \end{scope}
    
    \node[mainbox, below=0.5cm of algo5, minimum width=6.5cm, fill=blue!8] (sim)
    {Structure Formation Simulation\\[1pt]\scriptsize (Cosmological, zoom-in and idealized simulations)};
    
    \draw[arrow] (algos) -- (algo2);
    \draw[arrow] (algo5) -- (sim);
    
    \node[greenlabel, above right=-0.95cm and 0.1cm of ic.north east]
    {\normalsize Early Universe};
    
    \node[greenlabel, above left=-1.2cm and -0.1cm of nle.north west]
    {\normalsize \parbox{3cm}{\centering Late-time structure formation}};
    
    \end{tikzpicture}
\caption{Mapping particle-physics models onto the inputs required for structure-formation simulations. The particle-physics ingredients are introduced in Chapter~\ref{sec: sidm_pp_models}, while their mapping onto the corresponding simulation inputs is developed in Chapter~\ref{sec:pp_to_simulation_inputs}. The resulting DM algorithms are described in Chapter~\ref{sec: implementation} and validated in Chapter~\ref{sec: verification tests}. They are implemented in our new SIDM infrastructure, introduced in~\cite{Zier:2026}, and in our decaying-DM module. Flowcharts of the corresponding SIDM algorithms are shown in \cref{fig:sidm_flowchart}.}    \label{fig:pp_to_simulation_inputs}
    \end{figure*}
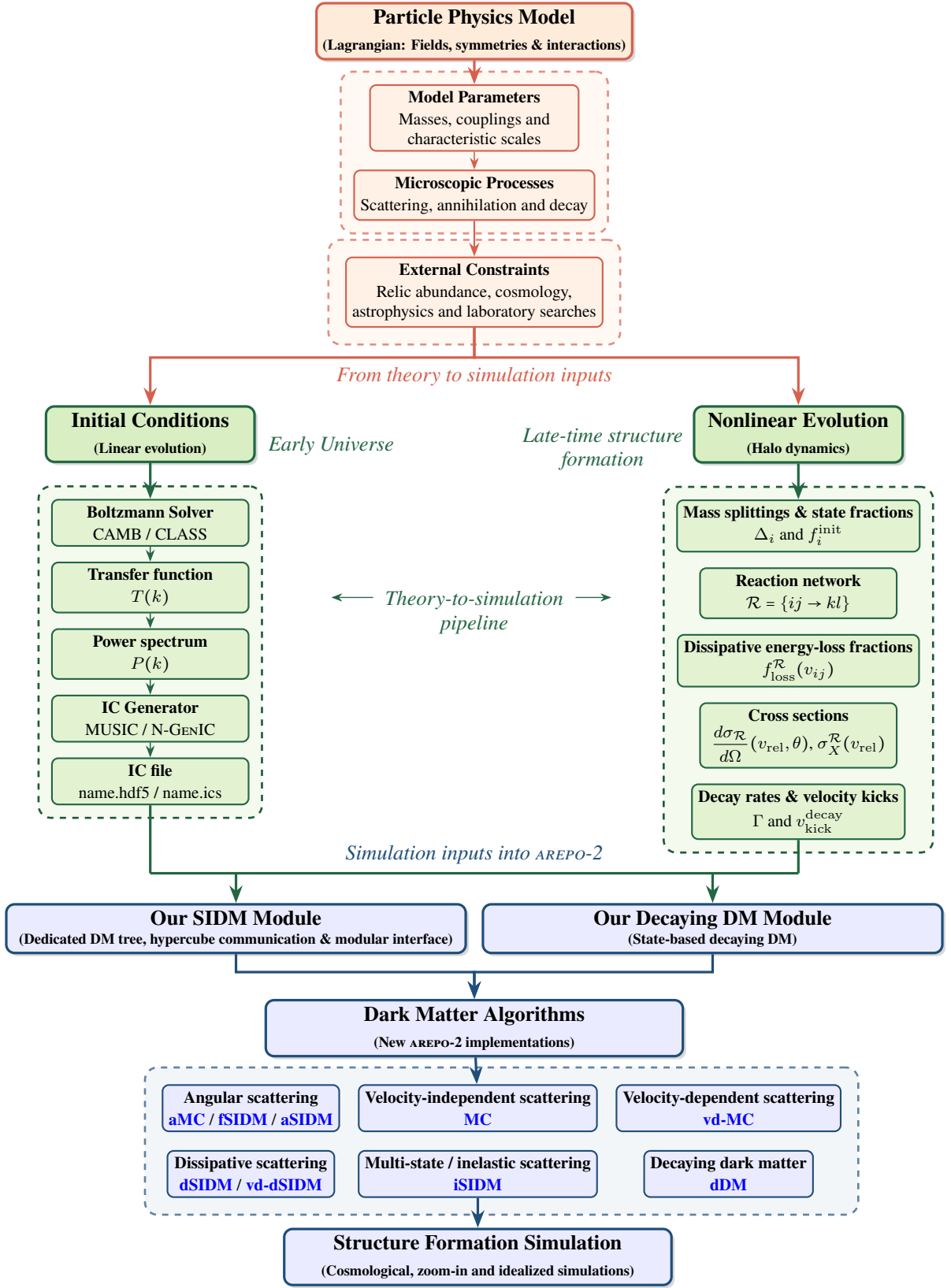

\medskip
\noindent
\textbf{Chapter overview.}
We first construct the general theory-to-simulation mapping through the initial conditions (ICs) (\cref{sec:initial_conditions}), internal-state structure (\cref{sec:internal_state_structure}), reaction network (\cref{sec:reaction_network}), dissipative energy loss (\cref{sec:dissipative_energy_loss}), differential and angle-weighted cross sections (\cref{sec:cross_sections}), and decay inputs (\cref{sec:decay_inputs}). We then apply the mapping to elastic single-state SIDM, including contact-like, light-mediator, resonant, vector and asymmetric realizations (\cref{sec: elastic_single_state,sec:contact-like,sec:light-mediator,sec:Resonant_SIDM,sec:Vector_SIDM,sec:Asymmetric_SIDM}), before extending it to multi-state inelastic SIDM (\cref{sec: inelastic_sidm_models}) and composite SIDM (\cref{sec:composite_SIDM}), including dark baryons, dark mesons, dark glueballs, and dark atoms and molecules (\cref{sec:dark_baryons,sec:dark_mesons,sec:dark_glueballs,sec:dark_atoms}). For each class, we identify the microscopic parameter space, derive the corresponding simulation inputs and connect them to the appropriate \arepoTwo\ implementation. We finally synthesize the mappings across model classes (\cref{sec:implemented_in_arepo}) and discuss the complementary astrophysical diagnostics through which their interaction properties can be tested (\cref{sec:simulation_diagnostics_DM}).

We begin by defining the common simulation-side parameterization. We organize $\mathcal{P}_{\rm SIM}$ into six ingredients:
\begin{enumerate}
\item \textbf{Initial conditions:} generated from the linear matter power spectrum $P(k)$ or transfer function $T(k)$, which set the amplitude of density fluctuations as a function of scale in the early Universe (see \cref{sec:initial_conditions}).

\item \textbf{Internal-state structure:} the macroparticle mass splittings $\Delta_i$ and initial state fractions $f_i^{\rm init}$ with $\sum_i f_i^{\rm init}=1$, encoding the spectrum and relative abundance of the internal states $i$ (see \cref{sec:internal_state_structure}).

\item \textbf{Reaction network:} $\mathcal{R} = \{ij \to kl\}$, enumerating all allowed elastic, inelastic and dissipative scattering channels between dark-sector states (see \cref{sec:reaction_network}).

\item \textbf{Dissipative energy-loss fractions:} $f_{\rm loss}^{\mathcal{R}}(v_{ij})$, specifying the fraction of the relative kinetic energy of the scattering pair lost in a dissipative scattering event (see \cref{sec:dissipative_energy_loss}).

\item \textbf{Differential and angle-weighted cross sections:} $\mathrm{d}\sigma_\mathcal{R}/\mathrm{d}\Omega(v_{\rm rel},\theta)$ and $\sigma_X^\mathcal{R}(v_{\rm rel})$, which determine the scattering rate and the angular structure of momentum transfer for each reaction channel (see \cref{sec:cross_sections}).

\item \textbf{Decay inputs:} the decay rates $\Gamma$ of unstable dark-sector states and the corresponding velocity kicks $v_{\rm kick}$ imparted to the daughter DM particles (see \cref{sec:decay_inputs}).
\end{enumerate}

These quantities constitute the direct inputs required to describe the relevant DM physics in structure-formation simulations. Energy exchange itself is generally not an independent simulation input, but is instead determined from these quantities and the kinematics of the corresponding scattering or decay event. For non-radiative inelastic scattering, energy is converted between internal and kinetic degrees of freedom according to the mass splittings $\Delta_i$ and the reaction $\mathcal{R}=\{ij\to kl\}$. For dissipative scattering, an additional energy loss is specified by $f_{\rm loss}^{\mathcal{R}}(v_{ij})$, while for decays the kinetic-energy injection into the daughter DM particle is encoded through $v_{\rm kick}$.

This mapping provides a unified interface between microscopic particle-physics models and structure-formation simulations: the particle physics enters through these quantities controlling the initial density field, internal-state evolution, reaction channels, interaction rates and event kinematics. These simulation ingredients separate naturally into two categories. The initial conditions (ingredient~1) encode the \emph{linear evolution} of density perturbations in the early Universe. The remaining ingredients (ingredients~2--6) determine the \emph{nonlinear evolution} of DM in haloes and are implemented directly in the simulation code, as discussed in \cref{sec: implementation}. In particular, the mass splittings and reaction network determine the kinematics of inelastic transitions, the dissipative energy-loss fractions specify non-conservative energy loss, the differential and angle-weighted cross sections determine scattering rates and angular momentum transfer, and the decay rates and velocity kicks determine the occurrence and kinematics of decays. \cref{fig:pp_to_simulation_inputs} gives an overview of this bridge between particle physics theories and structure-formation simulations. We now discuss each ingredient in turn, beginning with the initial conditions that set the stage for structure formation.

\subsection{Initial conditions}
\label{sec:initial_conditions}

Particle physics can influence structure formation already before the onset of nonlinear halo evolution by modifying the linear matter power spectrum used to generate cosmological initial conditions. A central framework for connecting microscopic dark-sector physics to these early-time signatures is ETHOS (Effective Theory of Structure Formation)~\cite{Cyr-Racine:2015ihg,Vogelsberger:2015gpr}.

A microscopic DM model can affect structure formation through two conceptually distinct stages. At early times, interactions within the dark sector or with additional relativistic species can alter the evolution of density perturbations and thereby modify the linear matter power spectrum $P(k)$. At later times, the same microscopic theory can modify the nonlinear evolution of haloes through self-scattering, inelastic transitions, dissipation and decays. ETHOS established a general mapping from microscopic dark-sector parameters to effective quantities describing both the linear evolution of perturbations and nonlinear DM self-interactions, with the latter characterized primarily through a velocity-dependent cross section~\cite{Cyr-Racine:2015ihg}. The present framework extends this nonlinear theory-to-simulation mapping to the broader interaction and decay structure represented by ingredients~2--6 above.

\medskip
\noindent
\emph{The linear matter power spectrum and transfer function.} The linear matter power spectrum $P(k)$ is defined through the two-point correlation of the Fourier-space matter density contrast,
\begin{equation}
\left\langle \delta(\mathbf{k})\,\delta^{*}(\mathbf{k}') \right\rangle =
(2\pi)^3\delta_{\rm D}(\mathbf{k}-\mathbf{k}')P(k),
\end{equation}
where $\delta(\mathbf{x})\equiv[\rho(\mathbf{x})-\bar{\rho}]/\bar{\rho}$ is the matter density contrast, $\rho(\mathbf{x})$ is the local matter density, $\bar{\rho}$ is the mean cosmic matter density, and $\delta_{\rm D}$ is the Dirac delta function. For a statistically homogeneous and isotropic density field, $P(k)$ depends only on the magnitude $k=|\mathbf{k}|$ and quantifies the amplitude of density fluctuations as a function of comoving scale, with larger $k$ corresponding to smaller scales. The linear matter power spectrum at redshift $z$ can be related to the primordial spectrum through
\begin{equation}
P(k,z)=T^2(k,z)\,P_{\rm primordial}(k),
\end{equation}
where $T(k,z)$ is the cosmological transfer function describing the linear evolution of primordial perturbations. When comparing a non-CDM model with a reference CDM cosmology, we instead define the relative transfer function
\begin{equation}
T_{\rm rel}(k)=
\left[
\frac{P_{\rm model}(k)}
{P_{\rm CDM}(k)}
\right]^{1/2},
\end{equation}
which isolates the scale-dependent modification relative to CDM.

In cosmological $N$-body simulations, the linear matter power spectrum $P(k,z_{\rm init})$ at the starting redshift provides the statistical input for generating the initial density and velocity fields. Starting from primordial perturbations, the density field is evolved linearly to a suitably chosen initial redshift $z_{\rm init}$, and the resulting $P(k,z_{\rm init})$ or $T(k,z_{\rm init})$ is used to construct the initial particle displacements and velocities. The starting redshift depends on the simulation box size, resolution and initial-condition method, and is commonly chosen such that the root-mean-square amplitude of the initial density fluctuations is $\sim0.1$--$0.2$~\cite{Knebe:2009start,Onorbe:2014zoom}. Starting too early can introduce numerical round-off and discreteness errors, while starting too late can compromise the validity of the perturbative initial conditions and subsequent structure formation. The simulation then follows the nonlinear evolution under gravity and any implemented DM physics. Particle-physics effects that modify the linear evolution must therefore be encoded in the initial conditions.

The linear power spectrum is the blueprint for structure formation. A cutoff at some scale $k_{\rm cut}$ suppresses structures below the corresponding mass scale. The shape of this cutoff, whether smooth and sharp as in warm DM or oscillatory as in models with \emph{dark acoustic oscillations}, further affects halo abundances and internal structure.

\medskip
\noindent
\emph{Why SIDM models can modify $P(k)$.} Not all SIDM models appreciably modify $P(k)$ on galaxy-formation scales. For sufficiently small primordial DM velocity dispersions, purely elastic self-scattering leaves the linear power spectrum approximately CDM-like on these scales, although modifications can arise at smaller scales~\cite{Egana-Ugrinovic:2021gnu}.
However, in multi-state SIDM, exothermic
conversions can inject kinetic energy and
modify linear density perturbations even
without coupling to relativistic dark-sector
species~\cite{Duan:2026zqj}.

Many motivated dark-sector models, however, contain additional relativistic species, such as dark photons, dark neutrinos or other light hidden-sector states~\cite{vandenAarssen:2012vpm,Bringmann:2013vra,Dasgupta:2013zpn,Cyr-Racine:2012tfp,Cyr-Racine:2013fsa,Schewtschenko:2014fca}. If DM remains thermally or kinetically coupled to a relativistic dark component in the early Universe, perturbations no longer evolve like collisionless CDM. Instead, DM participates in an effective dark plasma, analogous to the baryon--photon plasma before recombination. This can suppress small-scale perturbations, produce a cutoff in $P(k)$ and generate dark acoustic oscillations (DAO)~\cite{Buckley:2014hja,Vogelsberger:2015gpr}.

The physical origin is momentum exchange between DM and dark radiation (DR). Frequent DM--DR scattering produces a drag force opposing gravitational collapse, most efficiently on small scales and at early times. As the Universe expands, the interaction rate eventually drops below the Hubble rate, defining kinetic decoupling. After this point, DM perturbations grow freely. The time and efficiency of kinetic decoupling, together with diffusion or Silk damping, determine the scale and shape of the cutoff~\cite{Bringmann:2016ilk}.

\begin{figure}
    \centering    
    \includegraphics[width=0.9\columnwidth]{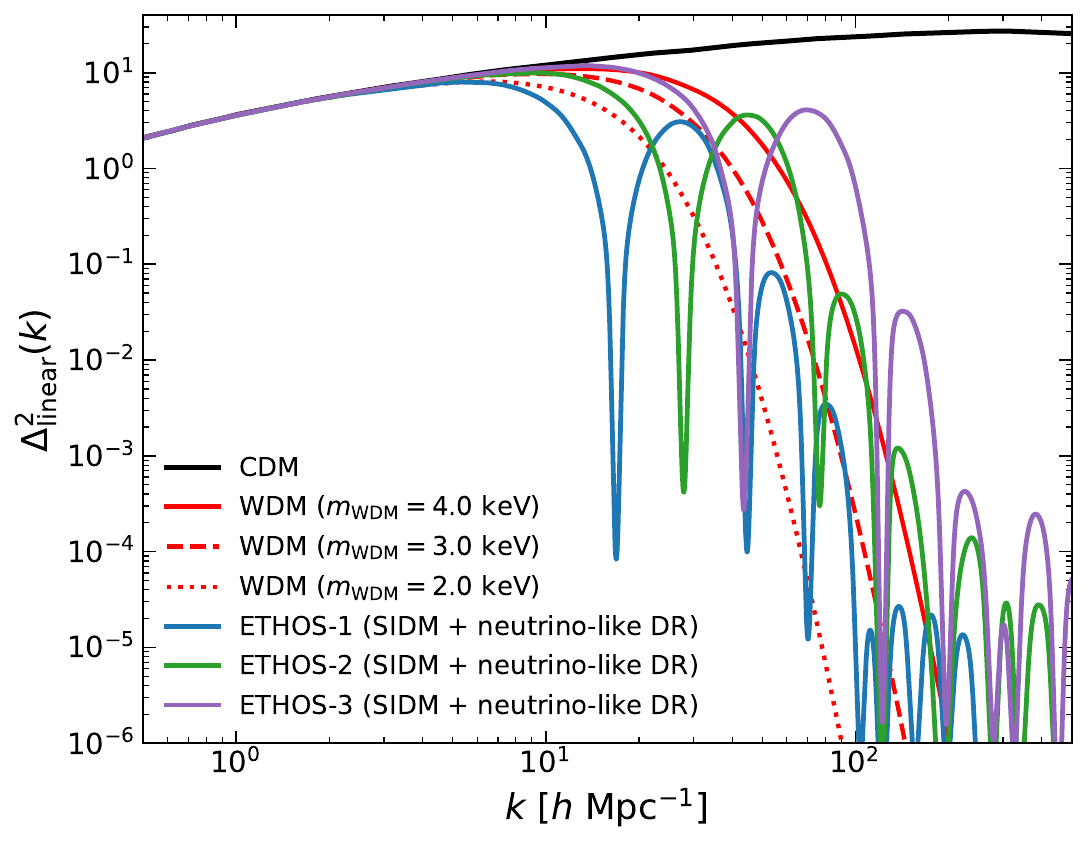}
\caption{Dimensionless linear matter power spectra, $\Delta^2_{\rm linear}(k)=k^3P(k)/(2\pi^2)$, for representative DM models. The black curve shows CDM. Red curves show thermal-relic WDM with $m_{\rm WDM}=2$, $3$ and $4\,{\rm keV}$, for which free streaming produces a smooth suppression of small-scale power. Blue, green and purple curves show the ETHOS-1, ETHOS-2 and ETHOS-3 benchmark models of~\cite{Cyr-Racine:2015ihg,Vogelsberger:2015gpr}, respectively, in which DM scattering with neutrino-like dark radiation generates dark acoustic oscillations and diffusion damping before kinetic decoupling. Unlike the smooth WDM cutoff, the ETHOS spectra retain oscillatory features from the coupled DM--dark-radiation fluid. The larger effective DM--DR opacity of ETHOS-1 shifts the suppression to larger scales and produces stronger damping than in ETHOS-3.}
\label{fig:power_spectra_examples}
\end{figure}

\medskip
\noindent
\emph{Particle-physics realizations that modify $P(k)$.} 
Several dark-sector mechanisms can modify the
linear matter power spectrum, including
DM--dark-radiation scattering, inelastic
state conversions, particle decays and
non-cold DM dynamics. For DM--dark-radiation
interactions, ETHOS provides a mapping onto
effective parameters controlling the interaction
opacity and transfer function~\cite{Cyr-Racine:2015ihg}.
Representative realizations are summarized below.

\begin{itemize}[leftmargin=*, labelsep=1.0em]
\item[(i)] \emph{ETHOS benchmark model: DM-DR scattering via a massive mediator.}
The ETHOS benchmark realization considered in~\cite{Vogelsberger:2015gpr} assumes a fermionic DM particle $\chi$ interacting with a massless neutrino-like fermion $\nu_D$ through a massive vector mediator $\phi_\mu$~\cite{vandenAarssen:2012vpm,Bringmann:2009vf,Dasgupta:2013zpn}. The interaction Lagrangian is
\begin{equation}
\mathcal{L}_{\rm int} 
\supset - g\, \bar{\chi}\gamma^\mu \chi\, \phi_\mu 
- g_\nu \bar{\nu}_D\gamma^\mu \nu_D\, \phi_\mu .
\end{equation}
This gives DM--DR scattering, $\chi\nu_D \leftrightarrow \chi\nu_D$, and hence a drag opacity (see Section II A of~\cite{Cyr-Racine:2015ihg}) that produces DAO and Silk damping in $P(k)$~\cite{Cyr-Racine:2015ihg}. ETHOS-1, ETHOS-2 and ETHOS-3 in~\cite{Vogelsberger:2015gpr} correspond to different choices of $(m_\chi, m_\phi, g, g_\nu)$, and therefore to different damping scales and oscillation amplitudes, as illustrated in Fig.~\ref{fig:power_spectra_examples}.

\item[(ii)] \emph{Hidden charged DM.}
If DM carries charge under an unbroken dark $U(1)_D$ gauge symmetry~\cite{Feng:2009mn}, it interacts with a massless gauge boson $A_{D\mu}$, commonly called the \emph{dark photon} and denoted $\gamma_D$. The interaction Lagrangian is
\begin{equation}
\mathcal{L}_{\rm int} \supset - (D_\mu\chi )^\dagger D^\mu\chi 
\quad \text{with} \quad 
D_\mu = \partial_\mu - ig_D A_{D\,\mu},
\end{equation}
which gives $\chi \gamma_D \leftrightarrow \chi \gamma_D$. Before kinetic decoupling, these interactions keep DM tightly coupled to dark radiation, forming a dark plasma analogous to the baryon--photon fluid. This generates DAOs and Silk-like damping in $P(k)$~\cite{Cyr-Racine:2012tfp}.

\item[(iii)] \emph{Non-Abelian dark radiation.}
If DM is charged under a non-Abelian gauge group, e.g.\ $SU(N)_D$~\cite{Buen-Abad:2015ova,Lesgourgues:2015wza}, the interaction Lagrangian is
\begin{equation}
\mathcal{L}_{\rm int} \supset - (D_\mu\chi )^\dagger D^\mu\chi  
\quad \text{with} \quad 
D_\mu = \partial_\mu - ig_D T^a A_{D\,\mu}^a.
\end{equation} Here $T^a$ are the generators of $SU(N)_D$ in the representation of $\chi$, and $A_{D\mu}^a$ are the corresponding dark gauge fields ($a=1,\dots,N^2-1$), analogous to gluon fields. Their quanta are dark gauge bosons, denoted by $\gamma_D^a$. The processes $\chi + \gamma_D^a \leftrightarrow \chi + \gamma_D^a$, together with dark-radiation self-interactions, keep the dark sector tightly coupled and behaving as a nearly perfect fluid. This induces strong drag on DM perturbations and modifies $T(k)$ through damping and the suppression of higher multipoles in the Boltzmann hierarchy~\cite{Cyr-Racine:2015ihg}.

\item[(iv)] \emph{Multi-state inelastic SIDM.}
Inelastic conversions between nearly degenerate
DM states can also modify the linear matter
power spectrum. Ref.~\cite{Duan:2026zqj}
showed that exothermic conversions can suppress
small-scale power and generate dark acoustic
oscillations even without dark radiation.
Their modified version of \textsc{CLASS}\footnote{
\url{https://github.com/xcduan-1/2cSIDM-CLASS}}
provides the corresponding linear power spectra.

\item[(v)] \emph{Atomic dark matter.}
In bound-state models, such as dark atoms~\cite{Goldberg:1986nk,Kaplan:2009de,Behbahani:2010xa,Kaplan:2011yj,Cline:2012is} discussed in \cref{sec:dark_atoms}, the dark sector contains charged constituents, e.g.\ $p_D$ and $e_D$, interacting through a dark photon $\gamma_D$. Before dark recombination, frequent Thomson-like scattering,
\begin{equation}
e_D + \gamma_D \leftrightarrow e_D + \gamma_D,
\end{equation}
keeps the dark plasma tightly coupled, producing DAO and Silk-like damping. After dark recombination, the coupling is strongly suppressed, setting a characteristic cutoff scale and imprinting oscillatory features in $P(k)$~\cite{Cyr-Racine:2013fsa}. 

\item[(vi)] \emph{Decaying dark matter.}
Decays arise in a broad range of dark-sector models, including gravitino DM with $R$-parity violation~\cite{Fayet:1981sq,Witten:1981nf,Pagels:1981ke,Khlopov:1984pf}, super-WIMP scenarios~\cite{Covi:1999ty,Feng:2003uy}, sterile-neutrino DM~\cite{Ma:2001ip,Dib:2011jh}, decaying gravitons or superheavy DM in string-theory constructions~\cite{Law-Smith:2023czn,Obied:2023clp,Allahverdi:2023nov}, and models with metastable excited DM states~\cite{Finkbeiner:2009mi,Brax:2021gpe}. Schematically, an unstable state $\chi_2$ may decay into a lighter daughter $\chi_1$ and relativistic dark-sector radiation $\gamma_D$,
\begin{equation}
\chi_2 \rightarrow \chi_1 + \gamma_D .
\end{equation}
If the decays and subsequent daughter-particle free streaming significantly modify the density perturbations before the simulation starting redshift, these effects must be encoded in the initial transfer function. For longer lifetimes, the initial conditions may remain approximately CDM-like, while the decays and recoil of the daughter particles must instead be evolved dynamically during the nonlinear simulation~\cite{Peter:2010jy,Wang:2014ina}. The corresponding simulation inputs, including the decay rate and daughter-particle velocity kick, are specified in \cref{sec:decay_inputs} and implemented in \arepoTwo\ as described in \cref{sec:imp_decaying_DM}.

\item[(vii)] \emph{Warm dark matter} (WDM) consists of particles with primordial velocities intermediate between the cold- and hot-DM limits~\cite{Bond:1980ha,Pagels:1981ke,Peebles:1982,Bond:1982PhRvL}. Representative candidates include sterile neutrinos~\cite{Dodelson:1993je,Berezhiani:1995am,Dolgov:2000ew,Abazajian:2001nj} and gravitinos~\cite{Fayet:1981sq,Witten:1981nf,Pagels:1981ke,Khlopov:1984pf}. Their primordial free streaming suppresses density perturbations below a characteristic scale and delays the formation of low-mass haloes~\cite{Bode:2000gq,Avila-Reese:2000nqd,Schneider:2011yu,Lovell:2013ola}. For thermal-relic WDM, this produces a smooth cutoff in $P(k)$, in contrast to the dark acoustic oscillations characteristic of interacting DM--radiation models. However, non-thermally produced sterile neutrinos can exhibit different cutoff shapes and additional small-scale features, depending on their primordial momentum distributions and the lepton asymmetry~\cite{Boyarsky:2018tvu,Lovell:2022vzx}. For thermal relics, Lyman-$\alpha$ forest measurements typically constrain the WDM mass to the $\mathcal{O}(\mathrm{few~keV})$ range or above, although the precise cutoff depends on the primordial phase-space distribution~\cite{Viel:2013fqw,Irsic:2017ixq,Irsic:2023equ}. The leading simulation effect is therefore encoded through the modified linear power spectrum, or equivalently the relative transfer function $T_{\rm rel}(k)$ defined above; depending on the WDM model and numerical setup, residual thermal velocities may additionally need to be included in the initial phase-space distribution. The thermal-relic WDM examples in \cref{fig:power_spectra_examples} illustrate the resulting smooth small-scale suppression without DAOs.

\item[(viii)] \emph{Light and ultra-light bosonic dark matter.}
Light and ultra-light bosonic DM candidates include axions and axion-like particles~\cite{Peccei:1977ur,Peccei:1977hh,Weinberg:1977ma,Wilczek:1977pj,Kim:2008hd}, hidden photons~\cite{Redondo:2008ec,Nelson:2011sf,Arias:2012az,Graham:2015rva,Long:2019lwl} and other ultra-light scalar fields~\cite{Antypas:2022asj}. Unlike fermionic DM, whose mass is constrained by phase-space bounds such as the Gunn--Tremaine limit~\cite{Tremaine:1979we,Alvey:2020xsk}, sufficiently light bosons can form highly occupied coherent fields with astrophysically relevant de Broglie wavelengths. Their wave dynamics suppress the growth of perturbations below a characteristic Jeans scale, producing a cutoff in $P(k)$. Ultra-light scalar DM with masses around $m_s\sim10^{-21}\,\mathrm{eV}$ is commonly referred to as fuzzy DM~\cite{Turner:1983he,Press:1989id,Sin:1992bg,Peebles:2000yy}; see also the reviews~\cite{Hui:2021tkt,Antypas:2022asj}. The corresponding modified linear power spectrum can be computed with a Boltzmann solver such as \textsc{AXION-CAMB}~\cite{Hui:2016ltb,Irsic:2017yje,Mocz:2017wlg}. Capturing the subsequent nonlinear evolution additionally requires a wave-based or equivalent dynamical treatment, as implemented with different numerical approaches in~\cite{Schive:2014dra,Mocz:2017wlg,Mocz:2019pyf}, including implementations within \arepo~\cite{Mocz:2017wlg,Mocz:2019pyf}.

\end{itemize}

\begin{table*}
\centering
\begin{tabular}{llllll}
\hline
& \textbf{Model} & \textbf{Key interactions} & \textbf{Effect on} $\textbf{\textit{P}(\textit{k})}$ & \textbf{References} & \textbf{Relevant model class} \\
\hline
(i) & SIDM + neutrino-like DR
& $\chi \nu_D \leftrightarrow \chi \nu_D$ 
& DAO + Silk damping 
& \cite{vandenAarssen:2012vpm,Bringmann:2009vf,Dasgupta:2013zpn} & E.g. light-mediator SIDM (\ref{sec:light-mediator}) \\

(ii) & Hidden charged DM 
& $\chi \gamma_D \leftrightarrow \chi \gamma_D$ 
& DAO + Silk damping 
& \cite{Feng:2009mn} & E.g. light-mediator SIDM (\ref{sec:light-mediator}) \\

(iii) & Non-Abelian DR 
& $\chi \gamma_D^a \leftrightarrow \chi \gamma_D^a$ 
& Strong damping 
& \cite{Buen-Abad:2015ova,Lesgourgues:2015wza} & E.g. vector SIDM (\ref{sec:Vector_SIDM}) \\

(iv) & Inelastic SIDM
& $\chi_2\chi_2 \leftrightarrow \chi_1\chi_1$
& Small-scale suppression + DAO (no DR)
& \cite{Duan:2026zqj}
& Inelastic SIDM (\ref{sec: inelastic_sidm_models}) \\

(v) & Atomic / molecular DM 
& $e_D \gamma_D \leftrightarrow e_D \gamma_D$ 
& DAO + sharp cutoff (recombination) 
& \cite{Goldberg:1986nk,Kaplan:2009de,Behbahani:2010xa,Kaplan:2011yj,Cline:2012is} & Atomic / molecular SIDM (\ref{sec:dark_atoms}) \\

(vi) & Decaying DM
& $\chi_2 \rightarrow \chi_1 + \gamma_D$
& Early decay / daughter free-streaming
& \cite{Peter:2010jy,Wang:2014ina}
& Decaying DM \\

(vii) & Warm DM 
& Free-streaming 
& Smooth cutoff (no oscillations) 
& \cite{Bode:2000gq,Viel:2005qj} & Warm DM \\

(viii) & Light / ultra-light DM
& Wave dynamics
& Small-scale cutoff
& \cite{Hui:2016ltb,Irsic:2017yje,Mocz:2017wlg}
& E.g. Fuzzy / axion-like DM \\

\hline
\end{tabular}
\caption{Representative particle-physics realizations that modify the linear matter power spectrum $P(k)$ and may therefore require modified initial conditions for structure-formation simulations. The mechanisms include DM interactions with
relativistic degrees of freedom, which generate
DAO and diffusion (Silk-like) damping;
exothermic inelastic DM conversions, which
can suppress small-scale power and generate
DAO even without dark radiation; and decaying DM, which can suppress small-scale power through early decays or daughter-particle free-streaming; WDM, which produces a smooth cutoff through primordial free-streaming; and light or ultra-light bosonic DM, which generates a cutoff through wave dynamics and quantum pressure. The impact of DM--DR interactions on the transfer function and structure formation is discussed in the ETHOS framework~\cite{Cyr-Racine:2015ihg,Vogelsberger:2015gpr}.}
\label{tab:pk_models}
\end{table*}

\noindent
For conventional cold SIDM, purely elastic self-scattering leaves the linear matter power spectrum nearly unchanged on scales relevant for galaxy formation. However, if the primordial DM velocity dispersion is appreciable, even elastic self-scattering can modify $P(k)$, producing small-scale suppression and acoustic oscillations without dark radiation~\cite{Egana-Ugrinovic:2021gnu}. The characteristic scales depend on the velocity dispersion and self-interaction rate, with smaller velocity dispersions shifting these effects to smaller scales. Additional modifications can arise from coupling to relativistic dark-sector species, as described by ETHOS~\cite{Cyr-Racine:2015ihg,Vogelsberger:2015gpr}, or from inelastic conversions between nearly degenerate DM states, which can generate small-scale suppression and acoustic features even without dark radiation~\cite{Duan:2026zqj}. Other mechanisms include primordial free streaming, particle decays and wave dynamics (see Table~\ref{tab:pk_models}).

\medskip
\noindent
\emph{How $P(k)$ is obtained in practice.} 
In principle, $P(k)$ for a given dark-sector model is obtained by solving the linearized Einstein--Boltzmann system for all relevant species. The microscopic particle physics enters through collision terms in the Boltzmann equations, determined by scattering amplitudes and interaction rates.

ETHOS provides an intermediate effective description in which the relevant early-Universe physics is encoded in a small set of effective parameters~\cite{Cyr-Racine:2015ihg}. These parameters describe how strongly DM couples to dark radiation, how the interaction rate scales with temperature, and how efficiently interactions damp higher angular moments of the perturbation hierarchy. The mapping is
\begin{equation}
\begin{aligned}
\mathcal{P}_{\rm PP}=\{m_{\mathrm{DM}},\{m_i\}, \{g_i\}, \xi \} &\to \{\omega_{\rm DR}, a_n, \alpha_l, \ldots\} \\ &\to T(k)\to P(k),
\end{aligned}
\end{equation}
where the left side contains the intrinsic particle physics parameters introduced in~\cref{sec: pp_terminology}: the DM mass $m_{\mathrm{DM}}$ (e.g., $m_s$, $m_\chi$, $m_{\phi_\mu}$), other masses $\{m_i\}$ such as mediator masses $m_{\mathrm{med}}$ (e.g., $m_\phi$, $m_{\phi_\mu}$), couplings $\{g_i\}$ including DM-mediator couplings ($\lambda$, $\mu$, $g$, $y$) and DM-DR couplings (e.g., to sterile neutrinos $g_\nu$ and dark gauge bosons $g_D$), and the dark sector temperature ratio $\xi \equiv (T_{\rm DR}/T_{\rm CMB})|_{z=0}$ at $z=0$.~\footnote{The dark sector temperature ratio at $z=0$, $\xi \equiv (T_{\rm DR}/T_{\rm CMB})|_{z=0}$, quantifies how much hotter or cooler the DR is compared to the cosmic microwave background today. This ratio is determined by the thermal history of the dark sector: if DR decouples from the SM bath before electron-positron annihilation, it does not share in the photon heating, yielding $\xi < 1$; if additional entropy is injected into the dark sector after decoupling -- for example, through late annihilations of heavy dark sector particles -- then $\xi > 1$. The value of $\xi$ is determined by the thermal history of the dark sector and, in a fully specified cosmological model, follows from the particle content, interactions, decoupling history and entropy evolution.} The right side contains the ETHOS effective parameters: the dark radiation energy density $\omega_{\rm DR} \equiv \Omega_{\rm DR}h^2$, interaction-rate coefficients $a_n$ describing the temperature or redshift dependence of the DM-DR drag opacity, angular damping coefficients $\alpha_l$ encoding the damping of higher multipoles in the Boltzmann hierarchy, and additional parameters describing model-dependent dark-sector thermodynamics. These parameters determine $T(k)$ and hence $P(k)$.

Once this mapping is performed, $T(k)$ can be computed with a modified Boltzmann solver such as CAMB~\cite{Seljak:1996is,2011ascl.soft02026L} or CLASS~\cite{2011arXiv11042932L,2011JCAP07034B,2011arXiv11042934L,2011JCAP09032L}, following~\cite{Cyr-Racine:2015ihg,Vogelsberger:2015gpr}. The resulting $P(k)$ is then used to generate simulation initial conditions with codes such as \textsc{MUSIC}~\cite{2011MNRAS.415.2101H} or \textsc{N-GenIC}~\cite{2005MNRAS3641105S,2015asclsoft02003S}.

Fig.~\ref{fig:power_spectra_examples} illustrates several possible modifications to the linear matter power spectrum, including SIDM models with dark acoustic oscillations from ETHOS and thermal-relic WDM models for comparison. WDM produces a smooth exponential cutoff from free-streaming, whereas SIDM coupled to dark radiation produces oscillatory features on top of an overall damping envelope. These features arise from the interplay of dark acoustic oscillations and Silk-like diffusion damping. Their scale and amplitude depend on the DM mass, DM--DR coupling strength and mediator properties. 
Models with similar cutoff scales can therefore exhibit different features in the linear power spectrum. However, nonlinear gravitational evolution couples different scales and can wash out these differences, making many DAO models nearly degenerate with thermal WDM in the nonlinear power spectrum on galaxy-formation scales. Ref.~\cite{Bohr:2020yoe} showed that, at high redshifts ($z\gtrsim5$), the nonlinear evolution of a broad class of ETHOS models can be characterized by two effective parameters, $k_{\rm peak}$ and $h_{\rm peak}$, describing the scale and amplitude of the first DAO peak. Strong DAO signatures can nevertheless survive nonlinear evolution, while halo abundances provide complementary information for distinguishing models.

\medskip
Once initial conditions are established, the subsequent nonlinear evolution of structure is governed by ingredients~2--6 in \cref{fig:pp_to_simulation_inputs}. We first introduce the internal-state structure and reaction network, which together determine the kinematics and associated energy exchange of inelastic transitions.

\subsection{Internal state structure}
\label{sec:internal_state_structure}

The internal-state composition of the dark sector is specified by the spectrum of mass splittings and the corresponding initial state fractions $f_i^{\rm init}$ satisfying
\begin{equation}
\sum_i f_i^{\rm init} = 1.
\end{equation}
The mass splittings define the hierarchy of internal states and determine the kinematics of inelastic transitions, while the state fractions encode the relative abundance of each state $i$ at initialization.

At the microscopic level, the mass of state $i$ is denoted by $m_i$, and we define the ground state as the lightest state with mass $m_0$. In an $N$-body simulation, each ground-state particle is a macroparticle of mass $M_0$ representing many microscopic particles, with $N_{\rm micro} \simeq M_0 / m_0$. The macroscopic mass of state $i$ is obtained by rescaling the microscopic mass, $M_i = M_0 (m_i/m_0)$. In practice, simulations use the dimensionless (relative) macroscopic mass splitting,
\begin{equation}\label{eq:relative_mass_splitting}
\Delta_i \equiv \frac{M_i - M_0}{M_0} = \frac{m_i - m_0}{m_0},
\end{equation}
which directly maps the microscopic mass differences onto simulation inputs. Together with the reaction network introduced below, these mass splittings determine the energy exchanged between internal and kinetic degrees of freedom during inelastic scattering, so no independent inelastic energy-exchange input is required.

The initial state fractions $f_i^{\rm init}$ are not free parameters, but are determined by the cosmological evolution of the particle physics model. In particular, they depend on the DM production mechanism in the early Universe (e.g.\ freeze-out or freeze-in, see \cref{sec:DM_production_mechanisms}), the mass splittings, and the rates for decays and inelastic scattering processes that interconvert the states in the early Universe. In many models, excited states are Boltzmann-suppressed prior to structure formation, leading to a hierarchy $f_0^{\rm init} \simeq 1$ and $f_{i>0}^{\rm init} \ll 1$. However, if decay rates are suppressed or inelastic transitions remain efficient at late times, a non-negligible population of excited states can persist.

More generally, the pair $(\Delta_i, f_i^{\rm init})$ fully characterizes the internal energy reservoir of the dark sector at initialization. This reservoir can subsequently be converted into kinetic energy through inelastic scattering, thereby playing a central role in the thermal and dynamical evolution of DM haloes.

\subsection{Reaction network}
\label{sec:reaction_network}

The reaction network $\mathcal{R} = \{ij \to kl\}$ specifies all possible two-body scattering processes between DM states. For a dark sector containing $N$ distinct particle species (e.g., ground and excited states, different spin configurations or multiple particle types), the reaction network enumerates every allowed channel through which particles $i$ and $j$ can scatter into final states $k$ and $l$.

Each reaction is characterized by a cross section
$\sigma_{ij\to kl}(v_{\rm rel})$ (with representative
examples introduced in \cref{sec:cross_sections}),
which determines the rate at which the process
occurs. For a particle of species $i$ moving
through a local mass density $\rho_j$ of species
$j$, the scattering rate is given by
\begin{equation}
R_{ij\to kl}
=
\frac{\rho_j}{m_j}
\left\langle
\sigma_{ij\to kl}(v_{\rm rel})v_{\rm rel}
\right\rangle,
\end{equation}
where $m_j=m_0(1+\Delta_j)$ is the microscopic
mass of species $j$, and the angle brackets
denote an appropriate velocity average.
In cosmological simulations, this rate determines
the probability of a scattering transition
during a timestep.

\medskip
\noindent
\textbf{Elastic scattering.} The simplest reaction network consists of a single elastic channel (e.g. $s + s \to s + s$ for scalar DM or $\chi + \chi \to \chi + \chi$ for fermionic DM), corresponding to standard elastic SIDM as shown in~\cref{fig:selfint_realizations_all}. 
For multiple DM species, elastic inter-species scattering, $i+j\to i+j$, can also exchange momentum and kinetic energy between species with different masses.

\medskip
\noindent
\textbf{Inelastic scattering.} More complex networks arise naturally in multi-state models (discussed in more detail in \cref{sec: inelastic_sidm_models}). For example, a two-state model with ground state $\chi_1$ and excited state $\chi_2$ separated by mass splitting $\delta$ includes five distinct channels:
\begin{align} 
\chi_1 + \chi_1 &\to \chi_1 + \chi_1 \quad \text{(elastic, ground)} \nonumber \\
\chi_2 + \chi_2 &\to \chi_2 + \chi_2 \quad \text{(elastic, excited)} \nonumber\\
\chi_1 + \chi_2 &\to \chi_1 + \chi_2 \quad \text{(elastic, mixed)} \label{eq:inelastic_channels} \\
\chi_1 + \chi_1 &\to \chi_2 + \chi_2 \quad \text{(inelastic, up-scattering)}\nonumber \\
\chi_2 + \chi_2 &\to \chi_1 + \chi_1 \quad \text{(inelastic, down-scattering)}. \nonumber
\end{align}

For the inelastic channels, energy is converted between internal and kinetic degrees of freedom. Once a reaction $\mathcal{R}=\{ij\to kl\}$ is selected, the corresponding internal-state energy change follows directly from the state masses,
\begin{equation} \label{eq:internal_state_energy_change}
\Delta E_{\rm int} = (M_i + M_j - M_k - M_l)c^2,
\end{equation}
and remains entirely within the DM sector. Importantly, $\Delta E_{\rm int}$ is not an additional simulation input: it is calculated internally from the mass splittings $\Delta_i$ and the reaction channel using energy--momentum conservation. Thus, apart from specifying the reaction network itself, the only additional quantities required to describe the internal-state composition are the mass splittings $\Delta_i$ and the initial state fractions $f_i^{\rm init}$; the subsequent inelastic event kinematics are determined internally by the code.

For a scattering reaction $ij\to kl$, the outgoing particle velocities are given by
\begin{equation}\label{eq:velocity_updates_energy_exchange}
\begin{aligned}
\mathbf{v}_i' &= \frac{M_i + M_j}{M_k + M_l} \mathbf{v}_{\rm cm}
+ \frac{M_l}{M_k + M_l}\,\varepsilon_{ij}\, v_{\rm rel} \hat{\mathbf{e}}, \\
\mathbf{v}_j' &= \frac{M_i + M_j}{M_k + M_l} \mathbf{v}_{\rm cm}
- \frac{M_k}{M_k + M_l}\,\varepsilon_{ij}\, v_{\rm rel} \hat{\mathbf{e}},
\end{aligned}
\end{equation}
where $v_{\rm rel}=|\mathbf{v}_i-\mathbf{v}_j|$ is the incoming relative speed, $\mathbf{v}_{\rm cm}=(M_i\mathbf{v}_i+M_j\mathbf{v}_j)/(M_i+M_j)$ is the incoming centre-of-momentum velocity, $\hat{\mathbf{e}}$ is the sampled outgoing direction and $\varepsilon_{ij}$ describes the change in relative speed due to energy exchange. For a non-radiative inelastic transition $ij\to kl$, energy conservation gives
\begin{equation}
\varepsilon_{ij}
=
\sqrt{
\frac{\mu_{ij}}{\mu_{kl}}
+
\frac{2\Delta E_{\rm int}}{\mu_{kl}v_{\rm rel}^2}
},
\end{equation}
where $\mu_{ij}=M_iM_j/(M_i+M_j)$ and $\mu_{kl}=M_kM_l/(M_k+M_l)$ are the initial- and final-state reduced masses, respectively. Hence, $\varepsilon_{ij}$ is likewise not supplied as an independent input, but is evaluated internally for each scattering event from the particle masses, relative velocity and selected reaction channel. Endothermic up-scattering gives $\varepsilon_{ij}<1$ and is allowed only when sufficient relative kinetic energy is available, whereas exothermic down-scattering gives $\varepsilon_{ij}>1$ and releases internal energy into kinetic motion.

For a two-state model with mass splitting $\delta$, exothermic down-scattering, $\chi_2 + \chi_2 \to \chi_1 + \chi_1$, provides a simple example. The released internal energy corresponds to the characteristic velocity kick
\begin{equation}\label{eq:velocity_kick_exothermic}
v_{\rm kick} = \sqrt{\frac{2\delta}{m_{\rm DM}}}\,c.
\end{equation}
Conversely, endothermic up-scattering ($\chi_1 + \chi_1 \to \chi_2 + \chi_2$)  requires $v_{\rm rel} > 2 v_{\rm kick}$ and is therefore kinematically suppressed at low velocities. This $v_{\rm kick}$ is again a derived quantity rather than an independent simulation input.

A crucial distinction is that these mass-splitting processes do \emph{not} remove energy from the system: they only redistribute energy between internal and kinetic degrees of freedom. Energy injection from exothermic internal transitions can nevertheless have dramatic dynamical effects. As shown by~\cite{Schutz:2014nka,vogelsberger2019evaporating}, down-scattering with $\delta = 10~{\rm keV}$ and $m_\chi = 10~{\rm GeV}$ produces velocity kicks $v_{\rm kick} \approx 424~{\rm km\,s^{-1}}$, sufficient to unbind particles from dwarf-scale haloes and efficiently reduce central densities in Milky Way-sized systems. Although no energy is lost from the dark sector, the redistribution of kinetic energy acts as an effective heating mechanism, with cumulative energy release comparable to that of large populations of supernovae.

\subsection{Dissipative energy-loss fractions}
\label{sec:dissipative_energy_loss}

Dissipative scattering differs from the non-radiative inelastic processes described above because part of the kinetic energy associated with the relative motion of the scattering pair is removed from the simulated DM system. Such processes can involve the emission of additional light dark-sector particles, for example
\begin{equation}
i+j \to k+l + \gamma_D,
\end{equation}
which carry energy away from the interacting DM particles.

For each dissipative reaction $\mathcal{R}$, the energy loss is specified through the dissipative energy-loss fraction $f_{\rm loss}^{\mathcal{R}}(v_{ij})$, defined as the fraction of the incoming relative kinetic energy of the scattering pair that is removed in the interaction. The corresponding dissipated energy is therefore
\begin{equation}\label{eq:diss_energy_loss_fraction}
E_{\rm diss}^{\mathcal{R}} = f_{\rm loss}^{\mathcal{R}}(v_{ij})\,\frac{1}{2}\mu_{ij}v_{\rm rel}^2,
\end{equation}
where $\mu_{ij}=M_iM_j/(M_i+M_j)$ is the reduced mass of the incoming pair. As with inelastic transitions, the post-scattering velocities are obtained using \cref{eq:velocity_updates_energy_exchange}, with $v_{ij}\equiv v_{\rm rel}$ denoting the incoming relative speed and $\varepsilon_{ij}$ accounting for the energy exchange. For a purely dissipative scattering event with unchanged particle masses, the reduction in relative kinetic energy implies
\begin{equation}
\varepsilon_{ij} = \sqrt{1-f_{\rm loss}^{\mathcal{R}}(v_{ij})},
\end{equation}
such that the outgoing relative speed is reduced according to $v_{\rm rel}'=\varepsilon_{ij}v_{\rm rel}$. Hence, neither the energy removed in an individual event nor $\varepsilon_{ij}$ needs to be supplied independently to the code; both are calculated internally from the input $f_{\rm loss}^{\mathcal{R}}(v_{ij})$ and the local scattering kinematics. More generally, if a dissipative reaction also changes the internal states, the mass-splitting contribution discussed in \cref{sec:reaction_network} is combined with the dissipative energy loss when determining $\varepsilon_{ij}$ through energy--momentum conservation.

\begin{figure}[t]
\centering

\begin{subfigure}[t]{0.48\columnwidth}
\centering
\begin{tikzpicture}
\begin{feynman}
  \vertex (a1) at (0, 0.5);
  \vertex (b1) at (0,-0.5);

  \vertex (i11) at (-1.8, 0.9) {$\chi$};
  \vertex (i12) at (-1.8,-0.9) {$\chi$};
  \vertex (f11) at ( 1.8, 0.9) {$\chi$};
  \vertex (f12) at ( 1.8,-0.9) {$\chi$};

  \vertex (c1) at (0.6, 0.9);
  \vertex (g1) at (1.85, 1.55) {$\phi$};

  \diagram*{
    (i11) -- [fermion] (a1) -- [fermion] (c1) -- [fermion] (f11),
    (i12) -- [fermion] (b1) -- [fermion] (f12),
    (a1)  -- [scalar, edge label'=$\phi$] (b1),
    (c1)  -- [scalar] (g1),
  };
\end{feynman}
\end{tikzpicture}
\caption{Emission of a dark scalar mediator $\phi$.}
\label{fig:brems_scalar}
\end{subfigure}
\hfill
\begin{subfigure}[t]{0.48\columnwidth}
\centering
\begin{tikzpicture}
\begin{feynman}
  \vertex (a3) at (0, 0.5);
  \vertex (b3) at (0,-0.5);

  \vertex (i31) at (-1.8, 0.9) {$\chi$};
  \vertex (i32) at (-1.8,-0.9) {$\chi$};
  \vertex (f31) at ( 1.8, 0.9) {$\chi$};
  \vertex (f32) at ( 1.8,-0.9) {$\chi$};

  \vertex (c3) at (0.6, 0.9);
  \vertex (g3) at (1.85, 1.55) {$\phi_\mu$};

  \diagram*{
    (i31) -- [fermion] (a3) -- [fermion] (c3) -- [fermion] (f31),
    (i32) -- [fermion] (b3) -- [fermion] (f32),
    (a3)  -- [photon, edge label'=$\phi_\mu$] (b3),
    (c3)  -- [photon] (g3),
  };
\end{feynman}
\end{tikzpicture}
\caption{Emission of a dark vector mediator $\phi_\mu$.}
\label{fig:brems_vector}
\end{subfigure}

\caption{Dissipative bremsstrahlung emission in fermionic DM self-interactions. Panel (a) shows the emission of a dark scalar mediator $\phi$, while panel (b) shows the emission of a dark vector mediator $\phi_\mu$.}
\label{fig:brems_scalar_vector}
\end{figure}
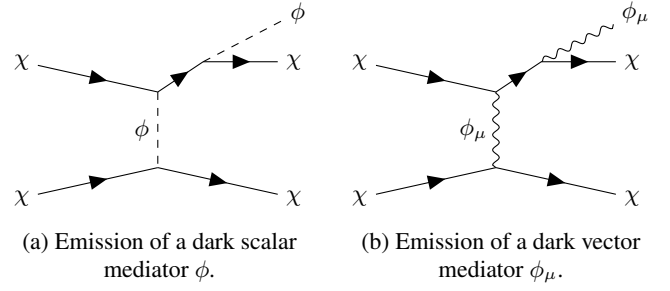

A concrete particle-physics realization of such energy loss is provided by dark bremsstrahlung. In \cite{Lankester--Broche:2025rkm}, the energy-differential cross sections and corresponding energy-loss rates were derived for a broad class of dSIDM models within the perturbative Born regime from the underlying particle physics. The analysis includes fermionic and scalar DM together with scalar or vector emission, and covers both long- and short-range mediated interactions. Of particular relevance here are long-range interactions mediated by a light dark-sector particle, for which the mediator itself can be radiated during the DM self-scattering process and thereby carry away part of the incoming kinetic energy. Representative examples with scalar and vector bremsstrahlung are illustrated in \cref{fig:brems_scalar_vector}. The resulting energy-differential cross section ${\rm d}\sigma/{\rm d}\omega$ determines the spectrum of radiated energies $\omega$, while the corresponding macroscopic energy-loss rate follows by weighting this spectrum by $\omega$ and integrating over the kinematics and DM velocity distribution. The analytical expressions derived in \cite{Lankester--Broche:2025rkm} are validated against a direct numerical integration of the full, unexpanded $2\rightarrow 3 $ scattering amplitudes, showing excellent agreement.

A dissipative scattering channel additionally requires the corresponding dissipative cross section, $\sigma_{\rm diss}^{\mathcal{R}}/m_{\rm DM}$, to determine the interaction probability. For anisotropic scattering, the differential cross section, ${\rm d}\sigma_{\rm diss}^{\mathcal{R}}/{\rm d}\Omega$, is further required to specify the angular distribution of the outgoing particles. Dissipative differential cross sections for several representative dissipative SIDM realizations, including the light-mediator scenarios discussed above, have been derived in \cite{Lankester--Broche:2025rkm}. These dissipative cross sections should be included as simulation inputs alongside the other differential and angle-weighted cross sections discussed in \cref{sec:cross_sections}. Another example of dissipative scattering is provided by the dark-atom SIDM model class presented in \cref{sec:dark_atoms}.

\medskip
The next ingredients we consider in \cref{fig:pp_to_simulation_inputs} are the differential and angle-weighted cross sections, whose emergence from the underlying microscopic interactions we derive below.

\subsection{\label{sec:cross_sections} Cross sections}

The microscopic interactions introduced above determine the two-body scattering amplitude $\mathcal{M}$ and hence the differential cross section $\mathrm{d}\sigma/\mathrm{d}\Omega(\theta,v_{\rm rel})$, as discussed briefly in \cref{sec:interactions_to_cross_sections}. Here we make this connection explicit for the representative contact and mediator-exchange realizations shown in \cref{fig:selfint_realizations_all}. The resulting differential cross sections can subsequently be used to construct the angle-weighted cross sections introduced in \cref{sec:relevant_cross_section}.

Throughout most of this section, we consider elastic scattering involving a single DM mass $m_{\rm DM}$, identified as $m_s$ for scalar DM and $m_\chi$ for fermionic DM. This provides the baseline relevant for single-state SIDM and keeps the expressions compact. For elastic scattering between particles with different masses $m_i$ and $m_j$, the relative two-body motion is governed by the reduced mass
\begin{equation}
\label{eq:reduced_mass}
\mu_{ij}\equiv\frac{m_i m_j}{m_i+m_j},
\end{equation}
with incoming relative momentum $k=\mu_{ij}v_{\rm rel}$. For equal masses, $\mu_{ij}=m_{\rm DM}/2$, so $k=m_{\rm DM}v_{\rm rel}/2$. Accordingly, expressions in which the DM mass enters through the relative two-body kinematics can be generalized to elastic unequal-mass scattering by replacing $m_{\rm DM}$ with $2\mu_{ij}$. More general scattering problems require additional information beyond this kinematic replacement. In particular, multi-state inelastic SIDM involves mass splittings, kinematic thresholds and coupled-channel scattering, as discussed in \cref{sec: inelastic_sidm_models}, while composite realizations such as dark atoms require cross sections derived from their internal bound-state dynamics, as discussed in \cref{sec:dark_atoms}.

The quartic interaction in \cref{fig:selfint_scalar_contact} corresponds to a local four-point vertex. For the non-derivative interaction considered here, the tree-level amplitude is momentum independent, resulting in short-range and isotropic scattering. By contrast, the mediator-exchange diagrams in \cref{fig:selfint_scalar_phi,fig:selfint_scalar_vec,fig:selfint_fermion_phi,fig:selfint_fermion_vec} introduce momentum dependence through the mediator propagator. In the non-relativistic regime relevant for DM structure formation and evolution, these interactions can be described by a Yukawa potential~\cite{Tulin:2013teo},
\begin{equation}
\label{eq:yukawa_potential_generic}
V(r)=\pm\frac{\alpha_{\rm DM}}{r}e^{-m_{\rm med}r},
\end{equation}
where $m_{\rm med}=m_\phi$ for scalar mediation and $m_{\rm med}=m_{\phi_\mu}$ for vector mediation. For a vector mediator, the potential is attractive ($-$) for particle--antiparticle scattering, such as $\chi\bar{\chi}$, and repulsive ($+$) for particle--particle and antiparticle--antiparticle scattering, such as $\chi\chi$ and $\bar{\chi}\bar{\chi}$. For a scalar mediator, the potential is purely attractive ($-$).

The dimensionless interaction strength $\alpha_{\rm DM}$ is determined by the microscopic couplings introduced in \cref{sec:renormalizable_interactions}: the scalar--scalar--mediator coupling $\mu$ in \cref{eq:scalar_mediator_couplings_mu}, the scalar--vector coupling $g$ in \cref{eq:vector_int}, the Yukawa coupling $y$ in \cref{eq:yukawa_coupling}, and the fermion--vector coupling $g$ in \cref{eq:vector_exchange}. For the representative realizations considered here,
\begin{equation}
\label{eq:alpha_DM_microscopic}
\alpha_{\rm DM}=
\begin{cases}
\dfrac{\mu^2}{16\pi m_s^2}, & \text{DM scalar, scalar mediator}, \\[6pt]
\dfrac{g^2}{4\pi}, & \text{DM scalar, vector mediator}, \\[6pt]
\dfrac{y^2}{4\pi}, & \text{DM fermion, scalar mediator}, \\[6pt]
\dfrac{g^2}{4\pi}, & \text{DM fermion, vector mediator}.
\end{cases}
\end{equation}
For vector mediation, these expressions assume unit dark charge. More generally, the interaction strength is $\alpha_{\rm DM}=g^2|Q_iQ_j|/(4\pi)$, while the relative sign of the dark charges determines whether the interaction is attractive or repulsive.

For equal-mass scattering, the momentum transfer is $q=2k\sin(\theta/2)=m_{\rm DM}v_{\rm rel}\sin(\theta/2)$. The scattering behaviour is then controlled by the mediator mass relative to the typical momentum transfer:
\begin{itemize}[leftmargin=*, labelsep=0.5em]
\item \emph{Light-mediator regime} ($m_{\rm med}\lesssim q$): the interaction range $m_{\rm med}^{-1}$ is comparable to or larger than the de Broglie wavelength. The scattering can therefore become strongly angle- and velocity-dependent and is typically forward-peaked.
\item \emph{Heavy-mediator limit} ($m_{\rm med}\gg q$): the interaction range is much shorter than the de Broglie wavelength. Mediator exchange reduces to an effective contact interaction, yielding approximately isotropic and velocity-independent scattering, consistent with the picture illustrated in \cref{fig:heavy_mediator_limit_both}.
\end{itemize}

In the following, we first consider mediator-induced elastic scattering in and beyond the Born regime, before turning to the contact-interaction limit and an effective-range description of composite scattering.

\medskip
\noindent
\textbf{Analytical cross sections in the Born regime.}
For mediator exchange as shown in \cref{fig:selfint_scalar_phi,fig:selfint_scalar_vec,fig:selfint_fermion_phi,fig:selfint_fermion_vec}, the scattering amplitude can be computed perturbatively in the non-relativistic Born regime~\cite{Tulin:2013teo},
\begin{equation}
\label{eq:born_condition}
\frac{\alpha_{\rm DM}m_{\rm DM}}{m_{\rm med}}\ll1 .
\end{equation}
For elastic scattering of unequal masses, this condition generalizes to $2\alpha_{\rm DM}\mu_{ij}/m_{\rm med}\ll1$ according to \cref{eq:reduced_mass}.

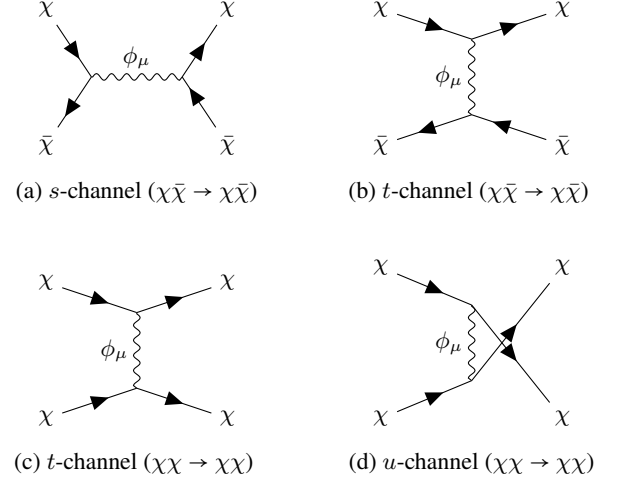
\begin{figure}[t]
\centering

\begin{subfigure}[t]{0.49\columnwidth}
\centering
\begin{tikzpicture}
\begin{feynman}
  \vertex (v1) at (-0.6,0);
  \vertex (v2) at ( 0.6,0);

  \vertex (i1) at (-1.2, 0.9) {$\chi$};
  \vertex (i2) at (-1.2,-0.9) {$\bar{\chi}$};
  \vertex (f1) at ( 1.2, 0.9) {$\chi$};
  \vertex (f2) at ( 1.2,-0.9) {$\bar{\chi}$};

  \diagram*{
    (i1) -- [fermion] (v1),
    (i2) -- [anti fermion] (v1),
    (v1) -- [photon, edge label=$\phi_\mu$] (v2),
    (v2) -- [fermion] (f1),
    (v2) -- [anti fermion] (f2),
  };
\end{feynman}
\end{tikzpicture}
\caption{$s$-channel ($\chi\bar{\chi}\rightarrow \chi\bar{\chi}$)}
\end{subfigure}
\hfill
\begin{subfigure}[t]{0.49\columnwidth}
\centering
\begin{tikzpicture}
\begin{feynman}
  \vertex (a) at (0, 0.5);
  \vertex (b) at (0,-0.5);

  \vertex (i1) at (-1.2, 0.9) {$\chi$};
  \vertex (i2) at (-1.2,-0.9) {$\bar{\chi}$};
  \vertex (f1) at ( 1.2, 0.9) {$\chi$};
  \vertex (f2) at ( 1.2,-0.9) {$\bar{\chi}$};

  \diagram*{
    (i1) -- [fermion] (a) -- [fermion] (f1),
    (i2) -- [anti fermion] (b) -- [anti fermion] (f2),
    (a)  -- [photon, edge label'=$\phi_\mu$] (b),
  };
\end{feynman}
\end{tikzpicture}
\caption{$t$-channel ($\chi\bar{\chi}\rightarrow \chi\bar{\chi}$) \label{fig:t-channel_dist}}
\end{subfigure}

\vspace{5mm}

\begin{subfigure}[t]{0.49\columnwidth}
\centering
\begin{tikzpicture}
\begin{feynman}
  \vertex (a) at (0, 0.5);
  \vertex (b) at (0,-0.5);

  \vertex (i1) at (-1.2, 0.9) {$\chi$};
  \vertex (i2) at (-1.2,-0.9) {$\chi$};
  \vertex (f1) at ( 1.2, 0.9) {$\chi$};
  \vertex (f2) at ( 1.2,-0.9) {$\chi$};

  \diagram*{
    (i1) -- [fermion] (a) -- [fermion] (f1),
    (i2) -- [fermion] (b) -- [fermion] (f2),
    (a)  -- [photon, edge label'=$\phi_\mu$] (b),
  };
\end{feynman}
\end{tikzpicture}
\caption{$t$-channel ($\chi\chi\rightarrow \chi\chi$) \label{fig:t-channel_identical}}
\end{subfigure}
\hfill
\begin{subfigure}[t]{0.49\columnwidth}
\centering
\begin{tikzpicture}
\begin{feynman}
  \vertex (a) at (0, 0.5);
  \vertex (b) at (0,-0.5);

  \vertex (i1) at (-1.2, 1.0) {$\chi$};
  \vertex (i2) at (-1.2,-1.0) {$\chi$};
  \vertex (f1) at ( 1.2, 1.0) {$\chi$};
  \vertex (f2) at ( 1.2,-1.0) {$\chi$};

  \diagram*{
    (i1) -- [fermion] (a) -- [fermion] (f2),
    (i2) -- [fermion] (b) -- [fermion] (f1),
    (a)  -- [photon, edge label'=$\phi_\mu$] (b),
  };
\end{feynman}
\end{tikzpicture}
\caption{$u$-channel ($\chi\chi\rightarrow \chi\chi$)\label{fig:u-channel_identical}}
\end{subfigure}

\caption{
Channel structure for vector-mediated fermionic DM scattering.
\emph{Top row}: distinguishable particles $\chi\bar{\chi} \to \chi\bar{\chi}$, where $\bar{\chi}$ denotes the antiparticle of $\chi$, allowing $s$- and $t$-channels.
\emph{Bottom row}: identical particles $\chi\chi \to \chi\chi$, allowing $t$- and $u$-channels. An analogous channel classification applies to all other mediator-induced self-interactions shown in \cref{fig:selfint_realizations_all}.}
\label{fig:stu_channels_consistent}
\end{figure}

For distinguishable DM particles, such as $ss^\ast\to ss^\ast$ and $\chi\bar{\chi}\to\chi\bar{\chi}$, the static Yukawa interaction arises from $t$-channel mediator exchange. For identical DM particles, including $ss\to ss$ and $\chi\chi\to\chi\chi$, the exchange-related $u$-channel contribution must also be included because the two outgoing particles are indistinguishable. These channel structures are illustrated in \cref{fig:stu_channels_consistent}. The $s$-channel topology shown for particle--antiparticle scattering does not generate the static Yukawa potential and is therefore not included in the potential-scattering expressions below; it can become relevant near an $s$-channel resonance or when the complete short-distance amplitude is required. 

\begin{figure*}
    \centering
    \includegraphics[width=2.0\columnwidth]{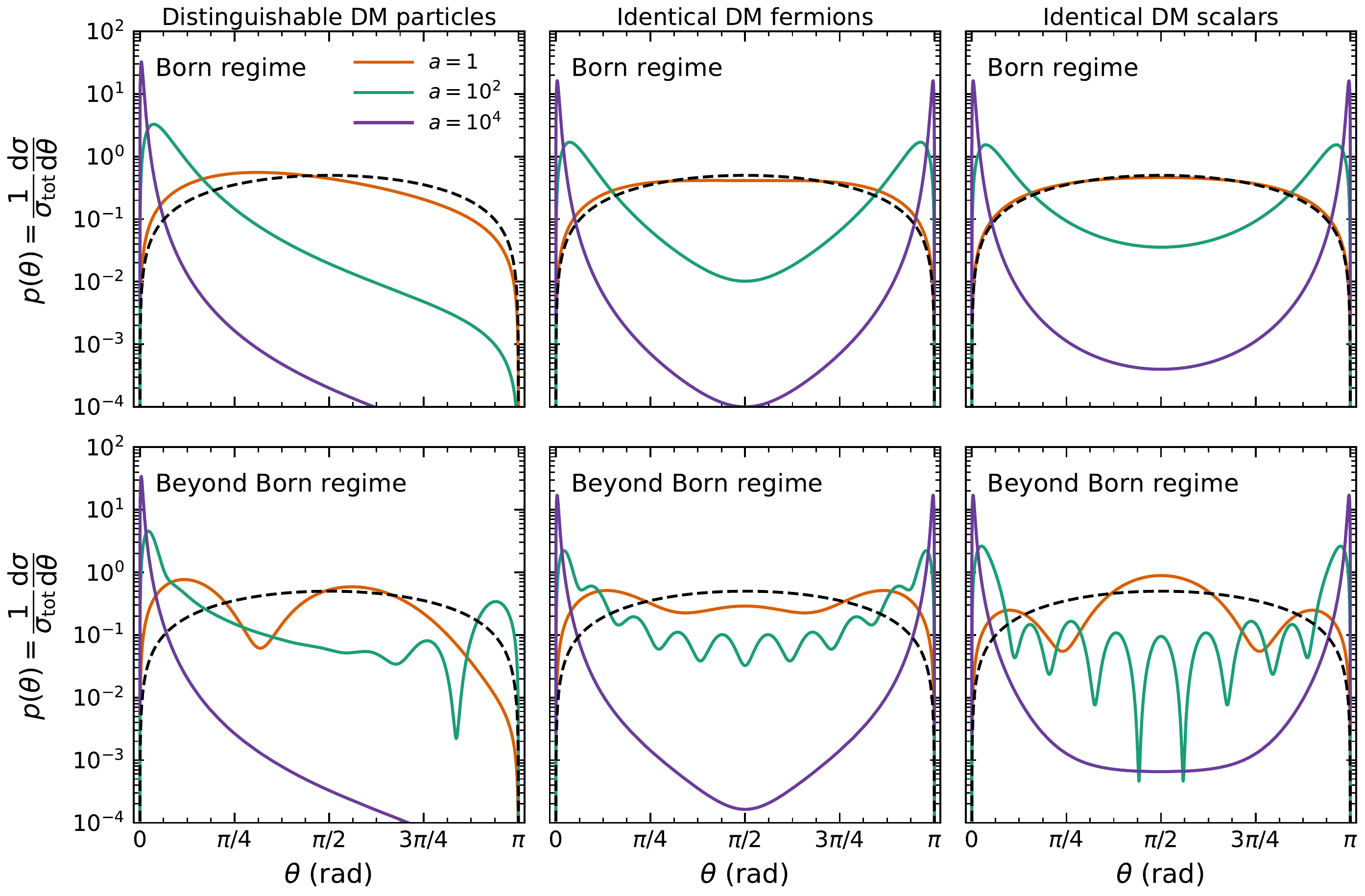}
\caption{
Normalized scattering-angle distributions for distinguishable DM particles (left column), identical DM fermions (middle column), and identical DM scalars (right column).
The top panels show the Born-regime results, $\alpha_{\rm DM}m_{\rm DM}/m_{\rm med}\ll 1$, based on the Rutherford-like and M{\o}ller-like approximations in \cref{eq:Rutherford_general,eq:Moeller_general}.
The bottom panels show the corresponding non-perturbative Yukawa results obtained from the numerical partial-wave treatment in \cref{eq:numerical_differential_cross_section,eq:differential_cross_distinguishable_general,eq:differential_cross_identical_bosons,eq:spin_averaged_identical_fermions}.
In both rows, the curves correspond to $a=1$, $10^2$, and $10^4$, while the dashed curve shows the isotropic reference distribution.
For the non-perturbative results, we fix $m_{\rm DM}=10~\mathrm{GeV}$, $m_{\rm med}=1~\mathrm{MeV}$ and $\alpha_{\rm DM}=10^{-2}$, corresponding to $\alpha_{\rm DM}m_{\rm DM}/m_{\rm med}=100$, and obtain the same three values of $a$ by varying $v_{\rm rel}$.
This allows a direct comparison between the Born-regime approximations and the full non-perturbative Yukawa calculation at the same values of the kinematic parameter $a$.
}
    \label{fig:rutherford_moller_angular_distribution}
\end{figure*}

In the distinguishable case, the Born differential cross section takes the Rutherford-like form~\cite{Tulin:2013teo}
\begin{equation}
\label{eq:Rutherford_general}
\frac{\mathrm{d}\sigma_{\rm R}}{\mathrm{d}\Omega}
=
\frac{\sigma_0}{4\pi}
\frac{1}{\left(1+a\sin^2\frac{\theta}{2}\right)^2},
\end{equation}
where
\begin{equation}
\label{eq:sigma0_w_definition}
\sigma_0
=
\frac{4\pi\alpha_{\rm DM}^2m_{\rm DM}^2}{m_{\rm med}^4},
\qquad
w\equiv\frac{m_{\rm med}c}{m_{\rm DM}},
\end{equation}
and
\begin{equation}
\label{eq:anisotropy_parameter}
a\equiv
\left(\frac{v_{\rm rel}}{w}\right)^2
=
\left(\frac{m_{\rm DM}v_{\rm rel}}{m_{\rm med}c}\right)^2.
\end{equation}
Here, $\sigma_0$ sets the overall cross-section normalization and $w$ the characteristic velocity scale~\cite{Yang:2022hkm}, while the dimensionless parameter $a$ determines the angular anisotropy at a given relative velocity. For $a\ll1$, the scattering is nearly isotropic, whereas $a\gg1$ corresponds to strong forward enhancement.

For identical DM particles, the two outgoing particles are indistinguishable, so the $t$- and $u$-channel diagrams shown in the bottom row of \cref{fig:stu_channels_consistent} correspond to the same physical final state. Their amplitudes must therefore be combined before computing the cross section, leading to interference between the two contributions. This gives a M{\o}ller-like differential cross section~\cite{Girmohanta:2022dog}
\begin{equation}
\label{eq:Moeller_general}
\begin{aligned}
\frac{\mathrm{d}\sigma_{\rm M}}{\mathrm{d}\Omega}
=
\frac{1}{2}\frac{\sigma_0}{4\pi}
\Bigg[
&\frac{1}{\left(1+a\sin^2\frac{\theta}{2}\right)^2}
+
\frac{1}{\left(1+a\cos^2\frac{\theta}{2}\right)^2}
\\
&+
\frac{c_{\rm int}}
{\left(1+a\sin^2\frac{\theta}{2}\right)
 \left(1+a\cos^2\frac{\theta}{2}\right)}
\Bigg],
\end{aligned}
\end{equation}
where
\begin{equation}
\label{eq:interference_coefficient}
c_{\rm int}
=
\begin{cases}
+2, & \text{identical DM scalars},\\
-1, & \text{identical DM fermions}.
\end{cases}
\end{equation}
The factor $1/2$ avoids double counting the identical final state. For identical DM scalars, the two contributions interfere constructively, resulting in $c_{\rm int}=2$, whereas for identical DM fermions the result depends on the symmetry of the spin state; after averaging over the possible spin configurations, one obtains $c_{\rm int}=-1$~\cite{Girmohanta:2022dog}. Because the two outgoing particles are indistinguishable, the resulting cross section is symmetric under $\theta\to\pi-\theta$ and is enhanced in both the forward and backward directions.

The Rutherford- and M{\o}ller-like differential cross sections are therefore conveniently parametrized by $(\sigma_0,w)$, with the angular dependence at a given relative velocity determined by the dimensionless parameter $a=(v_{\rm rel}/w)^2$.

The normalized scattering-angle distributions corresponding to \cref{eq:Rutherford_general,eq:Moeller_general} are shown in the top panels of \cref{fig:rutherford_moller_angular_distribution}. We define
\begin{equation}
\label{eq:normalized_scattering_angle_distribution}
p(\theta)\equiv\frac{1}{\sigma_{\rm tot}}\frac{\mathrm{d}\sigma}{\mathrm{d}\theta}
=
\frac{2\pi}{\sigma_{\rm tot}}\frac{\mathrm{d}\sigma}{\mathrm{d}\Omega}\sin\theta,
\end{equation}
with $\int_0^\pi p(\theta)\,\mathrm{d}\theta=1$. Increasing $a$ makes the Rutherford-like distribution increasingly forward-peaked, while the M{\o}ller-like distributions develop symmetric forward and backward peaks. The scalar--fermion difference is largest at intermediate angles, where $t$- and $u$-channel interference is most important.

\medskip
\noindent
\textbf{Numerical evaluation beyond the Born regime.}
For light-mediator SIDM models beyond the perturbative Born regime,
\begin{equation}
\label{eq:beyond_born}
\frac{\alpha_{\rm DM}m_{\rm DM}}{m_{\rm med}}\gtrsim1 ,
\end{equation}
multiple mediator exchanges become important and the Born expressions in \cref{eq:Rutherford_general,eq:Moeller_general} are no longer reliable. For elastic scattering of unequal masses, the corresponding condition is $2\alpha_{\rm DM}\mu_{ij}/m_{\rm med}\gtrsim1$, following the reduced-mass replacement discussed in \cref{eq:reduced_mass}. The elastic scattering problem can instead be treated non-perturbatively as a two-body potential-scattering problem, with the partial-wave phase shifts determined numerically from the Schr\"odinger equation for the Yukawa potential~\cite{Tulin:2013teo}.

The phase shifts obtained from the Schr\"odinger equation determine the scattering amplitude through the partial-wave expansion
\begin{equation}
\label{eq:numerical_differential_cross_section}
f(\theta)
=
\frac{1}{k}
\sum_{\ell=0}^{\infty}
(2\ell+1)e^{i\delta_\ell}\sin\delta_\ell
P_\ell(\cos\theta),
\end{equation}
where $P_\ell$ is the Legendre polynomial of order $\ell$, $\delta_\ell$ is the corresponding phase shift.

For distinguishable DM particles, the differential cross section follows directly from
\begin{equation}
\label{eq:differential_cross_distinguishable_general}
\frac{\mathrm{d}\sigma}{\mathrm{d}\Omega}
=
|f(\theta)|^2 .
\end{equation}
For identical particles, the final states at $\theta$ and $\pi-\theta$ are indistinguishable, so their amplitudes must be added or subtracted according to the exchange symmetry. For identical DM scalars,
\begin{equation}
\label{eq:differential_cross_identical_bosons}
\frac{\mathrm{d}\sigma}{\mathrm{d}\Omega}
=
\frac{1}{2}
\left|
f(\theta)+f(\pi-\theta)
\right|^2 ,
\end{equation}
whereas for unpolarized identical DM fermions,
\begin{equation}
\begin{aligned}
\label{eq:spin_averaged_identical_fermions}
\frac{\mathrm{d}\sigma}{\mathrm{d}\Omega}
=
\frac{1}{2}
\bigg[
\frac14\big|f(\theta)+f(\pi-\theta)\big|^2
+
\frac34\big|f(\theta)-f(\pi-\theta)\big|^2
\bigg].
\end{aligned}
\end{equation}
The symmetric and antisymmetric spatial states therefore contribute with weights $1/4$ and $3/4$ for unpolarized fermions. The factor $1/2$ avoids double counting identical final states when integrating over the full solid angle~\cite{Girmohanta:2022dog}. In the field-theory description, the same exchange symmetry arises from combining the $t$- and $u$-channel amplitudes shown in the bottom row of \cref{fig:stu_channels_consistent}.

The angle-weighted cross sections introduced in \cref{eq:weighted_cross_sections} can be obtained directly from the phase shifts. For distinguishable DM particles, the transfer and viscosity cross sections are~\cite{Tulin:2013teo}
\begin{align}
\label{eq:sigmaT_phaseshift}
\sigma_T
=&
\frac{4\pi}{k^2}
\sum_{\ell=0}^{\infty}
(\ell+1)
\sin^2\!\left(\delta_{\ell+1}-\delta_\ell\right),
\\
\label{eq:sigmaV_phaseshift}
\sigma_V
=&
\frac{6\pi}{k^2}
\sum_{\ell=0}^{\infty}
\frac{(\ell+1)(\ell+2)}{2\ell+3}
\sin^2\!\left(\delta_{\ell+2}-\delta_\ell\right).
\end{align}
The normalization of $\sigma_V$ follows the convention adopted in \cref{eq:weighted_cross_sections}, for which the angle-weighted cross sections reduce to $\sigma_{\rm tot}$ in the isotropic limit.

For identical particles, the modified transfer cross section can similarly be evaluated directly from the phase shifts,
\begin{equation}
\label{eq:sigmaT_tilde_phaseshift_general}
\begin{split}
\sigma_{\widetilde T}^{(\pm)}
=
\frac{2\pi}{k^2}
\sum_{\ell,\ell'=0}^{\infty}
&(2\ell+1)(2\ell'+1)
\left[1\pm(-1)^\ell\right]
\left[1\pm(-1)^{\ell'}\right]
\\
&\times
\cos\!\left(\delta_\ell-\delta_{\ell'}\right)
\sin\delta_\ell\sin\delta_{\ell'}
I_{\ell\ell'},
\end{split}
\end{equation}
where
\begin{equation}
\label{eq:I_llprime}
I_{\ell\ell'}
\equiv
\int_{-1}^{1}\mathrm{d}x\,
(1-|x|)
P_\ell(x)P_{\ell'}(x),
\qquad
x\equiv\cos\theta .
\end{equation}
The $+$ sign keeps only the symmetric even-$\ell$ partial waves, while the $-$ sign keeps only the antisymmetric odd-$\ell$ partial waves. Identical DM scalars therefore use the $+$ sector. For unpolarized identical DM fermions, the symmetric and antisymmetric spatial sectors are spin averaged according to
\begin{equation}
\label{eq:sigma_spin_average}
\sigma_{\widetilde T}^{\rm avg}
=
\frac14\sigma_{\widetilde T}^{(+)}
+
\frac34\sigma_{\widetilde T}^{(-)} ,
\end{equation}
where $\sigma_{\widetilde T}^{(+)}$ and $\sigma_{\widetilde T}^{(-)}$ are obtained from the even-$\ell$ and odd-$\ell$ sectors of \cref{eq:sigmaT_tilde_phaseshift_general}, respectively.

The phase shifts are obtained by solving the radial Schr\"odinger equation for the Yukawa potential in \cref{eq:yukawa_potential_generic}. Defining $\chi_\ell\equiv rR_\ell$ and introducing the dimensionless variables~\cite{Tulin:2013teo}
\begin{equation}
\label{eq:dimensionless_vars}
\begin{aligned}
x &\equiv \alpha_{\rm DM}m_{\rm DM}r,
&
a_{\rm DM} &\equiv \frac{v_{\rm rel}}{2\alpha_{\rm DM}},
\\
b_{\rm DM} &\equiv \frac{\alpha_{\rm DM}m_{\rm DM}}{m_{\rm med}},
&
\chi_\ell &\equiv rR_\ell ,
\end{aligned}
\end{equation}
the equal-mass radial equation becomes
\begin{equation}
\label{eq:dimensionless_radial}
\left[
\frac{\mathrm{d}^2}{\mathrm{d}x^2}
+a_{\rm DM}^2
-\frac{\ell(\ell+1)}{x^2}
\pm\frac{1}{x}e^{-x/b_{\rm DM}}
\right]
\chi_\ell(x)=0 .
\end{equation}
The plus sign in \cref{eq:dimensionless_radial} corresponds to the attractive Yukawa potential and the minus sign to the repulsive potential. This sign is opposite to that appearing explicitly in \cref{eq:yukawa_potential_generic} because the potential enters the radial Schr\"odinger equation through $-2\mu V$.

For each partial wave, \cref{eq:dimensionless_radial} is integrated outward from the regular small-radius solution,
\begin{equation}
\chi_\ell(x)\propto x^{\ell+1},
\end{equation}
and matched at sufficiently large radius, where the Yukawa interaction is negligible, to the free-wave solution
\begin{equation}
\label{eq:asymptotic_match}
\chi_\ell(x)
\propto
x
\left[
\cos\delta_\ell\,j_\ell(a_{\rm DM}x)
-
\sin\delta_\ell\,n_\ell(a_{\rm DM}x)
\right],
\end{equation}
where $j_\ell$ and $n_\ell$ are the spherical Bessel and Neumann functions. Matching the numerical and asymptotic solutions determines $\delta_\ell$. The calculation is repeated for increasing $\ell$ until the reconstructed differential or angle-weighted cross section converges; a detailed description of this numerical procedure is given in~\cite{Tulin:2013teo}.

Once the phase shifts have been determined, the full angular dependence can be reconstructed through \cref{eq:numerical_differential_cross_section} and converted to the physical differential cross section using \cref{eq:differential_cross_distinguishable_general} for distinguishable DM particles or \cref{eq:differential_cross_identical_bosons,eq:spin_averaged_identical_fermions} for identical DM particles. Alternatively, the angle-weighted cross sections can be evaluated directly from the phase shifts. The resulting quantities $\mathrm{d}\sigma/\mathrm{d}\Omega(\theta,v_{\rm rel})$ and $\sigma_X(v_{\rm rel})$ provide the simulation inputs required by the elastic SIDM algorithms discussed in \cref{sec: implementation}.

The corresponding normalized scattering-angle distributions are shown in the bottom panels of \cref{fig:rutherford_moller_angular_distribution}. They are obtained from the partial-wave amplitude in \cref{eq:numerical_differential_cross_section} using \cref{eq:differential_cross_distinguishable_general,eq:differential_cross_identical_bosons,eq:spin_averaged_identical_fermions} and normalized according to \cref{eq:normalized_scattering_angle_distribution}. We use the same values $a=1$, $10^2$ and $10^4$ as in the Born-regime examples above, fixing $m_{\rm DM}=10~\mathrm{GeV}$, $m_{\rm med}=1~\mathrm{MeV}$ and $\alpha_{\rm DM}=10^{-2}$, such that $\alpha_{\rm DM}m_{\rm DM}/m_{\rm med}=100$, and varying $v_{\rm rel}$ according to \cref{eq:anisotropy_parameter}. Compared with the Born expressions in \cref{eq:Rutherford_general,eq:Moeller_general}, the non-perturbative distributions can exhibit a more complex angular structure due to interference between multiple partial waves, while the distributions for identical DM particles remain symmetric under $\theta\rightarrow\pi-\theta$.

\medskip
\noindent
\textbf{Contact interactions.}
The self-scattering realizations shown in \cref{fig:selfint_realizations_all} reduce to short-range \emph{contact} interactions when the mediator mass is large compared to the typical momentum transfer in DM haloes (cf.~\cref{fig:heavy_mediator_limit_both}). In this heavy-mediator limit, the interaction range becomes negligible, the scattering amplitude becomes approximately momentum independent and the differential cross section is nearly isotropic and velocity independent over astrophysical scales.

In the contact limit, the differential cross section becomes independent of the scattering angle and therefore takes the isotropic form
\begin{equation}
\frac{\mathrm{d}\sigma}{\mathrm{d}\Omega}
\simeq
\frac{\sigma_{\rm tot}}{4\pi}.
\label{eq:contact_isotropic}
\end{equation}
The value of $\sigma_{\rm tot}$ depends on the scattering process. For the Rutherford-like interaction in \cref{eq:Rutherford_general}, the limit $a\to0$ gives $\sigma_{\rm tot}=\sigma_0$. For the M{\o}ller-like interaction in \cref{eq:Moeller_general}, the same limit gives $\sigma_{\rm tot}=2\sigma_0$ for identical DM scalars and $\sigma_{\rm tot}=\sigma_0/2$ for identical DM fermions, because the $t$- and $u$-channel amplitudes interfere constructively for DM scalars and destructively for DM fermions. Because the scattering is isotropic, all angle-weighted cross sections in \cref{eq:weighted_cross_sections} coincide up to trivial $\mathcal{O}(1)$ factors,
\begin{equation}
\sigma_{\rm tot}
\simeq
\sigma_{\rm T}
\simeq
\sigma_{\widetilde{\rm T}}
\simeq \sigma_{\rm V}.
\end{equation}

Below we summarize cross sections for the contact interactions corresponding to \cref{fig:selfint_scalar_contact,fig:heavy_mediator_limit_both}. For mediator-induced interactions, we consider the contact limit of the Yukawa potential-scattering contributions introduced above: $t$-channel exchange for distinguishable particles and the combined $t$- and $u$-channel contributions for identical particles. Possible additional short-distance $s$-channel contributions are not included, consistently with the potential-scattering treatment above.

\medskip
\noindent
\emph{Scalar DM contact self-interaction.}
The quartic interaction in \cref{eq:quartic_coupling} directly generates a momentum-independent $2\to2$ amplitude and hence isotropic scattering. For the complex and real scalar DM cases, respectively,
\begin{equation}
\begin{aligned}
\left.
\frac{\mathrm{d}\sigma}{\mathrm{d}\Omega}
\right|_{s s^\ast\to s s^\ast}
&=
\frac{\lambda^2}{16\pi^2m_s^2},
&
\sigma_{\rm tot}
&=
\frac{\lambda^2}{4\pi m_s^2},
\\[4pt]
\left.
\frac{\mathrm{d}\sigma}{\mathrm{d}\Omega}
\right|_{ss\to ss}
&=
\frac{\lambda^2}{512\pi^2m_s^2},
&
\sigma_{\rm tot}
&=
\frac{\lambda^2}{128\pi m_s^2}.
\end{aligned}
\label{eq:scalar_quartic_contact_cross_sections}
\end{equation}
The second line includes the symmetry factor $1/2$ for identical final-state particles.

\medskip
\noindent
\emph{Heavy-mediator interactions.}
For the mediator-induced interactions in \cref{fig:heavy_mediator_limit_both}, the contact-limit normalization follows directly from \cref{eq:anisotropy_parameter,eq:alpha_DM_microscopic},
\begin{equation}
\begin{aligned}
\sigma_0
&=
\frac{4\pi\alpha_{\rm DM}^2m_{\rm DM}^2}{m_{\rm med}^4}
\\[4pt]
&=
\begin{cases}
\dfrac{\mu^4}{64\pi m_s^2m_\phi^4},
& \text{DM scalar, scalar mediator}, \\[6pt]
\dfrac{g^4m_s^2}{4\pi m_{\phi_\mu}^4},
& \text{DM scalar, vector mediator}, \\[6pt]
\dfrac{y^4m_\chi^2}{4\pi m_\phi^4},
& \text{DM fermion, scalar mediator}, \\[6pt]
\dfrac{g^4m_\chi^2}{4\pi m_{\phi_\mu}^4},
& \text{DM fermion, vector mediator}.
\end{cases}
\end{aligned}
\label{eq:heavy_mediator_sigma0}
\end{equation}
The corresponding contact-limit differential and total cross sections are then obtained from
\begin{equation*}
\begin{array}{lll}
\text{Distinguishable DM particles:}
&
\dfrac{\mathrm{d}\sigma}{\mathrm{d}\Omega}
\simeq\dfrac{\sigma_0}{4\pi},
&
\sigma_{\rm tot}\simeq\sigma_0,
\\[10pt]
\text{Identical DM scalars:}
&
\dfrac{\mathrm{d}\sigma}{\mathrm{d}\Omega}
\simeq\dfrac{\sigma_0}{2\pi},
&
\sigma_{\rm tot}\simeq2\sigma_0,
\\[10pt]
\text{Identical DM fermions:}
&
\dfrac{\mathrm{d}\sigma}{\mathrm{d}\Omega}
\simeq\dfrac{\sigma_0}{8\pi},
&
\sigma_{\rm tot}\simeq\dfrac{\sigma_0}{2}.
\end{array}
\label{eq:heavy_mediator_contact_cross_sections}
\end{equation*}
Here the identical-particle expressions already include the symmetry factor for the final state and the interference between the $t$- and $u$-channel amplitudes. Although the scalar--vector coupling in \cref{eq:vector_int} contains derivatives, its leading non-relativistic contribution is not velocity suppressed, so it approaches the same momentum-independent contact form.

\medskip
\noindent
\textbf{Effective-range formalism.}
For short-range self-interactions at low relative momentum, it is convenient to replace the microscopic model with an effective parametrization of scattering phase shifts. This proves especially valuable for composite SIDM models (see \cref{sec:composite_SIDM}), where non-perturbative composite dynamics generate the interaction and may be far more complex than simple single-mediator potentials. Here, direct microscopic calculations of the full potential or angular dependence become difficult, whereas the effective-range formalism~\cite{Chu_2020} provides a compact, model-independent description of low-energy scattering relevant for structure formation.

The non-relativistic scattering amplitude expands in partial waves as~\begin{align}
\begin{aligned}
f(v_{\rm rel},\theta)
=&
\sum_{\ell=0}^{\infty}
(2\ell+1)\,f_\ell(k)\,P_\ell(\cos\theta),
\qquad \\ 
f_\ell(k)
=&
\frac{1}{k(\cot\delta_\ell-i)},
\label{eq:pw_effective_range}
\end{aligned}
\end{align}
where $\delta_\ell(k)$ is the phase shift of the $\ell$-th partial wave. For finite-range interactions, $k^{2\ell+1}\cot\delta_\ell(k)$ is analytic at $k=0$ and expands as~\cite{Bethe:1949yr,Blatt:1949zz}
\begin{equation}
k^{2\ell+1}\cot\delta_\ell(k) = -\frac{1}{a_\ell^{\,2\ell+1}} + \frac{1}{2}\,r_{e,\ell}\,k^2 + \mathcal{O}(k^4),
\label{eq:ere_general}
\end{equation}
where $a_\ell$ and $r_{e,\ell}$ are the generalized scattering length and effective range, respectively.

When a single partial wave dominates, this provides a compact, model-independent description for velocity dependence of the cross section. For s-wave dominated scattering ($\ell=0$), as in \cref{eq:effective_range_cross}, one obtains
\begin{equation}
\sigma_{0} = \frac{4\pi}{k^2}\sin^2\delta_0 \;\simeq\; \frac{4\pi a_0^2}{1+k^2\!\left(a_0^2-a_0 r_{e,0}\right)+\frac14 a_0^2 r_{e,0}^2 k^4}\,.
\label{eq:ere_swave}
\end{equation}
This form captures diverse low-energy SIDM scenarios, including Yukawa-induced scattering in the short-range regime and scattering enhanced by near-threshold bound states, virtual levels or resonances.\footnote{Near-threshold bound states are true two-body states with energies just below threshold, producing very strong low-velocity scattering. Virtual levels describe systems close to forming a bound state, still enhancing the cross section, while resonances are short-lived intermediate states at positive energies, producing cross-section peaks at specific velocities.} It proves particularly valuable for composite SIDM models, where detailed microscopic interactions are complicated but low-energy scattering can often be summarized by a few effective parameters such as $a_0$ and $r_{e,0}$.

The effective-range formalism extends beyond s-wave: keeping $\ell=1$ incorporates p-wave scattering and angular dependence through $P_1(\cos\theta)=\cos\theta$. When multiple partial waves contribute, the angular distribution follows from \cref{eq:pw_effective_range}, providing a systematic bridge between isotropic low-energy physics and treatments including higher partial waves and angular structure. However, the validity of the effective-range expansion is limited to finite-range interactions and sufficiently low momenta. When interactions become long-ranged or when $k$ is large enough that many partial waves contribute significantly, the truncated expansion in \cref{eq:ere_general} becomes insufficient. In such cases, one must return to full numerical phase-shift calculations described above and reconstruct the differential cross section from the complete partial-wave sum.

\subsection{Decay inputs}
\label{sec:decay_inputs}

The final ingredients required to complete the mapping from particle physics to simulation inputs are the decay rates $\Gamma$ of unstable dark-sector states and the corresponding velocity kicks $v_{\rm kick}$ imparted to the daughter DM particles. The decay rate $\Gamma$ sets the decay probability per unit time, and hence the lifetime and frequency of decay events, while $v_{\rm kick}$ determines the recoil imparted to the daughter DM particle and thereby encodes the associated kinetic-energy injection. While many SIDM models assume all DM species are stable on cosmological timescales, several well-motivated scenarios incorporate unstable states that can decay during or after structure formation.

\medskip
\noindent
\textbf{Physical origin of dark-sector decays.}
Decays arise whenever the dark sector contains unstable states, including excited states that can transition to lower-energy configurations. Their impact on structure formation depends on the decay lifetime relative to the relevant cosmological and dynamical timescales. Such decays can occur within SIDM models or in other DM scenarios in which particle decay affects structure formation. Representative examples include:

\begin{itemize}[leftmargin=*, labelsep=0.5em]

\item \emph{Excited-state DM:} Multi-state models can contain a stable ground state ${\rm DM}_1$ and a heavier state ${\rm DM}_2$, separated by a mass splitting $\delta$, with decays such as ${\rm DM}_2\to{\rm DM}_1+\gamma_{\rm D}$~\cite{Finkbeiner:2007kk,Arkani-Hamed:2008hhe}. Here ${\rm DM}_{1,2}$ may, for example, denote scalar states ($s_{1,2}$) or fermionic states ($\chi_{1,2}$), while $\gamma_{\rm D}$ represents a light dark-sector particle such as a scalar mediator $\phi$ or dark photon $\phi_\mu$. The microscopic masses and couplings determine both the decay rate and the recoil of the daughter DM state.

\item \emph{Dark atoms:} Atomic DM models can contain excited bound states that decay radiatively to lower-energy levels through dark-photon emission~\cite{Cyr-Racine:2013fsa,Cline_2014a}. The corresponding level splittings determine the released energy and recoil, while the transition rates determine the lifetime of the excited states; see \cref{sec:dark_atoms}.

\item \emph{Other dark-sector decay scenarios:} Decays also arise in a variety of other dark-sector realizations. Dark-sector mediators can themselves be unstable and decay into SM particles~\cite{Pospelov:2007mp}, while unstable DM appears in scenarios such as gravitino DM with $R$-parity violation~\cite{Fayet:1981sq,Witten:1981nf,Pagels:1981ke,Khlopov:1984pf}, super-WIMP models~\cite{Covi:1999ty,Feng:2003uy}, sterile-neutrino DM~\cite{Ma:2001ip,Dib:2011jh}, and models with decaying gravitons or superheavy DM~\cite{Law-Smith:2023czn,Obied:2023clp,Allahverdi:2023nov}.

\end{itemize}

The relevance of a decay for structure formation is controlled by the lifetime $\tau=1/\Gamma$ relative to the timescales of interest. Short-lived states decay before late-time halo evolution, whereas sufficiently long-lived states can survive into the structure-formation epoch and decay within evolving haloes.

\begin{figure}[t]
\centering

\begin{subfigure}[t]{0.49\columnwidth}
\centering
\begin{tikzpicture}
\begin{feynman}
  \vertex (i1) at (-1.2, 0) {$s_2$};
  \vertex (b) at ( 0, 0);
  \vertex (f1) at ( 1.2, 0.5) {$s_1$};
  \vertex (f2) at ( 1.2,-0.5) {$\phi$};

  \diagram*{
    (i1) -- [scalar] (b),
    (b)  -- [scalar] (f1),
    (b)  -- [scalar] (f2),
  };
\end{feynman}
\end{tikzpicture}
\caption{Scalar DM decay via scalar mediator: $s_2 \to s_1 + \phi$.}
\label{fig:decay_scalar_phi}
\end{subfigure}
\hfill
\begin{subfigure}[t]{0.49\columnwidth}
\centering
\begin{tikzpicture}
\begin{feynman}
  \vertex (i1) at (-1.2, 0) {$s_2$};
  \vertex (b) at ( 0, 0);
  \vertex (f1) at ( 1.2, 0.5) {$s_1$};
  \vertex (f2) at ( 1.2,-0.5) {$\phi_\mu$};

  \diagram*{
    (i1) -- [scalar] (b),
    (b)  -- [scalar] (f1),
    (b)  -- [photon] (f2),
  };
\end{feynman}
\end{tikzpicture}
\caption{Scalar DM decay via vector mediator: $s_2 \to s_1 + \phi_\mu$.}
\label{fig:decay_scalar_vec}
\end{subfigure}

\vspace{4mm}

\begin{subfigure}[t]{0.49\columnwidth}
\centering
\begin{tikzpicture}
\begin{feynman}
  \vertex (i1) at (-1.2, 0) {$\chi_2$};
  \vertex (b) at ( 0, 0);
  \vertex (f1) at ( 1.2, 0.5) {$\chi_1$};
  \vertex (f2) at ( 1.2,-0.5) {$\phi$};

  \diagram*{
    (i1) -- [fermion] (b),
    (b)  -- [fermion] (f1),
    (b)  -- [scalar] (f2),
  };
\end{feynman}
\end{tikzpicture}
\caption{Fermionic DM decay via scalar mediator: $\chi_2 \to \chi_1 + \phi$.}
\label{fig:decay_fermion_phi}
\end{subfigure}
\hfill
\begin{subfigure}[t]{0.49\columnwidth}
\centering
\begin{tikzpicture}
\begin{feynman}
  \vertex (i1) at (-1.2, 0) {$\chi_2$};
  \vertex (b) at ( 0, 0);
  \vertex (f1) at ( 1.2, 0.5) {$\chi_1$};
  \vertex (f2) at ( 1.2,-0.5) {$\phi_\mu$};

  \diagram*{
    (i1) -- [fermion] (b),
    (b)  -- [fermion] (f1),
    (b)  -- [photon] (f2),
  };
\end{feynman}
\end{tikzpicture}
\caption{Fermionic DM decay via vector mediator: $\chi_2 \to \chi_1 + \phi_\mu$.}
\label{fig:decay_fermion_vec}
\end{subfigure}

\caption{Feynman diagrams for DM two-body decays. \emph{Top row}: scalar DM ($s$) decaying via emission of a scalar mediator $\phi$ or vector mediator $\phi_\mu$. \emph{Bottom row}: fermionic DM ($\chi$) decaying via emission of a scalar mediator $\phi$ or vector mediator $\phi_\mu$.}
\label{fig:decay_diagrams}
\end{figure}
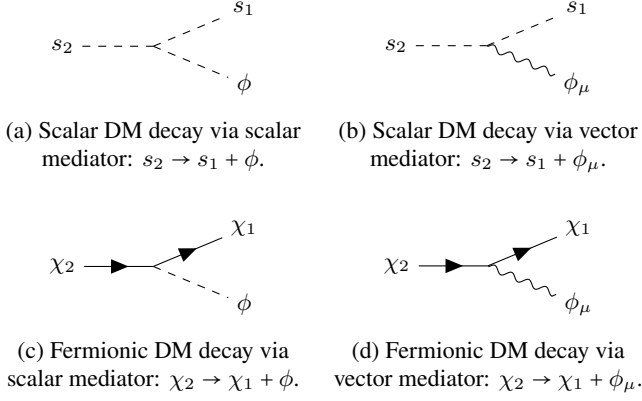

\medskip
\noindent
\textbf{Decay kinematics and velocity kick.}
For a two-body decay $a\to1+2$ in the rest frame of the parent particle $a$, energy--momentum conservation fixes the final-state momenta and hence the recoil of the daughter DM particle. We keep $c$ explicit to make the mapping to physical simulation velocities transparent. The two decay products are emitted back-to-back with equal and opposite momenta of magnitude
\begin{equation}
\label{eq:final_momenta_decay}
|\vec{p}_{1,2}|=
\frac{c}{2m_a}
\sqrt{\bigl[m_a^2-(m_1+m_2)^2\bigr]
\bigl[m_a^2-(m_1-m_2)^2\bigr]} .
\end{equation}
Their orientation is determined by the angular distribution of the decay and is sampled isotropically when the decay is isotropic.

If particle $1$ denotes the massive daughter DM state represented in the simulation and remains non-relativistic, its recoil speed in the parent rest frame defines the decay velocity kick,
\begin{equation}
v_{\rm kick}^{\rm decay}
\equiv
|\vec{v}_1^{\rm rest}|
\simeq
\frac{|\vec{p}_1|}{m_1}.
\end{equation}
The second decay product may instead be relativistic or massless, as in the emission of a light dark-sector particle such as a dark photon. If this state freely escapes the system and is not dynamically followed, its momentum is nevertheless fully accounted for through the recoil of the massive daughter. Thus, $v_{\rm kick}^{\rm decay}$ is not an independent microscopic quantity, but is fixed by the parent and daughter masses for a specified decay channel.

If the parent DM particle is non-relativistic and moves with velocity $\vec{v}_a$ in the simulation frame, the velocity of the massive daughter is
\begin{equation}
\vec{v}_1=
\vec{v}_a+\vec{v}_1^{\rm rest},
\end{equation}
where $\vec{v}_1^{\rm rest}$ has magnitude $v_{\rm kick}^{\rm decay}$ and is assigned a random isotropic direction in the parent rest frame. The microscopic masses therefore determine the two-body decay kinematics and map directly onto the velocity kick required by the simulation.

For decays into multiple particles ($a \to 1 + 2 + 3 + \ldots$), the final-state momenta must be sampled using Monte Carlo methods that account for both the available phase space and the decay dynamics~\cite{ParticleDataGroup:2024cfk,James:1968gu}. The sampled momenta are computed in the parent particle's rest frame and then boosted to the simulation frame. In such cases, the simple fixed-kick description is replaced by the corresponding sampled decay kinematics.

\medskip
\noindent
\textbf{Calculation of decay rates.}
For the field-theory expressions in this paragraph, we use natural units, $\hbar=c=1$. The decay width for a process $a \to 1+2+\dots$ is obtained from the matrix element $\mathcal{M}$ through~\citep{Peskin:1995ev}
\begin{equation}
\Gamma(a \to 1+2+\dots)
=
\frac{1}{2m_a}
\int \mathrm{d}\Pi_n\,|\mathcal{M}|^2,
\end{equation}
where
\begin{equation}
\mathrm{d}\Pi_n
=
(2\pi)^4
\delta^{(4)}
\!\left(
p_a-\sum_{i=1}^n p_i
\right)
\prod_{i=1}^n
\frac{\mathrm{d}^3p_i}{(2\pi)^3\,2E_i}
\end{equation}
is the $n$-body Lorentz-invariant phase-space element. The matrix element contains the dependence on the underlying dark-sector masses, couplings and interaction structure.

For a two-body decay $a\to1+2$ in the rest frame of particle $a$, the phase-space element reduces to
\begin{equation}
\mathrm{d}\Pi_2
=
\frac{|\vec p_1|}{16\pi^2m_a}\,
\mathrm{d}\Omega,
\end{equation}
where $|\vec p_1|$ is the final-state momentum given by \cref{eq:final_momenta_decay}. If the squared matrix element is isotropic after averaging over initial-state quantum numbers and summing over final-state polarizations, the decay products are emitted uniformly in solid angle and
\begin{equation}
\Gamma(a\to1+2)
=
\frac{1}{8\pi}
\frac{|\vec p_1|}{m_a^2}
|\mathcal M|^2.
\end{equation}

For decays such as those shown in \cref{fig:decay_fermion_vec}, the decay rate depends on the underlying interaction and hence on the particle masses and couplings. It must therefore be calculated for the specific microscopic model and supplied to the simulation.

\medskip
\noindent
\textbf{Decay probability.} When expressed in inverse-time units, the decay rate $\Gamma$ determines the probability for a particle $a$ to decay during a time interval $\Delta t$, \begin{equation}
P_{\rm decay} = 1 - \exp(-\Gamma \Delta t).
\label{eq:decay_probability}
\end{equation} This expression follows from the exponential decay law, where the survival probability decreases as $e^{-\Gamma \Delta t}$. If multiple decay channels are available, the total decay rate is $\Gamma_{\rm tot} = \sum_i \Gamma_i$ and the branching ratio to a specific channel is ${\rm BR}_i = \Gamma_i/\Gamma_{\rm tot}$.

\medskip
\noindent
\textbf{Impact on halo structure.} Decays can affect structure formation through several mechanisms:

\begin{itemize}[leftmargin=*, labelsep=0.5em]
\item \emph{Mass loss and disruption:} If recoil velocities are sufficiently large for daughter DM particles to escape the halo, the bound mass decreases and the gravitational potential becomes shallower, potentially disrupting low-mass haloes and reducing the abundance of subhaloes~\cite{Peter:2010jy,Wang:2014ina}.

\item \emph{Recoil heating:} Velocity kicks imparted to the massive daughter DM particles heat the halo, driving expansion and reducing its central density and concentration~\cite{Peter:2010jy,Abdelqader:2008wa}. At the simulation level, the magnitude of this effect is controlled by the velocity kick $v_{\rm kick}^{\rm decay}$.

\item \emph{Change in interaction properties:} If the daughter state has different self-interaction properties from the parent, decays change the relative state abundances and can thereby modify the effective scattering behaviour of the DM population~\cite{Schutz:2014nka}.

\end{itemize}

With this theory-to-simulation mapping in place, we have a general toolbox that makes explicit which microscopic ingredients are required to represent a given dark-sector model in structure-formation simulations. In the following, we use this toolbox to organize representative model classes and discuss, for each, their characteristic phenomenology, the relevant particle-physics, astrophysical and cosmological constraints, and how the corresponding physics can be incorporated into simulations.

\section*{SIDM model classes and their simulation mapping}

We now assemble the particle-physics ingredients developed above into representative SIDM model classes: contact-like and minimal light-mediator SIDM (Secs.~\ref{sec:contact-like} and \ref{sec:light-mediator}), resonant SIDM (\cref{sec:Resonant_SIDM}), vector SIDM (\cref{sec:Vector_SIDM}), asymmetric SIDM (\cref{sec:Asymmetric_SIDM}), multi-state inelastic SIDM (\cref{sec: inelastic_sidm_models}) and composite SIDM (\cref{sec:composite_SIDM}). Each name identifies the particle-physics feature used to organize the discussion rather than a property that uniquely characterizes the theory; individual models may therefore belong to several classes at once.

To make the connection between particle physics and structure formation transparent, we first give a \textit{microscopic description} of each model class, specifying the particle content, relevant interactions and resulting scattering or decay processes, followed by \textit{representative benchmarks} illustrating the parameter space of interest. Detailed cosmological, astrophysical and experimental constraints are deferred to App.~\ref{app:model_constraints}. At the end of each broader class, we then collect the \textit{mapping from particle physics to simulation inputs} and the corresponding \arepoTwo\ implementation, making explicit how the microscopic parameter space, $\mathcal{P}_{\rm PP}$, reduces to the simulation inputs, $\mathcal{P}_{\rm SIM}$, required for structure-formation simulations and which numerical schemes can realize them.

This delimitation of parameter space can also be approached from the simulation side, by constraining the simulation inputs directly against observations. Rather than simulating each model in full, we exploit the fact that different astrophysical systems act as distinct laboratories in which different DM properties dominate. Each model class is thus characterized by a distinct set of simulation diagnostics, allowing models to be constrained by targeting the dominant diagnostic in the system where its effect is strongest, as discussed in \cref{sec:simulation_diagnostics_DM}.

\subsection{Elastic single-state SIDM}
\label{sec: elastic_single_state}

We begin by considering the simplest classes of SIDM models in which the DM consists of a single stable DM particle species and the interactions are purely elastic.

\subsubsection{Contact-like (short-range) self-interactions}
\label{sec:contact-like}

Self-coupled (short-range interacting) DM constitutes the simplest class of SIDM models. In this class, the interactions are elastic, isotropic and velocity independent, providing a minimal and widely used benchmark for structure formation.

\medskip
\noindent
\textbf{Microscopic description:} The simplest realization of SIDM consists of a cosmologically stable scalar field $s$, which may be either real or complex~\cite{Bento:2000ah,McDonald:2001vt}. As discussed in \cref{eq:quartic_coupling}, self-interactions arise from renormalizable quartic terms,
\begin{equation}
\mathcal{L}_{\rm int} \supset
\begin{cases}
-\dfrac{\lambda}{4!}\, s^4 
& \text{(real scalar)} \\[6pt]
-\lambda\, (s^\dagger s)^2
& \text{(complex scalar)}
\end{cases}
\end{equation}
corresponding to a single four-point vertex (see \cref{fig:selfint_scalar_contact}) that induces elastic $ss \to ss$ or $ss^* \to ss^*$ scattering. For the non-derivative quartic contact interaction, the tree-level scattering amplitude is momentum independent, resulting in isotropic and velocity-independent scattering. As given in \cref{eq:scalar_quartic_contact_cross_sections}, the corresponding total cross sections for real and complex scalar DM are
\begin{equation}
\label{eq:sigma_tot_scalar_cases}
\sigma_{\rm tot} =
\begin{cases}
\dfrac{\lambda^2}{128\pi m_s^2}
& \text{(real scalar)}\\[6pt]
\dfrac{\lambda^2}{4\pi m_s^2}
& \text{(complex scalar)} 
\end{cases}
\end{equation}
where $m_s$ denotes the DM mass. Expressed in terms of the phenomenologically relevant quantity $\sigma/m_s$,
\begin{equation*}
m_s
\simeq
\lambda^{2/3}
\left(
\frac{1\,{\rm cm}^2/{\rm g}}{\sigma_\mathrm{tot}/m_s}
\right)^{1/3} \times \begin{cases}
16\,{\rm MeV}
& \text{(real scalar)}\\[6pt]
8\,{\rm MeV}
& \text{(complex scalar)} 
\end{cases}
\end{equation*}
implying MeV-scale masses for $\lambda \sim \mathcal{O}(1)$ and $\sigma/m_s \sim 0.1\text{--}10\,{\rm cm}^2/{\rm g}$.

Moreover, contact interactions arise for both scalar and fermionic DM when the mediator mass satisfies $m_{\rm med}^2 \gg q^2$. In this heavy-mediator limit, the $t$-channel exchange diagrams for fermionic DM with either a scalar mediator $\phi$ or a vector mediator $\phi_\mu$ (see \cref{fig:heavy_limit_fermion}) reduce to local four-fermion operators,
\begin{equation}
\mathcal{L}_{\rm eff} \supset
\begin{cases}
\dfrac{y^2}{m_\phi^2}
(\bar{\chi}\chi)(\bar{\chi}\chi)
& \text{(scalar mediator)} \\[6pt]
\dfrac{g^2}{m_{\phi_\mu}^2}
(\bar{\chi}\gamma^\mu \chi)(\bar{\chi}\gamma_\mu \chi)
& \text{(vector mediator)}
\end{cases}
\end{equation}
where $y$ and $g$ denote the mediator couplings, and $m_\phi$ and $m_{\phi_\mu}$ are the scalar and vector mediator masses, respectively. From \cref{eq:heavy_mediator_sigma0}, the corresponding total cross sections are
\begin{equation}
\sigma_{\rm tot} \simeq
\dfrac{m_\chi^2}{4\pi}\times
\begin{cases}
\dfrac{y^4}{m_\phi^4}
& \text{(scalar mediator)} \\[6pt]
\dfrac{g^4}{m_{\phi_\mu}^4}
& \text{(vector mediator)}.
\end{cases}
\end{equation}
For mediator masses comparable to the DM mass, $m_{\rm med} \sim m_\chi$ and couplings $y,g \sim \mathcal{O}(1)$, this again corresponds to MeV-scale DM masses with $\sigma/m_\chi \sim 0.1\text{--}10\,{\rm cm}^2/{\rm g}$. Several particle-physics models fall into this category, including scalar DM interacting via heavy scalar or vector mediators, as well as fermionic DM with analogous heavy mediators~\cite{Ko:2014nha,Kao:2020sqs,Khater:2021wcx}. In large regions of parameter space consistent with current constraints, these models predict approximately velocity-independent self-interactions; see, for example, the viable fermionic SIDM realizations discussed in~\cite{Hambye:2019tjt}.

The same effective description also emerges naturally in composite dark sectors below the confinement scale, where bound states (e.g.\ dark mesons or baryons) interact through short-range composite dynamics or residual forces, as illustrated in \cref{fig:selfint_composite_mesons,fig:selfint_composite_baryons}. These interactions can be effectively captured by local operators, as demonstrated for dark mesons~\cite{Hochberg:2014kqa}, in composite Higgs frameworks~\cite{Rosenlyst:2021elz} and for dark baryons~\cite{Cline:2022leq}.

These models are characterized by a momentum-independent matrix element, leading to an approximately isotropic and velocity-independent differential cross section in the non-relativistic regime. The microscopic parameter space can therefore be summarized by a small set of particle-physics parameters, e.g.
\begin{equation} \label{eq:PP_parameters_contact}
\mathcal{P}_{\rm PP} = \{ m_{\mathrm{DM}},\, \lambda \}
\quad \text{or} \quad
\{ m_{\mathrm{DM}},\, m_{\mathrm{med}},\, y \ \text{or}\ g \},
\end{equation}

\medskip
\noindent
\textbf{Representative benchmarks:}
For purely self-coupled scalar DM with a quartic
interaction, $\lambda\simeq1$ and
$m_s\simeq8~{\rm MeV}$ (real scalar) or
$m_s\simeq26~{\rm MeV}$ (complex scalar) yield
$\sigma_{\rm tot}/m_s\simeq1~{\rm cm}^2/{\rm g}$.
For fermionic DM with a heavy mediator, representative benchmarks include
\begin{equation}
m_\chi \simeq 25~{\rm MeV}, \qquad m_{\rm med} \sim m_\chi, \qquad g \sim 1,
\end{equation}
also yielding $\sigma/m_\chi \sim 1~{\rm cm}^2/{\rm g}$. These benchmarks yield cross sections consistent with astrophysical constraints from dwarf galaxies through clusters. For detailed discussion of the cosmological, astrophysical, and experimental viability of contact-like SIDM models, see App.~\ref{app:constraints_contact-like}.

\begin{figure*}
\centering    
    \includegraphics[width=1.4\columnwidth]{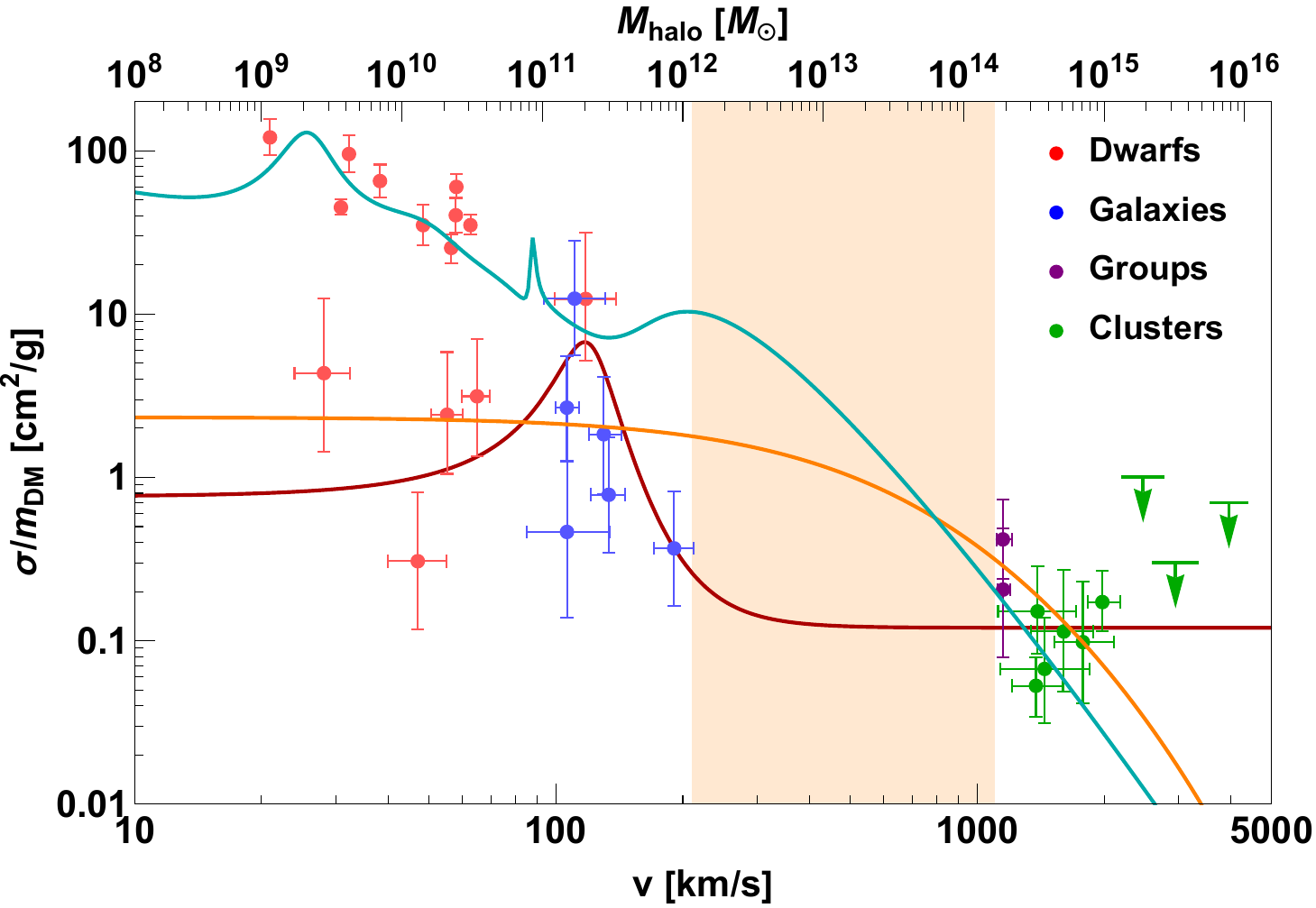}
    \caption{Elastic velocity-dependent cross sections: resonant (cyan), minimal light-mediated (orange) and dark-meson SIDM (dark red). The orange band highlights the weakly constrained intermediate-velocity regime, where the cross sections differ strongly. The resonant model can be simulated with either isotropic or anisotropic scattering, while the other models are treated using an isotropic approximation. Markers show inferred constraints from dwarf galaxies, low-surface-brightness galaxies, galaxy groups, clusters and cluster collisions~\cite{Kaplinghat:2015aga,Sagunski:2020spe,2021MNRAS503920C}, assuming elastic, isotropic scattering.}
    \label{fig:cross_sections}
\end{figure*}

\subsubsection{Minimal SIDM models with light mediator}
\label{sec:light-mediator}

Minimal SIDM models with a light mediator constitute one of the simplest and most widely studied classes of velocity-dependent SIDM scenarios. In these models, the DM consists of a single stable particle species, while self-interactions are mediated by an additional light scalar or vector particle. The light mediator generates a finite-range Yukawa interaction, which can lead to elastic but generally anisotropic and velocity-dependent scattering.

\medskip
\noindent
\textbf{Microscopic description:} At the microscopic level, light-mediator SIDM arises from the renormalizable Lagrangian interaction terms introduced in \cref{sec:renormalizable_interactions}. The simplest and most widely studied realizations involve either a light \textit{scalar mediator} $\phi$ or a light \textit{vector mediator} $\phi_\mu$~\cite{Buckley_2010,Feng:2009hw,Tulin:2013teo,Kaplinghat:2013yxa,Bernal:2015ova,Bringmann:2016din,Cirelli:2016rnw,Kahlhoefer:2017umn,Hufnagel:2017dgo,Hufnagel:2018bjp,Bernal:2019uqr}. For \textit{scalar DM} $s$, the relevant interactions in \cref{eq:scalar_mediator_couplings_mu,eq:vector_int} are \begin{equation}
\label{eq:scalar_DM_coplings}
\mathcal{L}_{\rm int} \supset
\begin{cases}
\mu (s^\dagger s)\phi
& \text{(scalar mediator)} \\[6pt]
ig\left(s^\dagger\partial^\mu s
-(\partial^\mu s^\dagger)s\right)\phi_\mu
& \text{(vector mediator)}
\end{cases}
\end{equation}
as illustrated in \cref{fig:scalar_vertices,fig:vertex_scalar_vector}. For \textit{fermionic DM} $\chi$, the corresponding interactions in \cref{eq:yukawa_coupling,eq:vector_exchange} are
\begin{equation} \label{eq:fermionic_DM_coplings}
\mathcal{L}_{\rm int} \supset
\begin{cases}
y\,\bar{\chi}\chi\,\phi,
& \text{(scalar mediator)} \\[6pt]
g\,\bar{\chi}\gamma^\mu\chi\,\phi_\mu
& \text{(vector mediator)}
\end{cases}
\end{equation}
corresponding to the vertices shown in \cref{fig:vertex_yukawa,fig:vertex_vector}. In both cases, the self-scattering process arises through mediator exchange between two DM particles, as shown in \cref{fig:selfint_realizations_all}, and is therefore fundamentally non-local, in contrast to the point-like contact interactions discussed in \cref{sec:contact-like}.

Although we focus here on scalar and fermionic DM, \textit{vector DM} is also possible and will be discussed separately in \cref{sec:Vector_SIDM}. Likewise, while scalar and vector mediators are the most commonly studied cases, light \textit{spin-2 tensor mediator} -- analogous to the hypothetical graviton as a force carrier of gravity -- has also been considered in the SIDM literature~\cite{Kang:2020yul,Kang:2020afi}.

This light-mediator SIDM scenario serves as our baseline model class and was used in Chapter~\ref{sec: sidm_pp_models} to introduce the main particle-physics ingredients relevant for SIDM model building. In the non-relativistic regime, mediator exchange is described by the Yukawa potential in \cref{eq:yukawa_potential_generic}. Vector exchange can generate either attractive or repulsive interactions depending on the dark charges, whereas scalar exchange is attractive. The interaction range is set by $m_{\mathrm{med}}^{-1}$, such that a sufficiently light mediator gives rise to finite-range, velocity-dependent and generally anisotropic self-scattering. The resulting scattering phenomenology is controlled by the mediator mass, DM mass and interaction strength. In the perturbative Born regime, $\alpha_{\mathrm{DM}}m_{\mathrm{DM}}/m_{\mathrm{med}}\ll1$, the differential cross sections admit simple analytic forms, given in \cref{eq:Rutherford_general} to (\ref{eq:interference_coefficient}). For $\alpha_{\mathrm{DM}}m_{\mathrm{DM}}/m_{\mathrm{med}}\gtrsim1$, the Yukawa potential must instead be treated non-perturbatively, with the differential cross section calculated numerically as described in \cref{eq:numerical_differential_cross_section} to (\ref{eq:asymptotic_match}); this regime can also give rise to resonant scattering. The contact-like limit is recovered when the mediator mass is large compared with the characteristic momentum transfer, whereas light mediators lead to increasingly forward-peaked scattering. The corresponding differential and angle-averaged cross sections, including the transfer and viscosity cross sections, are discussed in detail in \cref{sec:cross_sections}.

The particle-physics parameter space of this baseline light-mediator model can therefore be summarized as
\begin{equation}
\label{eq:PP_parameters_light_mediator}
\mathcal{P}_{\rm PP}
=
\left\{
m_{\mathrm{DM}},\,
m_{\mathrm{med}},\,
\alpha_{\mathrm{DM}},\,
\epsilon_{\rm DM}
\right\},
\end{equation}
where $m_{\mathrm{DM}}$, $m_{\mathrm{med}}$ and $\alpha_{\mathrm{DM}}$ determine the self-interaction strength and its velocity and angular dependence, while $\epsilon_{\rm DM}$ parametrizes the portal coupling to the SM (\cref{fig:portal_couplings}) and thereby affects DM production, mediator decay and the associated experimental constraints.

\medskip
\noindent
\textbf{Representative benchmarks:} A minimal Higgs-portal SIDM model with a scalar mediator provides a well-studied benchmark scenario:
\begin{equation}
m_{\rm med}\sim 1~{\rm MeV}, \, \, \, \, \, m_{\rm DM}\sim 0.5~{\rm GeV}, \, \, \, \, \, \alpha_\phi \sim 5 \times 10^{-5},
\end{equation} which produces the velocity-dependent self-interaction
cross section shown as the orange curve in
\cref{fig:cross_sections}. This benchmark is also
compatible with the observed relic abundance through
standard thermal freeze-out via $p$-wave annihilation
$\chi\bar{\chi}\to\phi\phi$, while Higgs mixing
connects the mediator to the SM
~\cite{Kaplinghat:2013yxa}.

A vector-mediator benchmark with comparable astrophysical signatures requires alternative DM-production mechanisms:
\begin{equation}
m_{\rm med} \sim 10~{\rm MeV}, \quad m_{\rm DM} \sim 1~{\rm GeV}, \quad \alpha_{\rm DM} \sim 10^{-3},
\end{equation}
where modified thermal histories or secluded dark sectors may help reconcile the model with stringent cosmological constraints on annihilation. For detailed discussion of the cosmological, astrophysical, and experimental viability of light-mediator SIDM models and the parameter space surrounding these benchmarks, see App.~\ref{app:constraints_light-mediator}.

\subsubsection{Resonant SIDM}
\label{sec:Resonant_SIDM}

Resonant SIDM constitutes a broad class of velocity-dependent self-interaction models in which the scattering cross section is enhanced by a near-threshold state in the two-body system. Unlike minimal light-mediator SIDM, this enhancement is not tied to a light mediator or to a finite-range Yukawa interaction. It can also arise from non-perturbative dynamics, short-range interactions or composite dark-sector physics. The resulting self-scattering is elastic in the simplest realization considered here (while it may also become inelastic, as discussed in \cref{sec: inelastic_sidm_models}) and exhibits a pronounced and sometimes sharply localized dependence on both the relative velocity and scattering angle.

\medskip
\noindent
\textbf{Microscopic description:} At the microscopic level, resonant SIDM arises when two DM particles can temporarily form a correlated intermediate state during a collision. If this state lies close in energy to the incoming two-particle state, the scattering probability can be strongly enhanced. This resonant enhancement is a generic feature of non-relativistic quantum scattering,\footnote{The relevant velocities are non-relativistic, $v \lesssim 10^{-2}c$, as typically encountered in DM haloes, justifying the use of non-relativistic quantum mechanics.} and can arise from a near-threshold bound state, a virtual level or an unstable resonance. In each case, the particles remain correlated for an unusually long time, increasing the probability of scattering.

In particle physics, unstable intermediate states are often described using the Breit--Wigner formalism~\cite{Breit:1936zzb}. This corresponds to an intermediate state that behaves like a particle with a well-defined mass, i.e. an \emph{on-shell} state, which can be formed during the collision and subsequently decay. However, resonant enhancements can also occur without producing a true on-shell particle. Near-threshold bound states and virtual levels describe systems that are close to forming a bound configuration; the particles therefore spend more time interacting, enhancing the scattering probability even without a real intermediate particle.

In the SIDM context, resonant behavior has been studied in many concrete models~\cite{Tulin:2013teo,Braaten:2013tza,Hochberg:2014dra,Duch:2017nbe,Tsai:2020vpi,Cline:2022leq,Csaki:2022xmu,Tang:2025orb,Lee:2025lko,Garcia-Cely:2025flv}, as well as in model-independent treatments based on the effective-range formalism~\cite{Chu:2018qrm,Chu:2018fzy,Chu:2019awd}. A key point is that strong velocity dependence can arise even without a light mediator~\cite{Chu:2018qrm,Chu:2018fzy}; depending on the underlying dynamics, the scattering may also acquire non-trivial angular structure.

Resonant SIDM can nevertheless occur within light-mediator models. The Yukawa interactions discussed in \cref{sec:light-mediator} enter a non-perturbative regime when $\alpha_{\rm DM} m_{\rm DM}/m_{\rm med} \gtrsim 1$. An attractive potential can then support near-threshold states, producing resonances and antiresonances. In this regime, the Born-level expressions in \cref{eq:Rutherford_general,eq:Moeller_general} are no longer valid, and the cross section must be computed using a partial-wave treatment, as discussed in \cref{sec:cross_sections}.

More generally, resonant behavior can arise when the interaction has a finite microscopic range but appears effectively point-like at halo energies. The low-energy dynamics can then be described by local effective operators, as discussed in \cref{sec: pp_ints_to_SM_self_interactions}. This perspective is closely related to the effective-range approach, originally developed for low-energy neutron--proton scattering~\cite{Bethe:1949yr,Blatt:1949zz}. For scalar DM $s$, the leading interactions may be written as
\begin{equation}
\mathcal{L}_{\rm eff} \supset 
C_4 (s^\dagger s)^2
+ \frac{C_6}{\Lambda^2} 
(\partial_\mu s^\dagger \partial^\mu s)(s^\dagger s),
\end{equation}
while for fermionic DM $\chi$,
\begin{equation}
\mathcal{L}_{\rm eff} \supset 
\frac{C_6}{\Lambda^2} (\bar{\chi}\chi)(\bar{\chi}\chi)
+ \frac{C_6'}{\Lambda^2} (\bar{\chi}\gamma^\mu\chi)(\bar{\chi}\gamma_\mu\chi),
\end{equation}
where $\Lambda$ denotes the microscopic scale above which the effective description breaks down. These are the leading terms in the expansion of \cref{eq:effective_Lagrangian_expansion}. Although point-like at low energies, their coefficients can encode strong dynamics, compositeness or near-threshold states.

A particularly important realization arises in composite dark sectors, discussed in \cref{sec:composite_dark_sectors_non_pert_ints}, with SIDM implementations reviewed in \cref{sec:composite_SIDM}. In these scenarios, the relevant low-energy degrees of freedom are bound states rather than elementary particles, and the scattering resembles that of nuclear systems. As illustrated in \cref{fig:selfint_composite_realizations}, dark mesons can scatter through short-range interactions set by the confinement scale $\Lambda_D$, while dark baryons can interact through residual forces mediated by lighter composite states, analogous to pion exchange in QCD. Such dynamics naturally generate large scattering lengths and near-threshold states. In the direct resonant case shown in \cref{fig:selfint_composite_resonance}, scattering proceeds through a short-lived intermediate state $R_D$, leading to a Breit--Wigner-type enhancement of the cross section.

\begin{figure*}[t]
\centering
\includegraphics[width=0.82\columnwidth]{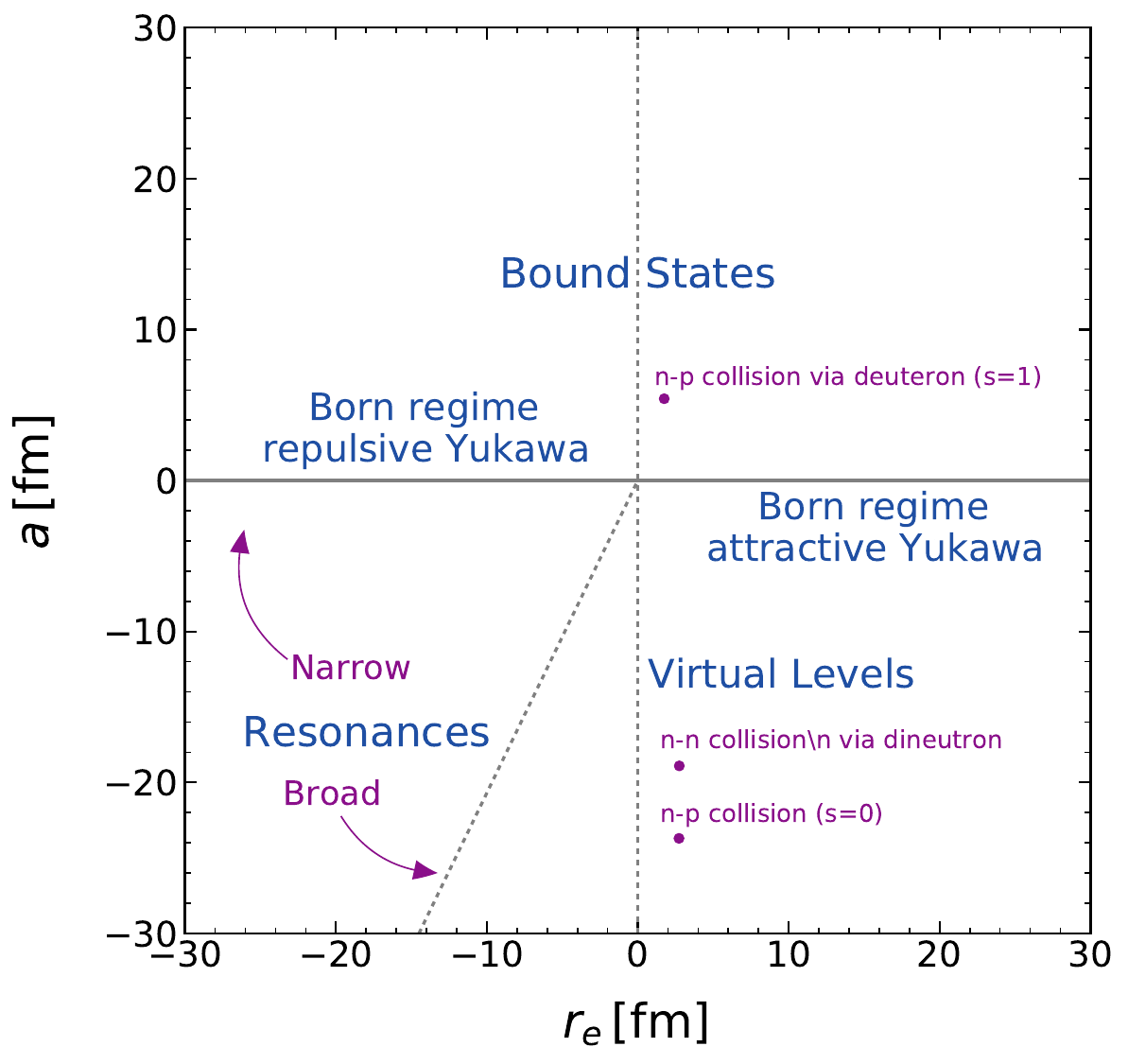}
\includegraphics[width=1.08\columnwidth]{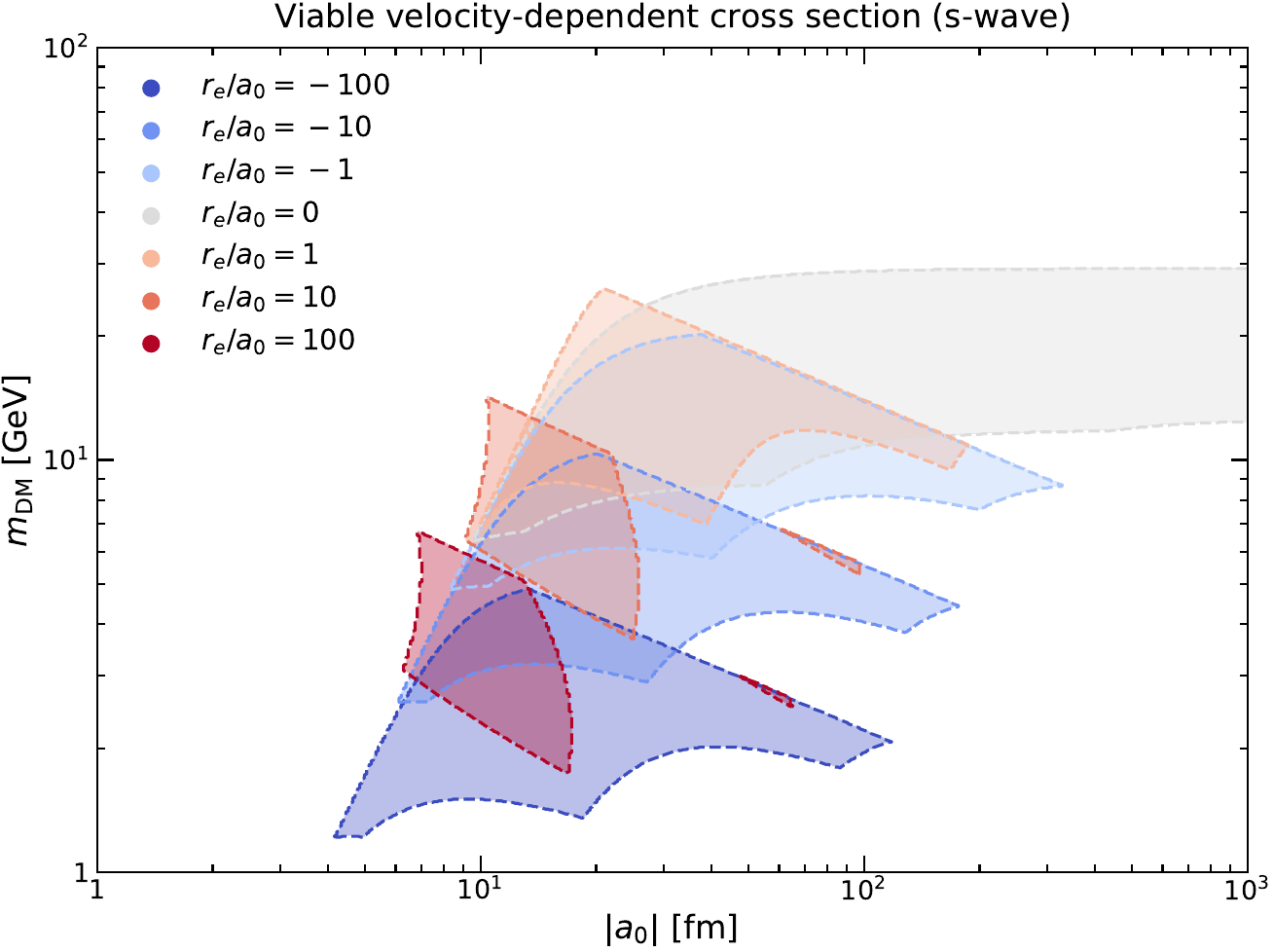}
\caption{
\emph{Left panel:} Effective-range parameter space in the $(r_e,a)$ plane, showing how bound states, virtual levels, Breit--Wigner resonances, and the Born regimes of attractive and repulsive Yukawa interactions map onto Eqs.~(\ref{eq:eff_range_formalism_1})--(\ref{eq:eff_range_formalism_2}). Purple points show nuclear-physics examples: proton--neutron scattering through the deuteron, neutron--neutron scattering through the dineutron, and spin-singlet proton--neutron scattering, with values from~\cite{PhysRevC.48.792}. Reproduced from~\cite{Chu:2019awd}.
\emph{Right panel:}
Allowed regions in the $(m_{\mathrm{DM}}, |a_0|)$ plane yielding velocity-dependent $s$-wave self-interaction cross sections consistent with the inferred astrophysical constraints in \cref{fig:cross_sections}. Each colored region corresponds to a fixed $r_{e,0}/a_0$, illustrating how the effective range changes the velocity dependence and shifts the viable parameter space. Positive and negative $r_{e,0}/a_0$ values lead to qualitatively different behaviors and occupy distinct regions.
}
\label{fig:effective_range_approx_parameters}
\end{figure*}

At low velocities, this behavior is described by the effective-range formalism. For distinguishable particles, the differential cross section is
\begin{equation}
\label{eq:eff_range_formalism_1}
\frac{d\sigma}{d\Omega} = |f(v_{\rm rel},\theta)|^2,
\end{equation}
where the scattering amplitude is given by the partial-wave expansion in \cref{eq:pw_effective_range}. For finite-range interactions, the phase shifts follow the effective-range expansion in \cref{eq:ere_general}, characterized by the generalized scattering lengths $a_\ell$ and effective ranges $r_{e,\ell}$. In the $s$-wave-dominated limit, \cref{eq:ere_swave} provides a compact, largely model-independent description of low-energy self-scattering.

The same framework also captures Breit--Wigner
resonances. Close to a resonance energy $E_R$,
the partial-wave cross section takes the form
\begin{equation}
\sigma_\ell(E)
=
\frac{4\pi(2\ell+1)}{m_{\rm DM}E}
\frac{\Gamma(E)^2/4}
{(E-E_R)^2+\Gamma(E)^2/4},
\end{equation}
where $E=k^2/m_{\rm DM}$ is the center-of-mass
kinetic energy for equal-mass DM particles,
and $\Gamma(E)$ is the energy-dependent width
of the intermediate state. Near threshold,
$\Gamma(E)=\Gamma_R(E/E_R)^{\ell+1/2}$,
with $\Gamma_R\equiv\Gamma(E_R)$
~\cite{Chu:2018fzy}. Matching the corresponding
Breit--Wigner phase shift to the effective-range
expansion in \cref{eq:ere_general} gives
\begin{equation}
\label{eq:eff_range_formalism_2}
\begin{aligned}
a_\ell &=
-\frac{1}{\sqrt{m_{\rm DM}E_R}}
\left(\frac{\Gamma_R}{2E_R}\right)^{
\frac{1}{2\ell+1}},\\[6pt]
r_{e,\ell} &=
-\frac{4(m_{\rm DM}E_R)^{\ell+1/2}}
{m_{\rm DM}\Gamma_R}.
\end{aligned}
\end{equation}
These relations apply to an isolated elastic
resonance when non-resonant background
contributions can be neglected. They have
the required dimensions
$[a_\ell]=L$ and
$[r_{e,\ell}]=L^{1-2\ell}$. It is important to distinguish such particle resonances from enhancements due to near-threshold bound states or virtual levels, where no on-shell particle is formed. For sufficiently short-range interactions, both cases are captured by the effective-range approximation. For long-range forces, where many partial waves contribute, this approximation breaks down and the cross section must be computed numerically, as discussed for light-mediator interactions in \cref{sec:light-mediator}. RSIDM can therefore be viewed as the general low-energy limit of non-perturbative scattering, encompassing both Yukawa resonances and more general short-range dynamics.

As shown schematically in the left panel of \cref{fig:effective_range_approx_parameters}, these resonant phenomena can be organized in the parameter space spanned by the scattering length $a_0$ and effective range $r_{e,0}$. Bound states, virtual levels, Breit--Wigner resonances and the Born regimes of Yukawa interactions occupy distinct regions of this plane, showing how different microscopic models can lead to similar low-energy scattering behavior. The figure also indicates well-known nuclear-physics examples relevant for composite dark-sector scenarios. Proton--neutron scattering in the spin-triplet channel, for kinetic energies up to a few MeV, is accurately described by the effective-range approximation with $a = 5.42\,\mathrm{fm}$ and $r_e = 1.75\,\mathrm{fm}$~\cite{Noyes_1972,PhysRevC.48.792}, reflecting the bound deuteron state. By contrast, neutron--neutron scattering has $a = -18.9\,\mathrm{fm}$ and $r_e = 2.75\,\mathrm{fm}$, corresponding to a virtual level rather than a bound state; a similar situation occurs for spin-singlet proton--neutron scattering. These examples illustrate how bound states and virtual levels, despite their different interpretations, both enhance low-energy scattering and are unified by the effective-range description.

Thus, resonant SIDM can be parametrized either by effective low-energy scattering parameters,
\begin{equation}\label{eq:PP_parameters_resonant_eff}
\mathcal{P}_{\rm PP}^{\rm eff}=\{m_{\rm DM}, \Lambda, a_\ell, r_{e,\ell}\},
\end{equation}
or by microscopic particle-physics parameters,
\begin{equation} \label{eq:PP_parameters_resonant}
\mathcal{P}_{\rm PP}=\{m_{\rm DM}, m_{\rm med}, \alpha_{\rm DM}\}.
\end{equation}

\medskip
\noindent
\textbf{Representative benchmarks:}
Resonant SIDM can generate large self-interaction cross sections at dwarf-galaxy velocities while remaining suppressed at the higher velocities characteristic of galaxy clusters. As a representative $s$-wave benchmark in the effective-range description, we consider
\begin{equation}
m_{\rm DM}\simeq10~{\rm GeV}, \, \, \, \, \,
a_0\simeq-500~\text{fm}, \, \, \, \, \,
r_{e,0}\simeq100~\text{fm},
\end{equation}
corresponding to the resonant parameter region shown in the right panel of \cref{fig:effective_range_approx_parameters}. Such effective parameters can arise naturally from the strong dynamics of composite dark sectors.

As a representative resonant light-mediator model, we adopt
\begin{equation}
m_{\rm DM}\simeq37.84~{\rm GeV}, \, 
m_{\rm med}\simeq4.39~{\rm MeV}, \, 
\alpha_{\rm DM}=10^{-2}.
\end{equation}
This benchmark produces the velocity-dependent resonant cross section shown by the cyan curve in \cref{fig:cross_sections}, consistent with the inferred astrophysical constraints. The broader cosmological, astrophysical and experimental viability of resonant SIDM and the surrounding parameter space is discussed in App.~\ref{app:constraints_resonant}.

\subsubsection{Vector SIDM}
\label{sec:Vector_SIDM}

Vector SIDM is another well-motivated class of elastic single-state SIDM models. As in the scalar and fermionic cases discussed above, the self-interactions may be approximately contact-like or mediated by a lighter force carrier, leading to velocity-dependent and generally anisotropic scattering. The distinguishing feature is the microscopic origin of the DM particle: in vector SIDM, DM is a massive gauge boson of a spontaneously broken dark gauge symmetry, resulting in a different interaction structure from scalar or fermionic models.

\medskip
\noindent
\textbf{Microscopic description:} In vector SIDM models, the DM particle is a massive spin-1 field, denoted by $\chi_\mu$. Such particles naturally arise as gauge bosons of a dark gauge symmetry that is spontaneously broken, analogous to how the SM $W^\pm$ and $Z$ bosons acquire mass through electroweak symmetry breaking.

At the Lagrangian level, these scenarios are generically described by a dark gauge theory of the form~\cite{Bohm:2001yx}
\begin{equation}
\mathcal{L}_{\rm VDM}
\supset
-\frac{1}{4} F^a_{\mu\nu} F^{a\,\mu\nu}
+ (D_\mu \Phi)^\dagger (D^\mu \Phi)
- V(\Phi,H),
\end{equation}
where $\Phi$ is the dark Higgs field, $H$ is the SM Higgs, and $V(\Phi,H)$ generates symmetry breaking and possible Higgs-portal interactions. The covariant derivative is
\begin{equation}
D_\mu \Phi
=
\left(
\partial_\mu - i g\, T^a \chi^a_\mu
\right)\Phi,
\end{equation}
where $g$ is the dark gauge coupling and $T^a$ are the generators of the dark gauge group. After symmetry breaking, this term generates the vector-boson masses and their couplings to the scalar sector. The field-strength tensor is
\begin{equation}
F^a_{\mu\nu}
=
\partial_\mu \chi^a_\nu
-
\partial_\nu \chi^a_\mu
+
g\, f^{abc} \chi^b_\mu \chi^c_\nu ,
\end{equation}
where $f^{abc}$ are the structure constants. For an $SU(2)$ dark gauge symmetry, $a=1,2,3$, and one may identify
\begin{equation}
\chi_\mu^\pm = \frac{\chi_\mu^1 \mp i \chi_\mu^2}{\sqrt{2}},
\qquad
\phi_\mu \equiv \chi_\mu^3 .
\end{equation}
In this basis, the charged massive gauge bosons $\chi_\mu^\pm$ can play the role of DM, while the neutral component $\phi_\mu$ can act as a light vector mediator.

The structure of the interactions depends crucially on whether the dark gauge symmetry is Abelian or non-Abelian. In Abelian theories, such as a $U(1)$ dark photon model, $f^{abc}=0$, and the field-strength tensor contains only derivative terms; Abelian vector bosons therefore do not self-interact at the renormalizable level. In non-Abelian theories, such as $SU(2)$ or $SU(3)$, the last term in $F^a_{\mu\nu}$ generates cubic and quartic vector interactions. After symmetry breaking, these induce couplings such as $\chi_\mu^+\chi_\mu^-\phi_\mu$ and $\chi_\mu^+\chi_\mu^-\phi_\mu\phi_\mu$, which contribute to annihilation and self-scattering. A concrete realization was presented in~\cite{Chaffey:2019fec}, where a Higgsed $SU(2)$ dark sector contains massive charged gauge boson DM stabilized by a residual global symmetry, with a lighter neutral gauge boson acting as a dark-photon-like mediator.

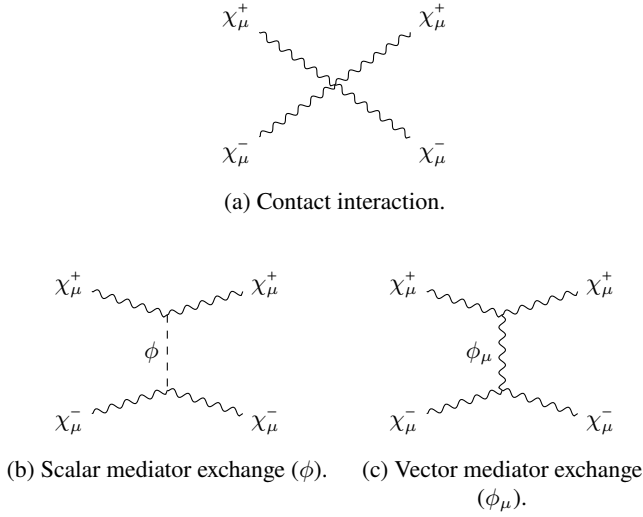
\begin{figure}[t]
\centering

\begin{subfigure}[t]{0.49\columnwidth}
\centering
\begin{tikzpicture}
\begin{feynman}
  \vertex (v) at (0,0);

  \vertex (i1) at (-1.3, 0.9) {$\chi_\mu^{+}$};
  \vertex (i2) at (-1.3,-0.9) {$\chi_\mu^{-}$};
  \vertex (f1) at ( 1.3, 0.9) {$\chi_\mu^{+}$};
  \vertex (f2) at ( 1.3,-0.9) {$\chi_\mu^{-}$};

  \diagram*{
    (i1) -- [photon] (v) -- [photon] (f1),
    (i2) -- [photon] (v) -- [photon] (f2),
  };
\end{feynman}
\end{tikzpicture}
\caption{Contact interaction.}
\label{fig:VDM_contact}
\end{subfigure}

\vspace{5mm}

\begin{subfigure}[t]{0.49\columnwidth}
\centering
\begin{tikzpicture}
\begin{feynman}
  \vertex (a) at (0,0.5);
  \vertex (b) at (0,-0.5);

  \vertex (i1) at (-1.3, 0.9) {$\chi_\mu^{+}$};
  \vertex (i2) at (-1.3,-0.9) {$\chi_\mu^{-}$};
  \vertex (f1) at ( 1.3, 0.9) {$\chi_\mu^{+}$};
  \vertex (f2) at ( 1.3,-0.9) {$\chi_\mu^{-}$};

  \diagram*{
    (i1) -- [photon] (a) -- [photon] (f1),
    (i2) -- [photon] (b) -- [photon] (f2),
    (a)  -- [scalar, edge label'=$\phi$] (b),
  };
\end{feynman}
\end{tikzpicture}
\caption{Scalar mediator exchange ($\phi$).}
\label{fig:VDM_scalar_med}
\end{subfigure}
\hfill
\begin{subfigure}[t]{0.49\columnwidth}
\centering
\begin{tikzpicture}
\begin{feynman}
  \vertex (a) at (0,0.5);
  \vertex (b) at (0,-0.5);

  \vertex (i1) at (-1.3, 0.9) {$\chi_\mu^{+}$};
  \vertex (i2) at (-1.3,-0.9) {$\chi_\mu^{-}$};
  \vertex (f1) at ( 1.3, 0.9) {$\chi_\mu^{+}$};
  \vertex (f2) at ( 1.3,-0.9) {$\chi_\mu^{-}$};

  \diagram*{
    (i1) -- [photon] (a) -- [photon] (f1),
    (i2) -- [photon] (b) -- [photon] (f2),
    (a)  -- [photon, edge label'=$\phi_\mu$] (b),
  };
\end{feynman}
\end{tikzpicture}
\caption{Vector mediator exchange ($\phi_\mu$).}
\label{fig:VDM_vector_med}
\end{subfigure}

\caption{Examples of vector SIDM self-interactions. The vector DM fields $\chi_\mu^\pm$ can scatter through a contact interaction or via the exchange of scalar or vector mediators.}
\label{fig:vector_sidm_selfint}
\end{figure}

Examples of vector-DM self-scattering are shown in \cref{fig:vector_sidm_selfint}. The fields $\chi_\mu^\pm$ may scatter through intrinsic non-Abelian gauge self-interactions, which effectively behave as contact interactions at low energies, or through scalar or vector mediator exchange. Mediator exchange typically produces a finite-range Yukawa potential, as in the light-mediator models of \cref{sec:light-mediator}. In explicit realizations such as~\cite{Chaffey:2019fec}, self-interactions are often dominated by vector-mediated processes as shown in \cref{fig:VDM_vector_med}, leading to velocity-dependent and anisotropic scattering. Vector exchange can be repulsive or attractive depending on the charges of the scattering particles: like charges repel, while particle--antiparticle pairs attract. Scalar exchange (\cref{fig:VDM_scalar_med}) is purely attractive. Consequently, vector SIDM can exhibit the same non-perturbative phenomena discussed in \cref{sec:Resonant_SIDM}, including near-threshold states and resonant enhancement of the cross section.

A wide range of vector-DM realizations has been proposed, including the Kaluza--Klein photon~\cite{Servant:2002aq}, hidden vector DM~\cite{Hambye:2008bq}, Abelian and non-Abelian vector DM models~\cite{Farzan:2012hh,Gross:2015cwa}, hidden-force scenarios~\cite{Davoudiasl:2013jma}, residual non-Abelian vector DM~\cite{Ko:2016fcd}, hidden-sector monopole and vector DM models~\cite{Baek:2013dwa,Khoze:2014woa}, non-Abelian hidden-sector SIDM~\cite{Boddy:2014yra,2014PhRvD}, multi-component vector DM~\cite{Elahi:2019jeo}, vector SIMP~\cite{Choi:2019zeb}, and more recent constructions~\cite{Frandsen:2022klh,Yuan:2024tfs,Bouzeraib:2026dbx,Shaikh:2026ixs}.

Thus, the particle-physics parameter space of vector SIDM can be summarized as
\begin{equation} \label{eq:PP_parameters_vector}
\mathcal{P}_{\rm PP}=\{m_{\rm DM}, m_{\rm med}, \alpha_{\rm DM},\epsilon_{\rm DM}\},
\end{equation}
where $\epsilon_{\rm DM}$ parametrizes the portal coupling to the SM (\cref{fig:portal_couplings}).

\medskip
\noindent
\textbf{Representative benchmarks:} Two representative vector SIDM benchmarks were
studied in~\cite{Chaffey:2019fec}. The light
benchmark has:
\begin{align}
\begin{aligned}
m_{\rm DM} \simeq 0.06~{\rm GeV},\, m_{\rm med} \simeq 0.095~{\rm MeV}, \,\alpha_{\rm DM} \simeq 1.35\times 10^{-6},
\end{aligned}
\end{align}
which approximately reproduces the inferred
astrophysical self-interaction cross sections
while remaining compatible with thermal
freeze-out in the model considered.
A heavier benchmark has
\begin{equation}
m_{\rm DM} \simeq 60~{\rm GeV}, \, \, \, \, \, m_{\rm med} \simeq 6~{\rm MeV}, \, \, \, \, \, \alpha_{\rm DM} \simeq 10^{-3},
\end{equation}
which reproduces the thermal relic abundance
but underpredicts the inferred cluster-scale
self-interaction cross sections. For detailed discussion of the cosmological, astrophysical, and experimental viability of vector SIDM models and the parameter space surrounding these benchmarks, see App.~\ref{app:constraints_vector}.

\subsubsection{Asymmetric SIDM}
\label{sec:Asymmetric_SIDM}

Asymmetric SIDM (ASIDM) denotes SIDM scenarios in which the DM relic abundance is primarily set by a particle--antiparticle asymmetry in the dark sector, while halo-scale self-interactions arise from the same mechanisms as in ordinary SIDM, e.g.\ light mediators or composite dark-sector dynamics. ASIDM is therefore not defined by a new scattering law, but by a different cosmological origin of the relic abundance, as discussed in \cref{sec:DM_production_mechanisms}. This setup naturally reconciles sizeable self-interactions with the observed DM density, suppresses late-time annihilation signals and can reopen parameter space that is strongly constrained in minimal symmetric light-mediator SIDM models.

\medskip
\noindent
\textbf{Microscopic description:} ASIDM combines two ingredients. First, as in ordinary light-mediator SIDM discussed in \cref{sec:light-mediator}, or in composite scenarios such as dark baryons discussed in \cref{sec:dark_baryons}, DM can have sizeable self-interactions mediated by light dark-sector states or by composite dynamics. This can produce velocity-dependent scattering, efficient in low-velocity systems such as dwarf haloes and weaker in galaxy clusters~\cite{Buckley:2009in,Feng:2009hw,Feng:2009mn,Loeb:2010gj,Bringmann:2016din,Kaplinghat:2015aga,vandenAarssen:2012vpm,Tulin:2013teo}. Second, as in asymmetric DM (ADM), the present-day DM abundance is set mainly by a primordial dark particle--antiparticle asymmetry, analogous to the baryon asymmetry of the visible Universe~\cite{Nussinov:1985xr,Kaplan:2009ag,Davoudiasl:2012uw,DuttaBanik:2020vfr,Barman:2021ost,Cui:2020dly,Falkowski:2011xh,Patel:2022xyv,Biswas:2018sib,Narendra:2018vfw,Nagata:2016knk,Arina:2011cu,Arina:2012fb,Narendra:2017uxl,Narendra:2019cyt,Mahapatra:2023dbr,Borah:2022ask}; see also our brief discussion in \cref{sec:DM_production_mechanisms} or the reviews~\cite{Petraki:2013wwa,Zurek:2013wia}.

A common motivation is cogenesis, where the dark and baryon asymmetries are generated by related early-Universe dynamics. This is motivated by the observed coincidence
\begin{equation}
\Omega_{\rm DM}\simeq 5\,\Omega_B ,
\end{equation}
as inferred from cosmological observations~\cite{Planck_2020b}. If the two asymmetries have a common origin, one naturally expects comparable number densities, $n_{\rm DM}\sim n_B$. Since $\Omega_X=n_Xm_X/\rho_c$, the similarity $\Omega_{\rm DM}\sim\Omega_B$ then points to DM masses near the baryonic scale, $m_{\rm DM}\sim m_B$, naturally motivating GeV-scale DM~\cite{Nussinov:1985xr,Kaplan:2009ag,Davoudiasl:2012uw,DuttaBanik:2020vfr,Barman:2021ost,Cui:2020dly,Falkowski:2011xh,Patel:2022xyv,Biswas:2018sib,Narendra:2018vfw,Nagata:2016knk,Arina:2011cu,Arina:2012fb,Narendra:2017uxl,Narendra:2019cyt,Mahapatra:2023dbr}. Many ADM frameworks are closely connected to baryogenesis~\cite{1979PhRvL42850W,1982NuPhB195542} or leptogenesis~\cite{Fukugita:1986hr}.\footnote{Baryogenesis refers to mechanisms in the early Universe that generate the observed excess of baryons over antibaryons, where baryons are particles such as protons and neutrons. Leptogenesis is a related mechanism in which an asymmetry is first generated in the lepton sector, i.e.\ among particles such as electrons and neutrinos, and then converted into a baryon asymmetry -- and in models with a coupled dark sector potentially also a dark asymmetry -- through electroweak sphaleron processes~\cite{Rubakov:1996vz}.}

A central feature of ASIDM is that the symmetric component can annihilate efficiently, leaving the asymmetric component as the observed relic density~\cite{Nussinov:1985xr,Kaplan:2009ag,Davoudiasl:2012uw,DuttaBanik:2020vfr,Barman:2021ost,Cui:2020dly,Falkowski:2011xh,Patel:2022xyv,Biswas:2018sib,Narendra:2018vfw,Nagata:2016knk,Arina:2011cu,Arina:2012fb,Narendra:2017uxl,Narendra:2019cyt,Mahapatra:2023dbr,Borah:2022ask}. This is particularly attractive for SIDM, because interactions that generate large self-scattering often also imply efficient DM--anti-DM annihilation in the early Universe. In symmetric light-mediator SIDM, this can make the relic abundance too small, especially for GeV-scale and heavier DM~\cite{Borah:2022ask}. The asymmetric setup may avoid this tension.

Microscopically, ASIDM can be realized with elementary DM and a light scalar or vector mediator, as in \cref{sec:light-mediator}, supplemented by an initial dark asymmetry. Another appealing possibility is composite DM, such as dark baryons in \cref{sec:dark_baryons}, whose masses arise from a confining gauge interaction analogous to QCD. If the visible and dark confinement scales are related, the similarity between visible and dark matter densities can emerge naturally. Such constructions can arise in mirror-symmetric models, either preserved~\cite{Hodges:1993yb,Foot:2003jt,Foot:2004pq,Foot:2004pa,Ibe:2018juk} or deliberately broken~\cite{An:2009vq,Cui:2011wk,Farina:2016ndq,Lonsdale:2017mzg,Lonsdale:2018xwd,Ibe:2019ena,Beauchesne:2020mih,Feng:2020urb,Ritter:2021hgu,Ibe:2021gil}. Twin Higgs models provide a notable example~\cite{Farina:2015uea,Craig:2015xla,GarciaGarcia:2015pnn,Barbieri:2016zxn}. Related structures can also arise from infrared fixed-point dynamics~\cite{Bai:2013xga,Newstead:2014jva,Ritter:2022opo} or unified visible--dark gauge sectors~\cite{Murgui:2021eqf,Foster:2022ajl}. These scenarios often predict strong self-interactions from the underlying dark dynamics, as discussed in \cref{sec:composite_SIDM}.

Composite ASIDM can also contain multiple nearly degenerate states~\cite{Rosenlyst:2023tyj}. The visible proton--neutron system provides a useful analogy: despite the small splitting $(m_n-m_p)/m_p \sim 10^{-3}$, the proton is cosmologically stable, while the neutron decays through weak interactions. Similar dark-sector multiplets with splittings $\delta m$ motivate multi-state DM models, where collisions can induce endothermic up-scattering or exothermic de-excitation, as discussed in \cref{sec: inelastic_sidm_models}. Viable scenarios require $\delta m$ to be small compared to typical halo kinetic energies to avoid kinematic suppression or excessive heating.

The asymmetry can also be generated in seesaw-based frameworks that address neutrino masses, including type-I~\cite{1977PhLB67421M,Gell-Mann:1979ijt,Gell-Mann:1979vob,1980PhRvL44912M,Schechter:1981bd,Schechter:1981cv}, type-II~\cite{Schechter:1981cv,1981PhRvD23.165M,Wetterich:1981bx,Lazarides:1980nt,Brahmachari:1997cq} and type-III seesaw scenarios~\cite{Foot:1988aq}. Recent ASIDM constructions based on these mechanisms have been explored in~\cite{Dutta:2022knf,Borah:2024wos,Dutta:2024bnt,Dutta:2025xlv,Ishida:2025xvr}. In such models, the visible-sector asymmetry can be transferred dynamically to the dark sector, for example through electroweak sphaleron processes if the dark sector couples to particles carrying electroweak charges~\cite{Rubakov:1996vz}. Such setups may simultaneously connect neutrino masses, baryogenesis and the DM relic abundance while retaining the desired SIDM phenomenology.

Taken together, the self-scattering sector of ASIDM is characterized by the same particle-physics parameters as in the corresponding symmetric SIDM models,
\begin{equation}
\mathcal{P}_{\rm PP}
=
\{m_{\rm DM},\, m_{\rm med},\, \alpha_{\rm DM}\},
\end{equation}
or, equivalently, by effective low-energy scattering parameters,
\begin{equation}
\mathcal{P}_{\rm PP}^{\rm eff}
=
\{m_{\rm DM},\, \Lambda,\, a_\ell,\, r_{e,\ell}\}.
\end{equation}
However, the relic-abundance constraint is different from that in symmetric SIDM, because the present-day DM density is set by a primordial particle--antiparticle asymmetry rather than by standard symmetric freeze-out. This changes which regions of the particle-physics parameter space are viable.

\medskip
\noindent
\textbf{Representative benchmark:}
A representative asymmetric SIDM model with repulsive vector-mediated self-interactions is defined by
\begin{equation}
m_{\rm DM}=5~\mathrm{GeV}, \, \, \, \, \,
m_{\rm med}=6~\mathrm{MeV}, \, \, \, \, \,
\alpha_{\rm DM}\simeq8\times10^{-4}.
\end{equation}
This benchmark yields a viable velocity-dependent self-interaction cross section. This benchmark demonstrates how asymmetric production can enlarge viable parameter space by decoupling the relic abundance from the self-interaction strength. Additional ASIDM benchmarks with different mediator types and mass hierarchies expand the range of viable models. For detailed discussion of the cosmological, astrophysical, and experimental viability of ASIDM models and the parameter space surrounding these benchmarks, see App.~\ref{app:constraints_asymmetric}. 

\subsubsection{Theory-to-simulation mapping: Elastic single-state SIDM}
\label{sec:mapping_elastic_SIDM}

Although the elastic single-state model classes above differ in their microscopic origin, they reduce to the same structure-formation inputs once the corresponding self-interaction cross section has been determined. The relevant distinction is whether the scattering is isotropic and velocity independent or instead velocity and/or angle dependent.

For contact-like SIDM, the scattering is elastic, isotropic and velocity independent, so the model maps onto a single constant cross section per unit mass,
\begin{equation}
\mathcal{P}_{\rm PP} \rightarrow \mathcal{P}_{\rm SIM}
=
\left\{
\frac{\sigma_{\rm tot}}{m_{\rm DM}}
\right\}.
\end{equation}
This applies to the contact-like scalar and fermionic realizations discussed in \cref{sec:contact-like}, as well as to vector SIDM when its self-interactions are dominated by contact interactions. These models are simulated with the isotropic Monte Carlo (MC) scattering scheme described in \cref{sec:mc_velocity_dep}, using $\sigma_{\rm tot}/m_{\rm DM}$ as the simulation input.

Minimal light-mediator SIDM, resonant SIDM, mediator-driven vector SIDM and elastic asymmetric SIDM instead generally produce velocity- and angle-dependent scattering. Their microscopic parameters therefore map onto the differential self-interaction cross section,
\begin{equation}
\mathcal{P}_{\rm PP}
\rightarrow
\mathcal{P}_{\rm SIM}
=
\left\{
\frac{1}{m_{\rm DM}}
\frac{{\rm d}\sigma}{{\rm d}\Omega}
(\theta,v_{\rm rel})
\right\},
\end{equation}
or, when the angular dependence is not resolved explicitly, onto an appropriate angle-weighted cross section $\sigma_X(v_{\rm rel})/m_{\rm DM}$ defined in \cref{eq:weighted_cross_sections}. The microscopic origin determines how this cross section is obtained: from mediator exchange for minimal light-mediator and mediator-driven vector SIDM, from numerical partial-wave calculations or the effective-range parametrization for resonant SIDM, and from the scattering properties of the surviving late-time species for asymmetric SIDM.

These velocity-dependent models can be simulated with the velocity-dependent Monte Carlo scheme (\hyperref[sec:mc_velocity_dep]{MC}) when the scattering is treated isotropically, or with the anisotropic Monte Carlo, frequent small-angle scattering and angular-hybrid schemes (\hyperref[sec:anisotropic_MC]{aMC}, \hyperref[sec: fsidm]{fSIDM} and \hyperref[sec:asidm]{aSIDM}) when the angular dependence is retained. Resonant cross sections can contain narrow velocity-dependent features and strong angular structure and may therefore require sufficiently fine velocity and angular tabulation; the interpolation scheme described in \cref{sec:imp_velocity_dep_SIDM} is used to resolve these features between tabulated points.

In all of these minimal elastic single-state realizations, the reaction network contains only elastic scattering, with no dissipative or decay channels. Standard CDM initial conditions apply unless additional early-time dark-sector interactions modify the linear matter power spectrum, in which case the corresponding non-CDM initial conditions must be used, as summarized in \cref{tab:pk_models}. Extensions involving dissipative processes, multiple internal states, inelastic transitions or decays require additional simulation inputs and are treated separately below.

\subsection{Multi-state SIDM (elastic and inelastic)}
\label{sec: inelastic_sidm_models}

We now consider SIDM scenarios containing multiple DM species or internal states. Their interactions can be purely elastic, allowing momentum and energy exchange between different DM species, or inelastic, involving transitions between internal states and the conversion of internal energy into kinetic energy or vice versa. Both cases extend the elastic, single-state framework introduced in \cref{sec: elastic_single_state} and require additional scattering channels, as discussed in \cref{sec:reaction_network}.

We first focus on \emph{two-state inelastic SIDM}. In this scenario, the dark sector contains two nearly degenerate states, conventionally denoted $\chi_1$ and $\chi_2$, whose self-interactions are mediated by a dark-sector force carrier, such as a light scalar mediator $\phi$ or vector mediator $\phi_\mu$. Because the mediator couples the different internal states, the resulting interaction structure generally permits both elastic and inelastic scattering channels. A variety of inelastic DM realizations have been proposed; see, e.g.,~\cite{Tucker-Smith:2001myb,Tucker-Smith:2004mxa,Chang:2008gd,Cui:2009xq}. We discuss the vector- and scalar-mediated cases together, since they exhibit the same basic coupled-channel scattering structure and differ primarily in their viable regions of parameter space. Elastic multi-component SIDM and composite realizations are briefly discussed at the end of this section.

\medskip
\noindent
\textbf{Microscopic description:} A simple realization consists of a ground state $\chi_1$ and an excited state $\chi_2$ separated by a small mass splitting
\begin{equation}
\delta \equiv m_{\chi_2}-m_{\chi_1} \ll m_{\rm DM}.
\end{equation}
Such a spectrum arises naturally when a single dark-sector particle is split into two states after symmetry breaking. A common example is a pseudo-Dirac fermion,\footnote{A Dirac fermion has distinct particle and antiparticle states, while a Majorana fermion is identical to its own antiparticle. A pseudo-Dirac fermion consists of two nearly degenerate Majorana states, which behave approximately as a single Dirac particle when the mass splitting is small.} which gives two nearly degenerate Majorana states at low energies.

For a vector mediator, the interaction is typically off-diagonal in the mass basis. Starting from a broken dark $U(1)$ gauge symmetry, the original interaction couples to the conserved charge of a Dirac fermion; after splitting into two Majorana states, the leading vector coupling connects $\chi_1$ and $\chi_2$,
\begin{equation}
\mathcal{L}_{\rm int} \supset g \phi_{\mu}\,\bar{\chi}_2\gamma^\mu \chi_1 + {\rm h.c.},
\end{equation}
while diagonal couplings such as $\phi_{\mu}\bar{\chi}_1\gamma^\mu\chi_1$ and $\phi_{\mu}\bar{\chi}_2\gamma^\mu\chi_2$, analogous to \cref{eq:vector_exchange}, are absent or strongly suppressed. Each interaction therefore changes the internal state, $\chi_1\leftrightarrow\chi_2$.

\begin{figure}[t]
\centering

\begin{subfigure}[t]{0.32\columnwidth}
\centering
\begin{tikzpicture}
\begin{feynman}
  \vertex (a) at (0,0.5);
  \vertex (b) at (0,-0.5);

  \vertex (i1) at (-1.2, 0.9) {$\chi_1$};
  \vertex (i2) at (-1.2,-0.9) {$\chi_1$};
  \vertex (f1) at ( 1.2, 0.9) {$\chi_2$};
  \vertex (f2) at ( 1.2,-0.9) {$\chi_2$};

  \diagram*{
    (i1) -- [fermion] (a) -- [fermion] (f1),
    (i2) -- [fermion] (b) -- [fermion] (f2),
    (a)  -- [photon, edge label'=$\phi_\mu$] (b),
  };
\end{feynman}
\end{tikzpicture}
\caption{$\chi_1\chi_1 \to \chi_2\chi_2$}
\label{fig:inelastic_vec_up}
\end{subfigure}
\hfill
\begin{subfigure}[t]{0.32\columnwidth}
\centering
\begin{tikzpicture}
\begin{feynman}
  \vertex (a) at (0,0.5);
  \vertex (b) at (0,-0.5);

  \vertex (i1) at (-1.2, 0.9) {$\chi_2$};
  \vertex (i2) at (-1.2,-0.9) {$\chi_2$};
  \vertex (f1) at ( 1.2, 0.9) {$\chi_1$};
  \vertex (f2) at ( 1.2,-0.9) {$\chi_1$};

  \diagram*{
    (i1) -- [fermion] (a) -- [fermion] (f1),
    (i2) -- [fermion] (b) -- [fermion] (f2),
    (a)  -- [photon, edge label'=$\phi_\mu$] (b),
  };
\end{feynman}
\end{tikzpicture}
\caption{$\chi_2\chi_2 \to \chi_1\chi_1$}
\label{fig:inelastic_vec_down}
\end{subfigure}
\hfill
\begin{subfigure}[t]{0.32\columnwidth}
\centering
\begin{tikzpicture}
\begin{feynman}
  \vertex (a) at (0,0.5);
  \vertex (b) at (0,-0.5);

  \vertex (i1) at (-1.2, 0.9) {$\chi_1$};
  \vertex (i2) at (-1.2,-0.9) {$\chi_2$};
  \vertex (f1) at ( 1.2, 0.9) {$\chi_2$};
  \vertex (f2) at ( 1.2,-0.9) {$\chi_1$};

  \diagram*{
    (i1) -- [fermion] (a) -- [fermion] (f1),
    (i2) -- [fermion] (b) -- [fermion] (f2),
    (a)  -- [photon, edge label'=$\phi_\mu$] (b),
  };
\end{feynman}
\end{tikzpicture}
\caption{$\chi_1\chi_2 \to \chi_2\chi_1$}
\label{fig:inelastic_vec_mixed}
\end{subfigure}

\vspace{0.35cm}

\begin{subfigure}[t]{0.49\columnwidth}
\centering
\begin{tikzpicture}
\begin{feynman}

  \vertex (a1) at (-0.32,0.5);
  \vertex (b1) at (-0.32,-0.5);
  \vertex (a2) at ( 0.32,0.5);
  \vertex (b2) at ( 0.32,-0.5);

  \vertex (i1) at (-1.25, 0.9) {$\chi_1$};
  \vertex (i2) at (-1.25,-0.9) {$\chi_1$};
  \vertex (f1) at ( 1.25, 0.9) {$\chi_1$};
  \vertex (f2) at ( 1.25,-0.9) {$\chi_1$};

  \diagram*{
    (i1) -- [fermion] 
    (a1) -- [fermion, edge label=$\chi_2$] (a2) -- [fermion] 
    (f1),
    (i2) -- [fermion] 
    (b1) -- [fermion, edge label'=$\chi_2$] (b2) -- [fermion] 
    (f2),
    (a1) -- [photon, edge label'=$\phi_\mu$] (b1),
    (a2) -- [photon, edge label=$\phi_\mu$] (b2),
  };
\end{feynman}
\end{tikzpicture}
\caption{$\chi_1\chi_1 \to \chi_1\chi_1$ (loop)}
\label{fig:inelastic_vec_loop_11}
\end{subfigure}
\hfill
\begin{subfigure}[t]{0.49\columnwidth}
\centering
\begin{tikzpicture}
\begin{feynman}
  \vertex (a1) at (-0.32,0.5);
  \vertex (b1) at (-0.32,-0.5);
  \vertex (a2) at ( 0.32,0.5);
  \vertex (b2) at ( 0.32,-0.5);

  \vertex (i1) at (-1.25, 0.9) {$\chi_2$};
  \vertex (i2) at (-1.25,-0.9) {$\chi_2$};
  \vertex (f1) at ( 1.25, 0.9) {$\chi_2$};
  \vertex (f2) at ( 1.25,-0.9) {$\chi_2$};

  \diagram*{
    (i1) -- [fermion] 
    (a1) -- [fermion, edge label=$\chi_1$] (a2) -- [fermion] 
    (f1),
    (i2) -- [fermion] 
    (b1) -- [fermion, edge label'=$\chi_1$] (b2) -- [fermion] 
    (f2),
    (a1) -- [photon, edge label'=$\phi_\mu$] (b1),
    (a2) -- [photon, edge label=$\phi_\mu$] (b2),
  };
\end{feynman}
\end{tikzpicture}
\caption{$\chi_2\chi_2 \to \chi_2\chi_2$ (loop)}
\label{fig:inelastic_vec_loop_22}
\end{subfigure}

\caption{
Self-scattering processes in a two-state inelastic SIDM model with off-diagonal vector interaction of the form
$g\,\phi_\mu\,\bar{\chi}_2\gamma^\mu\chi_1$.
\textit{Top panels}: tree-level inelastic and mixed elastic scattering channels.
\textit{Bottom panels}: loop-induced elastic scattering of identical states.
}
\label{fig:inelastic_vec_reactions_full}
\end{figure}
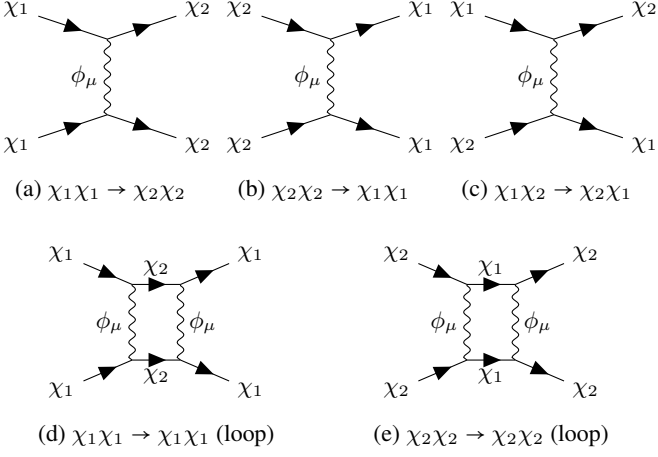

This structure determines the tree-level processes shown in the upper panels of \cref{fig:inelastic_vec_reactions_full}: the inelastic channels $\chi_1\chi_1 \to \chi_2\chi_2$ and $\chi_2\chi_2 \to \chi_1\chi_1$, and elastic scattering in the mixed channel $\chi_1\chi_2 \to \chi_2\chi_1$, equivalent to $\chi_1\chi_2 \to \chi_1\chi_2$ up to relabeling. Elastic scattering of identical states, $\chi_1\chi_1 \to \chi_1\chi_1$ and $\chi_2\chi_2 \to \chi_2\chi_2$, is absent at tree level but generated at higher order through virtual transitions, corresponding to the loop diagrams in the lower panels of \cref{fig:inelastic_vec_reactions_full}. In vector-mediated models these loop-induced elastic processes need not be negligible: in the non-relativistic regime, repeated mediator exchange and channel mixing, i.e.\ Sommerfeld-type effects, can enhance the low-velocity amplitude. The appropriate description is therefore a coupled-channel Schr\"odinger problem.

A related but phenomenologically distinct realization is scalar-mediated inelastic SIDM~\cite{Wang:2025xoq}. The dark sector again contains a pseudo-Dirac fermion split into $\chi_1$ and $\chi_2$, but the mediator is a light scalar $\phi$. This retains the kinematic threshold mechanism of inelastic SIDM while potentially avoiding some cosmological constraints of vector-mediated models, since annihilation into scalars can be $p$-wave suppressed, as discussed in \cref{sec:DM_search_experiments}. Avoiding these constraints may require tuned mass splittings or asymmetric DM scenarios; see, e.g.,~\cite{Kaplan:2009ag,Petraki:2013wwa} and \cref{sec:Asymmetric_SIDM}. In the benchmark of Ref.~\cite{Wang:2025xoq}, $\delta \sim 100\,{\rm eV}$, so up-scattering is suppressed in the coldest haloes while remaining efficient in dwarf galaxies.

Unlike the vector case, where the off-diagonal structure follows from the broken $U(1)$ gauge symmetry, a scalar mediator generically induces both diagonal and off-diagonal Yukawa couplings in the mass basis. To obtain inelastic-dominated scattering, one can impose a dark $\mathcal{Z}_2$ symmetry,
\begin{equation}
\chi_1 \to +\chi_1,
\qquad
\chi_2 \to -\chi_2,
\qquad
\phi \to -\phi,
\end{equation}
which enforces
\begin{equation}
\mathcal{L}_{\rm int} \supset -y\, \phi (\bar{\chi}_1 \chi_2 + \bar{\chi}_2 \chi_1),
\end{equation}
and forbids diagonal couplings $\phi \bar{\chi}_1\chi_1$ and $\phi \bar{\chi}_2\chi_2$, analogous to \cref{eq:yukawa_coupling}. The leading self-scattering processes then resemble the vector-mediated case in \cref{fig:inelastic_vec_reactions_full}, with $\phi_\mu \to \phi$.

A key difference is that, in the scalar-mediated case, elastic scattering of identical states can be dominated by virtual transitions through the closed $\chi_2\chi_2$ channel. This allows destructive interference, or an anti-resonance, where the elastic cross section is strongly suppressed. In vector-mediated models, additional non-perturbative contributions typically prevent a complete cancellation, so the elastic cross section remains non-zero even near an anti-resonance.

At the two-body level, both vector- and scalar-mediated two-state SIDM are described by a coupled-channel Schr\"odinger equation. Restricting to the basis $\ket{11}$ and $\ket{22}$,
\begin{equation} \label{eq:two_state_inelastic_cross}
\left[
-\frac{1}{2\mu_{\rm DM}}\nabla^2 + \mathbf{V}(r)
\right]\Psi(\mathbf{r})
= E\, \Psi(\mathbf{r}),
\end{equation}
where $\mu_{\rm DM}$ is the reduced mass. For off-diagonal mediator exchange, the potential matrix has the schematic form
\begin{equation}
\mathbf{V}(r)
=
\begin{pmatrix}
0 & V_{\rm off}(r) \\
V_{\rm off}(r) & 2\delta
\end{pmatrix},
\end{equation}
with
\begin{equation}
V_{\rm off}(r) \simeq -\frac{\alpha_{\rm DM}}{r}e^{-m_{\rm med} r},
\end{equation}
where $\alpha_{\rm DM}=g^2/(4\pi)$ for vector mediation and $\alpha_{\rm DM}=y^2/(4\pi)$ for scalar mediation in a Yukawa parametrization. Expanding in partial waves,
\begin{equation}
\Psi(\vec r)
=
\sum_{\ell=0}^{\infty}
R_{\ell,i}(r)\,
P_\ell(\cos\theta),
\end{equation}
and defining $\chi_{\ell,i}(r)\equiv r\,R_{\ell,i}(r)$,
the radial equations are
\begin{equation}
\label{eq:coupled_channel_Schrodinger_eq}
\left[
\frac{{\rm d}^2}{{\rm d}r^2}
-\frac{\ell(\ell+1)}{r^2}
+k^2
\right]\chi_{\ell,i}
=
2\mu_{\rm DM}
\sum_j V_{ij}(r)\chi_{\ell,j},
\end{equation}
where $k^2=2\mu_{\rm DM}E$, with $E$ measured
relative to the ground-state two-particle threshold.
The channel-dependent asymptotic momenta are
$k_1^2=2\mu_{\rm DM}E$ and
$k_2^2=2\mu_{\rm DM}(E-2\delta)$,
where the excitation threshold $2\delta$ is
already included in the diagonal potential
$V_{22}=2\delta$. A numerical procedure for solving this coupled-channel system is given in Appendix~B of
\cite{Blennow:2016gde}.

\begin{figure*}
    \centering
    \includegraphics[width=1.0\columnwidth]{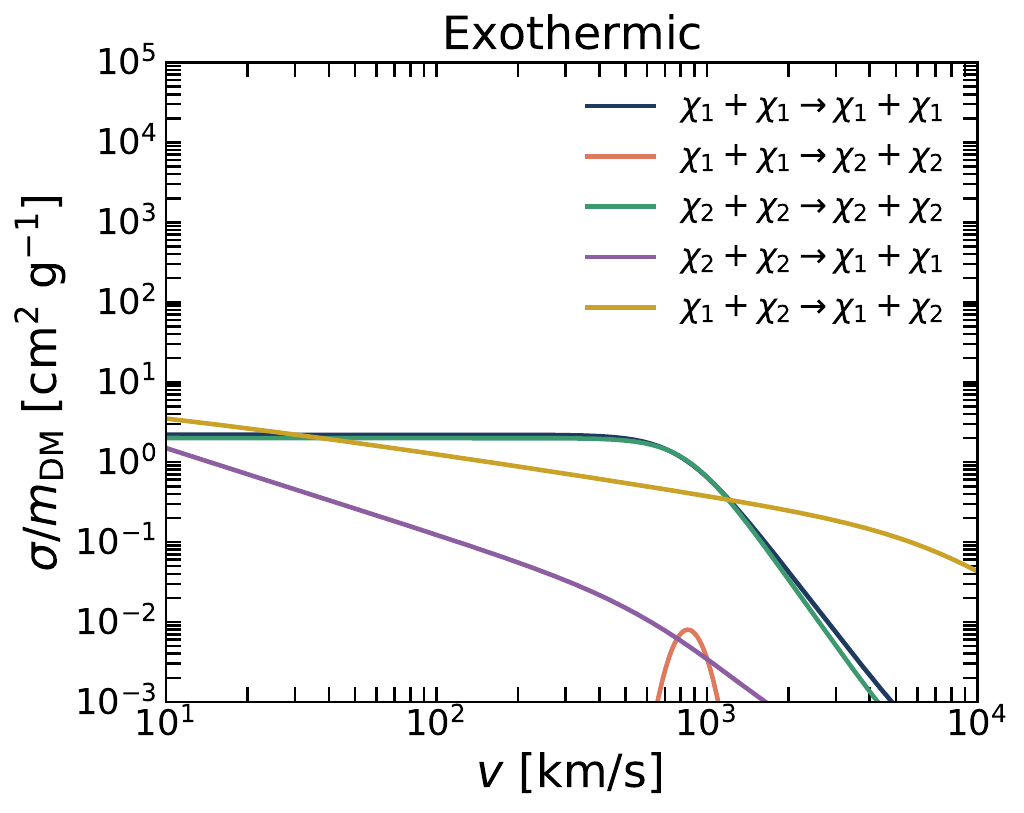}
    \includegraphics[width=1.0\columnwidth]{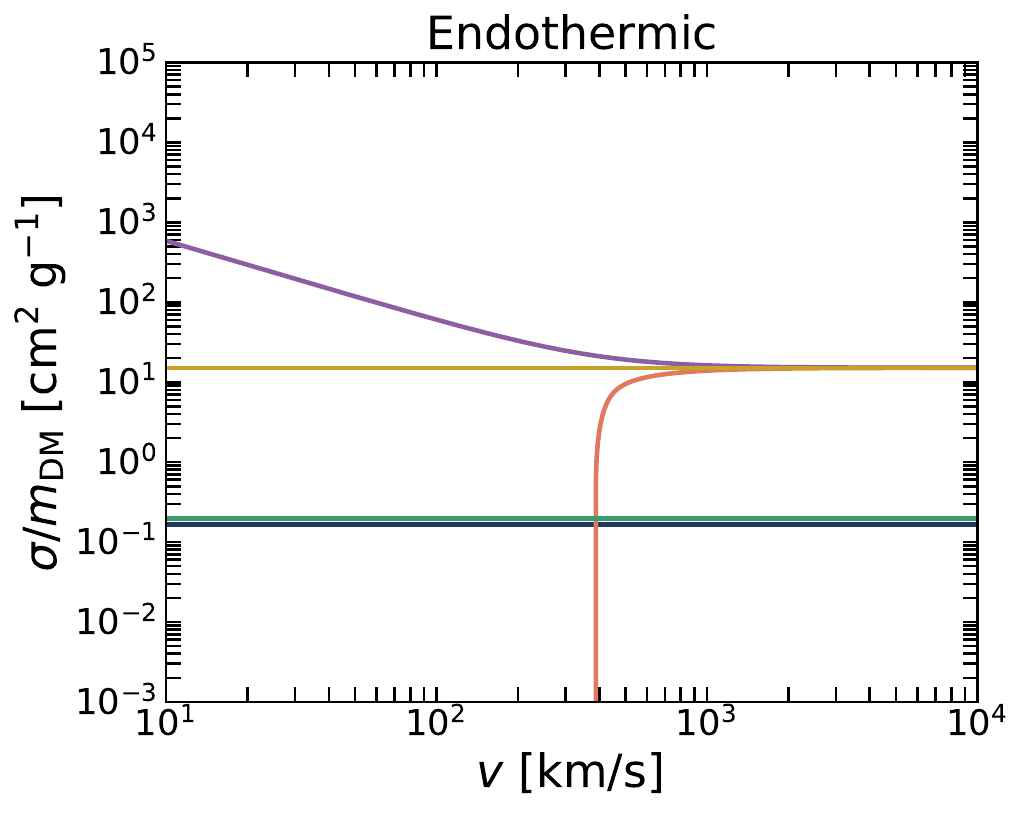}
\caption{
Velocity dependence of the self-interaction cross section per unit mass, $\sigma/m_{\rm DM}$, for representative two-state inelastic SIDM benchmarks with a vector mediator. 
\textit{Left panel:} Exothermic case, including elastic channels and down-scattering $\chi_2+\chi_2\to\chi_1+\chi_1$, which releases internal energy and can impart velocity kicks to the outgoing particles. 
\textit{Right panel:} Endothermic case, including elastic channels and up-scattering $\chi_1+\chi_1\to\chi_2+\chi_2$, which is suppressed below threshold and removes kinetic energy when active. 
The curves reproduce the benchmark cross sections of~\cite{ONeil:2022szc} using the same two-state SIDM setup and channel definitions, illustrating the threshold behaviour and channel dependence characteristic of multi-state SIDM.
}
    \label{fig:inelastic_cross_sections}
\end{figure*}

For relative velocity $v_{\rm rel}$, the incoming kinetic energy is $E=\mu_{\rm DM}v_{\rm rel}^2/2$. The excited channel opens only if
\begin{equation}
E > 2\delta
\;\;\Leftrightarrow\;\;
v_{\rm rel} > 2\sqrt{\frac{\delta}{\mu_{\rm DM}}},
\end{equation}
and is otherwise kinematically closed. This defines the threshold for endothermic up-scattering: below threshold, $\chi_1\chi_1\to\chi_2\chi_2$ is forbidden, while above threshold it can rapidly turn on and generate strong velocity dependence. Endothermic up-scattering removes kinetic energy and is inactive in cold systems, whereas exothermic down-scattering injects energy and can impart recoil kicks with magnitude (\cref{eq:velocity_kick_exothermic})
\begin{equation}
v_{\rm kick} = \sqrt{\frac{2\delta}{m_{\rm DM}}}\,c .
\end{equation}

Multi-component SIDM can also significantly modify halo structure through purely elastic scattering. In models with DM species of different masses, inter-species scattering transfers kinetic energy from heavier to lighter particles, driving mass segregation in which the heavier species becomes more centrally concentrated while the lighter species moves to larger radii. This can enhance central densities and modify gravothermal evolution, with implications for dense substructures and strong gravitational lensing~\cite{Yang:2025dgl,Yang:2025xsp}.

Composite dark sectors provide another natural realization of multi-state inelastic SIDM~\cite{Alves:2009nf}. In analogy with the visible sector, where protons and neutrons arise as bound states of quarks, confined dark sectors can contain dark baryons with large self-interactions and multiple nearly degenerate states, as discussed in \cref{sec:dark_baryons}. More generally, composite dark sectors are expected to produce a rich spectrum of states, providing a natural microscopic origin for the inelastic scenarios described above.

\medskip
\noindent
\textbf{Representative benchmarks:} An exothermic benchmark with vector mediation is
\begin{equation}
\begin{aligned}
m_{\chi_1} &= 10~{\rm GeV}, \qquad
m_{\phi_\mu} = 30~{\rm MeV}, \\
\delta &= 10~{\rm keV}, \qquad 
\alpha_{\rm DM} = 0.10,
\end{aligned}
\end{equation}
for which up-scattering is kinematically suppressed and the dynamics is dominated by elastic and down-scattering processes. The corresponding velocity-dependent self-interaction cross sections are shown in the left panel of \cref{fig:inelastic_cross_sections}.

An endothermic benchmark is
\begin{equation}
\begin{aligned}
m_{\chi_1} &= 2.3~{\rm MeV},  \qquad 
m_{\phi_\mu} = 7.3~{\rm MeV}, \\
\delta &= 0.48~{\rm eV},  \qquad  
\alpha_{\rm DM} = 0.17,
\end{aligned}
\end{equation}
for which up-scattering becomes kinematically allowed above
\begin{equation}
v \gtrsim 2\sqrt{\frac{2\delta}{m_{\chi_1}}}c \simeq 400~{\rm km/s},
\end{equation}
so that all channels can contribute. The corresponding cross sections are shown in the right panel of \cref{fig:inelastic_cross_sections}.

For a scalar-mediated model, a representative benchmark is
\begin{equation}
\begin{aligned}
m_\chi &\approx 40~{\rm GeV},  \qquad 
m_\phi \approx 20~{\rm MeV}, \\
\delta &\approx 100~{\rm eV}, \qquad
\alpha_{\rm DM} \approx 10^{-2},
\end{aligned}
\end{equation}
which suppresses self-scattering at the low relative velocities of ultra-faint Milky Way satellites ($v_{\rm rel}\sim20\,{\rm km\,s^{-1}}$), while allowing resonantly enhanced scattering in dwarf galaxies with higher characteristic velocities ($v_{\rm rel}\sim60\,{\rm km\,s^{-1}}$). For detailed discussion of the cosmological, astrophysical, and experimental viability of multi-state inelastic SIDM models and the parameter space surrounding these benchmarks, see App.~\ref{app:constraints_inelastic}.

\subsubsection{Theory-to-simulation mapping: Multi-state inelastic SIDM}
\label{sec:mapping_multi_inelastic_SIDM}

For two-state inelastic SIDM, the microscopic parameters
\begin{equation}
\label{eq:pp_parameters_inelastic}
\mathcal{P}_{\rm PP}
=
\{m_{\rm DM},\,m_{\rm med},\,\alpha_{\rm DM},\,\delta\}
\end{equation}
determine the internal-state structure, reaction kinematics and channel-dependent scattering cross sections. The mass splitting is encoded through the relative state masses, $\Delta_{\chi_1}=0$ and $\Delta_{\chi_2}=\delta/m_{\rm DM}$, while the cosmological history determines the initial state fractions $f_i^{\rm init}$. The reaction network $\mathcal R=\{ij\to kl\}$ contains the allowed elastic, endothermic and exothermic channels, as illustrated in \cref{fig:inelastic_vec_reactions_full}, with the associated energy exchange determined by the state masses and reaction kinematics. Solving the coupled-channel scattering problem in \cref{eq:two_state_inelastic_cross} gives the corresponding differential cross sections ${\rm d}\sigma_{\mathcal R}/{\rm d}\Omega(\theta,v_{\rm rel})$. The resulting mapping can therefore be summarized as
\begin{equation}
\label{eq:sim_inputs_inelastic_compact}
\mathcal{P}_{\rm PP}
\rightarrow
\mathcal{P}_{\rm SIM}
=
\left\{
\Delta_i,\,
f_i^{\rm init},\,
\mathcal R,\,
\frac{1}{m_{\rm DM}}
\frac{{\rm d}\sigma_{\mathcal R}}{{\rm d}\Omega}
\right\}.
\end{equation}
Only if one or more internal states are unstable and their decays, as illustrated in \cref{fig:decay_diagrams}, are dynamically relevant for structure formation are the decay rate $\Gamma$ and velocity kick $v_{\rm kick}$ required as additional simulation inputs, as described in \cref{sec:decay_inputs}. If the angular dependence is not resolved explicitly, the channel-dependent differential cross sections may instead be represented by the corresponding angle-weighted cross sections $\sigma_X^{\mathcal R}(v_{\rm rel})/m_{\rm DM}$ defined \cref{eq:weighted_cross_sections}. 
Standard CDM initial conditions apply unless
early-time dark-sector dynamics modify the
linear matter power spectrum. In particular,
exothermic inelastic conversions can suppress
small-scale power and generate dark acoustic
oscillations even without dark
radiation~\cite{Duan:2026zqj}.
If these effects are established before the
simulation starting redshift, they must be
included in the cosmological initial conditions,
as discussed in \cref{sec:initial_conditions}
and summarized in \cref{tab:pk_models}.

Multi-state inelastic SIDM is treated with the \hyperref[sec:imp_inelastic_SIDM]{iSIDM} implementation, in which each simulation particle carries an internal-state label and scattering events sample the allowed channels, enforce their kinematic thresholds and update both particle states and velocities. Depending on the velocity and angular dependence of the channel cross sections, iSIDM can be combined with the velocity-dependent Monte Carlo scheme (\hyperref[sec:mc_velocity_dep]{MC}), the anisotropic Monte Carlo scheme (\hyperref[sec:anisotropic_MC]{aMC}) or the angular-hybrid scheme (\hyperref[sec:asidm]{aSIDM}). If one or more internal states are unstable and their decays are dynamically relevant, these can additionally be treated with the state-based decaying-DM implementation (\hyperref[sec:imp_decaying_DM]{dDM}).

\subsection{Composite SIDM}
\label{sec:composite_SIDM}

Composite SIDM describes scenarios in which the DM particles are composite bound states rather than elementary particles. The relevant low-energy degrees of freedom may be dark baryons, dark mesons, dark glueballs, dark atoms or dark molecules, with self-interactions governed by non-perturbative dynamics, residual forces, bound-state formation or atomic/molecular scattering. By analogy with visible matter, these models provide a natural setting for large, velocity-dependent and potentially resonant self-interactions. They can also involve multiple internal states, inelastic transitions, dissipative processes and decays. We now discuss five representative composite SIDM model classes: dark-baryon SIDM (\cref{sec:dark_baryons}), dark-meson SIDM (\cref{sec:dark_mesons}), dark-glueball SIDM (\cref{sec:dark_glueballs}), and dark-atom or dark-molecule SIDM (\cref{sec:dark_atoms}).

\subsubsection{Dark-baryon SIDM}
\label{sec:dark_baryons}

Dark-baryon SIDM is a well-motivated composite DM scenario in which the DM consists of bound states formed by a confining dark-sector force. The resulting self-interactions are governed by non-perturbative, nuclear-like dynamics rather than elementary mediator exchange.

\medskip
\noindent
\textbf{Microscopic description:} A natural possibility is that DM is not elementary, but consists of composite bound states formed by a confining non-Abelian gauge interaction, analogous to QCD. In this class of models, the DM candidates are dark baryons, $N_D$, i.e.\ bound states of dark quarks $q_D$ charged under a dark gauge group such as $SU(N_D)$ or $SO(N_D)$~\cite{Cline:2013zca,Chu:2019awd,Cline:2022leq}, as discussed in \cref{sec:composite_dark_sectors_non_pert_ints}. 
For $SU(N_D)$, dark baryons are bound states
of $N_D$ fermionic dark quarks carrying a
conserved dark baryon number. They are
fermionic for odd $N_D$ and bosonic for even
$N_D$. For $SO(N_D)$, they are real, i.e.\ identical to their antiparticles, but can remain stable due to residual symmetries~\cite{Antipin:2015xia}. Their stability is often robust, since decay typically proceeds only through higher-dimensional operators, with dimension $d\geq 6$ for $N_D\geq 3$, making them naturally long-lived on cosmological timescales~\cite{Kribs:2016cew}.

The dynamics depend on the hierarchy between the dark quark mass $m_{q_D}$ and the confinement scale $\Lambda_{\rm dQCD}$. If $m_{q_D}\ll\Lambda_{\rm dQCD}$, the theory resembles QCD, with strongly coupled relativistic bound states whose properties must be inferred from lattice calculations~\cite{Kribs:2016cew}. If $m_{q_D}\gg\Lambda_{\rm dQCD}$, the dark quarks are non-relativistic and the bound states can be described using quantum-mechanical potential methods.
Recent studies of heavy dark baryons in confining $SU(N_D)$ theories have shown that chromoelectric polarizability induces van der Waals interactions, potentially generating large, velocity-dependent self-scattering cross sections~\cite{DallaValleGarcia:2026mis,DallaValleGarcia:2026aws}.

A minimal and phenomenologically relevant setup is a QCD-like dark sector with light dark quarks and strong gauge group $SU(N_D)$, analogous to the up and down quarks that form protons and neutrons. For $m_{q_D} \ll \Lambda_{\rm dQCD}$, the theory contains light composite dark pions, arising as pseudo-Nambu--Goldstone bosons of approximate chiral symmetry breaking as in \cref{eq:global_symmetry_breaking}. The dark baryon mass, which sets the DM mass scale, is generated dynamically by confinement,
\begin{equation}
m_{N_D} \sim {\cal O}(1)\,\Lambda_{\rm dQCD}.
\end{equation}
Guided by QCD, one expects scaling relations of the form
\begin{equation}
m_{N_D} \simeq 3.8\,\Lambda_{\rm dQCD},
\qquad
m_{\pi_D} \propto \sqrt{m_{q_D}\Lambda_{\rm dQCD}},
\end{equation}
where the baryon mass is controlled mainly by confinement, while the dark pion mass $m_{\pi_D}$ depends on both $m_{q_D}$ and $\Lambda_{\rm dQCD}$.

For simplicity, one often assumes equal dark quark masses and a QCD-like structure with $N_c=3$ and $N_f=2$. If $\Lambda_{\rm dQCD}$ is moderately larger than $\Lambda_{\rm QCD}\simeq 250~{\rm MeV}$, the resulting dark baryons naturally lie in the GeV mass range. This is particularly well motivated in asymmetric DM scenarios, discussed in \cref{sec:DM_production_mechanisms,sec:Asymmetric_SIDM}, which aim to explain $\Omega_{\rm DM}\simeq 5\,\Omega_B$~\cite{Planck_2020b}. If the dark and visible asymmetries are generated by related mechanisms~\cite{Nussinov:1985xr,Kaplan:2009ag,Davoudiasl:2012uw,DuttaBanik:2020vfr,Barman:2021ost,Cui:2020dly,Falkowski:2011xh,Patel:2022xyv,Biswas:2018sib,Narendra:2018vfw,Nagata:2016knk,Arina:2011cu,Arina:2012fb,Narendra:2017uxl,Narendra:2019cyt,Mahapatra:2023dbr}, one expects $n_{\rm DM}\sim n_B$. Since $\Omega_X=n_Xm_X/\rho_c$, this points to $m_{\rm DM}\sim m_B$ and hence GeV-scale dark baryons. With a conserved dark baryon number, the relic abundance is set by a primordial asymmetry rather than thermal freeze-out~\cite{Frandsen:2011kt,Cline:2022leq}.

A key feature of dark-baryon SIDM is that self-interactions are governed by strong, non-perturbative dynamics. At low energies, the scattering can be described using the effective-range expansion familiar from nuclear physics. Focusing on $s$-wave scattering in spin channel $s$, the amplitude follows from \cref{eq:pw_effective_range,eq:ere_general}. In the effective-range approximation one obtains~\cite{Chu:2019awd,Cline:2022leq}
\begin{equation} \label{eq:dark_baryon_cross}
\mathcal{A}_s=
\frac{4\pi}{m_N\left(-ip-a_s^{-1}+\frac12 r_{0,s}p^2+\mathcal{O}(p^4)\right)},
\end{equation}
where $m_N$ is the dark nucleon mass, $p=m_Nv_{\rm rel}/2$ is the center-of-mass momentum, and $a_s$ and $r_{0,s}$ are the scattering length and effective range in the given spin channel. The total cross section is obtained by summing over spin channels,
\begin{equation}
\sigma_{\rm tot} 
= 
\sum_s (2s+1)\,\frac{m_N^2|\mathcal{A}_s|^2}{16\pi}.
\end{equation}

Lattice QCD\footnote{Here, ``lattice'' refers to lattice gauge theory calculations, where the strongly coupled dynamics are solved numerically by discretizing spacetime~\cite{Gupta:1997nd}. These methods are essential for QCD-like theories, where perturbative techniques break down.} combined with nuclear effective field theory connects the strongly coupled dynamics to the low-energy scattering parameters. The scattering lengths can be expressed as functions of $m_{\pi_D}/\Lambda_{\rm dQCD}$, which controls the composite dynamics. Using fits to lattice QCD data~\cite{Beane:2008bt,Chen:2010yt}, the singlet and triplet scattering lengths can be parameterized as
\begin{equation}
a_s = \frac{0.71\,\Lambda_{\rm QCD}^{-1}}{m_{\pi}/\Lambda_{\rm QCD} - 0.58},
\qquad
a_t = \frac{0.96\,\Lambda_{\rm QCD}^{-1}}{m_{\pi}/\Lambda_{\rm QCD} - 0.42},
\end{equation}
for $\Lambda_{\rm QCD}=250~{\rm MeV}$. For a dark QCD sector, these expressions can be rescaled by replacing $\Lambda_{\rm QCD}\to\Lambda_{\rm dQCD}$ and $m_\pi\to m_{\pi_D}$, assuming the same dimensionless dependence on $m_{\pi_D}/\Lambda_{\rm dQCD}$. Since the effective ranges $r_{0,s}$ are parametrically smaller than the corresponding scattering lengths over the relevant parameter space, they can often be neglected to a good approximation~\cite{Chen:2010yt}.

The cross section is velocity dependent through
$p=m_Nv_{\rm rel}/2$. For comparison with
astrophysical systems, we average over the
relative-velocity distribution. Assuming a
Maxwellian distribution with one-dimensional
relative-velocity dispersion $v_0$, and neglecting
the effective range, $|r_{0,s}|\ll|a_s|$, the
average can be performed analytically:
\begin{equation}
\label{eq:dark_baryon_cross_end}
\langle \sigma v_{\rm rel} \rangle
=
\sum_s (2s+1)\,
\frac{a_s}{m_N}\,F(b_s),
\end{equation}
where
\begin{equation*}
F(b_s)
=
\sqrt{2\pi}
\left(
2b_s
+
b_s^3 e^{b_s^2/2}
\,\mathrm{Ei}(-b_s^2/2)
\right),
\end{equation*}
and
\begin{equation*}
b_s \equiv \frac{2}{m_N a_s v_0},
\qquad
v_0 =
\sqrt{\frac{\pi}{8}}\,
\langle v_{\rm rel}\rangle.
\end{equation*}
Here, $\mathrm{Ei}$ is the exponential integral. This form shows explicitly how the scattering rate depends on the velocity scale of the system. The model can therefore produce large self-interactions at dwarf-galaxy velocities while suppressing scattering in clusters~\cite{Kaplinghat:2015aga,Cline:2022leq}. Large cross sections arise near poles in the scattering lengths, corresponding to near-threshold bound or virtual states. Dark-baryon SIDM is thus a composite analogue of resonant SIDM, with the enhancement generated by nuclear-like dynamics rather than elementary mediator exchange.

The differential cross section may in general have angular dependence through the partial-wave expansion in \cref{eq:pw_effective_range}. In the simplest $s$-wave dominated case it is approximately isotropic, while higher partial waves introduce angular structure through $P_\ell(\cos\theta)$.

Thus, the particle-physics parameters of a QCD-like dark-baryon scenario can be summarized as
\begin{equation}
\mathcal{P}_{\rm PP}
=
\{ \Lambda_{\rm dQCD},\, m_{q_D},\, N_D \},
\end{equation}
which determine the low-energy quantities
\begin{equation}
\{ m_{N_D},\, m_{\pi_D},\, a_s,\, r_{0,s} \}.
\end{equation}

The resulting theory-to-simulation mapping and the corresponding \arepoTwo implementation are summarized in \cref{sec:mapping_composite_SIDM}.

\medskip
\noindent
\textbf{Representative benchmarks:} Two representative dark-baryon benchmarks compatible with astrophysical constraints are
\begin{equation}
m_{\pi_D}=0.5757\, \Lambda_{\rm dQCD},
\qquad
m_{\rm DM}\equiv m_{N_D}=5~\mathrm{GeV},
\end{equation}
compatible with constraints inferred from dwarf galaxies, low surface brightness galaxies, galaxy groups and clusters, shown in \cref{fig:cross_sections}. A second benchmark yields larger dwarf-galaxy cross sections:
\begin{equation}
m_{\pi_D}=0.5795\, \Lambda_{\rm dQCD},
\qquad
m_{\rm DM}\equiv m_{N_D}=6~\mathrm{GeV},
\end{equation} closer to the preferred values reported in~\cite{2021MNRAS503920C}. For detailed discussion of the cosmological, astrophysical, and experimental viability of dark-baryon SIDM models and the parameter space surrounding these benchmarks, see App.~\ref{app:constraints_dark_baryons}.

\subsubsection{Dark-meson SIDM}
\label{sec:dark_mesons}

Dark-meson SIDM is a broad class of composite DM models in which DM consists of mesonic bound states of a new confining dark gauge interaction. Unlike elementary SIDM models based on fundamental mediators, the DM particle is itself composite, and its self-interactions are governed by low-energy hadronic dynamics, chiral symmetry breaking and, in some cases, composite resonances or portal interactions.

\medskip
\noindent
\textbf{Microscopic description:} A generic starting point is a confining gauge theory with dark quarks, as introduced in \cref{sec:composite_dark_sectors_non_pert_ints}, for example
\begin{equation}
G_{\rm DC} = SU(N)_{\rm DC},\qquad SO(N)_{\rm DC},\qquad Sp(N)_{\rm DC},
\end{equation}
with fermionic matter fields $q_{D,i}$ transforming under $G_{\rm DC}$. At the renormalizable level, the dark sector can be written schematically as
\begin{equation}
\begin{aligned}
\mathcal{L}_{\rm DC}
=&
-\frac14\,G^a_{D\,\mu\nu}G_D^{a\,\mu\nu}
+\sum_i \bar q_{D,i}(i\gamma^\mu D_\mu-m_i)q_{D,i}
\\ &+\mathcal{L}_{\rm dark\text{-}int},
\end{aligned}
\label{eq:dark_meson_micro_lagrangian}
\end{equation}
where $G^a_{D\,\mu\nu}$ is the field-strength tensor of the dark gauge field, analogous to the gluon field-strength tensor in QCD. Here, $\mathcal{L}_{\rm dark\text{-}int}$ may contain additional portal or mediator interactions.

For light dark quarks, $m_{q_D}\ll\Lambda_D$, confinement induces spontaneous chiral symmetry breaking through $\langle \bar q_D q_D\rangle\neq0$, producing dark mesons as pseudo-Nambu--Goldstone bosons (pNGBs), $\pi_D\sim\bar q_Dq_D$. The precise symmetry-breaking pattern depends on the confining gauge group and matter representation, as summarized in \cref{eq:global_symmetry_breaking}.

At low energies, the pNGBs are described by a chiral effective theory with
\begin{equation}
\Sigma = \exp\!\left(i\pi_D/f\right),
\end{equation}
where $\pi_D=\pi_D^aT^a$, with $T^a$ denoting the generators of the broken flavor symmetry, and $f$ is the dark-pion decay constant. The leading effective Lagrangian is~\cite{Georgi:1984zwz}
\begin{equation}
\mathcal{L}_{\rm eff}
=
f^2\,{\rm tr}(\partial_\mu\Sigma^\dagger\partial^\mu\Sigma)
-
f^3\,{\rm tr}(M_q\Sigma+{\rm h.c.}),
\label{eq:dark_meson_chiral_lagrangian}
\end{equation}
giving
\begin{equation}
m_{\pi_D}^2 \sim f\,M_q .
\end{equation}
Thus, the dark-meson masses are controlled by explicit chiral-symmetry breaking, while their interaction scale is set by $f\sim\Lambda_D/(4\pi)$ up to order-one factors.

Dark-meson SIDM is an umbrella class that can be organized by both the microscopic realization and the relic-density mechanism. The most relevant realizations are:
\begin{itemize}
\item \emph{QCD-like dark mesons:} DM consists of one or more stable pNGB mesons of a confining dark sector~\cite{Bai:2010qg,Berlin:2018tvf}.
\item \emph{Composite-Higgs dark mesons:} DM is a pNGB arising from the same strong sector that triggers electroweak symmetry breaking and generates the composite Higgs~\cite{Ryttov:2008xe,Rosenlyst:2021elz,Rosenlyst:2024rsb}.
\end{itemize}
The relic abundance can be set by portal-driven $2\to2$ freeze-out, SIMP $3\to2$ freeze-out generated by the Wess--Zumino--Witten term, or by dynamics involving heavier vector resonances, including resonant enhancement and semi-annihilation~\cite{Bhattacharya:2013kma,Hochberg:2014dra,Hochberg:2014kqa,Bernal:2015xba,Choi:2018iit,Berlin:2018tvf}.

The resulting theory-to-simulation mapping, including the contact-like and resonant scattering regimes and their corresponding \arepoTwo implementations, is summarized in \cref{sec:mapping_composite_SIDM}.

\medskip
\noindent
\textbf{Representative benchmarks:}
For QCD-like dark mesons in a portal-coupled
realization, two representative scalar-like
$s$-wave resonances are
\begin{equation}
m_{\pi_D}\simeq22~{\rm GeV},\quad
v_R=120~{\rm km/s},\quad
\gamma=10^{-4.5},
\end{equation}
with $\sigma_0/m_{\pi_D}=0.12~{\rm cm}^2/{\rm g}$,
corresponding to the dark-red curve in
\cref{fig:cross_sections}, and
\begin{equation}
m_{\pi_D}\simeq16~{\rm GeV},\quad
v_R=5035~{\rm km/s},\quad
\gamma=10^{-1.1},
\end{equation}
with $\sigma_0/m_{\pi_D}=0.2~{\rm cm}^2/{\rm g}$,
representing a broader resonance not shown
in the figure. 
A third benchmark is a vector-like $p$-wave resonance:
\begin{equation}
m_{\pi_D} \simeq 400~{\rm MeV},\quad
v_R = 50~{\rm km/s},\quad
\gamma = 10^{-0.5},
\end{equation}
with $\sigma_0/m_{\pi_D}=0.2~{\rm cm}^2/{\rm g}$.

For composite-Higgs dark mesons, a representative benchmark is
\begin{equation}
m_{\rm DM}\simeq 100~{\rm MeV}, \qquad f\simeq 20~{\rm TeV}, \qquad \lambda_\Phi\simeq 1,
\end{equation}
which yields
\begin{equation}
\frac{\sigma_{\rm tot}}{m_{\rm DM}}\sim 0.5~{\rm cm}^2/{\rm g}.
\end{equation}
This benchmark combines naturalness considerations with astrophysical self-interaction requirements in a unified framework. For detailed discussion of the cosmological, astrophysical, and experimental viability of dark-meson SIDM models and the parameter space surrounding these benchmarks, see App.~\ref{app:constraints_dark_mesons}.

\subsubsection{Dark-glueball SIDM}
\label{sec:dark_glueballs}

Dark-glueball SIDM is one of the simplest composite DM scenarios. The dark sector is a pure confining non-Abelian gauge theory without dark quarks, so the low-energy spectrum consists entirely of glueball bound states. Unlike mediator-based SIDM, the self-interactions arise from non-perturbative strong dynamics in the confined phase rather than from exchange of a fundamental light particle.

\medskip
\noindent
\textbf{Microscopic description:} The minimal dark-glueball scenario contains only a new confining non-Abelian gauge force, for example based on $SU(N)_{\rm DC}$, with no dark quarks or other matter fields. Its microscopic Lagrangian is
\begin{equation}
\mathcal{L}_{\rm dark}
=
-\frac14\,G^a_{D\,\mu\nu}G_D^{a\,\mu\nu}
+
\frac{\theta_D g_D^2}{32\pi^2}\,
G^a_{D\,\mu\nu}\widetilde{G}_D^{a\,\mu\nu}.
\end{equation}
Here, $G^a_{D\,\mu\nu}$ is the field-strength tensor of the dark gauge field, analogous to the gluon field-strength tensor in QCD. The parameter $g_D$ is the dark gauge coupling, while $\theta_D$ is a CP-violating vacuum angle that is often neglected in the simplest scenarios.

As in QCD, the interaction becomes strong at low energies and confines at a characteristic scale $\Lambda_D$~\cite{Kribs:2016cew,Ochs:2013gi}; see \cref{fig:dark_sector_running_coupling} and the discussion in \cref{sec:composite_dark_sectors_non_pert_ints}. Below confinement, the dark gluons form glueball bound states. The lightest scalar glueball is associated with the operator
\begin{equation}
\mathcal{O}_{\rm GB} \sim \mathrm{Tr}(G_{D\,\mu\nu} G_D^{\mu\nu}),
\end{equation}
where $G_{D\,\mu\nu}=G^a_{D\,\mu\nu}T^a$.\footnote{Here, $G^a_{D\,\mu\nu}$ are the components of the dark-gluon field-strength tensor, while $T^a$ are the generators of the gauge group, encoding the internal dark-color structure.} Glueballs are classified by quantum numbers $J^{PC}$, denoting spin, parity and charge conjugation.

For $\theta_D=0$, lattice studies of pure $SU(3)$ and large-$N$ gauge theories\footnote{The large-$N$ limit considers $SU(N)$ gauge theories with many colors, allowing systematic expansions in powers of $1/N^2$ for quantities such as glueball masses.} show that the lightest state is typically the scalar $0^{++}$ glueball, followed by a tower of heavier states such as $0^{-+}$, $1^{+-}$ and $2^{++}$~\cite{Kribs:2016cew,Ochs:2013gi,Lucini:2010nv}. Its mass is set by the confinement scale,
\begin{equation}
m_{0^{++}} \equiv m_0 \sim {\cal O}(1)\,\Lambda_D,
\end{equation}
with the lattice-inspired pure-$SU(3)$ estimate $m_0 \simeq 6.7\,\Lambda_D$
obtained by rescaling pure-glue QCD results~\cite{Gockeler:2005rv}. More generally, in large-$N$ language one may write~\cite{Lucini:2010nv,Lucini:2001ej}
\begin{equation}
m_0 = \left(\alpha + \frac{\beta}{N^2}\right)\Lambda_D,
\end{equation}
with $\alpha$ and $\beta$ coefficients of order unity.

Thus, unlike dark-baryon or dark-meson models, the minimal glueball setup contains no constituent dark fermions, no dark flavor structure and no dark quark masses. Its low-energy parameters are mainly $\Lambda_D$, $N$ and, if relevant, $\theta_D$. If $\theta_D\neq0$, CP violation can mix the $0^{++}$ and $0^{-+}$ glueballs. In the following, we focus on the commonly considered $\theta_D=0$ case, where the lightest state is the scalar $0^{++}$ glueball~\cite{Soni:2016gzf}.

At energies well below confinement, the lightest glueball can be described by an effective scalar field $s$ representing the $0^{++}$ state. The effective Lagrangian contains
\begin{equation} \label{eq:eff_lagrangian_glueball}
\mathcal{L}_{\rm eff} \supset
-\frac12 m_0^2 s^2
-
\frac{\lambda_3}{3!} s^3
-
\frac{\lambda_4}{4!} s^4
-
\frac{\lambda_5}{5!} s^5
+\cdots,
\end{equation} where the couplings encode the underlying strong
dynamics. Combining large-$N$ counting with
naive dimensional analysis (NDA), the effective
couplings can be parametrized as~\cite{Cline:2022leq}
\begin{align}
\begin{aligned}
&\lambda_3 = a_3\frac{4\pi m_0}{N},
\qquad
\lambda_4 = a_4\left(\frac{4\pi}{N}\right)^2,
\qquad \\
&\lambda_5 = \frac{a_5}{m_0}
\left(\frac{4\pi}{N}\right)^3,
\end{aligned}
\end{align}
where $a_i=\mathcal{O}(1)$ are non-perturbative
coefficients. Without the NDA enhancement,
large-$N$ counting gives
$\sigma_{\rm tot}/m_0\sim1/(N^4m_0^3)$
~\cite{Soni:2016gzf}. Including the NDA
factors yields the estimate~\cite{Cline:2022leq}
\begin{equation}
\label{eq:dark_glueball_cross}
\frac{\sigma_{\rm tot}}{m_0}
\sim
\frac{12\pi^3}{N^4m_0^3}.
\end{equation}
The numerical prefactor is an NDA estimate
rather than an exact prediction, as the
self-scattering amplitude depends on the
underlying non-perturbative couplings. 

Requiring a self-interaction strength near
the Bullet Cluster bound,
$\sigma_{\rm tot}/m_0\sim
0.5~\mathrm{cm}^2/\mathrm{g}$, gives
\begin{equation}
m_{\rm DM}\equiv m_0
\sim130~\mathrm{MeV}
\left(\frac{3}{N}\right)^{4/3}.
\end{equation}
For a dark QCD-like $SU(3)_{\rm DC}$
gauge group, this corresponds to
$m_0\sim130~\mathrm{MeV}$ and
$\Lambda_D\sim20~\mathrm{MeV}$.

In the minimal glueball scenario, the interaction is short-range and approximately contact-like, so the cross section is nearly velocity independent. This makes pure glueball DM somewhat limited if one wants strong velocity dependence across astrophysical systems. Such velocity dependence can arise in extensions with additional dark-sector degrees of freedom. For example, adding a heavy adjoint fermion $X$ with $m_X\gg\Lambda_D$ can produce color-singlet glueballino states that interact through glueball exchange~\cite{Boddy:2014yra}. Their relic abundance can involve both perturbative annihilation, $X\bar X\to g_Dg_D$, and post-confinement annihilation of composite states~\cite{Contino:2018crt}. If glueballs decay into DM particles, the abundance can be further reduced. Remarkably, viable glueballino DM can occur for masses up to $10^6~{\rm GeV}$~\cite{Smirnov:2019ngs}, above the conventional perturbative unitarity bound~\cite{Griest:1989wd}.

At the level of particle-physics parameters, the minimal model can be summarized as
\begin{equation}
\begin{aligned}
\mathcal{P}_{\rm PP}=\{ \Lambda_D,\, N_D,\, \theta_D \}
&\;\rightarrow\; 
\{ m_0,\, \lambda_3,\, \lambda_4,\, \lambda_5,\ldots \} \\
&\;\rightarrow\;
\frac{\sigma_{\rm tot}}{m_{\rm DM}} \approx \text{constant}.
\end{aligned}
\end{equation}
Thus, minimal dark-glueball SIDM is best viewed as a composite realization of approximately contact-like SIDM. The resulting theory-to-simulation mapping and the corresponding \arepoTwo implementation are summarized in \cref{sec:mapping_composite_SIDM}.

\medskip
\noindent
\textbf{Representative benchmark:}
For $SU(3)_{\rm DC}$ dark-glueball SIDM, a representative benchmark with self-interactions relevant at the high relative velocities characteristic of merging galaxy clusters such as the Bullet Cluster is
\begin{equation}
m_{\rm DM}\sim130~\mathrm{MeV}, \qquad
\Lambda_D\sim20~\mathrm{MeV},
\end{equation}
corresponding to $\sigma/m_{\rm DM}\sim0.5~\mathrm{cm}^2/\mathrm{g}$. Different gauge groups and large-$N$ realizations extend this parameter space to other mass and interaction regimes. Further constraints on dark-glueball SIDM are summarized in App.~\ref{app:constraints_dark_glueballs}.

\subsubsection{Dark atoms and molecules}
\label{sec:dark_atoms}

Dark atoms constitute a composite class of SIDM models with rich elastic and dissipative dynamics. In these models, oppositely charged dark-sector constituents form neutral bound states through a new dark electromagnetic interaction. In parts of parameter space, these atoms can further bind into dark molecules, introducing additional scattering channels and cooling processes.

\medskip
\noindent
\textbf{Microscopic description:} In the simplest realization, the particle content consists of a dark proton $p_D$ with mass $m_{p_D}$, a dark electron $e_D$ with mass $m_{e_D}$ and a dark photon $\phi_\mu$, either massless or with mass $m_{\phi_\mu}$, all charged under a new gauge group $U(1)_D$~\cite{Kaplan:2009de,Kaplan:2011yj}. The low-energy Lagrangian may be written as
\begin{equation}
\begin{aligned}
\mathcal{L}_{\rm dark}
= & -\frac14 F'_{\mu\nu}F'^{\mu\nu}
+ \bar p_D(i\gamma^\mu D_\mu-m_{p_D})p_D \\ &
+ \bar e_D(i\gamma^\mu D_\mu-m_{e_D})e_D ,
\end{aligned}
\end{equation} where $F'_{\mu\nu}=\partial_\mu\phi_\nu-\partial_\nu\phi_\mu$ is the dark-photon field-strength tensor, $D_\mu=\partial_\mu+i g Q_D\phi_\mu$ is the covariant derivative, $Q_D$ denotes the dark charge of the corresponding particle, and $\alpha_D=g^2/(4\pi)$ is the dark fine-structure constant. The lightest neutral bound state $H_D=(p_D e_D)$ is the dark analogue of hydrogen. Its basic properties are controlled by
\begin{equation*}
\mu_{H_D}=\frac{m_{p_D}m_{e_D}}{m_{p_D}+m_{e_D}},
\quad
B_{H_D}=\frac12 \alpha_D^2\mu_{H_D},
\quad
a_{0,D}=\frac{1}{\alpha_D\mu_{H_D}},
\end{equation*}
where $\mu_{H_D}$ is the reduced mass, $B_{H_D}$ the atomic binding energy and $a_{0,D}$ the dark Bohr radius. The atomic mass is approximately
\begin{equation}
m_{H_D}\simeq m_{p_D}+m_{e_D}-B_{H_D}
\simeq m_{p_D}+m_{e_D},
\end{equation}
unless the theory is close to a strongly coupled regime. It is useful to define the mass ratio
\begin{equation}
R\equiv \frac{m_{p_D}}{m_{e_D}}\geq 1,
\end{equation}
which strongly affects the scattering spectrum and possible molecular dynamics~\cite{Cline:2013pca}. 
Requiring $m_{H_D}>0$ gives the formal bound
\begin{equation}
\alpha_D < \frac{\sqrt{2}(1+R)}{\sqrt{R}}.
\end{equation}
However, the non-relativistic approximation
underlying the binding-energy expression
breaks down well before this bound is reached.

The self-interactions of dark atoms, $H_D+H_D \to H_D+H_D$, are richer than those of point-like SIDM. Although the characteristic geometric scale is set by the Bohr radius,
\begin{equation}
\sigma_{\rm tot}\sim \pi a_{0,D}^2,
\end{equation}
the physical elastic and momentum-transfer cross sections can be much larger and exhibit nontrivial velocity dependence due to atomic potentials, spin channels, partial waves and resonances~\cite{Cline:2013pca}. Similar complexity is known from ordinary hydrogen scattering, where molecular potentials and quantum effects generate rich energy-dependent cross sections~\cite{KOLOS1974457,Silvera:1980zz,Wolniewicz1993}.

For structure formation, the relevant quantity is therefore not necessarily the total cross section, but an angle-weighted cross section such as $\sigma_{\rm T}$, $\sigma_{\rm \widetilde{T}}$ or $\sigma_{\rm V}$, as discussed in \cref{sec:relevant_cross_section}. For atom--atom scattering, these cross sections are obtained from non-relativistic potential scattering by solving the Schr\"odinger equation for the singlet and triplet molecular potentials $V_{s,t}(r)$. After decomposing the relative wave function into partial waves, the radial equation becomes
\begin{equation} \label{eq:dark_atom_SI_cross}
\begin{aligned} 
\Bigg[
\frac{d^2}{dr^2}
-\frac{\ell(\ell+1)}{r^2} 
+ f(R,\alpha_D)\,U_{s,t}(r)
\Bigg]\chi^{s,t}_{\ell}(r)=0,
\end{aligned}
\end{equation}
where $U_{s,t}(r)\equiv E-V_{s,t}(r)$, $\chi_\ell(r)=r\,R_\ell(r)$ and $\ell$ is the orbital angular momentum. The prefactor is
\begin{equation}
f(R,\alpha_D) 
\simeq R + 2 + R^{-1} - \frac{1}{2}\alpha_D^2.
\end{equation}
In practice, the small $\mathcal{O}(\alpha_D^2)$ correction can often be neglected, so that $f(R,\alpha_D)$ depends primarily on $R$. The potentials $V_{s}(r)$ and $V_{t}(r)$ denote the singlet and triplet Born--Oppenheimer molecular potentials between two dark atoms, corresponding to symmetric and antisymmetric electronic wave functions, respectively. They have been computed and fitted analytically in~\cite{KOLOS1974457,Silvera:1980zz,SILVERA1986139,Wolniewicz1993}; see~\cite{Cline:2013pca} for a convenient compilation. The phase shifts $\delta_\ell$ are obtained by numerically integrating \cref{eq:dark_atom_SI_cross} and matching to the asymptotic free-wave solution, following the procedure outlined after \cref{eq:dimensionless_radial}. The angle-weighted cross sections $\sigma_X(v_{\rm rel})$ are then constructed by summing over partial waves using \cref{eq:sigmaT_phaseshift,eq:sigmaV_phaseshift,eq:sigmaT_tilde_phaseshift_general}. The resulting cross sections exhibit resonances and threshold effects that can strongly affect halo structure.

\begin{figure*}
    \centering
    \includegraphics[width=1.1\columnwidth]{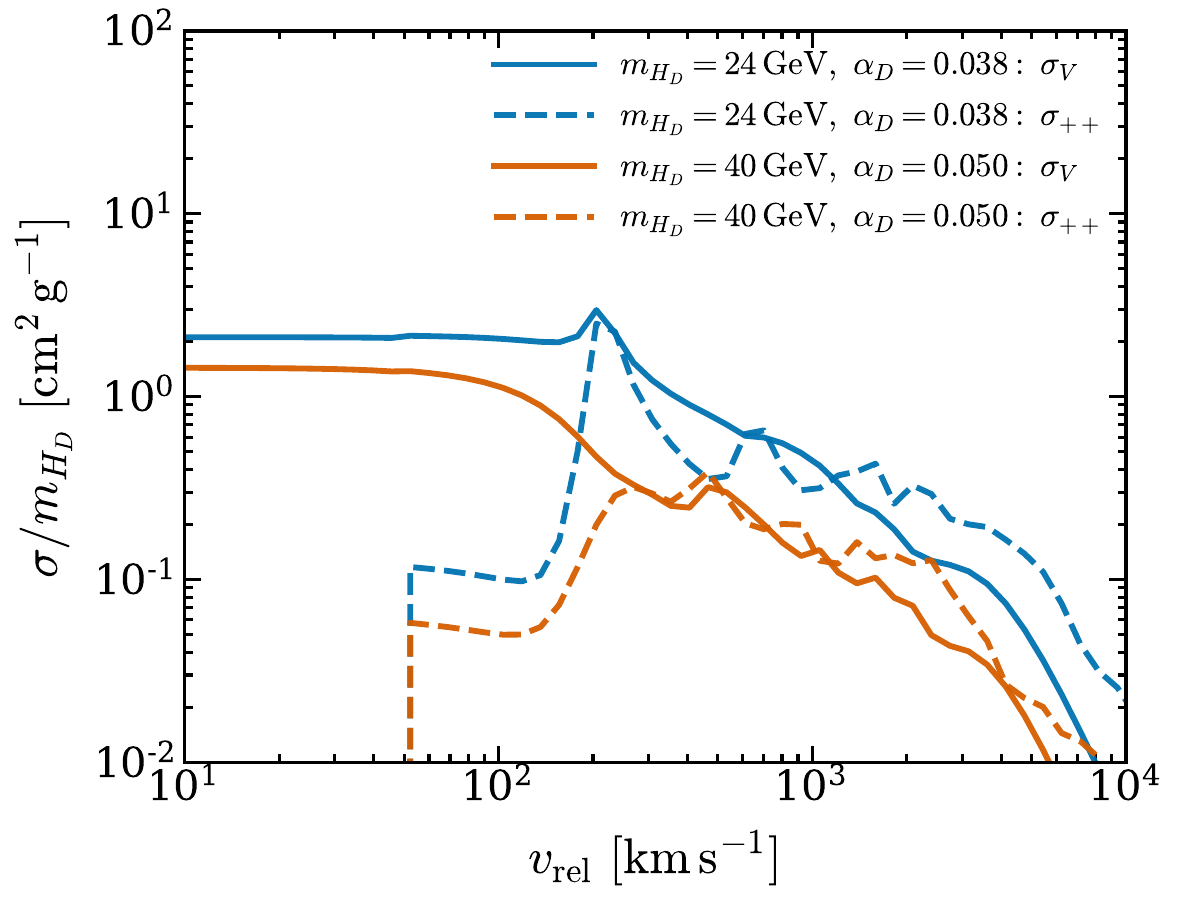}
    \caption{
    Velocity-dependent self-interaction cross sections for two atomic-DM benchmarks. Solid lines show the viscosity cross section $\sigma_V$ for elastic scattering, $H_D+H_D\to H_D+H_D$. Dashed lines show the inelastic cross section $\sigma_{++}$ for hyperfine up-scattering,
    $H_D+H_D\to H_D^\ast+H_D^\ast$, followed by the rapid decay $H_D^\ast \to H_D + \phi_\mu$. The dissipative channel has a kinematic threshold set by the hyperfine splitting, $v_{\rm rel}\geq v_{\rm th}=\sqrt{8E_{\rm hf}/m_{H_D}}$.
    }
    \label{fig:dark_atom_cross_sections}
\end{figure*}

Dark atoms can undergo a range of \emph{dissipative} processes, closely analogous to those of ordinary hydrogen. Dissipation occurs when kinetic or internal energy is converted into dark radiation, such as dark photons $\phi_\mu$, that escapes the halo. At low collision energies, an important cooling channel is hyperfine excitation followed by radiative de-excitation,
\begin{equation}
\begin{aligned}
H_D+H_D &\to H_D^\ast+H_D^\ast,\\
H_D^\ast &\to H_D+\phi_\mu ,
\end{aligned}
\end{equation}
where $H_D^\ast$ denotes a hyperfine excited state. For two identical atoms, single excitation, $H_D+H_D\to H_D^\ast+H_D$, is forbidden or strongly suppressed by angular-momentum selection rules, whereas double excitation is allowed. The corresponding dissipative cross section, denoted $\sigma_{++}(v_{\rm rel})$ and shown in \cref{fig:dark_atom_cross_sections}, therefore provides a dissipative scattering channel above the hyperfine excitation threshold~\cite{Boddy:2016bbu}.

For a given atomic model $(m_{H_D},\alpha_D,R)$, the characteristic atomic energy scale and hyperfine splitting are
\begin{equation}
E_0=\alpha_D^2\mu_{H_D},
\qquad
\mu_{H_D}\simeq m_{H_D}\frac{R}{(1+R)^2},
\end{equation}
and
\begin{equation}\label{eq:hyperfine_splitting_energy}
E_{\rm hf}
\simeq
\frac{8}{3}\,\alpha_D^2 E_0\,\frac{R}{(1+R)^2}.
\end{equation}
Since the dominant hyperfine process excites both incoming atoms, their subsequent radiative de-excitations remove approximately $2E_{\rm hf}$ from the scattering pair. Using the incoming centre-of-mass kinetic energy $E_{\rm kin}^{\rm in}=m_{H_D}v_{\rm rel}^2/4$ and the definition in \cref{eq:diss_energy_loss_fraction}, the corresponding energy-loss fraction is
\begin{equation}\label{eq:dark_atom_diss_energ_loss_fraction}
f_{\rm loss}(v_{\rm rel})
\simeq
\frac{8E_{\rm hf}}{m_{H_D}v_{\rm rel}^2} =
\left(\frac{v_{\rm th}}{v_{\rm rel}}\right)^2,
\qquad
v_{\rm rel}\geq v_{\rm th},
\end{equation}
where $v_{\rm th}=\sqrt{8E_{\rm hf}/m_{H_D}}$ is the threshold relative velocity for hyperfine excitation. The emitted dark photons are assumed to escape the halo. Thus, the microscopic parameters $(m_{H_D},\alpha_D,R)$ determine both the dissipative cross section $\sigma_{++}(v_{\rm rel})$ and the associated velocity-dependent energy-loss fraction $f_{\rm loss}(v_{\rm rel})$ required by the simulation.

At higher collision energies, electronic excitation can occur,
\begin{equation}
\begin{aligned}
H_D+H_D &\to H_D(n\geq2)+H_D,\\
H_D(n\geq2) &\to H_D+\phi_\mu ,
\end{aligned}
\label{eq:dark_electronic_excitation}
\end{equation}
removing an energy of order the atomic binding scale per event. In contrast to hyperfine excitation, single-atom electronic excitation generally dominates, while double excitation is kinematically and dynamically suppressed.

At still higher temperatures, collisional ionization can generate a partially ionized dark plasma,
\begin{equation}
H_D+H_D\to p_D+e_D+H_D,
\label{eq:collisional_ionization}
\end{equation}
after which the charged component can cool through processes such as
\begin{equation}
\begin{aligned}
p_D+e_D &\to H_D+\phi_\mu,\\
p_D+e_D &\to p_D+e_D+\phi_\mu ,
\end{aligned}
\label{eq:dark_atom_recombination}
\end{equation}
corresponding to recombination and bremsstrahlung, respectively. A residual ionized component can additionally exchange energy with the dark-radiation bath through processes such as Compton scattering~\cite{Cyr-Racine:2012tfp}.

Dark atoms may furthermore form molecular bound states, $H_{2,D}$, through radiative or three-body formation. The resulting molecular spectrum introduces additional dissipative channels, in particular rotational and vibrational excitation followed by radiative decay,
\begin{equation}
\begin{aligned}
H_{2,D}+H_{2,D} &\to H_{2,D}^{\ast}+H_{2,D}^{\ast},\\
H_{2,D}^{\ast} &\to H_{2,D}+\phi_\mu ,
\end{aligned}
\label{eq:dark_molecular_excitations}
\end{equation}
providing cooling mechanisms analogous to those of molecular hydrogen in the visible sector.

At the level of particle-physics parameters, the minimal atomic model can be summarized as
\begin{equation}
\begin{aligned}
\mathcal{P}_{\rm PP}
&=
\{m_{p_D},\,m_{e_D},\,\alpha_D,\,m_{\phi_\mu},\,\xi\}
\\
&\rightarrow
\{m_{H_D},\,R,\,\alpha_D,\,m_{\phi_\mu},\,\xi\},
\end{aligned}
\end{equation}
where the atomic self-interaction benchmarks are conveniently specified by the reduced set $\{m_{H_D},\alpha_D,R\}$ once a massless or sufficiently light dark photon and the dark-sector thermal history are fixed. These parameters determine the atomic spectrum and hence the elastic and dissipative scattering properties discussed above. The resulting theory-to-simulation mapping, including the elastic and dissipative simulation inputs and their corresponding \arepoTwo implementations, is summarized in \cref{sec:mapping_composite_SIDM}.

\medskip
\noindent
\textbf{Representative benchmarks:} Dark-atom self-interactions can be compatible with constraints~\cite{Cline:2013zca,Boddy:2016bbu}; a discussion of the cosmological, astrophysical and experimental viability of this parameter space is given in App.~\ref{app:constraints_dark_atoms}. Two representative atomic-DM benchmarks are
\begin{equation}\label{eq:dSIDM_benchmark_1}
m_{H_D}\simeq 24~{\rm GeV},
\qquad
\alpha_D\simeq 0.038,
\qquad
R\simeq 36,
\end{equation}
and
\begin{equation}\label{eq:dSIDM_benchmark_2}
m_{H_D}\simeq 40~{\rm GeV},
\qquad
\alpha_D\simeq 0.050,
\qquad
R\simeq 60,
\end{equation}
both with a massless or sufficiently light dark photon satisfying \cref{eq:dark_photon_constraint}. The corresponding elastic and dissipative cross sections are shown in \cref{fig:dark_atom_cross_sections}. At dwarf-galaxy velocities, the elastic viscosity cross section can reach $\sigma_V/m_{H_D}\sim\mathcal{O}(1),{\rm cm}^2/{\rm g}$ and decreases toward cluster velocities. The hyperfine dissipative cross section $\sigma_{++}$ is non-zero only above the threshold velocity $v_{\rm th}$, while the energy-loss fraction follows $f_{\rm loss}(v_{\rm rel})\propto v_{\rm rel}^{-2}$ as given in \cref{eq:dark_atom_diss_energ_loss_fraction}. Each benchmark therefore fixes both the velocity-dependent scattering rates and the energy removed per dissipative event.

\subsubsection{Theory-to-simulation mapping: Composite SIDM}
\label{sec:mapping_composite_SIDM}

The composite SIDM realizations discussed above span contact-like, resonant, multi-state and dissipative scattering. Despite their different microscopic origins, their particle-physics parameters map onto different subsets of the general simulation inputs introduced above, depending on the resulting low-energy scattering, internal-state structure and energy exchange.

\medskip
\noindent
\textbf{Dark-baryon SIDM:} As discussed in \cref{sec:dark_baryons}, the minimal realization contains a single elastic channel, $N_DN_D\to N_DN_D$, with the microscopic parameters $\mathcal{P}_{\rm PP}=\{\Lambda_{\rm dQCD},m_{q_D},N_D\}$ determining the dark-baryon spectrum and low-energy scattering parameters. The resulting mapping is
\begin{equation}
\mathcal{P}_{\rm PP}
\rightarrow
\mathcal{P}_{\rm SIM}
=
\left\{
\frac{1}{m_{N_D}}
\frac{{\rm d}\sigma}{{\rm d}\Omega}(\theta,v_{\rm rel})
\right\}.
\end{equation}
The differential cross section can be obtained from the effective-range description or from numerical calculations of the underlying composite dynamics. In the $s$-wave dominated regime, the scattering is approximately isotropic and can instead be represented by an appropriate angle-weighted cross section $\sigma_X(v_{\rm rel})/m_{N_D}$. Minimal dark-baryon SIDM can therefore be simulated with the velocity-dependent Monte Carlo scheme (\hyperref[sec:mc_velocity_dep]{MC}), while significant angular dependence can be treated with \hyperref[sec:anisotropic_MC]{aMC} or \hyperref[sec:asidm]{aSIDM}. Realizations with multiple nearly degenerate dark-baryon states and transitions between them instead map onto the multi-state framework of \cref{sec: inelastic_sidm_models} and can be treated with \hyperref[sec:imp_inelastic_SIDM]{iSIDM}.

\medskip
\noindent
\textbf{Dark-meson SIDM:} The dark-meson realizations described in \cref{sec:dark_mesons} can range from approximately contact-like scattering to strongly velocity-dependent resonant scattering. For approximately contact-like or $s$-wave dominated interactions, the mapping reduces to
\begin{equation}
\mathcal{P}_{\rm PP}
\rightarrow
\mathcal{P}_{\rm SIM}
=
\left\{
\frac{\sigma_{\rm tot}}{m_{\rm DM}}
\right\},
\end{equation}
whereas resonant dynamics or higher partial waves can generate velocity and angular dependence, in which case
\begin{equation}
\mathcal{P}_{\rm PP}
\rightarrow
\mathcal{P}_{\rm SIM}
=
\left\{
\frac{1}{m_{\rm DM}}
\frac{{\rm d}\sigma}{{\rm d}\Omega}(\theta,v_{\rm rel})
\right\}.
\end{equation}
The required cross sections may be obtained directly from the low-energy composite theory or through the effective-range or Breit--Wigner parametrizations discussed above. Approximately isotropic realizations can be simulated with \hyperref[sec:mc_velocity_dep]{MC}, while angular-dependent scattering can be treated with \hyperref[sec:anisotropic_MC]{aMC} or \hyperref[sec:asidm]{aSIDM}. Dark-meson realizations containing multiple states and transitions between them additionally map onto the \hyperref[sec:imp_inelastic_SIDM]{iSIDM} framework.

\medskip
\noindent
\textbf{Dark-glueball SIDM:} For the minimal dark-glueball realization of \cref{sec:dark_glueballs}, only the lightest $0^{++}$ glueball is dynamically relevant and the interaction is short-range and approximately velocity independent. The microscopic parameters $\mathcal{P}_{\rm PP}=\{\Lambda_D,N_D,\theta_D\}$ therefore map onto
\begin{equation}
\mathcal{P}_{\rm PP}
\rightarrow
\mathcal{P}_{\rm SIM}
=
\left\{
\frac{\sigma_{\rm tot}}{m_0}
\right\},
\end{equation}
with $\sigma_{\rm tot}/m_0$ given by the low-energy composite dynamics in \cref{eq:dark_glueball_cross}. The minimal realization can therefore be simulated with \hyperref[sec:mc_velocity_dep]{MC} using a constant cross section. More general glueball sectors containing additional long-lived states require the corresponding state populations, mass splittings and channel-dependent cross sections and can be treated with \hyperref[sec:imp_inelastic_SIDM]{iSIDM}. If unstable states decay on dynamically relevant timescales, the additional inputs $\Gamma$ and $v_{\rm kick}$ introduced in \cref{sec:decay_inputs} are required and can be evolved with \hyperref[sec:imp_decaying_DM]{dDM}.

\medskip
\noindent
\textbf{Dark atoms and molecules:} The atomic and molecular realizations discussed in \cref{sec:dark_atoms} combine elastic self-scattering with dissipative processes associated with internal excitations. For the benchmarks considered here, ordinary atom--atom scattering is accompanied by hyperfine excitation followed by prompt radiative de-excitation, which can be treated as a single dissipative channel because the emitted dark photons are assumed to escape the halo. The corresponding theory-to-simulation mapping is
\begin{equation}
\mathcal{P}_{\rm PP}
\rightarrow
\mathcal{P}_{\rm SIM}
=
\left\{
\frac{\sigma_{\rm el}(v_{\rm rel})}{m_{H_D}},\,
\frac{\sigma_{\rm diss}(v_{\rm rel})}{m_{H_D}},\,
f_{\rm loss}(v_{\rm rel})
\right\},
\end{equation}
where $\sigma_{\rm el}=\sigma_V$ and $\sigma_{\rm diss}=\sigma_{++}$ are shown in \cref{fig:dark_atom_cross_sections}, and $f_{\rm loss}(v_{\rm rel})$ is given in \cref{eq:dark_atom_diss_energ_loss_fraction}. The elastic channel can be simulated with \hyperref[sec:mc_velocity_dep]{MC}, while the dissipative channel can be treated with \hyperref[sec:imp_dissipative_SIDM]{vd-dSIDM}. More general atomic or molecular realizations may require explicit multi-state evolution, additional dissipative channels or decay inputs for long-lived excited states. Collisional ionization, recombination and the subsequent dynamics of a charged dark plasma, including Compton-like coupling to dark radiation, are not currently captured by the \arepoTwo implementations considered here.

For all of these composite realizations, standard CDM initial conditions apply when the dark-sector physics affects only late-time structure formation. If early-time dark-sector interactions modify the linear matter power spectrum, the corresponding non-CDM initial conditions must instead be supplied, as summarized in \cref{tab:pk_models}.

\begin{table*}[t]
\small
\setlength{\tabcolsep}{2pt}
\renewcommand{\arraystretch}{1.4}
\begin{tabular}{lcccccc}
\hline
\textbf{SIDM model class}
& \textbf{Reactions}
& \textbf{Cross sections}
& \textbf{Vel. dep.$\,$} 
& \textbf{Angular dep.$\,$}
& \textbf{Energy ex.}
& \textbf{Mod. ICs}\\

\hline

Contact-like (\ref{sec:contact-like})
& Single
& Eqs.~(\ref{eq:scalar_quartic_contact_cross_sections})-(\ref{eq:heavy_mediator_sigma0})
& No
& No
& Elastic
& -- \\

Minimal light-mediator (\ref{sec:light-mediator})
& Single
& Eqs.~(\ref{eq:Rutherford_general})--(\ref{eq:interference_coefficient}) / (\ref{eq:numerical_differential_cross_section})--(\ref{eq:asymptotic_match})
& Yes
& Yes
& Elastic (Dissipative)
& \hyperref[tab:pk_models]{(i)}--\hyperref[tab:pk_models]{(iii)} \\

Resonant (\ref{sec:Resonant_SIDM})
& Single
& Eqs.~(\ref{eq:numerical_differential_cross_section})--(\ref{eq:asymptotic_match}) /  (\ref{eq:eff_range_formalism_1})--(\ref{eq:eff_range_formalism_2})
& Yes (Strong)
& Yes (Strong)
& Elastic
& \hyperref[tab:pk_models]{(i)}--\hyperref[tab:pk_models]{(iii)} \\

Vector (\ref{sec:Vector_SIDM})
& Single
& Eqs.~(\ref{eq:numerical_differential_cross_section})--(\ref{eq:asymptotic_match})
& Yes
& Yes
& Elastic (Dissipative)
& \hyperref[tab:pk_models]{(i)}--\hyperref[tab:pk_models]{(iii)} \\

Asymmetric (\ref{sec:Asymmetric_SIDM})
& Single / multi
& Eqs.~(\ref{eq:numerical_differential_cross_section})--(\ref{eq:asymptotic_match})
& Yes
& Yes
& Elastic
& \hyperref[tab:pk_models]{(i)}--\hyperref[tab:pk_models]{(iii)} \\

Multi-state inelastic (\ref{sec: inelastic_sidm_models})
& Multi
& Eqs.~(\ref{eq:two_state_inelastic_cross})--(\ref{eq:coupled_channel_Schrodinger_eq})
& Yes
& Yes
& Inelastic (Decay)
& \hyperref[tab:pk_models]{(i)}--\hyperref[tab:pk_models]{(iii)} \\

Dark baryons (\ref{sec:dark_baryons})
& Single / multi
& Eqs.~(\ref{eq:dark_baryon_cross})--(\ref{eq:dark_baryon_cross_end})
& Yes
& Yes
& Elastic (Inelastic)
& \hyperref[tab:pk_models]{(iii)} \\

QCD-like dark mesons (\ref{sec:dark_mesons})
& Single / multi
& \cref{eq:s_wave_cross_QCD_dark_mesons}
& Yes
& Yes
& Elastic (Inelastic)
& \hyperref[tab:pk_models]{(iii)} \\

CH dark mesons (\ref{sec:dark_mesons})
& Single / multi
& \cref{eq:composite_Higgs_dark_meson_cross}
& No
& No
& Elastic (Inelastic)
& \hyperref[tab:pk_models]{(iii)} \\

Dark glueballs (\ref{sec:dark_glueballs})
& Single / multi
& \cref{eq:dark_glueball_cross}
& No
& No
& Elastic (Inelastic)
& \hyperref[tab:pk_models]{(iii)} \\

Dark atoms / molecules (\ref{sec:dark_atoms}) \ \
& Single / multi
& \cref{eq:dark_atom_SI_cross}
& Yes
& Yes
& Dissipative (Decay)
& \hyperref[tab:pk_models]{(iv)} \\
\hline

\\

\hline
\textbf{Non-SIDM model class}
& \textbf{Reactions}
& \textbf{Cross sections}
& \textbf{Vel. dep.$\,$} 
& \textbf{Angular dep.$\,$}
& \textbf{Energy ex.}
& \textbf{Mod. ICs} \\
\hline

Decaying DM
& Single / multi
& No
& No
& No
& Decay
& \hyperref[tab:pk_models]{(v)} \\

Warm DM 
& No
& No
& No
& No
& --
& \hyperref[tab:pk_models]{(vi)} \\

Fuzzy / axion-like DM 
& No
& No
& No
& No
& --
& \hyperref[tab:pk_models]{(vii)} \\
\hline
\end{tabular}

\vspace{0.3cm}
\centering
\caption{
Theory-to-simulation mapping for the DM model classes considered here, following \cref{sec:pp_to_simulation_inputs,fig:pp_to_simulation_inputs}. Columns list reaction structure, cross-section inputs, velocity and angular dependence, energy exchange, and possible initial-condition (IC) modifications. These are physical properties, mapped onto the \arepoTwo\ algorithms as follows: velocity-dependent scattering to \hyperref[sec:mc_velocity_dep]{MC}; anisotropic scattering to \hyperref[sec:anisotropic_MC]{aMC}, \hyperref[sec: fsidm]{fSIDM} or \hyperref[sec:asidm]{aSIDM}; multi-state inelastic transitions to \hyperref[sec:imp_inelastic_SIDM]{iSIDM}; dissipation to \hyperref[sec:imp_dissipative_SIDM]{dSIDM}; and decays to \hyperref[sec:imp_decaying_DM]{dDM}. In the energy-exchange column, the entry outside parentheses denotes the primary interaction type for each model class, while entries in parentheses indicate additional processes that may become relevant in particular realisations. The IC column indicates possible modifications to the initial matter power spectrum arising from the underlying particle physics; CDM-like ICs apply when these effects are negligible. Labels \hyperref[tab:pk_models]{(i)}--\hyperref[tab:pk_models]{(vii)} refer to the examples in \cref{tab:pk_models}. Our implementations support the listed scattering and decay processes, while full fuzzy-DM evolution requires an additional wave-dynamical solver~\cite{Schive:2014dra,Mocz:2017wlg,Mocz:2019pyf}. For dark atoms and molecules, elastic scattering, dissipative transitions and decays are included, but collisional ionisation, late-time recombination and charged-plasma dynamics are not.
}
\label{tab:sidm_model_overview}
\end{table*}

\subsection{Mapping model classes onto the \arepoTwo implementations}
\label{sec:implemented_in_arepo}

Table~\ref{tab:sidm_model_overview} brings together the theory-to-simulation mapping developed in Secs.~\ref{sec:initial_conditions} -- \ref{sec:decay_inputs} and illustrated in \cref{fig:pp_to_simulation_inputs}, linking the DM model classes reviewed in this chapter to the simulation inputs and \arepoTwo\ algorithms introduced in \cref{sec: implementation}. The representative benchmarks considered here are motivated by the microscopic models and their constraints, briefly summarised for each model class in App.~\ref{app:model_constraints}.

For elastic single-state SIDM, the required input is the scattering cross section, including its velocity and angular dependence. Isotropic scattering with constant or velocity-dependent cross sections is treated by the \hyperref[sec:mc_velocity_dep]{MC} scheme, while angular dependence is represented by \hyperref[sec:anisotropic_MC]{aMC} or, for strongly forward-peaked interactions, the \hyperref[sec: fsidm]{fSIDM} and \hyperref[sec:asidm]{aSIDM} schemes. Multi-state and inelastic models additionally require initial state fractions, mass splittings and reaction-dependent cross sections, with the \hyperref[sec:imp_inelastic_SIDM]{iSIDM} implementation accounting for the associated exchange between internal and kinetic energy. Dissipative channels require scattering cross sections and an energy-loss prescription, implemented through \hyperref[sec:imp_dissipative_SIDM]{dSIDM}, while \hyperref[sec:imp_decaying_DM]{dDM} requires decay rates and recoil kinematics. These ingredients can describe several particle-physics SIDM model classes, and a single model may require multiple ingredients.

The table also distinguishes late-time dynamics from modifications of the initial matter power spectrum, as can arise from free streaming in warm DM, wave effects in ultra-light bosonic DM, or coupling to dark radiation. Decaying DM additionally requires decay dynamics, while a full treatment of fuzzy DM requires a wave-dynamical solver~\cite{Schive:2014dra,Mocz:2017wlg,Mocz:2019pyf}. The IC labels \hyperref[tab:pk_models]{(i)}--\hyperref[tab:pk_models]{(vii)} refer to the representative modified-IC examples in \cref{tab:pk_models}, discussed in \cref{sec:initial_conditions}.

\subsection{Complementary diagnostics across DM model classes}
\label{sec:simulation_diagnostics_DM}

The mapping above provides a basis for using astrophysical systems as complementary laboratories for DM particle physics. Different model classes can produce similar signatures in one environment yet behave differently at other velocities, densities or dynamical states~\cite{Tulin:2013teo,Kaplinghat:2015aga,Boddy:2016bbu}. Testing the same underlying models across these regimes therefore provides a route to determining which interaction properties can be constrained and which remain degenerate.

Dwarf galaxies probe the low-velocity regime through core formation and gravothermal evolution~\cite{rocha2013cosmological,Balberg2002,Koda_2011}. Their density and velocity-dispersion profiles, core sizes and collapse times respond to heat transport~\cite{Outmezguine2023,Gad-Nasr:2023gvf} and to processes that inject or remove kinetic energy~\cite{vogelsberger2019evaporating,Essig2019}.
Contact-like SIDM (\cref{sec:contact-like}) provides the elastic, isotropic baseline, while light-mediator and resonant models (Secs.~\ref{sec:light-mediator} and \ref{sec:Resonant_SIDM}) introduce velocity and angular dependence that can modify heat transport and halo evolution~\cite{Tulin:2013teo,Outmezguine2023,Arido:2024hfk}. Inelastic transitions and dissipative radiative processes, including those in multi-state and dark atomic models (Secs.~\ref{sec: inelastic_sidm_models} and \ref{sec:dark_atoms}), can further modify the thermal evolution whenever their thresholds and state populations permit the relevant channels~\cite{Boddy:2016bbu,vogelsberger2019evaporating,Essig2019}.

Galaxies and galaxy groups connect dwarf and cluster scales, probing the intermediate-velocity behaviour of the same interactions~\cite{Kaplinghat:2015aga,Sagunski:2020spe}. Host-halo profiles and shapes, satellite orbits and mass loss, and the abundance and internal structure of subhaloes provide complementary diagnostics~\cite{Sagunski:2020spe,Vargya2022,Kummer_2018,Zeng_2022}. Satellites are particularly informative because their internal scattering samples lower velocities than their interactions with the host background, linking internal heat transport to interaction-driven evaporation and orbital evolution~\cite{Kummer_2018,Zeng_2022}. This sensitivity to multiple velocity scales makes galaxies and groups promising laboratories for distinguishing models with similar predictions at dwarf or cluster scales.

Galaxy clusters probe high-velocity scattering, while cluster mergers additionally expose its effects far from dynamical equilibrium~\cite{Sagunski:2020spe,Kahlhoefer_2014}. The preferred direction set by the collision axis makes mergers particularly sensitive to scattering anisotropy, alongside velocity dependence and energy exchange~\cite{Kahlhoefer_2014,Arido:2024hfk}. DM--galaxy offsets, evaporation, mass loss and remnant morphology trace the resulting redistribution of momentum and energy~\cite{Kahlhoefer_2014,Kim_2017b,Arido:2024hfk}. Their interpretation requires accounting for merger geometry, halo structure, viewing angle and the methods used to measure the observables~\cite{Robertson:2016xjh,Kim_2017b}.

Late-time decays and modified initial conditions affect structure formation in different ways. Decay-induced kicks (\cref{sec:decay_inputs}) can redistribute DM within an existing halo or allow daughter particles to escape, lowering its central density and bound mass. Low-mass haloes are especially sensitive because of their low escape velocities~\cite{Wang:2014ina}. Suppression or oscillations in the initial matter power spectrum instead change the density perturbations from which haloes form, altering their abundance and formation times~\cite{Schneider:2011yu,Cyr-Racine:2015ihg,Vogelsberger:2015gpr,Bohr:2021bdm}. Subhalo populations and strong-lensing perturbations probe the resulting halo abundance and structure~\cite{Lovell:2013ola,Gilman:2019nap}, while the Lyman-$\alpha$ forest probes small-scale matter clustering through absorption by intergalactic gas~\cite{Irsic:2017ixq,Irsic:2023equ,Bose:2018juc}. Halo density profiles and bound masses additionally trace the subsequent evolution driven by decays and self-interactions~\cite{Wang:2014ina,vogelsberger2019evaporating,Zeng_2022}.

Connecting these diagnostics to particle physics requires controlled simulations that isolate individual interaction properties, followed by cosmological and baryonic simulations that establish how their signatures change in realistic environments. Baryonic effects, including feedback, can substantially reshape the predicted halo structure~\cite{Sameie2021,Vargya2022}, while environmental evolution and numerical uncertainties affect the interpretation of core formation and collapse~\cite{Zeng_2022,Mace2024}. The cross-section constraints illustrated in \cref{fig:cross_sections} provide a useful reference, but depend on the assumed interaction physics, halo modelling and observational analysis~\cite{Kaplinghat:2015aga,Sagunski:2020spe,Robertson:2016xjh,vogelsberger2019evaporating}. A broader application of this framework could be to test whether a common microscopic model can explain observations across complementary astrophysical laboratories and to constrain the velocity dependence, angular structure and energy exchange of its interactions. The \arepoTwo\ implementations developed here provide the simulation framework needed to pursue this connection for a broad range of dark-sector scenarios.

\medskip
Having established the theory-to-simulation mapping and identified the simulation inputs for representative dark-sector model classes, we now turn to their numerical realizations in \arepoTwo and the validation of the resulting algorithms.

\section{\label{sec: implementation} Numerical implementation}

In this section, we describe the numerical implementation of the algorithms in the new \arepoTwo\ SIDM and decaying-DM physics modules. We first summarize the aspects of the \arepoTwo\ gravity solver and time integration that are relevant to these modules. We then introduce the new SIDM framework (\cref{sec:SIDM_module}) before presenting the implementations of isotropic (\cref{sec:imp_velocity_dep_SIDM}), anisotropic (\cref{sec:implementation_of_angular_SIDM}), multi-state inelastic (\cref{sec:imp_inelastic_SIDM}), and dissipative SIDM (\cref{sec:imp_dissipative_SIDM}), as well as decaying DM (\cref{sec:imp_decaying_DM}).

\subsection{\label{sec:arepo2_overview} The moving-mesh code \arepoTwo}

The massively parallel moving-mesh code \arepoTwo\ \citep{springel2010pur,pakmor2016improving,weinberger2020arepo} solves the Euler equations on a moving, unstructured Voronoi mesh using a second-order finite-volume method. The mesh-generating points move approximately with the local fluid velocity, yielding a quasi-Lagrangian and Galilean-invariant hydrodynamical scheme~\citep{vogelsberger2012moving}. In addition to solving the hydrodynamic equations for the gas, \arepoTwo\ solves the Poisson equation for the total matter distribution and evolves gas and collisionless particle species, including DM and stars, under their mutual gravity. Its modular code structure allows additional particle-physics processes to be implemented as largely self-contained physics modules.

Gravitational forces can be computed using a hierarchical octree, a particle--mesh (PM) solver, or their combination in the TreePM scheme~\citep{Barnes1986,Hockney1988nbody,Bagla2002,Bagla:2002kq,Bode2003}. In TreePM, long-range gravitational forces are evaluated using the PM solver, while short-range forces are calculated through the tree. Close gravitational interactions between simulation particles are regularized by gravitational softening, which suppresses artificial two-body gravitational scattering below the resolved spatial scale. For collisionless particle species, a fixed gravitational softening length $\epsilon_{\mathrm{soft}}$ is typically adopted.

The gravitational timestep of a collisionless particle is given by
\begin{equation}
    \Delta t_{\mathrm{grav},i} = \sqrt{\frac{2\,\eta\,\epsilon_{\mathrm{soft}}}{|\vec{a}_i|}},
    \label{eq:dt_grav}
\end{equation}
where $|\vec{a}_i|$ is the magnitude of the gravitational acceleration of simulation particle $i$, $\epsilon_{\mathrm{soft}}$ is the gravitational softening length and $\eta$ is a dimensionless accuracy parameter~\citep{weinberger2020arepo}. Additional physics modules can impose their own timestep constraints, and the timestep assigned to each simulation element is determined by the most restrictive of all applicable criteria. The SIDM timestep criteria introduced below therefore act as additional upper limits on the timesteps of interacting DM particles.

The particle equations of motion are integrated using a leapfrog kick--drift--kick scheme. By default, particles in \arepoTwo\ are advanced using individual, block-structured timesteps arranged in a power-of-two hierarchy~\citep{pakmor2016improving}. Alternatively, a single global timestep can be enforced, in which case all particles are advanced with the smallest timestep required by any of them. Individual timesteps allow particles in dynamically rapidly evolving regions to be integrated with short timesteps without requiring the entire particle population to advance at the same rate. At a given step, only particles in the currently active time bins are advanced, while particles assigned to longer timesteps remain inactive.

The gravitational integration employs the hierarchical scheme of~\cite{springel2021simulating}, in which the dynamics are separated into slow and fast subsystems at successive levels of the timestep hierarchy. Particles on shorter timesteps form the fast subsystem and are integrated more frequently while remaining coupled to particles on longer timesteps. A practical consequence is that the gravitational calculation can be restricted to the currently active particles, substantially reducing the computational cost when only a small fraction of the particle population occupies the shortest time bins.

Neighbour-based physics modules require particular care in combination with individual timesteps because an active particle can interact with neighbouring particles that are themselves inactive. The new \arepoTwo\ SIDM implementation introduced and validated in our companion work~\citep{Zier:2026} addresses this by decoupling the SIDM neighbour search from the gravity solver and constructing a dedicated DM-only neighbour tree. The neighbour search includes all relevant DM particles independently of their current activity state, while SIDM interactions are initiated for active target particles. This construction allows the SIDM module to remain compatible with the hierarchical gravity integration and provides the numerical framework for the interaction algorithms described below.

\subsection{\label{sec:SIDM_module} The new SIDM module in \arepoTwo}

The SIDM algorithms described below are implemented as physics extensions of the new \arepoTwo\ SIDM module introduced and validated in~\citep{Zier:2026}. That work presented a modern module for simulating SIDM in \arepoTwo, focusing on isotropic, elastic scattering with velocity-independent and velocity-dependent cross sections, while designing the framework to be straightforwardly extendable to more general interaction models.

Compared with the original \arepo\ SIDM implementation of \cite{Vogelsberger_2012}, the new module offers several advantages. At the heart of the design is a dedicated neighbour-search tree containing only DM particles, which decouples the SIDM solver from the gravity calculation and makes it compatible with the hierarchical gravity time-integration scheme of \arepoTwo~\citep{springel2021simulating}. The neighbour tree is used to identify the relevant DM neighbours of each active particle and to evaluate the corresponding pairwise scattering probability or effective interaction update.

The module also provides a parallel communication scheme that safely handles multiple scatterings per timestep. Accepted interactions are applied sequentially and self-consistently, even when a particle undergoes multiple interactions within a single timestep, including interactions between particles stored on different MPI tasks. This ensures that each interaction uses the most recent particle state and permits a less restrictive SIDM timestep criterion than schemes that require the total scattering probability per particle to remain small.

A further key feature is the extensible interaction interface. New two-body interactions can be implemented by specifying how particle states are read, how a pairwise interaction is evaluated, how the updated state is written back, and how repeated non-local updates are resolved. The neighbour search, communication, and parallelisation are handled by the module itself. For more details, read Ref.~\cite{Zier:2026}.

This infrastructure provides the basis for the physics extensions presented in the following sections: velocity-dependent Monte-Carlo (MC) scattering (\cref{sec:mc_velocity_dep}), implemented in Ref.~\cite{Zier:2026}; anisotropic MC scattering (\cref{sec:anisotropic_MC}); frequent small-angle SIDM scattering (\cref{sec: fsidm}); angular-hybrid SIDM (\cref{sec:asidm}); multi-state and inelastic SIDM (\cref{sec:imp_inelastic_SIDM}); constant and velocity-dependent dissipative SIDM (\cref{sec:imp_dissipative_SIDM}); and the separate decaying-DM module (\cref{sec:imp_decaying_DM}).

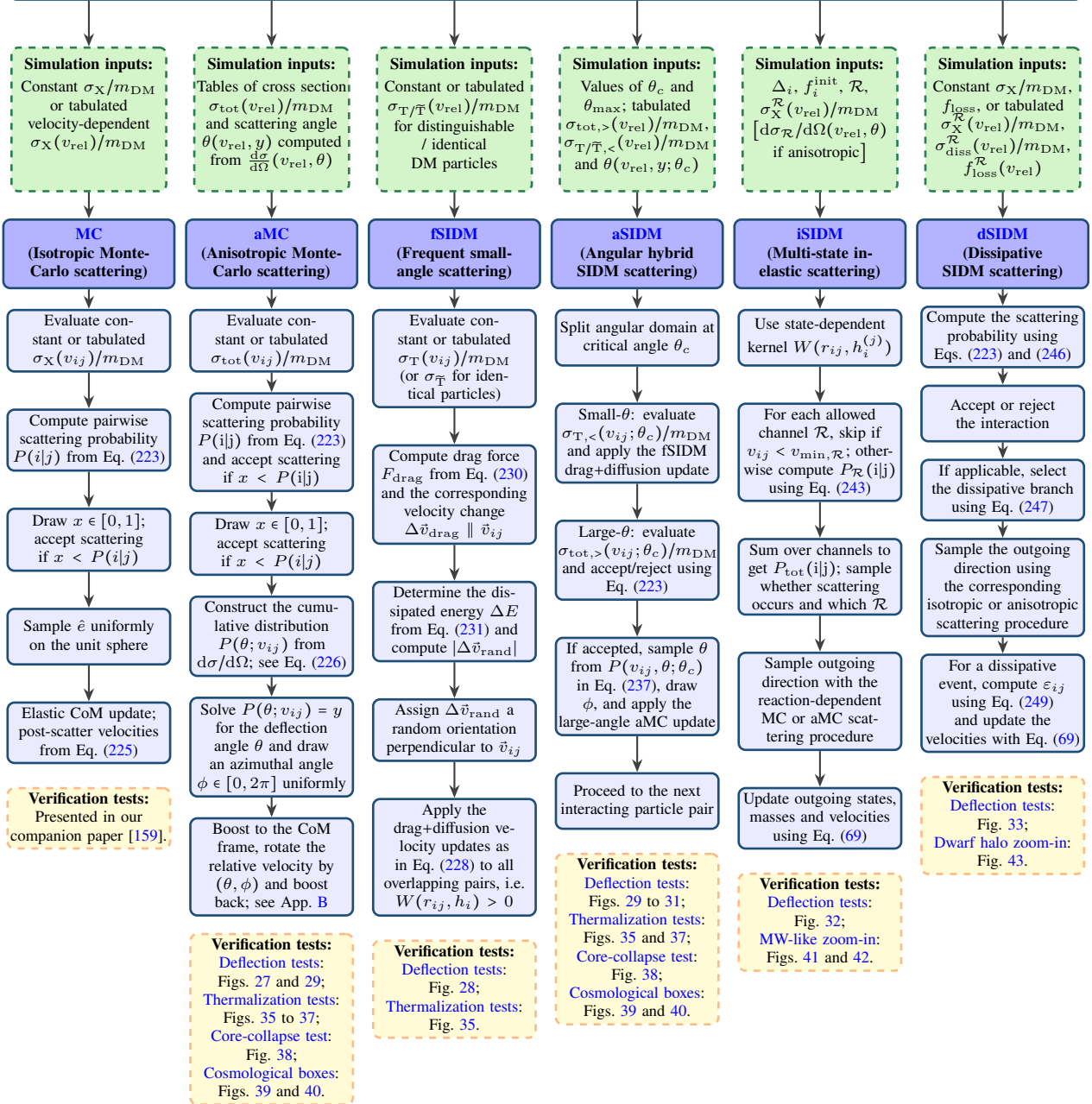
\begin{figure*}
\centering
\hspace*{-1.05cm}
\begin{tikzpicture}[
  >={Stealth[length=2mm, width=1.6mm]},
  arr/.style       ={->, thick, black!75},
  shared/.style    ={draw=darkblue, line width=1.0pt, rounded corners=3pt, fill=blue!8, align=left,
                     inner sep=6pt, font=\scriptsize},
  algotitle/.style ={draw=darkblue, line width=1.0pt, rounded corners=3pt, fill=blue!30, align=center,
                     text width=2.55cm, minimum height=1.05cm,
                     font=\scriptsize\bfseries, inner xsep=1pt, inner ysep=2pt},
  input/.style     ={draw=darkgreen, dashed, line width=1.0pt, rounded corners=3pt, fill={rgb,265:red,220; green,255; blue,210}, align=center,
                     text width=2.3cm, minimum height=2.2cm,
                     font=\scriptsize, inner xsep=1pt, 
                     inner ysep=2pt, text depth=1.55cm, text height=-0.4cm},
  algo/.style      ={draw=darkblue, line width=1.0pt, rounded corners=3pt, fill=blue!8,  align=center,
                     text width=2.42cm, minimum height=0.85cm,
                     font=\scriptsize, inner xsep=1pt, inner ysep=2pt},
  modhdr/.style    ={draw, thick, rounded corners=3pt, align=center,
                     font=\bfseries, minimum height=0.6cm, inner sep=4pt},
  verification/.style={draw=lightorange, dashed, line width=1.0pt, rounded corners=3pt, fill=yellow!20, align=center,
                     text width=2.42cm,
                     font=\scriptsize, inner xsep=1pt, inner ysep=2pt},
]

\node[modhdr, fill=blue!30, text width=16.5cm] (sidmHeader)
  {SIDM module \textnormal {\small (called after the second half-step kick)}};

\node[shared, below=0.15cm of sidmHeader, text width=16cm] (sharedSIDM) {\small
\textbf{Common steps used by all SIDM algorithms:}
\begin{enumerate}[leftmargin=*, labelsep=0.5em, itemsep=1pt, topsep=2pt]
    \item Refresh scale factor $a$ and Hubble rate $H(a)$; build the active DM target list.
    \item Local tree walk to find SIDM neighbours; export remote queries via MPI; bisect smoothing length $h_i$ until e.g. $N_{\rm ngb}\in [32{\pm}5]$.\\
    \emph{(Multi-state SIDM uses state-dependent smoothing lengths $h_i^{(j)}$, one per incoming-state population $j$).}
    \item For each candidate pair: compute relative velocity $v_{ij}=|\vec v_i-\vec v_j|$ and enforce the SIDM time-step if $\Delta t_{\mathrm{SIDM},i}< \Delta t_{\rm grav}$.
    \item Plan MPI peer ordering; execute primary (local) and secondary (remote) scatter phases; update tree bounds and release scratch buffers.
\end{enumerate}
};

\coordinate (branchBase) at ([yshift=-0.70cm]sharedSIDM.south);

\node[input, anchor=north] at ([xshift=-7.0cm]branchBase) (IMC)
  {\textbf{Simulation inputs:}\\[2pt]
   Constant $\sigma_{\rm X}/m_{\rm DM}$ or tabulated velocity-dependent $\sigma_{\rm X}(v_{\rm rel})/m_{\rm DM}$};

\node[input, anchor=north] at ([xshift=-4.2cm]branchBase) (IaMC)
  {\textbf{Simulation inputs:}\\[2pt]
Tables of cross section $\sigma_{\rm tot}(v_{\rm rel})/m_{\rm DM}$ and scattering angle $\theta(v_{\rm rel},y)$ computed from
$\frac{{\rm d}\sigma}{{\rm d}\Omega}(v_{\rm rel},\theta)$};

\node[input, anchor=north] at ([xshift=-1.4cm]branchBase) (IfS)
  {\textbf{Simulation inputs:}\\[2pt]
   Constant or tabulated $\sigma_{{\rm T}/\widetilde{{\rm T}}}(v_{\rm rel})/m_{\rm DM}$ for distinguishable \\ / identical DM particles};

\node[input, anchor=north] at ([xshift=1.4cm]branchBase) (IaS)
  {\textbf{Simulation inputs:}\\[2pt]
   Values of $\theta_c$ and $\theta_{\rm max}$; tabulated $\sigma_{\mathrm{tot},>}(v_{\rm rel})/m_{\rm DM}$, $\sigma_{{\rm T}/\widetilde{{\rm T}},<}(v_{\rm rel})/m_{\rm DM}$ and $\theta(v_{\rm rel},y;\theta_c)$};

\node[input, anchor=north] at ([xshift=4.2cm]branchBase) (IiS)
  {\textbf{Simulation inputs:}\\[2pt]
   $\Delta_i$, $f_i^{\rm init}$, $\mathcal R$, $\sigma_{\rm X}^{\mathcal R}(v_{\rm rel})/m_{\rm DM}$\\ $\big[{\rm d\sigma_{\mathcal R}}/{\rm d\Omega}(v_{\rm rel},\theta)$ $\text{if anisotropic}\big]$};

\node[input, anchor=north] at ([xshift=7.0cm]branchBase) (IdS)
  {\textbf{Simulation inputs:}\\[2pt]
   Constant $\sigma_{\rm X}/m_{\rm DM}$, $f_{\rm loss}$, or tabulated
   $\sigma_{\rm X}^{\mathcal R}(v_{\rm rel})/m_{\rm DM}$,
   $\sigma_{\rm diss}^{\mathcal R}(v_{\rm rel})/m_{\rm DM}$,
   $f_{\rm loss}^{\mathcal R}(v_{\rm rel})$};

\foreach \ii in {IMC, IaMC, IfS, IaS, IiS, IdS}{%
  \draw[arr] (sharedSIDM.south -| \ii.north) -- (\ii.north);
}

\node[algotitle, below=0.4cm of IMC] (TMC)
  {\hyperref[sec:mc_velocity_dep]{MC} \\ {\scriptsize (Isotropic Monte-Carlo scattering)}};

\node[algotitle, below=0.4cm of IaMC] (TaMC)
  {\hyperref[sec:anisotropic_MC]{aMC} \\ {\scriptsize (Anisotropic Monte-Carlo scattering)}};

\node[algotitle, below=0.4cm of IfS] (TfS)
  {\hyperref[sec: fsidm]{fSIDM} \\ {\scriptsize (Frequent small-angle scattering)}};

\node[algotitle, below=0.4cm of IaS] (TaS)
  {\hyperref[sec:asidm]{aSIDM} \\ {\scriptsize (Angular hybrid SIDM scattering)}};

\node[algotitle, below=0.4cm of IiS] (TiS)
  {\hyperref[sec:imp_inelastic_SIDM]{iSIDM} \\ {\scriptsize (Multi-state inelastic scattering)}};

\node[algotitle, below=0.4cm of IdS] (TdS)
  {\hyperref[sec:imp_dissipative_SIDM]{dSIDM} \\ {\scriptsize (Dissipative SIDM scattering)}};

\foreach \ii/\tt in {IMC/TMC, IaMC/TaMC, IfS/TfS, IaS/TaS,
                     IiS/TiS, IdS/TdS}{%
  \draw[arr] (\ii) -- (\tt);
}

\node[algo, below=0.30cm of TMC] (MC1) {Evaluate constant or tabulated $\sigma_{\rm X}(v_{ij})/m_{\rm DM}$};
\node[algo, below=0.55cm of MC1] (MC2) {Compute pairwise scattering probability $P(i|j)$ from Eq.~\eqref{eq:probabilityScatter_vdep_explicit}};
\node[algo, below=0.55cm of MC2] (MC3) {Draw $x\!\in\![0,1]$;\\ accept scattering if $x<P(i|j)$};
\node[algo, below=0.55cm of MC3] (MC4) {Sample $\hat e$ uniformly on the unit sphere};
\node[algo, below=0.55cm of MC4] (MC5)
  {Elastic CoM update;\\
   post-scatter velocities \\ from Eq.~\eqref{eq:isotropic_velocity_update_vdep}};

\node[algo, below=0.30cm of TaMC] (aMC1) {Evaluate constant or tabulated $\sigma_{\rm tot}(v_{ij})/m_{\rm DM}$};
\node[algo, below=0.3cm of aMC1] (aMC2) {Compute pairwise scattering probability $P({\rm i|j})$ from Eq.~\eqref{eq:probabilityScatter_vdep_explicit} and accept scattering if $x<P({\rm i|j})$};
\node[algo, below=0.3cm of aMC2] (aMC3) {Draw $x\!\in\![0,1]$;\\ accept scattering if $x<P(i|j)$};
\node[algo, below=0.3cm of aMC3] (aMC4) {Construct the cumulative distribution $P(\theta;v_{ij})$ from ${\rm d}\sigma/{\rm d}\Omega$; see \cref{eq:cdf_rsidm}};
\node[algo, below=0.3cm of aMC4] (aMC5) {Solve $P(\theta;v_{ij})=y$ for the deflection angle $\theta$ and draw an azimuthal angle $\phi\in[0,2\pi]$ uniformly};
\node[algo, below=0.3cm of aMC5] (aMC6) {Boost to the CoM frame, rotate the relative velocity by $(\theta,\phi)$ and boost back; see App.~\ref{app: anisotropic scattering calculation}};

\node[algo, below=0.30cm of TfS] (fS1) {Evaluate constant or tabulated $\sigma_{\rm T}(v_{ij})/m_{\rm DM}$ (or $\sigma_{\widetilde{\rm T}}$ for identical particles)};
\node[algo, below=0.55cm of fS1] (fS2)
  {Compute drag force $F_{\rm drag}$ from \cref{eq:drag} and the corresponding velocity change\\
   $\Delta\vec v_{\rm drag}\!\parallel\!\vec v_{ij}$};
\node[algo, below=0.55cm of fS2] (fS3) {Determine the dissipated energy $\Delta E$ from \cref{eq:energy_drag} and compute $|\Delta \vec{v}_\mathrm{rand}|$};
\node[algo, below=0.55cm of fS3] (fS4) {Assign $\Delta \vec{v}_\mathrm{rand}$ a random orientation perpendicular to $\vec{v}_{ij}$};
\node[algo, below=0.55cm of fS4] (fS5) {Apply the drag$+$diffusion velocity updates as in \cref{eq:velocity_updates_fsidm} to all overlapping pairs, i.e. $ W(r_{ij},h_i)>0$};

\node[algo, below=0.30cm of TaS] (aS1)
  {Split angular domain at\\ critical angle $\theta_c$};

\node[algo, below=0.55cm of aS1] (aS2)
  {Small-$\theta$: evaluate\\
   $\sigma_{\mathrm{T},<}(v_{ij};\theta_c)/m_{\mathrm{DM}}$\\
   and apply the fSIDM\\
   drag$+$diffusion update};

\node[algo, below=0.55cm of aS2] (aS3)
  {Large-$\theta$: evaluate\\
   $\sigma_{\mathrm{tot},>}(v_{ij};\theta_c)/m_{\mathrm{DM}}$\\
   and accept/reject using\\
   Eq.~\eqref{eq:probabilityScatter_vdep_explicit}};

\node[algo, below=0.55cm of aS3] (aS4)
  {If accepted, sample $\theta$ from
   $P(v_{ij},\theta;\theta_c)$ in \cref{eq:cdf_asidm},
   draw $\phi$, and apply
   the large-angle aMC update};

\node[algo, below=0.55cm of aS4] (aS5)
  {Proceed to the next\\
   interacting particle pair};

\node[algo, below=0.30cm of TiS] (iS1)
  {Use state-dependent kernel
   $W(r_{ij},h_i^{(j)})$};

\node[algo, below=0.55cm of iS1] (iS2)
  {For each allowed channel $\mathcal R$,
   skip if $v_{ij}<v_{\min,\mathcal R}$;
   otherwise compute $P_{\mathcal R}({\rm i|j})$
   using \cref{eq:imp_inelastic_probability}};

\node[algo, below=0.55cm of iS2] (iS3)
  {Sum over channels to get
   $P_{\rm tot}({\rm i|j})$; sample
   whether scattering occurs and which $\mathcal R$};

\node[algo, below=0.55cm of iS3] (iS4)
  {Sample outgoing direction with the reaction-dependent MC or aMC scattering procedure};

\node[algo, below=0.55cm of iS4] (iS5)
  {Update outgoing states, masses
   and velocities using \cref{eq:velocity_updates_energy_exchange}};

\node[algo, below=0.25cm of TdS] (dS1)
  {Compute the scattering probability using
   \cref{eq:probabilityScatter_vdep_explicit,eq:imp_vd_dsidm_combined_cross_section}};

\node[algo, below=0.25cm of dS1] (dS2)
  {Accept or reject the interaction};

\node[algo, below=0.25cm of dS2] (dS3)
  {If applicable, select the dissipative branch using
   \cref{eq:imp_vd_dsidm_branching_ratio}};

\node[algo, below=0.25cm of dS3] (dS4)
  {Sample the outgoing direction using the corresponding isotropic or anisotropic scattering procedure};

\node[algo, below=0.25cm of dS4] (dS5)
  {For a dissipative event, compute $\varepsilon_{ij}$ using
   \cref{eq:imp_dissipative_epsilon} and update the velocities with
   \cref{eq:velocity_updates_energy_exchange}};

\foreach \t/\first in {TMC/MC1, TaMC/aMC1, TfS/fS1, TaS/aS1,
                       TiS/iS1, TdS/dS1}{%
  \draw[arr] (\t) -- (\first);
}

\foreach \a/\b in {MC1/MC2, MC2/MC3, MC3/MC4, MC4/MC5,
                   aMC1/aMC2, aMC2/aMC3, aMC3/aMC4, aMC4/aMC5, aMC5/aMC6,
                   fS1/fS2,  fS2/fS3,  fS3/fS4,  fS4/fS5,
                   aS1/aS2,  aS2/aS3,  aS3/aS4,  aS4/aS5,
                   iS1/iS2,  iS2/iS3,  iS3/iS4,  iS4/iS5,
                   dS1/dS2,  dS2/dS3,  dS3/dS4,  dS4/dS5}{%
  \draw[arr] (\a) -- (\b);
}

\node[verification, below=0.35cm of MC5] (VMC)
  {\textbf{Verification tests:}\\
   Presented in our companion paper~\cite{Zier:2026}.};

\node[verification, below=0.25cm of aMC6] (VaMC)
  {\textbf{Verification tests:}\\
   \hyperref[sec: deflection tests]{Deflection tests}: \cref{fig:single_scatter,fig: Goudmit Saunderson check 100}; \hyperref[sec: thermalization tests]{Thermalization tests}: \\
\cref{fig:thermalization_anisotropic_iso_limit,fig:thermalization_table_lookup,fig:thermalization_vd_anisotropic}; \\
\hyperref[sec: core collapse tests]{Core-collapse test}: \\ \cref{fig:isolated_halo}; \\
\hyperref[sec:cosmological_DMO_simulations]{Cosmological boxes}: \cref{fig:subhalos_profiles_const_cross,fig:subhalos_profiles_velocity_cross}.};

\node[verification, below=0.35cm of fS5] (VfS)
  {\textbf{Verification tests:}\\
    \hyperref[sec: deflection tests]{Deflection tests}:\\ \cref{fig: Moliere vs fSIDM}; \\ \hyperref[sec: thermalization tests]{Thermalization tests}: \\ \cref{fig:thermalization_anisotropic_iso_limit}.};

\node[verification, below=0.35cm of aS5] (VaS)
  {\textbf{Verification tests:}\\
      \hyperref[sec: deflection tests]{Deflection tests}:\\ \cref{fig: Goudmit Saunderson check 100,fig:time_criterion_Rutherford,fig: cpu plot}; \hyperref[sec: thermalization tests]{Thermalization tests}: \\ \cref{fig:thermalization_anisotropic_iso_limit,fig:thermalization_vd_anisotropic}; 
       \\ \hyperref[sec: core collapse tests]{Core-collapse test}: \\ \cref{fig:isolated_halo}; \\
      \hyperref[sec:cosmological_DMO_simulations]{Cosmological boxes}: \\  \cref{fig:subhalos_profiles_const_cross,fig:subhalos_profiles_velocity_cross}.};

\node[verification, below=0.35cm of iS5] (ViS)
  {\textbf{Verification tests:}\\
      \hyperref[sec: deflection tests]{Deflection tests}: \\ \cref{fig:multi_sidm_deflection_test}; \\\hyperref[sec:MW_like_zoom_in_sims]{MW-like zoom-in}:  \cref{fig:main_and_subhalo_density_z0,fig:endothermic_state_fractions}.};

\node[verification, below=0.35cm of dS5] (VdS)
  {\textbf{Verification tests:}\\
      \hyperref[sec: deflection tests]{Deflection tests}: \\
      \cref{fig:vd_dsidm_energy_validation}; \\
      \hyperref[sec:dwarf_zoom_in_sims]{Dwarf halo zoom-in}:\\
      \cref{fig:dSIDM_test}.};

\begin{pgfonlayer}{background}
\node[rounded corners=6pt, inner sep=8pt,
      fit=(sidmHeader)(sharedSIDM)(IMC)(IdS)(TMC)(TdS)
          (VMC)(VaMC)(VfS)(VaS)(ViS)(VdS)] (sidmbox) {};
\end{pgfonlayer}

\end{tikzpicture}
\caption{Flowchart of the new SIDM module in \textsc{Arepo-2}. The SIDM module shares a common neighbour-search, smoothing-length, time-stepping and MPI-communication block (top), with the algorithm-specific branches below. Above each algorithm title, a dashed green box summarizes the simulation inputs required to run that algorithm, taken from the mapping between particle-physics parameters and simulation inputs in \cref{fig:pp_to_simulation_inputs}. Each branch then implements the common steps with its own scattering physics: MC (isotropic Monte Carlo), aMC (anisotropic Monte Carlo), fSIDM (frequent small-angle scattering), aSIDM (anisotropic hybrid), iSIDM (multi-state and inelastic), and dSIDM (dissipative SIDM, including both constant and velocity-dependent realizations). At the bottom of each algorithm column, yellow boxes indicate the verification tests performed.}
\label{fig:sidm_flowchart}
\end{figure*}

\subsection{\label{sec:imp_velocity_dep_SIDM} Implementation of velocity-dependent SIDM}

We now describe the implementation of elastic, isotropic SIDM with either a constant or velocity-dependent cross section in \arepoTwo. This section provides a short recap of the kernel choices (\cref{sec:kernel_velocity_dep}) and MC scattering scheme (\cref{sec:mc_velocity_dep}) implemented in \arepoTwo\ and validated with extensive tests in our recent work~\cite{Zier:2026}.

The velocity-independent MC scheme is implemented as a separate option in which the cross section per unit mass is a constant simulation input, $\sigma_{\rm X}/m_{\rm DM}$. The velocity-dependent extension replaces this constant by a function of the pairwise relative velocity, $\sigma_{\rm X}\rightarrow\sigma_{\rm X}(v_{\rm rel})$. Here, $\sigma_{\rm X}$ denotes an angle-weighted cross section as defined in \cref{eq:weighted_cross_sections}. Depending on the model and application, $\sigma_{\rm X}$ can be the total cross section $\sigma_{\rm tot}$, the transfer cross section $\sigma_{\rm T}$, the modified transfer cross section $\sigma_{\widetilde{\rm T}}$, or the viscosity cross section $\sigma_{\rm V}$, as defined and discussed in \cref{sec:relevant_cross_section}.

This captures SIDM models in which the effective scattering strength varies across astrophysical systems with different characteristic velocities, such as light-mediator, resonant, and composite scenarios with velocity-dependent low-energy scattering amplitudes (see the velocity-dependence column of \cref{tab:sidm_model_overview} for model classes that may require this velocity-dependent MC scattering algorithm). The angular structure of the differential cross section, ${\rm d}\sigma/{\rm d}\Omega$, is implemented separately in Sec.~\ref{sec:implementation_of_angular_SIDM}. Here, we focus on the isotropic case, where the particle-physics input is compressed into the effective angle-integrated cross section $\sigma_{\rm X}(v_{\rm rel})/m_{\rm DM}$.

\subsubsection{\label{sec:kernel_velocity_dep} Kernel estimator and local scattering rate}

A numerical scheme for SIDM scattering is derived from the Boltzmann equation,
\begin{equation}
    \dfrac{\mathrm{D}f(\vec{x}, \vec{v}, t)}{\mathrm{D}t}
    = \dfrac{\partial f}{\partial t}
    + \vec{v}\!\cdot\!\nabla_{\vec{x}} f
    - \nabla_{\vec{x}}\Phi \!\cdot\! \nabla_{\vec{v}} f
    = \Gamma[f, \sigma],
    \label{eq:CBE}
\end{equation} where $\mathrm{D}/\mathrm{D}t$ is the Lagrangian derivative, $f(\vec{x}, \vec{v}, t)$ the one-particle DM phase-space distribution function, $\Phi$ the gravitational potential and $\Gamma[f,\sigma]$ the collision operator encoding non-gravitational scattering. The left-hand side describes collisionless phase-space advection under gravity. 
For SIDM, the collision operator depends on the relative speed $v_{\rm rel}$ even for velocity-independent cross sections, while the differential cross section ${\rm d}\sigma/{\rm d}\Omega$ may itself depend on relative speed, scattering angle $\theta$, and the internal states and available scattering channels, including inelastic and dissipative processes.

Solving \cref{eq:CBE} on an Eulerian grid is impractical due to the six-dimensional phase space and highly inhomogeneous distribution. The $N$-body method instead evolves a coarse-grained sampling of the DM distribution~\citep{rocha2013cosmological,Jiang:2022aqw}, approximating it as \begin{equation}
    \hat{f}(\vec{x}, \vec{v}) =
    \dfrac{1}{\sum_i m_i}
    \sum_i M_i\,W(\vec{x}-\vec{x}_i, \epsilon_i)\,\delta(\vec{v}-\vec{v}_i),
    \label{eq:nbody}
\end{equation}
where $M_i$, $\vec{x}_i$ and $\vec{v}_i$ are the mass, position and velocity of the $i$-th simulation particle, respectively, while $W$ is the gravitational-softening kernel and $\epsilon_i$ the softening length.

This coarse-grained viewpoint is essential for any SIDM implementation. Microscopic DM scattering occurs between physical particles far below the resolution limit, so collisions cannot be modeled by literal pairwise scattering. Instead, SIDM schemes construct an effective prescription whose stochastic updates reproduce, in the continuum limit, the collision operator in \cref{eq:CBE}. The collision probability depends on the kernel weight, relative velocity, cross section, and simulation-particle mass. The velocity-space structure of \cref{eq:nbody} uses delta functions, neglecting unresolved internal velocity distributions. Numerical convergence requires that macroscopic observables -- core formation, heat transport, halo shapes -- become insensitive to resolution and kernel choice.

We model microscopic scattering by applying stochastic pairwise velocity updates to simulation particles at each timestep. The central challenge is computing the correct scattering rate while conserving momentum and energy, and recovering the continuum SIDM dynamics. Below we describe schemes implemented in \arepoTwo and review alternatives from the literature.

In Monte-Carlo (MC) schemes, the discretised collision operator is
\begin{equation}
    \dfrac{\mathrm{D}\hat{f}(\vec{x}, \vec{v}, t)}{\mathrm{D}t} = \Gamma[\hat{f}, \sigma] = \sum \hat{\Gamma}_{\rm ij}[\hat{f}, \sigma],
\end{equation}
where $\hat{\Gamma}_{\rm ij}$ is the numerical collision operator for pair $(i,j)$ and the sum runs over all DM particle pairs. From $\hat{\Gamma}_{\rm ij}$ we obtain the scattering probability; the scatter outcome is model-dependent and specified in later sections.

In the one-sided kernel formulation, we define the scattering-rate contribution obtained when particle $i$ evaluates neighbour $j$ as a target as
\begin{equation}
    \hat{\Gamma}({\rm i|j})
    =\frac{1}{2} \frac{\sigma_{\rm X}(v_{ij})}{m_{\rm DM}}\,
    v_{ij}\,
    M_j\,W_{ij},
    \label{eq:onesided_rate_vdep}
\end{equation}
where $v_{ij}\equiv|\vec{v}_i-\vec{v}_j|$. Here, $\hat{\Gamma}({\rm i|j})$ is a one-sided contribution and should be distinguished from the scattering rate $\hat{\Gamma}_{ij}$ associated with the particle pair. When both particles are active on the same timestep, the unordered pair $(i,j)$ is encountered through the two complementary evaluations $(i|j)$ and $(j|i)$. The factor $1/2$ therefore ensures the correct expected number of scatterings per pair for $P({\rm i|j})=\hat{\Gamma}({\rm i|j})\Delta t_i\ll1$, as enforced by the timestep criterion in \cref{sec: time-stepping criterion}.

In $N$-body codes, the phase-space distribution of DM is represented by numerical particles, each corresponding to a phase-space patch of many physical particles. A kernel is employed to smooth the matter distribution in configuration space and thereby estimate the local contribution entering $W_{ij}$ in \cref{eq:onesided_rate_vdep}. We use the standard cubic-spline kernel~\cite{Monaghan_Lattanzio_1985},
\begin{equation}
\label{eq:kernel}
W_{\rm CS}(r_{ij},h_i) = \frac{8}{\pi h_i^3}
\begin{cases}
1 - 6q^2 + 6q^3, & 0 \le q \le \frac{1}{2}, \\[2pt]
2(1-q)^3, & \frac{1}{2} < q \le 1, \\[2pt]
0, & q > 1,
\end{cases}
\end{equation}
where $q\equiv r_{ij}/h_i$. Here $r_{ij}\equiv|\vec{x}_i-\vec{x}_j|$ is the distance between particles $i$ and $j$. The kernel has compact support for $r_{ij} \le h_i$. The smoothing length $h_i$ is determined iteratively by requiring a target neighbour number $N_{\rm ngb}$, typically $N_{\rm ngb}=32\pm 5$. This choice is computationally efficient and was used in the original \arepoOne\ SIDM implementation~\cite{Vogelsberger_2012}.

Alternative kernel choices, including the top-hat and kernel-overlap formulations~\citep{Kochanek:2000pi,Robertson:2016xjh,rocha2013cosmological}, are also supported in \arepoTwo and are described and compared in detail in our companion work~\citep{Zier:2026}. In this work, we use the cubic-spline one-sided estimator throughout.

\subsubsection{\label{sec:mc_velocity_dep} Monte-Carlo scattering scheme (MC)}

The probability that particle $i$ scatters with neighbour $j$ during the timestep $\Delta t_i$ is given by $P({\rm i|j}) = \hat{\Gamma}({\rm i|j})\,\Delta t_i$, where $\hat{\Gamma}({\rm i|j})$ is the numerical scattering rate defined in \cref{eq:onesided_rate_vdep}. Explicitly,
\begin{equation}
    P({\rm i|j}) =
    \frac{1}{2}\frac{\sigma_{\rm X}(v_{ij})}{m_{\rm DM}}\,
    v_{ij}\,M_j\,W(r_{ij},h_i)\,\Delta t_i .
\label{eq:probabilityScatter_vdep_explicit}
\end{equation}
The factor $1/2$ avoids double counting the same physical particle pair through the complementary one-sided evaluations $(i|j)$ and $(j|i)$. For each candidate pair, a random number $x\in[0,1]$ is drawn and the scattering event is accepted if $x<P({\rm i|j})$. The timestep is restricted according to the general SIDM criterion in \cref{sec: time-stepping criterion}, with $(C,\sigma)=(C_{\rm MC},\sigma_{\rm X})$.

For an elastic isotropic scatter, the outgoing direction in the center-of-momentum (CoM) frame is sampled uniformly on the unit sphere. 
The CoM velocity is
\begin{equation}
    \vec{v}_{\rm CoM}
    =
    \frac{M_i\vec{v}_i+M_j\vec{v}_j}{M_i+M_j}.
\end{equation}
After an accepted scatter, the velocities are updated as
\begin{equation}
\begin{aligned}
    \vec{v}'_i  =& \vec{v}_{\rm CoM}
    +  \frac{M_j}{M_i+M_j}\,  v_{ij}\,\hat{e},
    \\
    \vec{v}'_j =& \vec{v}_{\rm CoM}
    -  \frac{M_i}{M_i+M_j}\,
    v_{ij}\,\hat{e},
    \label{eq:isotropic_velocity_update_vdep}
\end{aligned}
\end{equation}
where $\hat{e}$ is a randomly sampled unit vector.  This update conserves total momentum and kinetic energy exactly for elastic scattering. The only role of the velocity-dependent cross section is to modify the probability that the event occurs; once a scatter has been accepted, the isotropic elastic velocity update is identical to the constant-cross-section case.

The MC branch of the SIDM module is summarized in \cref{fig:sidm_flowchart}. 
After the common SIDM neighbour-search and pair-construction steps, the algorithm evaluates either a constant or velocity-dependent cross section $\sigma_{\rm X}(v_{ij})/m_{\rm DM}$, computes the pairwise scattering probability using Eq.~\eqref{eq:probabilityScatter_vdep_explicit}, and probabilistically accepts or rejects the event. 
For accepted events, the post-scatter velocities are assigned using the isotropic elastic CoM update in Eq.~\eqref{eq:isotropic_velocity_update_vdep}.

This MC algorithm therefore captures both constant and velocity-dependent SIDM scattering strengths across different astrophysical systems.

\subsubsection{\label{sec:cross_section_tables} Simulation inputs: velocity-dependent SIDM}

Velocity-dependent cross sections are supplied to \arepoTwo\ as tabulated functions of the relative velocity, $\sigma_{\rm X}(v_{\rm rel})/m_{\rm DM}$, over the velocity range relevant for the simulation. For single-state elastic SIDM, a single table is required, whereas models with multiple reaction channels $\mathcal{R}=\{ij\rightarrow kl\}$ require one table for each channel. The particle-physics model determines the corresponding velocity dependence; see \cref{tab:sidm_model_overview} for representative expressions. The simulation inputs required by each algorithm are summarized in the dashed green boxes above the corresponding branches in \cref{fig:sidm_flowchart}, following the mapping between particle-physics parameters and simulation inputs in \cref{fig:pp_to_simulation_inputs}.

\subsection{\label{sec:implementation_of_angular_SIDM} Implementation of angular SIDM}

In this section, we describe the implementation of elastic SIDM with anisotropic scattering in \arepoTwo. We extend the isotropic MC scattering framework described in \cref{sec:mc_velocity_dep} to account for the full angular dependence of the differential cross section ${\rm d}\sigma/{\rm d}\Omega$. This is essential for capturing SIDM particle-physics model classes with strong angular structure, especially resonant scattering in both elementary and composite SIDM models as described in \cref{sec:Resonant_SIDM} (see the angular dependence column of \cref{tab:sidm_model_overview} for model classes that may require anisotropic scattering algorithms). Anisotropic SIDM scattering has previously been implemented in several $N$-body codes, beginning with the implementation of~\cite{Robertson_2017b}; see~\cite{Fischer:2026lwx} for a recent overview of subsequent implementations. The present work provides such an implementation in \arepoTwo.

Anisotropic scattering introduces significant computational challenges when the scattering is strongly forward-dominated, as the total cross section $\sigma_{\rm tot}$ can greatly exceed the momentum-transfer cross sections $\sigma_{\rm T}$ and $\sigma_{\widetilde{\rm T}}$. To address this, we implement three complementary algorithms in \arepoTwo:
\begin{itemize}[leftmargin=*, labelsep=0.5em]
    \item The anisotropic Monte Carlo (aMC) scheme (\cref{sec:anisotropic_MC}) \citep{Robertson:2016qef}, which treats all scattering events as discrete collisions with deflection angles sampled from ${\rm d}\sigma/{\rm d}\Omega$;
    \item The frequent small-angle drag-diffusion scheme (fSIDM) (\cref{sec: fsidm}) \citep{Fischer_2021a}, which treats frequent forward scattering as a continuous drag-diffusion process;
    \item The anisotropic hybrid scheme (aSIDM) (\cref{sec:asidm}), developed by~\cite{Arido:2024hfk}, which combines aMC for rare large-angle scattering with fSIDM for frequent small-angle scattering, achieving computational efficiency without loss of physical accuracy.
\end{itemize}
In what follows, we describe each algorithm in detail.

\subsubsection{\label{sec:anisotropic_MC} Anisotropic Monte Carlo scattering scheme (aMC)}

The anisotropic Monte Carlo (aMC) scheme extends the MC scattering algorithm introduced in \cref{sec:mc_velocity_dep} by incorporating the full angular dependence of the microscopic interaction~\citep{Robertson:2016qef}. The outgoing direction in the centre-of-momentum frame is sampled from the angular distribution determined by the differential cross section ${\rm d}\sigma/{\rm d}\Omega$, rather than uniformly over the unit sphere. Explicit sampling of anisotropic deflection angles has previously been considered in SIDM simulations~\citep{Robertson_2017b,Banerjee_2020}.

The neighbour search, kernel estimator, and MC acceptance procedure are otherwise identical to the isotropic MC scheme of \cref{sec:mc_velocity_dep}. The scattering probability retains the form of \cref{eq:probabilityScatter_vdep_explicit}, with the replacement $\sigma_{\rm X}(v_{ij}) \rightarrow \sigma_{\rm tot}(v_{ij})$. Here $\sigma_{\rm tot}$ determines the overall scattering rate, while the differential cross section ${\rm d}\sigma/{\rm d}\Omega$ determines the angular distribution of accepted events. The timestep follows the general SIDM criterion in \cref{sec: time-stepping criterion}, with $(C,\sigma)=(C_{\rm aMC},\sigma_{\rm tot})$.

The corresponding cumulative probability distribution is~\citep{Robertson:2016qef}
\begin{equation}
\label{eq:cdf_rsidm}
\hspace*{-0.15cm}
P(<\theta;v_{ij})
=
\frac{1}{\sigma_{\rm tot}(v_{ij})}
\int_{0}^{\theta}
2\pi \sin\theta'\,
\frac{{\rm d}\sigma}{{\rm d}\Omega'}
(v_{ij},\theta')\,
{\rm d}\theta' .
\end{equation}
which gives the probability that the scattering angle is smaller than or equal to $\theta$. 

For \textit{distinguishable particles}, the angular domain is $\theta\in[0,\pi]$, and the normalization is chosen such that $P(\pi;v_{ij})=1$.
This is the appropriate angular domain for distinguishable-particle Rutherford scattering in the Born approximation, as in \cref{eq:Rutherford_general}, and for the corresponding numerical differential cross section in \cref{eq:differential_cross_distinguishable_general}. 

For \textit{identical particles}, final states related
by $\theta\rightarrow\pi-\theta$ are physically
indistinguishable. The differential cross sections
in \cref{eq:Moeller_general,eq:differential_cross_identical_bosons,eq:spin_averaged_identical_fermions}
already include the corresponding symmetry factor
$1/2$. We therefore sample the full angular domain
$\theta\in[0,\pi]$, using the same total cross
section $\sigma_{\rm tot}$ that enters the
scattering probability. The cumulative distribution
satisfies $P(\pi;v_{ij})=1$ and, by symmetry,
$P(\pi/2;v_{ij})=1/2$, consistent with the
single-scattering deflection test in
\cref{fig:single_scatter}.

The actual scattering angle is obtained by inverse-transform sampling.
We draw a second random number $y\in[0,1]$ and solve
\begin{equation}
    P(\theta;v_{ij}) = y
    \label{eq:inverse_sampling_theta}
\end{equation}
for $\theta$.
For analytic differential cross sections, this inversion can sometimes be performed analytically.
In the general case, and in particular for numerically tabulated differential cross sections, the cumulative distribution is tabulated, and the inverse relation is obtained by interpolation.

The scattering kinematics are evaluated in the CoM frame. We begin by boosting the incoming particle velocities into the CoM frame, where the relative velocity vector is rotated by the sampled scattering angle $\theta$ and an azimuthal angle $\phi \in [0,2\pi]$, drawn uniformly to ensure rotational symmetry around the initial relative velocity. After applying this rotation, the updated velocities are transformed back to the laboratory frame by reversing the initial boost. The detailed velocity update procedure is provided in App.~\ref{app: anisotropic scattering calculation}.

The aMC branch of the SIDM module is summarized in \cref{fig:sidm_flowchart}, including the evaluation of $\sigma_{\rm tot}(v_{ij})/m_{\rm DM}$, probabilistic event selection, angular sampling, and the subsequent CoM-frame velocity update.

The main drawback of the aMC scheme is that, for strongly forward-dominated scattering, $\sigma_{\rm tot}$ can be much larger than the momentum-transfer cross sections, $\sigma_{\rm T}$ or $\sigma_{\widetilde{\rm T}}$. Since the timestep criterion in \cref{eq:general_timestep} scales inversely with $\sigma_{\rm tot}$, this can make the pure aMC algorithm prohibitively expensive. This computational challenge motivates the anisotropic hybrid scattering algorithm introduced by~\cite{Arido:2024hfk}, which we have implemented in \arepoTwo\ as the aSIDM scheme described in \cref{sec:asidm}.

\subsubsection{\label{sec: fsidm} Frequent small-angle scattering scheme (fSIDM)}

The frequent small-angle scattering scheme (fSIDM) models the cumulative effect of frequent forward-dominated SIDM interactions through an effective drag--diffusion description~\citep{Fischer_2021a}. In contrast to MC and aMC scattering, the interaction update is applied to every pair of numerical particles within the interaction range, corresponding to a non-zero kernel weight, $W(r_{ij},h_i)>0$. The pairwise interaction consists of a deterministic drag term describing momentum transfer and a stochastic transverse contribution that restores the kinetic energy removed by the drag. The \arepoTwo\ SIDM module handles the parallelisation required when particles undergo multiple such interactions within a single timestep, as described in our companion work~\citep{Zier:2026}.

For a pair of numerical particles $i$ and $j$ with
masses $M_i$ and $M_j$, we formulate the velocity
update in terms of equal and opposite momentum
transfers, following
\citep{Fischer_2021a,Fischer:2026lwx}.
Defining the relative velocity as
$\vec{v}_{ij}=\vec{v}_i-\vec{v}_j$, with
$v_{ij}=|\vec{v}_{ij}|$, the post-interaction
velocities are
\begin{equation}
\begin{aligned}
\label{eq:velocity_updates_fsidm}
\vec{v}_i' &=
\vec{v}_i
-\frac{\Delta\vec{p}_{\rm drag}}{M_i}
+\frac{\Delta\vec{p}_{\rm rand}}{M_i},
\\
\vec{v}_j' &=
\vec{v}_j
+\frac{\Delta\vec{p}_{\rm drag}}{M_j}
-\frac{\Delta\vec{p}_{\rm rand}}{M_j}.
\end{aligned}
\end{equation}
Here, $\Delta\vec{p}_{\rm drag}$ is the
deterministic momentum transfer induced by the
effective drag interaction, while
$\Delta\vec{p}_{\rm rand}$ is a stochastic
transverse momentum transfer that restores the
kinetic energy removed by the drag.
The drag impulse is directed along the
relative velocity,
\begin{equation}
\label{eq:vdrag_time_int}
\Delta\vec{p}_{\rm drag}
=
F_{\rm drag}\Delta t_i
\frac{\vec{v}_{ij}}{v_{ij}},
\end{equation}
where $\Delta t_i$ is the timestep of particle $i$.
The equal and opposite momentum changes ensure
exact conservation of the total linear momentum,
independently of the particle mass ratio. The effective drag force~\citep{Fischer_2021a} is
\begin{equation}
\label{eq:drag}
    F_{\mathrm{drag}} = \frac{1}{2} v_{ij}^2 \frac{\sigma_{\mathrm{T}}(v_{ij})}{m_{\mathrm{DM}}} \frac{M_i M_j}{1+M_j/M_i} W(r_{ij},h_i)\,,
\end{equation} where $\sigma_{\mathrm{T}}/m_{\mathrm{DM}}$
is the transfer cross section for distinguishable
particles, as defined in
\cref{eq:weighted_cross_sections}.
For identical particles, it is replaced by
$\sigma_{\widetilde{\mathrm{T}}}$.
This normalization follows~\cite{Arido:2024hfk}
and is twice that adopted in~\cite{Fischer_2021a}
for the modified transfer cross section.

The kinetic energy removed by the drag interaction
is restored through an equal and opposite stochastic
momentum transfer perpendicular to the relative
velocity. Introducing the reduced mass of the
numerical particle pair,
$\mu_{ij}=M_iM_j/(M_i+M_j)$, the kinetic energy
removed by the drag impulse is
\begin{equation}
\label{eq:energy_drag}
\Delta E_{\rm drag}
=
v_{ij}|\Delta\vec{p}_{\rm drag}|
-\frac{|\Delta\vec{p}_{\rm drag}|^2}
{2\mu_{ij}}.
\end{equation}
Energy conservation then fixes the magnitude
of the stochastic transverse impulse,
\begin{equation}
\label{eq:velocity_rand}
|\Delta\vec{p}_{\rm rand}|
=
\sqrt{
|\Delta\vec{p}_{\rm drag}|
\left(
2\mu_{ij}v_{ij}
-|\Delta\vec{p}_{\rm drag}|
\right)
}.
\end{equation}
Its direction is chosen randomly in the plane
perpendicular to $\vec{v}_{ij}$. Applying the
resulting equal and opposite momentum changes
conserves both the total linear momentum and
kinetic energy of the interacting pair. In the equal-mass limit, $M_i=M_j$, these
expressions reduce to the standard fSIDM
velocity update. For unequal-mass scattering,
the numerical particle mass ratio must match
the corresponding physical DM particle mass
ratio for the pairwise scattering kinematics
to represent the microscopic interaction
consistently~\citep{Fischer:2026lwx}. The timestep follows the general SIDM criterion in \cref{sec: time-stepping criterion}, with $(C,\sigma)=(C_{\rm fSIDM},\sigma_{\rm T})$, or $\sigma_{\widetilde{\rm T}}$ for identical particles.

The fSIDM branch of the SIDM module is summarized in \cref{fig:sidm_flowchart}. For each overlapping pair, the algorithm evaluates the relevant transfer cross section, applies the drag contribution in \cref{eq:drag}, and restores the removed kinetic energy through the stochastic transverse contribution in \cref{eq:velocity_rand}, with the final velocities given by \cref{eq:velocity_updates_fsidm}.

\subsubsection{\label{sec:asidm} 
Anisotropic hybrid scattering scheme (aSIDM)} 

To efficiently treat strongly anisotropic scattering, we implement the anisotropic hybrid scattering algorithm (aSIDM) in \arepoTwo, developed by~\cite{Arido:2024hfk}. The scheme combines the rare large-angle Monte Carlo treatment of \cref{sec:anisotropic_MC} with the frequent small-angle drag-diffusion treatment of \cref{sec: fsidm}. The differential cross section is separated at a transition angle into rare and frequent components, which are evolved with aMC and fSIDM, respectively.

The key idea is to partition the angular domain into two regimes separated by a critical angle $\theta_c$:
\begin{itemize}[leftmargin=*, labelsep=0.5em]
    \item \textit{Small-angle scattering} ($\theta < \theta_c$): Treated using the fSIDM drag-diffusion approach. These scatterings are frequent but produce only small momentum transfers per event.
    \item \textit{Large-angle scattering} ($\theta \geq \theta_c$): Treated using the anisotropic MC sampling approach. These scatterings are rare but produce significant momentum transfers.
\end{itemize}
The critical angle $\theta_c$ is a technical parameter that separates the two regimes. Physical observables must be independent of $\theta_c$ within a reasonable range, which we verify in our tests in \cref{sec: verification tests}. In our tests in Chapter~\ref{sec: verification tests}, values $\theta_c \sim 0.1$--$0.3$ radians provide a suitable balance between accuracy and computational efficiency.

For each particle pair $(i,j)$ with non-zero kernel weight $W(r_{ij},h_i)>0$, the aSIDM algorithm applies the small- and large-angle contributions sequentially within the same timestep, as summarized in \cref{fig:sidm_flowchart}. The two components reuse the fSIDM and aMC updates described in \cref{sec: fsidm,sec:anisotropic_MC}, respectively, with the cross sections restricted to the corresponding angular sub-domains. Since both updates conserve momentum and energy, the combined pair update preserves these conservation properties before the algorithm proceeds to the next pair.

\textit{Small-angle contribution (fSIDM component):} We apply the fSIDM drag force and transverse kick exactly as in \cref{sec: fsidm}, but with the transfer cross section in \cref{eq:drag} restricted to angles below $\theta_c$,
\begin{equation}
    \sigma_{\mathrm{T}}(v_{ij}) \;\rightarrow\; \sigma_{\mathrm{T},<}(v_{ij};\theta_c),
\end{equation}
where the angle-restricted transfer cross section is defined as
\begin{equation}
    \sigma_{\mathrm{T},<}(v_{ij};\theta_c) 
    \equiv \int_{0}^{\theta_c} (1-\cos\theta)\,\frac{{\rm d}\sigma}{{\rm d}\Omega}(v_{ij},\theta)\, {\rm d}\Omega.
\end{equation}
For identical particles, $\sigma_{\mathrm{T},<}$ is replaced by the corresponding modified transfer cross section $\sigma_{\widetilde{\mathrm{T}},<}(\theta_c)$. The velocity updates then follow \cref{eq:velocity_updates_fsidm} unchanged.

\textit{Large-angle contribution (aMC component):} We use aMC sampling exactly as in \cref{sec:anisotropic_MC}, but restrict the angular domain to $\theta \geq \theta_c$. The scattering probability retains the form of \cref{eq:probabilityScatter_vdep_explicit}, now with the angle-restricted total cross section
\begin{equation}
    \sigma_{\mathrm{tot}}(v_{ij}) \;\rightarrow\; \sigma_{\mathrm{tot},>}(v_{ij};\theta_c),
\end{equation} where
\begin{equation}
    \sigma_{\mathrm{tot},>}(v_{ij};\theta_c)
    \equiv
    \int_{\theta_c}^{\theta_{\rm max}}
    \frac{{\rm d}\sigma}{{\rm d}\Omega}
    (v_{ij},\theta)\,{\rm d}\Omega,
\end{equation}
with $\theta_{\rm max}=\pi$ for distinguishable
particles (Rutherford scattering) and
$\theta_{\rm max}=\pi-\theta_c$ for identical
particles (M{\o}ller scattering). If an event is accepted, the deflection angle is sampled from the cumulative distribution of \cref{eq:cdf_rsidm} restricted to the large-angle domain,
\begin{equation}
    P(v_{ij},\theta;\theta_c)
    =
    \frac{1}{\sigma_{\rm tot,>}(v_{ij};\theta_c)}
    \int_{\theta_c}^{\theta}
    \frac{{\rm d}\sigma}{{\rm d}\Omega'}(v_{ij},\theta')
    \,
    {\rm d}\Omega',
    \label{eq:cdf_asidm}
\end{equation} The sampled angle satisfies
$\theta\in[\theta_c,\theta_{\max}]$,
with $P(v_{ij},\theta_{\max};\theta_c)=1$. The scattering angle is obtained by inverse-transform sampling as in \cref{eq:inverse_sampling_theta}, and the velocity update proceeds as described in \cref{sec:anisotropic_MC} and App.~\ref{app: anisotropic scattering calculation}. The aSIDM timestep combines separate constraints from the rare large-angle and frequent small-angle components, as described in \cref{sec: time-stepping criterion}.

The key advantage of aSIDM is that the computationally expensive quantity $\sigma_{\rm tot}$ is replaced by two much smaller cross sections: $\sigma_{\mathrm{T},<}(\theta_c)$ for the drag force and $\sigma_{\rm tot,>}(\theta_c)$ for the MC acceptance probability. In strongly forward-dominated models, $\sigma_{\rm tot,>}(\theta_c) \ll \sigma_{\rm tot}$, which dramatically relaxes the timestep constraint. Our tests in \cref{sec: deflection tests} show that the aSIDM algorithm achieves speed-ups of up to $\sim 100\times$ compared to the pure aMC scheme (see \cref{fig: cpu plot}), consistent with the findings of~\cite{Arido:2024hfk}. At the same time, several tests in \cref{sec: verification tests} show that the aSIDM algorithm correctly captures both the cumulative effect of frequent small-angle scattering through drag-diffusion and the discrete rare large-angle scattering events, in agreement with the fully anisotropic MC scheme in \cref{sec:anisotropic_MC}. Notably, we find this agreement even for relatively large values of $\theta_c$, which are favourable for a less restrictive timestep criterion.

\subsubsection{\label{sec:angular_cross_section_tables} Simulation inputs: angular SIDM} 

For angle-dependent SIDM, the required cross sections and angular distributions are supplied to \arepoTwo\ as tabulated functions of the relative velocity. The aMC scheme requires the total cross section $\sigma_{\rm tot}(v_{ij})/m_{\rm DM}$ together with the inverse cumulative angular distribution $\theta(v_{ij},y)$ used to sample the deflection angle. The fSIDM scheme requires the transfer cross section $\sigma_{\rm T}(v_{ij})/m_{\rm DM}$ for distinguishable particles, or the modified transfer cross section $\sigma_{\widetilde{\rm T}}(v_{ij})/m_{\rm DM}$ for identical particles. For the aSIDM scheme, the corresponding quantities are restricted by the critical angle $\theta_c$: the angle-restricted total cross section $\sigma_{\rm tot,>}(v_{ij};\theta_c)/m_{\rm DM}$ for rare large-angle scattering, the small-angle transfer cross section $\sigma_{\rm T,<}(v_{ij};\theta_c)/m_{\rm DM}$ (or $\sigma_{\widetilde{\rm T},<}(v_{ij};\theta_c)/m_{\rm DM}$ for identical particles), and the inverse cumulative angular distribution $\theta(v_{ij},y;\theta_c)$. The critical angle $\theta_c$ and maximum scattering angle $\theta_{\rm max}$ are specified separately, with $\theta_{\rm max}=\pi$ for distinguishable particles and $\theta_{\rm max}=\pi/2$ for identical particles.

For models with multiple reaction channels $\mathcal{R}=\{ij\rightarrow kl\}$, the corresponding cross-section and angular-distribution tables are supplied separately for each channel, analogous to the velocity-dependent inputs described in \cref{sec:cross_section_tables}. In addition, the mass splittings, $\Delta_i$, between the internal states and the initial state fractions, $f_i$, must be specified. The simulation inputs required by each algorithm are summarized in \cref{fig:sidm_flowchart}.

\subsection{\label{sec:imp_inelastic_SIDM} Implementation of multi-state and inelastic SIDM}

We now describe the implementation of multi-state and inelastic SIDM in \arepoTwo. This extends the elastic scattering schemes discussed in the previous sections by allowing DM particles to occupy different internal states and by supporting multiple reaction channels between them. Each reaction channel can be assigned its own cross section, while transitions that change the total rest mass of the interacting pair exchange energy between internal and kinetic degrees of freedom.

This implementation is particularly relevant for the multi-state inelastic SIDM model class described in \cref{sec: inelastic_sidm_models}, including models with dark-vector or dark-scalar mediators that permit transitions between internal DM states. Such models are characterized by multiple reaction channels (see the \textit{Reactions} column of \cref{tab:sidm_model_overview}) and energy exchange in inelastic transitions (see the \textit{Energy exchange} column). Other examples include composite DM scenarios with excited states, as discussed in \cref{sec:composite_SIDM}.

\subsubsection{\label{sec:imp_multi_states} Multi-state representation and state-dependent smoothing lengths}

We first describe how the internal-state structure is represented at the macroparticle level and how the neighbour search is generalized to obtain the appropriate local density for each incoming-state population.

\medskip
\noindent
\textbf{Internal states and reaction network.}
At initialization, each SIDM macroparticle is assigned an internal state according to the specified initial fractions $f_i^{\rm init}$, as described in \cref{sec:internal_state_structure}. The macroparticle mass of state $i$ is then calculated internally as
\begin{equation}
    M_i = (1+\Delta_i) M_0 ,
    \label{eq:imp_state_mass}
\end{equation}
where $\Delta_0=0$ for the ground state and the relative mass splitting $\Delta_i$ is determined from the corresponding particle-physics masses $m_i$ through \cref{eq:relative_mass_splitting}. For an interacting pair in states $i$ and $j$, the reaction network introduced in \cref{sec:reaction_network} specifies the available channels $\mathcal{R}=\{ij\rightarrow kl\}$.

\medskip
\noindent
\textbf{State-dependent smoothing lengths.}
In multi-state SIDM, the scattering rate for a given channel depends on the local density of particles in the corresponding incoming state. A smoothing length determined from all SIDM neighbours can therefore give a biased estimate when the state populations are unequal. For example, in a two-state system, a state-0 particle may scatter with either state-0 or state-1 particles. If one state is subdominant, a common smoothing length can reach the target neighbour number before enough particles of that state are enclosed, leading to an under-resolved kernel estimate of its local density and hence an incorrect scattering rate. State-dependent smoothing lengths are therefore required to resolve each interacting population independently.

To avoid this, \arepoTwo\ assigns a separate SIDM smoothing length to each relevant neighbour state. For an active particle in state $i$, interactions with particles in state $j$ are evaluated using
\begin{equation}
W(r_{ij},h_i)
\rightarrow
W(r_{ij},h_i^{(j)}),
\end{equation}
where $h_i^{(j)}$ is determined from the state-$j$ neighbour population. If state $j$ is sufficiently well sampled, $h_i^{(j)}$ is iterated independently until the enclosed number of state-$j$ neighbours lies within $N_{\rm ngb}\pm\Delta N_{\rm ngb}$. The resulting kernel therefore estimates the local density of the relevant incoming state rather than the total SIDM density. In the single-state limit, this reduces to the standard SIDM smoothing-length determination.

The state-dependent smoothing lengths $h_i^{(j)}$ characterize the local population of partner state $j$ and are therefore independent of the current internal state of particle $i$. If particle $i$ changes state during a timestep, subsequent reactions are evaluated using its updated state, while the state-dependent smoothing lengths remain those determined during the current neighbour-search calculation and are updated at the next search.

This procedure requires special treatment when an interacting state is absent or sparsely sampled. Enforcing $N_{\rm ngb}$ neighbours of a rare state would otherwise drive its state-specific smoothing length to increasingly large, and eventually non-local, scales. At each state-dependent neighbour-search calculation, we therefore determine the global population of every SIDM state. If no particles of the required partner state are available, the corresponding search is omitted. For a non-empty partner state $j$, we identify the state as sparsely sampled if fewer than $N_{\rm ngb}-\Delta N_{\rm ngb}$ partners exist globally or if
\begin{equation}
N_{\rm ngb}\frac{N_j^{\rm avail}}{N_{\rm ref}^{\rm avail}} < 1,
\label{eq:sparse_state_hsml}
\end{equation}
where $N_j^{\rm avail}$ is the available population of state $j$ and $N_{\rm ref}^{\rm avail}$ is that of the most abundant interacting state that is itself sufficiently sampled. The second condition corresponds to a global state abundance for which fewer than one state-$j$ particle is expected per $N_{\rm ngb}$ particles of the reference population. In this regime, $h_i^{(j)}$ is not enlarged further. Instead, we adopt the smoothing length $h_i^{(\rm ref)}$ of the well-sampled reference state as the local interaction scale and perform a single search for state-$j$ neighbours. This retains a local kernel estimator, at the expense of increased sampling noise for very rare states, rather than suppressing that noise by extending the interaction kernel to non-local scales. Once state $j$ becomes sufficiently populated, the algorithm automatically reverts to the independently iterated state-dependent smoothing length. Consistently, the predictor updates $h_i^{(j)}$ only when at least $N_{\rm ngb}-\Delta N_{\rm ngb}$ state-$j$ neighbours are available. 

This treatment differs from the multi-state implementation in \arepoOne~\cite{ONeil:2022szc}, where a common smoothing length is determined from all SIDM neighbours irrespective of their internal state. This provides a good approximation when one state strongly dominates the local DM population, since the total SIDM density then closely traces the density of the dominant state. This is the case for much of the endothermic and exothermic SIDM evolution considered in Ref.~\cite{ONeil:2022szc}, where the populations are strongly asymmetric, with approximately $f_0\simeq1$, $f_1\simeq0$ and $f_1\simeq1$, $f_0\simeq0$, respectively. The approximation becomes less accurate once both states contribute appreciably to the local density, because a common smoothing length no longer independently resolves the density of the specific incoming state entering each reaction channel. The \arepoTwo\ implementation removes this approximation by determining separate smoothing lengths for sufficiently sampled states, while reverting to a well-resolved local interaction scale when a state becomes too sparsely sampled to determine its own smoothing length robustly.

\subsubsection{\label{sec:imp_inelastic_scattering_scheme} Reaction-dependent scattering and inelastic kinematics}

The multi-state scattering algorithm generalizes the elastic MC scheme by allowing several reaction channels to compete for a given pair of incoming states. For each channel, the code determines the corresponding kinematic threshold, evaluates the reaction probability and, if a scattering event occurs, updates the internal states and particle velocities according to the selected reaction.

\medskip
\noindent
\textbf{Energy exchange and kinematic threshold.}
For each reaction $\mathcal{R}=\{ij\rightarrow kl\}$, the incoming and outgoing state masses are obtained from \cref{eq:imp_state_mass}. The corresponding change in internal-state energy is then calculated internally as
\begin{equation}
    \Delta E_{\rm int}
    =
    \left(M_i+M_j-M_k-M_l\right)c^2 .
    \label{eq:internal_state_energy_change}
\end{equation}
With this convention, $\Delta E_{\rm int}>0$ corresponds to an exothermic reaction that releases internal energy into kinetic energy, whereas $\Delta E_{\rm int}<0$ corresponds to an endothermic reaction that requires kinetic energy.

For endothermic channels, the incoming pair must have sufficient centre-of-momentum kinetic energy to produce the heavier final state. The corresponding threshold relative velocity is
\begin{equation}
    v_{\rm min,\mathcal{R}}
    =
    \sqrt{
    \frac{-2\Delta E_{\rm int}}{\mu_{\rm in}}
    },
    \qquad
    \mu_{\rm in}
    =
    \frac{M_i\,M_j}{M_i+M_j},
    \label{eq:imp_inelastic_vmin}
\end{equation}
for $\Delta E_{\rm int}<0$. Elastic and exothermic channels have $v_{\rm min,\mathcal{R}}=0$, while endothermic reactions with $v_{ij}<v_{\rm min,\mathcal{R}}$ are kinematically forbidden.

\medskip
\noindent
\textbf{Reaction-dependent scattering probability.}
For each candidate pair, consisting of an active particle in state $i$ and a neighbouring particle in state $j$, the algorithm identifies all reaction channels with incoming states $i+j$. For each kinematically allowed reaction $\mathcal{R}=\{ij\rightarrow kl\}$, the scattering probability is obtained by generalizing \cref{eq:probabilityScatter_vdep_explicit} to a reaction-dependent cross section and a state-dependent kernel estimate,
\begin{equation}
    P_{\mathcal{R}}({\rm i|j})
    =
    \frac{\sigma_{\rm X}^{\mathcal{R}}(v_{ij})}{m_0}
    \frac{v_{ij}}{2}
    \frac{1}{1+\Delta_j} M_j \, W(r_{ij},h_i^{(j)})\,
    \Delta t_i ,
    \label{eq:imp_inelastic_probability}
\end{equation}
where $m_0$ is the microscopic ground-state mass and $\sigma_{\rm X}^{\mathcal{R}}$ is the reaction-specific cross section entering the scattering rate. For $v_{ij}<v_{\rm min,\mathcal{R}}$, the corresponding probability is set to zero. The factor $M_j/(1+\Delta_j)$ provides the number-density normalization relative to the microscopic ground-state mass $m_0$, while $h_i^{(j)}$ is the state-dependent smoothing length defined above. The factor $1/2$ accounts for the pairwise interaction in the same way as in the elastic single-state schemes.

For angle-dependent scattering, \cref{eq:imp_inelastic_probability} retains the same form, with $\sigma_{\rm X}^{\mathcal{R}}$ replaced by the reaction-specific total cross section $\sigma_{\rm tot}^{\mathcal{R}}$. The differential cross section ${\rm d}\sigma_{\mathcal{R}}/{\rm d}\Omega$ instead determines the angular distribution of the outgoing particles after the reaction channel has been selected, following the aMC procedure described in \cref{sec:anisotropic_MC}.

\medskip
\noindent
\textbf{Reaction selection.}
The total scattering probability for a given incoming-state pair is
\begin{equation}
    P_{\rm tot}({\rm i|j})
    =
    \sum_{\mathcal{R}\in ij\rightarrow kl}
    P_{\mathcal{R}}({\rm i|j}) .
    \label{eq:imp_inelastic_total_probability}
\end{equation}
A random number $x\in[0,1]$ is drawn. If $x>P_{\rm tot}({\rm i|j})$, no scattering occurs. Otherwise, a scattering event is accepted and the reaction channel is selected from the cumulative distribution of the individual probabilities $P_{\mathcal{R}}({\rm i|j})$. This allows multiple elastic and inelastic channels to compete for the same incoming-state pair. The multi-state timestep accounts for the summed interaction probability over all allowed reaction channels and the state-dependent kernel estimator, as described in \cref{sec: time-stepping criterion}.

\medskip
\noindent
\textbf{Inelastic reaction update.}
Once a reaction $\mathcal{R}=\{ij\rightarrow kl\}$ has been accepted, the internal states of the interacting particles are updated from $(i,j)$ to $(k,l)$ and their macroparticle masses are set according to \cref{eq:imp_state_mass}. Energy conservation determines the change in the magnitude of the relative velocity through
\begin{equation}
    \varepsilon_{ij}
    =
    \sqrt{
    \frac{\mu_{\rm in}}{\mu_{\rm out}}
    +
    \frac{2\Delta E_{\rm int}}
    {\mu_{\rm out}v_{ij}^2}
    },
    \qquad
    \mu_{\rm out}
    =
    \frac{M_kM_l}{M_k+M_l},
    \label{eq:imp_inelastic_epsilon}
\end{equation}
where $\mu_{\rm in}$ is defined in \cref{eq:imp_inelastic_vmin}. Thus, $\varepsilon_{ij}=1$ for elastic scattering with unchanged masses, $\varepsilon_{ij}<1$ for endothermic reactions and $\varepsilon_{ij}>1$ for exothermic reactions.

The direction of the outgoing relative velocity is sampled according to the scattering model. For isotropic scattering, it is drawn uniformly on the unit sphere. For anisotropic reactions, the deflection angle is sampled from the reaction-dependent differential cross section using the aMC procedure described in \cref{sec:anisotropic_MC} and App.~\ref{app: anisotropic scattering calculation}.

Once the outgoing direction $\hat{\mathbf{e}}$ has been sampled, the particle velocities are updated using \cref{eq:velocity_updates_energy_exchange}, with $v_{ij}\equiv v_{\rm rel}=|\mathbf{v}_i-\mathbf{v}_j|$. The centre-of-momentum term accounts for the change in total macroparticle mass between the incoming and outgoing states, while the relative-motion term incorporates the energy exchange. This update conserves the total momentum of the interacting pair while accounting for the conversion between internal and kinetic energy.

The complete implementation flow, from the assignment of internal states to reaction selection and the subsequent state and velocity updates, is summarized in \cref{fig:sidm_flowchart}.

\subsubsection{\label{sec:inputs_inelastic} Simulation inputs: inelastic (multi-state) SIDM}

The multi-state and inelastic SIDM implementation requires the relative mass splittings $\Delta_i$, the initial state fractions $f_i^{\rm init}$ and the allowed reaction channels $\mathcal{R}=\{ij\rightarrow kl\}$. For each reaction channel, the corresponding reaction-specific cross section is supplied as a tabulated function of the relative velocity, $\sigma_{\rm X}^{\mathcal{R}}(v_{\rm rel})/m_0$. For anisotropic reactions, the corresponding angular information required by the chosen scattering algorithm is supplied separately for each channel, following the tabulation procedure described in \cref{sec:angular_cross_section_tables}. The state-dependent macroparticle masses $M_i$, internal-state energy changes $\Delta E_{\rm int}$, kinematic thresholds $v_{\rm min,\mathcal{R}}$ and post-scattering kinematics are calculated internally from these simulation inputs and are therefore not additional simulation inputs. 

\subsection{\label{sec:imp_dissipative_SIDM} Implementation of dissipative SIDM}

We extend the scattering framework described above to dissipative
self-interactions, in which part of the relative kinetic energy of an
interacting DM pair is transferred to degrees of freedom that are not
evolved explicitly, such as freely escaping dark radiation. The
corresponding energy-loss fraction and its mapping from microscopic
dark-sector models are defined in \cref{sec:dissipative_energy_loss}.
Here we describe how this energy loss is incorporated into the
\arepoTwo\ scattering algorithms.

The implementation supports both a constant dSIDM model, specified by
a velocity-independent cross section and a constant energy-loss
fraction $f_{\rm loss}$, and a general velocity-dependent formulation.
In the latter, conservative and dissipative contributions may coexist
within the same reaction channel and may be combined with the
multi-state and anisotropic scattering schemes introduced above. Such
channels occur, for example, in radiative light-mediator interactions
and dark-atomic scattering, as discussed in
\cref{sec:light-mediator,sec:Vector_SIDM,sec:dark_atoms}.

\subsubsection{\label{sec:imp_dissipative_model}
Dissipative event selection and kinematics}

\medskip
\noindent
\textbf{Event selection.}
For the constant dSIDM model, interactions are sampled using the
velocity-independent cross section $\sigma_{\rm X}/m_{\rm DM}$, and
every accepted interaction is dissipative. The event probability is
otherwise identical to that of the corresponding MC or aMC scheme
described in \cref{sec:mc_velocity_dep,sec:anisotropic_MC}.

More generally, consider a reaction
$\mathcal{R}=\{ij\rightarrow kl\}$ for which both a conservative and a
dissipative process contribute. At relative velocity $v_{ij}$, the
total rate is determined by the combined cross section
\begin{equation}
    \sigma_{\rm comb}^{\mathcal{R}}(v_{ij})
    =
    \sigma_{\rm X}^{\mathcal{R}}(v_{ij})
    +
    \sigma_{\rm diss}^{\mathcal{R}}(v_{ij}) .
    \label{eq:imp_vd_dsidm_combined_cross_section}
\end{equation}
Accordingly, $\sigma_{\rm comb}^{\mathcal{R}}$ replaces
$\sigma_{\rm X}^{\mathcal{R}}$ in the reaction probability of
\cref{eq:imp_inelastic_probability}; for a single-state model, the
analogous replacement is made in
\cref{eq:probabilityScatter_vdep_explicit}.

Conditional on an accepted interaction, the dissipative branch is
selected with probability
\begin{equation}
    P_{\rm diss}^{\mathcal{R}}(v_{ij})
    =
    \frac{\sigma_{\rm diss}^{\mathcal{R}}(v_{ij})}
         {\sigma_{\rm comb}^{\mathcal{R}}(v_{ij})}.
    \label{eq:imp_vd_dsidm_branching_ratio}
\end{equation}
A second random draw therefore selects between the conservative and
dissipative processes without modifying the total interaction rate. The corresponding timestep constraint is described in \cref{sec: time-stepping criterion}.

\medskip
\noindent
\textbf{Kinematic update.}
Once the dissipative branch has been selected, a fraction
$f_{\rm loss}^{\mathcal{R}}(v_{ij})$ of the incoming relative kinetic
energy is removed from the simulated DM pair. For the dissipative
reactions considered here, the internal states and particle masses are
unchanged, so that
\begin{equation}
    K_{\rm rel,out}
    =
    \left[1-f_{\rm loss}^{\mathcal{R}}(v_{ij})\right]
    K_{\rm rel,in}.
    \label{eq:imp_dissipative_energy_balance}
\end{equation}
The factor $\varepsilon_{ij}$ entering the velocity update in
\cref{eq:velocity_updates_energy_exchange} is therefore
\begin{equation}
    \varepsilon_{ij}
    =
    \sqrt{1-f_{\rm loss}^{\mathcal{R}}(v_{ij})}.
    \label{eq:imp_dissipative_epsilon}
\end{equation}
The outgoing relative velocity is thus reduced according to the
fraction of kinetic energy lost in the dissipative interaction. The
constant dSIDM limit is recovered by setting
$f_{\rm loss}^{\mathcal{R}}(v_{ij})=f_{\rm loss}$. After sampling the outgoing direction $\hat{\mathbf e}$, the particle
velocities are updated using
\cref{eq:velocity_updates_energy_exchange} with the value of
$\varepsilon_{ij}$ above. Physical dissipative reactions require
$0\leq f_{\rm diss}^{\mathcal{R}}(v_{ij})\leq1$.

For anisotropic interactions, the outgoing direction is sampled from
the angular distribution of the selected branch. In particular, a
dissipative channel may be assigned its own differential cross
section,
${\rm d}\sigma_{\rm diss}^{\mathcal{R}}/{\rm d}\Omega$, which is
sampled with the aMC procedure of \cref{sec:anisotropic_MC}. The same
event-selection and kinematic machinery therefore applies to
isotropic and anisotropic, and constant- or velocity-dependent,
dissipative interactions. The same kinematic framework can in principle be generalized to reactions that simultaneously change the internal states by including the corresponding internal-energy change $\Delta E_{\rm int}$.

\subsubsection{\label{sec:inputs_dissipative}
Simulation inputs: dissipative SIDM}

The constant dSIDM model requires the velocity-independent cross
section $\sigma_{\rm X}/m_{\rm DM}$ and the energy-loss fraction
$f_{\rm loss}$. For the velocity-dependent implementation, each
reaction channel $\mathcal{R}=\{ij\rightarrow kl\}$ is specified by
\begin{equation}
    \frac{\sigma_{\rm X}^{\mathcal{R}}(v_{\rm rel})}{m_{\rm DM}},
    \qquad
    \frac{\sigma_{\rm diss}^{\mathcal{R}}(v_{\rm rel})}{m_{\rm DM}},
    \qquad
    f_{\rm loss}^{\mathcal{R}}(v_{\rm rel}),
\end{equation}
supplied as tabulated functions of relative velocity. For anisotropic
dissipative scattering, the corresponding angular information is
provided following \cref{sec:angular_cross_section_tables}. The
branching probabilities, energy changes, and post-scattering
kinematics are then calculated internally during the simulation. The
complete dissipative branch is summarized in
\cref{fig:sidm_flowchart}.

\subsection{\label{sec:imp_decaying_DM} Implementation of dark-sector decays}

While decay is a one-body process rather than a DM self-interaction, SIDM models may contain unstable or excited dark-sector states whose decays modify the state populations and dynamics relevant for structure formation.

We implement dark-sector decays as stochastic single-particle processes in which a DM particle receives a recoil associated with the production of one non-relativistic daughter and one or more relativistic or otherwise dynamically unresolved decay products. Only the non-relativistic daughter is evolved explicitly; the remaining decay products are assumed to escape or free-stream without further interactions. The decay is therefore represented numerically by a stochastic transition followed by an isotropic recoil kick. The mapping from microscopic decay kinematics to the decay rate and recoil velocity used by the simulation is described in \cref{sec:decay_inputs}.

We implement two variants. The standalone scheme provides an effective recoil-kick model for a single DM state and can be evolved independently of SIDM. The state-based scheme is integrated with the multi-state SIDM framework and allows unstable internal states to decay to lower-mass states. This includes, for example, decays such as $\chi_2\rightarrow\chi_1+\gamma_{\rm D}$ in the two-state SIDM models of \cref{sec: inelastic_sidm_models}, as well as cascades through multiple unstable states such as KK-graviton towers~\cite{Law-Smith:2023czn,Obied:2023clp,Allahverdi:2023nov}. We restrict the implementation to decays with a single dynamically evolved, non-relativistic daughter particle and do not consider decays that produce multiple massive daughters.

\subsubsection{\label{sec:imp_decaying_DM_model} Decay probability and recoil update}

For a particle $i$ in internal state $a$, the state-based scheme assigns a decay rate $\Gamma_a$ and recoil velocity $v_{{\rm kick},a}$. The probability that the particle decays during its timestep $\Delta t_i$ is
\begin{equation}
    P_i^{\rm decay}
    =
    1-\exp\left(-\Gamma_a\Delta t_i\right).
    \label{eq:ddm_decay_probability}
\end{equation}
A uniform random number $x\in[0,1]$ is drawn and the decay is accepted if $x<P_i^{\rm decay}$. The recoil direction $\hat{\mathbf{n}}$ is then sampled isotropically in the parent-particle rest frame, and the particle velocity is updated as
\begin{equation}
    \vec{v}_i'
    =
    \vec{v}_i
    +
    v_{{\rm kick},a}\hat{\mathbf{n}} .
    \label{eq:ddm_velocity_update_code}
\end{equation}

In the standalone scheme, the same expressions apply with $\Gamma_a$ and $v_{{\rm kick},a}$ replaced by the global values $\Gamma$ and $v_{\rm kick}$. The particle mass and identity remain unchanged after an accepted event, and no flag records whether the particle has previously decayed. A numerical particle may therefore receive multiple recoil kicks, so this scheme should be interpreted as an effective stochastic recoil model rather than as the explicit evolution of a finite unstable-state population.

In the state-based scheme, stable states have $\Gamma_a=0$, with $\Gamma_0=v_{{\rm kick},0}=0$ for the ground state. Following an accepted decay, the particle transitions as $a\rightarrow a-1$ and its macroparticle mass is updated according to the multi-state prescription of \cref{sec:imp_multi_states}, with $M_a=(1+\Delta_a)M_0$. Initial states are drawn from the specified fractions $f_a^{\rm init}$, and a particle may therefore undergo successive decays until it reaches a stable state. Since decays are local single-particle operations, no neighbour search is required.

\subsubsection{\label{sec:inputs_decaying_DM} Simulation inputs: decaying dark matter}

The standalone scheme requires the decay rate $\Gamma$ and recoil velocity $v_{\rm kick}$. The state-based scheme instead specifies $\Gamma_a$ and $v_{{\rm kick},a}$ for each unstable internal state, together with the mass splittings $\Delta_a$ and initial state fractions $f_a^{\rm init}$. These decay quantities are determined by the underlying particle-physics model as discussed in \cref{sec:decay_inputs}; the decay probabilities, state transitions, mass updates, and recoil kicks are calculated internally.

\subsection{\label{sec: time-stepping criterion} Time-stepping criterion}

As described in \cref{sec:arepo2_overview}, \arepoTwo\ advances particles using individual hierarchical timesteps, with the timestep of each particle determined by the most restrictive of the applicable numerical and physical criteria. For DM particles subject to self-interactions and/or decays, the timestep is therefore set by
\begin{equation}
    \Delta t_i
    =
    \min\!\left(
    \Delta t_{\mathrm{grav},i},
    \Delta t_{\mathrm{SIDM},i},
    \Delta t_{\mathrm{DDM},i}
    \right),
    \label{eq:total_sidm_timestep}
\end{equation}
where $\Delta t_{\mathrm{grav},i}$ controls the temporal accuracy of the gravitational integration, $\Delta t_{\mathrm{SIDM},i}$ that of the self-interaction update, and $\Delta t_{\mathrm{DDM},i}$ that of the decaying DM (DDM) treatment. Only the criteria applicable to a given particle are imposed.

\medskip
\noindent
\textbf{Gravitational timestep.}
The gravitational timestep $\Delta t_{\mathrm{grav},i}$ is defined in \cref{eq:dt_grav}. Its dimensionless accuracy parameter $\eta$ controls the temporal resolution of the gravitational integration, with smaller values imposing shorter timesteps, particularly in regions of large gravitational acceleration. Although the default choice $\eta=0.025$ is sufficient for many cosmological structure-formation simulations, substantially smaller values can be required for accurate studies of late-time SIDM core collapse. Ref.~\cite{Mace2024} showed that, for isolated haloes evolved well beyond a Hubble time, $\eta=0.025$ does not yield a converged core-collapse time and that even $\eta=0.005$ can remain insufficient at high resolution; for some configurations, convergence is not fully established even at $\eta=0.001$, despite the substantially increased computational cost. The gravitational timestep parameter should therefore be treated as an explicit numerical convergence parameter in core-collapse simulations, alongside the mass and spatial resolution and the SIDM accuracy parameters introduced below.

\medskip
\noindent
\textbf{SIDM timestep.}
The SIDM timestep is chosen to keep the pairwise interaction update sufficiently small. For discrete scattering schemes, this requires the scattering probability of each candidate pair to remain small, while for drag--diffusion schemes the momentum transfer per pairwise update must remain a small fraction of the relative momentum.

For the single-state SIDM schemes, we use the general criterion
\begin{equation}
    \Delta t_{\mathrm{SIDM},i}
    =
    \frac{2\,C}
    {v_{ij}^{\rm max}\,M_j\,W(0,h_i)\,
    \left[\sigma(v_{ij})/m_{\mathrm{DM}}\right]^{\rm max}},
    \label{eq:general_timestep}
\end{equation}
where $C$ is a dimensionless accuracy parameter, $\sigma$ is the relevant cross section for the given scheme, $v_{ij}^{\rm max}$ is the maximum pairwise relative velocity, and $W(0,h_i)$ is the maximum kernel value. For MC (\cref{sec:imp_velocity_dep_SIDM}), aMC (\cref{sec:anisotropic_MC}) and fSIDM (\cref{sec: fsidm}), respectively,
\begin{equation}
    (C,\sigma)
    =
    (C_{\rm MC},\sigma_{\rm X}),
    \,
    (C_{\rm aMC},\sigma_{\rm tot}),
    \,
    (C_{\rm fSIDM},\sigma_{\rm T}),
\end{equation}
where $\sigma_{\rm T}$ is replaced by $\sigma_{\widetilde{\rm T}}$ for identical particles.

For aSIDM (\cref{sec:asidm}), the rare large-angle and frequent small-angle components impose separate constraints,
\begin{equation}
    \Delta t_{{\rm SIDM},i}
    =
    \min\!\left(
        \Delta t_{\mathrm{rare},i},
        \Delta t_{\mathrm{freq},i}
    \right),
    \label{eq:asidm_timestep}
\end{equation}
where $\Delta t_{\mathrm{rare},i}$ and $\Delta t_{\mathrm{freq},i}$ are each given by \cref{eq:general_timestep} with
$(C,\sigma)=(C_{\rm rare},\sigma_{\mathrm{tot},>})$ and
$(C,\sigma)=(C_{\rm freq},\sigma_{\mathrm{T},<})$, respectively. For identical particles, $\sigma_{\mathrm{T},<}$ is replaced by $\sigma_{\widetilde{\rm T},<}$.

For multi-state (inelastic) SIDM (\cref{sec:imp_inelastic_SIDM}), the presence of several reaction channels and state-dependent smoothing lengths requires the generalized criterion
\begin{equation}
    \Delta t_{\mathrm{SIDM},i}
    =
    \frac{2\,C_{\mathrm{iSIDM}}}
    {
        v_{ij}^{\rm max}
        \rho_{i,\rm eff}^{\rm max}
        \left[
        \sum_{\mathcal{R}\in ij\rightarrow kl}
        \sigma_{\rm X}^{\mathcal{R}}(v_{ij})/m_0
        \right]^{\rm max}
    },
    \label{eq:iSIDM_timestep}
\end{equation}
where
\begin{equation}
    \rho_{i,\rm eff}^{\rm max}
    =
    \max_j
    \left[
    \frac{M_j}{1+\Delta_j}
    W(0,h_i^{(j)})
    \right]
\end{equation}
is the maximum effective kernel-density factor entering \cref{eq:imp_inelastic_probability}. The maximum is taken over the relevant partner states, relative velocities and reaction channels. For angle-dependent scattering,
$\sigma_{\rm X}^{\mathcal R}=\sigma_{\rm tot}^{\mathcal R}$.
This criterion ensures that the summed interaction probability remains small for every candidate pair.

Constant and velocity-dependent dissipative SIDM use the corresponding single- or multi-state timestep criterion. When conservative and dissipative processes contribute to the same reaction channel, $\sigma_{\rm X}^{\mathcal R}$ is replaced by the combined cross section $\sigma_{\rm comb}^{\mathcal R}$ defined in \cref{eq:imp_vd_dsidm_combined_cross_section}.

We use accuracy parameters of $C=0.1$ for the SIDM schemes considered here, for which the numerical tests presented below are converged.

\medskip
\noindent
\textbf{DDM timestep.}
For stochastic dark-sector decays, an additional timestep constraint can be imposed to resolve recoil kicks and changes in internal states. Although the exponential decay probability in \cref{eq:ddm_decay_probability} exactly reproduces the survival probability for a constant decay rate, the finite timestep limits the temporal resolution of individual decay events. We therefore adopt
\begin{equation}
    \Delta t_{{\rm DDM},i}
    =
    \frac{C_{\rm DDM}}{\Gamma_a},
    \label{eq:ddm_timestep}
\end{equation}
where $C_{\rm DDM}$ is a dimensionless accuracy parameter and $\Gamma_a$ is the decay rate of the particle's current state. This limits the decay probability per timestep to $P_i^{\rm decay}\leq 1-e^{-C_{\rm DDM}}$. For standalone DDM, $\Gamma_a$ is replaced by the global decay rate $\Gamma$, while stable states impose no decay timestep constraint. The total timestep is determined by the minimum of the applicable gravitational, SIDM and DDM constraints.

\medskip
After describing the new SIDM module, the SIDM algorithms implemented within it, and the separate decaying-DM implementations, we now turn to their numerical verification.

\section{\label{sec: verification tests} Verification tests}

We verify the implementations introduced in \cref{sec: implementation} using controlled tests followed by isolated-halo and cosmological simulations. We begin with beam-deflection tests of aMC, fSIDM and aSIDM, comparing the scattering-angle distributions with analytical predictions. The same setup tests the state-dependent smoothing lengths
of the multi-state SIDM implementation and the energy loss in dSIDM and vd-dSIDM. A separate uniform-box test verifies the decay probabilities, state transitions and velocity kicks of the state-based dDM implementation. Thermalization tests then examine the approach to thermal equilibrium with aMC,
fSIDM and aSIDM, including the velocity-dependent aMC and aSIDM implementations. We next compare aMC and aSIDM during gravothermal core collapse in an isolated halo. Cosmological box simulations test MC, aMC and aSIDM with both velocity-independent and velocity-dependent cross sections. Finally, cosmological zoom-in simulations benchmark iSIDM and dSIDM against previously published results. \Cref{tab:test_overview} summarizes the tests, the algorithms examined and the
corresponding simulation setups.

\begin{table*}
\centering
\renewcommand{\arraystretch}{1.3}
\begin{tabular}{l c l l c c}
\hline\hline
\textbf{Physical process} & \textbf{Vel. dep.} & \textbf{Algorithm(s)} & \textbf{Test configuration} & $\textbf{N}_{\textbf{DM}}$ & \textbf{Figure(s)} \\
\hline

Anisotropic & No & \hyperref[sec:anisotropic_MC]{aMC} 
& Deflection (single scatter) 
& $80^3$ 
& \cref{fig:single_scatter} \\

Anisotropic & No & \hyperref[sec: fsidm]{fSIDM} 
& Deflection (multiple scatter) 
& $80^3$ 
& \cref{fig: Moliere vs fSIDM} \\

Anisotropic & No & \hyperref[sec:anisotropic_MC]{aMC}, \hyperref[sec:asidm]{aSIDM} 
& Deflection (multiple scatter) 
& $80^3$ 
& \cref{fig: Goudmit Saunderson check 100} \\

Anisotropic & No & \hyperref[sec:asidm]{aSIDM}
& Deflection ($t_{\rm min}$ criterion) 
& $80^3$
& \cref{fig:time_criterion_Rutherford} \\

Anisotropic & No & \hyperref[sec:asidm]{aSIDM} vs.\ \hyperref[sec:anisotropic_MC]{aMC} 
& Deflection (CPU cost) 
& $80^3$ 
& \cref{fig: cpu plot} \\

Multi-state & No & \hyperref[sec:imp_inelastic_SIDM]{multi-SIDM}
& Deflection (multiple scatter) 
& $80^3$
& \cref{fig:multi_sidm_deflection_test} \\

Dissipative & No & \hyperref[sec:imp_dissipative_SIDM]{dSIDM} 
& Deflection (single scatter) 
& $80^3$ 
& \cref{fig:vd_dsidm_energy_validation} (left) \\

Dissipative & Yes & \hyperref[sec:imp_dissipative_SIDM]{vd-dSIDM} 
& Deflection (single scatter) 
& $80^3$
& \cref{fig:vd_dsidm_energy_validation} (right) \\

Decay & No & \hyperref[sec:imp_decaying_DM]{dDM}
& Uniform box 
& $10^{6}$ 
& \cref{fig:state_based_dDM_validation} \\

Anisotropic & No & \hyperref[sec:anisotropic_MC]{aMC}, \hyperref[sec:asidm]{aSIDM}, \hyperref[sec: fsidm]{fSIDM} 
& Thermalization 
& $10^{4}$ 
& \cref{fig:thermalization_anisotropic_iso_limit} \\

Anisotropic & Yes & \hyperref[sec:anisotropic_MC]{aMC} 
& Thermalization (table lookup) 
& $10^{4}$ 
& \cref{fig:thermalization_table_lookup} \\

Anisotropic & Yes & \hyperref[sec:anisotropic_MC]{aMC}, \hyperref[sec:asidm]{aSIDM} 
& Thermalization 
& $10^{4}$ 
& \cref{fig:thermalization_vd_anisotropic} \\

Anisotropic & Yes & \hyperref[sec:anisotropic_MC]{aMC}, \hyperref[sec:asidm]{aSIDM}
& Isolated core collapse 
& $10^{6}$ 
& \cref{fig:isolated_halo} \\

Anisotropic & No & \hyperref[sec:mc_velocity_dep]{MC}, \hyperref[sec:anisotropic_MC]{aMC}, \hyperref[sec:asidm]{aSIDM} 
& Cosmological box (const.\ $\sigma$) 
& $512^{3}$ 
& \cref{fig:subhalos_profiles_const_cross} \\

Anisotropic & Yes & \hyperref[sec:mc_velocity_dep]{MC}, \hyperref[sec:anisotropic_MC]{aMC}, \hyperref[sec:asidm]{aSIDM} 
& Cosmological box (vel.-dep.\ $\sigma$) 
& $512^{3}$ 
& \cref{fig:subhalos_profiles_velocity_cross} \\

Inelastic & Yes & \hyperref[sec:imp_inelastic_SIDM]{iSIDM} 
& Zoom-in (MW-like galaxy) 
& 55,717,200 
& \cref{fig:main_and_subhalo_density_z0,fig:endothermic_state_fractions} \\

Dissipative & No & \hyperref[sec:imp_dissipative_SIDM]{dSIDM} 
& Zoom-in (dwarf galaxy) 
& 15,686,824 
& \cref{fig:dSIDM_test} \\
\hline\hline
\end{tabular}
\caption{\label{tab:test_overview}
Overview of the verification and benchmark simulations.
The velocity-dependence column indicates whether the interaction
inputs vary with relative velocity. For deflection tests,
$N_{\rm DM}$ denotes the number of beam particles, excluding
target particles and any additional non-interacting population.
For the isolated-halo test, the quoted count is
$N_{200}$, the number initially enclosed within $r_{200}$.
For cosmological boxes it is the total particle count, and for
zoom-in simulations it is the number of high-resolution DM
particles.
}
\end{table*}

\subsection{\label{sec: deflection tests} Deflection by a uniform lattice}

Following \cite{rocha2013cosmological,Robertson2017SIDM,Fischer_2021a},
we consider a beam-deflection test in which particles with a common
initial velocity scatter through a uniform target medium. The
distribution of beam-particle deflection angles, measured relative
to the initial beam axis, provides a direct test of the angular
scattering algorithms.

We first describe the setup in \cref{sec:test_setup} and the
analytical reference solutions in
\cref{sec:analytical_solutions_angular_distribution}.
We then test aMC, fSIDM and aSIDM in
\cref{sec:deflection_tests_angular_SIDM}, including the dependence
of the aSIDM results on the critical angle $\theta_{\rm c}$.
The same setup is used to test the state-dependent smoothing
lengths of the multi-state SIDM implementation in
\cref{sec:deflection_test_multi_state_SIDM}, and the energy loss
in dSIDM and vd-dSIDM in
\cref{sec:deflection_test_dSIDM_scattering_algorithms}.

\subsubsection{Test setup}
\label{sec:test_setup}

The initial conditions consist of $80^3 = 512{,}000$ target particles placed on a uniform Cartesian grid and initially at rest, together with $512{,}000$ beam particles that are randomly distributed within the same volume and all move along a common axis with an initial speed of $v = 2.5\,\kms$. Placing the target particles on a regular lattice enforces a genuinely constant target density and avoids artefacts (e.g.\ spurious low-density regions) that would arise from a random Poisson sampling of the target medium. Each particle, in both the beam and the target population, has the same mass $10^{5}\,\Msun$. Both populations uniformly fill a cubic volume of side length $25\,\kpc$, so that the target particles define a uniform target density
$\rho_{\rm target} = 3.28 \times 10^{6}\,\Msun\,\kpc^{-3}$.
Beam particles scatter off target particles according to the specified differential cross section $\mathrm{d}\sigma/\mathrm{d}\Omega$, with a cross section per unit DM mass fixed to $\sigma_\mathrm{T}/m_\mathrm{DM}$, while beam--beam and target--target scatterings are explicitly disabled and gravity is switched off for this test.

For comparisons with the analytical fixed-target solution, we use the no-recoil limit. The target particles remain at
rest, and the beam-particle velocity update is evaluated in the limit of infinite target inertia. For elastic scattering, this preserves the beam-particle speed and changes only its
direction. The finite target macroparticle masses continue
to determine $\rho_{\rm target}$ and the scattering rate; the infinite-inertia limit is used only in the scattering
kinematics. This test-specific update must be applied consistently to both the aMC and fSIDM components. It is distinct from an equal-mass scattering update in which the target velocity change is simply discarded. The main multiple-scattering comparison uses the Rutherford differential cross section, while the additional angular distributions tested below are specified individually.

\subsubsection{Analytical solutions for the angular distribution}
\label{sec:analytical_solutions_angular_distribution}

Following the multiple-scattering deflection-test
treatment of~\cite{Arido:2024hfk}, we consider
a homogeneous target in the no-recoil limit,
for which the beam-particle speed and scattering
rate remain constant. We define the angular distribution $f(t,\theta)$ such that $f(t,\theta)\sin\theta\,\mathrm{d}\theta$ is the fraction of beam particles with deflection angles between $\theta$ and $\theta+\mathrm{d}\theta$. The probability density per unit polar angle is therefore $p(t,\theta)=f(t,\theta)\sin\theta$. The corresponding Boltzmann equation admits the Goudsmit--Saunderson solution \cite{Goudsmit_1940a,Goudsmit_1940b}, written in the compact form of \cite{Bethe_1953} as
\begin{equation}
\label{eq:Goudsmit-Saunderson}
\begin{split}
    f(t,\theta) &= \sum_{\ell=0}^{\infty}\left(\ell+\frac12\right)P_\ell(\cos\theta)\\
    &\times\exp\!\left[-\frac{\rho v t}{m_{\rm DM}}\int\mathrm{d}\Omega'\,\frac{\mathrm{d}\sigma}{\mathrm{d}\Omega'}\left(1-P_\ell(\cos\theta')\right)\right].
\end{split}
\end{equation}
Here, $\rho\equiv\rho_{\rm target}$, $v$ is the constant beam-particle speed, and $P_\ell$ is the Legendre polynomial of order $\ell$. The angle $\theta'$ describes an individual scattering relative to the incoming direction, whereas $\theta$ is the accumulated deflection relative to the initial beam axis. The differential cross section is evaluated at the fixed relative speed $v$. The total opacity,
\begin{equation}
    \tau_{\rm tot}(t)=\rho v t\,\frac{\sigma_{\rm tot}}{m_{\rm DM}}=\frac{\rho v t}{m_{\rm DM}}\int\mathrm{d}\Omega\,\frac{\mathrm{d}\sigma}{\mathrm{d}\Omega},
\end{equation}
is the mean number of scatterings experienced by a beam particle by time $t$. For a finite total cross section, the unscattered fraction is $\exp[-\tau_{\rm tot}(t)]$. The exact angular distribution therefore retains a forward delta-function contribution with this weight at finite times. 

Several structural properties of Eq.~\eqref{eq:Goudsmit-Saunderson} are noteworthy. One obtains\footnote{Using the identities
$\int_{0}^{\pi} \mathrm{d}\theta \sin\theta P_\ell(\cos\theta) = 0$ for $\ell>0$ and $\int_{0}^{\pi} \mathrm{d}\theta \sin\theta P_0(\cos\theta) = 2$, one can show that the distribution function in Eq.~\eqref{eq:Goudsmit-Saunderson} remains normalized.}
\begin{equation}
  \int_{0}^{\pi}
  \mathrm{d}\theta\,\sin\theta\,
  f(t,\theta) =1,
\end{equation}
indicating that the distribution remains normalized at all times. Furthermore, the completeness relation
\begin{equation}
  \sum_{\ell=0}^{\infty}
  \left(\ell+\frac12\right)
  P_\ell(\cos\theta)
  =
  \delta(\cos\theta-1)
\end{equation} guarantees recovery of the initial forward-peaked configuration in the limit $t\rightarrow 0$. Finally, we have\footnote{Because $P_\ell(\cos\theta)\leq 1$, with strict inequality for $\ell>0$ and $\theta>0$, the multipole terms with $\ell>0$ decay exponentially at late times.}
\begin{equation}
  f(t\rightarrow\infty,\theta)
  \rightarrow
  \frac12,
\end{equation}
demonstrating the approach toward an isotropic angular distribution after many scatterings.

The time evolution thus interpolates between a sharply forward-directed initial state and an isotropic configuration, with the details of the intermediate regime determined by the angular structure of the differential cross section.

\paragraph{Isotropic scattering.}

If the cross section is angle-independent, all multipoles with $\ell>0$ decay with the same rate set by $\tau_{\rm tot}(t)$. The solution reduces to \begin{align}
  f(t,\theta)\big|_{\rm iso}
  &=
  \frac12
  +
  \sum_{\ell>0}
  \left(\ell+\frac12\right)
  P_\ell(\cos\theta)\,
  e^{-\tau_{\rm tot}(t)}
  \nonumber\\
  &=
  \frac12\!\left(1-e^{-\tau_{\rm tot}(t)}\right)
  +
  e^{-\tau_{\rm tot}(t)}
  \delta (\cos\theta-1).
\end{align} The distribution can therefore be interpreted as a mixture of an isotropic component and a forward $\delta$-function contribution, whose relative weights are controlled by the total opacity.

\paragraph{Strongly forward-peaked scattering.}

For interactions dominated by small deflection angles, as in the fSIDM scenario, the Legendre polynomials can be expanded as
\begin{equation}
  1 - P_\ell(\cos\theta)
  \simeq
  \frac12\,\ell(\ell+1)\,(1-\cos\theta),
\end{equation} which leads to \begin{equation}
\begin{aligned}
  f(t,\theta)\big|_{\rm forward}
  =
  \sum_{\ell=0}^{\infty} &
  \left(\ell+\frac12\right)
  P_\ell(\cos\theta)\,
  \\ & \exp\!\left[
    -\frac12\,\ell(\ell+1)\,
    \tau_{\rm T}(t)
  \right].
\end{aligned}
\end{equation}
In this regime, higher multipoles are suppressed increasingly strongly, and the evolution is dictated by the transfer opacity,\begin{equation} \label{eq:transfer_opacity}
  \tau_{\rm T}(t)
  =
   \rho\, v\, t \frac{\sigma_\mathrm{T}}{m_\mathrm{DM}}
  =
  \frac{\rho\, v\, t}{m_\mathrm{DM}}
  \int \mathrm{d}\Omega\,
  (1-\cos\theta)\,
  \frac{\mathrm{d}\sigma}{\mathrm{d}\Omega}.
\end{equation} Hence, it is the transfer cross section $\sigma_{\rm T}$, rather than the total cross section, that controls the dynamics for forward-dominated interactions.

When $\tau_{\rm T}$ is sufficiently small, the discrete sum over multipoles may be approximated by a continuous integral. In this limit one recovers the well-known Moli\`ere form~\cite{Moliere_1948},
\begin{equation}
\label{eq: Moliere solution}
  f(t,\theta)
  \approx
  \frac{1}{\tau_{\rm T}(t)}
  \exp\!\left(
    -\frac{\theta^2}{2\,\tau_{\rm T}(t)}
  \right),
\end{equation} which corresponds to a Gaussian broadening of the beam induced by the cumulative effect of many small-angle scatterings.

This diffusive angular spreading directly mirrors the effective fSIDM description, in which forward-dominated interactions are represented by transverse momentum diffusion. Both approaches therefore yield the same characteristic Gaussian broadening at early times. As the system evolves toward isotropy, the fSIDM formulation continues to provide a valid effective description even beyond the strict domain of applicability of the Moli\`ere approximation.

Because the parameters of the test configuration are chosen somewhat arbitrarily, we express all timescales in units of a reference time of the system $t_\mathrm{sys}$. This reference time is defined as the characteristic dynamical timescale $t_\mathrm{dyn}$ over which the angular distribution changes appreciably. Specifically, $t_\mathrm{sys}$ is determined from the condition that the opacity
\begin{equation}
  \tau(t)
  \equiv
  \rho\, v\, t\, \frac{\sigma}{m_\mathrm{DM}}
\end{equation}
reaches unity at $t=t_\mathrm{sys}$ for a representative cross section $\sigma/m_\chi $, i.e. $\tau(t_{\rm sys}) = 1$. This definition is independent of the particular type of cross section $\sigma$ adopted in the model.

\begin{figure}
    \centering
\includegraphics[width=1.00\columnwidth]{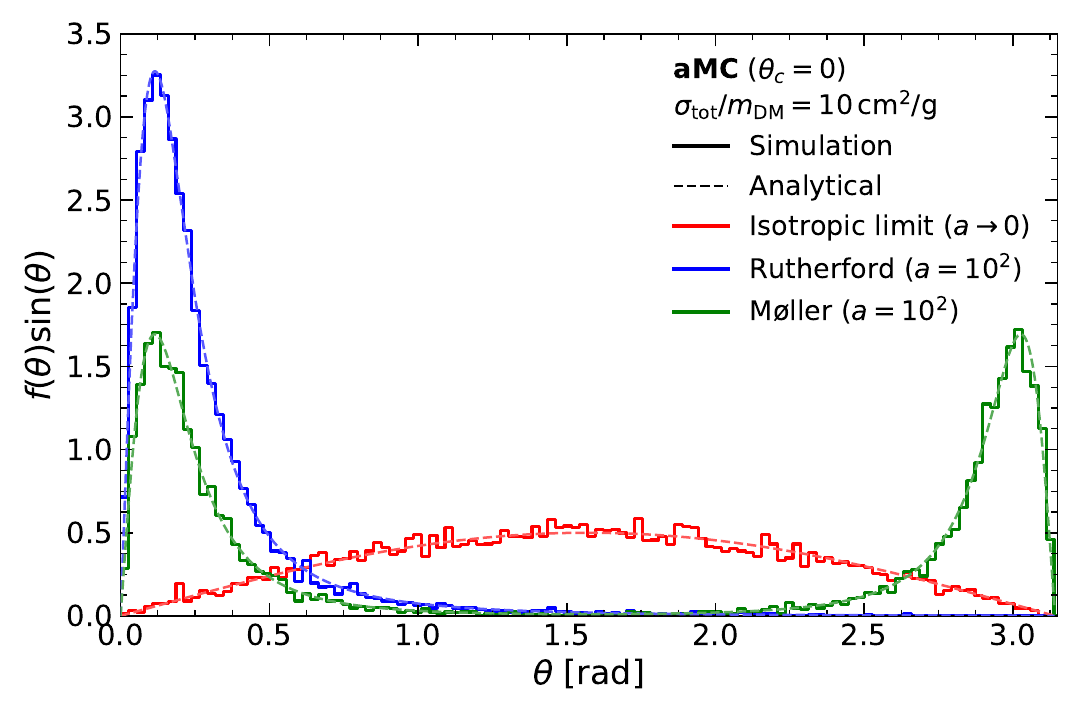}
\caption{Single-scattering test of the aMC angular sampler (\cref{sec:anisotropic_MC}). Each beam particle scatters once before the distribution is measured. Solid curves show the sampled angular distributions, and dashed curves show \cref{eq:analytical_probability_distribution}. The cases are Rutherford scattering (blue) and M{\o}ller scattering (green), both with $a=10^2$, and isotropic scattering (red). All use $\sigma_{\rm tot}/m_{\rm DM}=10\,\cpm$. The agreement confirms that the aMC implementation correctly samples the corresponding differential cross sections.}
    \label{fig:single_scatter}
\end{figure}

\subsubsection{Deflection tests of angular SIDM scattering algorithms}
\label{sec:deflection_tests_angular_SIDM}

We first test the angular sampling of aMC (\cref{sec:anisotropic_MC}). Each beam particle is allowed to scatter at most once, after which its interactions are disabled. The angular distribution is evaluated after all beam particles have scattered. For an azimuthally symmetric differential cross section, the normalized single-scattering distribution is
\begin{equation}
\label{eq:analytical_probability_distribution}
    p(\theta)=\frac{1}{\sigma_{\rm tot}}\frac{\mathrm{d}\sigma}{\mathrm{d}\theta}=\frac{2\pi}{\sigma_{\rm tot}}\frac{\mathrm{d}\sigma}{\mathrm{d}\Omega}\sin\theta,
\end{equation}
with $\int_0^\pi p(\theta)\,\mathrm{d}\theta=1$. Here, $\theta$ denotes the sampled scattering angle, which coincides with the beam-particle deflection in the no-recoil limit. We use the Rutherford and M{\o}ller differential cross sections in \cref{eq:Rutherford_general,eq:Moeller_general} with $a=10^2$, together with an isotropic reference for which $p(\theta)=(1/2)\sin\theta $. All three cases have $\sigma_{\rm tot}/m_{\rm DM}=10\,\cpm$. This normalization sets the scattering rate but cancels from the normalized angular distribution. The agreement in \cref{fig:single_scatter} verifies that aMC samples the prescribed angular distributions.

\begin{figure}
    \centering
\includegraphics[width=1.00\columnwidth]{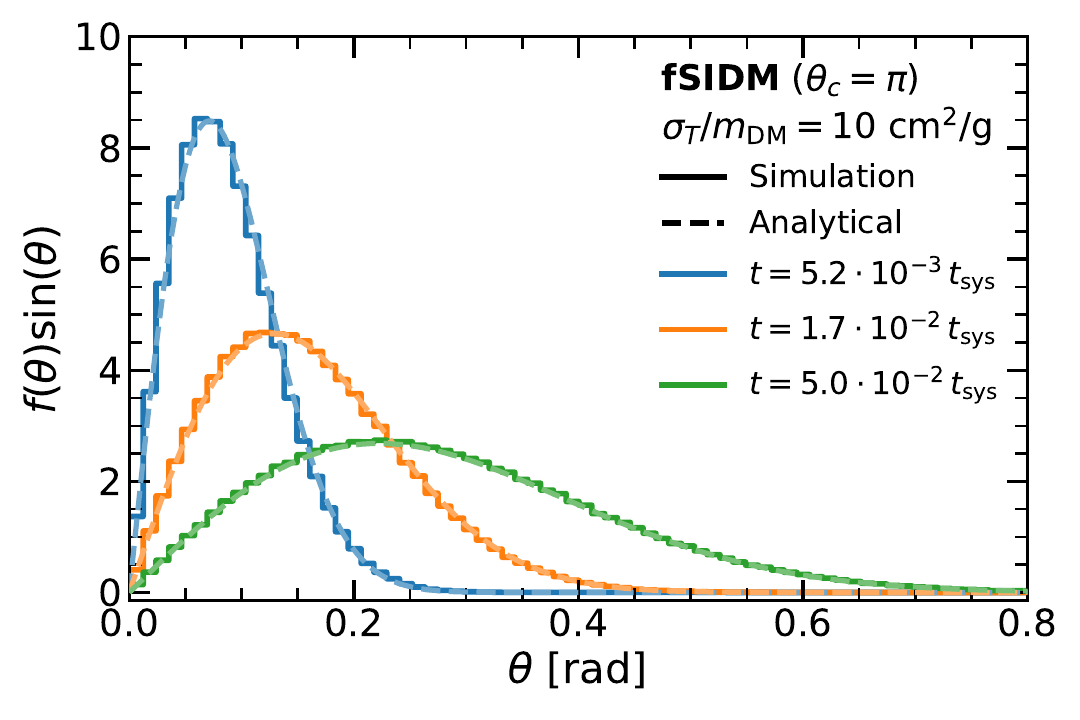}
    \caption{Validation of the fSIDM implementation, based on the effective drag-force and momentum-diffusion scheme, described in \cref{sec: fsidm} using the deflection test. The plot shows the angular distribution of beam particles in the deflection-test setup for a transfer cross section of $10~\mathrm{cm}^2/\mathrm{g}$ at three different times (solid lines), compared to the Gaussian Moli\`ere distribution with a width set by the transfer cross section (dashed lines; see \cref{eq: Moliere solution}). We assume beam--target scattering without recoil.
    }
    \label{fig: Moliere vs fSIDM}
\end{figure}

We next validate the fSIDM implementation described in \cref{sec: fsidm} using the multiple-scattering deflection test. The numerical angular distributions obtained with the effective drag-force and momentum-diffusion scheme are compared directly with the analytical Moli\`ere solution in \cref{eq: Moliere solution}. The results are shown in \cref{fig: Moliere vs fSIDM}. Each panel shows the angular distribution of the beam particles at three different times (solid curves), together with the corresponding Gaussian Moli\`ere predictions (dashed curves), whose widths are determined by the transfer opacity $\tau_{\rm T}$. The test is performed in the no-recoil limit, such that only the propagation directions of the beam particles change, while their speeds remain constant. The agreement between the numerical fSIDM results and the Gaussian Moli\`ere solution confirms that the effective treatment of strongly forward-dominated scattering through a drag term and transverse momentum diffusion correctly reproduces the expected small-angle broadening of the beam.

\begin{figure}
    \centering
    \includegraphics[width=1.00\columnwidth]{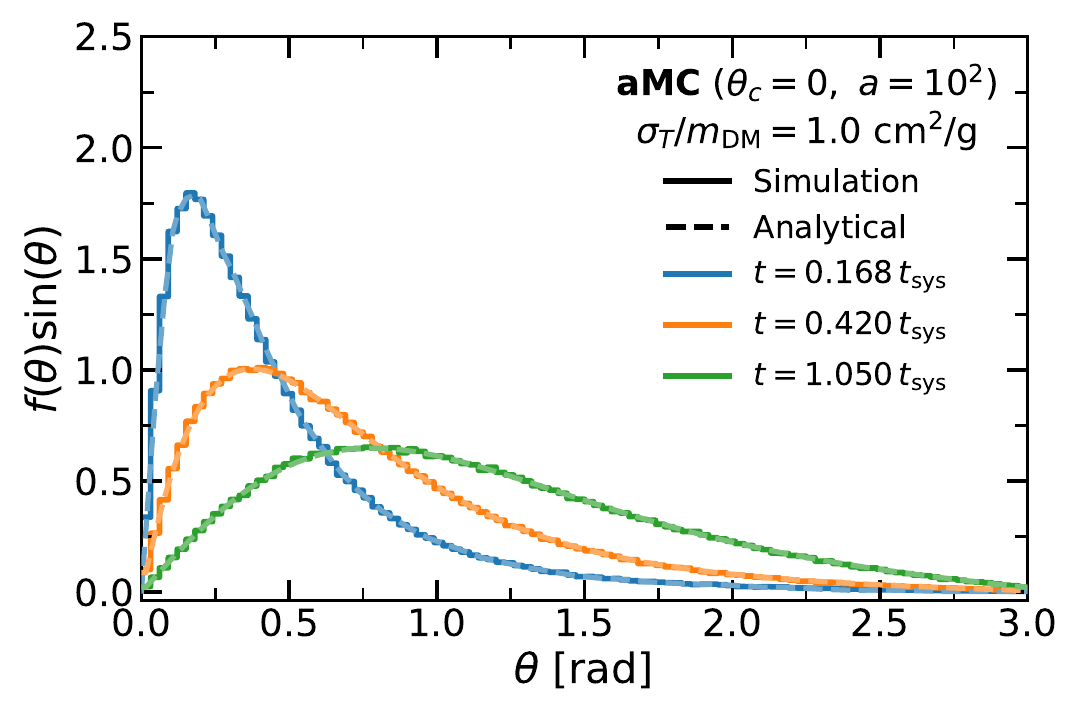}
    \includegraphics[width=1.00\columnwidth]{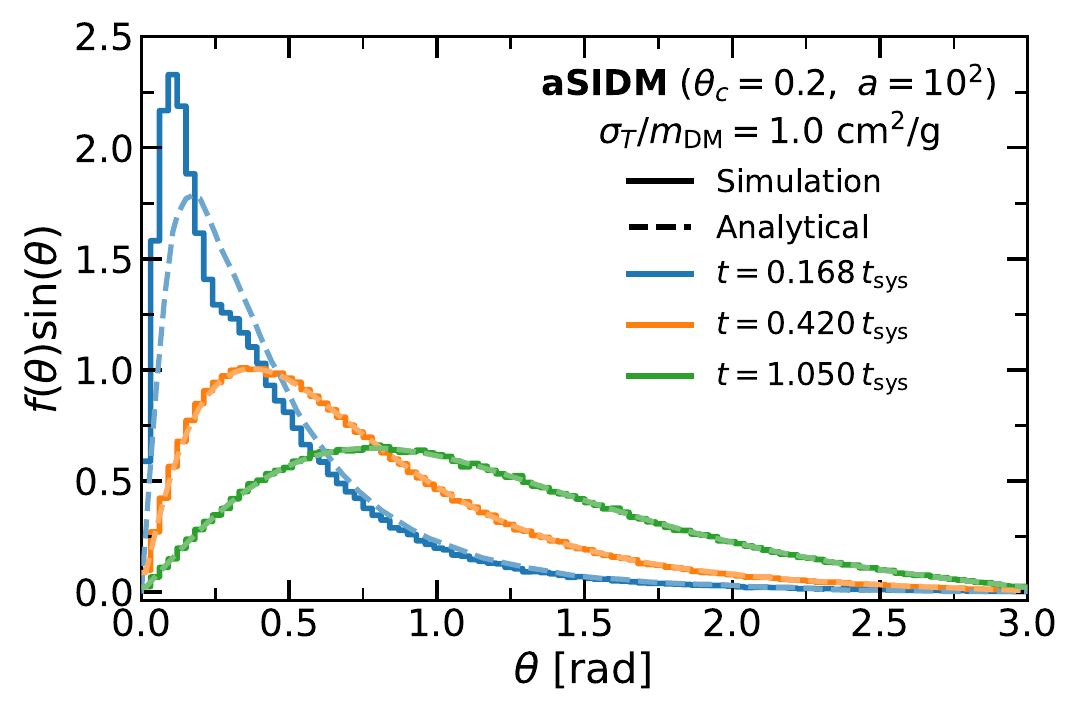}
    \caption{
    Validation of the anisotropic MC implementation described in \cref{sec:anisotropic_MC} with $\theta_{\mathrm{c}}=0$ (upper panel) and the aSIDM implementation described in \cref{sec:asidm} with $\theta_{\mathrm{c}}=0.2$ (lower panel). The deflection test assumes the Rutherford differential cross section in \cref{eq:Rutherford_general}, with the anisotropy parameter $a=10^2$ defined in \cref{eq:anisotropy_parameter}, a transfer cross section per unit DM mass of $\sigma_{\rm T}/m_{\rm DM}=1~\cpm$, and beam--target scattering without recoil. The solid curves show the numerical angular distributions, $f(t,\theta)\sin\theta$, at three times expressed in units of $t_{\mathrm{sys}}$, while the dashed curves show the exact multiple-scattering solution in \cref{eq:Goudsmit-Saunderson}. The aMC results agree with the analytical solution at all times shown. For aSIDM, agreement is reached for $t\gtrsim t_{\rm min}(0.2)\simeq0.3\,t_{\rm sys}$, after sufficient small-angle deflections have accumulated for the fSIDM component to reproduce its angular redistribution.
    }
    \label{fig: Goudmit Saunderson check 100}
\end{figure}

Following the hybrid-scheme validation strategy
of~\cite{Arido:2024hfk}, we compare our
\arepoTwo\ aMC and aSIDM implementations
(\cref{sec:anisotropic_MC,sec:asidm}) with
the exact multiple-scattering solution in
\cref{eq:Goudsmit-Saunderson}. The benchmark uses the Rutherford differential cross section in \cref{eq:Rutherford_general}, with $a=10^2$, $\sigma_{\rm T}/m_{\rm DM}=1\,\cpm$ and no target recoil. In \cref{fig: Goudmit Saunderson check 100}, the upper panel shows aMC, corresponding to $\theta_{\rm c}=0$, while the lower panel shows aSIDM with $\theta_{\rm c}=0.2$. The latter uses fSIDM for $\theta<\theta_{\rm c}$ and aMC for the remaining angles. The aMC distributions agree with the analytical prediction at all times shown. For aSIDM, the small-angle approximation reproduces the distribution after sufficient small deflections have accumulated. Using the agreement criterion discussed below, we obtain $t_{\rm min}(0.2)\simeq0.3\,t_{\rm sys}$ for $a=10^2$ and $t_{\rm min}(0.2)\simeq0.03\,t_{\rm sys}$ for $a=10^4$. At fixed $\sigma_{\rm T}$, the more strongly forward-peaked Rutherford interaction accumulates small-angle scatterings more rapidly, allowing agreement to be reached earlier.

\begin{figure*}
    \centering
    \includegraphics[width=1.00\columnwidth]{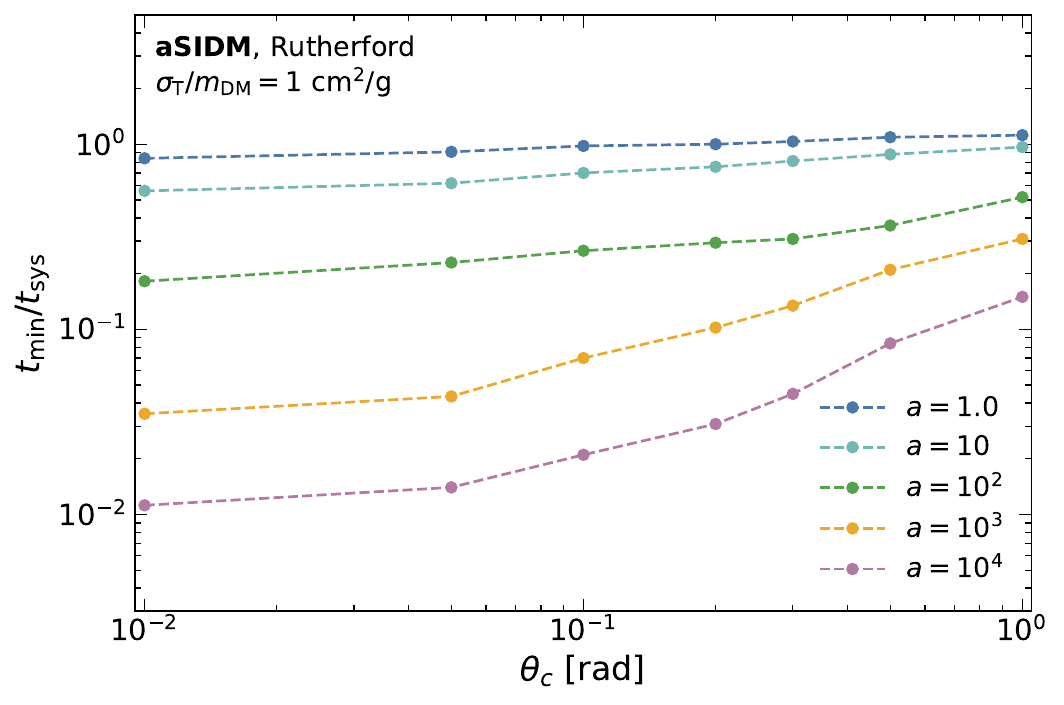}
    \includegraphics[width=1.00\columnwidth]{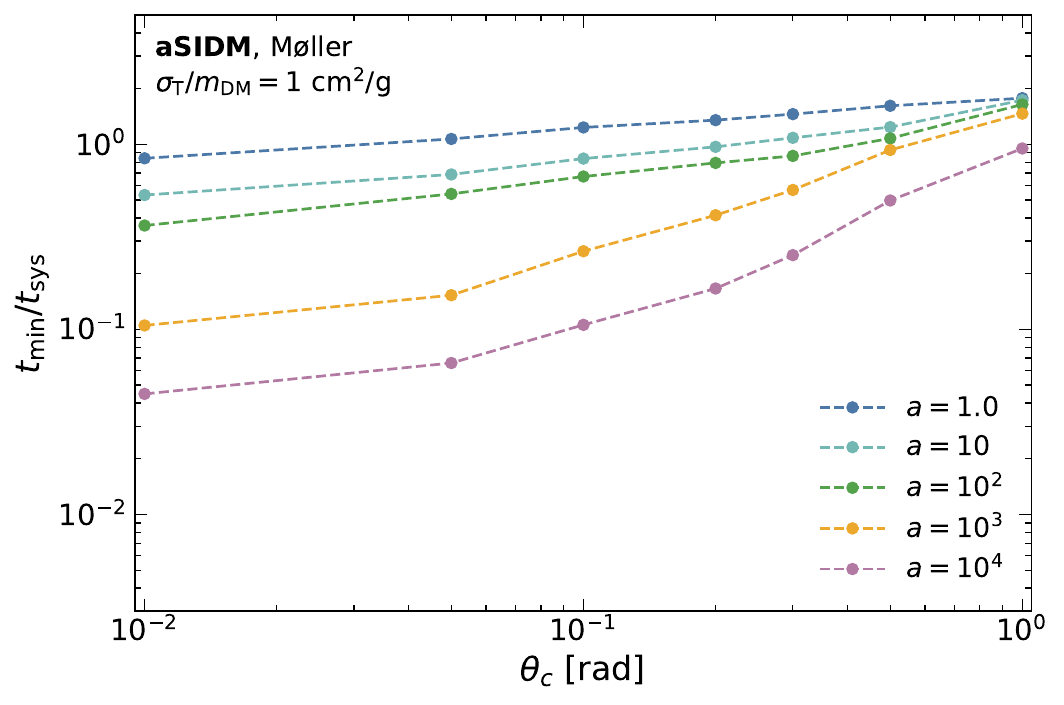}
\caption{Minimum agreement time $t_{\rm min}(\theta_{\rm c})$ for aSIDM in the deflection test, shown for Rutherford (left) and M{\o}ller (right) scattering. For Rutherford scattering, the reference is the exact multiple-scattering solution in \cref{eq:Goudsmit-Saunderson}, while the M{\o}ller results are compared with the corresponding fully anisotropic aMC calculation. We consider $\sigma_{\rm T}/m_{\rm DM}=1\,\cpm$. The agreement time generally increases with $\theta_{\rm c}$ and decreases towards the strongly forward-dominated regime at large $a$. An example of the Rutherford comparison is shown in \cref{fig: Goudmit Saunderson check 100}.}
    \label{fig:time_criterion_Rutherford}
\end{figure*}

\begin{figure*}
    \centering
    \includegraphics[width=1.0\columnwidth]{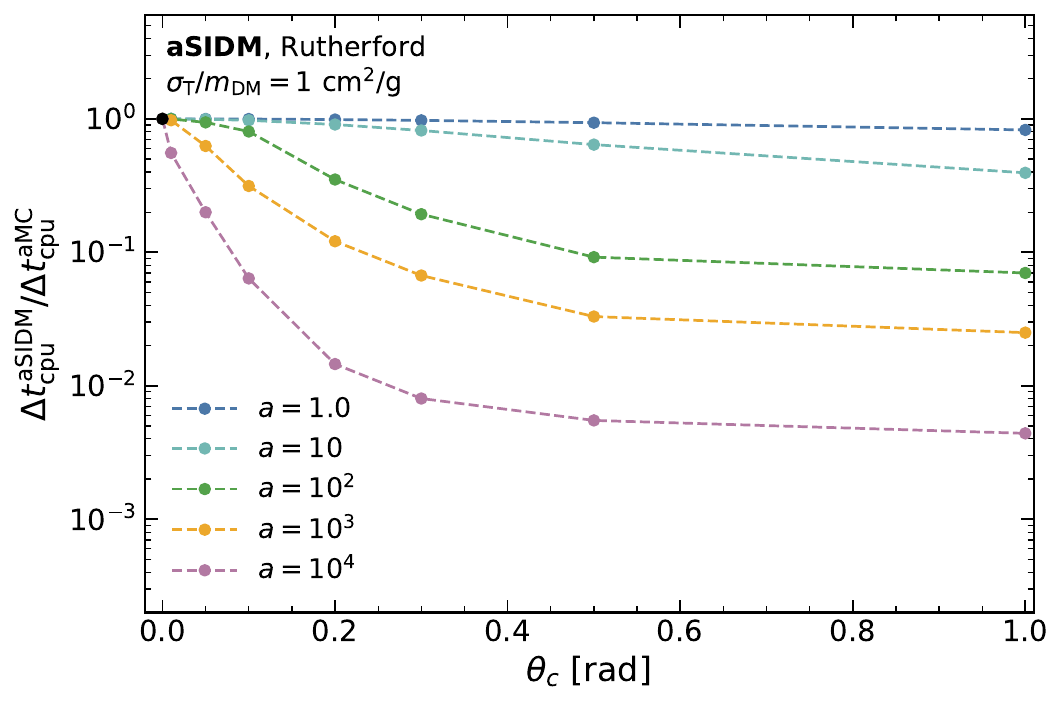}
    \includegraphics[width=1.0\columnwidth]{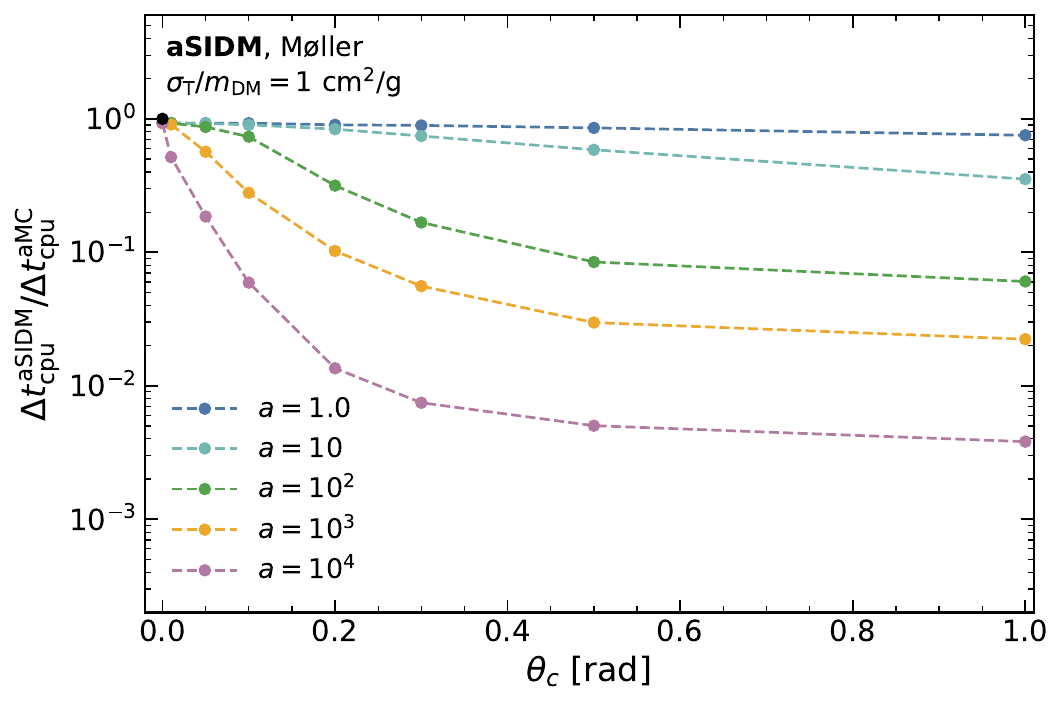}
\caption{Computational cost of aSIDM relative to fully anisotropic aMC in the deflection test, evolved to $5\,t_{\rm sys}$, for Rutherford (left) and M{\o}ller (right) scattering. We consider $\sigma_{\rm T}/m_{\rm DM}=1\,\cpm$. The single-core CPU time is normalized to the corresponding aMC simulation with $\theta_{\rm c}=0$. Increasing $\theta_{\rm c}$ transfers a larger fraction of the forward scattering to the fSIDM drag--diffusion treatment and relaxes the timestep constraint, yielding speed-ups of up to $\mathcal{O}(100)$ in the strongly anisotropic cases.}
    \label{fig: cpu plot}
\end{figure*}

Following the critical-angle validity tests
of~\cite{Arido:2024hfk}, we quantify the
regime of validity of our aSIDM implementation
in \cref{fig:time_criterion_Rutherford}. For each critical angle $\theta_{\mathrm{c}}$ and anisotropy parameter $a$, we compare both the fully aMC scheme and the aSIDM scheme against the exact multiple-scattering solution in \cref{eq:Goudsmit-Saunderson}. The aMC scheme explicitly samples the full differential cross section at every scattering event and agrees with the analytical solution throughout the evolution. By contrast, the aSIDM scheme agrees with the analytical solution only after a minimum time $t_{\rm min}(\theta_{\mathrm{c}})$, defined as the time beyond which the absolute difference between the simulated and analytical angular distributions, normalized by the square root of the particle count in each angular bin, remains below a threshold of $0.5$ in every bin, following~\cite{Arido:2024hfk}. This comparison is illustrated in the lower panel of \cref{fig: Goudmit Saunderson check 100} for $\theta_{\mathrm{c}}=0.2$ and $a=10^2$. 
\cref{fig:time_criterion_Rutherford} shows results for both the Rutherford differential cross section in \cref{eq:Rutherford_general} (left panel) and the Møller differential cross section in \cref{eq:Moeller_general} (right panel). In both cases, $t_{\rm min}$ increases monotonically with $\theta_{\mathrm{c}}$, because a larger critical angle causes a greater fraction of the deflections to be treated using the effective small-angle approximation. This approximation becomes accurate only after sufficiently many small-angle scatterings have accumulated. For strongly anisotropic cross sections with large $a$, the minimum time decreases because the enhanced rate of forward scatterings accelerates the onset of the drag--diffusion regime. In all cases, $t_{\rm min}$ remains below or of the order of the characteristic system timescale $t_{\mathrm{sys}}$, indicating that the aSIDM scheme becomes applicable early in the evolution. For weakly anisotropic cases, such as $a=1$, the scattering is already very close to isotropic, and the standard isotropic MC algorithm is therefore the more natural and efficient choice.  

The computational advantage of the aSIDM scheme relative to the fully aMC implementation is illustrated in \cref{fig: cpu plot}. We use the deflection-test setup with a fixed transfer cross section per unit DM mass of $\sigma_{\rm T}/m_{\rm DM}=1\,\cpm$ and compare both the Rutherford differential cross section, shown in the left panel, and the Møller differential cross section, shown in the right panel. For each critical angle $\theta_{\mathrm{c}}$ and for anisotropy parameters $a$, we measure the single-core CPU time $\Delta t_{\rm cpu}$ required to advance the simulation by the same fixed physical time interval and normalize it to the corresponding CPU time of the aMC scheme at $\theta_{\mathrm{c}}=0$. Increasing $\theta_{\mathrm{c}}$ moves a larger fraction of the strongly forward-peaked scatterings into the effective drag-force and momentum-diffusion treatment, thereby relaxing the timestep constraint. The speed-up primarily results from the less restrictive timestep criterion of the aSIDM scheme in \cref{eq:asidm_timestep}. In the strongly forward-dominated regime, the total cross section $\sigma_{\rm tot}$ can be much larger than the effective cross sections entering the hybrid treatment: $\sigma_{{\rm T},<}$ for deflections with $\theta<\theta_{\mathrm{c}}$, which are treated using fSIDM, and $\sigma_{{\rm tot},>}$ for deflections with $\theta>\theta_{\mathrm{c}}$, which are sampled explicitly using aMC. Because the timestep of the fully aMC scheme scales inversely with $\sigma_{\rm tot}$, it becomes increasingly restrictive for strongly anisotropic cross sections. By contrast, the aSIDM timestep is controlled by the smaller effective cross sections associated with the two angular regimes, allowing substantially larger timesteps.
For $\theta_{\mathrm{c}}\sim 0.5$, the computational cost is reduced by a factor of $\sim 10$ and up to $\mathcal{O}(100)$ in the anisotropic cases with $a\sim 10^2$--$10^4$. We find no substantial difference between the Rutherford and Møller cases, nor for other fixed values of $\sigma_{\rm T}/m_{\rm DM}$ than the one used in the figure.

The results demonstrate that aSIDM provides a considerable computational improvement while retaining the correct angular evolution for $t\gtrsim t_{\rm min}(\theta_{\mathrm{c}})$. This makes the scheme particularly well suited for large-scale $N$-body simulations of strongly anisotropic DM self-interactions.

\subsubsection{Verification of the multi-state SIDM scheme}
\label{sec:deflection_test_multi_state_SIDM}

We verify the state-dependent smoothing-length scheme introduced in \cref{sec:imp_multi_states} using the deflection-test setup. This scheme is a key component of the multi-state SIDM implementation: the local interaction rate for each reaction channel must be estimated from the particles in the relevant incoming states, rather than from the total DM particle distribution. We retain the beam and target initial conditions of the fSIDM test in \cref{fig: Moliere vs fSIDM} and add $10^6$ non-interacting state-2 particles, approximately twice the number of particles in the interacting target state. Because the smoothing length is determined separately for each state, this additional population must leave both the smoothing length and the interaction rate of the beam--target channel unchanged.

\begin{figure}
    \centering
    \includegraphics[width=1.00\columnwidth]{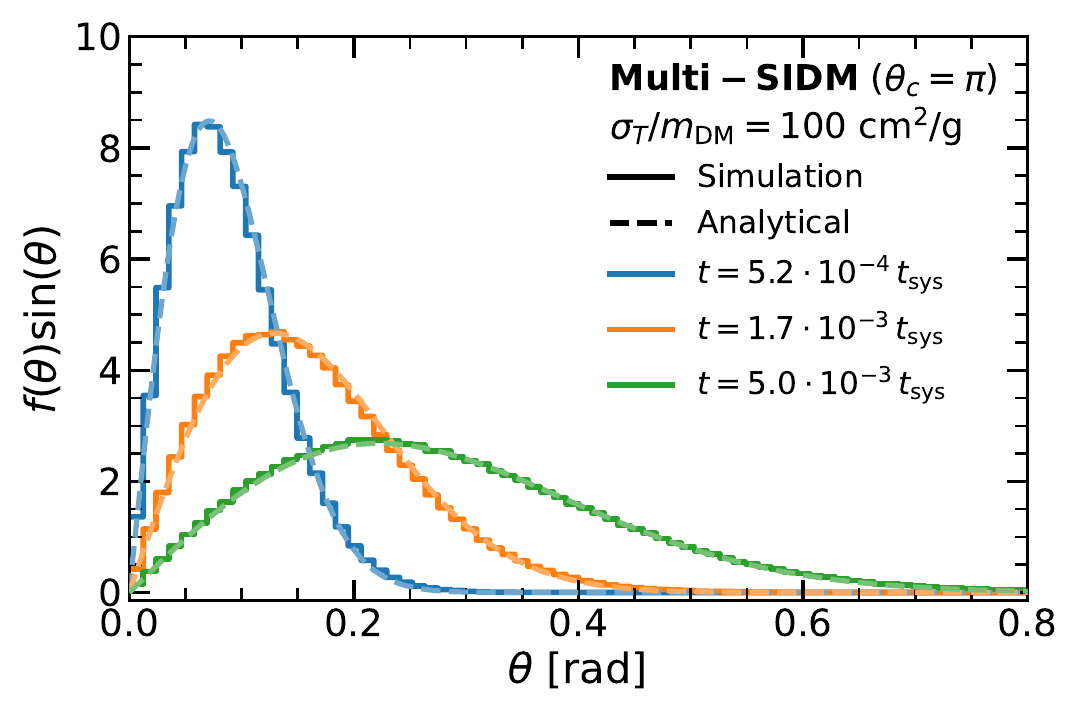}
    \caption{
    Validation of the multi-state SIDM implementation with state-dependent smoothing lengths described in \cref{sec:imp_multi_states}. The deflection-test setup is identical to that used in \cref{fig: Moliere vs fSIDM}, but includes $10^6$ additional non-interacting state-2 particles, corresponding to twice the number of particles in the interacting target state. The solid curves show the angular distributions of the beam particles at three different times for $\sigma_{\rm T}/m_{\rm DM}=100~\mathrm{cm}^2\,\mathrm{g}^{-1}$, while the dashed curves show the corresponding Gaussian Moli\`ere solutions. The agreement confirms that the state-dependent smoothing-length scheme correctly isolates the interacting target population and remains unaffected by the additional non-interacting state.
    }
    \label{fig:multi_sidm_deflection_test}
\end{figure}

\cref{fig:multi_sidm_deflection_test} compares the numerical angular distributions of the beam particles at three times with the corresponding Gaussian Moli\`ere solutions in \cref{eq: Moliere solution}, using $\sigma_{\rm T}/m_{\rm DM}=100~\mathrm{cm}^2\,\mathrm{g}^{-1}$. The agreement demonstrates that the state-dependent smoothing-length scheme correctly isolates the interacting target population even when it is outnumbered by particles in the additional non-interacting state. The comparison also tests the expected scaling with transfer opacity: the cross section adopted here is ten times larger than that used in \cref{fig: Moliere vs fSIDM}. In this fixed-target fSIDM setup with gravity disabled, the angular evolution depends only on the transfer opacity defined in \cref{eq:transfer_opacity}. Increasing $\sigma_{\rm T}$ by a factor of ten therefore yields the same angular distribution after one-tenth of the evolution time. The results in \cref{fig:multi_sidm_deflection_test,fig: Moliere vs fSIDM} reproduce this scaling at equal transfer opacity.

\begin{figure*}
    \centering
    \includegraphics[width=1.00\columnwidth]{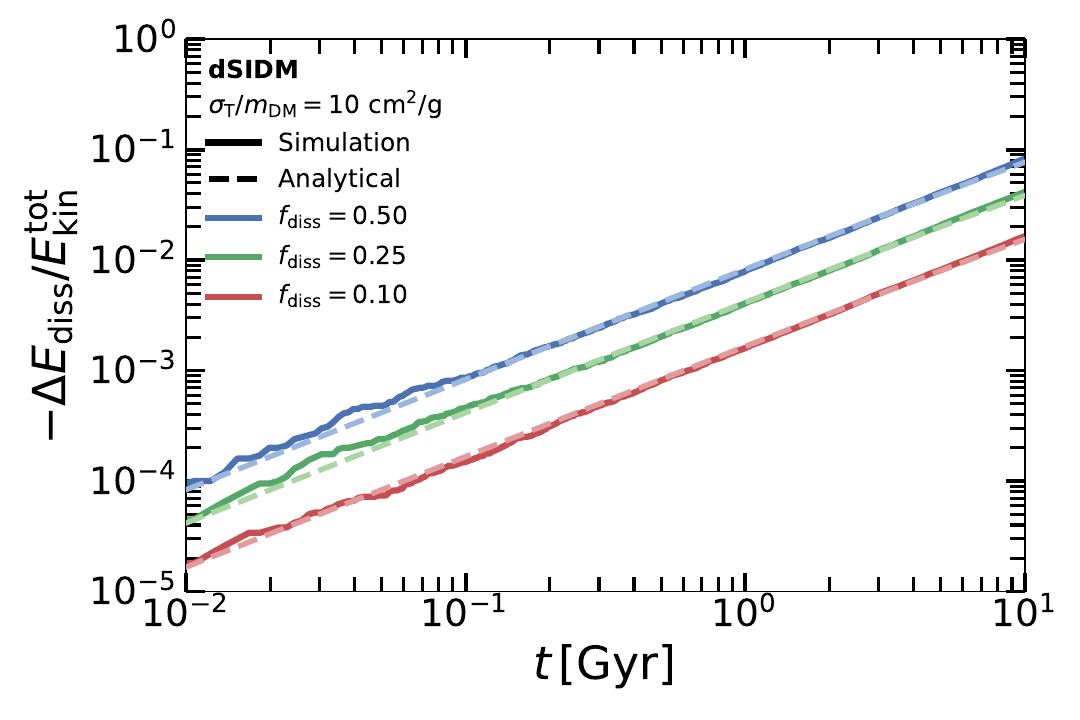}
    \includegraphics[width=1.00\columnwidth]{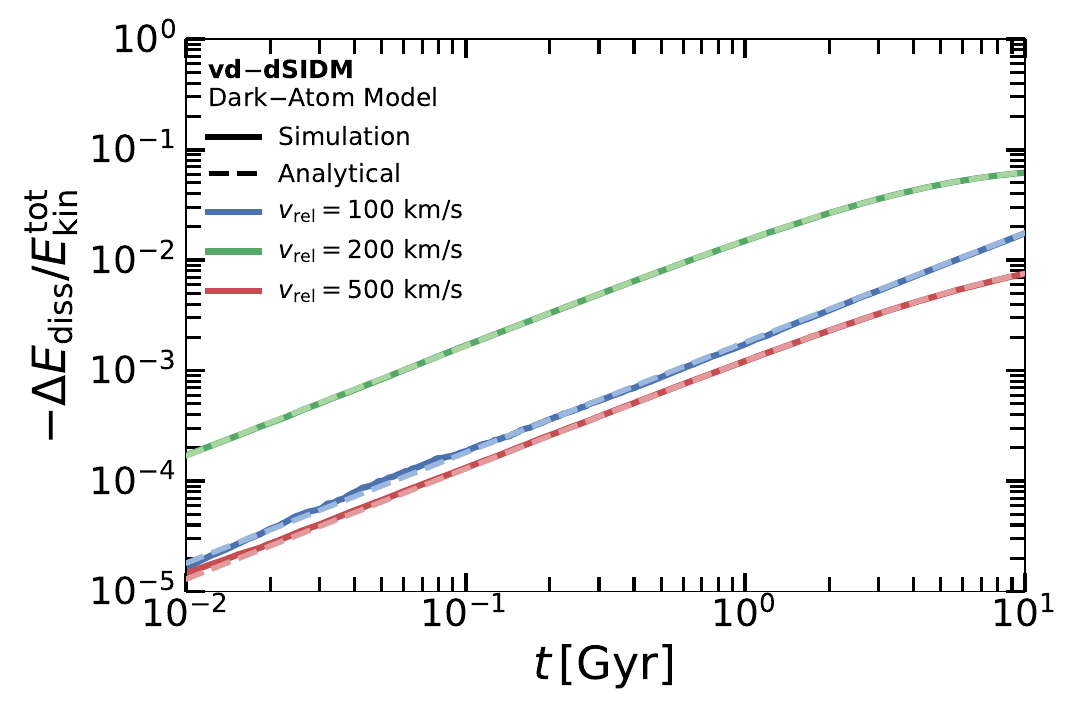}
\caption{Validation of the constant dSIDM (left) and velocity-dependent dSIDM (vd-dSIDM; right) implementations described in \cref{sec:imp_dissipative_SIDM} using the single-scattering-test setup. Solid curves show the cumulative dissipated energy measured in the simulations, normalized by the initial kinetic energy of the beam, while dashed curves show the analytical predictions from \cref{eq:dissipative_energy_loss}. The constant dSIDM runs use $\sigma_{\rm T}/m_{\rm DM}=10\,\cpm$ with $f_{\rm loss}=0.10$, $0.25$ and $0.50$. The vd-dSIDM runs use the first dark-atom benchmark defined in \cref{eq:dSIDM_benchmark_1}, with the velocity-dependent dissipative cross section shown in \cref{fig:dark_atom_cross_sections} and the corresponding energy-loss fraction from \cref{eq:dark_atom_diss_energ_loss_fraction}, evaluated at $v_{\rm rel}=100$, $200$ and $500\,\kms$. The agreement with the analytical predictions validates both the scattering rates and dissipative energy updates in the constant and velocity-dependent implementations.}
    \label{fig:vd_dsidm_energy_validation}
\end{figure*}

\subsubsection{Validation of the dissipative SIDM scattering algorithms}
\label{sec:deflection_test_dSIDM_scattering_algorithms}

We validate the constant dissipative SIDM (dSIDM) and velocity-dependent dissipative SIDM (vd-dSIDM) implementations described in \cref{sec:imp_dissipative_SIDM} using the controlled deflection-test setup of \cref{sec:test_setup}. In these tests, the primary diagnostic is the cumulative kinetic energy removed by dissipative interactions, rather than the angular distribution of the scattered beam. This provides a direct test of both the scattering rate and the energy loss applied in each accepted interaction.

A beam of state-0 particles propagates through an initially stationary population of state-1 target particles. Gravity and same-state scatterings are disabled, and beam--target interactions are treated in the no-recoil limit. Each beam particle is allowed to scatter at most once, preventing repeated energy losses from the same particle. The resulting cumulative dissipated energy can therefore be compared directly with an analytical prediction for both the constant and velocity-dependent implementations.

With the single-scattering condition, each beam particle undergoes a dissipative interaction with the constant rate $\Gamma=\rho\,(\sigma/m_{\rm DM})\,v_{\rm rel}$, giving an expected scattered fraction $1-\exp(-\Gamma t)$. Here, $\sigma/m_{\rm DM}$ denotes the constant cross section in the dSIDM test or the velocity-dependent dissipative cross section evaluated at the fixed relative velocity of a vd-dSIDM run. In the no-recoil limit, the target is treated as having effectively infinite inertia, so the reduced mass approaches the beam-particle mass $M_0$. Each interaction therefore removes a fraction $f_{\rm loss}$ of the relative kinetic energy, $\Delta E_{\rm diss}^{\rm pair}=-M_0v_{\rm rel}^2f_{\rm loss}/2$, entirely from the state-0 beam particle. Normalizing the cumulative energy loss by the initial beam kinetic energy, $E_{\rm kin}^{\rm tot}=N_0^{\rm init}M_0v_{\rm rel}^2/2$, gives
\begin{equation}
\label{eq:dissipative_energy_loss}
\frac{\Delta E_{\rm diss}(t)}{E_{\rm kin}^{\rm tot}}
=
-f_{\rm loss}
\left[
1-\exp\left(-\Gamma t\right)
\right].
\end{equation}
The dashed curves in \cref{fig:vd_dsidm_energy_validation} show this analytical prediction. Since $\Delta E_{\rm diss}<0$, the figure displays the positive removed-energy fraction, $-\Delta E_{\rm diss}/E_{\rm kin}^{\rm tot}$. The approach to the asymptotic value therefore tests the scattering rate, while the saturation level directly tests the energy loss applied per interaction.

For the constant dSIDM test in the left panel of \cref{fig:vd_dsidm_energy_validation}, we adopt $\sigma_{\rm T}/m_{\rm DM}=10\,\cpm$ and consider $f_{\rm loss}=0.10$, $0.25$ and $0.50$. The three runs have the same scattering rate but different energy losses per interaction, and therefore approach different asymptotic dissipated-energy fractions on the same characteristic timescale. The agreement with \cref{eq:dissipative_energy_loss} verifies both the velocity-independent scattering probability and the prescribed energy loss.

For the vd-dSIDM test in the right panel, we use the first dark-atom benchmark model defined in \cref{eq:dSIDM_benchmark_1}, together with the corresponding elastic and dissipative cross sections shown by the blue curves in \cref{fig:dark_atom_cross_sections}. We consider $v_{\rm rel}=100$, $200$ and $500~\mathrm{km/s}$, sampling distinct parts of the velocity-dependent dissipative cross section: the $200~\mathrm{km/s}$ case lies close to its maximum, while the $100$ and $500~\mathrm{km/s}$ cases probe lower-cross-section regimes on either side. For each run, the scattering rate is $\Gamma_{\rm vd\text{-}dSIDM}(v_{\rm rel})=\rho\,[\sigma_{\rm diss}(v_{\rm rel})/m_{\rm DM}]\,v_{\rm rel}$, while the energy removed per interaction is set by the velocity-dependent fraction $f_{\rm loss}(v_{\rm rel})$ in \cref{eq:dark_atom_diss_energ_loss_fraction}. The three velocities therefore test the velocity-dependent interaction rate and dissipative energy update simultaneously. Their agreement with the analytical predictions demonstrates that both $\sigma_{\rm diss}(v_{\rm rel})$ and $f_{\rm loss}(v_{\rm rel})$ are evaluated and applied consistently by the vd-dSIDM implementation.

\begin{figure*}
    \centering
    \includegraphics[width=1.00\columnwidth]{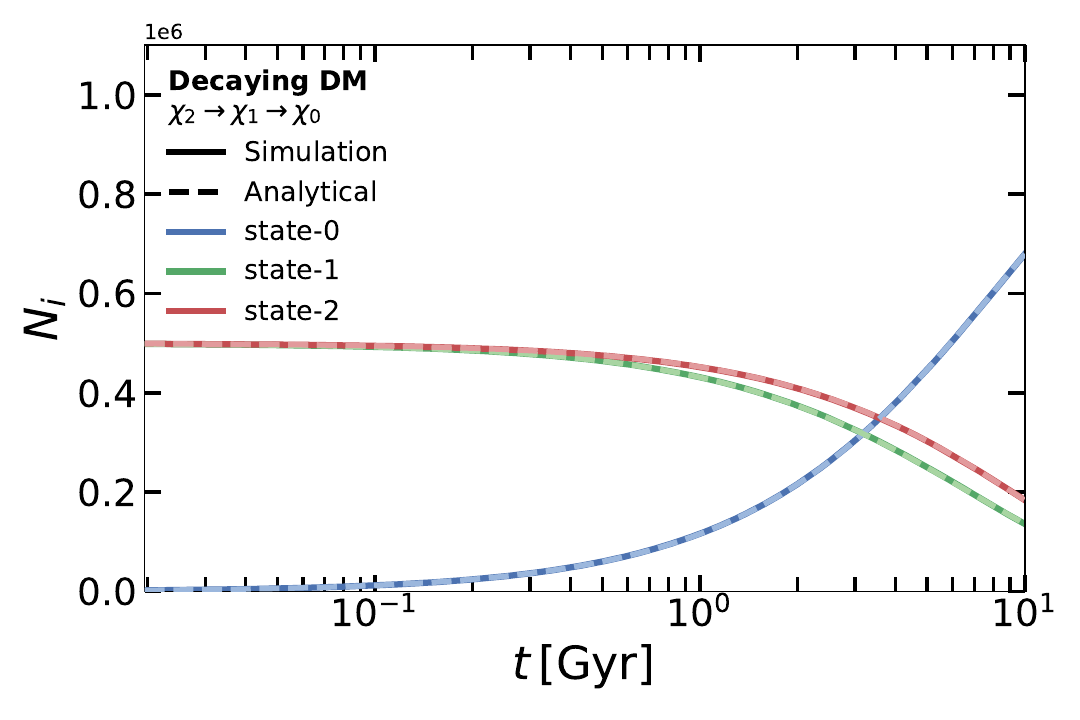}
    \includegraphics[width=1.00\columnwidth]{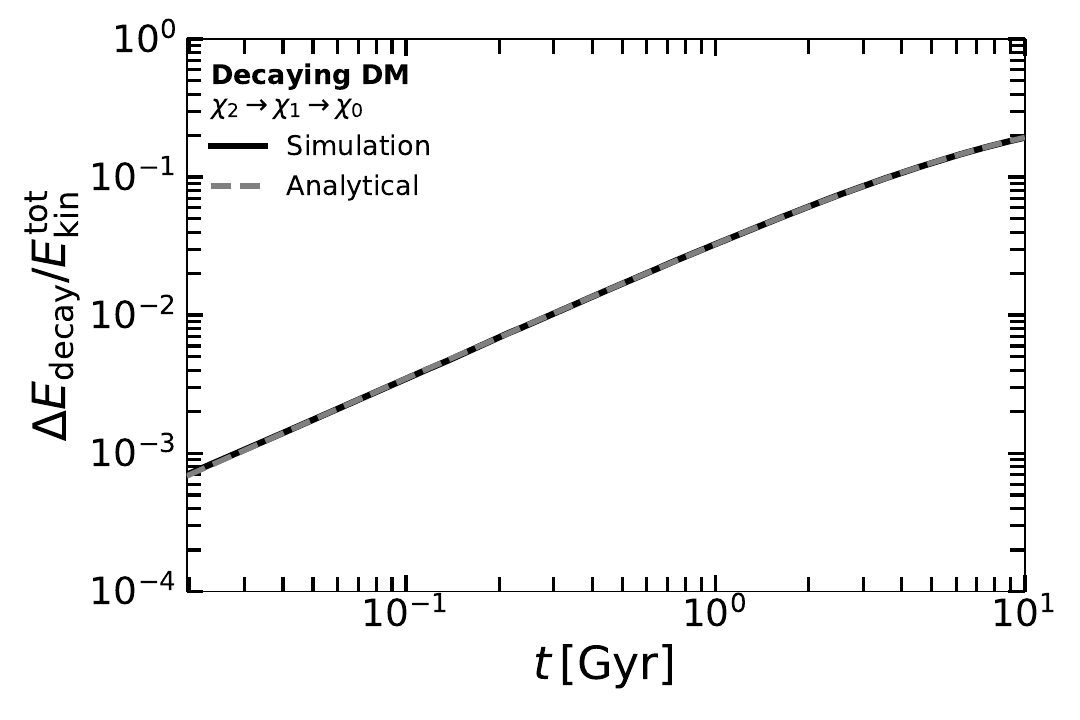}
\caption{Uniform-box validation of the state-based decaying-DM implementation described in \cref{sec:imp_decaying_DM}. The initial conditions contain $500{,}000$ particles in each of the unstable states~1 and~2 and no particles in the stable state~0. 
Particles undergo the sequential decay chain
$\chi_2\rightarrow\chi_1\rightarrow\chi_0$,
with decay rates
$\Gamma_{21}=0.10\,\mathrm{Gyr}^{-1}$
and $\Gamma_{10}=0.25\,\mathrm{Gyr}^{-1}$,
and isotropic recoil kicks
$v_{{\rm kick},21}=1.0\,\mathrm{km\,s^{-1}}$
and $v_{{\rm kick},10}=2.0\,\mathrm{km\,s^{-1}}$. \textit{Left:} evolution of the populations in the three states. \textit{Right:} cumulative kinetic energy injected by the recoil kicks, normalized by the initial total kinetic energy, $\Delta E_{\rm decay}/E_{\rm kin}^{\rm tot}(0)$. Solid curves show the simulation and dashed curve the corresponding analytical prediction from \cref{eq:state_number_evolution_decay,eq:kinetic_energy_injection_decays}. The agreement validates the prescribed decay rates, sequential state transitions and recoil-energy updates.}
    \label{fig:state_based_dDM_validation}
\end{figure*}

\subsection{\label{sec:dDM_box_test} Decaying DM in a uniform box}

We validate the state-based decaying-DM implementation described in \cref{sec:imp_decaying_DM} using a controlled uniform-box test of the decay probabilities, sequential state transitions and recoil kicks. The initial conditions contain $500{,}000$ particles in each of the unstable states~1 and~2 and no particles in the stable state~0. The initial particle speeds are $v_1=2.5~\mathrm{km\,s^{-1}}$ and $v_2=5.0~\mathrm{km\,s^{-1}}$. We consider the decay chain
\begin{align}
\begin{aligned}
&\chi_2 \xrightarrow{\Gamma_{21}} \chi_1
\xrightarrow{\Gamma_{10}} \chi_0,
\qquad \\ 
\Gamma_{21}=0.10&~\mathrm{Gyr}^{-1},
\quad
\Gamma_{10}=0.25~\mathrm{Gyr}^{-1},
\end{aligned}
\end{align} with isotropically oriented recoil kicks
$v_{{\rm kick},21}=1.0~\mathrm{km\,s^{-1}}$
and $v_{{\rm kick},10}=2.0~\mathrm{km\,s^{-1}}$. For constant decay rates, the state populations and ensemble-averaged kinetic energy injected by the recoil kicks have closed-form analytical solutions, which are derived in App.~\ref{app:ddm_validation}.

\Cref{fig:state_based_dDM_validation} compares these predictions with the simulations. The left panel shows the evolution of the three state populations, directly testing the decay probabilities and sequential transitions $\chi_2\rightarrow\chi_1\rightarrow\chi_0$. The right panel shows the cumulative recoil-energy injection normalized by the initial total kinetic energy, providing an independent test of the velocity-kick update. In both cases, the simulations closely follow the analytical predictions, validating the state-based decay implementation.

\begin{figure*}
    \centering
    \includegraphics[width=0.995\columnwidth]{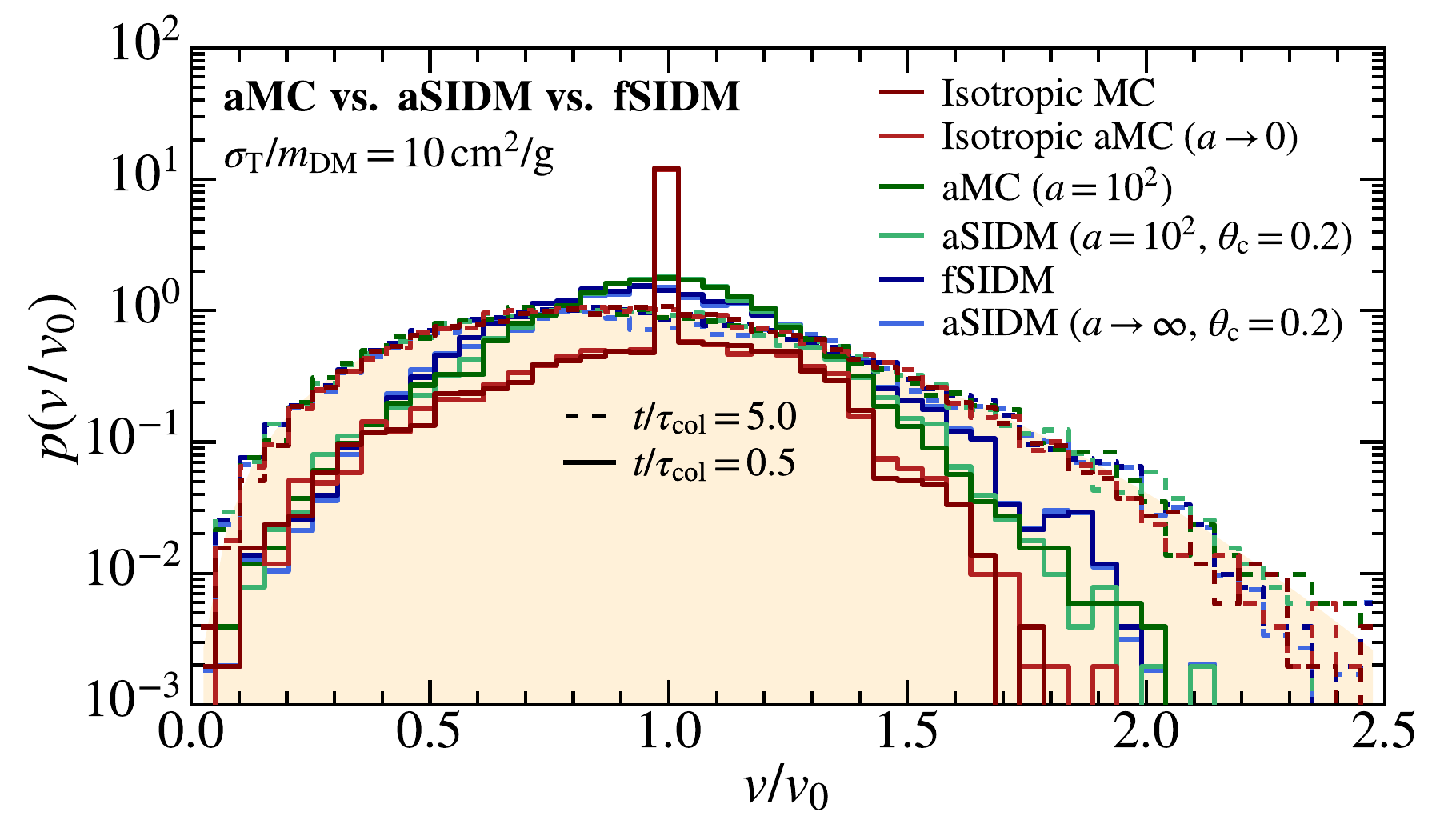}
    \includegraphics[width=1.005\columnwidth]{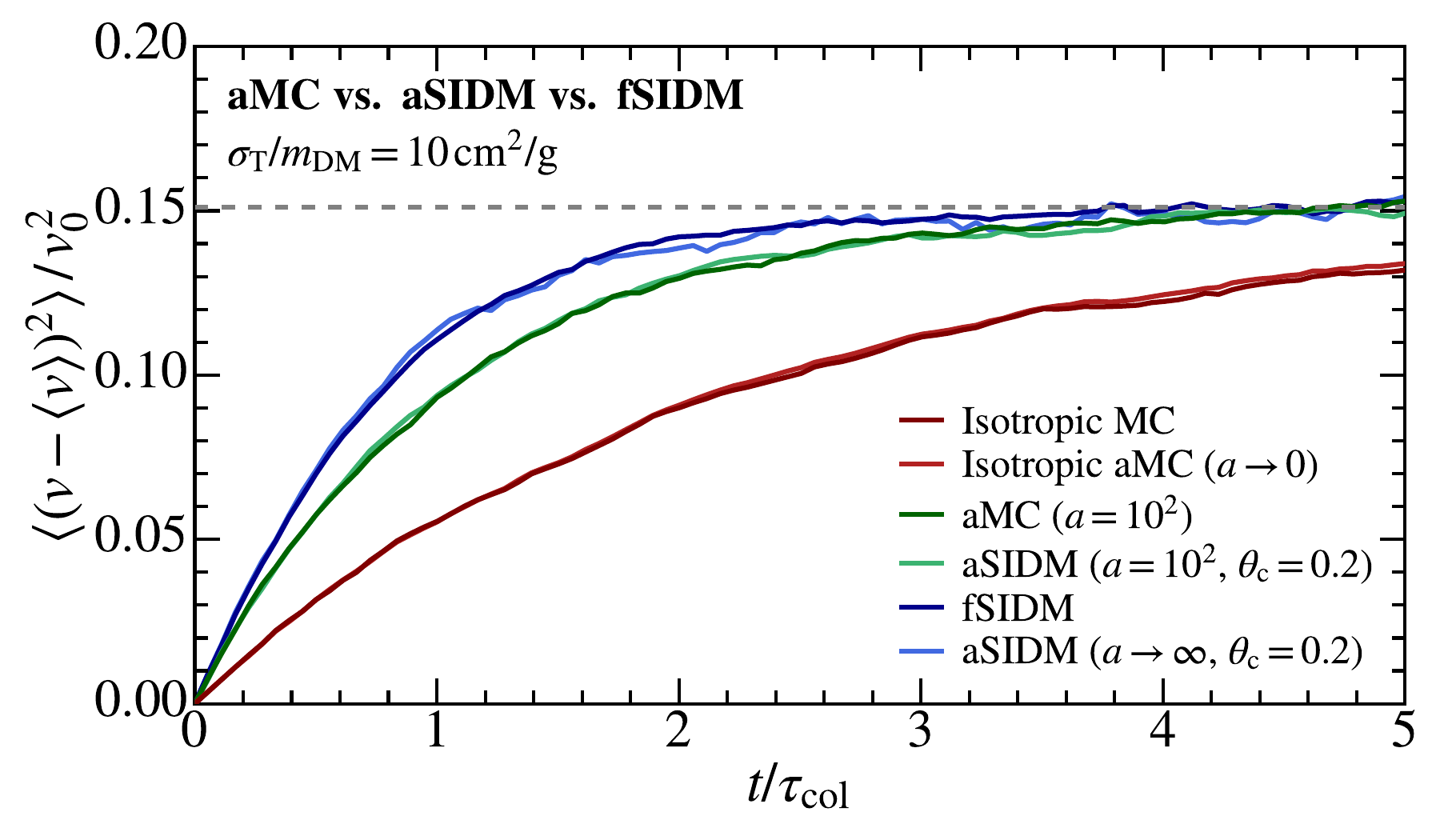}
\caption{Thermalization test of the aMC, aSIDM and fSIDM implementations described in \cref{sec:anisotropic_MC,sec:asidm,sec: fsidm}. The simulations use $N_{\rm part}=10^4$, $N_{\rm ngb}=32$, $v_0=2\,\kms$ and $\sigma_{\rm T}/m_{\rm DM}=10\,\cpm$. \textit{Left:} speed distributions at $t=0.5\,\tau_{\rm col}$ (solid) and $t=5.0\,\tau_{\rm col}$ (dashed), with $\tau_{\rm col}$ defined in \cref{eq:thermalize_box_collision_time}; the yellow shaded region shows the Maxwell--Boltzmann equilibrium distribution from \cref{eq:MB_distribution}. \textit{Right:} corresponding evolution of the squared speed dispersion, with the gray dashed line marking the Maxwell--Boltzmann expectation. The three comparisons test the isotropic limit, where aMC with $a=10^{-3}$ reproduces isotropic MC (red); finite anisotropy, where aSIDM with $\theta_{\rm c}=0.2$ reproduces fully anisotropic aMC with $\theta_{\rm c}=0$ at $a=10^2$ (green); and the strongly forward-dominated limit, where aSIDM approaches fSIDM at $a=10^{10}$ (blue). All schemes relax towards the same thermal equilibrium, validating their consistency across the isotropic, anisotropic and frequent-scattering limits.}
\label{fig:thermalization_anisotropic_iso_limit}
\end{figure*}

\begin{figure}
    \centering
\includegraphics[width=1.00\columnwidth]{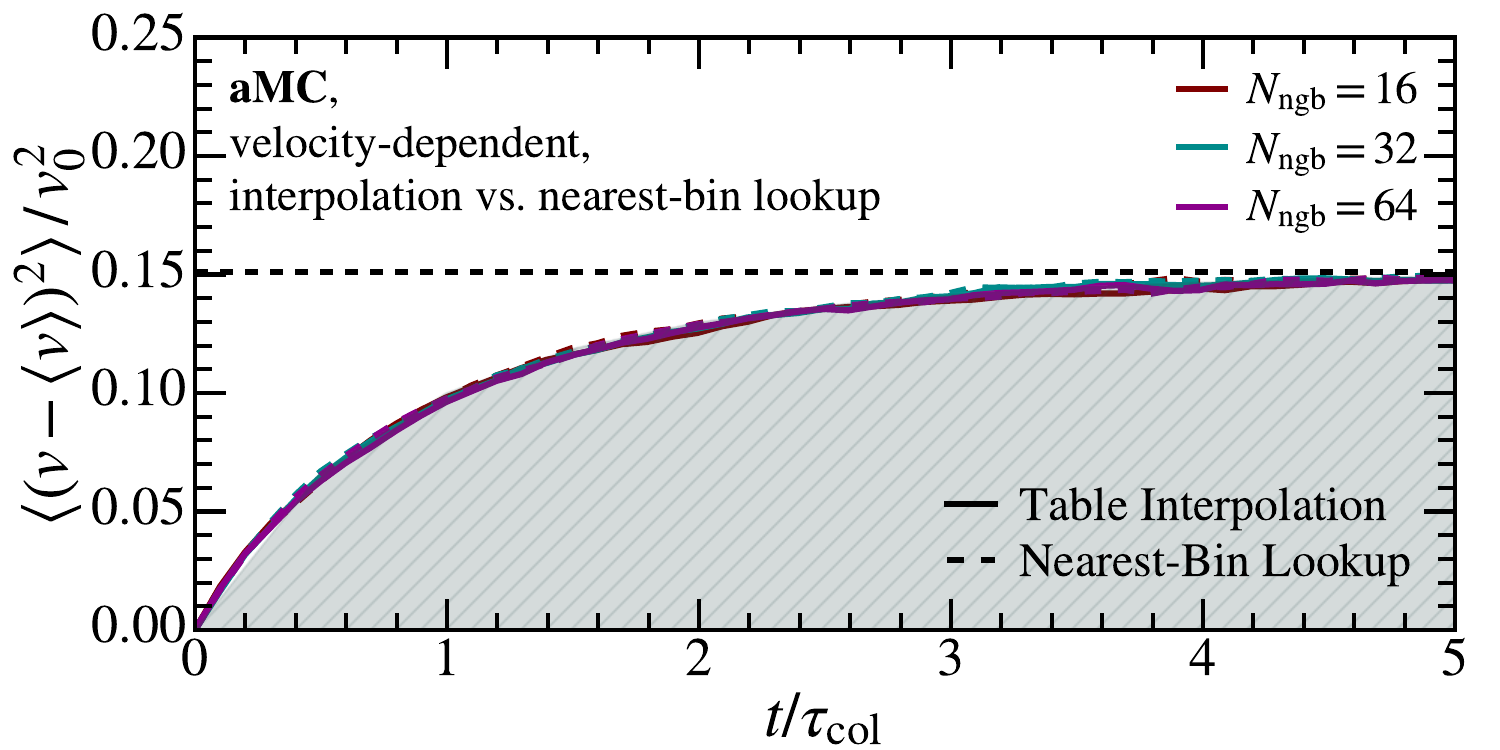}
\caption{Thermalization test of the velocity-dependent
aMC implementation described in
\cref{sec:anisotropic_MC}, using an
angle- and velocity-dependent light-mediator
SIDM test model with
$\sigma_0/m_{\rm DM}=300\,\cpm$
and $w=20\,\kms$. The figure shows the evolution of the normalized squared speed dispersion for $N_{\rm ngb}=16$, $32$ and $64$. Solid curves use interpolation between neighbouring bins of the tabulated cross sections and angular distributions, while dashed curves use nearest-bin lookup. The simulations contain $N_{\rm part}=10^4$ particles in a periodic box of side length $100\,\kpc$, with $\rho=10^6\,\Msun\,\kpc^{-3}$ and $v_0=40\,\kms$. The gray dashed line marks the Maxwell--Boltzmann expectation and the shaded band the semi-analytic Boltzmann prediction described in App.~B1 of~\cite{Zier:2026}. The nearly identical numerical curves demonstrate that the velocity-dependent aMC thermalization is insensitive to both the neighbour number and the tabulated-data lookup scheme.}
\label{fig:thermalization_table_lookup}
\end{figure}

\begin{figure}
    \centering  \includegraphics[width=1.00\columnwidth]{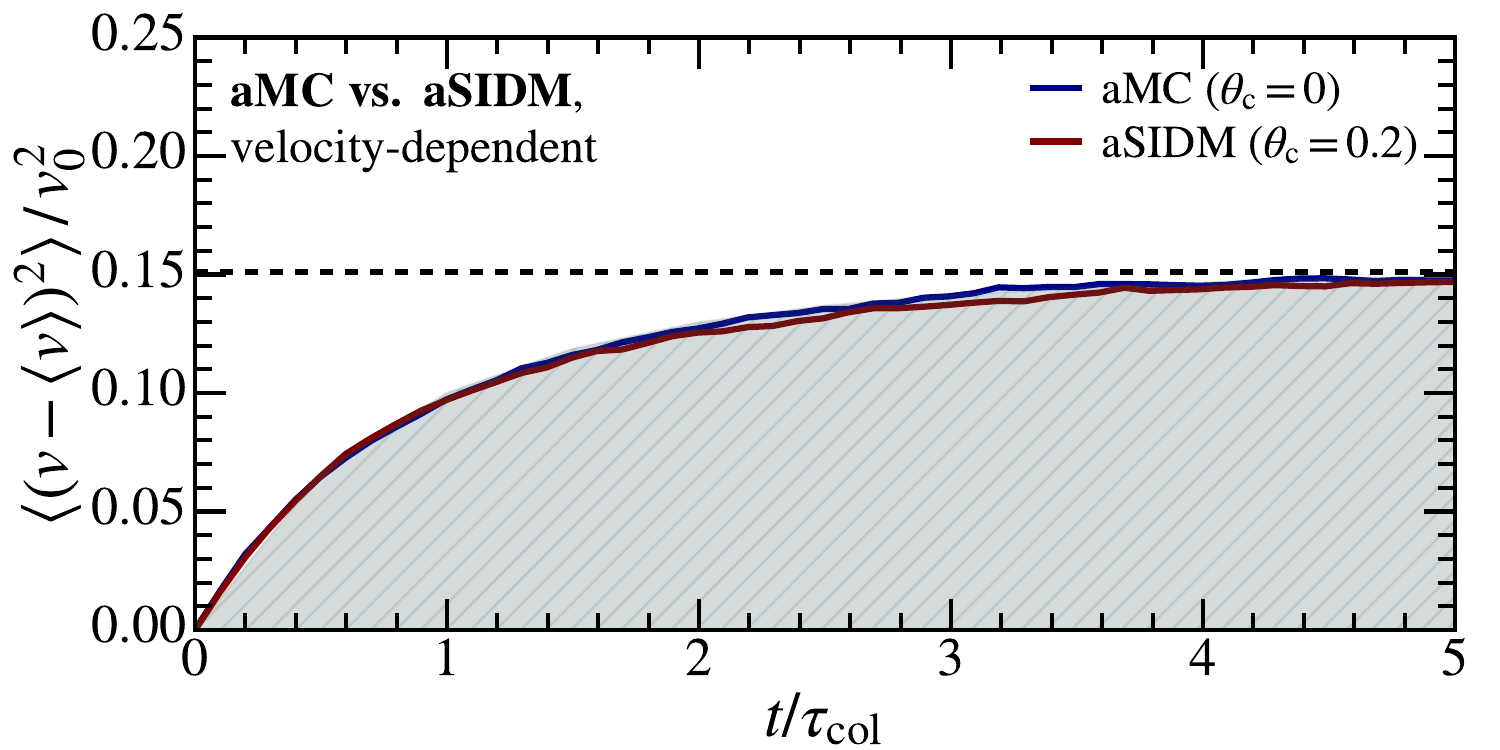}
\caption{Thermalization comparison of the velocity-dependent aMC and aSIDM implementations described in
\cref{sec:anisotropic_MC,sec:asidm}, using
the same test model with $\sigma_0/m_{\rm DM}=300\,\cpm$ and $w=20\,\kms$ as in
\cref{fig:thermalization_table_lookup}. The blue curve shows fully anisotropic aMC with $\theta_{\rm c}=0$, while the purple curve shows aSIDM with $\theta_{\rm c}=0.2$. The gray dashed line marks the Maxwell--Boltzmann expectation from \cref{eq:MB_distribution}, and the shaded band shows the semi-analytic Boltzmann prediction described in App.~B1 of~\cite{Zier:2026}. The two implementations exhibit indistinguishable relaxation towards thermal equilibrium and closely follow the semi-analytic solution, demonstrating that aSIDM reproduces the thermalization rate of the fully anisotropic velocity-dependent aMC treatment.}    \label{fig:thermalization_vd_anisotropic}
\end{figure}

\subsection{\label{sec: thermalization tests} Thermalization in a uniform box}

We use uniform-box thermalization tests, following the setup of~\cite{Fischer_2021a}, to validate the aMC, fSIDM and aSIDM scattering algorithms described in \cref{sec:anisotropic_MC,sec: fsidm,sec:asidm}, together with their velocity-dependent aMC and aSIDM extensions. The tests follow the relaxation of an initially mono-energetic particle distribution towards thermal equilibrium and thereby probe the cumulative effect of repeated self-interactions, complementing the angular deflection tests above. Gravity is disabled, so the evolution is driven entirely by elastic DM self-scattering. All particles are initialized with the same speed $v_0$ and isotropically distributed velocity directions. Self-interactions subsequently redistribute kinetic energy among the particles while conserving the total kinetic energy, driving the speed distribution towards the Maxwell--Boltzmann (MB) equilibrium distribution
\begin{equation}
\label{eq:MB_distribution}
f_{\rm MB}(v)
=
4\pi v^2
\left(
\frac{3}{2\pi v_0^2}
\right)^{3/2}
\exp\left(
-\frac{3v^2}{2v_0^2}
\right),
\end{equation}
which satisfies $\langle v^2\rangle=v_0^2$ and therefore has the same mean kinetic energy as the initial distribution.

Although the problem is formally scale-free, we adopt a periodic box of side length $10\,\kpc$ containing $N_{\rm part}=10^4$ particles with total mass $M=10^{10}\,\Msun$, corresponding to a uniform density $\rho=10^7\,\Msun\,\kpc^{-3}$. For the thermalization tests of the anisotropic SIDM algorithms described in \cref{sec:anisotropic_MC,sec: fsidm,sec:asidm}, all particles are initialized with speed $v_0=2\,\kms$, and we adopt $\sigma_{\rm T}/m_{\rm DM}=10\,\cpm$. We characterize the relaxation using the reference collision timescale
\begin{equation}
\label{eq:thermalize_box_collision_time}
\tau_{\rm col}
=
\frac{\sqrt{3\pi}}
{4 v_0 \left(\sigma_{\rm T}/m_{\rm DM}\right)\rho}
=
17.97\,\Gyr .
\end{equation}
Here, $4v_0/\sqrt{3\pi}$ is the mean relative speed of the equilibrium MB distribution with $\langle v^2\rangle=v_0^2$. This differs only slightly from the mean relative speed, $4v_0/3$, of the initial isotropic mono-speed distribution. We therefore use a constant $\tau_{\rm col}$ to characterize the evolution from the initial state to thermal equilibrium. All runs are evolved to $t_{\rm max}=100\,\Gyr$.

\Cref{fig:thermalization_anisotropic_iso_limit} shows the thermalization tests for the anisotropic SIDM scattering algorithms. The left panel compares the speed distributions at $t=0.5\,\tau_{\rm col}$ and $t=5.0\,\tau_{\rm col}$. At the earlier time, the distributions remain only partially thermalized and retain memory of the initial mono-energetic state, whereas by $5.0,\tau_{\rm col}$ they have broadened towards the Maxwell--Boltzmann prediction in \cref{eq:MB_distribution}, indicated by the yellow shaded region. The figure provides three consistency checks. Near the isotropic limit, $a=10^{-3}$, the aMC result agrees with isotropic MC (red). At finite anisotropy, $a=10^2$, aSIDM with $\theta_{\rm c}=0.2$ reproduces the corresponding fully anisotropic aMC calculation with $\theta_{\rm c}=0$ (green). In the strongly forward-dominated regime, $a=10^{10}$, aSIDM approaches the fSIDM result (blue), as expected when relaxation is dominated by frequent small-angle deflections. The right panel shows the corresponding evolution of the squared speed dispersion. In all three regimes, the different scattering treatments converge towards the same Maxwell--Boltzmann equilibrium (the gray dashed line) while conserving the total kinetic energy, demonstrating consistent thermalization in the isotropic, anisotropic and frequent-scattering limits.

\Cref{fig:thermalization_table_lookup} tests the
numerical robustness of the velocity-dependent
aMC implementation using an independent
light-mediator SIDM test model parametrized
by $\sigma_0/m_{\rm DM}=300\,\cpm$
and $w=20\,\kms$, as defined in
\cref{eq:sigma0_w_definition}. Because the differential cross section depends on both relative velocity and scattering angle, the implementation relies on tabulated velocity-dependent cross sections and their corresponding angular distributions. For a sufficiently resolved table, different lookup prescriptions should converge to the same underlying functions. As a robustness check at the adopted table resolution, we compare interpolation between neighbouring velocity bins, introduced in \arepoTwo, with nearest-bin lookup, each tested with $N_{\rm ngb}=16$, $32$ and $64$. The simulations use a periodic box of side length $100\,\kpc$ containing a total mass $M=10^{12}\,\Msun$, corresponding to a uniform density $\rho=10^6\,\Msun\,\kpc^{-3}$, with initial particle speed $v_0=40\,\kms$. We define a reference collision timescale by evaluating the Rutherford total cross section at the mean relative speed $4v_0/\sqrt{3\pi}$, yielding $\tau_{\rm col}\simeq2.33\,\Gyr$. The evolution of the normalized squared speed dispersion is essentially insensitive to both the lookup prescription and $N_{\rm ngb}$: all simulations closely follow the semi-analytic Boltzmann solution described in App.~B1 of~\cite{Zier:2026}\footnote{The solver is publicly available at \url{https://github.com/vinh-qtran/ThermBoxSIDM}} and approach the Maxwell--Boltzmann expectation at late times. This agreement shows that, for the adopted tabulation, the velocity-dependent aMC thermalization is insensitive to the lookup prescription and to the neighbour number over the range tested.

Building on this result, \cref{fig:thermalization_vd_anisotropic} compares the velocity-dependent aMC and aSIDM implementations for the same angle- and velocity-dependent light-mediator model as mentioned above. The fully anisotropic aMC simulation with $\theta_{\rm c}=0$ and the aSIDM simulation with $\theta_{\rm c}=0.2$ show indistinguishable relaxation towards thermal equilibrium and closely follow the semi-analytic prediction. This agreement demonstrates that aSIDM reproduces the thermalization rate of the fully anisotropic velocity-dependent aMC treatment.

\subsection{\label{sec: core collapse tests} Isolated core collapse}

\begin{figure}
    \centering
\includegraphics[width=1.00\columnwidth]{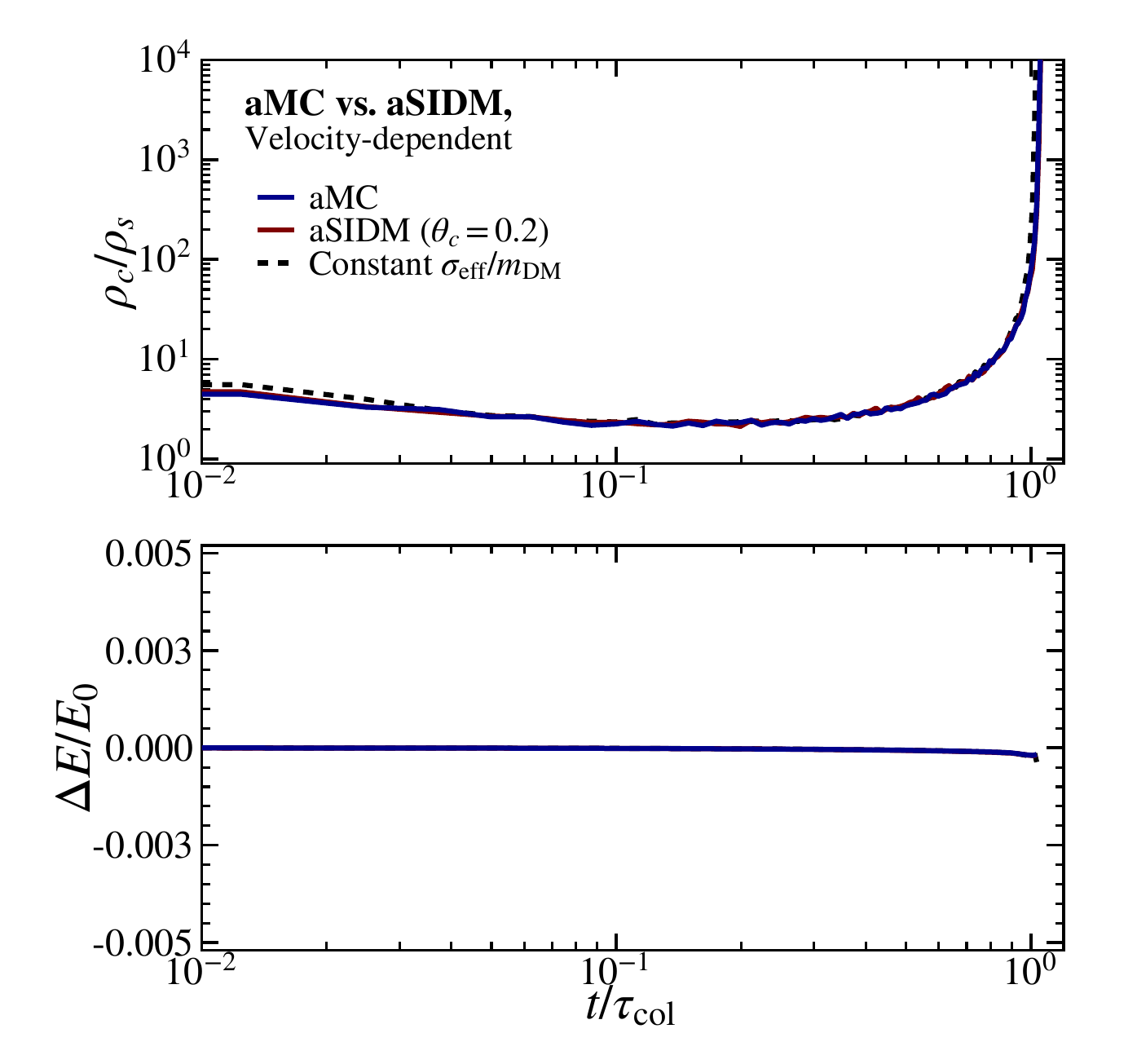}
    \caption{The evolution of the core density (top) and fractional energy deviation (bottom) of an isolated halo with mass $M_{200} = 10^{10} \, \Msun$ and concentration $c = 10$. The core density $\rho_{\rm c}$ is normalized to the NFW scale density $\rho_{\rm s}$, while the physical time $t$ is normalized to the collapse timescale $\tau_{\rm col}$ given in \cref{eq:collapse_timescale}. The halo evolutions using the aMC ($\theta_{\rm c} = 0$, red) and aSIDM ($\theta_{\rm c} = 0.2$, blue) implementations show good agreement. The collapse times of these haloes also agree well with that of the halo evolved under a velocity-independent cross section of $\sigma_{\rm eff} = 48.18\,\cpm$ (\cref{eq:sigma_eff}). The energy is well conserved in all simulations, with small oscillations around zero until core collapse. Such short-term oscillations are expected for leapfrog integration, which primarily suppresses long-term secular energy drift. After core collapse, the deviation is smaller than $0.1\%$.}
    \label{fig:isolated_halo}
\end{figure}

\begin{table*}
    \centering
    \setlength{\tabcolsep}{18pt}
    \renewcommand{\arraystretch}{1.25}

    \begin{tabular}{@{}lcl@{}}
        \hline\hline
        \textbf{Run}
        & \textbf{SIDM algorithm}
        & \textbf{Scattering treatment}
        \\
        \hline

        CDM
        & ---
        & Collisionless
        \\

        \hline
        \multicolumn{3}{c}{
        \textit{Velocity-independent SIDM:}
        $\sigma_{\rm T}/m_{\rm DM}=1\,{\rm cm^2\,g^{-1}}$
        }
        \\
        \hline

        MC
        & \hyperref[sec:mc_velocity_dep]{MC}
        & Isotropic
        \\

        aMC-iso
        & \hyperref[sec:anisotropic_MC]{aMC}
        & Isotropic limit ($a=10^{-3}$)
        \\

        aMC
        & \hyperref[sec:anisotropic_MC]{aMC}
        & Anisotropic ($a=10^{2}$)
        \\

        aSIDM
        & \hyperref[sec:asidm]{aSIDM}
        & Anisotropic ($a=10^{2}$, $\theta_c=0.2$)
        \\

        \hline
        \multicolumn{3}{c}{
        \textit{Velocity-dependent SIDM: resonant benchmark}
        }
        \\
        \hline

        vd-MC
        & \hyperref[sec:mc_velocity_dep]{MC}
        & Isotropic
        \\

        vd-aMC-iso
        & \hyperref[sec:anisotropic_MC]{aMC}
        & Isotropic limit
        \\

        vd-aMC
        & \hyperref[sec:anisotropic_MC]{aMC}
        & Anisotropic
        \\

        vd-aSIDM
        & \hyperref[sec:asidm]{aSIDM}
        & Anisotropic ($\theta_c=0.2$)
        \\

        \hline\hline
    \end{tabular}

    \caption{
    Overview of the nine DM-only cosmological box simulations used in
    \cref{fig:subhalos_profiles_const_cross,fig:subhalos_profiles_velocity_cross}.
    All runs contain $512^3$ DM particles with particle mass
    $m_{\rm DM}=4.6\times10^{6}\,\Msun$ and use a comoving
    Plummer-equivalent gravitational softening of $\epsilon=600\,{\rm cpc}$.
    The velocity-independent suite compares isotropic MC with the near-isotropic
    aMC limit, and fully anisotropic aMC with aSIDM. The velocity-dependent suite
    performs the corresponding comparisons for the resonant SIDM benchmark
    represented by the cyan curve in \cref{fig:cross_sections}. The prefix
    ``vd-'' denotes the velocity-dependent runs, while CDM provides the common
    collisionless reference.
    }
    \label{tab:cosmological_box_simulations}
\end{table*}

To test our various SIDM physics implementations in a fully self-gravitating astrophysical context, we simulate an isolated DM halo. The halo is initialized as a stable CDM configuration following \cite{Tran2024_1}, with the modifications of \cite{Tran2025_3} that ensure stability in the presence of finite gravitational softening. The corresponding initial-condition (IC) generator, \textsc{SoftIsoICs}, is publicly available\footnote{\url{https://github.com/vinh-qtran/SoftIsoICs}}. Within the virial radius $r_{200}$, i.e. the radius within which the average density is $200$ times the critical density of the Universe, the halo follows a classical NFW profile \citep{navarro1997universal}, and it transitions to a modified exponential cut-off \citep{Springel1999} beyond $r_{200}$,
\begin{align}
    \label{eq:nfw_inner_density}
    \rho_{r < r_{200}} (r) &= \frac{\rho_{\rm{s}}}{\left(r/r_{\rm s}\right) \left(1 + \left(r/r_{\rm s}\right)\right)^2}, \\
    \label{eq:nfw_outer_density}
    \rho_{r > r_{200}} (r) &= \frac{\rho_{\rm s}}{c \left(1 + c\right)^2} \left(\frac{r}{r_{200}}\right)^{\epsilon_{\rm d}} \exp\left(-\frac{r-r_{200}}{r_{\rm d}}\right).
\end{align}
We adopt a scale density $\rho_{\rm{s}}=5.69\times 10^{6}\,\Msun\,\kpc^{-3}$ and scale radius $r_{\rm{s}}=4.55\,\kpc$, corresponding to a virial mass (the mass enclosed within $r_{200}$) of $M_{200}=10^{10}\,\Msun$, and a concentration $c=r_{200}/r_s=10$; for simplicity we set the decay scale $r_{\rm d}=r_{200}$. Continuity of the logarithmic density slope at $r_{200}$ then fixes the decay index $\epsilon_{\rm d}$,
\begin{equation}
    \label{eq:nfw_decay_exponential}
    \epsilon_{\rm d} = \frac{r_{200}}{r_{\rm d}} - \frac{1+3c}{1+c}.
\end{equation}
We sample $N_{200}=10^6$ particles within $r_{200}$ and extend the sampling out to $r_{\rm max}=5\,r_{200}$. The resulting Plummer-equivalent gravitational softening, $\epsilon=0.054\,\kpc$, satisfies the convergence criteria of \cite{Mace2024}. 

The halo is evolved with the Rutherford-like cross section in \cref{eq:Rutherford_general}, using $\sigma_0/m_{\rm DM} = 80\,\cpm$ and $w = 120\,\kms$, where $\sigma_0$ and $w$ are determined from the underlying particle-physics parameters through \cref{eq:sigma0_w_definition}. We compare the core density evolutions in the aMC implementation with that in the aSIDM implementation with $\theta_{\rm c} = 0.2$. The core density is measured following \cite{Tran2025_2}. At each snapshot, we determine the halo centre using the shrinking-sphere method \cite{Power2003} and partition the halo into shells with log-linearly spaced radial bins $r_i$, merging adjacent shells as needed to ensure each contains a minimum bin count of $400$ particles. Beginning from the innermost radius $r_i$ at which the enclosed particle count reaches $N_i \ge 2500$, i.e.\ a minimum core count of $2500$, we compute the mean enclosed density $\bar\rho_i \pm \Delta\bar\rho_i$ under Poisson statistics. We identify $\bar\rho_i$ as the core density $\rho_{\rm c}$ at the first radius satisfying
\begin{equation}
    \label{eq:core_density}
    \lvert \bar\rho_i - \rho_{i+1} \rvert \, \geq 2 \left[\left(\Delta{\bar\rho_i}\right)^2+\left(\Delta{\rho_{i+1}}\right)^2\right]^{1/2},
\end{equation}
where $\rho_{i+1}$ and $\Delta{\rho_{i+1}}$ are the mean density of the shell between $r_i$ and $r_{i+1}$ and its associated Poisson uncertainty. If this criterion is not met, we advance to $r_{i+1}$ and repeat. A profile fit \citep[e.g.,][]{Yang2023,Tran2025_1} yields very similar results. 

We quote times in units of the characteristic gravothermal collapse time \citep{Essig2019,Yang2024},
\begin{equation}
    \label{eq:collapse_timescale}
    \tau_{\rm col} = \frac{150}{C_{\rm grav}} \frac{1}{\sigma_{\rm eff}/m} \frac{1}{\rho_{\rm{s}}} \left(\frac{1}{4 \pi \, G \, \rho_{\rm{s}} \, r_{\rm{s}}^2}\right)^{1/2}.
\end{equation} We take $C_{\rm grav}=0.8$ following the fluid model
calibration in~\cite{Zier:2026}. Following
\cite{Yang+2022}, we define the heat-conductivity-weighted effective cross section using the viscosity cross section,
\begin{equation}
\label{eq:sigma_eff}
\frac{\sigma_{\rm eff}}{m_{\rm DM}}
=
\frac{\left\langle
(\sigma_{\rm V}/m_{\rm DM})v_{\rm rel}^5
\right\rangle}
{\left\langle v_{\rm rel}^5\right\rangle}
\simeq 48.18\,\cpm.
\end{equation}
Here, $\sigma_{\rm V}$ is defined using the normalized
angular weight $w_{\rm V}(\theta)=3\sin^2\theta/2$
in \cref{eq:weighted_cross_sections}, and the brackets
denote an average over the Maxwellian relative-speed
distribution, with an effective one-dimensional velocity
dispersion of
$\sigma_{\rm 1D,eff}
=1.1\sqrt{G\rho_{\rm s}r_{\rm s}^2}
\simeq24.76\,\kms$ \citep{Tran2025_2}.
For comparison, using $\sigma_{\rm tot}$ instead gives
$49.39\,\cpm$. We also evolve the halo with the
velocity-independent effective cross section
$\sigma_{\rm eff}/m_{\rm DM}=48.18\,\cpm$
as a benchmark for comparison.

Figure \ref{fig:isolated_halo} shows the core density and fractional energy deviation evolutions of the three haloes evolved using the aMC and aSIDM implementations, as well as under the velocity-independent cross section of 
$\sigma_{\rm eff}/m_{\rm DM}=48.18\,\cpm$. We observe that the two anisotropic cases yield nearly identical results, and that the halo collapse times remain consistent with that of the velocity-independent cross section run. Energy conservation also remains at the $\sim 0.01\%$ level across all three cases until core collapse, when the deviation increases to $\sim 0.1\%$.

\begin{figure*}
    \centering
    \includegraphics[width=0.93\textwidth]{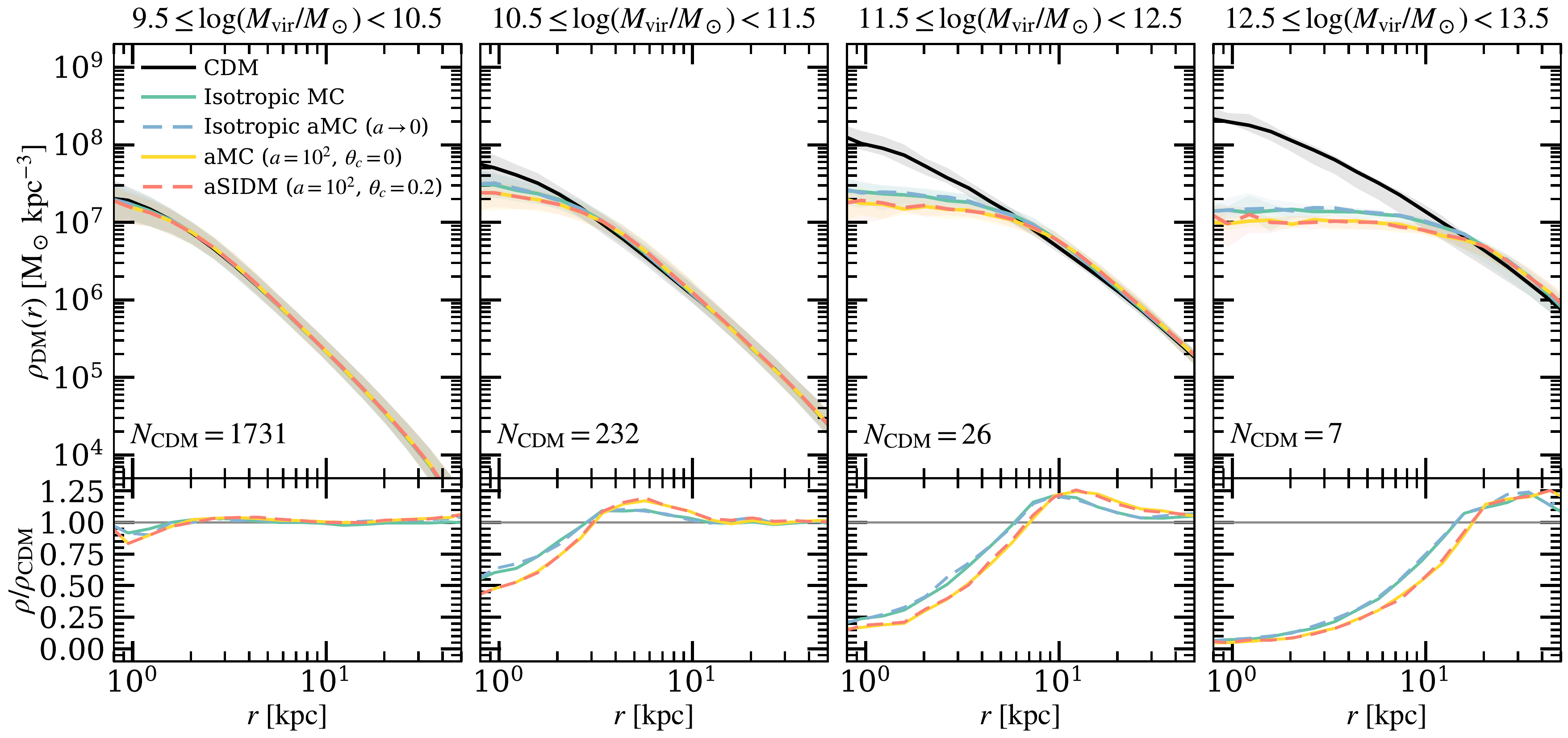} 
\caption{Median DM density profiles at $z=0$ in five virial-mass bins of width $\Delta\log(M_{\rm vir}/\Msun)=1$ from the DM-only $512^3$ simulations. Each panel is labelled by the corresponding mass bin and the number of CDM haloes, $N_{\rm CDM}$. We show CDM (black) and compare the isotropic MC calculation (light green solid) with the near-isotropic aMC result for $a=10^{-3}$ (light blue dashed), as well as the fully anisotropic aMC calculation with $a=10^2$ and $\theta_{\rm c}=0$ (yellow solid) with the corresponding aSIDM calculation using $\theta_{\rm c}=0.2$ (light red dashed). The upper panels show the median profiles with 16th--84th percentile bands, while the lower panels show the ratio to CDM. All SIDM runs use the velocity-independent normalization $\sigma_{\rm T}/m_{\rm DM}=1\,{\rm cm^2\,g^{-1}}$. The isotropic MC and near-isotropic aMC profiles agree closely, as do the fully anisotropic aMC and aSIDM profiles, while resolved central density reductions relative to CDM are visible in the higher-mass bins.}
\label{fig:subhalos_profiles_const_cross}
\end{figure*}

\begin{figure*}
    \centering
    \includegraphics[width=0.93\textwidth]{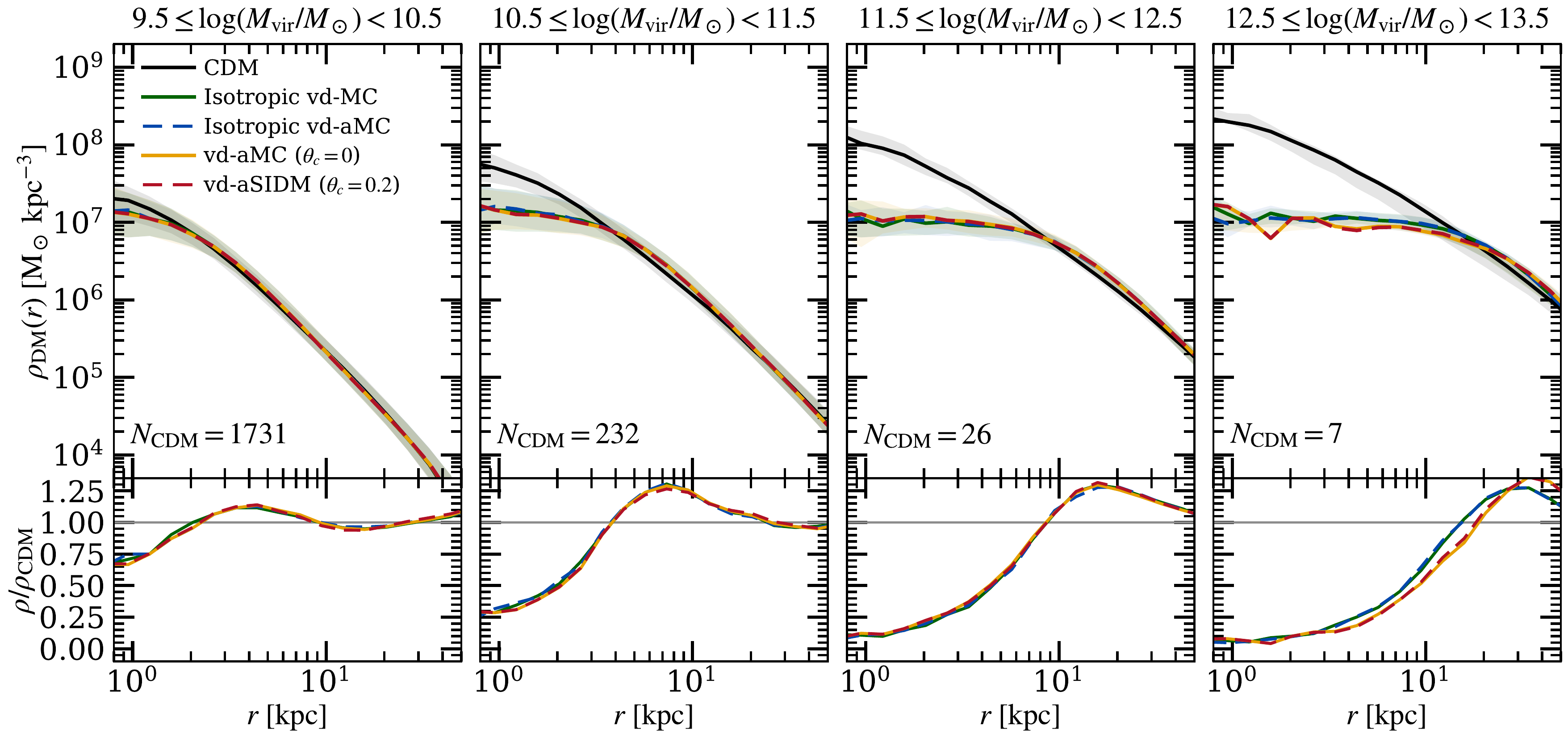}  
    \caption{Median DM density profiles at $z=0$ for the velocity-dependent resonant SIDM benchmark corresponding to the cyan curve in \cref{fig:cross_sections}, using the same halo-mass bins and simulation setup as in \cref{fig:subhalos_profiles_const_cross}. We show CDM (black) and compare the isotropic vd-MC simulation (green) with the near-isotropic vd-aMC result (blue dashed), as well as the fully anisotropic vd-aMC calculation (orange) with the corresponding vd-aSIDM calculation using $\theta_{\rm c}=0.2$ (dark red dashed). The upper panels show the median profiles with 16th--84th percentile bands, while the lower panels show the ratio to CDM. The isotropic vd-MC and near-isotropic vd-aMC profiles agree closely, demonstrating that vd-aMC recovers the isotropic limit for a velocity-dependent interaction, while the fully anisotropic vd-aMC and vd-aSIDM profiles likewise agree across the resolved halo-mass range. The more massive haloes develop central cores relative to CDM, and in the highest-mass bin the anisotropic profiles show a modest shift relative to the isotropic results, where the benchmark reaches $a\sim\mathcal{O}(10)$ at characteristic relative velocities of $v_{\rm rel}\sim\mathcal{O}(100)\,\kms$.} \label{fig:subhalos_profiles_velocity_cross}
\end{figure*}

\subsection{\label{sec:cosmological_DMO_simulations} Cosmological DM-only box simulations}

As a cosmological validation test, we use a suite of DM-only periodic-box simulations to compare halo density profiles across velocity-independent and velocity-dependent SIDM models with both isotropic and anisotropic scattering. These simulations test the velocity-dependent scattering implementation described in \cref{sec:imp_velocity_dep_SIDM} together with the angular scattering algorithms introduced in \cref{sec:implementation_of_angular_SIDM}. In particular, they allow us to verify that physically equivalent isotropic treatments produce consistent halo profiles, that the fully anisotropic MC (aMC) and angular-hybrid SIDM (aSIDM) schemes agree when applied to the same anisotropic interaction, and that the implementations recover the expected formation of SIDM cores in cosmological haloes.

\begin{figure*}
    \centering
    \includegraphics[width=1.00\columnwidth]{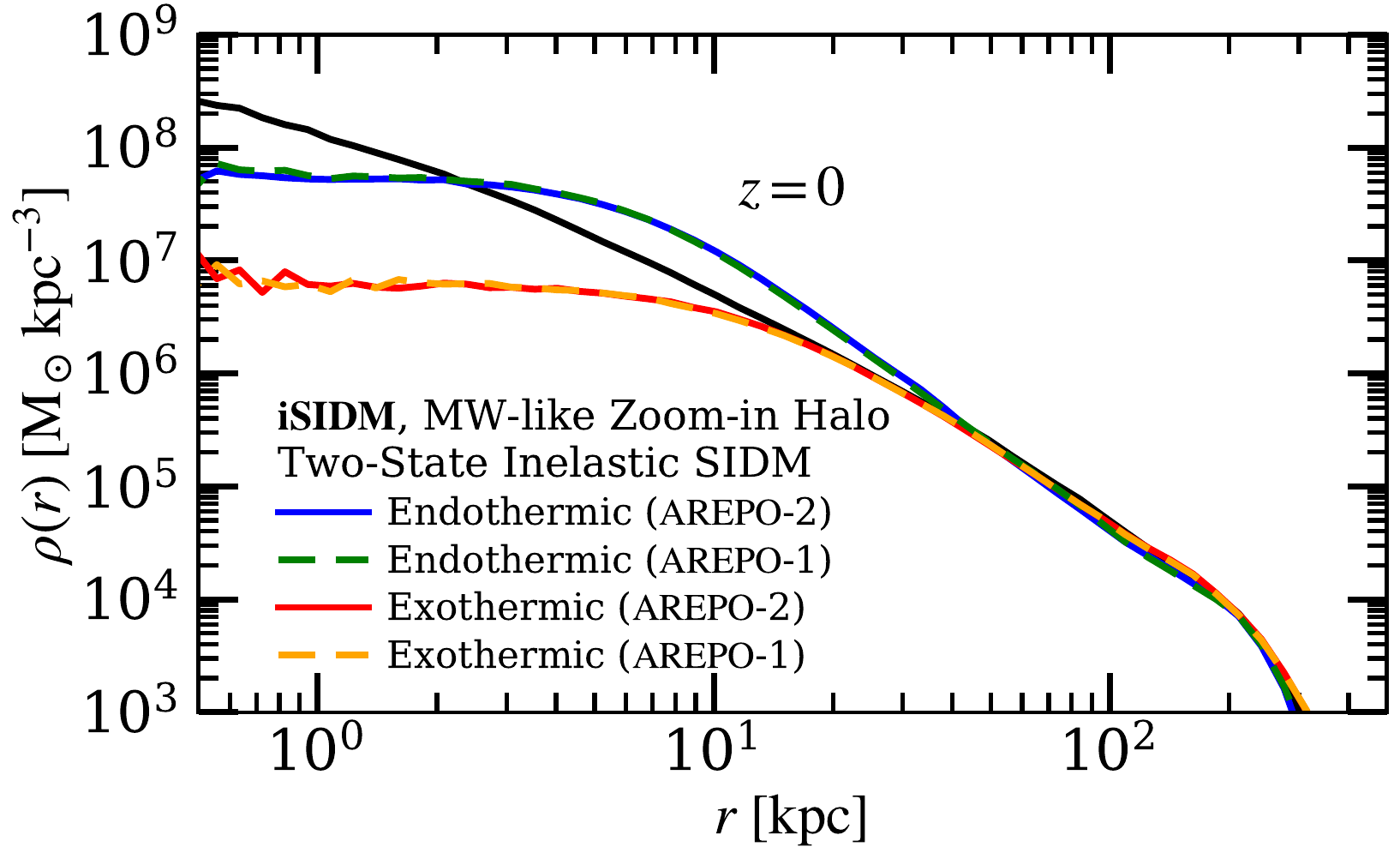}  
        \includegraphics[width=1.00\columnwidth]{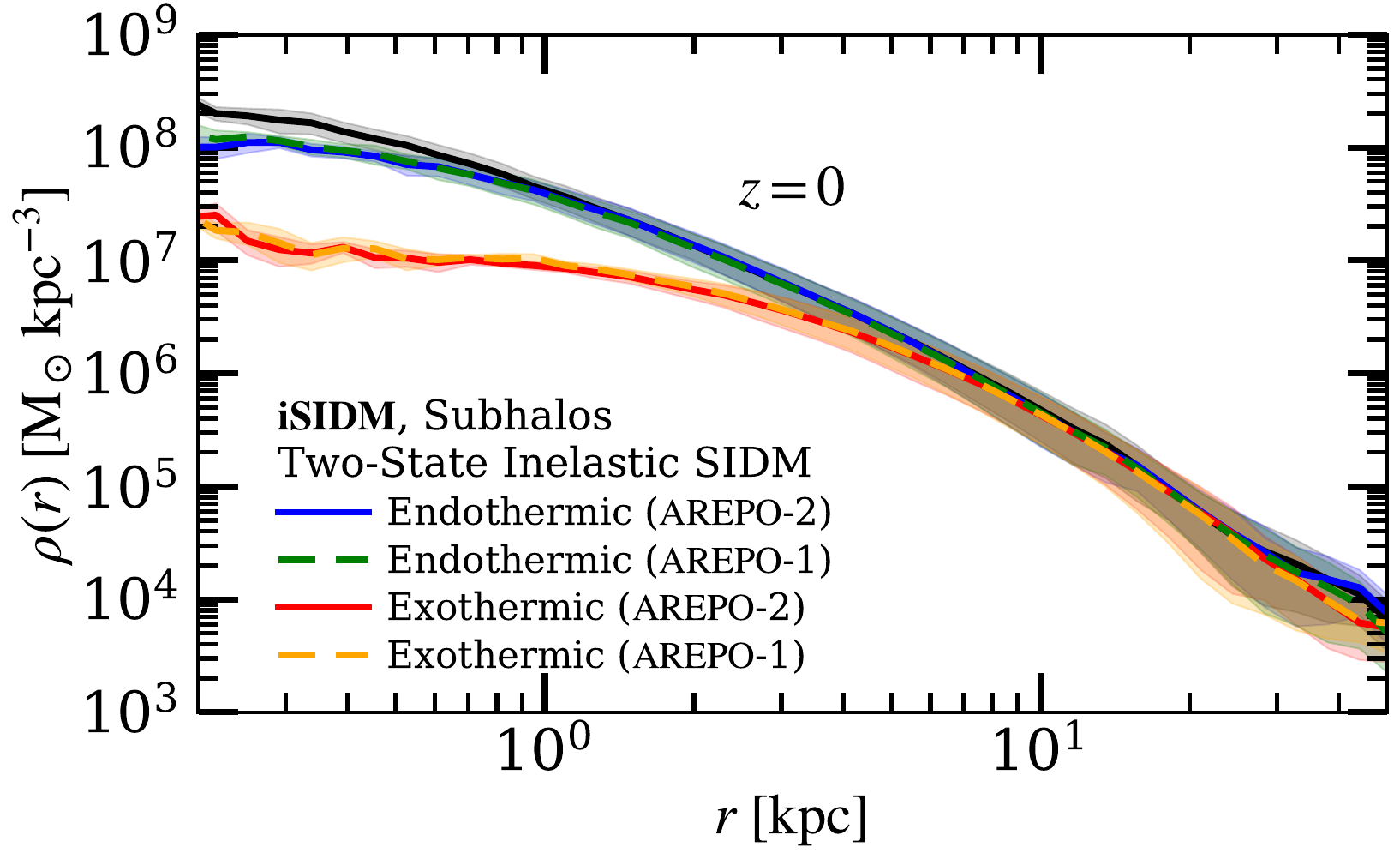}  
\caption{DM density profiles at $z=0$ from the Milky Way-like DM-only zoom-in simulations of the two-state inelastic SIDM benchmarks described in \cref{sec: inelastic_sidm_models}. \textit{Left:} density profile of the main halo for CDM (black line) and the endothermic and exothermic SIDM models. \textit{Right:} median density profile of the 15 most massive subhaloes, with shaded regions showing the 16th--84th percentiles. Solid curves show the new \arepoTwo iSIDM implementation described in \cref{sec:imp_inelastic_SIDM}, while dashed curves show the corresponding \arepoOne benchmark results of~\cite{ONeil:2022szc}, using the same velocity-dependent inelastic cross sections shown in \cref{fig:inelastic_cross_sections}. The exothermic model produces lower central densities through energy release in down-scattering, whereas the endothermic profiles remain closer to CDM, particularly for the subhaloes. The close agreement between \arepoOne and \arepoTwo demonstrates that the new iSIDM implementation reproduces the benchmark density profiles of both the main halo and its massive subhaloes.}
\label{fig:main_and_subhalo_density_z0}
\end{figure*}

\begin{figure}
    \centering
\includegraphics[width=0.975\columnwidth]{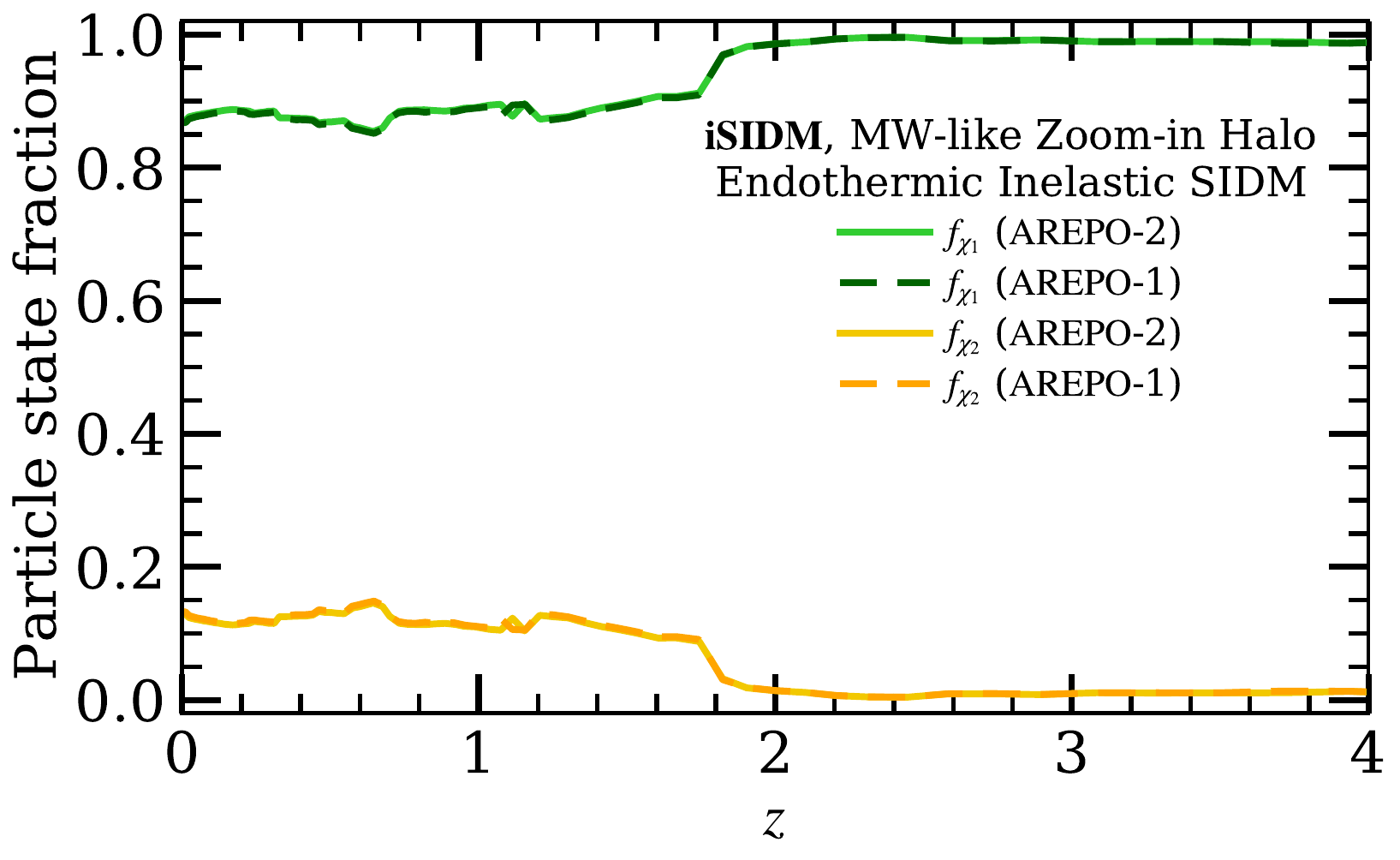} 
\caption{Evolution of the internal-state fractions in the main halo for the endothermic two-state SIDM benchmark described in \cref{sec: inelastic_sidm_models}. Green and yellow curves show the ground-state fraction, $f_{\chi_1}$, and excited-state fraction, $f_{\chi_2}$, respectively. Solid curves show the new \arepoTwo iSIDM implementation, while dashed curves show the corresponding \arepoOne benchmark results of~\cite{ONeil:2022szc}. Up-scattering progressively populates the excited state, with more rapid changes during periods of enhanced density and velocity dispersion, most prominently during the major merger at $z\simeq2$. The close agreement between \arepoOne and \arepoTwo demonstrates that the new iSIDM implementation reproduces the benchmark evolution of the internal-state populations.}
\label{fig:endothermic_state_fractions}
\end{figure}

\begin{figure}
    \centering
    \includegraphics[width=0.975\columnwidth]{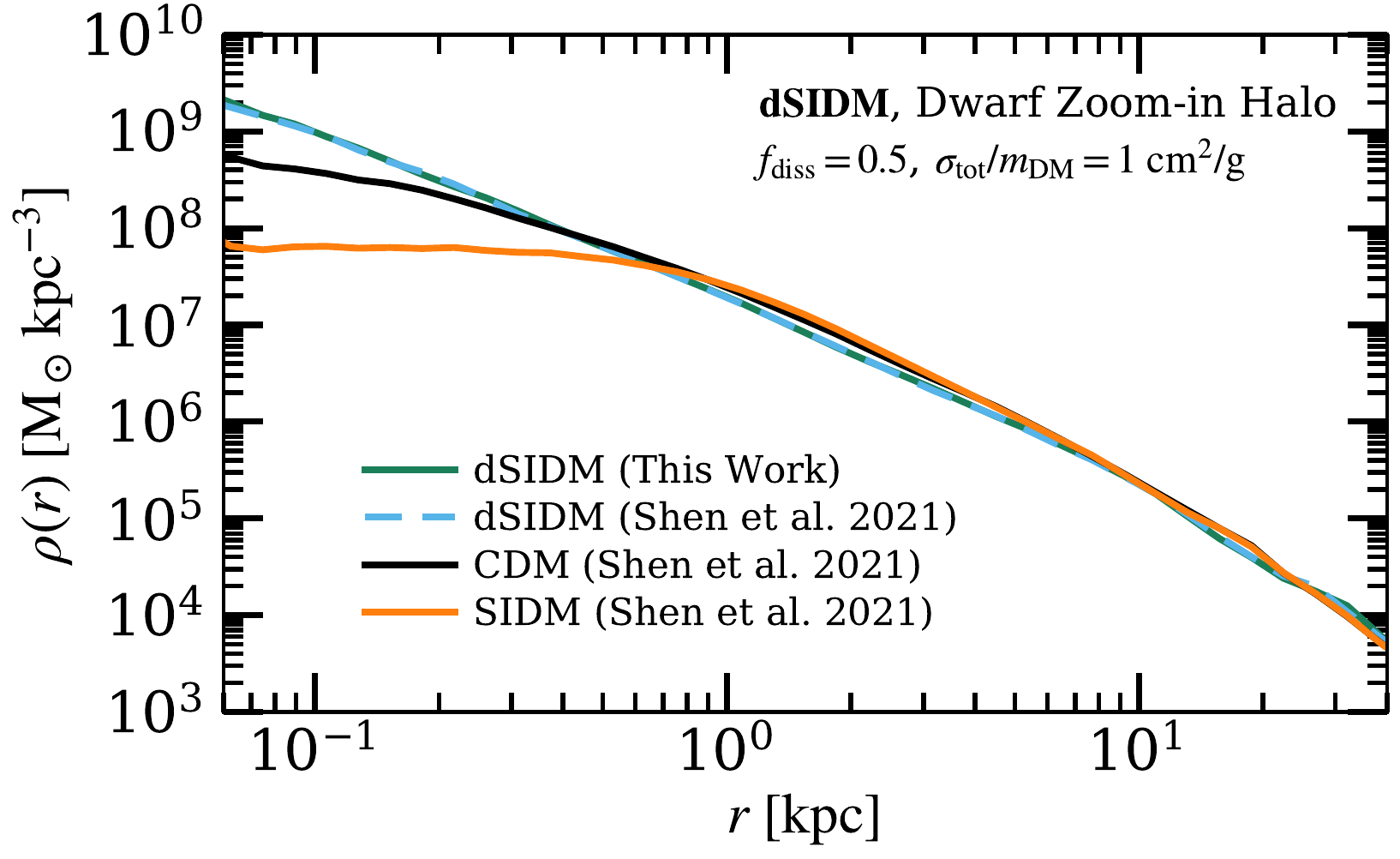}
\caption{DM density profiles at $z=0$ from DM-only cosmological zoom-in simulations of the m10q dwarf halo of~\cite{Shen:2021frv}. The solid and dashed green curves show the constant dSIDM model with $\sigma/m_{\rm DM}=1\,{\rm cm^2\,g^{-1}}$ and $f_{\rm loss}=0.5$, evolved with the new \arepoTwo implementation and the original \gizmo implementation, respectively. The corresponding CDM (black) and elastic SIDM (orange) profiles from~\cite{Shen:2021frv} are shown for reference. Elastic SIDM produces a central core through collisional heat transport, whereas dissipative scattering removes kinetic energy and drives halo contraction, resulting in a substantially steeper central density profile. The close agreement between the \arepoTwo and \gizmo dSIDM profiles demonstrates that the new constant dSIDM implementation reproduces the published benchmark halo structure.}
    \label{fig:dSIDM_test}
\end{figure}

We use the same simulation setup as in the performance study of the new \arepoTwo\ SIDM module in our companion work~\cite{Zier:2026}.
We fix the comoving box size to $L=25\,{\rm cMpc}$ and evolve all runs to $z=0$. Initial conditions are generated at $z=49$ with \textsc{N-GenIC}, distributed with \textsc{GADGET-4}~\citep{springel2021simulating}, for a \textsc{Planck}\,2018 cosmology~\citep{Planck2018Parameters}: $\Omega_\Lambda=0.69035856$, $\Omega_{\rm m}=0.30964144$, $\Omega_{\rm b}=0.04897468$, $h=0.6766$, $\sigma_8=0.8102$ and $n_{\rm s}=0.9655$. We use second-order Lagrangian perturbation theory together with the fixed-and-paired variance-suppression technique~\citep{Angulo2016} to reduce cosmic variance. All runs share the same input linear power spectrum and the same random seed, and therefore have identical large-scale phases. This isolates the impact of the SIDM scattering model and its numerical implementation. We neglect any back-reaction of SIDM on the linear power spectrum at these redshifts. The nine simulations and their corresponding scattering algorithms and model configurations are summarized in \cref{tab:cosmological_box_simulations}.

As a verification test, we compute the median DM density profiles of haloes at $z=0$ in five virial-mass bins of width $\Delta\log(M_{\rm vir}/\Msun)=0.5$. The profiles are measured from the $512^3$ runs and are shown relative to the corresponding CDM result. This comparison tests whether the different SIDM implementations produce the expected core formation in cosmological haloes and whether physically equivalent setups yield consistent profiles.

\Cref{fig:subhalos_profiles_const_cross} shows the velocity-independent comparison for $\sigma_{\rm T}/m_{\rm DM}=1\,{\rm cm^2\,g^{-1}}$. The isotropic MC calculation agrees closely with the near-isotropic aMC result, while the fully anisotropic aMC and aSIDM calculations likewise produce consistent profiles across the resolved halo-mass range. These comparisons demonstrate that aMC recovers the isotropic limit and that aSIDM reproduces the corresponding fully anisotropic MC evolution in cosmological haloes. Relative to CDM, the three highest-mass bins develop resolved central cores, whereas no comparable modification is visible in the lower-mass bins. This trend reflects the dependence of the local scattering rate on both halo density and characteristic relative velocity, together with the different dynamical evolution of haloes across the mass range.

\Cref{fig:subhalos_profiles_velocity_cross} shows the corresponding comparison for the velocity-dependent resonant SIDM model with the cross section shown by the cyan curve in \cref{fig:cross_sections}. The isotropic vd-MC simulation agrees closely with the near-isotropic vd-aMC result across the resolved radial and halo-mass range, demonstrating that vd-aMC recovers the isotropic limit also for a velocity-dependent interaction. Likewise, the fully anisotropic vd-aMC and vd-aSIDM calculations produce consistent density profiles, showing that the hybrid treatment reproduces the corresponding fully anisotropic evolution when both the scattering rate and angular distribution depend on relative velocity. The more massive haloes develop central cores relative to CDM, while in the highest-mass bin the anisotropic profiles are shifted slightly relative to the isotropic results. This follows the same qualitative trend as in \cref{fig:subhalos_profiles_const_cross}, but the difference is smaller because the SIDM benchmark model is less strongly anisotropic at the characteristic velocities of these haloes, with $a\sim10$ at $v_{\rm rel}\sim100\,{\rm km\,s^{-1}}$, compared with $a=100$ in the velocity-independent benchmark.

\subsection{\label{sec: cosmological zoom-in simulations} Cosmological zoom-in simulations}

We further validate the new implementations using cosmological DM-only zoom-in simulations by reproducing two previously published SIDM benchmarks. We first test the multi-state inelastic SIDM implementation using the Milky Way-like zoom-in simulations of~\cite{ONeil:2022szc}, comparing directly with the original \arepoOne results. We then test the constant dissipative SIDM implementation using the dwarf-galaxy zoom-in benchmark of~\cite{Shen:2021frv}, comparing with the corresponding \gizmo simulation. These tests extend the controlled and periodic-box validations above to fully self-gravitating cosmological haloes with realistic assembly histories and resolved substructure, providing direct cross-code benchmarks of the inelastic and dissipative implementations.

\subsubsection{\label{sec:MW_like_zoom_in_sims} Milky Way-like zoom-in halo simulations}

We validate the multi-state inelastic SIDM (iSIDM) implementation described in \cref{sec:imp_inelastic_SIDM} using the DM-only cosmological zoom-in initial conditions of~\cite{ONeil:2022szc}. The simulations follow a Milky Way-like halo with $M_{200}\simeq2\times10^{12}\,\Msun$ at $z=0$. The high-resolution region contains the main halo and its surrounding substructure, while lower-resolution particles sample the large-scale environment and interact only gravitationally. We rerun the endothermic and exothermic two-state SIDM benchmarks with \arepoTwo using the same velocity-dependent inelastic cross sections shown in \cref{fig:inelastic_cross_sections}, enabling a direct comparison with the original \arepoOne calculations.

\Cref{fig:main_and_subhalo_density_z0} compares the $z=0$ density profile of the main halo and the median profile of its 15 most massive subhaloes. The \arepoTwo results closely reproduce the corresponding \arepoOne profiles for both the endothermic and exothermic benchmarks, for the host halo as well as its substructure. The exothermic model develops a lower-density central region, consistent with down-scattering converting internal-state energy into kinetic energy. The endothermic model retains a higher central density than the exothermic case, while its subhalo profiles remain closer to CDM, consistent with the reduced efficiency of threshold-suppressed up-scattering at the lower characteristic velocities of subhaloes. The agreement between the two codes demonstrates that the new iSIDM implementation reproduces the benchmark halo structure across both host- and subhalo scales.

We further test the evolution of the internal-state populations using the endothermic benchmark. \Cref{fig:endothermic_state_fractions} shows the ground- and excited-state fractions of particles in the main halo as a function of redshift. The state fractions remain nearly constant until up-scattering becomes sufficiently efficient to populate the excited state. Periods of enhanced density and velocity dispersion increase the interaction rate and drive more rapid changes in the state populations, most notably during the major merger at $z\simeq2$. The close agreement between the \arepoTwo and \arepoOne results demonstrates that the new implementation reproduces not only the resulting halo density structure, but also the underlying evolution of the internal-state populations.

\subsubsection{\label{sec:dwarf_zoom_in_sims} Dwarf galaxy zoom-in halo simulations}

We validate the constant dissipative SIDM (dSIDM) implementation described in \cref{sec:imp_dissipative_SIDM} using the DM-only cosmological zoom-in initial conditions of the \texttt{m10q} dwarf-halo benchmark of~\cite{Shen:2021frv}. This isolated, early-forming dwarf halo reaches a mass of $M_{\rm halo}\simeq7.5\times10^{9}\,\Msun$ at $z=0$. To enable a direct comparison with the published benchmark, we adopt the same constant cross section, $\sigma/m_{\rm DM}=1\,{\rm cm^2\,g^{-1}}$, and dissipative energy-loss fraction, $f_{\rm loss}=0.5$.

\Cref{fig:dSIDM_test} compares the resulting $z=0$ DM density profile from the new \arepoTwo implementation with the original \gizmo result of~\cite{Shen:2021frv}. The corresponding CDM and elastic SIDM profiles are shown for reference. Elastic self-interactions redistribute heat within the halo and produce a central core, whereas dissipative scattering removes kinetic energy from the DM and drives contraction, resulting in a substantially steeper and more centrally concentrated density profile. The close agreement between the \arepoTwo and \gizmo dSIDM profiles demonstrates that the new constant dSIDM implementation reproduces the published benchmark halo structure in a fully cosmological, self-gravitating dwarf-halo simulation.

\section{Discussion}
\label{sec:discussion}

We discuss the advances over \arepoOne, the scope and validity of the implemented algorithms, and the degeneracies in the theory-to-simulation mapping (\cref{sec:pp_to_simulation_inputs}). We then examine numerical convergence and the interplay between cosmological initial conditions, late-time DM interactions and baryonic physics, and discuss the prospects for inferring microscopic DM properties from astrophysical observations.

\textit{Advances relative to \arepoOne.} The original SIDM implementation in \arepoOne\ couples the SIDM neighbour search to the gravitational tree~\citep{Vogelsberger_2012,vogelsberger2019evaporating}. Our companion paper~\citep{Zier:2026} introduced a dedicated SIDM neighbour tree and a communication scheme supporting multiple interactions per timestep. Building on this infrastructure, the present work implements anisotropic Monte Carlo scattering (aMC), frequent small-angle drag--diffusion (fSIDM) and the angular-hybrid scheme (aSIDM), together with state-dependent multi-state interactions, generalized dissipative scattering and a dedicated DM decay module. The angular schemes build on earlier anisotropic and hybrid approaches~\citep{Robertson_2017b,Fischer_2021a,Arido:2024hfk}; for strongly forward-peaked interactions, aSIDM achieves speed-ups of up to $\sim100\times$ over aMC in our deflection tests (\cref{fig: cpu plot}). The mapping in \cref{sec:pp_to_simulation_inputs} provides common simulation inputs for these implementations and links them to the model classes illustrated in \cref{fig:map_pp_to_effective} and summarized in \cref{tab:sidm_model_overview}. The applicability to any given particle-physics model depends on whether all its relevant interactions and initial-condition effects are represented.

\textit{Inverting the mapping.} The inverse mapping $\mathcal{P}_{\rm SIM}\rightarrow\mathcal{P}_{\rm PP}$ is generally non-unique, since different microscopic DM models can produce similar effective interaction properties. Furthermore, different interaction properties can lead to similar astrophysical signatures, introducing additional degeneracies when constraining DM physics from observations. Observations of dwarf galaxies, satellite systems, galaxies, groups and clusters, including galaxy and cluster mergers, probe a broad range of relative velocities and dynamical environments (\cref{sec:simulation_diagnostics_DM}). Combining these observations may help distinguish models with similar predictions at one scale but different velocity dependences at others (\cref{fig:cross_sections}). Our mapping allows these predictions to be derived consistently from the same microscopic model. Independent constraints from early-Universe cosmology and particle-physics experiments probe additional aspects of the underlying theory and may help resolve degeneracies that remain in structure-formation observables, as discussed in \cref{sec:interplay_constraints,app:model_constraints}.

\textit{Anisotropic scattering and effective transport cross sections.} Whether anisotropic DM scattering can be approximated by isotropic scattering with an angle-weighted cross section depends on both the dynamical regime and the observable. For elastic scattering in approximately relaxed haloes, matching to the effective cross section can reproduce gravothermal evolution more accurately than matching to the transfer cross section as shown in \cref{fig:isolated_halo}. However, a single transport cross section discards angular information (\cref{sec:relevant_cross_section}) that can affect momentum transfer and evaporation in satellite--host interactions and galaxy or cluster mergers~\citep{Kahlhoefer_2014,Robertson_2017b,Arido:2024hfk}. The agreement between aMC and aSIDM in our tests (\cref{sec: verification tests}) supports the accuracy of the hybrid treatment for the differential cross sections and regimes tested, but does not validate their replacement by isotropic scattering. Systematic comparisons at fixed $\sigma_{\rm eff}(v_{\rm rel})$ but different angular distributions are therefore needed to assess the limits of transport approximations and the sensitivity of observables such as DM--galaxy offsets in mergers and satellite mass loss to the full ${\rm d}\sigma/{\rm d}\Omega$.

\textit{Choice of the critical angle in aSIDM.} The critical angle $\theta_{\rm c}$ is a numerical parameter separating the drag--diffusion and aMC treatments, rather than a property of the microscopic interaction. Physical predictions should therefore be insensitive to its value within the accuracy of the hybrid approximation. Increasing $\theta_{\rm c}$ shifts more scattering into the drag--diffusion treatment and can reduce computational cost, but our deflection tests also show an increase in the minimum agreement time $t_{\rm min}(\theta_{\rm c})$ for the angular distributions (\cref{fig:time_criterion_Rutherford,fig: cpu plot}). Values $\theta_{\rm c}\sim0.1$--$0.3$ radians provide a useful balance between accuracy and efficiency in the tested configurations, but do not define a universal range for astrophysical applications. The choice must account for the differential cross section, local interaction rate and physical timescales relevant to the observables, with accuracy checked through convergence tests in $\theta_{\rm c}$.

\textit{Multi-state and inelastic SIDM.}
Multi-state SIDM couples the evolution of internal-state populations to halo dynamics: the local abundances determine the rates of the available reaction channels, while inelastic transitions modify these populations and exchange energy between internal and kinetic degrees of freedom. This coupling cannot generally be captured by a single effective velocity-dependent cross section. Our state-dependent smoothing lengths provide separate estimates of the relevant incoming-state densities, as verified by the multi-state deflection test (\cref{sec:imp_multi_states,fig:multi_sidm_deflection_test}). The cosmological zoom-in simulations reproduce previously published endothermic and exothermic halo density profiles and the evolution of internal-state fractions, including enhanced conversion during merger activity~\citep{ONeil:2022szc} (\cref{fig:main_and_subhalo_density_z0,fig:endothermic_state_fractions}). An important next step is to investigate how inelastic scattering, evolving state populations and decays of unstable DM states jointly influence halo structure and evolution across different masses and dynamical environments.

\textit{Velocity-dependent dissipation.}
Dissipative SIDM modifies halo evolution through the competition between collisional heat transport and energy loss. For identical DM particles of mass $m_{\rm DM}$ undergoing dissipative two-body scattering, the local volumetric cooling rate is $\mathcal{C}=\rho^2\Lambda_{\rm eff}$, with $\Lambda_{\rm eff}=\frac{1}{8}\langle[\sigma_{\rm diss}(v_{\rm rel})/m_{\rm DM}]f_{\rm loss}(v_{\rm rel})v_{\rm rel}^3\rangle$, averaged over the local relative-velocity distribution~\cite{Essig2019,Schmidt:2026hmn}. Here, $f_{\rm loss}$ denotes the fraction of the incoming relative kinetic energy removed per dissipative collision. For constant cross sections and energy-loss fractions, the cooling rate scales with $\langle v_{\rm rel}^3\rangle$, whereas velocity dependence and kinematic thresholds can substantially modify this behaviour. For example, dark-atomic hyperfine excitation produces an energy-loss fraction $f_{\rm loss}\propto v_{\rm rel}^{-2}$ above threshold (\cref{eq:dark_atom_diss_energ_loss_fraction}), corresponding to an approximately fixed energy loss per collision when followed by radiative de-excitation. The competition between dissipative cooling and heat transport can alter the velocity dispersion structure and accelerate central contraction and gravothermal collapse~\cite{Essig2019,Shen:2021frv,Schmidt:2026hmn}. Our constant and velocity-dependent dissipative implementations reproduce analytical scattering and energy-loss predictions in controlled tests (\cref{fig:vd_dsidm_energy_validation}), while the constant-dSIDM cosmological zoom-in reproduces a published benchmark~\citep{Shen:2021frv} (\cref{fig:dSIDM_test}). These implementations provide a basis for investigating how velocity- and channel-dependent dissipation modifies gravothermal evolution.

\textit{Scope and limitations of the implementations.} 
Although our \arepoTwo implementations cover a broad range of SIDM interactions and DM decay processes, they do not capture all the physics of the dark-sector models considered in this work. For dark atoms and molecules, selected scattering, dissipative and decay processes are supported, but collisional ionisation, late-time recombination and charged-plasma
dynamics are not (\cref{fig:map_pp_to_effective,tab:sidm_model_overview}). The scattering algorithms treat two dynamically evolved DM particles per event, with radiative losses represented through an effective energy-loss prescription. Processes that change the number of
dynamically evolved particles, such as annihilation into untracked products or radiative bound-state formation, require additional numerical treatment. Likewise, modified initial conditions alone
do not capture the nonlinear wave dynamics of fuzzy DM, which require a wave-based or equivalent dynamical treatment
~\citep{Schive:2014dra,Mocz:2017wlg,Mocz:2019pyf}.
The applicability of the present implementations therefore depends on which processes are dynamically relevant in the model and astrophysical environment under consideration.

\textit{Numerical convergence.} Reliable predictions of nonlinear structure formation require systematic convergence studies across the different SIDM algorithms and interaction regimes. The relevant numerical parameters include particle mass resolution, gravitational softening $\epsilon_{\rm soft}$, the gravitational timestep accuracy parameter $\eta$ (\cref{eq:dt_grav}), the scheme-dependent SIDM timestep parameters $C$ (\cref{sec: time-stepping criterion}), SIDM neighbour number $N_{\rm ngb}$, kernel choice and gravitational tree-force accuracy. Additional considerations include the critical angle $\theta_{\rm c}$ in aSIDM and the resolution and interpolation of tabulated velocity- and angle-dependent cross sections (\cref{sec:cross_section_tables,sec:angular_cross_section_tables}). Gravothermal core collapse provides a particularly stringent convergence test, as increasing central densities and scattering rates place increasingly demanding requirements on gravitational integration and collisional heat transport~\cite{Mace2024}. Extending convergence studies to velocity-dependent, anisotropic, multi-state and dissipative SIDM is therefore essential. Convergence should be assessed through the evolution of central density, velocity dispersion and core-collapse time, together with appropriate energy-budget diagnostics. For multi-state SIDM, the convergence of state-dependent scattering rates and internal-state populations requires particular attention when some states are sparsely sampled. Our isolated-halo tests demonstrate good global energy conservation throughout gravothermal core collapse (\cref{fig:isolated_halo}), providing an important numerical consistency check, although energy conservation alone does not establish convergence. Ultimately, the numerical uncertainties must be sufficiently small to resolve the physical differences between SIDM models in the observables used for inference.

\textit{Coupling between initial conditions and late-time interactions.} Our cosmological benchmarks use matched initial conditions to isolate the effects of late-time DM interactions (\cref{sec:cosmological_DMO_simulations,sec:MW_like_zoom_in_sims,sec:dwarf_zoom_in_sims}). Predictions for particular microscopic models must also include any changes that their early-Universe physics induces in the initial matter distribution (\cref{tab:pk_models}). DM--dark-radiation coupling can produce dark acoustic oscillations and diffusion damping~\citep{Cyr-Racine:2015ihg,Vogelsberger:2015gpr}; exothermic inelastic conversions can suppress small-scale power even without dark radiation~\citep{Duan:2026zqj}; and sufficiently early decays can alter the transfer function through daughter-particle free streaming. These effects can change halo abundances, assembly histories and concentrations, thereby affecting subsequent heat transport and gravothermal evolution. Likewise, the initial internal-state fractions $f_i^{\rm init}$ depend on the earlier thermal and reaction history and set the populations available for inelastic transitions. Self-consistent model predictions therefore require deriving both the cosmological initial conditions and the late-time interaction inputs from the same microscopic theory and cosmological history, following \cref{sec:pp_to_simulation_inputs}.

\textit{Baryonic physics.} A necessary step toward robust particle-physics inference is to establish which SIDM signatures remain distinguishable in full-physics cosmological simulations. While collisionless CDM haloes typically develop centrally cuspy density profiles~\cite{Navarro_1996b}, baryonic processes, including gas cooling, star formation and stellar and AGN feedback, can substantially reshape their gravitational potentials and internal structure. Feedback-driven potential fluctuations can generate central cores even in CDM haloes~\cite{Governato_2010,PontzenGovernato2012CuspCore}, whereas centrally concentrated baryons can induce contraction and accelerate gravothermal evolution in SIDM haloes~\cite{robles2017sidm,Despali2019InterplaySIDMBaryons,Sameie2021,Robertson2018}. Consequently, baryonic physics can mimic SIDM effects, or affect or obscure the differences between CDM and SIDM predictions for halo density profiles, shapes, kinematics and satellite evolution~\cite{Vargya2022,Robertson2021,Despali2025AIDATNG}. Matched DM-only and full-physics simulations are therefore essential for quantifying this interplay and identifying observables that retain sensitivity to DM microphysics. Existing hydrodynamical SIDM simulations, including EAGLE-SIDM~\cite{Robertson2018,Robertson2021,Forouhar2022}, BAHAMAS-SIDM~\cite{Robertson_2019}, TANGO-SIDM~\cite{Correa2022,Correa2025}, AIDA-TNG~\cite{Despali2025AIDATNG}, and high-resolution FIRE and Lyra simulations~\cite{robles2017sidm,Sameie2021,Vargya2022,Gutcke2025}, provide a foundation for quantifying the interplay between baryonic physics and DM self-interactions. Because the SIDM and decay modules developed here are decoupled from the gravity solver and compatible with the hierarchical time integration of \textsc{arepo-2}, they can be combined with its hydrodynamics and galaxy-formation physics, providing a basis for systematic full-physics studies of how distinct DM properties shape structure formation across astrophysical scales.

\textit{Towards particle-physics inference.} The longer-term goal is to connect the mapping $\mathcal{P}_{\rm PP}\rightarrow\mathcal{P}_{\rm SIM}$ to joint inference from complementary astrophysical systems (\cref{sec:simulation_diagnostics_DM}).
In such a framework, predictions derived from the same $\mathcal{P}_{\rm PP}$ through $\mathcal{P}_{\rm SIM}$ are confronted jointly with observations of dwarf galaxies, satellites and stellar streams, strong-lensing perturbers, galaxies and groups, and galaxy clusters and their mergers, while accounting for baryonic and environmental effects, numerical systematics and observational uncertainties. Exploring the parameter space $\mathcal{P}_{\rm SIM}$ will likely require targeted simulation suites and machine-learning-assisted inference. Such analyses could constrain the effective DM properties in $\mathcal{P}_{\rm SIM}$ and help identify which combinations of microscopic parameters in $\mathcal{P}_{\rm PP}$ can be constrained by astrophysical observations, providing a quantitative connection between nonlinear structure formation and fundamental DM physics.

\section{Conclusions}
\label{sec: conclusions}

We have established a unified theory-to-simulation framework for SIDM, combining a comprehensive treatment of microscopic dark-sector physics, an explicit mapping to nonlinear structure-formation simulation inputs, and new numerical implementations in \arepoTwo. The SIDM implementations build on the infrastructure introduced in our companion paper~\citep{Zier:2026}. Our main contributions are:

\begin{itemize}[leftmargin=*]

\item \textbf{A simulation-oriented SIDM particle-physics framework.}
We develop a systematic particle-physics reference framework for SIDM (Chapter~\ref{sec: sidm_pp_models}), organizing the microscopic ingredients, phenomenology and constraints of representative contact-like, light-mediator, resonant, vector, asymmetric, multi-state inelastic and composite dark-sector model classes.

\item \textbf{A generalized mapping from particle physics to simulation inputs.}
We establish an explicit mapping from microscopic dark-sector physics to simulation inputs and apply it to representative SIDM model classes (Chapter~\ref{sec:pp_to_simulation_inputs}). The mapping specifies cosmological initial conditions, internal-state masses and populations, reaction networks, velocity- and angle-dependent cross sections, dissipative energy-loss fractions, and decay rates and recoil kinematics. Inelastic energy exchange follows from the state masses and reaction channels. This extends the ETHOS approach~\citep{Cyr-Racine:2015ihg,Vogelsberger:2015gpr} to a broader range of dark-sector processes.

\item \textbf{Anisotropic SIDM in \arepoTwo.}
We implement anisotropic Monte Carlo scattering (aMC; \cref{sec:anisotropic_MC}), frequent small-angle drag--diffusion (fSIDM; \cref{sec: fsidm}) and an angular-hybrid scheme (aSIDM; \cref{sec:asidm}). The hybrid scheme combines fSIDM with explicit aMC sampling of rare large-angle events, achieving speed-ups of up to $\sim100\times$ over aMC for strongly forward-peaked interactions in our deflection tests. These implementations extend the previous isotropic SIDM treatment in \arepo, building on anisotropic, frequent-scattering and hybrid approaches developed in earlier studies~\citep{Robertson_2017b,Fischer_2021a,Arido:2024hfk}.

\item \textbf{State-dependent neighbour estimates for multi-state SIDM.}
Building on the \arepoOne treatment of reaction networks and inelastic kinematics~\citep{vogelsberger2019evaporating,2021MNRAS5001531C,ONeil:2022szc}, we introduce state-dependent smoothing lengths in \arepoTwo to estimate the local partner-state densities entering each reaction channel (\cref{sec:imp_inelastic_SIDM}). Sparsely sampled states use a local fallback, while searches for absent states are omitted. The multi-state deflection test verifies that an additional non-interacting population does not alter the scattering of the interacting states, while cosmological zoom-in benchmarks reproduce halo density profiles and internal-state evolution.

\item \textbf{Velocity-dependent and multi-state dissipative SIDM.}
We extend the dissipative-scattering framework in \arepo (\cref{sec:imp_dissipative_SIDM}) to allow dissipative and non-dissipative processes within the same reaction network, with velocity- and reaction-dependent cross sections and energy-loss fractions, and combinations with multi-state and anisotropic scattering. Compared with the previous \arepo treatment, which assumed a single velocity-independent cross section and a constant energy-loss fraction, the new implementation allows the dissipative physics to depend on relative velocity, reaction channel and internal state.

\item \textbf{Dark-matter decays in \arepoTwo.}
We implement a dedicated DM decay module (\cref{sec:imp_decaying_DM}) supporting state-based decay chains with state-dependent rates and recoil velocities, alongside a standalone effective recoil scheme. The state-based implementation follows sequential transitions and mass updates and is integrated with the multi-state SIDM framework.

\item \textbf{Validation from controlled tests to cosmological simulations.}
We validate the implementations through controlled deflection, thermalization and decay tests, isolated-halo gravothermal-collapse simulations, and cosmological box and zoom-in simulations. The results reproduce analytical predictions, demonstrate agreement between numerical schemes in their common regimes of validity, and recover published benchmarks for multi-state inelastic and constant dissipative SIDM~\citep{ONeil:2022szc,Shen:2021frv}.

\end{itemize}

Together, these advances establish a quantitative, numerically validated connection between microscopic dark-sector physics and nonlinear structure formation, substantially broadening the range of SIDM and DM decay processes accessible with \arepoTwo.

\begin{acknowledgments}
We thank Stephanie O’Neil for providing the initial conditions used in \cref{sec:MW_like_zoom_in_sims}. MR acknowledges support from the Carlsberg Foundation, grant CF24-1996. OZ acknowledges support from Harvard University through the Institute for Theory and Computation Fellowship.
\end{acknowledgments}

\section*{Data Availability} 
The initial-condition generator and the Boltzmann reference solver used in the validation tests are publicly available. The SIDM and DDM implementations, as well as the data underlying this article, are available from the corresponding author (MR) upon reasonable request.

\appendix

\section{Constraints on SIDM model classes}
\label{app:model_constraints}

This appendix summarizes the cosmological, astrophysical and particle-physics constraints on the SIDM model classes introduced in \cref{sec:pp_to_simulation_inputs}. Their general origin, including relic-abundance requirements and constraints from cosmology, astrophysics and laboratory searches, is discussed in \cref{sec:DM_production_mechanisms,sec:DM_search_experiments}, with their interplay and resulting viable parameter space summarized in \cref{sec:DM_search_experiments}. Here we apply these considerations to the individual SIDM model classes and extend the representative benchmarks of \cref{sec:pp_to_simulation_inputs} to a broader discussion of their viable particle-physics parameter space.

\subsection{Contact-like (short-range) self-interactions}
\label{app:constraints_contact-like}

In the simplest realization, scalar DM $s$ has only quartic self-interactions and no additional mediators. If $s$ couples to the SM, for example through a Higgs portal $\lambda_{sh} s^2 h^2 /4$ in \cref{eq:higgs_portal_dm}, the observed relic abundance can be obtained via standard thermal freeze-out~\cite{Burgess:2000yq}. However, such Higgs-portal scenarios are strongly constrained by precision Higgs measurements (set by collider experiments discussed in \cref{sec:DM_search_experiments}), in particular bounds on invisible Higgs decays~\cite{CMS:2016dhk}. In purely dark-sector realizations, the relic density can instead be set through $3\to2$ ``cannibalization'' processes~\cite{Carlson:1992fn,Hochberg:2014kqa}, such as $s s s \to s s$, which arise naturally in the presence of cubic and quartic self-interactions of the form $s^3$ and $s^4$, respectively.

In the heavy-mediator scenarios, the same interactions that generate approximately contact-like self-scattering also open annihilation channels that determine the relic abundance (see \cref{sec:DM_production_mechanisms}). A common possibility is standard WIMP freeze-out, where the relic density is set by the thermally averaged annihilation cross section in \cref{eq:relic_abundance_freezeout_explicit}, $\Omega_{\rm DM} h^2 \propto 1/\langle\sigma_{\rm ann} v\rangle$ (see, e.g.,~\cite{Kao:2020sqs}). In portal realizations, annihilation proceeds directly into SM states via the mediator, as in Higgs-portal scalar DM or vector-portal fermionic DM~\cite{Ko:2014nha}. In more secluded scenarios, annihilation into dark-sector states such as $\chi\chi \to \phi\phi$ or $ss \to \phi\phi$ can dominate, with the mediator subsequently decaying to SM particles through a small portal coupling. In this case the relic density is primarily controlled by dark-sector interactions, while the portal mainly governs mediator decay and experimental signatures. Variants of this setup include coannihilation with nearby dark states that enhance the effective annihilation rate, as well as asymmetric realizations in which the relic density is set by an initial particle--antiparticle asymmetry rather than thermal freeze-out. Such mechanisms arise in viable SIDM models consistent with current observational constraints~\cite{Hambye:2019tjt}.

Composite SIDM realizations, discussed further in \cref{sec:composite_SIDM}, generically give rise to analogous interaction structures, as illustrated in~\cite{Hochberg:2014kqa,Rosenlyst:2021elz,Cline:2022leq}.

\subsection{Minimal SIDM models with light mediator}
\label{app:constraints_light-mediator}

At the microscopic level, minimal light-mediator SIDM models are specified by the DM mass, mediator mass, DM--mediator coupling and portal couplings to the SM, such as the photon, $Z$-boson and Higgs portals in \cref{eq:photon_portal,eq:Z_portal,eq:higgs_portal_mediator}. These parameters determine the self-interaction strength, relic abundance, mediator lifetime and experimental signatures.

A viable model must reproduce the observed DM relic abundance, generate astrophysically relevant self-interactions, and satisfy cosmological, astrophysical and particle-physics constraints. The interplay between self-interactions and DM production (discussed in \cref{sec:DM_production_mechanisms}) is illustrated in the upper-right and lower-left panels of \cref{fig:effective_force_vector} for vector- and scalar-mediated interactions, respectively. Additional constraints arise from cosmology, including BBN, the CMB and $\Delta N_{\rm eff}$; astrophysics, such as supernova cooling; and particle-physics searches, including direct detection, indirect detection and colliders, as discussed in \cref{sec:DM_search_experiments}. In particular, the same portal couplings that control mediator decay and DM production also determine the experimental signatures. This creates a tension: large couplings are needed for viable DM production and prompt mediator decay, while small couplings are required to evade experimental bounds.

In the simplest Higgs-portal SIDM model with a scalar mediator, this tension leaves only a very small viable region. A narrow range around $m_{\rm med}\sim 1~{\rm MeV}$ and $m_{\rm DM}\sim 0.5~{\rm GeV}$ remains consistent with all constraints, with $\alpha_\phi \sim 5 \times 10^{-5}$ producing the desired velocity-dependent self-interactions; see the lower-left panel of \cref{fig:effective_force_vector} and the corresponding benchmark cross section, shown as the orange curve in \cref{fig:cross_sections}. By contrast, minimal vector-mediator models are typically strongly constrained or effectively ruled out by a combination of CMB and indirect-detection bounds, as discussed in \cref{sec:DM_search_experiments}.

The relic abundance need not be set by standard thermal freeze-out. Alternative mechanisms include freeze-in, reannihilation, asymmetric DM, non-thermal production from decays and modified dark-sector thermal histories, as briefly reviewed in \cref{sec:DM_production_mechanisms}. These scenarios can significantly change the relation between microscopic parameters and the present-day DM density. As discussed at the end of \cref{sec:DM_search_experiments}, extensions beyond the minimal setup -- including modified thermal histories, suppressed portal couplings, modified annihilation channels, inelastic DM, asymmetric DM and extended dark sectors -- can reopen substantial regions of parameter space. The strongest exclusions therefore apply mainly to the simplest minimal light-mediator SIDM models, while more general realizations remain viable.

\subsection{Resonant SIDM}
\label{app:constraints_resonant}

At the microscopic level, RSIDM requires a state sufficiently close to the two-particle threshold to affect non-relativistic halo scattering,
\begin{equation}
m_R - 2m_{\rm DM} \ll m_{\rm DM},
\end{equation}
with the phenomenology controlled by the resonance width and the dominant partial wave. Many viable scenarios are $s$-wave dominated, naturally producing large low-velocity cross sections and admitting the effective-range description discussed above. The right panel of \cref{fig:effective_range_approx_parameters} shows the corresponding viable parameter space in the $(m_{\rm DM}, |a_0|)$ plane for different values of $r_{e,0}/a_0$. The shaded regions indicate models with velocity-dependent self-interaction cross sections consistent with the astrophysical constraints in \cref{fig:cross_sections}: large cross sections at dwarf-galaxy velocities, $v \sim 10\,\mathrm{km/s}$, and suppressed values at cluster velocities, $v \sim 10^3\,\mathrm{km/s}$. This behavior is characteristic of near-threshold $s$-wave resonant scattering.

As in other SIDM scenarios, a viable RSIDM model must reproduce the observed DM relic abundance, generate astrophysically relevant self-interactions, and satisfy cosmological, astrophysical and particle-physics constraints. In RSIDM, however, the relic density and self-scattering signal can be closely linked: the same near-threshold state that enhances self-scattering may also enhance annihilation. This can invalidate the standard velocity expansion of $\langle \sigma_{\rm ann} v\rangle$ near the resonance and lead to strong temperature dependence, early kinetic decoupling and sizable corrections to the relic density~\cite{Duch:2017nbe}. The abundance may therefore depend on resonance-enhanced annihilation, early kinetic decoupling, freeze-in, reannihilation, asymmetric production, inverse decays or other non-thermal processes, as briefly reviewed in \cref{sec:DM_production_mechanisms}.

Concrete realizations differ in how the near-threshold state is generated. In non-perturbative Yukawa models, it arises dynamically from an attractive potential and is tied to the light-mediator interactions discussed in \cref{sec:light-mediator}. In composite dark sectors, it corresponds to a bound or excited hadron-like state close to threshold, as discussed in \cref{sec:composite_SIDM}. Examples include dark QCD RSIDM~\cite{Tsai:2020vpi}, super-resonant SIDM in supersymmetric QCD~\cite{Csaki:2022xmu}, extra-dimensional realizations~\cite{Lee:2025lko}, dark meson SIMPs~\cite{Hochberg:2014dra} and self-interacting dark baryons~\cite{Cline:2022leq}. The constraints depend on the origin of the resonance. If it is tied to elementary mediator dynamics, the model is subject to the usual cosmological and experimental bounds, including BBN, CMB, $\Delta N_{\rm eff}$, indirect detection, direct detection and collider probes through portal couplings, as discussed in \cref{sec:DM_search_experiments}. If the resonance belongs to a secluded or composite sector, the constraints are typically weaker and more model dependent.

Overall, RSIDM provides a natural mechanism for large self-interactions in low-velocity systems and suppressed scattering at high velocities, without necessarily requiring an elementary light mediator. Its near-threshold origin also makes both the self-scattering signal and relic-density calculation more sensitive to the underlying dynamics than in simpler SIDM scenarios.

\subsection{Vector SIDM}
\label{app:constraints_vector}

As in the scalar and fermionic SIDM cases discussed in \cref{sec:light-mediator}, a viable vector SIDM model must ensure cosmological stability of the DM particle, reproduce the observed relic abundance, generate astrophysically relevant self-interactions and satisfy experimental constraints. In gauge-theory realizations, stability is typically guaranteed by residual symmetries after symmetry breaking, such as discrete symmetries ($\mathbb{Z}_2$ or KK parity)~\cite{Servant:2002aq,Hambye:2008bq,Farzan:2012hh}, conserved gauge symmetries~\cite{Gross:2015cwa,Davoudiasl:2013jma,Ko:2016fcd,Boddy:2014yra,2014PhRvD}, custodial or global symmetries~\cite{Hambye:2008bq,Choi:2019zeb,Chaffey:2019fec,Yuan:2024tfs}, accidental global $U(1)$ symmetries~\cite{Frandsen:2022klh,Shaikh:2026ixs}, topological stability~\cite{Baek:2013dwa,Khoze:2014woa} or multi-component charge assignments~\cite{Elahi:2019jeo,Bouzeraib:2026dbx}; see \cref{sec:DM_mass_stability_abundance}.

The observed relic abundance can be obtained through standard thermal freeze-out, discussed in \cref{sec:DM_production_mechanisms}. The dominant annihilation channels are into light vector mediators, $\chi^+\chi^- \to \phi_\mu\phi_\mu$, and scalar states, $\chi^+\chi^- \to \Phi\Phi$, where $\Phi$ denotes the scalar multiplet. Parametrically,
\begin{equation}
\langle \sigma v \rangle \sim \frac{\alpha_{\rm DM}^2}{m_{\rm DM}^2},
\end{equation}
which, using \cref{eq:relic_abundance_freezeout_explicit}, gives the approximate thermal target
\begin{equation}
\alpha_{\rm DM}^{\rm th} \simeq 0.017 \left(\frac{m_{\rm DM}}{\mathrm{TeV}}\right).
\end{equation}
For $m_{\rm DM}\simeq60~{\rm MeV}$, this gives
$\alpha_{\rm DM}^{\rm th}\simeq10^{-6}$,
consistent with the light vector SIDM benchmark
of~\cite{Chaffey:2019fec}. At weak-scale masses,
the required coupling increases correspondingly,
reaching $\alpha_{\rm DM}^{\rm th}\sim10^{-3}$
for $m_{\rm DM}\simeq60~{\rm GeV}$.

Astrophysically relevant self-interactions arise from exchange of the light vector mediator $\phi_\mu$, generating an effective Yukawa potential analogous to the minimal light-mediator models in \cref{sec:light-mediator}. This produces velocity-dependent scattering cross sections. Parameters close to the light benchmark can reproduce the required cross sections on dwarf and galaxy scales, whereas heavier DM masses generally fail to reach the preferred cluster-scale range.

Consistency with experimental constraints requires suppressed but non-zero interactions between the dark and visible sectors. In the kinetic-mixing scenario, maintaining thermal equilibrium with the SM during freeze-out requires
\begin{equation}
\epsilon_{\rm DM}^2 \left(\frac{m_{\rm med}}{10~\mathrm{MeV}}\right)
\gtrsim
10^{-11}
\left(\frac{m_{\rm DM}}{50~\mathrm{GeV}}\right)^2,
\end{equation}
while direct-detection searches impose upper bounds on $\epsilon_{\rm DM}$. Viable regions therefore require kinetic mixing small enough to evade experimental limits but large enough to thermalize the dark sector. More generally, if the vector mediator couples to the SM through kinetic mixing or a Higgs portal, the model is constrained by BBN, CMB, direct and indirect detection, collider searches and dark-photon experiments. Although~\cite{Chaffey:2019fec} mainly focuses on relic abundance and self-interactions, these probes further restrict realistic parameter space. Higgs-portal realizations are already tightly constrained by direct detection, leaving only restricted viable regions~\cite{Frandsen:2022klh}. Additional theoretical constraints can arise from ultraviolet consistency: embedding vector DM models into asymptotically safe quantum gravity can impose non-trivial restrictions on the couplings and in simple Higgs-portal realizations may significantly reduce or even exclude otherwise phenomenologically viable regions~\cite{deBrito:2023ydd}.

As in other minimal light-mediator SIDM scenarios, viable parameter space can reopen in more general constructions. The discussion above assumes thermal freeze-out, but vector SIDM often arises from richer gauge-theory structures that allow alternative production mechanisms, such as freeze-in or decays, as well as multi-component or inelastic sectors and more secluded realizations with suppressed portal couplings. Recent work such as~\cite{Shaikh:2026ixs} shows that ultraviolet completions can substantially modify the phenomenology: additional states and interactions can open new annihilation channels and weaken direct-detection bounds, reopening regions excluded in simpler Higgs-portal descriptions.

\subsection{Asymmetric SIDM}
\label{app:constraints_asymmetric}

As in all viable SIDM scenarios, ASIDM models must reproduce the observed DM relic abundance, generate astrophysically relevant self-interactions and satisfy cosmological, astrophysical and particle-physics constraints as discussed in \cref{sec:DM_production_mechanisms,sec:DM_search_experiments}. The key difference from symmetric SIDM is that the relic abundance is set by the surviving asymmetric component rather than by thermal freeze-out. This may resolve the tension in minimal light-mediator SIDM, where large self-interactions often imply overly efficient annihilation and an under-abundant relic density~\cite{Borah:2022ask}, while also offering a dynamical explanation of the relation $\rho_{\rm DM}\sim 5\rho_B$.

At the same time, the symmetric component must be efficiently depleted. In practice, this requires sizeable annihilation cross sections into light mediators or other dark-sector states, which is generically satisfied in light-mediator models. In asymmetric scenarios, this efficient annihilation is not problematic, since the relic density is instead set by the primordial asymmetry. A crucial consequence is that late-time DM--anti-DM annihilation signals are strongly suppressed, as the anti-DM population is depleted. This allows ASIDM models to evade stringent indirect-detection and CMB bounds that typically constrain symmetric light-mediator scenarios~\cite{Slatyer:2015jla,Lin:2011gj,Baldes:2017gzu}.

The lower-right panel of \cref{fig:effective_force_vector} illustrates this structure for an asymmetric SIDM scenario with a repulsive vector-mediated force. The relic abundance is dominated by a primordial particle--antiparticle asymmetry, while the surviving DM population is mostly one species. The symmetric component must be efficiently removed, for example through annihilations mediated by a kinetic-mixing portal, requiring parameters outside the clock-shaped relic-density contours for the corresponding $m_{\mathrm{DM}}$ in the upper-left panel of \cref{fig:effective_force_vector}. Thus, unlike in symmetric freeze-out, the relic abundance is not directly tied to the same parameter combination that controls the present-day halo self-interaction strength, substantially enlarging the viable parameter space.

Direct-detection constraints arise from portal interactions between the dark and visible sectors. For a vector mediator, these are typically kinetic-mixing interactions with SM gauge bosons, as in \cref{eq:photon_portal,eq:Z_portal}. Current bounds require very small mixing, $\epsilon_{\rm DM} \lesssim 10^{-8}$ for MeV-scale mediators, which suppresses the direct-detection cross section sufficiently for the benchmark to remain viable. Cosmological constraints are also satisfied in this regime: the small portal coupling prevents the dark and visible sectors from thermalizing after asymmetry generation, keeping interaction rates far below the Hubble rate. Contributions to $\Delta N_{\rm eff}$ are therefore negligible, BBN constraints are satisfied, and the depleted anti-DM population ensures compatibility with CMB and indirect-detection bounds from gamma rays, cosmic rays and neutrinos. In explicit cogenesis scenarios, the final DM abundance depends on particle masses, asymmetries, washout effects and the transfer of asymmetry between the visible and dark sectors. In the models considered here, these effects may be controlled if washout processes are either forbidden by conserved quantum numbers or remain out of equilibrium, thereby preserving both the baryon and dark asymmetries after generation~\cite{Dutta:2022knf,Borah:2024wos,Dutta:2024bnt,Dutta:2025xlv,Ishida:2025xvr}.

Thus, ASIDM can enlarge the viable parameter space compared to symmetric SIDM in suitable realizations. By partly decoupling the relic-abundance mechanism from the present-day self-interaction strength, it may allow strong, velocity-dependent self-interactions while remaining consistent with cosmological, astrophysical and particle-physics constraints.

\subsection{Multi-state inelastic SIDM}
\label{app:constraints_inelastic}

For vector-mediated models, the light mediator that generates large self-interactions induces strong cosmological constraints, as discussed in \cref{sec:DM_search_experiments}. Annihilation is typically $s$-wave and can be Sommerfeld-enhanced at late times, leading to stringent CMB bounds. Mediator decay and portal couplings must also satisfy BBN and direct-detection constraints. As we will see, viable parameter space can nevertheless open through mechanisms such as inelasticity and modified portal structures.

This interplay is illustrated by the benchmark models of~\cite{ONeil:2022szc}, who performed cosmological zoom-in simulations of Milky Way-like haloes with inelastic SIDM. Their simulations showed that the relative importance of elastic, endothermic and exothermic scattering strongly affects halo structure and subhalo survival, depending on the mass splitting and mediator parameters.

Astrophysically, one typically requires $\sigma/m_\chi \sim 1\text{--}10~{\rm cm}^2/{\rm g}$ on dwarf-galaxy scales while keeping the cross section suppressed in ultra-faint satellites and clusters. Inelastic models can achieve this through their threshold structure: endothermic scattering is suppressed at low velocities but can turn on at dwarf-galaxy scales. A particularly important requirement is suppressing low-velocity elastic scattering, $\chi_1\chi_1 \to \chi_1\chi_1$, which is crucial for reconciling satellite survival with core formation in dwarf galaxies~\cite{Vogelsberger_2012}. If elastic and inelastic scattering were comparable, steep velocity dependences such as $\sigma \propto v^{-4}$ could produce excessively large cross sections in satellites, conflicting with dwarf-galaxy constraints such as those inferred for Draco under non-collapsing SIDM halo models~\cite{Read:2018pft}.

Scalar-mediated models can alleviate these tensions. In Ref.~\cite{Wang:2025xoq}, annihilation into scalars is $p$-wave suppressed, avoiding strong CMB constraints. The mediator is chosen to be leptophilic,\footnote{A leptophilic mediator couples predominantly to leptons, such as electrons and neutrinos, rather than to quarks. This suppresses interactions with nuclei and weakens conventional direct-detection constraints based on nuclear recoils; see the discussion of direct detection in \cref{sec:DM_search_experiments}.} allowing rapid decay before BBN while suppressing direct-detection bounds. The model also permits the radiative decay
$\chi_2 \to \chi_1+\gamma_{\rm D}$. The excited-state
population is primarily depleted in the early
Universe through exothermic self-interactions
$\chi_2\chi_2 \to \chi_1\chi_1$.
Consequently, the abundance of $\chi_2$ becomes
negligible, and elastic processes involving the
excited state, such as
$\chi_1\chi_2 \to \chi_1\chi_2$,
are strongly suppressed.

In summary, vector- and scalar-mediated two-state SIDM share the same basic coupled-channel scattering structure but differ in their viable parameter space. Vector-mediated models are strongly constrained by $s$-wave annihilation and mediator couplings, whereas scalar models provide additional freedom through $p$-wave annihilation and portal structure, allowing a more robust realization of the inelastic SIDM mechanism.

\subsection{Dark-baryon SIDM}
\label{app:constraints_dark_baryons}

For dark baryons to make up the dominant DM component, four broad consistency conditions are typically required~\cite{Cline:2022leq}. First, dark mesons, especially dark pions, must be unstable or sufficiently depleted. If the dark sector contains a dark $U(1)_D$ interaction, charged dark pions can annihilate efficiently through $\pi_D^+\pi_D^-\to \gamma_D \gamma_D$, and for moderate dark gauge couplings their relic abundance can become negligible~\cite{Cline:2022leq}. Neutral dark pions can also decay to dark photons if kinematically allowed, while vector mesons and glueballs are typically unstable in the parameter range of interest.

Second, dark-atom formation must be avoided if the dark sector contains oppositely charged states and a dark photon. If dark electrons are present to ensure $U(1)_D$ charge neutrality, dark proton--dark electron bound states may form and dominate the phenomenology. This can be prevented if the dark photon is sufficiently massive that bound states are not supported,
\begin{equation}
m_{\gamma_D} \gtrsim \alpha_{\rm DM} \frac{m_{e_D}m_{N_D}}{m_{e_D}+m_{N_D}},
\end{equation}
with $\alpha_{\rm DM}=g^2/(4\pi)$, or more generally if recombination is inefficient~\cite{Petraki:2016cnz,Cline:2021itd,Cline:2022leq}.

Third, dark nucleosynthesis into larger nuclei must be suppressed. If dark deuteron-like or dinucleon bound states form efficiently, the DM may end up in dark nuclei rather than individual nucleons, substantially changing the scattering phenomenology~\cite{Krnjaic:2014xza,Redi:2018muu,Cline:2022leq}. In the parameter regions that give the desired self-interaction cross sections, the scattering lengths are often close to a pole, implying that the relevant two-body bound state is marginally bound or absent. Dark big-bang nucleosynthesis can also be kinematically blocked if the dark photon mass exceeds the would-be binding energy~\cite{Cline:2022leq}.

Fourth, residual dark-photon-mediated long-range forces between dark baryons should remain subdominant to the strong composite interaction. Otherwise, the model effectively reverts to Yukawa-type light-mediator SIDM rather than genuinely dark-baryonic SIDM. This typically imposes a lower bound on the dark photon mass. For benchmark values such as $\alpha_{\rm DM}\sim10^{-3}$ and $m_{N_D}\sim5~{\rm GeV}$, taking
\begin{equation}
m_{\gamma_D}\gtrsim{\cal O}(10)~{\rm MeV}
\end{equation}
is often sufficient to keep the dark-photon contribution subdominant over the relevant velocity range~\cite{Tulin:2013teo,Cline:2022leq}.

Additional cosmological and portal-related constraints arise if dark photons are unstable through kinetic mixing $\epsilon_\gamma$; see \cref{eq:photon_portal,fig:portal_photon}. Their decays into SM states, such as $e^+e^-$, help avoid overclosure but are constrained by beam-dump experiments, supernova cooling, BBN, the CMB and direct detection~\cite{Kaplinghat:2013yxa,Fabbrichesi:2020wbt,Chang:2016ntp,Li:2020roy,XENON:2018voc,PandaX-II:2018xpz,Billard:2021uyg}. Residual annihilation or bound-state formation in a subdominant mesonic component can also inject energy during or after recombination and is constrained by the CMB and related probes~\cite{Slatyer:2015jla,Liu:2018uzy,Depta:2019lbe,Mahbubani:2019pij}. The viable parameter space is therefore restricted, but explicit analyses show that consistent benchmark regions exist~\cite{Cline:2022leq}.

\subsection{Dark-meson SIDM}
\label{app:constraints_dark_mesons}

\noindent
\textbf{Model constraints: QCD-like dark mesons.} In the simplest QCD-like dark-meson SIDM scenario, dark mesons are lighter than dark baryons and glueballs and dominate the low-energy spectrum. With a single dark quark, the lightest pseudoscalar resembles the QCD $\eta'$ and is not parametrically light. With several light dark quarks, the mesons can instead be pNGBs and therefore natural DM candidates. Stability is typically ensured by residual global or discrete symmetries, such as a dark analogue of $G$-parity, so that the lightest $G$-odd meson is stable while heavier states decay~\cite{Bai:2010qg}. Without such a symmetry, dark mesons generically decay through portal interactions and cannot constitute the DM abundance~\cite{Berlin:2018tvf}.

In portal-coupled dark-meson models, the dark sector remains thermally connected to the SM through Higgs, vector or axion-like portals. At low energies, portal interactions can arise from terms such as~\cite{Bhattacharya:2013kma}
\begin{equation}
\begin{aligned}
&\lambda_h \left(|H|^2-v^2\right){\rm tr}(\partial_\mu\Sigma^\dagger\partial^\mu\Sigma)
+
\\
&\frac{\lambda_v}{f}\,B_{\mu\nu}
\,{\rm tr}(M_q\Sigma\,\partial^\mu\Sigma^\dagger\partial^\nu\Sigma + {\rm h.c.}) .
\end{aligned}
\end{equation}
These induce annihilation channels such as $\pi_D\pi_D \to WW,\ HH$ and, in the presence of a dark vector mediator, also $\pi_D\pi_D\to\phi_\mu\phi_\mu$. Achieving the correct relic density while satisfying CMB, BBN, direct-detection and laboratory bounds is nontrivial, especially when the same portal thermalizes the dark sector and induces late-time annihilation into visible states. Heavier vector resonances can also modify freeze-out through semi-annihilation channels such as $\pi_D\pi_D \to \pi_D V_D$, followed by $V_D$ decay into SM particles. Here $V_D$ is a heavier spin-1 bound state, analogous to the QCD $\rho$ meson. Such channels can affect the relic abundance and lead to signatures in beam-dump, fixed-target and collider experiments~\cite{Berlin:2018tvf}.

A second important possibility is SIMP dark meson DM, where the relic abundance is set by number-changing $3\to2$ reactions within the dark sector~\cite{Hochberg:2014dra,Hochberg:2014kqa}. In chiral perturbation theory, these processes are generated by the Wess--Zumino--Witten term,
\begin{equation*}
\mathcal{L}_{\rm WZW}
=
\frac{2N_D}{15\pi^2 f_{\pi_D}^5}\,
\epsilon^{\mu\nu\rho\sigma}
\,\mathrm{Tr}
\!\left(
\pi_D \partial_\mu \pi_D \partial_\nu \pi_D \partial_\rho \pi_D \partial_\sigma \pi_D
\right).
\end{equation*}
This term is present only for suitable symmetry-breaking patterns, for example $SU(N)_{\rm DC}$ and $SO(N)_{\rm DC}$ with $N_f\ge3$, and $Sp(N)_{\rm DC}$ for even $N$ with $N_f\ge2$~\cite{Hochberg:2014kqa}. The corresponding thermally averaged cross section can be written as
\begin{equation}
\langle \sigma v^2\rangle_{3\to2}
\equiv
\frac{\alpha_{\rm eff}^3}{m_{\pi_D}^5}
\sim
\frac{N^2 N_F^5\,m_{\pi_D}^5}{f_{\pi_D}^{10}},
\end{equation}
up to order-one and group-theory factors. Viability requires a large enough $3\to2$ rate to reproduce the relic abundance while remaining within the validity of chiral perturbation theory, typically requiring $m_{\pi_D}/f_{\pi_D}\lesssim{\cal O}(2\pi)$ rather than approaching the naive cutoff~\cite{Hochberg:2014kqa,Bernal:2015xba}. Together with self-interaction requirements, this typically favors
\begin{equation}
m_{\pi_D}\sim 30~{\rm MeV} - 1~{\rm GeV},
\end{equation}
with the precise range depending on the gauge group, number of flavors and flavor-symmetry breaking.

SIMP models also require kinetic equilibrium with the SM bath during freeze-out. Otherwise, $3\to2$ reactions heat the dark sector, leading to a cannibal phase and potentially excessive small-scale suppression. Viable models therefore require a portal that maintains kinetic contact without allowing ordinary $2\to2$ annihilation into SM states to dominate~\cite{Hochberg:2014dra,Hochberg:2014kqa}. Examples include dark-photon and axion-like mediator portals~\cite{Lee:2015gsa,Hochberg:2015vrg}, constrained by CMB energy injection, BBN, beam-dump and collider searches and Lyman-$\alpha$ structure-formation bounds~\cite{Viel:2013fqw,Lee:2015gsa,Hochberg:2015vrg}.

At low energies, elastic dark-meson scattering can be computed in chiral perturbation theory. The leading $s$-wave contribution gives
\begin{equation}
\frac{\sigma_{\rm tot}}{m_{\pi_D}}
=
\frac{m_{\pi_D}}{64\pi f_{\pi_D}^4}
\,
\frac{2N_F^4-25N_F^2+90-65/N_F^2}{N_F^2-1},
\end{equation}
so large self-interactions naturally favor MeV--GeV dark mesons and sufficiently small decay constants.

At low velocities, the effective-range formalism in \cref{eq:pw_effective_range,eq:ere_general} provides a model-independent description of short-range self-interactions. In composite dark sectors, such as those illustrated in \cref{fig:selfint_composite_realizations}, the scattering may be dominated by a near-threshold composite resonance $R_D$, shown in panel (c). This resonance corresponds to a short-lived bound or quasi-bound two-particle state, e.g.\ a meson--meson or baryon--baryon resonance, with mass
\begin{equation}
m_R \simeq 2 m_{\rm DM} + \delta m,
\end{equation}
where $\delta m \ll m_{\rm DM}$ controls the proximity to threshold and hence the strength of the velocity dependence. When a single partial wave dominates the scattering amplitude in \cref{eq:pw_effective_range}, the cross section can be written in Breit--Wigner form, \begin{align}
\begin{aligned}
\label{eq:s_wave_cross_QCD_dark_mesons}
\sigma(v_{\rm rel})
\simeq
\sigma_0 +&
\frac{16\pi(2\ell+1)}
{m_{\pi_D}^2v_{\rm rel}^2}
\\&
\times\frac{\Gamma(v_{\rm rel})^2/4}
{\big(E(v_{\rm rel})-E_R\big)^2
+\Gamma(v_{\rm rel})^2/4}. \end{aligned} 
\end{align} 
Here $E(v_{\rm rel})=m_{\pi_D}v_{\rm rel}^2/4$
is the centre-of-mass kinetic energy,
$E_R=m_R-2m_{\pi_D}$ is the resonance energy,
and $v_R=2\sqrt{E_R/m_{\pi_D}}$ is the
corresponding resonance velocity. The energy-dependent width $\Gamma(v_{\rm rel})$ measures the coupling of the two-particle scattering state to the composite resonance: a small width gives a narrow peak, while a large width gives a broad enhancement. We parameterize it as
\begin{equation}
\Gamma(v_{\rm rel}) = \gamma\, m_R\, v_{\rm rel}^{2\ell+1},
\end{equation}
where $\gamma$ encodes the resonance coupling to the scattering channel. The term $\sigma_0$ denotes the non-resonant short-range background, while $\ell$ is the dominant partial wave, with $\ell=0$ corresponding to an $s$-wave resonance and $\ell=1$ to a $p$-wave resonance.

\noindent 
\textbf{Model constraints: Composite-Higgs dark mesons.} A predictive realization arises in composite-Higgs models where both the Higgs and the DM are pNGBs of a common confining sector~\cite{Rosenlyst:2021elz}, with related constructions in~\cite{Ryttov:2008xe,Rosenlyst:2024rsb}. This framework is motivated by electroweak naturalness and triviality problems, which can be alleviated if the Higgs is composite~\cite{Terazawa:1976xx,Terazawa:1979pj,Kaplan:1983fs,Kaplan:1983sm}. It may simultaneously provide a dynamical origin for electroweak symmetry breaking, the DM relic abundance and large self-interactions.

The dark-meson DM candidate $\Phi$ in \cite{Rosenlyst:2021elz} is stabilized by an unbroken global $U(1)$ symmetry of the strong sector. Interactions between the Higgs and DM sectors are generated through partial compositeness~\cite{Kaplan:1991dc}, which also generates SM fermion masses. At low energies, the relevant interactions are dominated by
\begin{equation}
\mathcal{L}_{\rm eff}
\supset
-\,\lambda_\Phi (\Phi^\dagger \Phi)^2
-\lambda_{\Phi\Phi\Theta\Theta}\,\Phi^\dagger\Phi\,\Theta^2,
\end{equation} where $\Theta$ is a light pNGB singlet. The second
term induces $\Phi\bar{\Phi}\to\Theta\Theta$,
which sets the relic abundance through thermal
freeze-out, with~\cite{Rosenlyst:2021elz}
\begin{equation}
\langle \sigma v \rangle
\simeq
\frac{\lambda_{\Phi\Phi\Theta\Theta}^2}
{32\pi m_{\rm DM}^2}
\sqrt{1-\frac{m_\Theta^2}{m_{\rm DM}^2}}.
\end{equation}
The annihilation coupling is suppressed by the
vacuum misalignment in the composite-Higgs sector,
\begin{equation}
\lambda_{\Phi\Phi\Theta\Theta}
\simeq
\lambda_0\left(\frac{v_{\rm SM}}{f}\right)^2,
\qquad
\lambda_0\sim\mathcal{O}(1),
\end{equation}
where $v_{\rm SM}\simeq246\,{\rm GeV}$ is the
electroweak vacuum expectation value and $f$
is the pNGB decay constant of the composite
sector. For $m_\Theta\ll m_{\rm DM}$, reproducing
the thermal relic abundance requires approximately
\begin{equation}
m_{\rm DM}\simeq
0.9\,{\rm GeV}\,
\lambda_0
\left(\frac{10\,{\rm TeV}}{f}\right)^2.
\end{equation}

The same strong dynamics generates approximately contact-like elastic self-interactions~\cite{Rosenlyst:2021elz},
\begin{equation}\label{eq:composite_Higgs_dark_meson_cross}
\frac{\sigma_{\rm tot}}{m_{\rm DM}}
=
\frac{9\lambda_\Phi^2}{\pi} \left(\frac{\mathrm{GeV}}{m_{\rm DM}}\right)^3
\times 2\times10^{-4}\ {\rm cm}^2/{\rm g}.
\end{equation}
Since these arise from short-range quartic interactions rather than light-mediator exchange, they are approximately velocity independent. Large self-interactions require $\lambda_\Phi\sim{\cal O}(1)$ and sub-GeV DM masses to obtain $\sigma/m\sim0.1$--$1~{\rm cm}^2/{\rm g}$. Collider and Higgs measurements impose $f\gtrsim{\cal O}(10)~{\rm TeV}$, while cosmology requires the light pNGB $\Theta$ to decay before BBN, typically into SM fermions. 
Viable regions therefore occur for
$m_{\rm DM}\sim10~{\rm MeV}$--$1~{\rm GeV}$
and $f\sim10$--$100~{\rm TeV}$,
with $\lambda_0\sim\lambda_\Phi\sim\mathcal{O}(1)$.

\subsection{Dark-glueball SIDM}
\label{app:constraints_dark_glueballs}

Although dark-glueball DM is theoretically simple, its viable parameter space is strongly constrained. The first requirement is to reproduce the observed relic density. This is non-trivial because glueballs do not carry a conserved particle number. The same effective interactions that generate elastic $2\to2$ scattering in \cref{eq:eff_lagrangian_glueball} also induce number-changing processes such as $3s\to2s$. If the dark sector is thermally populated, these processes can remain efficient after the glueballs become non-relativistic, leading to a cannibalization regime in which rest mass is converted into kinetic energy and the dark-sector temperature redshifts more slowly than in an ordinary non-relativistic gas~\cite{Soni:2016gzf,Carlson_1992,Boddy:2014yra}. The relic abundance is therefore not set by conventional $2\to2$ freeze-out. In fact, simple thermal freeze-out through a portal to the SM is typically not viable: a portal strong enough to deplete the abundance generally also induces glueball decay at an unacceptable rate~\cite{Soni:2016gzf,Boddy:2014yra,Cline:2021itd}.

Viable glueball cosmologies therefore often require a secluded dark sector with a lower temperature than the visible sector, parameterized by $\xi=T_{\rm DM}/T_{\rm VM}\ll1$. In this case, the initial abundance is suppressed at confinement, and subsequent $3\to2$ processes may or may not reach equilibrium depending on $\xi$~\cite{Soni:2016gzf}. Other possibilities include late entropy injection, preferential reheating into the visible sector, gravitational production or additional hidden-sector ingredients that reduce the final abundance~\cite{Boddy:2014yra,Halverson:2016nfq,Acharya:2017szw,Soni:2017nlm,Jo:2020ggs,Redi:2020ffc}. Thus, glueball DM is generically overproduced unless the dark sector is sufficiently cold or diluted.

Portal interactions are also tightly constrained. At low energies, one may couple the glueball field $s$ to SM singlet operators,
\begin{equation}
{\cal L}_{\rm portal}
\supset
M^{3-n}\,s\,{\cal O}_{\rm SM},
\end{equation}
or, above confinement, write operators such as
\begin{equation}
\begin{aligned}
\frac{1}{\Lambda^4}B_{\mu\nu}B^{\mu\nu}\,\mathrm{tr}&(G_D^2),
\qquad
\frac{1}{\Lambda^4}B_{\mu\nu}\,\mathrm{tr}(G_D^3)^{\mu\nu},
\\
&\frac{1}{\Lambda^2}|H|^2\,\mathrm{tr}(G_D^2),
\end{aligned}
\end{equation}
where $B_{\mu\nu}$ is the hypercharge field strength and $H$ is the Higgs doublet. These operators can induce glueball decays into photons or other SM states, constrained by BBN, the CMB and indirect detection~\cite{Soni:2016gzf,Forestell:2016qhc,Forestell:2017wov}. If the portal is too strong, the glueball is cosmologically unstable; if it is too weak, the dark sector is typically overabundant.

The excited glueball spectrum can also be cosmologically relevant. Some states, such as $1^{+-}$, may be long-lived if accidental symmetries suppress their decays~\cite{Forestell:2016qhc,Forestell:2017wov}. Even if subdominant, their late decays can inject energy into the visible sector, so a consistent model must account for the full spectrum rather than only the lightest glueball.

Finally, glueballs always decay gravitationally, but this effect is typically negligible. Parametrically,
\begin{equation}
\Gamma \sim \frac{m_0^5}{M_{\rm Pl}^4},
\end{equation}
where $M_{\rm Pl} \simeq 1.22\times10^{19}~\mathrm{GeV}$ is the Planck mass. Glueballs are therefore cosmologically stable provided $m_0 \ll 10^7~\mathrm{GeV}$~\cite{Boddy:2014yra}.

\subsection{Dark atoms and molecules}
\label{app:constraints_dark_atoms}

The relic abundance of dark atoms is most naturally generated in an \emph{asymmetric DM} scenario; see \cref{sec:DM_production_mechanisms,sec:Asymmetric_SIDM}. An excess of $p_D$ over $\bar p_D$ and $e_D$ over $\bar e_D$, subject to overall $U(1)_D$ charge neutrality, remains after the symmetric component annihilates through
\begin{equation}\label{eq:annihilations_dark_atoms}
p_D\bar p_D\to\phi_\mu\phi_\mu,
\qquad
e_D\bar e_D\to\phi_\mu\phi_\mu,
\end{equation}
thereby setting the DM abundance analogously to the baryon asymmetry in the visible sector. UV-complete realizations\footnote{A \emph{UV-complete} model remains valid at arbitrarily high energies, in contrast to an effective field theory defined only below a cutoff scale.} have been proposed in~\cite{Cline:2013pca,Choquette:2015mca}. Without an asymmetry, the relic abundance may instead arise from thermal freeze-out, with
\begin{equation}
\langle\sigma v\rangle\simeq\frac{\pi\alpha_D^2}{m_{e_D}^2},
\end{equation}
for $m_i\lesssim10^3~\mathrm{GeV}$, above which Sommerfeld corrections may become important~\cite{Agrawal:2016quu}. However, simultaneously obtaining the correct relic abundance and efficient recombination is difficult in the symmetric case unless the dark sector is substantially colder than the visible sector, so asymmetric realizations are generally preferred~\cite{Agrawal:2016quu}.

After recombination, the residual ionized fraction $f_i\equiv n_{e_D}/n_{H_D}$ controls the strength of long-range dark Coulomb interactions. Useful estimates are~\cite{Cline_2014a,Cyr-Racine_2012}
\begin{equation}
f_i\simeq
\min\left[
1,\,
10^{-10}\,\xi\,\alpha_D^{-4}R^{-1}
\left(\frac{m_{H_D}}{\mathrm{GeV}}\right)^2
\right],
\end{equation}
and
\begin{equation}
f_i\sim
2\times10^{-16}\,
\xi\,\alpha_D^{-6}
\left(\frac{m_{H_D}}{\mathrm{GeV}}\right)
\left(\frac{B_{H_D}}{\mathrm{keV}}\right),
\end{equation}
where $\xi\equiv T_D/T_{\rm VM}$. Since the ionized component interacts through long-range forces, Bullet Cluster and halo-shape constraints favor efficient recombination, $f_i\ll1$, unless atomic DM constitutes only a subdominant fraction of the total DM~\cite{Randall:2008ppe,Markevitch:2003at,Feng:2009mn}.

The dark photon also contributes to the radiation density. For $T_D=\xi T_{\rm VM}$,
\begin{equation}
\Delta N_{\rm eff}
=
\frac{4}{7}
\left(\frac{11}{4}\right)^{4/3}
g_{\ast,D}\,\xi^4
\lesssim0.45,
\end{equation}
which for a single light dark photon, $g_{\ast,D}=2$, gives approximately $\xi\lesssim0.57$~\cite{Planck:2018vyg,Fan:2013yva}. Viable atomic-DM models therefore often require a colder dark sector.

Dark atoms may be reionized during halo formation if the virial temperature approaches the binding energy. Reionization is negligible for approximately $T_{\rm vir}/B_{H_D}\lesssim0.1$, which for Milky-Way-like systems corresponds roughly to~\cite{Ghalsasi:2017jna}
\begin{equation}
\alpha_D\gtrsim1.4\times10^{-3}\sqrt{R}.
\end{equation}
Below this limit, shock heating can regenerate an ionized component and enhance dissipative plasma effects.

For a massive dark photon, the interaction becomes Yukawa-suppressed,
\begin{equation}
V(r)=-\alpha_De^{-m_{\phi_\mu}r}/r,
\end{equation}
and atomic bound states require approximately~\cite{Petraki:2016cnz}
\begin{equation}\label{eq:dark_photon_constraint}
m_{\phi_\mu}\lesssim1.2\,\alpha_D\mu_{H_D}.
\end{equation}
If $m_{\phi_\mu}\sim B_{H_D}$, dark recombination can furthermore become kinematically suppressed, increasing $f_i$~\cite{Kahlhoefer:2020rmg}. Kinetic mixing with the SM photon through $\epsilon_\gamma$, as in \cref{eq:photon_portal}, introduces additional interactions with ordinary matter. These are suppressed for neutral atoms by charge neutrality and atomic form factors but can be substantially stronger for any residual ionized component, leading to constraints from direct detection, beam-dump experiments, stellar cooling, BBN and the CMB~\cite{Cline:2021itd}.

Finally, sufficiently efficient dissipation can drive atomic or molecular DM into dark disks, compact clumps or other dense structures, constrained by Gaia measurements of dark-disk surface densities, MACHO-like object searches and galactic mass-to-light ratios~\cite{Ghalsasi:2017jna,Schutz:2017tfp,Randall:2014lxa,Fan:2013tia,McCullough:2013jma,Kramer:2016dqu,Buch:2018qdr}. Dissipation may also accelerate gravothermal collapse and has been considered as a mechanism for forming early supermassive-black-hole seeds~\cite{Choquette:2018lvq,Mortlock:2011va,Banados:2017unc,Pollack:2014rja,DAmico:2017lqj}, while molecular cooling and dark chemistry introduce additional model dependence~\cite{Rosenberg:2017qia,Ryan:2021dis}.

\section{\label{app: anisotropic scattering calculation} Velocity update for anisotropic Monte-Carlo scattering}

For large-angle anisotropic scattering, the deflection
is defined in the center-of-momentum (CoM) frame,
where elastic scattering rotates the relative-velocity
vector without changing its magnitude. Following
the three-dimensional scattering-vector construction
of~\cite{Arido:2024hfk}, we describe the corresponding
laboratory-frame velocity update for arbitrary particle
mass ratios, ensuring momentum and energy conservation
and isotropic sampling of the azimuthal angle.
Throughout this derivation, $\vec{u}_a$ and $\vec{u}_b$
denote physical peculiar velocities. Conversion to
and from the cosmological velocity variables used
by \arepoTwo\ is performed consistently outside
the physical-frame scattering update.

We first construct the relative velocity, which is the only combination of the two velocities that changes under an elastic collision (the CoM velocity is conserved),
\begin{equation}
\vec{V} = \vec{u}_a - \vec{u}_b\,.
\end{equation}
Projecting onto the CoM frame, the velocity of particle $a$ is a fixed fraction of the relative velocity set by the reduced mass,
\begin{equation}
\vec{v}_a = \frac{m_b}{m_a+m_b}\,\vec{V},
\end{equation}
with magnitude $v = |\vec{v}_a|$. Because the collision is elastic, $v$ is unchanged by the scattering; only the \emph{direction} of $\vec{v}_a$ is rotated.

The polar deflection angle $\theta$ is drawn from the cumulative distribution corresponding to the differential cross section, as described in App.~\ref{app:Anisotropic_large_angle_aSIDM}. In order to apply the deflection in a controlled way, we introduce an auxiliary coordinate system in which the incoming relative velocity is aligned with the $x$-axis and the scattering is taken to occur in the $x$–$y$ plane. This choice makes the rotation trivial to write down, since the deflection then only mixes the $x$- and $y$-components. In this frame, the post-scattering velocity increments (already expressed as laboratory-frame increments relative to particle $b$, hence the $m_a/m_b$ factors that enforce momentum conservation) are constructed as
\begin{equation}
\vec{V}'_{a} =
\begin{pmatrix}
v\!\left(\cos\theta + \frac{m_a}{m_b}\right) \\
v\sin\theta \\
0
\end{pmatrix},
\quad
\vec{V}'_{b} =
\begin{pmatrix}
\frac{m_a}{m_b}\,v(1-\cos\theta) \\
-\frac{m_a}{m_b}\,v\sin\theta \\
0
\end{pmatrix}.
\end{equation}
One can verify directly that $m_a\vec{V}'_{a}+ \ldots$ reproduces the correct momentum balance and that in the equal-mass limit $m_a=m_b$ these expressions reduce to the familiar symmetric result.

The construction so far confines the scattering to a single plane, which does not yet respect the full rotational symmetry of the problem. To recover three-dimensional isotropy about the incoming direction, we draw an azimuthal angle $\varphi \in [0,2\pi]$ uniformly and rotate the increment vectors about the $x$-axis (the incoming-beam axis),
\begin{equation}
R_x(\varphi) =
\begin{pmatrix}
1 & 0 & 0 \\
0 & \cos\varphi & -\sin\varphi \\
0 & \sin\varphi & \cos\varphi
\end{pmatrix},
\end{equation}
\begin{equation}
\vec{V}'_{a,\mathrm{rot}} = R_x(\varphi)\,\vec{V}'_{a},
\qquad
\vec{V}'_{b,\mathrm{rot}} = R_x(\varphi)\,\vec{V}'_{b}.
\end{equation}
Since $\theta$ and $\varphi$ together parametrise the deflection into the full solid angle, this step guarantees that the scattered directions are correctly distributed in three dimensions.

The vectors are still expressed in the auxiliary frame whose $x$-axis points along $\vec{V}$. We must therefore rotate back to the original laboratory orientation. Let $\vec{V}=(V_x,V_y,V_z)$ be the incoming relative velocity. The two angles needed to undo the alignment are obtained from
\begin{equation}
\begin{aligned}
\beta &= \operatorname{atan2}(-V_z,V_y),
 \\
\alpha &= \operatorname{atan2}\!\left(-\cos\beta\,V_y+\sin\beta\,V_z,\,V_x\right),
\end{aligned}
\end{equation}
where the two-argument $\operatorname{atan2}$ is used to resolve the correct quadrant for arbitrary orientations of $\vec{V}$. The inverse rotation matrix that restores the original orientation is then
\begin{equation}
R^{-1}(\alpha,\beta) =
\begin{pmatrix}
\cos\alpha & \sin\alpha & 0 \\
-\sin\alpha\cos\beta & \cos\alpha\cos\beta & \sin\beta \\
\sin\alpha\sin\beta & -\cos\alpha\sin\beta & \cos\beta
\end{pmatrix}.
\end{equation}
Applying this transformation maps the scattered increments from the auxiliary frame back into the laboratory frame,
\begin{equation}
\begin{aligned}
\vec{V}'_{a,\mathrm{lab}} &= R^{-1}(\alpha,\beta)\,\vec{V}'_{a,\mathrm{rot}},
\\
\vec{V}'_{b,\mathrm{lab}} &= R^{-1}(\alpha,\beta)\,\vec{V}'_{b,\mathrm{rot}}.
\end{aligned}
\end{equation}

Finally, the updated physical peculiar velocities
are obtained by adding back the initial
laboratory-frame velocity of particle $b$,
\begin{equation}
\vec{u}'_a
=
\vec{u}_b+\vec{V}'_{a,\mathrm{lab}},
\qquad
\vec{u}'_b
=
\vec{u}_b+\vec{V}'_{b,\mathrm{lab}}.
\end{equation}
Since all velocities are expressed as physical
peculiar velocities, no additional scale-factor
conversion is required within the scattering
update. In the zero-deflection limit,
$\theta=0$, we recover
$\vec{u}'_a=\vec{u}_a$ and
$\vec{u}'_b=\vec{u}_b$, as expected.

By construction, this procedure ensures exact momentum conservation, allows for arbitrary mass ratios $m_a/m_b$, and correctly samples the anisotropic differential cross section in three dimensions. It therefore provides an unbiased Monte-Carlo realisation of individual large-angle scattering events.

\section{Technical details of the aSIDM scheme}
\label{app:technical_details_aSIDM_scheme}

The angular hybrid SIDM (aSIDM) scheme splits scattering at a critical angle $\theta_{\mathrm c}$ into rare large-angle events, treated with anisotropic Monte Carlo sampling (\cref{app:Anisotropic_large_angle_aSIDM}), and frequent small-angle deflections, modeled through drag and transverse diffusion (\cref{app:small_angle_aSIDM}). The angle-restricted cross sections and cumulative angular distributions for Rutherford scattering and spin-averaged identical fermions follow the analytical expressions of~\cite{Arido:2024hfk}. We additionally derive the corresponding expressions for identical DM scalars, providing a unified analytical basis for constructing the angular input tables for our aSIDM implementation in \arepoTwo.

The cross sections are parametrized by $\sigma_0$ and $w$, as defined in \cref{eq:sigma0_w_definition}, with the velocity-dependent anisotropy parameter $a(v_{\rm rel})=(v_{\rm rel}/w)^2$. For compactness, we use $a$ in the analytical expressions below, while the corresponding simulation inputs are tabulated as functions of $v_{\rm rel}$.

\subsection{Large-angle scattering in the aSIDM scheme}
\label{app:Anisotropic_large_angle_aSIDM}

In the aSIDM framework, large-angle deflections are treated stochastically by sampling directly from the differential cross section, restricted to scattering angles $\theta \ge \theta_{\mathrm c}$. To turn the differential cross section into a probability distribution from which $\theta$ can be drawn, the cumulative distribution must be normalized by the partial cross section that collects all events above the critical angle, \begin{equation}
\sigma_{\mathrm{tot},>}(\theta_{\mathrm c})
=
2\pi\int_{\theta_{\mathrm c}}^{\theta_{\max}}
\sin\theta\,
\frac{\mathrm d\sigma}{\mathrm d\Omega}\,
\mathrm d\theta,
\end{equation}
where $\theta_{\max}=\pi$ for distinguishable particles and $\theta_{\max}=\pi-\theta_{\mathrm c}$ for identical particles. Inverting the resulting cumulative distribution $P(\theta)$ for a uniform random number then yields a properly sampled deflection angle.

For identical DM scalars ($c_{\rm int}=+2$), the M{\o}ller cumulative probability reads
\begin{equation}
P(\theta)
=
\frac{\sigma_0}{\sigma_{\mathrm{tot},>}(\theta_{\mathrm c})}
\left[S(\theta_{\mathrm c})-S(\theta)\right],
\end{equation}
where the auxiliary function is
\begin{align}
\begin{aligned}
S(\theta)
=&
\frac{2\cos\theta}
{(2+a)^2-a^2\cos^2\theta}
\\ & \qquad+
\frac{1}{a(2+a)}
\ln\!\left(
\frac{2+a+a\cos\theta}
     {2+a-a\cos\theta}
\right).
\end{aligned}
\end{align}
The corresponding large-angle total cross section that normalises $P(\theta)$ is
\begin{align}
\sigma_{\mathrm{tot},>}(\theta_{\mathrm c})
&=
\sigma_0
\Bigg[
\frac{4\cos\theta_{\mathrm c}}
{(2+a)^2-a^2\cos^2\theta_{\mathrm c}}
\nonumber\\
&\qquad
+\frac{2}{a(2+a)}
\ln\!\left(
\frac{2+a+a\cos\theta_{\mathrm c}}
     {2+a-a\cos\theta_{\mathrm c}}
\right)
\Bigg].
\end{align}

For spin-averaged identical fermions, corresponding to $c_{\rm int}=-1$ in \cref{eq:Moeller_general}, the M{\o}ller cumulative probability reads
\begin{equation}
P(\theta)
=
\frac{\sigma_0}{4\,\sigma_{\mathrm{tot},>}(\theta_{\mathrm c})}
\frac{A_>(\theta)}{B_>(\theta)},
\end{equation}
where the numerator and denominator have been grouped into compact auxiliary functions to keep the inversion tractable,
\begin{align}
A_>(\theta) &=
4a(2+a)C_{>,2}(\theta_{\mathrm c})\cos\theta
\nonumber\\
&
- C_{>,3}(\theta)
\Big[
8a(2+a)\cos\theta_{\mathrm c}
\\ &\qquad\qquad\quad+ C_{>,2}(\theta_{\mathrm c})\ln C_{>,1}(\theta)
\Big],\\[4pt]
B_>(\theta) &=
a(2+a)C_{>,3}(\theta)C_{>,3}(\theta_{\mathrm c}),
\end{align}
and the auxiliary functions, which encode the dependence on the velocity-dependent anisotropy parameter $a$, are
\begin{align}
C_{>,1}(\theta) &=
\frac{(2+a-a\cos\theta)(2+a+a\cos\theta_{\mathrm c})}
     {(2+a+a\cos\theta)(2+a-a\cos\theta_{\mathrm c})},\\
C_{>,2}(\theta_{\mathrm c}) &=
-8-a(8+a)+a^2\cos(2\theta_{\mathrm c}),\\
C_{>,3}(\theta) &=
-(2+a)^2+a^2\cos^2\theta.
\end{align} For the same spin-averaged fermionic case, the angular integral gives the large-angle total cross section that normalises $P(\theta)$,
\begin{align}
\sigma_{\mathrm{tot},>}(\theta_{\mathrm c})
&=
\sigma_0
\Bigg[
\frac{4\cos\theta_{\mathrm c}}
     {(2+a)^2-a^2\cos^2\theta_{\mathrm c}}
\nonumber\\
&\qquad
+\frac{1}{a(2+a)}
\ln\!\left(
\frac{2(2+a)}{2+a+a\cos\theta_{\mathrm c}}-1
\right)
\Bigg].
\end{align}

For Rutherford scattering, Eq.~\eqref{eq:Rutherford_general}, the corresponding expressions are considerably simpler, since the cross section does not contain the identical-particle interference terms of the M{\o}ller case. The cumulative probability is \begin{equation}
\hspace*{-0.15cm}
P(\theta)
=
\frac{2\sigma_0}{\sigma_{\mathrm{tot},>}(\theta_{\mathrm c})}
\frac{\cos\theta_{\mathrm c}-\cos\theta}
     {(2+a-a\cos\theta_{\mathrm c})(2+a-a\cos\theta)}.
\end{equation} with the normalising large-angle cross section
\begin{equation}
\sigma_{\mathrm{tot},>}(\theta_{\mathrm c})
=
\sigma_0
\frac{1+\cos\theta_{\mathrm c}}
     {(1+a)(2+a-a\cos\theta_{\mathrm c})}.
\end{equation}
In this case $P(\theta)$ can be inverted analytically, which makes the large-angle sampling particularly efficient.

\subsection{Small-angle scattering in the aSIDM scheme}
\label{app:small_angle_aSIDM}

Deflections below the critical angle $\theta_{\mathrm c}$ are far too numerous to resolve individually. Instead they are treated through an effective drag term combined with transverse diffusion, following \cite{Fischer_2021a}: the drag reproduces the systematic momentum loss along the direction of motion, while the diffusion reproduces the random transverse kicks. Only the small-angle contribution to the transfer cross section enters this channel, so that the large- and small-angle treatments together recover the full momentum-transfer cross section without double counting.

For identical DM scalars ($c_{\rm int}=+2$), the modified transfer cross section restricted to $\theta\leq\theta_{\mathrm c}$ and, by symmetry, $\theta\geq\pi-\theta_{\mathrm c}$ is
\begin{align}
\sigma_{\widetilde T,<}(\theta_{\mathrm c})
&=
\sigma_0
\Bigg[
\frac{4}{a(2+a)}
\ln\!\left(
\frac{(1+a)(2+a-a\cos\theta_{\mathrm c})}
     {2+a+a\cos\theta_{\mathrm c}}
\right)
\nonumber\\
&\qquad
-\frac{8\cos\theta_{\mathrm c}
(1-\cos\theta_{\mathrm c})}
{(2+a)^2-a^2\cos^2\theta_{\mathrm c}}
\Bigg].
\end{align}

For spin-averaged identical fermions ($c_{\rm int}=-1$), the modified transfer cross section restricted to $\theta\leq\theta_{\mathrm c}$ and, by symmetry, $\theta\geq\pi-\theta_{\mathrm c}$ is
\begin{equation}
\sigma_{\widetilde T,<}(\theta_{\mathrm c})
=
2\sigma_0
\frac{A_<(\theta_{\mathrm c})}{B_<(\theta_{\mathrm c})},
\end{equation}
where the numerator collects the various logarithmic and polynomial terms that arise from integrating the forward-peaked cross section,
\begin{equation}
\begin{aligned}
\hspace*{-0.25cm}
A_<(&\theta) =
24\ln C_{<,1}(\theta)
-2a
\Big[
-4\ln C_{<,2}(\theta)
\\
&
+a\Big(
2(2+a)\cos\theta
-(7+a)\ln(2+a-a\cos\theta)
\\
&
-(11+2a)\ln C_{<,3}(\theta)
+\cos^2\theta
\big(
-4+3\ln C_{<,4}(\theta)
\\
&+a(-2+\ln C_{<,5}(\theta))
-6\ln2
\big)
+3(6+a)\ln2
\Big)
\Big],
\end{aligned}
\end{equation}
and the denominator is
\begin{equation}
B_<(\theta)
=
a^2(2+a)(2+a-a\cos\theta)(2+a+a\cos\theta),
\end{equation}
with the auxiliary functions
\begin{align}
C_{<,1}(\theta) &=
\frac{(2+a-a\cos\theta)(2+a+a\cos\theta)}
     {4(1+a)},\\
C_{<,2}(\theta) &=
\frac{(2+a-a\cos\theta)^4(2+a+a\cos\theta)^5}
     {512(1+a)^5},\\
C_{<,3}(\theta) &=
\frac{2+a+a\cos\theta}{1+a},\\
C_{<,4}(\theta) &=
\frac{(2+a-a\cos\theta)(2+a+a\cos\theta)}
     {1+a},\\
C_{<,5}(\theta) &=
\frac{(2+a-a\cos\theta)(2+a+a\cos\theta)^2}
     {8(1+a)^2}.
\end{align}
These closed-form M{\o}ller expressions apply to spin-averaged identical fermions ($c_{\rm int}=-1$). For identical scalars ($c_{\rm int}=+2$), the corresponding angle-restricted cross sections must instead be evaluated using the differential cross section in \cref{eq:Moeller_general}.

For Rutherford scattering, which is not symmetric under $\theta\to\pi-\theta$, only the forward region $\theta\leq\theta_{\mathrm c}$ contributes to the small-angle channel, yielding the more compact result
\begin{equation}
\sigma_{T,<}(\theta_{\mathrm c})
=
\sigma_0
\frac{
C_{{\rm R},1}(\theta_{\mathrm c})
+2\ln C_{{\rm R},2}(\theta_{\mathrm c})
-2
}{a^2},
\end{equation}
with
\begin{align}
C_{{\rm R},1}(\theta_{\mathrm c}) &=
\frac{4}{2+a-a\cos\theta_{\mathrm c}},\\
C_{{\rm R},2}(\theta_{\mathrm c}) &=
\frac{2+a-a\cos\theta_{\mathrm c}}{2}.
\end{align}
Together with the large-angle cross sections of the previous subsection, these expressions fully specify how the total cross section is partitioned between the discrete-scattering and drag-diffusion channels for a given choice of $\theta_{\mathrm c}$.

\section{\label{app:ddm_validation}Analytical solution for the decaying-DM verification test}

Here we give the analytical predictions used for the uniform-box verification of the state-based decaying-DM implementation in \cref{sec:dDM_box_test}. The test follows the sequential decay chain $\chi_2\rightarrow\chi_1\rightarrow\chi_0$, where states~2 and~1 are unstable and state~0 is stable. The decay rates are constant, and the initial conditions contain $500{,}000$ particles in each of states~1 and~2 and no particles in state~0.

For constant decay rates, the evolution of the state populations follows
\begin{equation}
\begin{aligned}\label{eq:state_number_evolution_decay}
    \dot{N}_2 &= -\Gamma_{21}N_2, \\
    \dot{N}_1 &= \Gamma_{21}N_2-\Gamma_{10}N_1, \\
    \dot{N}_0 &= \Gamma_{10}N_1 ,
\end{aligned}
\end{equation}
with the analytic solution, for $\Gamma_{10}\neq\Gamma_{21}$,
\begin{equation}
\begin{aligned}
    N_2(t) &= N_2(0)e^{-\Gamma_{21}t}, \\
    N_1(t) &= N_1(0)e^{-\Gamma_{10}t}
    +\frac{\Gamma_{21}N_2(0)}
    {\Gamma_{10}-\Gamma_{21}}
    \left(e^{-\Gamma_{21}t}-e^{-\Gamma_{10}t}\right), \\
    N_0(t) &= N_{\rm tot}-N_1(t)-N_2(t),
\end{aligned}
\end{equation}
where $N_{\rm tot}=N_0+N_1+N_2$ is conserved. The cumulative numbers of decays through the two channels are as follows
\begin{equation}
\begin{aligned}
    D_{21}(t)&=N_2(0)-N_2(t),
    \\
    D_{10}(t)&=N_0(t)-N_0(0).
\end{aligned}
\end{equation} For the benchmark values
$\Gamma_{21}=0.10~\mathrm{Gyr}^{-1}$
and $\Gamma_{10}=0.25~\mathrm{Gyr}^{-1}$,
these expressions provide the analytical
state-population evolution compared with
the numerical results in
\cref{fig:state_based_dDM_validation}.

Each decay additionally imparts an isotropically oriented velocity kick to the daughter particle. For equal simulation-particle masses, $M\equiv M_0=M_1=M_2$, the ensemble-averaged kinetic energy injected by the decay kicks is therefore
\begin{equation}
\label{eq:kinetic_energy_injection_decays}
    \Delta E_{\rm decay}(t)
    =
    \frac{M}{2}
    \left[
        v_{{\rm kick},21}^{2}D_{21}(t)
        +
        v_{{\rm kick},10}^{2}D_{10}(t)
    \right].
\end{equation}
For isotropically oriented kicks, the cross term between the pre-decay velocity and the kick vanishes upon ensemble averaging. We normalize the injected energy by the initial total kinetic energy,
\begin{equation}
    E_{\rm kin}^{\rm tot}(0)
    =
    \frac{M}{2}
    \sum_{s=0}^{2}N_s(0)v_s^2,
\end{equation} Together, the analytical state populations and cumulative energy injection provide independent tests of the state-dependent decay probabilities, sequential state transitions and recoil updates discussed in \cref{sec:dDM_box_test}.

\clearpage
\pagestyle{empty}

\newpage

\bibliographystyle{apsrev4-2}
\bibliography{main} 

\end{document}